\documentclass [12pt, twoside] {uwthesis}

\pdfoutput=1

\usepackage[pdftex]{graphicx}
\usepackage{subfig}
\usepackage{dcolumn}
\usepackage{bm}
\usepackage{amssymb}
\usepackage{amsmath}
\usepackage{amsfonts}
\usepackage{multirow}
\usepackage[usenames, dvipsnames]{color}
\usepackage{verbatim}
\usepackage{enumerate}
\usepackage{xspace}
\usepackage{slashed}
\usepackage[figuresright]{rotating}
\usepackage{listings}
\usepackage{feynmp}
\usepackage{hyperref}
\hypersetup{linkbordercolor=1 1 1, citebordercolor=1 1 1, urlcolor=black}

\usepackage{alltt}  %

\newcommand{\kt}[0]{\text{k}_{\text{T}}}
\newcommand{\cut}[0]{\text{cut}}

\newcommand{\eq}[0]{\text{eq}}

\newcommand{\beq}[0]{\begin{equation}}
\newcommand{\eeq}[0]{\end{equation}}
\newcommand{\ee}[0]{e^+e^-}

\newcommand{\prog}[1]{{\sc #1}\xspace}
\newcommand{\code}[1]{{\tt #1}\xspace}
\newcommand{\codes}[1]{{\tt #1}s\xspace}
\newcommand{\SJ}[0]{\prog{SpartyJet}}
\newcommand{\FJ}[0]{\prog{FastJet}}
\newcommand{\FP}[0]{\prog{FastPrune}}

\DeclareMathOperator{\tr}{Tr}

\def\Dslash{D\!\!\!\!\slash}

\def\nslash{n\!\!\!\slash}
\def\bnslash{\bar n\!\!\!\slash}

\newcommand{\nn}{\nonumber} 
\newcommand{\bn}{{\bar n}}
\newcommand{\mcdot}{\!\cdot\!}
\newcommand{\vect}[1]{\mathbf{#1}}
\newcommand{\abs}[1]{\left\lvert #1\right\rvert}
\newcommand{\bra}[1]{\left\langle #1\right\rvert}
\newcommand{\ket}[1]{\left\lvert #1\right\rangle}
\newcommand{\Lqcd}{\Lambda_{\text{QCD}}}

\renewcommand{\eq}[1]{Eq.~\eqref{#1}}
\newcommand{\eqs}[2]{Eqs.~\eqref{#1} and \eqref{#2}}

\renewcommand{\sec}[1]{Sec.~\ref{#1}}

\newcommand{\fig}[1]{Fig.~\ref{#1}}
\newcommand{\tab}[1]{Table~\ref{#1}}
\DeclareMathOperator{\Tr}{Tr}

\DeclareMathOperator{\Disc}{Disc}

\newcommand{\CF}{C_F}

\newcommand{\CA}{N_C}

\newcommand{\as}{\alpha_s}

\newcommand{\cO}{\mathcal{O}}

\newcommand{\xcone}{x_{\rm cone}}

\newcommand{\incl}{ {\rm incl}}

\newcommand{\qjetnaive}{\tilde{J}}

\renewcommand{\dotplus}[0]{+}

\newcommand{\tabruleA}{\rule{-1pt}{2ex} \rule[-1ex]{0pt}{0pt}}
\newcommand{\tabruleB}{\rule{-2pt}{3.25ex} \rule[-1.5ex]{-2pt}{0pt}}

\begin{document}
 
 
 \definecolor{light-gray}{gray}{0.9}
 \lstloadlanguages{C++, [gnu]make, Python}
 \lstset{language=C++,
 	breaklines,
 	basicstyle=\scriptsize\ttfamily,
	keywordstyle=\color{magenta}\bfseries,
	commentstyle=\color{blue},
	stringstyle=\ttfamily,
	showstringspaces=false,
	tabsize=2
  }

\prelimpages
 
%
%
%
\Title{Jet Substructure at the Large Hadron Collider: \\ Harder, Better, Faster, Stronger}
\Author{Christopher K. Vermilion}
\Year{2010}
\Program{Department of Physics}
\titlepage

%
%

\Chair{Stephen D. Ellis}{Professor}{Department of Physics}

\Signature{Stephen D. Ellis}
\Signature{Laurence G. Yaffe}
\Signature{Gordon T. Watts}
\signaturepage

 \doctoralquoteslip

%
%

\setcounter{page}{-1}
\abstract{%
I explore many aspects of jet substructure at the Large Hadron Collider, ranging from theoretical techniques for jet calculations, to phenomenological tools for better searches with jets, to software for implementing and comparing such tools.  I begin with an application of soft-collinear effective theory, an effective theory of QCD applied to high-energy quarks and gluons.  This material is taken from \cite{SCETLetter}, in which we demonstrate factorization and logarithmic resummation for a certain class of observables in electron-positron collisions.  I then explore various phenomenological aspects of jet substructure in simulated events.  After observing numerous features of jets at hadron colliders, I describe a method --- jet pruning --- for improving searches for heavy particles that decay to one or more jets.  This material is a greatly expanded version of \cite{Pruning2}.  Finally, I give an overview of the software tools available for these kinds of studies, with a focus on \SJ, a package for implementing and comparing jet-based analyses I have collaborated on.  Several detailed calculations and software examples are given in the appendices.  Sections with no new content are \textit{italic} in the Table of Contents.
}
 
%
%
\tableofcontents
\listoffigures
\listoftables  
 
%
%
\chapter*{Glossary}      
\addcontentsline{toc}{chapter}{Glossary}
\thispagestyle{plain}
\begin{glossary}

\item[asymptotic freedom] The property of QCD that the strong coupling is weak at high energies.  This means that high-energy processes can be calculated perturbatively, and that partons within hadrons will appear as weakly-bound constituents if probed at high energy.

\item[color] The QCD charge, analogous to electric charge.  Quarks carry one of three fundamental colors (red, green, and blue); gluons can be thought of as carrying a color-anticolor pair, such as (green-antired).  Particles that do not carry color charge are color singlets.

\item[confinement] The reverse of asymptotic freedom: at low energies (below $\Lambda_\text{QCD} \approx 200$ MeV) quarks and gluons are bound into hadrons because the strength of the coupling in this regime.

\item[factorization] The division of a physics process into subprocesses which can be calculated separately, and which typically depend on fewer energy scales than the full process.  A final cross section can then be expressed as a product of several functions, each of which depends on a subset of the relevant variables characterizing the event.  For example, factorization is what allows the non-perturbative evolution of incoming protons, and the likelihood to find a parton of given momentum in them, to be treated separately from the perturbative hard scattering.

\item[final-state radiation (FSR)] Radiation from outgoing particles produced in the hard scattering.

 \item[hadron] A bound state of a quark and an antiquark (meson) or of three quarks/antiquarks (baryon).  Hadrons are the relevant particles in QCD at low energies (compared to $\Lambda_\text{QCD} \approx 200$ MeV).

\item[hard scattering]  The central high-energy process at a hadron collider, where two quarks or gluons collide to produce 2 or more other high-energy particles.  The outgoing particles then typically decay to produce the particles seen in the detector.  The hard scattering is to be contrasted to the subsequent final-state radiation, previous initial-state radiation, and the underlying event.

\item[initial-state radiation (ISR)]  Radiation from incoming particles in the hard scattering.

\item[jet] A mostly collimated spray of hadrons produce by the showering of one or more quarks or gluons at a particle collider. 
 
 \item[jet algorithm] A procedure for constructing jets from initial objects such as particles or calorimeter cells.
 
 \item[leading log] A cross section is correct to leading logarithm accuracy if it includes all terms of order $\alpha_s^n L^{2n}$, where $L$ is some large logarithm.  See Sec.~\ref{sec:qcd:example:resum}.
 
 \item[Monte Carlo X] X uses, or was produced using, random numbers, as in a ``Monte Carlo event generator'' or a ``Monte Carlo data set''.  Refers to the famous casino in Monaco.
 
 \item[parton] A quark or gluon.  The term comes from the ``parton model'', a phenomenological model of the strong interaction that predates QCD.
 
 \item[parton shower] The process whereby a high-energy quark or gluon repeatedly radiates soft and collinear gluons, which subsequently radiate themselves, producing a multiplicity of low-energy partons, which will later hadronize.  A ``parton shower Monte Carlo'' such as \prog{Pythia} is a program that simulates this process, typically only including the leading-log portion of the gluon emission matrix element (i.e., the double soft/collinear singularity).
 
 \item[Perfect Strangers] An American sitcom that ran from 1986 to 1993 on ABC.  It ``chronicles the rocky coexistence of Larry Appleton (Mark Linn-Baker) and his distant cousin Balki Bartokomous (Bronson Pinchot)'' \cite{PerfectStrangers}.  Notable for producing the spin-off series \textit{Family Matters} in 1989.
 
 \item[pile-up (PU)] The effect of multiple proton collisions occurring at once at the LHC.  The expected energy exchanged in a proton collision has a sharply falling distribution, so most pile-up interactions are much less energetic than the main interaction, which is selected to have very large momentum transfer (e.g., having several high-$p_T$ jets).  Pile-up collisions are completely independent of the principal interaction.

\item[QCD jet] In contrast to a ``heavy particle jet'', which includes the shower from multiple quarks and/or gluons, which were produced in the decay of a massive particle.  A QCD jet includes the shower of one or more partons from the hard scattering, but has no intrinsic mass scale.

\item[splash-in/splash-out] Splash-in is radiation included in a jet that did not come from the showering of the initial parton(s).  Splash-out is the reverse: radiation that came from the initial parton(s) but is not included in that jet.  Note that after hadronization splash-in and splash-out cannot be unambiguously defined unless the initial partons form a color singlet.

\item[underlying event (UE)] The combined effect of beam remnants and their potential multiple interactions.  Beam remnants are what remains of the colliding protons after one parton each is involved in the hard scattering.  At minimum, they must combine with other parts of the events to create color singlet hadrons for the final state.  The beam remnants can also produce secondary collisions, known as multiple parton interactions (MPI).  This typically produces additional low-$p_T$ jets in the event as well as soft radiation throughout the detector.  The underlying event is approximately independent of the hard scattering, but is typically color-connected and thus impossible to separate completely.
 
\newpage
 
\item[\textbf{Hadron collider variables}] Kinematic variables at a hadron collider are chosen to have simple behavior under Lorentz boosts along the beam axis, since the center-of-momentum frame of the initial parton collision is only known up to such a boost.  Here are the most important variables:
\begin{itemize}
\item[$\phi$] Azimuthal angle about the beam axis.
\item[$y$] Rapidity, $y \equiv \frac{1}{2}\left( \frac{E+p_z}{E-p_z} \right)$.  Under a longitudinal boost $\gamma \equiv \cosh y_b$, $y \to y + y_b$.
\item[$\eta$] Pseudorapidity, equal to $y$ for massless particles; maps directly to polar angle: $\eta \equiv - \ln \tan(\theta/2)$.
\item[$p_T$] Momentum transverse to the beam axis, $p_T^2 = p_x^2 + p_y^2$.
\item[$\Delta R(p_1, p_2)$] Longitudinal boost-invarant angle between two particles: $\Delta R^2(p_1, p_2) \equiv (\phi_1 - \phi_2)^2 + (y_1 - y_2)^2$.
\item[$z(p_1, p_2)$] Minimum transverse momentum fraction for a merging/splitting from/to $p_1$ and $p_2$: $z \equiv \frac{\min(p_1^T, p_2^T)}{p_{1+2}^T}$, where $p_{1+2} = p_1 + p_2$.
\item[NOTE:] $z$ and $\Delta R$ are useful in describing the twin soft and collinear singularities of QCD radiation.  An emission with small $z$ is soft; an emission with small $\Delta R$ is collinear.
\end{itemize}

\item[\textbf{Note on Conventions}]  In this thesis I attempt to follow the conventions of Peskin and Schroeder \cite{Peskin} where possible.  In particular, this includes a ``West Coast metric'', $g \equiv \text{diag} (1,-1,-1,-1)$, so $p^2 = m^2$ for an on-shell particle of mass $m$.  The gamma matrices $\gamma^\mu$ are defined in the Weyl basis,
\[
\gamma^0 \equiv \begin{pmatrix} 0 & 1 \\  1 & 0 \end{pmatrix}; \qquad
\gamma^i \equiv \begin{pmatrix} 0 & \sigma^i \\ - \sigma^i & 0 \end{pmatrix}.
\]
``Natural'' units, where $\hbar = c = 1$, are used throughout.

\end{glossary}
 
%
%
\acknowledgments{
The author wishes to acknowledge foremost his collaborators in the work this thesis is based on: Steve Ellis, Jon Walsh, Andrew Hornig, Chris Lee, Joey Huston, Brian Martin, and Pierre-Antoine Delsart.  He would like to thank Dr. Walsh for significantly raising the average competence of their two-man team, and Prof. Ellis for his advice, wisdom, and friendly mentorship.  A better research group is hard to imagine.

Thanks are also in order to collaborators at earlier stages of this work, in particular Matt Strassler and Jacob Miner, and in general Kyle Armour.

There is no way to know if the support, friendship, distraction, encouragement, discouragement, inspiration, and disillusion provided by fellow graduate students at the University of Washington helped or hindered the completion of this thesis, but every bit of it is gratefully acknowledged.  The members of the Nuclear Theory Journal Club especially have the author's appreciation, but preferably not his forwarding address.

The author wishes to thank a long string of wonderful teachers, in particular Dion Terwilliger, Dave Weiner, Andy Cohen, Rob Carey, Ann Nelson, and Larry Yaffe.

Family in Seattle and elsewhere have been incredibly supportive in the design and execution of this graduate adventure and their love is appreciated and returned.

Finally, words cannot express the author's love and gratitude for his wife, Allison, who has been there every step of the way.  But just in case, he has included 61,452 of them.

}

%
%
\dedication{

\begin{center}

To Allison, the love of my life.  This is for you --- but if you'd rather have something else, I kept the receipt.

\end{center}
}

%

 
%
%

\textpages

 
\chapter {Introduction}
\label{sec:intro}

At the dawn of the LHC era, the prospects for high-energy particle physics are bright.  The Large Hadron Collider \cite{LHC, LHCWebsite} will almost certainly resolve some of the outstanding questions in particle physics.  How is electroweak symmetry, central to the remarkably successful Standard Model (SM), broken --- as it necessarily must be?  What is the nature of dark matter, which makes up a quarter of the mass of the universe?  Why is the Planck mass, the only ``natural'' scale in the universe, so much bigger than everything else?  And why are there so many particles in the Standard Model, anyway?  The possibilities for new discoveries are endless.

And yet --- prospects for \emph{easy} discovery are bleak.  Almost any physical effect observable at the LHC will require either deeply sophisticated analysis techniques, patient accumulation of vast statistics, or both.  A quick and easy discovery, with a few exceptions \cite{Supermodels}, would already have been made at earlier experiments at the Tevatron \cite{DZero, CDF} or LEP \cite{LEP}.

The chief difficulty in discovering new physics at the LHC is that new particles created will almost certainly exist for times much too short to ever interact directly with the detectors surrounding the point of collision.  Most new particle searches, then, are concerned with observing \emph{decay products}.  To be observed in the detector, these decay products must be stable enough to get there and interact strongly enough to be noticed.  These two requirements ensure that the most likely candidates are simply SM particles; certainly anything that can be produced by the collision of protons must be able to decay back to SM particles.  The unfortunate upshot is that the signature of new physics --- the signal we want to see --- will almost necessarily have a substantial overlap with the signatures of well-known SM processes that can produce the same decay products.

One of the most difficult types of signals will involve decays to quarks and gluons, which are subsequently observed as the phenomena known as jets.  The SM cross sections for basic processes involving jets, even when a $W$ or $Z$ boson is involved, typically dwarf any new physics signals with similar signature.  The common supersymmetric signature of a lepton, jets, and missing energy is easily faked by the $W+$jets background.  To have any hope of extracting these kinds of signals, we will need to advance our understanding and usage of jets.

Fortunately, many such advances have been made in recent years.  Theoretical advances have extended the precision with which we can predict cross sections involving jets, both through brute force calculations to higher order in perturbation theory and through new effective theories that make these calculations more tractable.  In the latter category, soft/collinear effective theory (SCET) \cite{Bauer:2000ew, Bauer:2000yr, Bauer:2001ct, Bauer:2001yt} has shown great potential to improve our ability to calculate jet-based observables by factorizing the relevant calculations into separate pieces involving single energy scales.  These improved theory tools will help us to better characterize the backgrounds to interesting new signals.

Another theoretical development in the run-up to the LHC has been increased interest in jet \emph{substructure}.  Whereas jets at previous experiments were typically thought of as corresponding to a single initial quark or gluon, this will not always be a good model at the LHC.  In particular, if heavy particles that decay to multiple quarks or gluons are highly boosted, the jets corresponding to the multiple decay products will move closer together and eventually appear as a single jet.  For example, while a top quark decaying $t \to Wb \to u\bar d b$ would be observed as three jets at the Tevatron, it is now common to imagine ``top jets'' in LHC analyses.  Finding these jets, as well as the single jets arising from decays of new particles, requires a new way of thinking about jets.  A jet corresponding to a top quark can be expected to have a mass, as well as substructure related to the two-step decay $t \to W b \to q \bar q' b$.  Separating top jets from their QCD doppelg\"angers requires understanding the substructure of both kinds of jets.

Beyond understanding the physics of parton showers and decays, we must consider the experimental environment in which we observe these phenomena.  The LHC will be a phenomenally noisy experiment.  We must account for radiation from the incoming protons, the interactions of the ``beam remnants'' (components of the protons not involved in the largest-energy collision), and even the effect of more than one pair of protons colliding at once.  All of these are sources of additional radiation in LHC events, and hence will contribute to the characteristics of observed jets.  An important development in the last few years has been the variety of ideas related to ``filtering'' jets to remove many of these contributions \cite{FilteringHiggs, FilteringNeutralino, FatJets, TopTagging, Pruning1, Pruning2, Trimming}.

As the theoretical tools to find, measure, and modify jets proliferate, the need for software to easily implement them grows.  The \FJ package \cite{FastJet, FastJetWebsite} has provided efficient implementations of nearly all common jet algorithms, as well as facilities for user-defined plugins and tools.  More recently, the \SJ package \cite{FamousJetReview, SpartyJetWebsite} has emerged as an analysis package that extends the capabilities of \FJ with support for a variety of input and output methods, simple chains of jet measurement and modification tools, and an increasingly powerful graphical interface for quickly comparing and exploring different analyses.

This thesis is divided into four main sections.  Chapter \ref{sec:qcd} provides background to the rest of the thesis, surveying QCD, effective theories of QCD like SCET, and basic jet physics.  Chapter \ref{sec:scet} demonstrates the ability of SCET to factorize jet observables in $\ee$ collisions.  Its content is essentially the same as \cite{SCETLetter}; further details are given in the companion paper \cite{Ellis:2010rw} and in Jonathan Walsh's thesis \cite{Jon}.  Chapter \ref{sec:sub} discusses predictions and (Monte Carlo) observations of jet substructure in heavy particle decays and their QCD background.  The theoretical discussion is taken from \cite{Pruning2}; the demonstrative plots and accompanying discussions are new.  Chapter \ref{sec:prune} describes and explores a method for improving heavy particle searches using ``pruned'' jet substructure.  This chapter is also largely drawn from \cite{Pruning2}, but the example plots and accompanying discussion in Secs.~\ref{sec:prune:ee} and \ref{sec:prune:pp}, as well as the discussions in Secs.~\ref{sec:prune:others} and \ref{sec:prune:using}, are new.  Chapter \ref{sec:tools} surveys the available software tools for studying jet substructure, with emphasis on the tools implementing jet pruning developed by the author, and the \SJ package, to which the author has made significant contributions.  The text is entirely new.  Finally, in Chapter \ref{sec:conc}, these various strands are tied together and the Future of the Jet is considered thoughtfully.


 
 \graphicspath{{chapter2/graphics/}}
 
\chapter{QCD Phenomenology}
\label{sec:qcd}
 
Quantum Chromodynamics (QCD) is well established as our best theory of the strong interaction, governing the behavior of hadrons such as protons and neutrons as well as their constituents, quarks and gluons.  QCD is a gauge quantum field theory, similar to quantum electrodynamics (QED), Feynman's ``strange theory of light and matter''.  As I will review in this chapter, there are several important differences between QCD and QED, which lead to a theory at once richer and more challenging.

I will begin with a review of the QCD Lagrangian, the running of the strong coupling, and the twin features of asymptotic freedom and infrared slavery (``confinement'', if you prefer).  I then give an example of a perturbative QCD calculation of the cross section for $\ee$ annihilation into hadrons, on the way encountering many of the fundamental issues that appear in perturbative QCD.  This will include a discussion of several systematic approaches to improving the precision of such calculations.  In Sec.~\ref{sec:qcd:eft}, I discuss soft collinear effective theory, an effective theory of high-energy quarks and gluons at particle colliders.  Finally, in Sec.~\ref{sec:qcd:jets} I discuss ``jets'', the chief QCD phenomenon observed and studied in collider experiments.  Sec.~\ref{sec:qcd:eft} is intended as background to Chapter \ref{sec:scet}; Sec.~\ref{sec:qcd:jets} is intended as background to Chapters \ref{sec:sub} and \ref{sec:prune}.

\section{Basics}
\label{sec:qcd:basics}

That this review of QCD phenomenology is incomplete is too obvious to belabor.  What follows is a list of references and reviews, themselves incomplete but collectively comprehensive.  A considerably more exhaustive survey of the QCD literature can be found in \cite{QCDResourceLetter}.

The basic features of QCD and gauge theories are discussed in the standard textbooks, e.g. \cite{Peskin, Zee, Srednicki, Burgess}.  A more focused resource, geared toward collider physics, is \cite{PinkBook}, known universally as ``the pink book''.  \cite{PinkBook} also contains a broad set of citations to the theoretical and experimental literature.  An extensive review of perturbative QCD is given in \cite{pQCD}.  An extensive review of the non-perturbative aspects of QCD (and field theories in general) is given in \cite{npQCD}.  References specifically relevant to the following sections and chapters will be given therein.


\subsection{The QCD Lagrangian}
\label{sec:qcd:basics:lagrangian}

The Lagrangian of QCD, omitting for now gauge-fixing terms, is

\beq
\mathcal{L}_\text{QCD} = -\frac{1}{4} F^A_{\alpha \beta} F_A^{\alpha \beta} + \sum_\text{flavors} \bar q_a ( i \slashed D - m )_{a b} q_b .
\label{eq:LQCD}
\eeq

The second term represents a set of spin-1/2 quarks, interacting with a gauge field hiding in the covariant derivative $\slashed D$ (Dirac indices have been suppressed).  The gauge interaction corresponding to QCD is $SU(3)$, with the gauge charge conventionally referred to as ``color''.  Quarks (antiquarks) live in the fundamental (antifundamental) representation of $SU(3)$, so a quark field carries a color index: $q_a$, where $a$ runs from 1 to 3.  The gauge fields, called gluons, live in the adjoint (dimension 8) representation.  Note that $SU(3)$ is non-Abelian, so the field strength term $-\frac{1}{4} F^A_{\alpha \beta} F_A^{\alpha \beta}$ contains self-interaction terms:
\beq
F^A_{\alpha \beta} = \left[ \partial_\alpha \mathcal{A}^A_\beta - \partial_\beta \mathcal{A}^A_\alpha - g f^{ABC} \mathcal{A}^B_\alpha \mathcal{A}^C_\beta \right].
\label{eq:F}
\eeq
The indices $\{A, B, C\}$ run over the eight color degrees of freedom for the gluon.  The interaction terms mean that the gluons themselves carry color charge.  This is the key distinguishing feature between the QCD and QED Lagrangians: gluons interact with each other where photons do not.

The sum over flavors in Eq.~\ref{eq:LQCD} runs over the six quark flavors.  As far as the strong interaction is concerned these flavors are identical except for their different masses.  In the electroweak sector the quarks are grouped into three pairs termed ``generations''.  The quark flavors and their approximate masses are given in Table \ref{table:quarkmasses}.

\begin{table}[htbp]
\begin{center}
\begin{tabular}{|l||c|c|}
\hline
Name (symbol) & Electric charge & Mass \\
\hline
Up ($u$) & $\frac{2}{3}$ & 1.5--3.3 MeV \\
\hline
Down ($d$) & -$\frac{1}{3}$ & 3.5--6.0 MeV \\
\hline
\hline
Charm ($c$) & $\frac{2}{3}$ & $1.27^\text{\tiny{+0.07}}_\text{\tiny{-0.11}}$ GeV \\
\hline
Strange ($s$) & -$\frac{1}{3}$ & $104^\text{\tiny +26}_\text{\tiny -34}$ MeV \\
\hline
\hline
Top ($t$) & $\frac{2}{3}$ & $171.2 \pm 2.1$ GeV \\
\hline
Bottom ($b$) & -$\frac{1}{3}$ & $4.20^\text{\tiny +0.17}_\text{\tiny-0.07}$ GeV \\
\hline
\end{tabular}
\end{center}
\caption[The six flavors of quarks]{The six flavors of quarks.  Masses are taken from \cite{PDG}.  Note that there is some ambiguity in defining a ``quark mass'', since quarks do not propagate as free particles.  See \cite{PDG} and the references therein for further discussion of this subtle point, e.g.:  ``The estimates of $u$ and $d$ masses are not without controversy and remain under active investigation.''}
\label{table:quarkmasses}
\end{table}

It is worth making the color structure of Eqs.~\ref{eq:LQCD} and \ref{eq:F} explicit.  The covariant derivatives, acting on quark (color triplet) and gluon (color octet) fields, are
\beq
\begin{split}
(D_\alpha)_{ab} &= \partial_\alpha \delta_{ab} + i g \left(t^C \mathcal{A}^C_\alpha \right)_{ab}, \\
(D_\alpha)_{AB} &= \partial_\alpha \delta_{AB} + i g \left(T^C \mathcal{A}^C_\alpha \right)_{AB}.
\end{split}
\eeq
The $\mathcal{A}^C$ are the eight gluon fields, multiplying the fundamental (adjoint) generators $t^C$ ($T^C$).  The generators obey the standard $SU(N)$ relations:
\beq
\begin{split}
[t^A, t^B] &= i f^{ABC} t^C, \\
[T^A, T^B] &= i f^{ABC} T^C, \\
(T^A)_{BC} &= - i f^{ABC},
\end{split}
\eeq
where $f^{ABC}$ are the structure constants of $SU(N)$.  The normalization of the fundamental generator matrices is chosen such that
\beq
\tr t^A t^B = T_R \delta^{AB} = \frac{1}{2} \delta^{AB} .
\eeq
An explicit form of the $t^A$ is not usually necessary, but the following relations are useful:
\beq
\begin{split}
t^A_{ab} t^A_{bc} &= C_F \delta_{ac} = \frac{N^2-1}{2 N} \delta_{ac}, \\
\tr (T^CT^D) &= f^{ABC}f^{ABD} = C_A \delta^{CD} = N \delta^{CD}.
\end{split}
\eeq
For $SU(3)$, the color factors are $C_F = 4/3$ and $C_A = 3$.

\subsubsection{Gauge fixing and ghosts}

The QCD Lagrangian is invariant under the gauge transformation
\beq
\begin{split}
q_a(x) &\to e^{(i t \cdot \theta(x))_{ab}} q_b(x) \equiv U(x)_{ab} q_b (x), \\
\mathcal{A}_\alpha &\to U(x) \mathcal{A}_\alpha U^{-1}(x) + \frac{i}{g} (\partial_\alpha U(x)) U^{-1}(x) .
\end{split}
\label{eq:gauge}
\eeq
defined by the matrix-valued function $(t \cdot \theta(x))_{ab}$.  Before we can define Feynman rules for QCD, we must choose a specific gauge to work in.  In the absence of a gauge choice, the gluon propagator would not be well defined.\footnote{The quadratic term for the gauge field is $\frac{1}{2} \int \frac{d^4 k}{(2 \pi)^4} \mathcal{A}_\alpha (k)(-k^2 g^{\alpha \beta} + k^\alpha k^\beta) \mathcal{A}_\beta (-k)$.  The integrand vanishes for a large space of gauge configurations, and hence the quadratic operator does not have a well-defined inverse.}  In terms of the functional integral,
\beq
Z = \int \mathcal{D A} \mathcal{D} \bar q \mathcal{D} q \exp \left\{ - \int d x \mathcal{L}_\text{QCD} \right\} ,
\eeq
choosing a gauge corresponds to ``factoring out'' the integration over the redundant space of gauge-equivalent field configurations.  The standard procedure is to introduce a gauge-fixing term to the functional integral:
\beq
1 = \int \mathcal{D} \theta \delta \left( G(\mathcal{A}^\theta) \right) \det \left( \frac{\delta G(\mathcal{A}^\theta)}{\delta \theta} \right),
\label{eq:FaddeevPopov}
\eeq
where $G(\mathcal{A^\theta})$ is some function of the (transformed) gauge field (Eq.~\ref{eq:gauge}).  For linear $G$, $\delta G(\mathcal{A^\theta}) / \delta \theta$ is independent of $\theta(x)$, so the functional integral $(\int \mathcal{D} \theta)$ factors out.  We have isolated the integration over different gauge configurations at the expensive of introducing a new term to the Lagrangian, the Faddeev-Popov determinant \cite{FaddeevPopov} in Eq.~\ref{eq:FaddeevPopov}.  With some manipulation, it can be shown\footnote{See, e.g., \cite{Peskin} Sections 9.4 and 16.2.} that the $\delta$ function and the determinant terms can be represented as a functional integral over two additional terms in the Lagrangian:
\beq
\begin{split}
\mathcal{L}_\text{gauge-fixing} &= -\frac{1}{2 \lambda}(\partial^\alpha \mathcal{A}_\alpha^A)^2, \\
\mathcal{L}_\text{ghost} &= \partial_\alpha \chi^{A \dagger} (D^\alpha_{AB} \chi^B).
\end{split}
\eeq

The gauge-fixing term modifies the gluon propagator.  Any value of $\lambda$ is allowed, different values corresponding to different gauge choices.  The ghost term introduces a pair of complex, scalar, anti-commuting ``fields'' $\{\bar \chi, \chi\}$, which come with their own functional integral.  These ``Faddeev-Popov ghosts'' are not physical particles --- they do not even exist in certain gauges! --- but they can be treated as such in the calculation of Feynman diagrams.  In practice, ghosts only appear in certain loop diagrams, since they are never external legs.

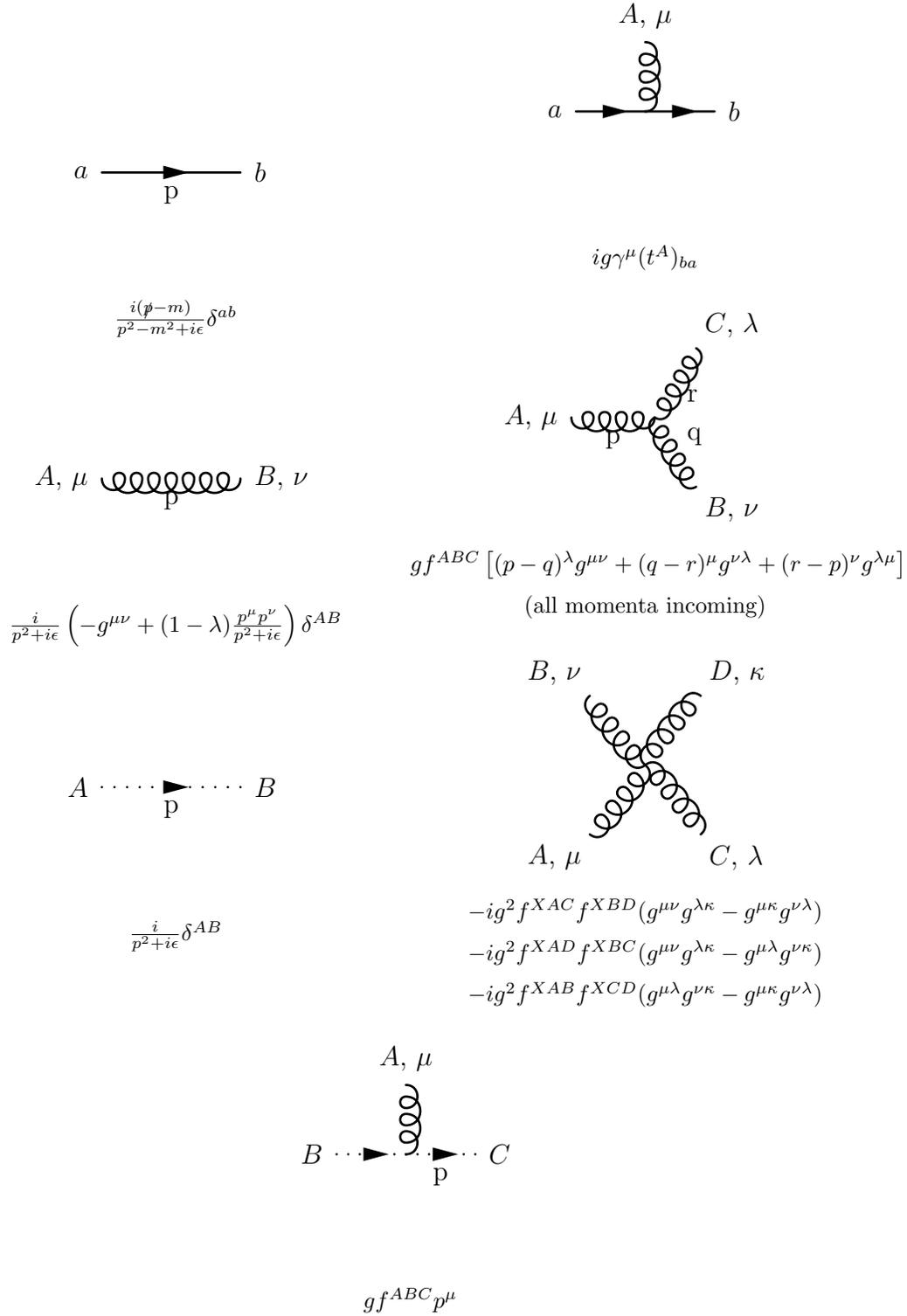
\begin{figure}[htbp]
\centering
\captionsetup[subfigure]{labelformat=empty,justification=centering,margin=0pt}

\def\feynheight{35mm} 
\vfill
\begin{minipage}{0.45\textwidth} \centering
\subfloat[][$\frac{i(\slashed p -m)}{p^2 - m^2 + i\epsilon} \delta^{ab}$] {
	\vbox to \feynheight{
	\vfil
\begin{fmffile}{qprop}
\begin{fmfgraph*}(60,30)
	\fmfleft{in}
	\fmflabel{$a$}{in}
	\fmfright{out}
	\fmflabel{$b$}{out}
	\fmf{quark,label=p}{in,out}
\end{fmfgraph*}
\end{fmffile}
	\vfil}
}

\subfloat[][$\frac{i}{p^2 + i\epsilon} \left( -g^{\mu\nu}  + (1-\lambda)\frac{p^\mu p^\nu}{p^2 + i\epsilon}\right)\delta^{AB}$] {
	\vbox to \feynheight{
	\vfil
\begin{fmffile}{gprop}
\begin{fmfgraph*}(60,30)
	\fmfleft{in}
	\fmflabel{$A$, $\mu$}{in}
	\fmfright{out}
	\fmflabel{$B$, $\nu$}{out}
	\fmf{gluon,label=p}{in,out}
\end{fmfgraph*}
\end{fmffile}
	\vfil}
}

\subfloat[][$\frac{i}{p^2 + i\epsilon} \delta^{AB}$] {
	\vbox to \feynheight{
	\vfil
\begin{fmffile}{xprop}
\begin{fmfgraph*}(60,30)
	\fmfleft{in}
	\fmflabel{$A$}{in}
	\fmfright{out}
	\fmflabel{$B$}{out}
	\fmf{ghost,label=p}{in,out}
\end{fmfgraph*}
\end{fmffile}
	\vfil}
}

\end{minipage}
\begin{minipage}{.1\textwidth}\end{minipage}
\begin{minipage}{0.45\textwidth} \centering
\subfloat[][$i g \gamma^{\mu} (t^{A})_{ba}$] {
	\vbox to \feynheight{
	\vfil
\begin{fmffile}{qqg}
\begin{fmfgraph*}(60,60)
	\fmfleft{in}
	\fmflabel{$a$}{in}
	\fmfright{out}
	\fmflabel{$b$}{out}
	\fmftop{gluon}
	\fmf{quark}{in,v}
	\fmf{quark}{v,out}
	\fmffreeze
	\fmf{gluon}{v,gluon}
	\fmflabel{$A$, $\mu$}{gluon}
\end{fmfgraph*}
\end{fmffile}%
	\vfil}
}

\subfloat[][$g f^{ABC} \left\lbrack (p-q)^\lambda g^{\mu\nu} + (q-r)^\mu g^{\nu \lambda} + (r-p)^\nu g^{\lambda \mu} \right\rbrack $

(all momenta incoming)] {
	\vbox to \feynheight{
	\vfil
\begin{fmffile}{ggg}
\begin{fmfgraph*}(60,60)
	\fmfleft{one}
	\fmfright{two,three}
	\fmf{gluon,label=p}{one,v}
	\fmf{gluon,label=q}{v,two}
	\fmf{gluon,label=r}{v,three}
	\fmflabel{$A$, $\mu$}{one}
	\fmflabel{$B$, $\nu$}{two}
	\fmflabel{$C$, $\lambda$}{three}
\end{fmfgraph*}
\end{fmffile}
	\vfil}
}

\centering
\subfloat[][$-i g^2 f^{XAC} f^{XBD} (g^{\mu\nu} g^{\lambda\kappa} - g^{\mu\kappa}g^{\nu\lambda})$\\
$-i g^2 f^{XAD} f^{XBC} (g^{\mu\nu} g^{\lambda\kappa} - g^{\mu\lambda}g^{\nu\kappa})$\\
$-i g^2 f^{XAB} f^{XCD} (g^{\mu\lambda} g^{\nu\kappa} - g^{\mu\kappa}g^{\nu\lambda})$] {
	\vbox to \feynheight{
	\vfill
\begin{fmffile}{gggg}
\begin{fmfgraph*}(60,60)
	\fmfleft{g1,g2}
	\fmfright{g3,g4}
	\fmf{gluon}{g1,v}
	\fmf{gluon}{g2,v}
	\fmf{gluon}{g3,v}
	\fmf{gluon}{g4,v}
	\fmflabel{$A$, $\mu$}{g1}
	\fmflabel{$B$, $\nu$}{g2}
	\fmflabel{$C$, $\lambda$}{g3}
	\fmflabel{$D$, $\kappa$}{g4}
\end{fmfgraph*}
\end{fmffile}%
	\vfill}
}
\vfill
\end{minipage}
\vfill
\subfloat[][$g f^{ABC} p^\mu $] {
	\vbox to \feynheight{
	\vfill
\begin{fmffile}{xxg}
\begin{fmfgraph*}(60,60)
	\fmfleft{in}
	\fmflabel{$B$}{in}
	\fmfright{out}
	\fmflabel{$C$}{out}
	\fmftop{gluon}
	\fmf{ghost}{in,v}
	\fmf{ghost,label=p}{v,out}
	\fmffreeze
	\fmf{gluon}{v,gluon}
	\fmflabel{$A$, $\mu$}{gluon}
\end{fmfgraph*}
\end{fmffile}
	\vfill}
}

\vfill

\caption[Feynman rules for QCD]{Feynman rules for QCD.  The ``ghost'' fields (dotted lines) can be treated as anti-commuting scalars that only propagate internally.}
\label{fig:QCDFeynmanRules}
\end{figure}

The Feynman rules arising from the gauge-fixed QCD Lagrangian are given in Fig.~\ref{fig:QCDFeynmanRules}.  Note the appearance of the free parameter $\lambda$ in the gluon propagator.  Any value of $\lambda$ can be used; any gauge-invariant calculation will be independent of the choice.  $\lambda = 1 (0)$ is the \emph{Feynman-'t Hooft (Landau) gauge}.

\subsection{Running of \texorpdfstring{$\alpha_s$}{alpha\_s}}
\label{sec:qcd:basics:running}

The most important difference in the phenomenology of QED and QCD is in the renormalization flow of the couplings.  At lowest order in perturbation theory, we find for both theories a result of the form \cite{PinkBook}:
\beq
\label{eq:alpha}
\frac{1}{\alpha(\mu_1)} = \frac{1}{\alpha(\mu_2)} + b_0 \ln \left( \frac{\mu_2^2}{\mu_1^2} \right),
\eeq
where $\alpha = \frac{e^2}{4 \pi}$ for QED and $\alpha = \frac{g^2}{4 \pi}$ for QCD.  The crucial difference lies in the sign of $b_0$, which for QED is positive and for QCD negative.  At small energies the QED coupling asymptotes to a small value, $1/\alpha(\mu) \sim 1/\alpha_0 \approx 137$, ($m_e \neq 0$ cuts off the running at $\mu \lesssim m_e$), growing logarithmically at larger energies: $1/\alpha(m_Z) \approx 128$.
For QCD however, the coupling grows logarithmically smaller at large energies and diverges at small energies.  At the scale of $m_Z$,
$\alpha_s(m_Z) \approx 0.118$
is small enough to calculate interactions perturbatively.  However, at scales $\mu \sim \Lambda_\text{QCD} \approx 200$ MeV, the perturbative result (Eq.~\ref{eq:alpha}) diverges.  This does not mean the coupling itself is becoming infinite, only that it is becoming large enough that perturbation theory is breaking down.  We observe that at low energies quarks are bound together in hadronic states, and the perturbative breakdown of QCD is this regime indicates that quarks and gluons are not the appropriate degrees of freedom at low energies.  In fact the mass scale of the lightest hadrons is about 200 MeV, confirming that this is the relevant scale for low-energy QCD. That quarks are observed only as bound states is known as ``confinement''; that the coupling becomes small at large energies is known as ``asymptotic freedom''.


\subsection{Confinement vs. asymptotic freedom and factorization}
 \label{sec:qcd:basics:confinement}
 
The twin phenomena of confinement and asymptotic freedom have important consequences for QCD phenomenology.  Confinement implies that quarks and gluons are not well-defined ``particles'' in the sense of asymptotic states that propagate freely.  The coupling binding quarks together in hadrons is so strong that individual quarks can never be removed.  In particular, the binding energy between quarks is $\cO$(200 MeV), but the lightest quark masses are $\cO$(5 MeV).  As two quarks in a meson are pulled apart, creating an additional $q \bar q$ pair from the vacuum becomes energetically favorable, resulting in two mesons.  At low energies, or equivalently large distance scales, hadrons --- not quarks or gluons --- are the relevant degrees of freedom.

Asymptotic freedom, meanwhile, means that at high energies the quarks and gluons in hadrons will behave like free particles.  Probed at high energies, a proton will appear to be a collection of weakly interacting quarks and gluons, or ``partons''.  For example, in a fast-moving proton, partons can only exchange large amounts of \emph{longitudinal} momentum: the relevant scale is the invariant mass of the exchanged gluon, which is small if the exchanged momentum is longitudinal and large if it is transverse.  Large transverse momentum fluctuations involve a factor of $\alpha_s(p_T)$ and are therefore suppressed.  This leads to the picture of a high-momentum proton as a collection of partons, all moving in the same direction, each carrying some fraction of the total momentum.

Similarly, a collision involving large transverse momentum exchange will ``resolve'' the parton structure of the proton; interactions involving more than one parton are suppressed.  The cross section for the process $pp \to q' \bar q'$ can be related to the partonic cross section $\sigma(q \bar q \to \gamma^* \to q' \bar q')$, which can be calculated perturbatively.  Explicitly, we can factorize a proton collision cross section into a partonic cross section convolved with functions that give the probability to find partons with specific momenta inside a proton:
\[
\sigma(p(k_1) p'(k_2) \to X) \approx \int d x_1 d x_2 \sigma \left((q(x_1 k_1) q'(x_2 k_2) \to X) \right) f_q (x_1, \mu), f_{q'}(x_2, \mu)
\]
The ``parton distribution functions'' $f_q$ depend on the parton flavor, momentum fraction $x$, and some energy scale $\mu$ --- the ``factorization scale'' --- which is not well defined but is generally taken to be related to some scale characteristic of the $qq' \to X$ process.  The parton distribution functions characterize the low-energy, non-perturbative interaction of partons within a proton and cannot be predicted using perturbative QCD.  They are however universal across a broad class of processes, and can therefore be measured once and used as an input to other analyses.

The largeness of the coupling at low energies makes it inevitable that the incoming and outgoing quarks will radiate energy away in the form of lower-energy gluons, that the gluons will themselves radiate and split into $q \bar q$ pairs, and that many low-energy partons will result.  At a ``hadronization scale'' $\mathcal{O}(\Lambda_\text{QCD})$, these partons arrange themselves into color singlets --- hadrons like pions and protons.  The basic QCD observable at high-energy colliders are ``jets'' of hadrons, about which much more will be said in Sec.~\ref{sec:qcd:jets}.


\section{Perturbative QCD example: \texorpdfstring{$\ee \to$}{ee ->} hadrons}
\label{sec:qcd:example}

We now consider an example calculation in perturbative QCD which although simple will exhibit many of the features of QCD relevant to collider experiments.  The simplest collider process that involves QCD in a fundamental way is the production of jets at electron-positron colliders.  The presence of strongly-interacting particles in the initial state at $ep$ or $pp(p\bar p)$ colliders introduces additional complications we will consider in Sec.~\ref{sec:qcd:jets}.

The simplest calculation in QED is the scattering cross section $\ee \to \mu^+ \mu^-$.  The QCD analog is the process $\ee \to q \bar q$, the annihilation of an $\ee$ pair into a quark and an anti-quark.  The tree-level Feynman diagrams for each process are shown in Fig.~\ref{fig:eeTreeLevel}.  Of course, whereas muons propagate for distances comparable to the size of a particle detector and thus can be directly detected, quarks cannot.  With a quark-antiquark pair produced initially, we know that they must radiate additional colored partons, all of which eventually organize into hadrons at a lower energy scale.  We might worry that trying to calculate $\sigma(\ee \to \text{hadrons})$ in terms of $\sigma(\ee \to q\bar q)$ is hopeless.

\begin{figure}[htbp]
\begin{center}

\subfloat[]{
\begin{fmffile}{eemm}
\begin{fmfgraph*}(100,60)
	\fmfleft{ep,em}
	\fmfright{qbar,q}
	\fmflabel{$e^{-}$}{em}
	\fmflabel{$e^{+}$}{ep}
	\fmflabel{$\mu^-$}{q}
	\fmflabel{$\mu^+$}{qbar}
	\fmf{fermion}{em,v1,ep}
	\fmf{photon}{v1,v2}
	\fmf{fermion}{qbar,v2,q}
\end{fmfgraph*}
\end{fmffile}
}
\qquad \qquad
\subfloat[]{
\begin{fmffile}{eeqq}
\begin{fmfgraph*}(100,60)
	\fmfkeep{eeqq}
	\fmfleft{ep,em}
	\fmfright{qbar,q}
	\fmflabel{$e^{-}$}{em}
	\fmflabel{$e^{+}$}{ep}
	\fmflabel{$q$}{q}
	\fmflabel{$\bar q$}{qbar}
	\fmf{fermion}{em,v1,ep}
	\fmf{photon}{v1,v2}
	\fmf{fermion}{qbar,v2,q}
\end{fmfgraph*}
\end{fmffile}
}

\end{center}
\caption[Feynman diagrams for $\ee \to \mu^+ \mu^-$ and $\ee \to q \bar q$]{Feynman diagrams for (a) $\ee \to \mu^+ \mu^-$ and (b) $\ee \to q \bar q$.}
\label{fig:eeTreeLevel}
\end{figure}

We are saved, however, by asymptotic freedom.  For a high-energy $\ee$ collision, with $(p_{e^+} + p_{e^-})^2 \equiv s \gg \Lambda_\text{QCD}$, the ``parton-level'' $\ee \to q \bar q$ process and the subsequent radiation and hadronization \emph{factorize}.  This is the first example we will see of a much more general phenomenon in QCD that relies on the running of the strong coupling and a wide separation of energy scales.  At the scale of the parton-level process --- known as the ``hard scattering'' due to the large energy scale involved --- $\alpha_s(s) \ll 1$ and perturbation theory is useful.  Corrections to the tree-level process involving high-energy gluons are perturbative and can be treated as a correction.  Low-energy radiation and hadronization, while non-perturbative, occur at a lower energy scale.  \cite{PinkBook} gives a nice picture of factorization in this case: consider the process as a function of time.  The $\ee$ pair comes together and first annihilates into an off-shell photon or $Z$ boson.  The uncertainty principle dictates that this intermediate particle can only propagate for a time (or distance) inversely proportional to its energy: $t \sim x \sim (\sqrt{s})^{-1}$.  If $\sqrt{s}$ is much larger than the energy scale for radiation and hadronization, then those processes occur over a much longer time scale and do not resolve the effectively instantaneous annihilation.  We can then assume that whatever happens subsequent to the hard scattering occurs with probability 1, so that the total cross section is simply the parton-level cross section:
\beq
\sigma(\ee \to \text{hadrons}) = \sigma(\ee \to q\bar q) + \text{perturbative corrections}.
\eeq
In the following subsections, we will explore the perturbative calculation of this cross section as well as systematic methods of improvement.  A much more detailed version of this calculation is given in Appendix \ref{app:eeNLO}.

\subsection{Tree-level prediction}
\label{sec:qcd:example:tree}

The tree-level diagram for $\ee \to q\bar q$ is given in Fig.~\ref{fig:eeTreeLevel}.  The intermediate boson can be a photon or a $Z$, but we will only consider the case of a photon.  At leading order in the electroweak coupling, including the $Z$ amplitude only contributes an overall factor to the total cross section.  A simple calculation yields the differential cross section
\beq
\frac{d\sigma}{d \cos \theta} = \frac{ \pi \alpha^2 Q^2_q}{2 s} (1+\cos^2\theta),
\eeq
which can be integrated to yield the total cross section
\beq
\sigma_\text{tree} = \frac{4 \pi \alpha^2}{3 s} Q_q^2 \equiv \sigma_0 Q_q^2.
\eeq

It is common to define the ratio $R$ of the total cross section for annihilation to hadrons versus muons:
\beq
R \equiv \frac{\sigma(\ee \to \text{hadrons})}{\sigma(\ee \to \mu^+ \mu^-)} = \frac{\sum_q \sigma(\ee \to q\bar q)}{\sigma(\ee \to \mu^+ \mu^-)} = 3 \sum_q Q_q^2.
\eeq
The sum is over quark flavors; a sum over quark colors has already been performed to yield the factor of 3.


\subsection{Next-to-leading order}
\label{sec:qcd:example:nlo}
	
At the first non-trivial order in $\alpha_s$, five additional diagrams contribute to the processes $\ee \to q\bar q$ and $\ee \to q\bar q g$, shown in Fig.~\ref{fig:eeNLO}.  If we are measuring the \emph{inclusive} cross section $\sigma(\ee \to \text{hadrons})$ we must include both of these processes.  If we wish to calculate a differential cross section in the three body phase space, only the $q\bar q g$ process contributes to $\mathcal{O}(\alpha_s)$, but as we will see we must be careful to restrict ourselves to regions of phase space where a perturbative expansion in $\alpha_s$ is well behaved.

\begin{figure}[htbp]
\begin{center}

\subfloat[]{
\begin{fmffile}{eeqqNLO1}
\begin{fmfgraph*}(100,60)
\fmfkeep{eeqqNLO1}
	\fmfleft{ep,em}
	\fmfright{qbar,q}
	\fmflabel{$e^{-}$}{em}
	\fmflabel{$e^{+}$}{ep}
	\fmflabel{$q$}{q}
	\fmflabel{$\bar q$}{qbar}
	\fmf{fermion,tension=0.5}{em,v1,ep}
	\fmf{photon}{v1,v2}
	\fmf{plain}{v3,v2,v4}
	\fmf{fermion}{v3,q}
	\fmf{fermion}{qbar,v4}
	\fmffreeze
	\fmf{gluon,tension=0.7}{v4,v3}
\end{fmfgraph*}
\end{fmffile}
\qquad \qquad
\begin{fmffile}{eeqqNLO2}
\begin{fmfgraph*}(100,60)
\fmfkeep{eeqqNLO2}
	\fmfleft{ep,em}
	\fmfright{qbar,q}
	\fmflabel{$e^{-}$}{em}
	\fmflabel{$e^{+}$}{ep}
	\fmflabel{$q$}{q}
	\fmflabel{$\bar q$}{qbar}
	\fmf{fermion,tension=0.5}{em,v1,ep}
	\fmf{photon}{v1,v2}
	\fmf{plain}{v5,v3,v2}
	\fmf{fermion}{v5,q}
	\fmf{fermion,tension=1/3}{qbar,v2}
	\fmffreeze
	\fmf{gluon,right=.7}{v3,v5}
\end{fmfgraph*}
\end{fmffile}
\qquad \qquad
\begin{fmffile}{eeqqNLO3}
\begin{fmfgraph*}(100,60)
\fmfkeep{eeqqNLO3}
	\fmfleft{ep,em}
	\fmfright{qbar,q}
	\fmflabel{$e^{-}$}{em}
	\fmflabel{$e^{+}$}{ep}
	\fmflabel{$q$}{q}
	\fmflabel{$\bar q$}{qbar}
	\fmf{fermion,tension=0.5}{em,v1,ep}
	\fmf{photon}{v1,v2}
	\fmf{plain}{v5,v3,v2}
	\fmf{fermion}{qbar,v5}
	\fmf{fermion,tension=1/3}{v2,q}
	\fmffreeze
	\fmf{gluon,right=.7}{v3,v5}
\end{fmfgraph*}
\end{fmffile}
}

\subfloat[]{
\begin{fmffile}{eeqqg1}
\begin{fmfgraph*}(100,60)
\fmfkeep{eeqqg1}
	\fmfleft{ep,em}
	\fmfright{qbar,g,q}
	\fmflabel{$e^{-}$}{em}
	\fmflabel{$e^{+}$}{ep}
	\fmflabel{$q$}{q}
	\fmflabel{$\bar q$}{qbar}
	\fmf{fermion}{em,v1,ep}
	\fmf{photon}{v1,v2}
	\fmf{fermion,tension=1/2}{qbar,v2}
	\fmf{plain}{v2,v3}
	\fmf{fermion}{v3,q}
	\fmffreeze
	\fmf{gluon}{v3,g}
\end{fmfgraph*}
\end{fmffile}
\qquad\qquad
\begin{fmffile}{eeqqg2}
\begin{fmfgraph*}(100,60)
\fmfkeep{eeqqg2}
	\fmfleft{ep,em}
	\fmfright{qbar,g,q}
	\fmflabel{$e^{-}$}{em}
	\fmflabel{$e^{+}$}{ep}
	\fmflabel{$q$}{q}
	\fmflabel{$\bar q$}{qbar}
	\fmf{fermion}{em,v1,ep}
	\fmf{photon}{v1,v2}
	\fmf{fermion,tension=1/2}{v2,q}
	\fmf{plain}{v2,v3}
	\fmf{fermion}{qbar,v3}
	\fmffreeze
	\fmf{gluon}{v3,g}
\end{fmfgraph*}
\end{fmffile}
}
\end{center}
\caption[Feynman diagrams for $\ee \to q\bar q$ and $\ee \to q \bar q g$]{Feynman diagrams for (a) $\ee \to q\bar q$ and (b) $\ee \to q \bar q g$.}
\label{fig:eeNLO}
\end{figure}

As I discuss in more detail in Appendix \ref{app:eeNLO}, the squared matrix elements for the real emission ($q\bar q g$) and virtual ($q \bar q$) diagrams are separately divergent in the infrared.  Calculationally, these divergences come from internal quark or gluon propagators going on shell.  


We can see the divergence explicitly if we consider the differential cross section in the energies of the two quarks.  Writing $x_i \equiv 2 k_i \cdot q/q^2$, we have:
\beq
\label{eq:diffXsec}
\frac{d^2 \sigma}{dx_1 dx_2} = \frac{4 \pi \alpha^2}{3 s}\frac{\alpha_s Q_q^2 C_F}{2 \pi}  \left[ \frac{(x_1^2 + x_2^2)}{(1-x_1)(1-x_2)} \right].
\eeq
While this differential cross section is well behaved for large $x_1, x_2$, it diverges for $x_1\to 1$ and/or $x_2 \to 1$.  Physically, these are the regions of phase space where the gluon is either collinear with one quark or the other, or the gluon is soft.

With a suitable infrared regulator, we find that the \emph{sum} of the real and virtual diagrams is finite.  The divergences only arise because of our insistence on describing the event in terms of quarks and gluons, which are not sensible degrees of freedom over all of phase space.  Performing the calculation requires a choice of regulator; in this thesis I use dimensional regularization (see, e.g., \cite{Peskin}).  In $d = 4 - 2\epsilon$ dimensions, the real and virtual contributions to the total cross section are given be Eqs.~\ref{eq:xsec_real} and \ref{eq:xsec_virt}:
\beq
\begin{split}
\sigma_\text{real}  &= \sigma_0 \left(\sum_q Q_q^2 \right) C_F N_c \frac{\alpha_s}{2 \pi} H(\epsilon) \left[ \frac{2}{\epsilon^2} + \frac{3}{\epsilon} + 19/2 - \pi^2 \right], \\
\sigma_\text{virt} &= \sigma_0 \left(\sum_q Q_q^2\right) C_F N_c \frac{\alpha_s}{2 \pi} H(\epsilon)\left[ -\frac{2}{\epsilon^2} - \frac{3}{\epsilon} - 8 + \pi^2 \right].
\end{split}
\eeq
I have performed the sum over colors and left a sum over flavors.  $H(\epsilon)$ is defined in Eq.~\ref{eq:Heps} and is equal to $1+\mathcal{O}(\epsilon)$.  Adding these to the tree-level contribution yields the finite final answer (Eq.~\ref{eq:final})
\beq
\begin{split}
\sigma(\ee \to \text{hadrons}) &= \sigma_0 \left(\sum_q Q_q^2 \right) N_c \left[1 + \frac{\alpha_s}{\pi} \frac{3 C_F}{4} \right] \\
&= \sigma_0 \left(\sum_q 3 Q_q^2 \right) \left[1 + \frac{\alpha_s}{\pi} \right].
\end{split}
\eeq

We can now see the contrast between two types of QCD calculations.  Some perturbative calculations will yield finite answers; some will not.  The distinction will be whether the calculation adds together contributions that contribute to the same observable phase space --- sometimes described as whether the calculation is ``suitably inclusive''.  The cross section for $(\ee \to q\bar q)$ is not finite beyond tree-level due to an infinite virtual correction.  The total cross section for $(\ee \to \text{hadrons})$ on the other hand is finite because there are canceling divergences in the two- and three-parton cross sections.  Of course, the total cross section is not the only quantity we can calculate that is well-defined.  Provided we group the \emph{singular pieces} of the real contribution with the virtual part, the resulting ``two-body'' and ``three-body'' calculations will be separately finite --- e.g. the differential cross section Eq.~\ref{eq:diffXsec} for large $x_1, x_2$.  This leads to the idea of jets, which we will discuss further in Sec.~\ref{sec:qcd:jets}.

\subsection{Logarithmic resummation}
\label{sec:qcd:example:resum}
	
Calculations in perturbative QCD that involve multiple scales will typically depend on logarithms of ratios of those scales.  We will see an explicit example of this in the calculation in Chapter \ref{sec:scet}.  If the lower scale is regulating an infrared divergence in the differential cross section, up to two powers of the logarithm will appear at every order in perturbation theory, corresponding to the double singularity for gluon emission seen in Eq.~\ref{eq:diffXsec}:
\beq
\begin{split}
\int^\mu \frac{d\sigma}{dX} = C_0 +& C_{12} \alpha_s L^2 + C_{11} \alpha_s L + C_{10} \alpha_s\\
+& C_{24} \alpha_s^2 L^4 + C_{23} \alpha_s^2 L^3 \ldots\\
+& \ldots 
\end{split}
\eeq

Often the logarithmic dependence will exponentiate, meaning that the cross section can be written:
\beq
\begin{split}
\int^\mu \frac{d\sigma}{dX} = C_0 \exp [ &C'_{12} \alpha_s L^2 + C'_{11} \alpha_s L + C'_{10} \alpha_s\\
+& C'_{23} \alpha_s^2 L^3 + C'_{22} \alpha_s^2 L^2 \ldots\\
+& \ldots ]
\end{split}
\eeq
All of the $\alpha_s^n L^{2 n}$ terms in the expansion are captured by the $\alpha_s L^2$ term in the exponent, which only contains logarithms up to $\alpha_s^n L^{n+1}$. In the terminology of Chapter \ref{sec:scet}, terms of order $\alpha_s^n L^{n+1}$ are ``leading logarithmic (LL)'', terms of order $\alpha_s^n L^n$ are ``next-to-leading logarithmic (NLL)'', etc.  In general, perturbation theory including logarithmic resummation exhibits greater convergence, particularly in the regions of phase space where the logarithms are large.


\section{Effective theories of QCD}
\label{sec:qcd:eft}

In the previous section we saw hints that seemingly straightforward calculations in perturbative QCD can be difficult to perform and subject to large corrections due to logarithms of ratios of scales.  These issues can at least partly be addressed by using \emph{effective theories} of QCD.  Effective theories, in the ``top-down'' approach where we know the full theory already, are simply field theories in which some modes have been integrated out, leaving a different set of operators in an effective Lagrangian for the remaining modes.  The classic example is the Fermi theory of the weak interaction where the $W$ boson is integrated out, leaving (non-renormalizable) four-fermion interactions.  In general, an effective theory removes particles above some mass or energy scale in order to simplify the description of physics below that scale.  By construction it must reproduce the low-energy physics of the full theory, up to corrections $\mathcal{O}(p/\Lambda)$, where $p$ is some relevant scale for the problem and $\Lambda$ is ``cutoff'' scale that delineates what has been integrated out.  In the case of Fermi theory $\Lambda \sim m_W$.

In this thesis I will consider a particular effective theory of QCD, soft-collinear effective theory (SCET) \cite{Bauer:2000ew,Bauer:2000yr,Bauer:2001ct,Bauer:2001yt}, an effective theory relevant to radiation from high-energy quarks and gluons.  High-energy, large-angle --- perturbative --- emissions are integrated out, leaving only low-energy (soft) and small-angle (collinear) degrees of freedom.  In Chapter \ref{sec:scet} we will see that this formulation, after suitable field-redefinitions, decouples the soft and collinear modes from each other.  This allows jet-based cross sections to be factorized into several pieces, each of which depends on a single momentum scale and hence contains no large logarithms.

The remainder of this section will be a review of SCET.  At the end of the next section (Sec.~\ref{sec:qcd:jets:shapes}), I will briefly review the class of observable considered in Chapter \ref{sec:scet}.

\subsection[\textit{Review of SCET}]{Review of SCET}
\label{sec:qcd:eft:scet}

SCET is the effective field theory for QCD with all degrees of freedom integrated out, other than those traveling with large energy but small virtuality along a light-like trajectory $n$, and those with small momenta in all components.\footnote{This subsection is taken, with small edits, from Sec.~4.1 of \cite{Ellis:2010rw}.}  A particularly useful set of coordinates is light-cone coordinates, which uses light-like directions $n$ and $\bar{n}$, with $n^2 = \bar{n}^2 = 0$ and $n\cdot\bar{n} = 2$.  In Minkowski coordinates, we take $n = (1,0,0,1)$ and $\bn = (1,0,0,-1)$, corresponding to collinear particles moving in the $+z$ direction. A generic four-vector $p^{\mu}$ can be decomposed into components
\[
p^{\mu} = \bar{n}\cdot p\frac{n^{\mu}}{2} + n\cdot p\frac{\bar{n}^{\mu}}{2} + p_{\perp}^{\mu}  .
\]
In terms of these components, $p = (\bar n\cdot p, n\cdot p, p_\perp)$, collinear and soft momenta scale with some small parameter $\lambda$ as
\begin{equation}
p_n = E(1,\lambda^2,\lambda),\quad p_s\sim E(\lambda^2,\lambda^2,\lambda^2),
\end{equation}
where $E$ is a large energy scale, for example, the center-of-mass energy in an $e^+ e^-$ collision.  $\lambda$ is then the ratio of the typical transverse momentum of the constituents of the jet to the total jet energy. Quark and gluon fields in QCD are divided into collinear and soft effective theory fields with these respective momentum scalings:
\begin{equation}
\label{QCDsplit}
q(x) = q_n(x) + q_s(x),\quad A^\mu(x) = A_n^\mu(x) + A_s^\mu(x).
\end{equation}
We factor out a phase containing the largest components of the collinear momentum from the fields $q_n,A_n$. Defining the ``label'' momentum $\tilde p_n^\mu = \bar n\cdot\tilde p_n \frac{n^\mu}{2} + \tilde p_\perp^\mu$, where $\bar n\cdot \tilde p_n$ contains the $\mathcal{O}(1)$ part of the large light-cone component of the collinear momentum $p_n$, and $\tilde p_\perp$ the $\mathcal{O}(\lambda)$ transverse component, we can partition the collinear fields $q_n,A_n$ into their labeled components,
\begin{equation}
q_n(x) = \sum_{\tilde p\not = 0} e^{-i\tilde p\cdot x}q_{n,p}(x),\quad A_{n}^\mu(x) = \sum_{\tilde p \not=0} e^{-i\tilde p\cdot x} A_{n,p}^\mu(x).
\end{equation}
The sums are over a discrete set of $\mathcal{O}(1,\lambda)$ label momenta into which momentum space is partitioned. The bin $\tilde p = 0$ is omitted to avoid double-counting the soft mode in \eq{QCDsplit} \cite{Manohar:2006nz}. The labeled fields $q_{n,p},A_{n,p}$ now have spacetime fluctuations in $x$ which are conjugate to ``residual'' momenta $k$ of order $E\lambda^2$, describing remaining fluctuations within each labeled momentum partition \cite{Bauer:2001ct,Manohar:2006nz}. It will be convenient to define label operators $\mathcal{P}^\mu = \bn\cdot \mathcal{P} n^\mu/2 + \mathcal{P}_\perp^\mu$  which pick out just the label components of momentum of a collinear field:
\begin{equation}
\mathcal{P}^\mu \phi_{n,p}(x) = \tilde p^\mu \phi_{n,p}(x).
\end{equation}
Ordinary derivatives $\partial^\mu$ acting on effective theory fields $\phi_{n,p}(x)$ are of order $E\lambda^2$.

The final step to construct the effective theory fields is to isolate the two large components of the Dirac spinor $q_{n,p}$ for a fermion with lightlike momentum along $n$. The large components $\xi_{n,p}$ and the small $\Xi_{n,p}$ can be separated by the projections
\begin{equation}
\xi_{n,p} =  \frac{\nslash\bnslash}{4}q_{n,p},\quad \Xi_{n,p} = \frac{\bnslash\nslash}{4}q_{n,p},
\end{equation}
and we have $q_{n,p} = \xi_{n,p} + \Xi_{n,p}$. One can show, substituting these definitions into the QCD Lagrangian, that the fields $\Xi_{n,p}$ have an effective mass of order $E$ and can be integrated out of the theory. The effective theory Lagrangian at leading order in $\lambda$ is \cite{Bauer:2000yr,Bauer:2001ct,Bauer:2001yt}
\begin{equation}
\label{SCETLag}
\mathcal{L}_{\text{SCET}} = \mathcal{L}_{\xi} + \mathcal{L}_{A_n} + \mathcal{L}_s,
\end{equation}
where the collinear quark Lagrangian $\mathcal{L}_\xi$ is
\begin{equation}
\label{xiLag}
\mathcal{L}_\xi = \bar\xi_{n} (x) \left[in\cdot D + i\Dslash_\perp^{\; c} W_n(x) \frac{1}{i\bar n\cdot \mathcal{P}} W_n^\dag(x) i\Dslash_\perp^{\; c}\right] \frac{\bnslash}{2}\xi_n(x),
\end{equation}
where $W_n$ is the Wilson line of collinear gluons,
\begin{equation}
\label{Wdef}
W_n(x) = \sum_{\text{perms}} \exp\left[-g\frac{1}{\bar n\cdot{\mathcal{P}}}\bar n\cdot A_n(x)\right] ;
\end{equation}
the collinear gluon Lagrangian $\mathcal{L}_{A_n}$ is
\begin{equation}
\label{AnLag}
\begin{split}
\mathcal{L}_{A_n} &= \frac{1}{2g^2}\Tr\biggl\{\Bigl[ i\mathcal{D}^\mu + gA_n^\mu, i\mathcal{D}^\nu + gA_n^\nu \Bigr]\biggr\}^2  \\
& \quad + 2\Tr\biggl\{\bar c_n \Bigl[i\mathcal{D}_\mu, \Bigl[i\mathcal{D}^\mu + g A_n^\mu,c_n\Bigr]\Bigr]\biggl\} + \frac{1}{\alpha} \Tr\biggl\{ \Bigl[i\mathcal{D}_\mu , A_n^\mu\Bigr]\biggr\},
\end{split}
\end{equation}
where $c_n$ is the collinear ghost field and $\alpha$ the gauge-fixing parameter; and the soft Lagrangian $\mathcal{L}_s$ is
\begin{equation}
\mathcal{L}_s = \bar q_s i\Dslash_s q_s(x) - \frac{1}{2}\Tr G_s^{\mu\nu} G_{s\mu\nu}(x),
\end{equation}
which is identical to the form of the full QCD Lagrangian (the usual gauge-fixing terms are implicit). In the collinear Lagrangians, we have defined several covariant derivative operators,
\begin{equation}
D^\mu = \partial^\mu - ig A_n^\mu - ig A_s^\mu,\quad iD_c^{\mu} = \mathcal{P}^\mu + g A_n^\mu ,\quad i\mathcal{D}^\mu  = \mathcal{P}^\mu + in\mcdot D\frac{\bar n^\mu}{2}.
\end{equation}
In addition, there is an implicit sum over the label momenta of each collinear field and the requirement that the total label momentum of each term in the Lagrangian be zero.

Note the soft quarks do not couple to collinear particles at leading order in $\lambda$. Meanwhile, the coupling of the soft gluon field to a collinear field is in the component $n\mcdot A_s$ only, according to \eqs{xiLag}{AnLag}, which makes possible the decoupling of  such interactions through a field redefinition of the soft gluon field given in \cite{Bauer:2001yt}. We will utilize this soft-collinear decoupling to simplify the proof of factorization in Chapter \ref{sec:scet}.

The SCET Lagrangian  \eq{SCETLag} may be extended to include collinear particles in more than one direction \cite{Bauer:2002nz}. One adds multiple copies of the collinear quark and gluon Lagrangians \eqs{xiLag}{AnLag} together. The collinear fields in each direction $n_i$ constitute their own independent set of quark and gluon fields, and are governed in principle by different expansion parameters $\lambda$ associated with the transverse momentum of each jet, set either by the angular cut $R$ in the jet algorithm or by the measured value of the jet shape $\tau_a$. Each collinear sector may be paired with its own associated soft field $A_s$ with momentum of order $E\lambda^2$ with the appropriate $\lambda$. For the purposes of keeping the notation tractable while proving the factorization theorem in this section, we will for simplicity take all $\lambda$'s to be the same, with a single soft gluon field $A_s$ coupling to collinear modes in all sectors. In \cite{Ellis:2010rw} we discuss how to ``refactorize'' the soft function further into separate soft functions each depending only on one of the various possible soft scales.

The effective theory containing $N$ collinear sectors and the soft sector is appropriate to describe QCD processes with strongly-interacting particles collimated in $N$ well-separated directions. Thus, in addition to the power counting in the small parameter $\lambda$ within each sector, guaranteeing that the particles in each direction are well collimated, we will find in calculating an $N$-jet cross section the need for another parameter that guarantees that the different directions $n_i$ are well separated. This latter condition requires $t_{ij} \gg 1$, where $t_{ij}$ is defined for jets $i$ and $j$ in \eq{tijdef}.\footnote{This condition is a consequence of our insistence on using operators with exactly $N$ directions to create the final state. We could move away from the large-$t$ limit and account for corrections to it by using a basis of operators with arbitrary numbers of jets and properly accounting for the regions of overlap between an $N$ jet operator and $(N\pm 1)$-jet operators. This is outside the scope of the present work, where we limit ourselves to kinematics well described by an $N$-jet operator, and thus, limit ourselves to the large-$t$ limit.}


\section{Jet physics and collider phenomenology}
\label{sec:qcd:jets}

In Sec.~\ref{sec:qcd:example} we saw that perturbative QCD predictions were finite when we combined cross sections in such a way that we included all processes leading to the same observable final state.  This observation is the basis of jet physics.  Whereas the ``two-parton'' and ``three-parton'' cross sections were both infinite at NLO, the ``two-jet'' cross section, where we combine the two-parton cross section with the soft/collinear parts of the three-parton cross section, was finite.  Likewise, the ``three-jet'' cross section, where we restrict the three partons to be well separated by some metric, will also be finite.  To make this more precise, we need a ``jet algorithm'', which I will discuss more carefully in Sec.~\ref{sec:qcd:jets:algorithm}.  In terms of a perturbative calculation, the role of a ``jet algorithm'' is to combine different final states such that the appropriate real and virtual diagrams have canceling singularities.  An algorithm that does this in all cases is said to be ``infrared safe''; one that does not, at least for some configurations, is said to be ``infrared unsafe'', or perhaps more accurately ``infrared sensitive''.  With this goal in mind, we now review some of the basic collider physics relevant to the production and reconstruction of jets.

\subsection{The parton shower and hadronization}
\label{sec:qcd:jets:basics}

We can see the need for something like jets by considering two- and three-parton final states in $\ee$ collisions, but the same effects are present at every order in perturbation theory.  A final state with $n$ partons will have, at tree level, real singularities that must cancel against virtual singularities in all $m$-parton processes for $m<n$.  A jet algorithm combines these canceling singularities by re-arranging $n$-parton phase space into $N$-jet phase space where cross sections are individually finite.  The ``parton shower'' is the process by which a high-energy quark or gluon radiates many more gluons, which themselves can radiate and (for gluons) split into $q\bar q$ pairs.  The radiation is dominated by the soft/collinear singularities in the gluon emission cross section seen in Sec.~\ref{sec:qcd:example}.  The jet algorithm can be thought of as trying to reverse this process.

An additional complication arises once the $n$-parton final state hadronizes.  Whereas a partonic final state can in principal be grouped such that there is a one-to-one mapping of jets to initial partons (ignoring interference), this is no longer possible after hadronization.  A jet algorithm acting on hadrons must produce groups of hadrons, necessarily color singlets, which can never be mapped unambiguously to colored initial partons.\footnote{An exception to this rule is the ARCLUS dipole clustering algorithm \cite{ARCLUS}, which proceeds via $3\to2$ recombinations and does not assign hadrons to specific jets.  Of course, this does not solve the problem of ambiguity so much as accept it as unavoidable.}  This means that the standard language of equating jets with initial partons is always subject to corrections, expected to be $\cO(\Lambda_\text{QCD}/Q)$, where $Q$ is some relevant hard scale.  An important consideration in the evaluation of a jet algorithm is the size of hadronization corrections (see, e.g., the discussion in \cite{ARCLUS}).

\subsection{Observing jets}
\label{sec:qcd:jets:detector}

Every event at an $\ee$ collider that produces strongly interacting particles, and every event at a hadron collider, involves jets in a fundamental way.  The ability to measure and understand jets is therefore central to collider physics.  Modern detector experiments observe jets primarily as energy depositions in a calorimeter: the set of energetic hadrons produced in the collision is seen as a two- or three-dimensional distribution of energy.  Information from a tracking system, where the paths of individual particles can be observed, is also increasingly being used in the study of jets.

At the LHC, the principal detectors are ATLAS \cite{ATLAS} and CMS \cite{CMS}.  As far as jet measurements are concerned, they share a few essential features.  Both detectors are roughly cylindrical and surround the point of interaction, providing full coverage out to $|\eta| \equiv  |-\ln \tan(\theta/2)| \sim 5$.\footnote{See the glossary item \textsc{Hadron Collider Variables} for definitions of the various kinematic variables used at hadron colliders, and the reasons for their use.}  The innermost layers are tracking layers, which pinpoint locations where charged particles pass.  With multiple tracking layers, the paths of individual particles can be reconstructed with high precision.  The entire system is placed within a magnetic field, so measuring the curvature of a particle's path determines its momentum.  Beyond the tracking system are two levels of calorimetry: and electromagnetic and hadronic calorimeters.  Calorimeters absorb and measure the energy of particles entering them.  The electromagnetic calorimeter is thick enough to absorb essentially all of the energy contained in electron or photon showers, but high energy hadrons like nucleons and pions will only deposit some of their energy in this layer and must be stopped by the hadronic calorimeter.  The hadronic calorimeters are larger and less finely segmented than the electromagnetic calorimeters.  The segmentation for both ATLAS and CMS hadronic calorimeters is approximately $\Delta \eta \times \Delta \phi = 0.1 \times 0.1$.

A typical event at the LHC will have many calorimeter cells with significant ($p_T \gtrsim 1$ GeV) energy deposition, which must be organized into jets for analysis.  One possible input to a jet algorithm is simply the set of calorimeter cells, each having some measured energy and associated with an direction.  Assuming that an individual cell corresponds either to a single particle or multiple essentially collinear particles, we can assign it a four-momentum by assuming that the corresponding mass is zero.  We can imagine an ``ideal calorimeter'' that only combined nearby particles in this way (but did not have any uncertainty on the total four-momentum).  A reasonable jet algorithm should at minimum be insensitive to this kind of initial merging of nearby particles.

Two interesting possibilities exist to supplement the information from the hadronic calorimeter in defining the inputs to a jet algorithm.  First, particles in a jet also deposit energy in the electromagnetic calorimeter, which has higher spatial resolution.  Using information from the electromagnetic calorimeter could allow the resolution of smaller-scale features in jet physics.  This could be particularly useful in the case of jets from heavy particle decays at very large transverse momentum, where the decay products become boosted very close together.

A second possibility is the use of tracking information in describing jets.  In principle, tracks can identify single particles and measure their momentum more precisely than the calorimeters measure their energy.  CMS, for example, uses a ``jets-plus-tracks'' algorithm \cite{JetsPlusTracks} that improves jet energy resolution by using the tracking system to measure the momentum of charged particles in the jet (including particles that are bent out of the jet cone by the magnetic field).  CMS also uses a ``particle flow'' algorithm  \cite{ParticleFlow} that attempts to distinguish electrons, photons, charged hadrons, neutral hadrons, and muons based on their activity in multiple detector layers --- identified particles can then be individually calibrated.  Both methods significantly improve the final jet energy resolution \cite{CMSJetsEarlyData}.

\subsection{The event environment at the LHC}
\label{sec:qcd:jets:collision}

An event at the LHC is considerably more complicated than the simple $\ee \to q\bar q$ events imagined in Sec.~\ref{sec:qcd:example}.  Most of the complications arise from the simple difference that the LHC will collide protons, which are composite objects.  Rather than collide quarks or gluons (which would be ideal), the LHC will collide bags of them --- protons.  The asymptotic freedom of QCD means that high-energy proton interactions can be viewed as perturbative interactions between relatively free quarks and gluons, with the remainder of the protons acting as spectators.  Unfortunately, asymptotic freedom also means that partonic (quark-on-quark, say) collisions involving large transverse momentum transfer --- and hence involving $\alpha_s(\mu)$ evaluated at a large scale --- are rare relative to the overall inelastic (proton-breaking) cross section.

In a hadron collider, the strongly-interacting incoming partons can radiate prior to the hard interaction (initial state radiation, ISR).  This adds to the radiation from any outgoing colored partons (final state radiation, FSR).  Moreover, the remnants of the proton can also interact with each other (multiple parton interactions, MPI; also known as the underlying event (UE)).  In principle this \emph{must} happen to some degree because the beam remnants are not color singlets and must interact at least enough to hadronize.  Likewise, initial state radiation and the underlying event are not in general independent from the hard scattering final state due to color connections.  If the final state is colored (a $g \to t\bar t$ event, say), there is not even a unique assignment of outgoing hadrons to FSR, ISR or UE.  Moreover, quantum mechanics allows interference between these processes.  Of the three, the underlying event is the most difficult to model and measure; for an extensive selection of recent work on this subject see \cite{UE}.

One final contribution adds to hadronic activity in an LHC event: pile-up (PU).  The LHC is designed to collide bunches of many protons at once to increase the likelihood of a high-transverse-momentum interaction.  In the background to these events, however, are much lower-energy collisions between other proton pairs.  At full design luminosity the LHC will observe approximately 25 collisions at once!  While pile-up, unlike ISR and UE, is truly independent of the final state physics, at large luminosities it grows in importance.

All of these effects of the hadronic environment make it more difficult to predict and observe phenomena at the LHC.  We will see in Chapter \ref{sec:prune} that techniques that reduce these effects can significantly improve the performance of LHC searches.

\subsection[\textit{Jets and jet algorithms}]{Jets and jet algorithms} 
\label{sec:qcd:jets:algorithm}

\begin{figure}[htbp]
\begin{center}
\includegraphics[width = \columnwidth]{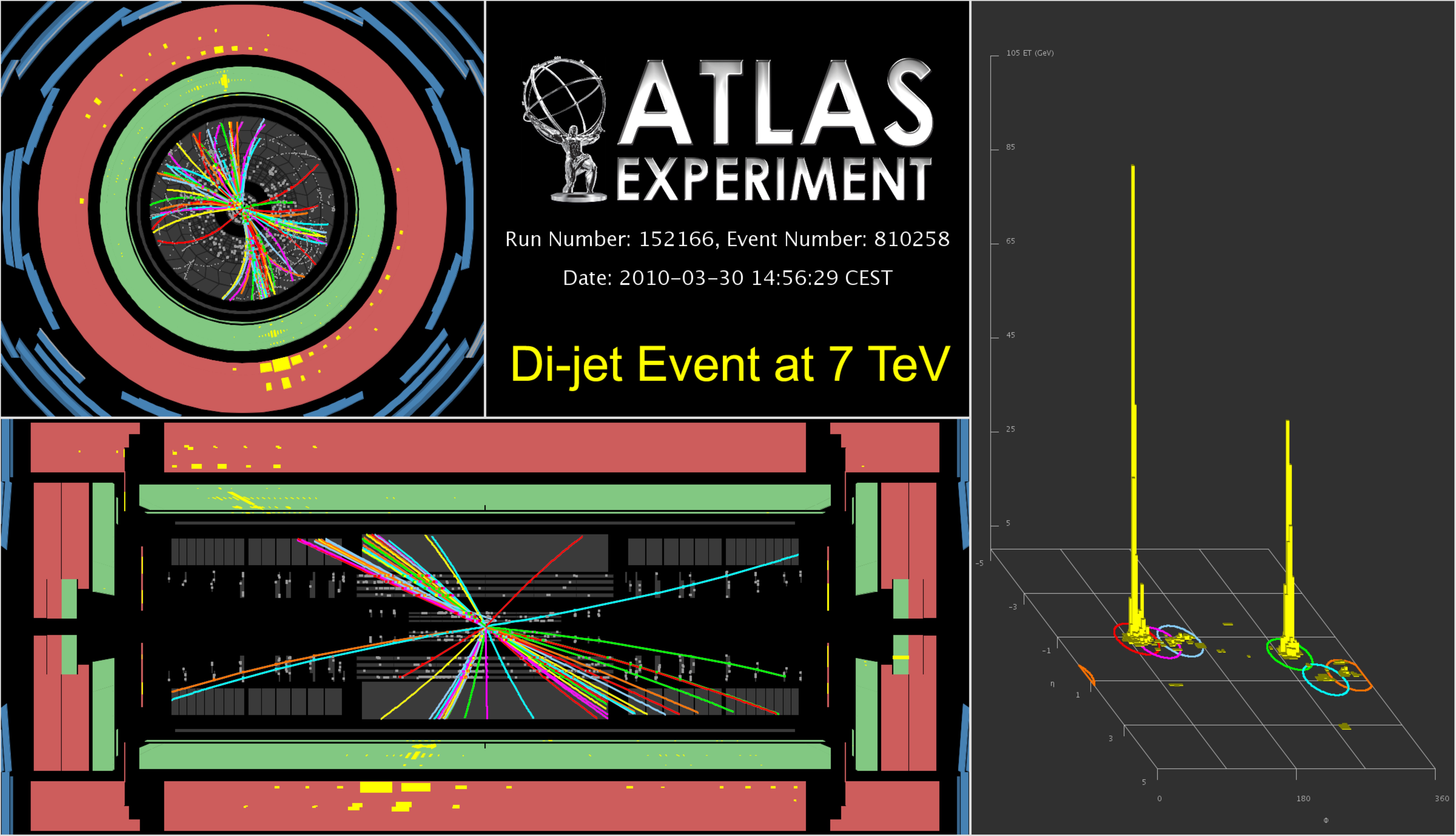}
\end{center}
\caption[Several event views for an event at ATLAS]{Several event views for an event at ATLAS.  Two high-energy ``jets'' have been identified, along with several much lower-energy jets clustered around them (colored circles in right plot).  Taken from the ATLAS public website \cite{ATLAS}.}
\label{fig:dijet_lego}
\end{figure}

To make sense of the multiplicity of hadrons produced in collisions with final-state quarks or gluons, we group them into jets (for two good reviews, see \cite{FamousJetReview} and \cite{Jetography}).\footnote{This subsection, with small modifications, is taken from Sec.~II of \cite{Pruning2}.}  High-energy quarks and gluons radiate many more gluons and $q\bar q$ pairs, but that the radiation is dominantly soft and/or collinear.  This means that most of the energy of the initial parton will be located in a small angular area in the detector, plus low-energy deposits at larger angle.  Large-energy, large-angle radiation can only come from perturbative emission, and therefore tends to happen with probability $\sim \alpha_s (p_{T_J}) \sim 0.1$.  An ATLAS event with two jets is shown in Fig.~\ref{fig:dijet_lego}.

\subsubsection{Recombination algorithms}


To identify jets we need a \emph{jet algorithm}.  Jet algorithms can be broadly divided into two categories, recombination algorithms and cone algorithms.  Both types of algorithms form jets from protojets, which are initially generic objects such as calorimeter towers, topological clusters%
\footnote{In addition to single cells, ATLAS also uses three-dimensional ``topological clusters'' of calorimeter cells as inputs to jet analyses.  Topological clustering is a method of combining nearby cells into an object significant and well enough measured to be locally calibrated.}%
, or final state particles.  Cone algorithms fit protojets within a fixed geometric shape, the cone, and attempt to find stable configurations of those shapes to find jets.  In the cone-jet language, ``stable'' means that the direction of the total four-momentum of the protojets in the cone matches the direction of the axis of the cone. Recombination algorithms, on the other hand, give a prescription to \emph{pairwise} (re)combine protojets into new protojets, eventually yielding a jet.  For the recombination algorithms studied in this work, this prescription is based on an understanding of how the QCD shower operates, so that the recombination algorithm attempts to undo the effects of showering and approximately trace back to objects coming from the hard scattering.  The anti-$\kt$ algorithm \cite{AntiKT} functions more like the original cone algorithms, and its recombination scheme is not designed to backtrack through the QCD shower.  Cone algorithms have been the standard in collider experiments, but recombination algorithms are finding more frequent use.  Analyses at the Tevatron \cite{Aaltonen:2008eq} have shown that the most common cone and recombination algorithms agree in measurements of jet cross sections.  In this work we are most interested in jet \emph{substructure}, and we therefore consider only recombination algorithms, which define substructure in a natural way.

A general recombination algorithm uses a distance measure $\rho_{ij}$ between protojets to control how they are merged.  A ``beam distance'' $\rho_i$ determines when a protojet should be promoted to a jet.  The algorithm proceeds as follows:

\begin{itemize}
\item[0.] Form a list $L$ of all protojets to be merged.

\item[1.] Calculate the distance between all pairs of protojets in $L$ using the metric $\rho_{ij}$, and the beam distance for each protojet in $L$ using $\rho_i$.

\item[2.] Find the smallest overall distance in the set $\{\rho_i, \rho_{ij}\}$.

\item[3.] If this smallest distance is a $\rho_{ij}$, merge protojets $i$ and $j$ by adding their four vectors.  Replace the pair of protojets in $L$ with this new merged protojet.  If the smallest distance is a $\rho_i$, promote protojet $i$ to a jet and remove it from $L$.

\item[4.] Iterate this process until $L$ is empty, i.e., all protojets have been promoted to jets.\footnote{This defines an \emph{inclusive} algorithm.  For an \emph{exclusive} algorithm, there are no promotions, but instead of recombining until $L$ is empty, mergings proceed until all $\rho_{ij}$ exceed a fixed $\rho_\text{cut}$.}
\end{itemize}

For the $\kt$ \cite{KT1, KT2, KT3}, Cambridge-Aachen (CA) \cite{CambridgeAachen}, and anti-$\kt$ \cite{AntiKT} recombination algorithms the metrics are
\beq
\begin{split}
\kt: \rho_{ij} \equiv \min(p_{Ti},p_{Tj})\Delta R_{ij}/D, & \qquad \rho_i \equiv p_{Ti}; \\
\text{CA}: \rho_{ij} \equiv \Delta R_{ij}/D, & \qquad \rho_i \equiv 1.\\
\text{anti-}\kt: \rho_{ij} \equiv \min(p^{-1}_{Ti},p^{-1}_{Tj})\Delta R_{ij}/D, & \qquad \rho_i \equiv p^{-1}_{Ti}.
\end{split}
\label{eq:algo}
\eeq
Note that all three are specific instances of the general metric with parameter $\alpha$:
\beq
\text{generic} \, \kt: \rho_{ij} \equiv \min(p^{\alpha}_{Ti},p^{\alpha}_{Tj})\Delta R_{ij}/D,  \qquad \rho_i \equiv p^{\alpha}_{Ti}.
\eeq
Here $p_{Ti}$ is the transverse momentum of protojet $i$ and $\Delta R_{ij} \equiv \sqrt{(\phi_i - \phi_j)^2 + (y_i - y_j)^2}$ is a measure of the angle between two protojets that is invariant under boosts along and rotations around the beam direction.  $\phi$ is the azimuthal angle around the beam direction, $\phi = \tan^{-1} {p_y}/{p_x}$, and $y$ is the rapidity, $y = \tanh^{-1} {p_z}/{E}$, with the beam along the $z$ axis.  The angular parameter $D$ governs when protojets should be promoted to jets: it determines when a protojet's beam distance is less than the distance to other objects.  $D$ provides a rough measure of the typical angular size (in $y$--$\phi$) of the resulting jets.

The recombination metric $\rho_{ij}$ determines the \emph{order} in which protojets are merged in the jet, with recombinations that minimize the metric performed first.  From the definitions of the recombination metrics in Eq.~(\ref{eq:algo}), it is clear that the $\kt$ algorithm tends to merge low-$p_T$ protojets earlier, while the CA algorithm merges pairs in strict angular order. This distinction will be very important in our subsequent discussion.  Anti-$\kt$, meanwhile, tends to cluster protojets around the hardest protojet, producing cone-like jets with less interesting substructure.

These definitions are all appropriate for finding jets at a hadron collider, where invariance under longitudinal boosts is desired.  At an $\ee$ collider, $p_T$ is replaced by $E$, and $\Delta R^2$ is typically replaced by $(1-\cos \theta)$.  Moreover, the beam metric $\rho_i$ is not used; instead, merging proceeds until all $\rho_{ij}$ exceed some (usually dimensionful) value $y_\text{cut}$ which depends on the center-of-mass energy $Q^2$.

\subsubsection{Jet Substructure}


A recombination algorithm naturally defines substructure for the jet.  The sequence of recombinations tells us how to construct the jet in step-by-step $2\to1$ mergings, and we can unfold the jet into two, three, or more subjets by undoing the last recombinations.  The jet algorithm begins and ends with physically meaningful information (starting at calorimeter cells, for example, and ending at jets), so we might expect that the intermediate (subjet) information to have physical significance as well.   In particular, we expect the earliest recombinations to approximately reconstruct the QCD shower, while the last recombinations in the algorithm, those involving the largest-$p_T$ degrees of freedom, may indicate whether the jet was produced by QCD alone or a heavy particle decay plus QCD showering.  This will be true for the CA and $\kt$ algorithms, where the metric reflects the soft ($\kt$) and collinear (CA and $\kt$) dynamics of the parton shower.  To discuss the details of jet substructure, we begin by defining relevant variables.

\subsubsection{Variables Describing Branchings and Their Kinematics}


Whereas the jet algorithm can be thought of as a sequence of \emph{mergings}, the parton shower, possibly preceded by a decay, can be thought of as a sequence of \emph{branchings}.  In studying the substructure produced by jet algorithms, it will be useful to describe branchings using a set of kinematic variables.  Since we will consider the substructure of (massive) jets reconstructing kinematic decays and of QCD jets, there are two natural choices of variables.  Jet--rest-frame variables are useful to understand decays because the decay cross section takes a simple form.  Lab-frame variables are useful because jet algorithms are formulated in the lab frame, so algorithm systematics are most easily understood there.  The QCD soft/collinear singularity structure is also easy to express in lab frame variables.

Naively, there are twelve variables completely describing a $1 \to 2$ splitting.  Here we will focus on the top branching (the last merging) of the jet splitting into two daughter subjets, which we will label $J \to 1,2$.  Imposing the four constraints from momentum conservation to the branching leaves eight independent variables.  The invariance of the algorithm metrics under longitudinal boosts and azimuthal rotations removes two of these (they are irrelevant).  For simplicity we will use this invariance to set the jet's direction to be along the $x$ axis, defining the $z$ axis to be along the beam direction.  Therefore there are six relevant variables needed to describe a $1 \to 2$ branching.  Three of these variables are related to the three-momenta of the jet and subjets, and the other three are related to their masses.

Of the six variables, only one needs to be dimensionful, and we can describe all other scales in terms of this one.  We choose the mass $m_J$ of the jet.  In addition, we use the masses of the two daughter subjets scaled by the jet mass:
\beq
a_1 \equiv \frac{m_1}{m_J} \quad \text{and} \quad a_2 \equiv \frac{m_2}{m_J} .
\label{eq:a1def}
\eeq
We choose the particle labeled by `1' to be the heavier particle, $a_1 > a_2$.  The three masses, $m_J$, $a_1$, and $a_2$, will be common to both sets of variables.  Additionally, we will typically want to fix the $p_{T}$ of the jet and determine how the kinematics of a system change as $p_{T_J}$ is varied.  For QCD, a useful dimensionless quantity is the ratio of the mass and $p_T$ of the jet, whose
square we call $x_J$:
\beq
x_J \equiv \frac{m_J^2}{p_{T_J}^2} .
\label{eq:xJdef}
\eeq
For decays, we will opt instead to use the familiar magnitude $\gamma$ of the boost of the heavy particle from its rest frame to the lab frame, which is related to $x_J$ by
\beq
\gamma = \sqrt{\frac{1}{x_J} + 1},\quad x_J=\frac{1}{\gamma^2-1}.
\eeq
The remaining two variables, which are related to the momenta of the subjets, will differ between the rest-frame and lab-frame descriptions of the splitting.

Unpolarized $1\to2$ decays are naturally described in their rest frame by two angles.  These angles are the polar and azimuthal angles of one particle (the heavier one, say) with respect to the direction of the boost to the lab frame, and we label them $\theta_0$ and $\phi_0$ respectively.  Since we are choosing that the final jet be in the $\hat{x}$ direction, $\theta_0$ is measured from the $\hat{x}$ direction while $\phi_0$ is the angle in the $y$--$z$ plane, which we choose to be measured from the $\hat{y}$ direction.  Putting these variables together, the set that most intuitively describes a heavy particle decay is the ``rest-frame'' set
\beq
\{m_J,\ a_1,\ a_2,\ \gamma,\ \cos\theta_0,\ \phi_0\} .
\label{eq:decayvars}
\eeq

In the lab frame, we want to choose variables that are invariant under longitudinal boosts and azimuthal rotations.  The angle $\Delta R_{12}$ between the daughter particles is a natural choice, as is the ratio of the minimum daughter $p_T$ to the parent $p_T$, which is commonly called $z$:
\beq
z \equiv \frac{\min(p_{T_1},p_{T_2})}{p_{T_J}} .
\label{eq:zdef}
\eeq
These variables make the recombination metrics for the $\kt$ and CA algorithms simple:
\beq
\rho_{12}(\kt) = p_{T_J}z\Delta R_{12} \quad \text{and} \quad \rho_{12}(\text{CA}) = \Delta R_{12} .
\label{eq:reconCAKT}
\eeq
Note that for a generic recombination, the momentum factors in the denominator of Eq.~(\ref{eq:zdef}) and in the $\kt$ metric in Eq.~(\ref{eq:reconCAKT}) should be $p_{Tp}$, the momentum of the the parent or combined subjet of the $2\to1$ recombination.

From these considerations we choose to describe recombinations in the lab frame with the set of variables
\beq
\{m_J,\ a_1,\ a_2,\ x_J,\ z,\ \Delta R_{12}\} .
\label{eq:labvars}
\eeq

\begin{figure*}[htbp]
\begin{center}
\subfloat[$a_1=a_2=0$] {\label{fig:thetaphicontours1} \includegraphics[width = .45\textwidth] {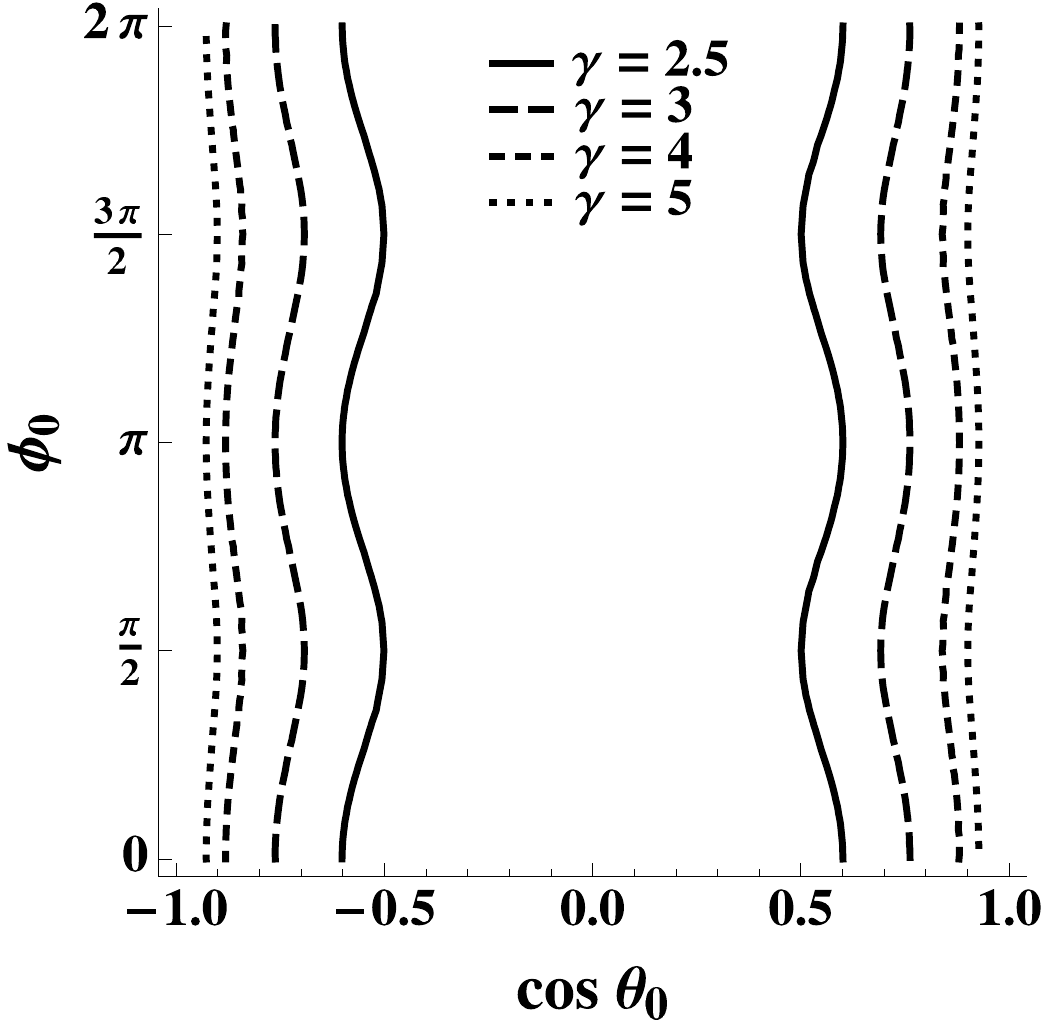}}
\subfloat[$a_1=0.46,\ a_2=0$] {\label{fig:thetaphicontours2} \includegraphics[width = .45\textwidth] {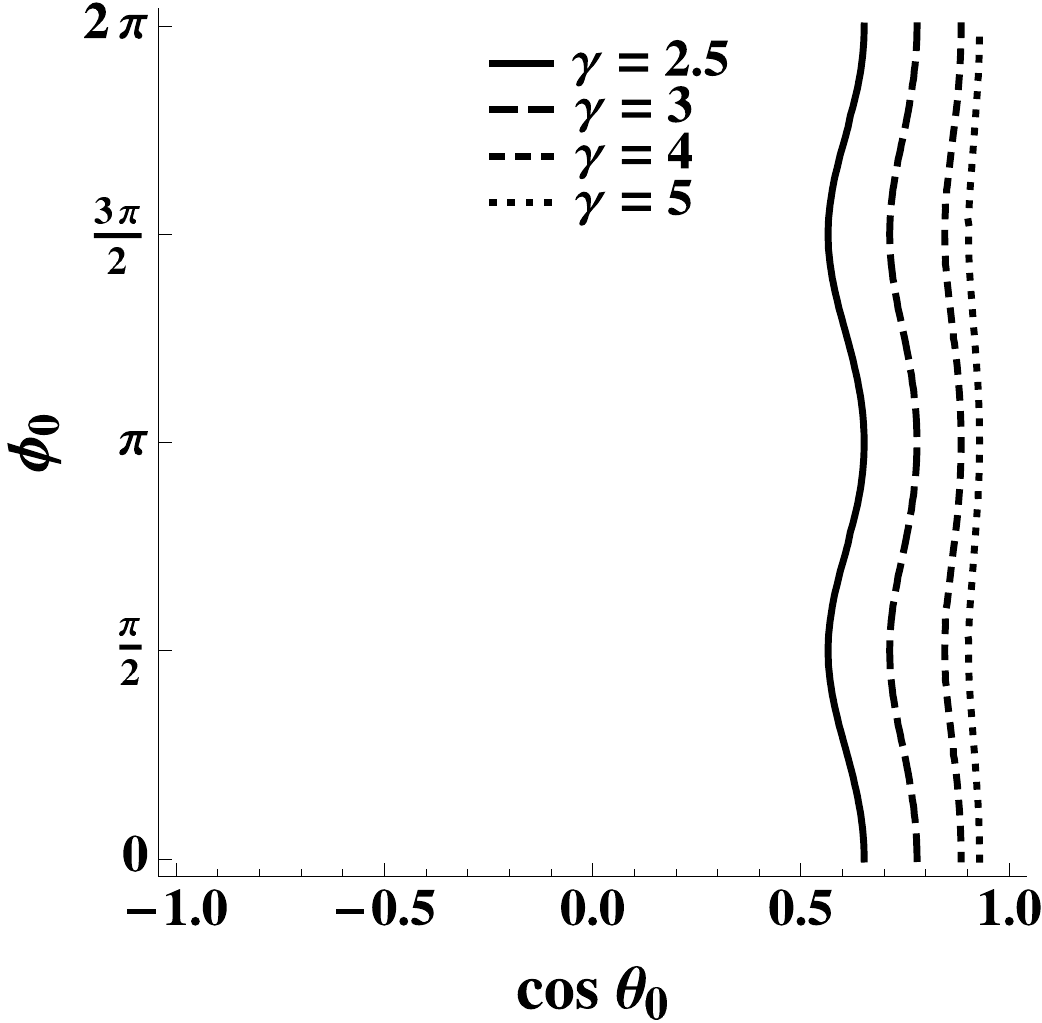}}

\subfloat[$a_1=0.9,\ a_2=0$] {\label{fig:thetaphicontours3} \includegraphics[width = .45\textwidth]{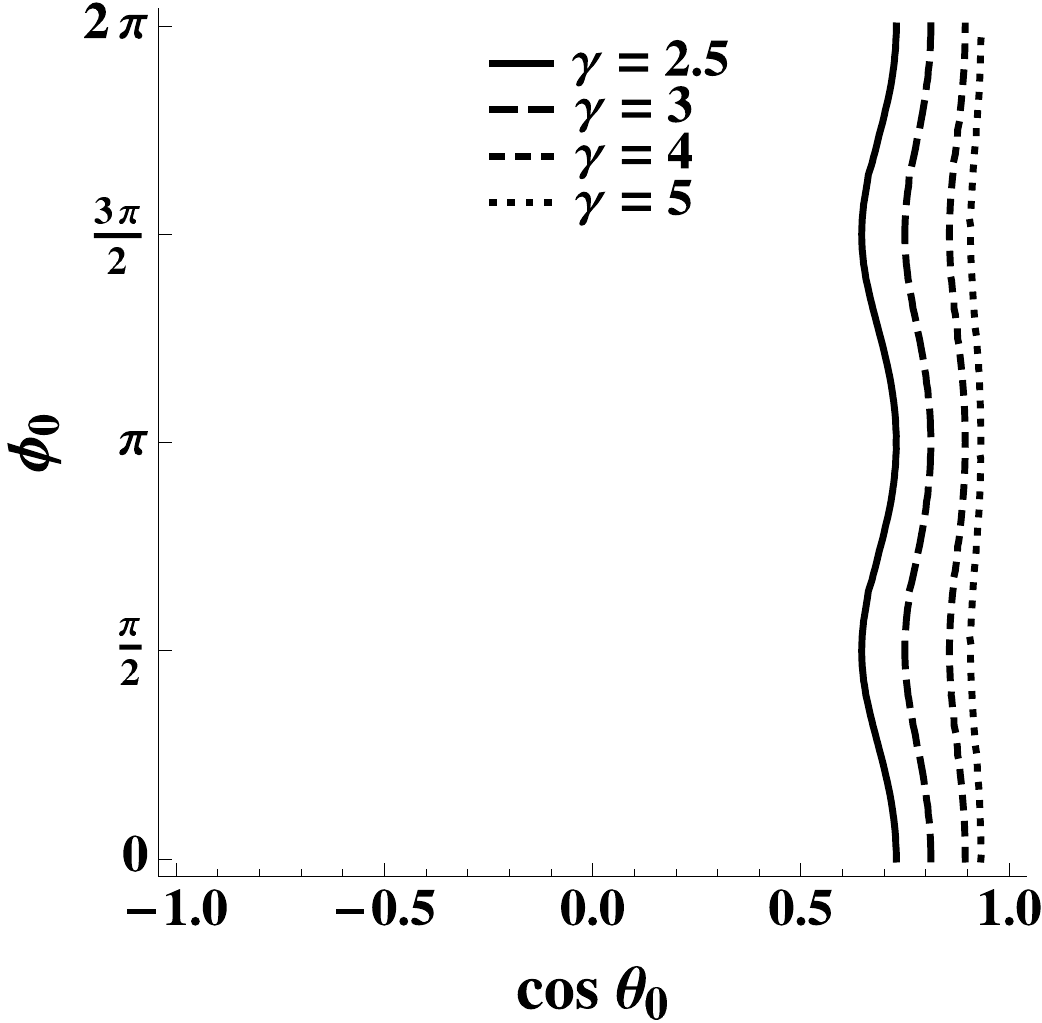}}
\subfloat[$a_1=0.3,\ a_2=0.1$] {\label{fig:thetaphicontours4} \includegraphics[width = .45\textwidth] {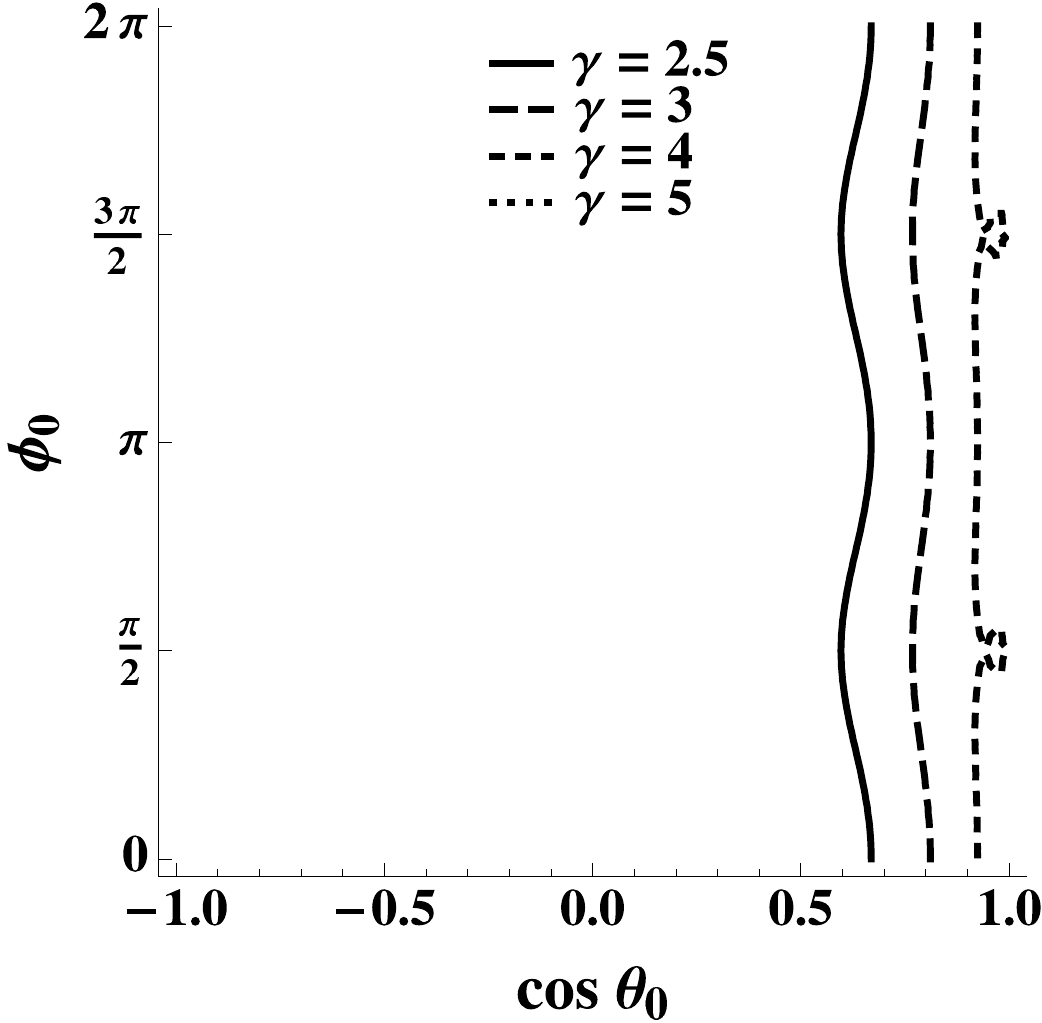}}
\end{center}
\caption[Boundaries in the $\cos\theta_0$--$\phi_0$ plane]{Boundaries in the $\cos\theta_0$--$\phi_0$ plane for a recombination step to fit in a jet of size $D=1.0$, for several values of the boost $\gamma$ and the subjet masses $\{a_1,\ a_2\}$.  The ``interior'' region has $\Delta R_{12} < D$.}
\label{fig:thetaphicontours}
\end{figure*}

In using these variables it is essential to understand the structure of the corresponding phase space, especially for the last two variables in both sets.  If we require that the decay ``fits'' in a jet, constraints and correlations appear.  These are clearest in terms of the lab frame variables $\Delta R_{12}$ and $z$.  As a first step in understanding these correlations, we plot in Fig.~\ref{fig:thetaphicontours} the contour $\Delta R_{12} = D(=1.0)$ in the $(\cos\theta_0, \phi_0)$ phase space for different values of $\gamma$ and over different choices for $a_1$ and $a_2$.  These specific values of $a_1$ and $a_2$ correspond to a variety of interesting processes:  $a_1 = a_2 = 0$ gives the simplest kinematics and is therefore a useful starting point; $a_1 = 0.46, a_2 = 0$ gives the kinematics of the top quark decay; $a_1 = 0.9, a_2 = 0$ and $a_1 = 0.3, a_2 = 0.1$ are reasonable values for subjet masses from the CA and $\kt$ algorithms respectively.  The contour $\Delta R_{12} = D$ defines the boundary in phase space where a $1\to2$ process will no longer fit in a jet, with the interior region corresponding to splittings with $\Delta R_{12} < D$.  Note that the contour is nearly vertical, increasingly so for larger $\gamma$.  This is a reflection of the fact that $\Delta R_{12}$ is nearly independent of $\phi_0$, up to terms suppressed by $\gamma^{-2}$.

\begin{figure*}[htbp]
\begin{center}
\subfloat[$a_1=a_2=0$] {\label{fig:zRcontours1} \includegraphics[width = .45\textwidth] {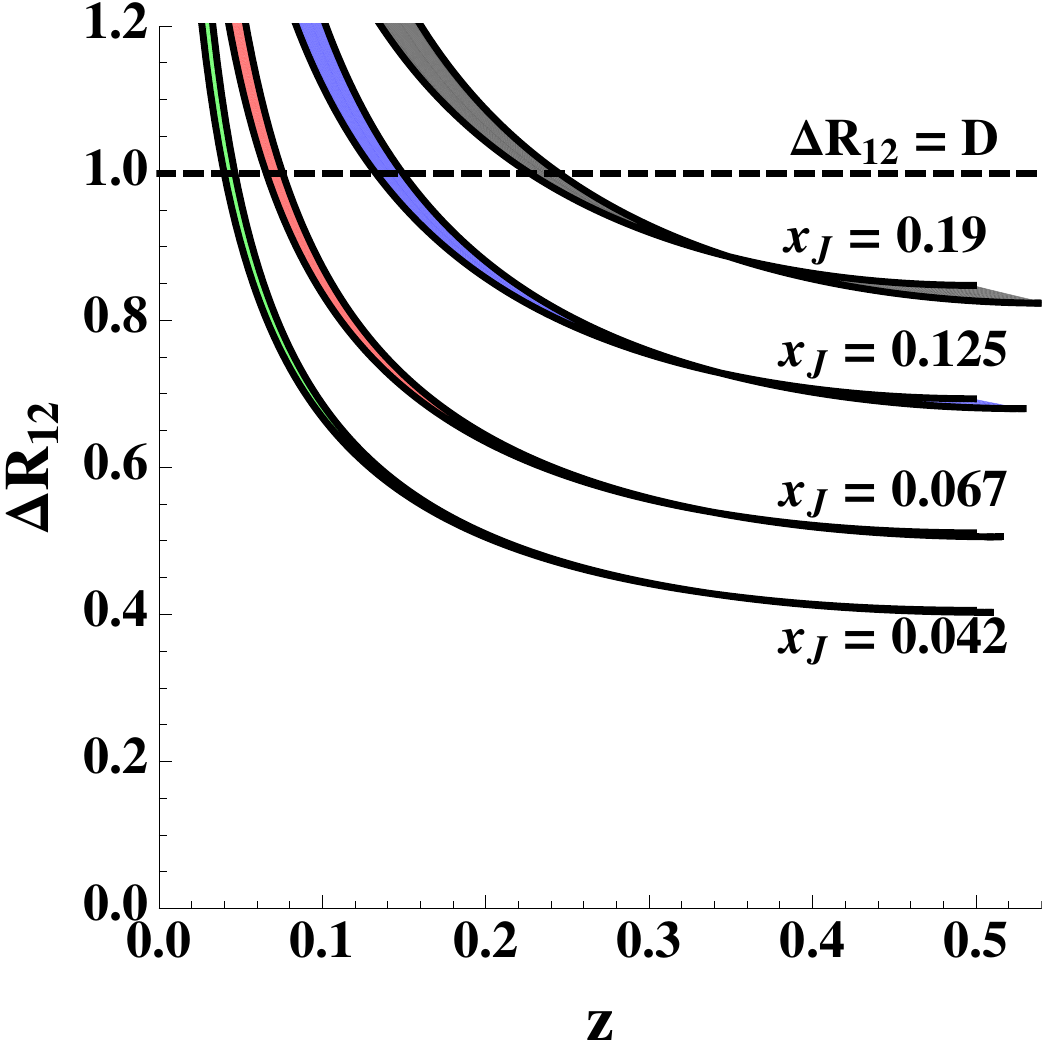}}
\subfloat[$a_1=0.46,\ a_2=0$] {\label{fig:zRcontours2} \includegraphics[width = .45\textwidth] {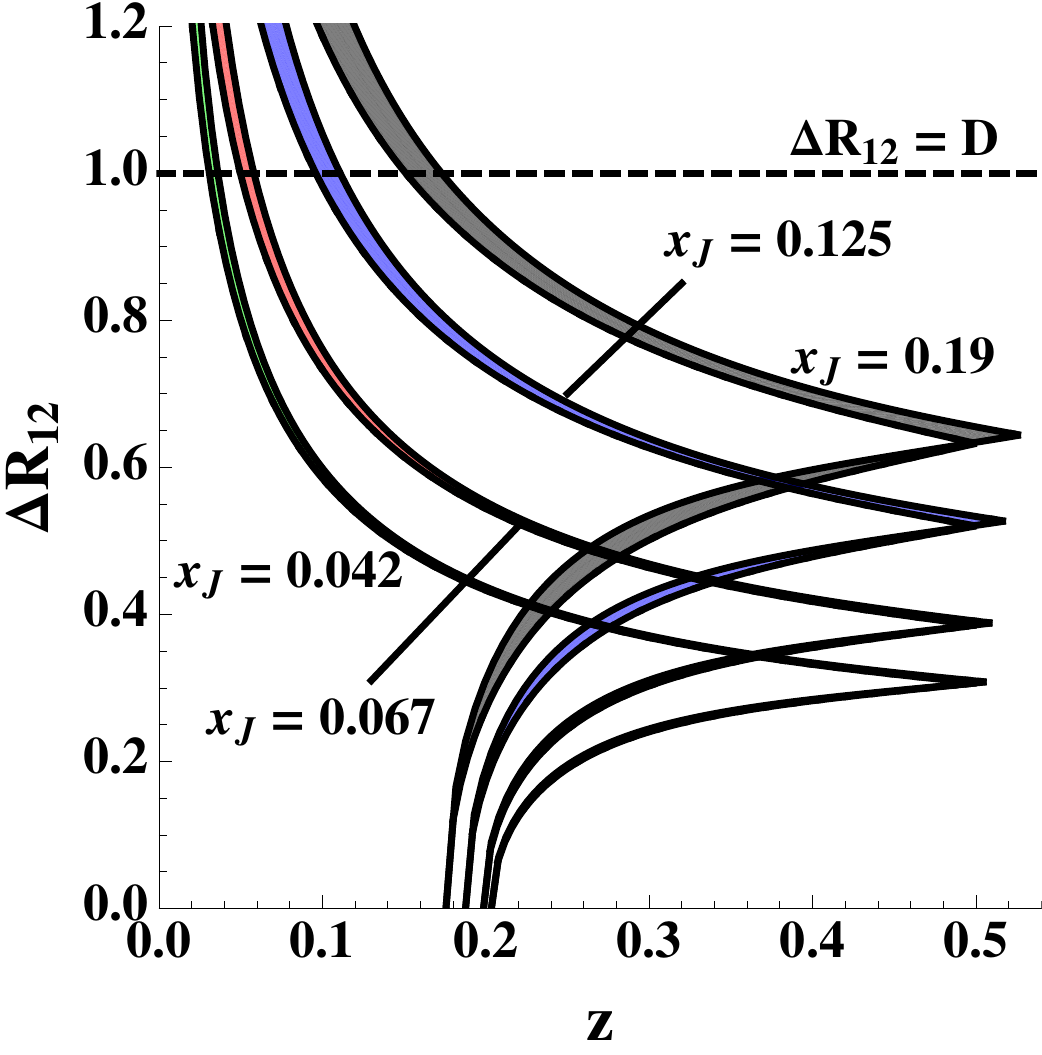}}

\subfloat[$a_1=0.9,\ a_2=0$] {\label{fig:zRcontours3} \includegraphics[width = .45\textwidth] {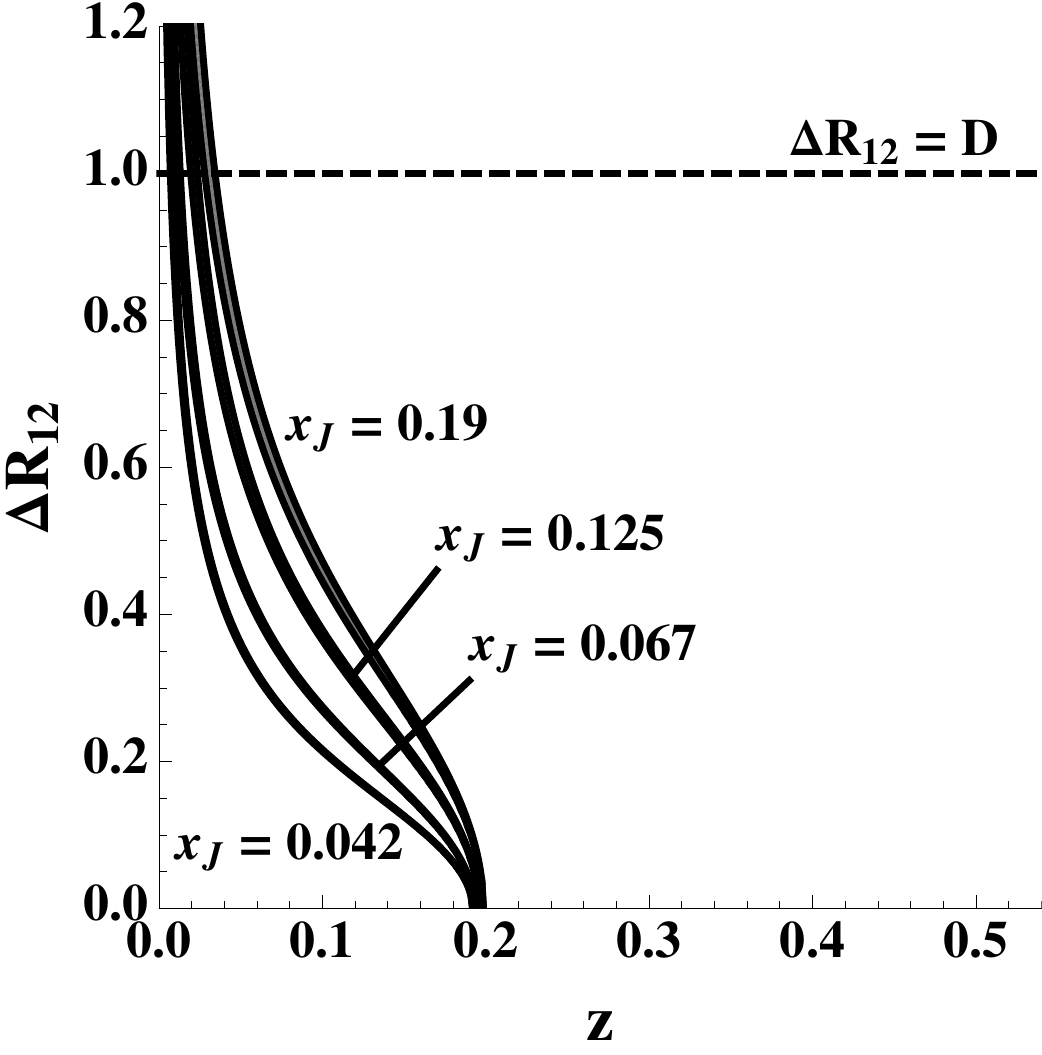}}
\subfloat[$a_1=0.3,\ a_2=0.1$] {\label{fig:zRcontours4} \includegraphics[width = .45\textwidth]{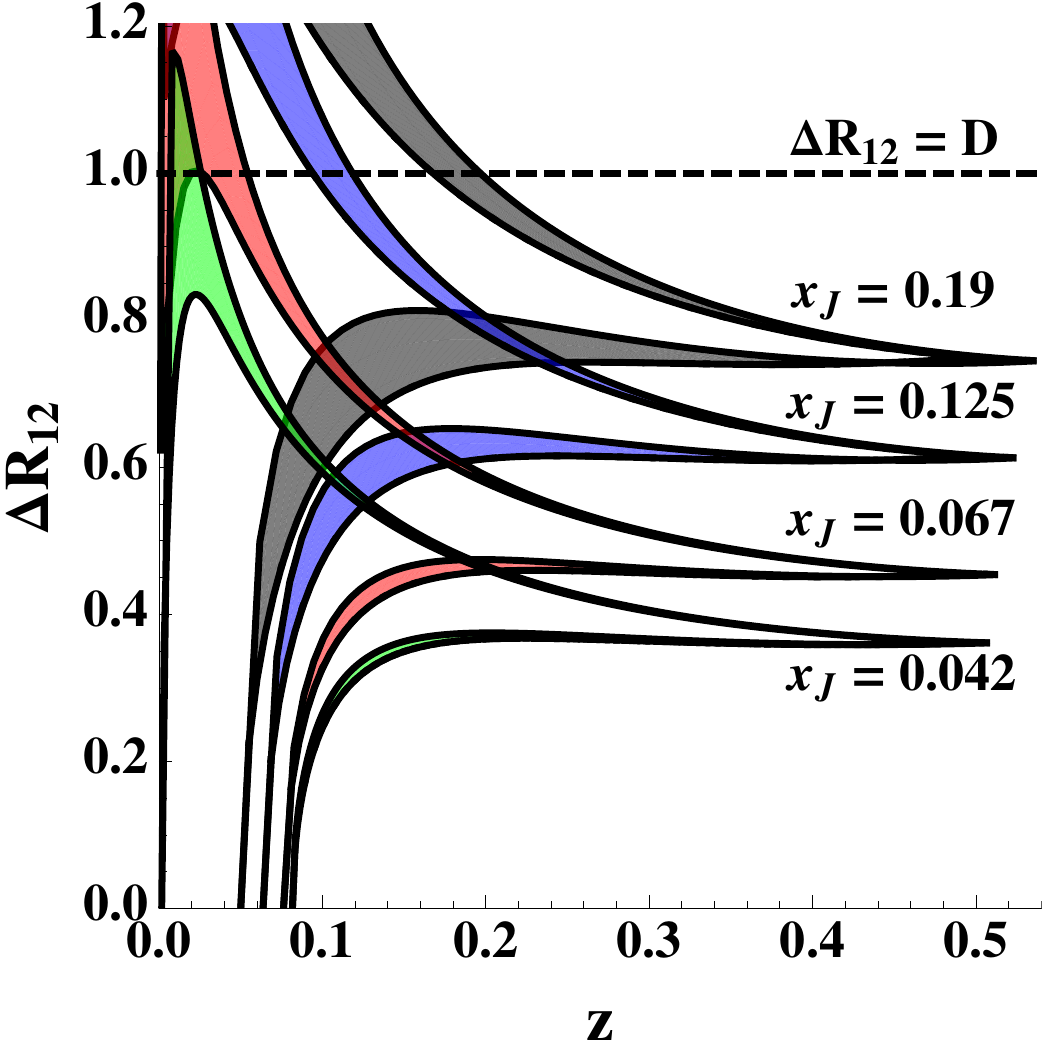}}
\end{center}
\caption[Boundaries in the $z$--$\Delta R_{12}$ plane]{Boundaries in the $z$--$\Delta R_{12}$ plane for a recombination step of fixed $\{a_1,\ a_2,\ x_J\}$, for various values of $x_J$ and the subjet masses $\{a_1,\ a_2\}$.  Configurations with $\Delta R_{12} < D$ fit in a jet; $D = 1.0$ is shown for example.  }
\label{fig:zRcontours}
\end{figure*}

While the constraint $\Delta R_{12} < D$ becomes simpler in the $(z, \Delta R_{12})$ phase space, the boundaries of the phase space become more complex.  In Fig.~\ref{fig:zRcontours}, we plot the available phase space in $(z, \Delta R_{12})$ for the same values of $x_J$, $a_1$, and $a_2$ as in Fig.~\ref{fig:thetaphicontours}, translating the value of $\gamma$ into $x_J$.  The most striking feature is that for fixed $x_J$, $a_1$, and $a_2$, the phase space in ($z$, $\Delta R_{12}$) is nearly one-dimensional; this is again due to the fact that $\Delta R_{12}$ and also $z$ are nearly independent of $\phi_0$.  In particular, for $a_1 = a_2 = 0$ (as in Fig.~\ref{fig:zRcontours1}), the phase space approximates the contour describing fixed $x_J$ for small $\Delta R_{12}$, which takes the simple form
\beq
x_J \equiv \frac{m_{J}^{2}}{p_{T_J}^{2}} \approx z \left(  1-z \right) \Delta R_{12}^{2}.
\label{eq:simplejmass}
\eeq
This approximation is accurate even for larger angles, $\Delta R_{12} \approx 1$, at the $10\%$ level. Note also that the width of the band about the contour described by Eq.~(\ref{eq:simplejmass}) is itself of order $x_J$.  As we decrease $x_J$ the band moves down and becomes narrower as indicated in Fig.~\ref{fig:zRcontours1}).

As illustrated in Figs.~\ref{fig:zRcontours2} and \ref{fig:zRcontours4}, we can also see a double-band structure to the $(z, \Delta R_{12})$ phase space.  The upper band corresponds to the case where the lighter daughter is softer (smaller-$p_T$) than the heavier daughter (and determines $z$), while the lower band corresponds to the case where the heavier daughter is softer.  This does not occur in Fig.~\ref{fig:zRcontours1} because $a_1 = a_2$ (the single band is double-covered), or in Fig.~\ref{fig:zRcontours3} because the heavier particle is never the softer one for the chosen values of $x_J$.

We have said nothing about the density of points in phase space for either pair of variables.  This is because the weighting of phase space is set by the dynamics of a process, while the boundaries are set by the kinematics.  Decays and QCD splittings weight the phase space differently, as we will see in Sec. \ref{sec:sub:parton}.

\subsubsection{Ordering in Recombination Algorithms}


Having laid out variables useful to describe $1\to2$ processes, we can discuss how the jet algorithm orders recombinations in these variables.  Recombination algorithms merge objects according to the pairwise metric $\rho_{ij}$.  The sequence of recombinations is almost always monotonic in this metric: as the algorithm proceeds, the value increases.  Only certain kinematic configurations will decrease the metric from one recombination to the next, and the monotonicity violation is small and rare in practice.

This means it is straightforward to understand the typical recombinations that occur at different stages of the algorithm.  We can think in terms of a phase space boundary: the algorithm enforces a boundary in phase space at a constant value of the recombination metric that evolves to larger values as the recombination process proceeds.  If a recombination occurs at a certain value of the metric, $\rho_0$, then subsequent recombinations are very unlikely to have $\rho_{ij} < \rho_0$, meaning that region of phase space is unavailable for further recombinations.

\begin{figure}[htbp]
\begin{center}
\subfloat[CA] {\label{fig:algboundariesCA} \includegraphics[width=0.45\textwidth] {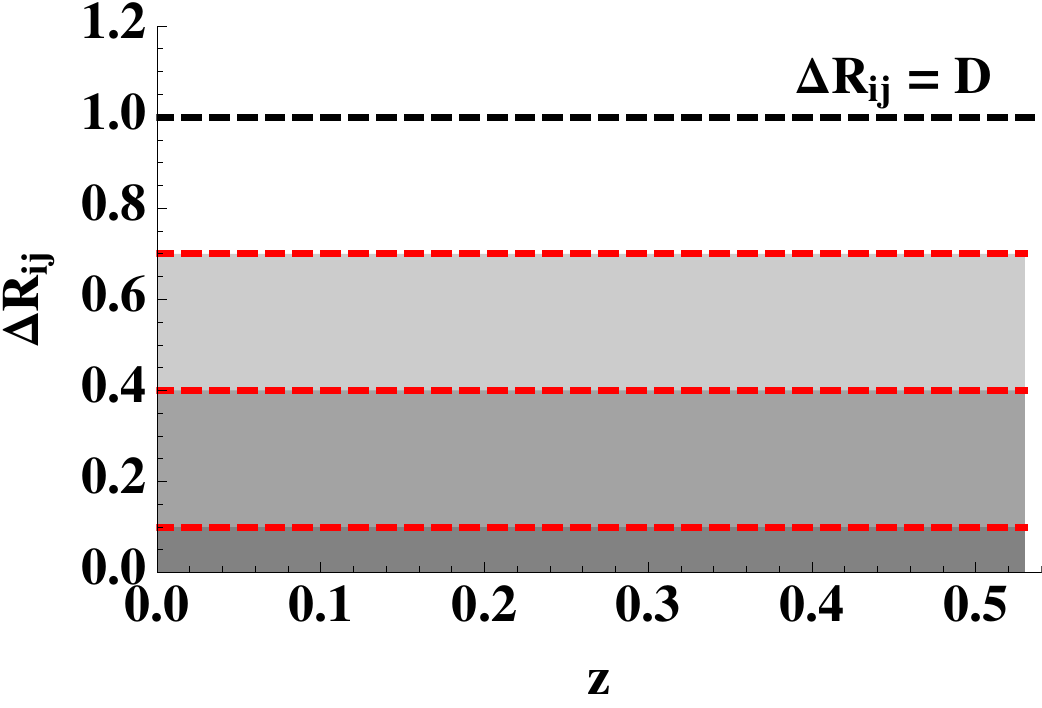}}
\subfloat[$\kt$] {\label{fig:algboundariesKT} \includegraphics[width=0.45\textwidth] {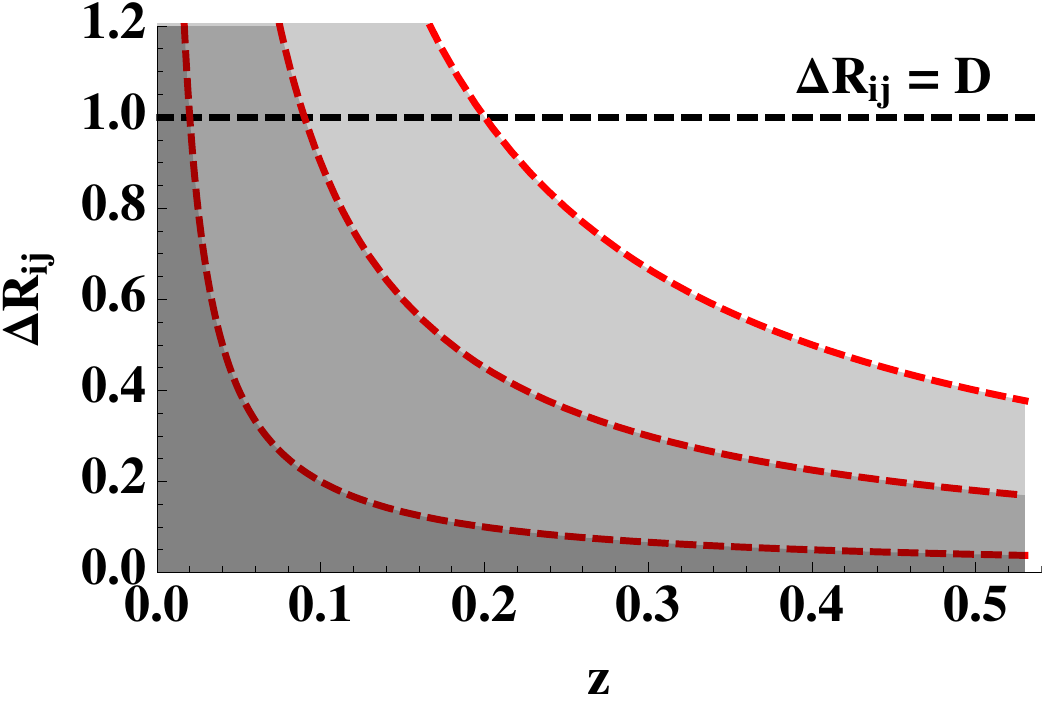}}
\end{center}
\caption[Typical boundaries on phase space due to ordering in the CA and $\kt$ algorithms]{Typical boundaries (red, dashed lines) on phase space due to ordering in the CA and $\kt$ algorithms.  The shaded region below the boundaries is cut out, and the more heavily shaded regions correspond to earlier in the recombination sequence.  The cutoff $\Delta R_{ij} = D = 1.0$ is shown for reference (black, dashed line).}
\label{fig:algboundaries}
\end{figure}

In Fig.~\ref{fig:algboundaries}, we plot typical boundaries for the CA and $\kt$ algorithms in the $(z, \Delta R_{12})$ phase space.  For CA, these boundaries are simply lines of constant $\Delta R_{12}$, since the recombination metric is $\rho_{ij}(\textrm{CA}) = \Delta R_{ij}$.  For $\kt$, these boundaries are contours in $z\Delta R_{12}$, and implicitly depend on the $p_T$ of the parent particle in the splitting.  Because the $\kt$ recombination metric for $i,j\to p$ is $\rho_{ij}(\kt) = z\Delta R_{ij}p_{Tp}$, increasing the value of $p_{Tp}$ will shift the boundary in to smaller $z\Delta R_{ij}$.  These algorithm-dependent ordering effects will be important in understanding the restrictions on the kinematics of the last recombinations in a jet.  For instance, we expect to observe no small-angle late recombinations in a jet defined by the CA algorithm.
%


\subsection[\textit{Event shapes and jet shapes}]{Event shapes and jet shapes}
\label{sec:qcd:jets:shapes}

An alternative characterization of hadronic activity is an \emph{event shape}.  Event shapes, such as thrust, characterize events based on the distribution of energy in the final state by assigning differing weights to events with differing energy distributions.\footnote{This subsection is taken, with small modifications, from Sec.~2 of \cite{Ellis:2010rw}.}  Events that are two-jet--like, with two very collimated back-to-back jets, produce values of the observable at one end of the distribution, while spherical events with a broad energy distribution produce values of the observable at the other end of the distribution.  While event shapes can quantify the global geometry of events, they are not sensitive to the detailed structure of jets in the event.  Two classes of events may have similar values of an event shape but characteristically different structure in terms of number of jets and the energy distribution within those jets.

Jet shapes, which are event shape-like observables applied to single jets, are an effective tool to measure the structure of individual jets.  Just as event shapes are an alternative to jets in characterizing an \emph{event}, jet shapes are an alternative to subjet descriptions of \emph{jet substructure}.  These observables can be used to not only quantify QCD-like events, but study more complex, non-QCD topologies, as illustrated  for light quark vs. top quark and $Z$ jets in \cite{Almeida:08.1,Almeida:2008tp}.  Broad jets, with wide-angle energy depositions, and very collimated jets, with a narrow energy profile, take on distinct values for jet shape observables.  In Chapter \ref{sec:scet}, we consider the example of the class of jet shapes called angularities, defined in \eq{angularitydefn} and denoted $\tau_a$.  Every value of $a$ corresponds to a different jet shape.  As $a$ decreases, the angularity weights particles at the periphery of the jet more, and is therefore more sensitive to wide-angle radiation.  Simultaneous measurements of the angularity of a jet for different values of $a$ can be an additional probe of the structure of the jet.


 
 \graphicspath{{chapter3/graphics/}}
 
\chapter[\textit{Consistent Factorization of Jet Observables in Exclusive Multijet Cross Sections}]{Consistent Factorization of Jet Observables in Exclusive Multijet Cross Sections}
\label{sec:scet}

\section[\textit{Introduction}]{Introduction}

Final states that contain several jets are important Standard Model backgrounds to many new physics processes in high-energy colliders, in addition to serving as sensitive probes of Quantum Chromodynamics (QCD) itself over a wide range of energy scales.\footnote{This chapter, with small modifications, is taken from \cite{SCETLetter}.} The structure of jet-like final states contains signatures of the hard scattering of parton-like degrees of freedom, the branching and showering at ever lower energies, and hadronization at the lowest scale $\Lqcd$. Probing the structure of jets both teaches us about QCD and can help us to distinguish jets of Standard Model origin from those that are truly signatures for new physics. 

The presence of multiple scales governing jets is at once an opportunity to probe many aspects of their physics and also a challenge due to the generation of large logarithms of ratios of these scales spoiling the behavior of perturbation theory. A powerful framework to separate physics at different scales and  to improve the behavior of perturbation series is effective field theory (EFT). EFTs aid in factorizing an observable dependent on multiple scales into pieces  each sensitive to a single energy scale. Renormalization group (RG) evolution of these pieces in  EFT achieves resummation of large logarithms to all orders in perturbation theory. Factorization also allows the disentangling of perturbative and non-perturbative physics \cite{Collins:1989gx,Sterman:1995fz}.

Soft-Collinear Effective Theory (SCET) \cite{Bauer:2000ew,Bauer:2000yr,Bauer:2001ct,Bauer:2001yt} has  had considerable success in applications to many hard-scattering cross sections \cite{Bauer:2002nz} and jet cross sections. SCET separates degrees of freedom in QCD  into distinct soft and collinear modes, expanding the full theory in a parameter $\lambda$ that characterizes the size of collinear momenta transverse to the jet direction, and provides a framework to factorize cross sections into separate pieces coming from interactions at hard, collinear, and soft scales. This was done in SCET for event shape variables using hemisphere jet algorithms in $e^+e^-$ colliders ~\cite{Fleming:2007xt,Hornig:2009vb} and for ``isolated Drell-Yan'' (where central jets are vetoed) in hadron colliders \cite{Stewart:2009yx}. 
In addition, there has been progress in understanding how to implement jet algorithms other than the simple hemisphere jet algorithm in SCET. 
In \cite{Bauer:2003di, Trott:2006bk}, total two-jet rates where the jets are defined by Sterman-Weinberg jet algorithms were computed at NLO. These results were extended to the cases of the exclusive $\kt$ and JADE algorithms in \cite{Cheung:2009sg}. 

In most applications of SCET to exclusive jet cross sections considered to date, there are two back-to-back jets. (Recently Ref.~\cite{Becher:2009th} considered direct photon production in hadron collisions, involving three collinear directions.) In this work we consider for the first time exclusive $N$-jet final states with arbitrary $N\ge 2$ for the SISCone \cite{SISCone}, Snowmass \cite{Huth:1990mi}, inclusive $\kt$ \cite{KT3}, anti-$\kt$ \cite{AntiKT}, and Cambridge-Aachen \cite{CambridgeAachen} jet algorithms. We find that a new feature that arises when more than two jets are present is that the parameter $\lambda$ is not in itself sufficient to ensure factorization. In particular, 
factorization is valid to leading order in $\lambda$ \emph{and} in a jet separation measure $1/t$, where 
 \beq
 \label{tijdef}
 t = \frac{\tan(\psi/2)}{\tan(R/2)}
 ,\eeq 
 with $R$ the angular size of a jet as defined by a jet algorithm and $\psi$ the minimum angle between two jets. This is due to the fact that jets need to be both well-collimated ($\lambda \ll 1$) and well-separated ($t \gg 1$). The latter requirement is trivial for back-to-back jets since $1/t = 0$ for $\psi = \pi$.

Our analysis applies not only to the total $N$-jet cross section, but also in the case that 
jet observables are measured on some number $M \le N$ of the jets. We will illustrate the measurement of  angularities $\tau_a$ (cf. \cite{Berger:2003iw,Almeida:08.1}), defined by
\begin{equation}
\label{angularitydefn}
\tau_a(J) = \frac{1}{2E_J}\sum_{i\in J} \abs{\vect{p}_T^i} e^{-\eta_i(1-a)},
\end{equation}
where $E_J$ is the energy of the jet $J$, the sum is over particles $i$ in the jet, and $p_T^i$ and $\eta_i$ are the transverse momentum and (pseudo-)rapidity of particle $i$ with respect to the jet axis.  However, most of our results do not depend on this choice of observable, and we organize the calculation such that  other observables can be easily implemented. 

Distributions of jet shapes such as angularities contain logarithms of $\tau_a$ that become large in the limit $\tau_a \to 0$, of the form $(\alpha_s^n \ln^{k-1}\tau_a)/\tau_a$ with $k\leq 2n$. 
The factorization theorem we present provides the basis for resummation of sets of these logarithms to all orders in $\alpha_s$. In the exponent, $\ln R(\tau_a)$, of the ``radiator''   $R(\tau_a) = (1/\sigma_0)\int_0^{\tau_a}d\tau'_a (d\sigma/d\tau'_a)$, these appear in the form $\alpha_s^n \ln^m\tau_a$ with $m\leq n+1$ \cite{Catani:1992ua,Berger:2003pk}. Our results here allow us to sum to leading-logarithmic (LL) ($m=n+1$) and next-to-leading-logarithmic (NLL) ($m=n$) accuracy in this exponent.

The set of jet shapes $\tau_a$ contain similar information as the ``original'' jet shape $\Psi(r/R)$ \cite{Ellis:1991vr,Ellis:1992qq,Abe:1992wv}, the fraction of energy of a jet of size $R$ in a sub-cone of size $r$. Distributions in this jet shape in hadron collisions were resummed to so-called ``modified LL'' accuracy (which includes the $k=2n$ and $k=2n-1$ terms as enumerated for the distribution above) in  \cite{Seymour:1997kj}.

Factorization of event shape distributions in SCET was proven in \cite{Fleming:2007qr,Bauer:2008dt}, and factorization for multijet observables defined with arbitrary algorithms was considered in \cite{Bauer:2008jx}. The extension to the more general case that we consider involves the straightforward combination of the techniques developed in these papers and will be derived in detail in \cite{Ellis:2010rw}. In this work we demonstrate that, after intricate cancellations among the various contributions to the jet and soft functions, consistency of the factorization theorem is satisfied at NLL accuracy. In order for the factorization theorem to be consistent, the hard, jet, and soft functions defined must satisfy a strong condition on their anomalous dimensions:
\begin{equation}
\label{consistency}
\begin{split}
0 = &\left(\gamma_H + \sum_{i=M+1}^N \gamma_{J_i}\right) \delta(\tau_a^1) \cdots \delta(\tau_a^M) \\
&+ \sum_{i=1}^M\gamma_{J_i}(\tau_a^i) \prod_{\substack{j=1 \\ j\not = i}}^M \delta(\tau_a^j)  + \gamma_S(\tau_a^1,\dots,\tau_a^M),
\end{split}
\end{equation} 
for any number $N$ of total jets and $M$ of measured jets, and any color representation of each jet. This consistency condition is made even more nontrivial by the potential dependence of the jet and soft anomalous dimensions on the jet algorithm parameters.  In this chapter we demonstrate that \eq{consistency} does in fact hold for arbitrary numbers, types, and sizes of jets in the final state, up to certain power corrections we are able to identify.

Observables measuring jet shapes like $\tau_a$, while also restricting the phase space into which soft gluons can be emitted, can be plagued by ``non-global'' logarithms \cite{Dasgupta:2001sh} beginning at NLL order that may not resummed by our methods. In particular there can be logarithms in our jet shape distributions generated by the energy cut $\Lambda$ that we place on soft radiation outside jets \cite{Ellis:2010rw}. Ref.~\cite{Dokshitzer:2003uw} demonstrated the factorization of similar distributions into global and non-global parts. Our results here allow the resummation of logarithms of $\tau_a$ in the global part. More simply, the non-global logarithms can be removed by  choosing $\Lambda\sim E_J\tau_a$  \cite{Berger:2003iw}. In \cite{Ellis:2010rw} we address resummation in the case that these scales remain disparate. Despite these potential complications, which deserve additional study, our demonstration of a consistent factorization theorem for jet shapes defined with a jet algorithm provides a key advance towards the resummation of any such jet shape distributions.

We begin in \sec{sec:cuts} by defining the phase space cuts needed to implement our choice of jet algorithms.  In \sec{sec:fact} we then present the factorization theorem for $N$-jet events and define the hard, jet, and soft functions, and identify power corrections to the factorization. In \sec{sec:RGE} we give the form of the RG evolution equations obeyed by the factorized functions.  In \sec{sec:anom} we summarize the results of all the anomalous dimensions needed for NLL running and demonstrate how they intricately satisfy the consistency condition \eq{consistency}. This requires calculating only the infinite parts of the bare functions. We give the finite pieces of the jet and soft functions (which are not needed at NLL) in \cite{Ellis:2010rw}. In \sec{sec:resum} as an example we calculate quark and gluon angularity jet shapes in 3-jet final states with logarithms of $\tau_a$ resummed to NLL accuracy.

\section[\textit{Phase Space Cuts and the Jet Algorithm}]{Phase Space Cuts and the Jet Algorithm}
\label{sec:cuts}

Two general categories of jet algorithms, cone algorithms and recombination ($\kt$-type) algorithms, are commonly used to find jets.  For a jet composed of two particles, as in a next-to-leading order description, the phase space constraints implied by each type of algorithm become very simple.  In this work we deal with the common forms of cone and (inclusive) $\kt$-type algorithms; our cone algorithms include the Snowmass and SISCone algorithms, and our recombination algorithms include the inclusive $\kt$, Cambridge-Aachen, and anti-$\kt$ algorithms.  Cone algorithms require each particle to be within an angle $R$ of the jet axis, while recombination algorithms require the angle between the two particles to be within an angle $D$ of each other.  If we label the jet axis as $\vect{n}$ and its constituent particles as 1 and 2, then the algorithm constraints for a two-particle jet are:
\begin{equation}
\label{injetconstraint}
\begin{split}
\textrm{cone type: }&\theta_{1\vect{n}} < R \textrm{ and } \theta_{2\vect{n}} < R  ,\\
\textrm{$\kt$ type: }&\theta_{12} < D  .
\end{split}
\end{equation}
For the parts of the jet and soft functions that we give in this work, we find that the functional form is the same for cone-type and $\kt$-type algorithms in terms of the angular parameter $R$ or $D$.  Therefore, we will use the more common $R$ in writing down the jet and soft functions, but we note here that the functional form is the same for $\kt$ with the replacement $R\to D$.

Note that, while all algorithms that we consider fall into one of the two constraints in \eq{injetconstraint} at NLO, at higher orders the various algorithms will behave differently. Without taking this into account, we have no guarantee that we can resum all logarithms of jet algorithm parameters correctly.\footnote{The $\kt$ algorithm, for example, is known to spoil naive exponentiation \cite{Banfi:2005gj}.}  This is not a problem we solve in this paper.
In this paper, we resum logarithms of  jet observables in the presence of phase space cuts due to an algorithm, demonstrate that the factorization theorem and NLL running are valid and consistent, and identify the power corrections to this statement.

At the hard scale, we match an $N$-leg amplitude in QCD onto an $N$-jet operator in SCET, meaning we must enforce that the number of jets is fixed to be $N$. To enforce that we have no more than $N$ jets, we require that the total energy of particles that do not enter jets to be less than a cutoff $\Lambda$. To enforce that we have at least $N$ jets, we need that pairwise each jet is well separated from every other jet. The requirement of consistency of NLL running will give a quantitative measure  of this separation requiring that $t \gg1$.

\section[\textit{Factorized Jet Shapes in \texorpdfstring{$N$}{N}-Jet Production}]{Factorized Jet Shapes in \texorpdfstring{$N$}{N}-Jet Production}
\label{sec:fact}

The cross section for $e^+ e^-$ annihilation to $N$ jets  at center-of-mass energy $Q$, differential in the jet three-momenta $\vect{P}_i$ of the jets and in the shapes of $M$ of these jets, is given in QCD by
\begin{equation}
\label{QCDcs}
\begin{split}
&\frac{d\sigma}{d\tau_a^1 \cdots d\tau_a^M  d^3\vect{P}_1\cdots d^3\vect{P}_N} \\
&\quad\quad = \frac{1}{2Q^2}\sum_X  (2\pi)^4\delta^4(Q-p_X)\abs{\bra{X}j^\mu(0)\ket{0}L_\mu}^2 \\
&\quad\qquad\times\delta_{n(\mathcal{J}(X))-N} \prod_{i=1}^M \delta(\tau_a^i - \tau_a(J_i)) \prod_{j=1}^N\delta^3(\vect{P}_j - \vect{P}(J_j)),
\end{split}
\end{equation}
where $J_{i}$ is the $i$th jet in $X$ identified by the jet algorithm $\mathcal{J}$. The Kronecker delta restricts the sum over states to those that are identified as having $N$ jets by the algorithm. The final state is produced by the QCD current $j^\mu = \bar q \gamma^\mu q$, and $L_\mu$ is the leptonic part of the amplitude for $e^+ e^-\to \gamma^*$. 

To factorize the cross section \eq{QCDcs}, we begin by matching the QCD current $j^\mu$ onto a set of $N$-jet operators in SCET.
These operators are built from quark and gluon jet fields,
\begin{equation}
\label{jetfields}
\chi_n = W_n^\dag\xi_n\,,\quad B_n^\perp = \frac{1}{g}W_n^\dag(\mathcal{P}_\perp + A_n^\perp)W_n,
\end{equation}
where $\xi_n,A_n$ are collinear quark and gluon fields in SCET, and $W_n$ is a Wilson line of the $\mathcal{O}(1)$ component $\bn\cdot A_n$ of collinear gluons,
\begin{equation}
W_n(x) = \sum_{\text{perms}}\exp \left[-\frac{g}{\bn\cdot\mathcal{P}}\bn\cdot A_n(x)\right].
\end{equation}
We have made use of the label operator $\mathcal{P}^\mu$ which picks out the large $\mathcal{O}(1)$ $\bn\cdot \tilde p$ and $\mathcal{O}(\lambda)$ $\tilde p_\perp$ components of the label momentum $\tilde p$ of collinear field in SCET. We will not need to construct the $N$-jet operators explicitly, but bases of $2,3,4$  jet operators have been given in \cite{Bauer:2002nz,Bauer:2006mk,Marcantonini:2008qn}, respectively.

To describe an $N$-jet cross section, we construct an effective theory Lagrangian by adding $N$ copies of the collinear Lagrangian in SCET (in $N$ different light-cone directions $n_i$) together with one soft Lagrangian. In each collinear sector, we redefine collinear fields by multiplying by Wilson lines of soft gluons to eliminate the coupling of soft gluons to collinear modes in the leading-order SCET Lagrangian \cite{Bauer:2001yt},
$
\xi_n = Y_n^\dag \xi_n^{(0)}$ and $ A_n = \mathcal{Y}_n A_n^{(0)} ,
$
where 
\begin{equation}
\label{Yndef}
Y_n(x) = P\exp\left[ig\int_0^\infty ds\,n\cdot A_s(ns+x)\right],
\end{equation}
with $A_s$ in the fundamental representation, and $\mathcal{Y}$ similarly defined but in the adjoint representation.

Performing the above steps in \eq{QCDcs} for the jet shape distribution, the details of which we report in \cite{Ellis:2010rw}, we obtain the factorized form in SCET,
\begin{equation}
\label{SCETcs}
\begin{split}
&\frac{d\sigma}{\prod_{i=1}^M d\tau_a^i \prod_{k=1}^N d^3\vect{P}_k} = \frac{d\sigma^{(0)}}{\prod_{k=1}^Nd^3\vect{P}_k} H(\vect{P}_1,\dots,\vect{P}_N)\!\!\prod_{j=M+1}^{N}\!\! J_{n_j,\omega_j}^{f_j}\\
&\times \prod_{i=1}^M \int\!d\tau_J^i \, d\tau_S^i \, \delta(\tau_a^i - \tau_J^i - \tau_S^i) \,J_{n_i,\omega_i}^{f_i}(\tau_J^i) S(\tau_S^1,\dots,\tau_S^M),
\end{split}
\end{equation}
where $\sigma^{(0)}$ is the Born cross section for $e^+ e^-\to N\text{ partons}$, $H = 1+\mathcal{O}(\alpha_s)$ is the hard coefficient given by the matching coefficient of the SCET $N$-jet operator, and $J$ and $S$ are jet and soft functions. The superscripts $f_{i}$ denote the color representation (corresponding to a quark, antiquark, or gluon) of the jet corresponding to the $i$th leg in the $N$-jet operator. We number the legs so that $i=1,\dots,M$ are the jets whose shapes we measure, and the remainder $j=M+1,\dots,N$ are left unmeasured. 

The quark  and gluon jet functions for jets whose shapes are measured are defined by\footnote{The normalization of \eq{quark} has been changed by a factor of $1/2$ to agree with the definition in \cite{Ellis:2010rw}, where $J^q_\omega(\tau_a) = 1 + \mathcal{O}(\as)$.}
\begin{subequations}
\label{jetfuncs}
\begin{align}
\begin{split}
\label{quark}
&J^q_{n,\omega}(\tau_J) =  \frac{1}{2\CA} \Tr\sum_{X_n}\int\frac{dn\mcdot k}{2\pi} \int d^4 x \, e^{-ik\cdot x} \frac{\bnslash}{2}  \delta_{n(\mathcal{J}(X_n)) - 1}  \\
&\quad\times\bra{0} \chi_{n,\omega}(x)\ket{X_n}\bra{X_n}\bar\chi_{n,\omega}(0)\ket{0} \delta(\tau_J - \tau_a(J(X_n))),
\end{split} \\
\label{gluon}
\begin{split}
&J^g_{n,\omega}(\tau_J) =  \frac{\omega}{2\CA \CF} \Tr\sum_{X_n}\int\frac{dn\mcdot k}{2\pi} \int d^4 x \, e^{-ik\cdot x}  \delta_{n(\mathcal{J}(X_n)) - 1}  \\
&\times\frac{1}{D-2}\bra{0} gB_{n,\omega}^{\perp\mu}(x)\ket{X_n}\bra{X_n}gB_{n,\omega\mu}^{\perp}(0)\ket{0}\delta(\tau_J - \tau_a(J(X_n))),
\end{split}
\end{align}
\end{subequations}
where the traces are over color and spinor indices, and $D$ is the number of dimensions. The sums are over states in the $n$-collinear sector. The label direction and energy $n,\omega$ are chosen to match the jet momentum $\vect{P}$. We have factored the Kronecker delta in the full cross section \eq{QCDcs} restricting the sum over states to those with $N$ jets according to the algorithm $\mathcal{J}$ into individual restrictions that there is precisely one jet in each collinear sector. The delta functions of $\tau_J$ restrict the angularity of the jet $J$ identified in the state $X_n$ by the jet algorithm. The jet functions $J_{n_j,\omega_j}^{f_j}$ for jets whose shapes are left unmeasured are given by \eq{jetfuncs} without the delta functions of $\tau_J$. 

The soft function, meanwhile, is given by matrix elements of  $N$ soft Wilson lines in each of the collinear directions $n_i$ and color representations $r_i$ of the $i$th jet. For arbitrary $N$, multiple color structures may appear, and if so there is an implicit sum over multiple hard functions $H$ and soft functions $S$ in \eq{SCETcs}. An $N$-jet soft function takes the general form
\begin{equation}
\label{softfunc}
\begin{split}
S_{N}&(\{\tau_S^i\}) = \frac{1}{\mathcal{N}} \sum_{X_s}\delta_{n(\mathcal{J}(X_s))} \prod_{i=1}^M \delta(\tau_S^i - \tau_a^{i}(X_s)) \\
& \times\bra{0} Y_{n_N}^{r_N\dag}\cdots Y_{n_1}^{r_1\dag}(0)\ket{X_s} \bra{X_s} Y_{n_1}^{r_1} \cdots Y_{n_N}^{r_N}(0)\ket{0},
\end{split}
\end{equation} 
where $\mathcal{N}$ normalizes the soft function to $\delta(\tau_a^1)\cdots\delta(\tau_a^M)$ at tree level. There is an implicit contraction of color indices which we have left unspecified. The whole soft function is color singlet.
Note that the sum over soft states is restricted so that soft particles do not create an additional jet when the jet algorithm is run on $X_s$. $\tau_a^{i}(X_s)$ is the contribution to the jet shape from soft particles which are actually in the jet $J_i$. 

The factorization of the cross section \eq{SCETcs} is valid in the following limits of QCD:
\begin{enumerate}
\item The SCET expansion parameter $\lambda$, determined either by the jet shape $\tau_a$ for measured jets or the jet radius $R$ for unmeasured jets, must be small. In other words, each jet must be \emph{well collimated}.

\item The separation between any pair of jets must be large. We will find that the natural measure for this separation is the variable $t = \tan(\psi/2)/\tan(R/2)$, where $\psi$ is the minimum angle between two jet directions. $t$ must be large, that is, jets must be \emph{well separated} in order for us to factor the $N$-jet condition in the full cross section \eq{QCDcs} into $N$ individual 1-jet conditions in each collinear sector as in \eq{jetfuncs} and a no-jet condition in the soft sector as in \eq{softfunc}. 
This approximation is inevitable because each jet function $J_i$  already approximates all radiation emitted by other jets as coming from a Wilson line $W_{n_i}$ along the exactly back-to-back direction $\bar n_i$, whereas the hard and soft functions  know the directions of all $N$ jets exactly.

\item The energy of all particles not included in a jet must be of the order of soft momenta. This is so that setting the label energy on each of the jet fields in \eq{jetfuncs} to be equal to the total jet energy is correct at leading order in $\lambda$. In particular, the energy cut parameter $\Lambda$ on energy outside of all jets is required to be soft, $\Lambda\sim \lambda^2 E_J$. 

\item Power corrections associated with the jet algorithm are small. For instance, setting the jet axis equal to the label direction $n$ is valid up to $\mathcal{O}(\lambda^2)$ corrections, which induce corrections to the jet shape $\tau_a^J$ which are subleading for $a<1$ \cite{Berger:2003iw,Bauer:2008dt,Lee:2006nr}. Similarly, assuming soft particles know only about the total collinear jet momentum by the time they are included or excluded from a jet  induces power corrections to $\tau_a^J$ that are power suppressed for sufficiently large $R$.
\end{enumerate}
We go into greater detail about these approximations in \cite{Ellis:2010rw}.

\section[\textit{Renormalization Group Evolution}]{Renormalization Group Evolution}
\label{sec:RGE}

The functions that we consider either renormalize multiplicatively or through convolutions in $\tau$. The multiplicative form of a renormalization group equation (RGE) obeyed by a function $F$ is
\begin{equation}
\label{RGEF}
\mu\frac{d}{d\mu}F(\mu) = \gamma_F(\mu)F(\mu),
\end{equation}
with the anomalous dimension of the form
\begin{equation}
\label{gammaF}
\gamma_F(\mu) = \Gamma_F[\alpha] \ln\frac{\mu^2}{\omega^2} + \gamma_F[\alpha].
\end{equation}
This RGE has the solution
\begin{equation}
F(\mu) = U_F(\mu,\mu_0)F(\mu_0),
\end{equation}
where
\begin{equation}
\label{UF}
U_F(\mu,\mu_0) = e^{K_F(\mu,\mu_0)}\left(\frac{\mu_0}{\omega}\right)^{\omega_F(\mu,\mu_0)},
\end{equation}
where we define $\omega_F,K_F$ below in \eq{omegaK}.
The convolved form of an RGE obeyed by functions $F$ that depend on the observable is
\begin{equation}
\label{RGEFtau}
\mu\frac{d}{d\mu}F(\tau;\mu) = \int d\tau' \gamma_{F}(\tau-\tau';\mu) F(\tau';\mu),
\end{equation}
where to all orders in $\alpha$ \cite{Fleming:2007xt,Grozin:1994ni}
\begin{equation}
\label{gammaFtau}
\gamma_{F}(\tau;\mu) = \left(\Gamma_F[\alpha]\ln\!\frac{\mu^2}{\omega^2} + \gamma_F[\alpha]\right)\delta(\tau) - \frac{2}{j_F}\Gamma_F[\alpha]\left[\frac{\theta(\tau)}{\tau}\right]_+ \!\! .
\end{equation}
The solution to this RGE is \cite{Fleming:2007xt,Becher:2006mr,Korchemsky:1993uz,Balzereit:1998yf,Neubert:2005nt}
\begin{equation} 
F(\tau;\mu) = \int d\tau' U_F(\tau-\tau';\mu,\mu_0)F(\tau';\mu_0),
\end{equation}
where
\begin{equation}
\label{UFtau}
U_F(\tau;\mu,\mu_0) = \frac{e^{K_F + \gamma_E\omega_F}}{\Gamma(-\omega_F)} \left(\frac{\mu_0}{\omega}\right)^{j_F\omega_F}\left[\frac{\theta(\tau)}{\tau^{1+\omega_F}}\right]_+.
\end{equation}
We note that the anomalous dimensions $\gamma_F(\mu)$ and $\gamma_F(\tau; \mu)$ in general also depend on the jet algorithm parameters $R$ and $\Lambda$ which we have made implicit.

The part of the anomalous dimensions in \eqs{gammaF}{gammaFtau} multiplying $\ln(\mu^2/\omega^2)$ is proportional, to all orders in $\alpha_s$, to the \emph{cusp anomalous dimension} $\Gamma(\alpha_s)$, given to $\mathcal{O}(\alpha_s)$ by $\Gamma(\alpha_s) = \alpha_s/\pi$.  With one-loop results for the anomalous dimensions, and using the two-loop form of the cusp anomalous dimension, the RGE solutions are accurate to NLL order.  In \eqs{UF}{UFtau}, $\omega_F,K_F$ are given by
\begin{subequations}
\label{omegaK}
\begin{align}
\label{omegaF}
\omega_F(\mu,\mu_0) &= \frac{2}{j_F}\int_{\alpha_s(\mu_0)}^{\alpha_s(\mu)}\frac{d\alpha}{\beta[\alpha]}\Gamma_F[\alpha] \\
\label{KF}
\begin{split}
K_F(\mu,\mu_0) &= \int_{\alpha_s(\mu_0)}^{\alpha_s(\mu)}\frac{d\alpha}{\beta[\alpha]}\gamma_F[\alpha] \\
&\quad + 2\int_{\alpha_s(\mu_0)}^{\alpha_s(\mu)}\frac{d\alpha}{\beta[\alpha]}\Gamma_F[\alpha] \int_{\alpha_s(\mu_0)}^{\alpha}\frac{d\alpha}{\beta[\alpha]},
\end{split}
\end{align}
\end{subequations}
where $\beta[\alpha]$ is the beta function of QCD. 
We define $j_F = 1$ for RGEs of the form \eq{gammaF}.

We will find that the hard function can be written as a sum over functions that each obey a multiplicative renormalization group equation.  The unmeasured jet function also obeys a multiplicative RGE, while the measured jet function obeys a RGE with a convolution over $\tau$.  The soft function, whose structure we will discuss in detail, can be decomposed into terms which obey multiplicative RGEs and terms which obey convolved RGEs.  

In the next section we outline the calculations necessary to obtain all the above anomalous dimensions to $\mathcal{O}(\alpha_s)$.

\section[\textit{Anomalous Dimensions and Consistency of Factorization}]{Anomalous Dimensions and Consistency of Factorization}
\label{sec:anom}

In this section we discuss the calculation of the one-loop hard, jet, and soft anomalous dimensions and the form of the anomalous dimensions in Table~\ref{table:gammas} and demonstrate that the consistency condition, \eq{consistency}, is satisfied to one-loop order, to leading order in the approximations we enumerated above. This is already an intricate test whose satisfaction turns out to be highly nontrivial. Having verified this condition, we proceed at the end of the Letter to give an application of NLL resummation of the jet shape distribution made possible by  our one-loop calculation of the anomalous dimensions. 

\subsection[\textit{Hard Function}]{Hard Function}
\label{sec:hard}

The hard function $H$ in the factorized cross section \eq{SCETcs} is given by the square of the Wilson coefficient in the matching of the $N$-parton amplitude in QCD onto an $N$-jet operator in SCET,
\begin{equation}
\bra {N} \bar q \Gamma q\ket{0} = \bra{N}C_N\mathcal{O}_N\ket{0},
\end{equation}
where the right-hand side is actually a sum over many possible $N$-jet operators built from the jet fields in \eq{jetfields} and soft Wilson lines \eq{Yndef}. The allowed basis of operators $\mathcal{O}_N$ is determined by gauge and Lorentz symmetry. If there is only one operator, the hard function is simply $H = \abs{C_N}^2$.

The one-loop anomalous dimension of the $N$-jet matching coefficient $C_N$ can be determined from calculations existing in the literature, for example, Table III of \cite{Chiu:2009mg}. For an operator with $N$ legs with color charges $\vect{T}_i$, the anomalous dimension of the matching coefficient $C_N$ is
\begin{equation}
\begin{split}
\gamma_{C_N}  (\alpha_s) = & - \sum_{i=1}^N\left[ \vect{T}_i^2 \Gamma(\alpha_s) \ln\frac{\mu}{\omega_i} + \frac{1}{2}\gamma_i(\alpha_s)\right] \\
& - \frac{1}{2}\Gamma(\alpha_s) \sum_{i\not = j}\vect{T}_i\cdot \vect{T}_j\ln\left(\frac{-n_i\cdot n_j - i0^+}{2}\right)
\end{split}
\end{equation}
where $\gamma_i$ is given to $\mathcal{O}(\alpha_s)$ for quarks and gluons by
\begin{equation}
\label{noncusp}
\gamma_q = \frac{3\alpha_sC_F}{2\pi}\,,\quad \gamma_g = \frac{\alpha_s}{\pi}\frac{11 C_A - 4 T_R n_f}{6}.
\end{equation}
The anomalous dimension of the hard function itself is then given by $\gamma_H = \gamma_{C_N} +\gamma_{C_N}^*$ and can be written as
\begin{equation}
\label{hardanomdim}
\gamma_H(\mu) = \sum_{i = 1}^N \gamma_H^i(\mu) + \gamma_H^{\text{pair}}(\mu).
\end{equation}
Because the hard function obeys a multiplicative RGE, each term in the hard function obeys a multiplicative RGE, and so each term in \eq{hardanomdim} has the form \eq{gammaF}. Each $H^i$ has $\omega = \omega_i$, while $\Gamma[\alpha] = 0$ for $H^{\text{pair}}$, as listed in Table~\ref{table:gammas}.

\subsection[\textit{Jet Functions}]{Jet Functions}

The quark and gluon jet functions are given by \eqs{quark}{gluon} and are calculated from cutting all possible diagrams at a given order in $\alpha_s$ correcting a collinear propagator  with label momentum $\omega n$.  The jet functions include phase space restrictions on the final-state particles from the cut requiring that only one jet is produced.  When we cut through a single propagator, the solitary parton in the final state is automatically in the jet, but these diagrams turn out to be scaleless and thus zero in dimensional regularization. For the cuts through loops, two collinear particles are created in the final state, and both particles are in the jet if \eq{injetconstraint} is satisfied.  If \eq{injetconstraint} is not satisfied, we require one of the particles to have energy $E < \Lambda$, so that only one jet is produced by the final state.  Additionally, for jets whose shapes are measured, we include a delta function, $\delta(\tau_J - \tau_a(J(X)))$, measuring the jet shape for the particles in the jet.  The restrictions on unmeasured jet functions are the same as the measured jets except for this delta function.

We report here the results of calculating only the infinite parts of the relevant loop graphs in dimensional regularization, in $D = 4-2\epsilon$ dimensions, in the $\overline{\text{MS}}$ scheme. We give the finite parts in \cite{Ellis:2010rw}. Our calculations give anomalous dimensions for quark and gluon jets $\gamma_J^i$ of the form \eq{gammaF} for unmeasured jets and $\gamma_{J}^k(\tau_a)$ of the form \eq{gammaFtau} for measured jets, with the values given in \tab{table:gammas}. 

In the measured jet function, we find that the zero-bin subtraction plays a key role.  The zero-bin subtraction removes doubly-counted regions of phase space from the ``naive'' contributions to the jet function \cite{Manohar:2006nz}.  For the measured jet functions, the naive contributions to the anomalous dimension only depend on $\delta(\tau_a)$ and do not contain $(1/\tau_a)_+$ distributions.  However, the zero-bin contribution to the anomalous dimension contains non-trivial $\tau_a$ dependence away from $\tau_a=0$, and it is only by performing the zero-bin subtraction that we obtain the correct running of the measured jet function.  

When the final-state particles in the jet function do not pass the cuts in \eq{injetconstraint}, only one particle is in a jet.  In this case the contribution to the jet function is power suppressed by $\cO(\Lambda/\omega)$, since a collinear parton must have $E<\Lambda$ to be outside of the jet.  This power contribution is not  power suppressed in the na\"{\i}ve contribution alone, but only after the zero-bin subtraction.  Additionally, the zero-bin removes the dependence of the measured jet function anomalous dimension on the jet algorithm parameter $R$.  For unmeasured jets, the zero-bin is a scaleless integral, and the $R$ dependence remains in the unmeasured jet function. 

Tabulating the results, we find the anomalous dimensions are 
\begin{equation}
\label{gammaJ}
\gamma_{J_i} = \Gamma(\alpha_s)\vect{T}_i^2 \ln\frac{\mu^2}{\omega_i^2\tan^2\frac{R}{2}} + \gamma_i,
\end{equation}
for unmeasured jet functions, and
\begin{equation}
\label{gammaJM}
\begin{split}
\gamma_{J_i}(\tau_a^i) &= \vect{T}_i^2\left[\Gamma(\alpha_s)\frac{2-a}{1-a}\ln\frac{\mu^2}{\omega_i^2} + \gamma_i\right]\delta(\tau_a^i) \\
&\quad - 2\Gamma(\alpha_s)\vect{T}_i^2 \frac{1}{1-a}\left[\frac{\theta(\tau_a)}{\tau_a}\right]_+
\end{split}
\end{equation}
for measured jet functions.

\begin{table}
\begin{center}
\begin{tabular}{||c | ccccc ||}
\hline\hline
 & \tabruleA $\Gamma_F[\alpha]$ & & $\gamma_F[\alpha]$ & & $j_F$  \\ \hline
\tabruleA $H^i$ & $-\Gamma \vect{T}_i^2$  && $-\gamma_i$ && 1 \\
\tabruleA $H^{\text{pair}}$ & 0  && $ - \Gamma\sum_{i\not=j}\vect{T}_i\cdot \vect{T}_j\ln\frac{ n_i\cdot n_j}{2}$ && 1 \\
\tabruleA $J^i$ & $\Gamma \vect{T}_i^2$ &&$ \gamma_i - \Gamma \vect{T}_i^2\ln\tan^2\frac{R}{2}$ && 1 \\
\tabruleA $J^k(\tau_a^k)$ &$ \Gamma \vect{T}_k^2\frac{2-a}{1-a}$ && $ \gamma_k$ && $2-a$ \\
\tabruleA $S^k(\tau_a^k)$ &$ - \Gamma\vect{T}_k^2 \frac{1}{1-a} $ && 0 && 1 \\
\tabruleA $S^i$ & 0 && $\Gamma\vect{T}_i^2 \ln\tan^2\frac{R}{2}$ && 1 \\
\tabruleA $S^{\text{pair}}$&  0 && $\Gamma\sum_{i\not=j}\vect{T}_i\cdot\vect{T}_j\ln\frac{n_i\cdot n_j}{2}$ && 1 \\
\hline\hline
\end{tabular}
\end{center}
\caption[Anomalous dimensions of hard, jet, and soft functions]{Anomalous dimensions of hard, jet, and soft functions. The cusp parts $\Gamma_F$ and non-cusp parts $\gamma_F$ of the anomalous dimensions for hard, unmeasured jet, measured jet, and soft functions are given, along with the constant $j_F$ appearing in \eqs{gammaFtau}{omegaF}. $\Gamma$ is the cusp anomalous dimension, given to one-loop by $\Gamma = \alpha_s/\pi$. The pieces $\gamma_{i}$ for quarks and gluons are given by \eq{noncusp}. The three rows for the soft anomalous dimensions are organized to correspond to the three groups of evolution factors given in \eq{softsolution} and are given in the limit $1/t^2\to 0$.}
\label{table:gammas}
\end{table}

\subsection[\textit{Soft Function}]{Soft Function}

\begin{figure}[b]
\begin{center}
\resizebox{\columnwidth}{!}{\includegraphics{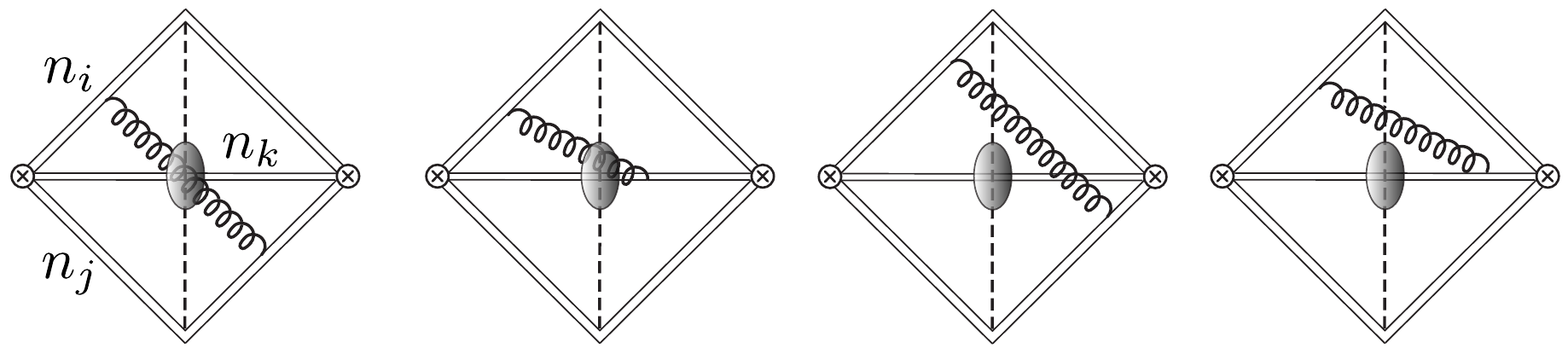}}
\end{center}
\vspace{-2em}
\caption[Soft function diagrams]{Soft function diagrams. A gluon exchanged between jets $i$ and $j$ crosses the cut which imposes phase space restrictions due to the jet algorithm. The blob represents the jet in direction $k$, which the gluon may enter or not.}
\label{fig:soft}
\end{figure}

The soft function in an $N$-jet cross section is given by  \eq{softfunc}, containing matrix elements of $N$ soft Wilson lines in the $N$ jet directions, with each Wilson line in the color representation of the corresponding jet.  At $\mathcal{O}(\alpha_s)$, this soft function is given by a sum over cut diagrams represented in \fig{fig:soft}.  The blob represents the jet in direction $n_k$, and we leave implicit the phase space cuts needed for each diagram.  We use Feynman gauge, in which each diagram is proportional to $n_i\cdot n_j$. (Note this allows us to drop graphs with $i=j$ or $i=k$ since $n_i^2 = 0$.)  

To calculate the soft function, we must implement phase space cuts on the soft gluon in the final state requiring that it either be in a jet or not produce a new jet (i.e., it has energy less than $\Lambda$).  The soft function is a sum over contributions from all pairs of directions $i$ and $j$ that exchange the soft gluon, and we calculate the total contribution with $i$ and $j$ fixed before summing over directions.  A natural way to organize the phase space of the soft gluon in the final state is as follows:
\begin{enumerate}[(1)]
\item The gluon enters a measured jet and contributes to $\tau_a^{k}(X_s)$. 
\item The gluon enters an unmeasured jet and has any energy. 
\item The gluon is not in any jet and has energy $E<\Lambda$. 
\end{enumerate}
We name contribution (1) $S_{ij}^\text{meas}(\tau_a^k)$, where the subscript $ij$ denotes that the gluon goes from $i$ to $j$. Regions (2) and (3) do not contribute to the angularity of any jet and just give an additive contribution $S_{ij}^{\text{non-meas}}$ to the coefficient of $\delta(\tau_a^1)\cdots \delta(\tau_a^M)$ in the full soft function $S(\tau_a^1,\dots,\tau_a^M)$. Contribution (3), however, is very awkward to calculate, as we must integrate over a phase space with many ``holes'' (corresponding to the jets) removed, resembling Swiss cheese. It is easier to reorganize contributions (2) and (3) into the following form:
\begin{enumerate}
\item[(A)] $S_{ij}^{\text{incl}}$: the gluon is anywhere with energy $E<\Lambda$.
\item[(B)] $S_{ij}^{{k}}$: the gluon is in jet $k$ with energy $E>\Lambda$.
\item[(C)] $\bar S_{ij}^{{k}}$: the gluon is in jet $k$ with energy $E<\Lambda$. 
\end{enumerate}
Then, the unmeasured soft gluon contribution $S_{ij}^{\text{unmeas}}$ (the sum of (2) and (3) in the original list) is given by the combination
\begin{equation}
\label{Snonmeas}
S_{ij}^{\text{unmeas}} = S_{ij}^{\text{\incl}} + \sum_{k=M+1}^N S_{ij}^k - \sum_{k = 1}^M \bar S_{ij}^k .
\end{equation}
In the first term, coming from region (A), we filled in the holes in the Swiss cheese-like region (3) in the original list, allowing the soft gluon to go anywhere with energy $E<\Lambda$. We compensated by adding the second term given by region (B) containing gluons  with energy $E>\Lambda$ inside unmeasured jets (part of the original region (2)) and subtracting the third term from region (C), removing gluons with $E<\Lambda$ inside measured jets, which are already correctly accounted for in $S_{ij}^{\text{meas}}(\tau_a^k)$.

The total soft function at $\mathcal{O}(\alpha_s)$ is then given by
\begin{equation}
\label{Stotal}
\begin{split}
S(\tau_a^1,\dots,\tau_a^M) =  \sum_{i\not =j}&\left[ \sum_{k=1}^M S_{ij}^{\text{meas}}(\tau_a^k)\prod_{\substack{l=1 \\ l\not = k}}^M\delta(\tau_a^l) \right. \\
&\quad \left. + S_{ij}^{\text{unmeas}}\prod_{l=1}^M\delta(\tau_a^l) \right] .
\end{split}
\end{equation}
Note that the second line is independent of the jet shape. This contribution is universal and will appear in any $N$-jet cross section in which some of the jets defined by a particular jet algorithm are not measured.

The contributions of the measured jet piece $S_{ij}^{\text{meas}}(\tau_a^k)$ to the anomalous dimension of the soft function are given in \tab{table:soft} separately in the cases that $k=i$ or $j$ and $k\not = i,j$.  These contributions are given by the form \eq{gammaFtau}, with the values given in \tab{table:soft}.
The results  are given in terms of the distance measure $t_{ij}=\tan(\psi_{ij}/2)/\tan(R/2)$ between jets of size $R$ separated by an angle $\psi_{ij}$,
and the angle $\beta_{ij}$ between the $ik$ and $jk$ planes. For well-separated jets, the contributions to the non-cusp part of the anomalous dimension are suppressed by $1/t^2$. 

The ``inclusive'' contribution $S_{ij}^{\text{incl}}$ for a soft gluon going anywhere with energy $E<\Lambda$ contributes a term to the soft anomalous dimension given by the general form \eq{gammaF}, with values  given in \tab{table:soft}.

Finally, for the contributions of soft gluons entering jets with $E>\Lambda$  or $E<\Lambda$ in (B) and (C) in the list above, we can combine the last two terms  in \eq{Snonmeas} using the following observation. The sum $S_{ij}^k + \bar S_{ij}^k$ is the contribution of a soft gluon entering jet $k$ with any energy. The phase space integral for this contribution contains a scaleless integral (of energy from 0 to $\infty$), and so this sum is zero in pure dimensional regularization. Thus we can set $\bar S_{ij}^k = -S_{ij}^k$, and the last two terms in \eq{Snonmeas} add up to the contribution of a soft gluon entering \emph{any} jet with energy $E>\Lambda$. 
These contributions can again be split up into those with $k=i$ or $j$ and $k\not=i,j$. They contribute parts to the soft anomalous dimension falling into the form \eq{gammaF}, with values in \tab{table:soft}. The non-cusp pieces are again suppressed by $1/t^2$ for well-separated jets.  

Using the contributions described above, we sum over directions $i$ and $j$ and obtain the anomalous dimensions for $S^{\text{meas}}(\tau_a^k)$ and $S^{\text{unmeas}}$, which we record in \tab{table:soft}.

The soft function obeys the renormalization group equation
\begin{equation}
\label{RGES}
\begin{split}
\mu\frac{d}{d\mu}S(\tau_1,\dots,\tau_M;\mu) = & \int d\tau_1'\cdots d\tau_M'  S(\tau_1',\dots,\tau_M';\mu) \\
&\times \gamma_S(\tau_1-\tau_1',\dots,\tau_M-\tau_M';\mu).
\end{split}
,\end{equation}
Because the soft function at $\mathcal{O}(\alpha_s)$ in \eq{Stotal} is a sum of terms that depend non-trivially on at most one jet shape, the anomalous dimension can be decomposed as
\begin{equation}
\label{Sanomdimparts}
\begin{split}
\gamma_S(\tau_1,\dots,\tau_M;\mu) &= \gamma_S^{\text{unmeas}}(\mu)\,\delta(\tau_1)\cdots\delta(\tau_M) \\
& \quad + \sum_{k=1}^M\gamma_S^{\text{meas}}(\tau_k;\mu)\prod_{\substack{j=1 \\ j\not=k}}^M \delta(\tau_j) ,
\end{split}
\end{equation}
The non-cusp parts of the anomalous dimension of $S^{\text{meas}}$ and $S^{\text{unmeas}}$ share the same dependence on $\tau$, and therefore we are free to shift non-cusp terms freely between anomalous dimensions.  While this does not change the physics, it allows us to organize the anomalous dimensions to match the contributions in \tab{table:gammas}, which we find more convenient for assembling the solution to the soft RGE \eq{RGES}.  By making the non-cusp part of $S^{\text{meas}}(\tau_a^k)$ zero, we find that the shifted $S^{\text{meas}}(\tau_a^k)$ is equal to $S^k(\tau_a^k)$ from \tab{table:gammas}, and that the shifted $S^{\text{unmeas}}$ is equal to $S^{\text{pair}} + \sum_i S^i$.

\begin{table}
\begin{center}
\begin{small}
\begin{tabular}{||c | cc @{}c ||}
\hline\hline
 & \tabruleA  $\Gamma_F[\alpha]$ &  $\gamma_F[\alpha]$ &  \\ 
 \hline
\tabruleB   $S_{ij}^{\text{meas}}(\tau_a^i)$ & $\frac{1}{2}\Gamma \vect{T}_i\cdot\vect{T}_j\frac{1}{1-a}$  & $\frac{1}{2}\Gamma\vect{T}_i\cdot\vect{T}_j \ln\frac{t_{ij}^2\tan^2(R/2)}{t_{ij}^2-1} $  &\\
\tabruleB  $S_{ij}^{\text{meas}}(\tau_a^k)$ & 0  & $\frac{1}{2}\Gamma\vect{T}_i\cdot\vect{T}_j \ln\frac{t_{ik}^2t_{jk}^2- 2t_{ik}t_{jk}\cos\beta_{ij}+1}{(t_{ik}^2 - 1)(t_{jk}^2 - 1)}$ &  \\
\tabruleB $S_{ij}^{\text{incl}}$ & $-\Gamma\vect{T}_i\cdot\vect{T}_j$  & $\Gamma\vect{T}_i\cdot\vect{T}_j \,\Big( \ln(n_i\mcdot n_j/2) + \ln \frac{\omega_i^2}{4 \Lambda^2}\Big)$ & \\
\tabruleB $S_{ij}^i$ & $\frac{1}{2}\Gamma\vect{T}_i\cdot\vect{T}_j$  & $-\frac{1}{2}\Gamma\vect{T}_i\cdot\vect{T}_j \, \Big( \ln\frac{t_{ij}^2\tan^2(R/2)}{t_{ij}^2-1} +  \ln \frac{\omega_i^2}{4 \Lambda^2}\Big)$ & \\
\rule{-2pt}{3ex} \rule[-2ex]{0pt}{0pt} $S_{ij}^k$ & 0   & $-\frac{1}{2}\Gamma\vect{T}_i\cdot\vect{T}_j \ln\frac{t_{ik}^2t_{jk}^2- 2t_{ik}t_{jk}\cos\beta_{ij}+1}{(t_{ik}^2 - 1)(t_{jk}^2 - 1)}$   &\\ 
\hline
\rule{-2pt}{3ex} \rule[-1.5ex]{0pt}{0pt} $S^{\text{meas}}(\tau_a^k)$ & $-\Gamma\frac{1}{1-a}\vect{T}_k^2$ & $-\Gamma \vect{T}_k^2\ln\tan^2\frac{R}{2} + \mathcal{O}(1/t^2)$  &\\
\tabruleB $S^{\text{unmeas}}$ & 0 & $\Gamma \sum_{i\not =j}\vect{T}_i\mcdot\vect{T}_j \ln(n_i\cdot n_j/2) $  &\\
\tabruleB & & $+ \Gamma \sum_{i=1}^N \vect{T}_i^2\ln\tan^2(R/2)+ \mathcal{O}(1/t^2)$ & \\
\hline\hline
\end{tabular}
\end{small}
\end{center}
\caption[Soft anomalous dimensions]{Soft anomalous dimensions. Contributions to the anomalous dimension of the soft function are given for soft gluons emitted by jet $i$ or $j$ and entering jet $k$ (with $k=i$ or $j$ in the first row and $k\not= i,j$ in the second) and being measured with angularity $\tau_a^k$; soft gluons emitted by jet $i$ or $j$ in any direction with energy $E<\Lambda$ in the third row; and soft gluons emitted by jet $i$ or $j$ and entering jet $k$ and angularity unmeasured in the fourth ($k=i$ or $j$) and fifth ($k\not=i,j$) rows. In the second-to-last row we summed the first two rows over all pairs of jets $i,j$ to obtain the measured contribution for a specific $\tau_a^k$, and in the last row, we summed all unmeasured soft gluon contributions. In the last two rows, we have taken the large $t$ limit. $j_F = 1$ in all cases.}
\label{table:soft}
\end{table}

Finally, we can give the soft function anomalous dimension.  Omitting terms which are suppressed by $\mathcal{O}(1/t^2)$, the soft function anomalous dimension is
\begin{equation}
\label{gammaS}
\begin{split}
\gamma_S &(\tau_a^1,\dots,\tau_a^M)  = \Gamma(\alpha_s)\biggl[ - \frac{1}{1-a}\sum_{k=1}^M \vect{T}_k^2\ln\frac{\mu^2}{\omega_k^2} \\
& + \sum_{i=M+1}^N \vect{T}_i^2\ln\tan^2\frac{R}{2} + \sum_{i\not = j}\vect{T}_i\mcdot\vect{T}_j\ln\frac{n_i\cdot n_j}{2}\biggr] \\
&\qquad\times \delta(\tau_a^1)\cdots \delta(\tau_a^M) \\
& + 2\Gamma(\alpha_s)\frac{1}{1-a}\sum_{k=1}^M \vect{T}_k^2 \left[\frac{\theta(\tau_a^k)}{\tau_a^k}\right]_+\prod_{\substack{j=1 \\ j\not=k}}^M\delta(\tau_a^j),
\end{split}
\end{equation}

The solution of the RGE is
\begin{equation}
\label{softsolution}
\begin{split}
&S(\tau_1,\dots,\tau_M;\mu) = \int\!d\tau_1'\cdots d\tau_M' \,S(\tau_1',\dots,\tau_M';\mu_0) \\
&\quad\times   U_S^{\text{pair}}(\mu,\mu_0)\prod_{k=1}^M U_S^k(\tau_k-\tau_k';\mu,\mu_0) \! \prod_{i=M+1}^N\! U_S^i(\mu,\mu_0) ,
\end{split}
\end{equation}
where $U_S^k(\tau_k)$ is an evolution kernel of a convoluted RGE  and is of the form in \eq{UFtau}, and $U_S^i$ and $U_S^{\text{pair}}$ are evolution kernels of multiplicative RGEs and are of the form in \eq{UF}.  The evolution kernels $U_S^k(\tau_k)$, $U_S^i$, and $U_S^{\text{pair}}$ correspond to the soft anomalous dimensions from $S^k(\tau_a^k)$, $S^i$, and $S^{\text{pair}}$ in \tab{table:gammas}.

\subsection[\textit{Consistency of Factorization}]{Consistency of Factorization}
\label{sec:consistency}

Adding together all jet and soft anomalous dimensions, we find, miraculously, the $R$ dependence cancels between the unmeasured jet anomalous dimension \eq{gammaJ} and sum over unmeasured jets in the soft function \eq{gammaS}, and the $\tau_a \not = 0$ dependence cancels between the measured jet anomalous dimension \eq{gammaJM} and the sum over measured jets in the soft function. The remaining pieces precisely match the hard anomalous dimension $\gamma_H$ given in \sec{sec:hard} such that the consistency condition \eq{consistency} is satisfied. Note, however, that satisfying \eq{consistency} exactly required that we drop corrections of $\mathcal{O}(1/t^2)$ in the soft function. Requiring consistency of the anomalous dimensions at one loop has provided the measure $t^2\gg1$  to quantify the condition we used in justifying the factorization theorem in \sec{sec:fact} that jets be ``well separated''.

\section[\textit{Application: Jet Shapes in \texorpdfstring{$e^+ e^-\to$}{e+e- to} 3 Jets}]{Application: Jet Shapes in \texorpdfstring{$e^+ e^-\to$}{e+e- to} 3 Jets}
\label{sec:resum}

As an example of using the above results to calculate a jet observable in an exclusive multijet final state, we give the resummed angularity jet shape distribution for a single measured quark or gluon jet in a three-jet final state in $e^+ e^-$ annihilation.  The techniques to derive and solve the RGEs to resum logarithms in jet shape distributions in SCET  are essentially identical to those for event shape distributions as performed in 
 \cite{Fleming:2007xt,Hornig:2009vb,Schwartz:2007ib,Becher:2008cf}.
  
We assemble the appropriate RG-evolved hard function, measured  jet function, two unmeasured jet functions, and  soft function given in Secs.~\ref{sec:RGE} and \ref{sec:anom}. Evolving these from their tree-level values at initial scales $\mu_H,\mu_J^i,\mu_S$ to the scale $\mu$ with NLL running, we obtain the distribution in the shape $\tau_a$ of jet 1 with jets 2, 3 unmeasured. Written as the derivative of the radiator,
\begin{equation}
\label{3jetshape}
\begin{split}
&\frac{1}{\sigma^{(0)}_{\vect{P}_1\vect{P}_2\vect{P_3}}}\frac{d\sigma_{\vect{P}_1\vect{P}_2\vect{P_3}}}{d\tau_a} = \frac{dR(\tau_a)}{d\tau_a} \\
&= -\frac{d}{d\tau_a}\Biggl\{ \frac{\exp\bigl[\mathcal{K}(\mu;\mu_{H},\mu_J^{1,2,3},\mu_S) + \gamma_E\bigl( \omega_J^1(\mu,\mu_J^1)  + \omega_S^1(\mu,\mu_S)\bigr)\bigr]}{\Gamma(1-\omega_J^1(\mu,\mu_J^1)  - \omega_S^1(\mu,\mu_S)\bigr)}  \\
&\qquad\times\left(\frac{\mu_H}{\bar\omega_H}\right)^{\omega_H(\mu,\mu_H)}  \left(\frac{\mu_J^1}{\omega_1}\right)^{(2-a)\omega_J^1(\mu,\mu_J^1)}\left(\frac{\mu_J^2}{\omega_2}\right)^{\omega_J^2(\mu,\mu_J^2)} \left(\frac{\mu_J^3}{\omega_3}\right)^{\omega_J^3(\mu,\mu_J^3)} \\
&\qquad\times \left(\frac{\mu_S}{\omega_1}\right)^{\omega_S^1(\mu,\mu_S)}\left[\frac{1}{\tau_a^{\omega_J^1(\mu,\mu_J^1)+\omega_S^1(\mu,\mu_S)}}\right]_+ \Biggr\},
\end{split}
\end{equation}
where $\sigma_{\vect{P}_1\vect{P}_2\vect{P_3}}$ is the cross section differential in the three jet momenta $\vect{P}_i = \omega_i \vect{n}_i$, the effective hard scale $\bar\omega_H = (\omega_1^{\vect{T}_1^2}\omega_2^{\vect{T}_2^2}\omega_3^{\vect{T}_3^2})^{\frac{1}{\vect{T}^2}}$ where $\vect{T}^2 = \vect{T}_1^2 +  \vect{T}_2^2 +  \vect{T}_3^2$, and $\mathcal{K}$ is the sum of the hard, jet, and soft evolution factors,
\begin{equation}
\begin{split}
\mathcal{K}&= K_H(\mu,\mu_H) + \sum_{i=1}^3 [K_J^i(\mu,\mu_J^i) + K_S^i(\mu,\mu_S)] + K_S^{\text{pair}}(\mu,\mu_S).
\end{split}
\end{equation}
Inspection of \eq{3jetshape} suggests the reasonable choices for initial scales to minimize large logarithms,\footnote{There are also phase-space logarithms $\ln(\mu_S/\Lambda)$ in the finite part of the soft function \cite{Ellis:2010rw} which are not resummed by the choices \eq{scalechoices}. These logarithms can be minimized by choosing $\Lambda\sim\omega_1\tau_a$ or, when these scales are disparate, by performing a further factorization of the soft function as we explain in  \cite{Ellis:2010rw}.}
\begin{equation}
\label{scalechoices}
\mu_H = \bar\omega_H , \ \mu_J^1 = \omega_1\tau_a^{1/(2-a)} ,\  \mu_J^{2,3} = \omega_{2,3}\tan\frac{R}{2} , \ \mu_S = \omega_1\tau_a.
\end{equation}
For the unmeasured jet scales $\mu_{J}^{2,3}$ we kept in mind the factor of $\ln\tan^2\frac{R}{2}$ present in $K_J^2$ (see \tab{table:gammas}).  To obtain the shape of a quark or gluon jet from \eq{3jetshape} we designate jet 1 as either quark or gluon and plug in the appropriate color factors and anomalous dimensions from \tab{table:gammas} into $\omega_F$ and $K_F$ appearing in \eq{3jetshape}. We report on a more detailed phenomenological study of these jet shapes in \cite{Ellis:2010rw} and their application to the discrimination of quark vs. gluon jets  in future work.

\section[\textit{Summary}]{Summary}

We have demonstrated the intricate fashion in which the factorized cross section to produce exclusive $N$-jet final states when $M \le N$ are measured with a jet observable remains consistent for NLL running. We identified sources of power corrections to this factorization theorem and the consistency condition. Up to these corrections, the factorization theorem remains consistent independently of the number of measured and unmeasured jets and number of quark and gluon jets.

One novel power correction that explicitly manifested itself in our calculation is in the separation parameter $t$. Since $1/t$ is identically zero for all jet sizes when jets are back-to-back, this parameter has not been identified in the literature before.

We find that, when a jet measurement is performed, the NLL resummed result has no dependence on the jet algorithm across the algorithms we considered (the Snowmass and SISCone cone algorithms and the inclusive $\kt$, anti-$\kt$, and the Cambridge-Aachen $\kt$-type algorithms). In addition, for unmeasured jets the dependence on the jet algorithm parameter $R$ (or $D$) is universal across these algorithms at NLL.

Jet shapes such as angularities can be used to describe the substructure of a jet, and can be used, for instance, to distinguish quark jets from gluon jets. In a future publication we will develop and describe a strategy to do so. We presented our calculations in such a way that  allows for straightforward adaptation to other measurements as well, as we separated those parts of the jet and soft function that depend only on the jet algorithm and not the choice of jet observable. In addition, the ideas we discussed such as the power corrections that arise in the factorization formula and the method of calculating the soft and jet functions, will carry over to a calculation involving jet algorithms at hadron colliders, essentially amounting to having algorithm parameters that are invariant under boosts along the beam axis.

\graphicspath{{chapter4/graphics/}}

\chapter{Jet Substructure, Theory and Practice}
\label{sec:sub}

\begin{figure}[htbp]
\begin{center}
\includegraphics[width = \columnwidth]{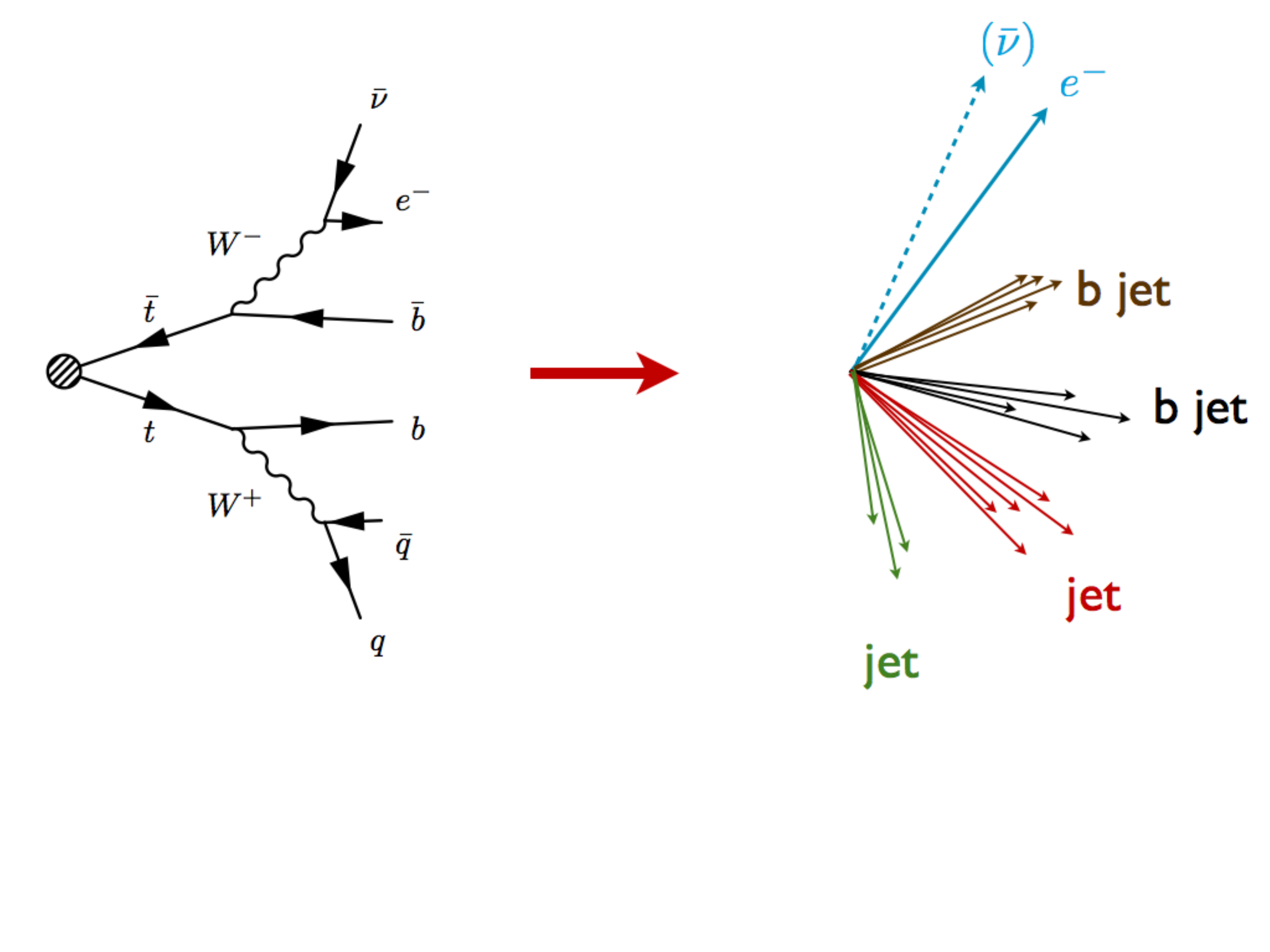}

\end{center}
\caption[Parton-level top-antitop quark event]{A parton-level description of the production of a top-antitop pair of quarks.  Each top quark decays to a $W$ boson and a bottom quark.  The $W$ can then decay either to a pair of quarks or to a charged lepton-neutrino pair.  In this case one $W$ decays leptonically, and one decays to quarks.  The result is described as an event with an electron, missing energy (from the invisible neutrino), and ``four jets''.}
\label{fig:ttbar_partons}
\end{figure}

A jet traditionally has been thought of as a proxy for a high-energy parton, e.g., a quark or gluon produced in a high-energy proton collision.  As an example of this approach, consider a measurement of the top quark mass at the Tevatron.  At the parton level, the production of a top-antitop pair looks like Fig.~\ref{fig:ttbar_partons}.  Each top quark decays to a $W$ boson and a bottom quark.  The $W$ can then decay either to a pair of quarks or to a charged lepton-neutrino pair.  In this case one $W$ decays leptonically, and one decays to quarks.  At this level of description, the outgoing particles include two bottom quarks and two other quarks (a $u$ and a $\bar d$, say).  These quarks will shower and hadronize, leading to jets.  A reconstruction analysis forms jets and matches jets to partons.  To the extent that the showers from each quark are independent and well separated, the total four-momenta of these jets will correspond to the four-momenta of the partons.

Consider, however, the case where the top quarks are produced with energies much larger than their mass.  They will be highly boosted, and their decay products will move closer together in the lab frame.  As the angular distance between partons becomes comparable to the characteristic size of their shower, they will not  in general appear as distinct jets.  A hadronic top quark decay might appear as two or even one jet instead of three.

When a top quark decay can be modeled as producing three jets, we can search for three jets, ``assign them'' to the partons of the decay, and then proceed with the analysis as if we are talking about partons instead of jets.  All of the subtleties of the QCD shower, hadronization, etc.~are hidden in the jet-to-parton matching step.  But when a top quark decay appears as a single jet, simply searching for one jet and calling it a ``top quark'' --- the same way we assign a light jet to a light quark --- throws away information.  In addition to the four-momentum of the top quark, we also have information about its decay, for example that a real $W$ boson should be present.  To search for a top quark jet, then, we should use our knowledge of the top quark's decay to look at the \emph{substructure} of the jet we think may be a top quark.

In principle, any heavy particle that decays to light quarks or gluons can be sufficiently boosted to be observed as a single jet.  To identify these decays, we should look for jets with the appropriate substructure.  The most important background to jets from heavy particle decays will be pure QCD jets.  Although QCD jets tend to be light, the tail of their mass distribution combined with their enormous production cross section mean that they will be a background to essentially any jet signature.  Separating heavy particle jets from this background will require a thorough understanding of the substructure we expect from both types of jets.  In the next section, we will take some first steps in this direction by working out parton-level predictions for the substructure of jets arising from pure QCD as well as the decays of heavy particles.  In subsequent sections, we will consider how showering, jet reconstruction, and splash-in modify these predictions and constrain our ability to distinguish different types of jets.

\section[\textit{Parton-level predictions}]{Parton-level predictions}
\label{sec:sub:parton}


Understanding the detailed substructure of jets presents an interesting challenge.\footnote{This section, with small modifications, is taken from Sections III and IV of \cite{Pruning2}.}  QCD jets are typically characterized by the soft and collinear kinematic regimes that dominate their evolution, but QCD populates the entire phase space of allowed kinematics.  Due to its immense cross section relative to other processes, small effects in QCD can produce event rates that still dominate other signals, even after cuts. Furthermore, the full kinematic distributions in QCD jet substructure currently can only be approximately calculated, so we will focus on understanding the key features of jets and the systematic effects that arise from the algorithms that define them. Note that even when an on-shell heavy particle is present in a jet, the corresponding kinematic decay(s) will contribute to only a few of the branchings within the jet. QCD will still be responsible for bulk of the complexity in the jet substructure, which is produced as the colored partons shower and hadronize, leading to the high multiplicity of color singlet particles observed in the detector.

It is a complex question to ask whether the jet substructure is accurately reconstructing the parton shower, and somewhat misguided, as the parton shower represents colored particles while the experimental algorithm only deals with color singlets. A more sensible question, and an answerable one, is to ask whether the algorithm is faithful to the dynamics of the parton shower. This is the basis of the metrics of the $\kt$ and CA recombination algorithms --- the ordering of recombinations captures the dominant kinematic features of branchings within the shower. In particular, the cross section for an extra real emission in the parton shower contains both a soft ($z$) and a collinear ($\Delta R$) singularity:
\beq
d\sigma_{n+1}\sim d\sigma_{n}\frac{dz}{z}\frac{d\Delta R}{\Delta R}.
\label{eq:socol}
\eeq
While these singularities are regulated (in perturbation theory) by virtual corrections, the enhancement remains, and we expect emissions in the QCD parton shower to be dominantly soft and/or collinear.  Due to their different metrics, the $\kt$ and CA algorithms will recombine these emissions differently, producing distinct substructure.  In the rest of this section, we will consider some generic features of jets and jet substructure.  We will elaborate on this discussion in following subsections, where we will contrast the features of jets arising from heavy particle decays with those from pure QCD showering.

\subsection[\textit{A simple model for QCD substructure}]{A simple model for QCD substructure}
\label{sec:sub:parton:qcd}

\begin{figure}[htbp]
\begin{center}
\includegraphics[width = .5\columnwidth]{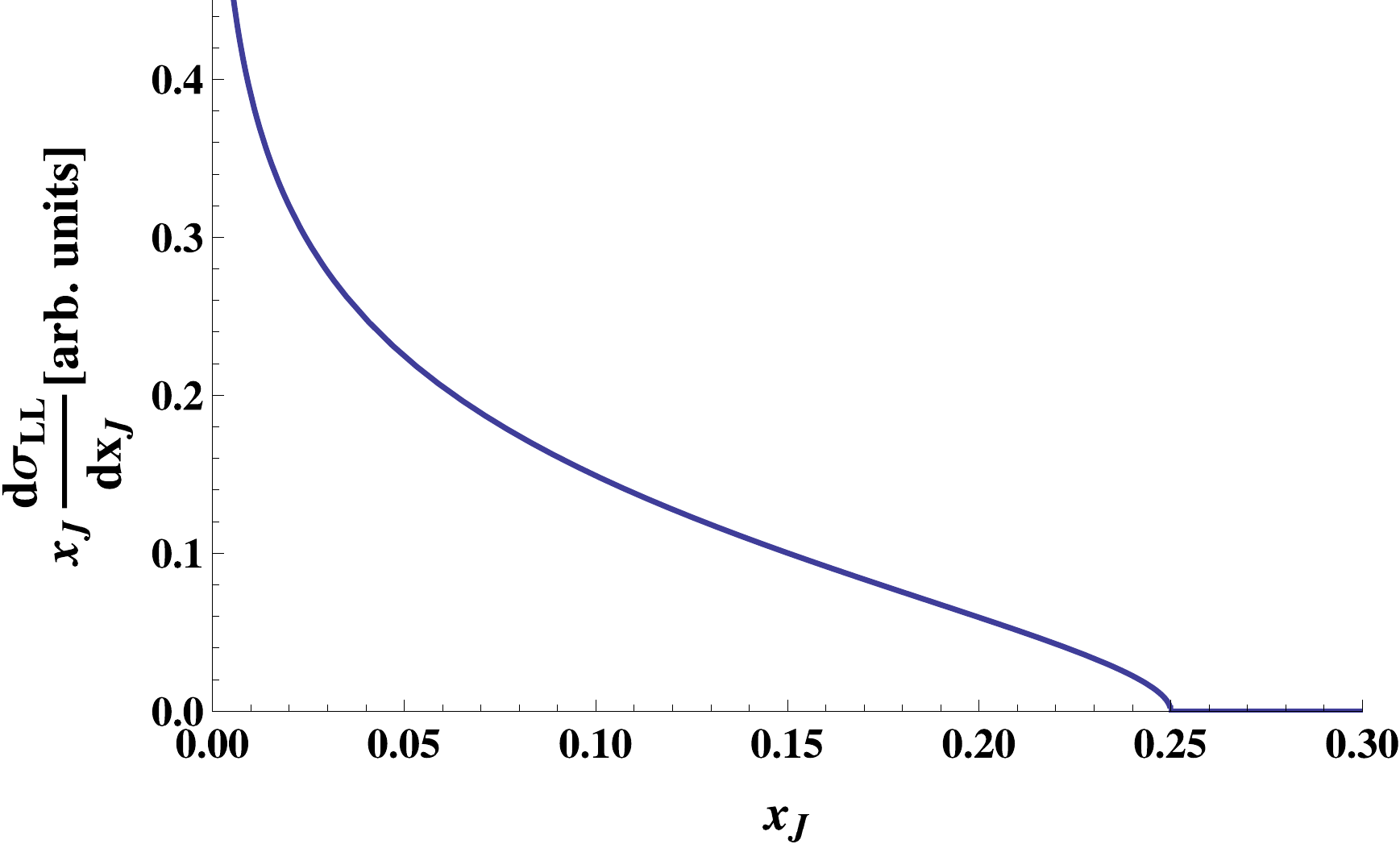}
\end{center}
\caption[Distribution in $x_J$ for a simple LL toy model]{Distribution in $x_J$ for a simple LL toy model with D = 1.0.}
\label{fig:LLmjet}
\end{figure}

To establish an intuitive level of understanding of jet substructure in QCD we consider a toy model description of jets in terms of a single branching and the kinematic variables $x_J$, $z$, and $\Delta R_{12}$ (introduced in \sec{sec:qcd:jets:algorithm}).  We take the jet to have a fixed $p_{T_J}$.  We combine the leading-logarithmic dynamics of of Eq.~(\ref{eq:socol}) with the approximate expression for the jet mass in Eq.~(\ref{eq:simplejmass}), and we label this combined approximation as the ``LL'' approximation.  Recall that this approximation for the jet mass is useful for small subjet masses and small opening angles.  From Section \ref{sec:qcd:jets:algorithm}, recall that fixing $x_J$ provides lower bounds on both $z$ and $\Delta R_{12}$ and ensures finite results for the LL approximation. This approach leads to the following simple form for the $x_J$ distribution,
\begin{eqnarray}
&&\frac{1}{\sigma}\frac{d\sigma_{LL}}{d(m_J^2/p_{T_J}^2)}\equiv\frac{1}{\sigma}
\frac{d\sigma_{LL}}{d x_J}\nonumber\\
&&\sim \int^{1/2}_0 \int^{D}_0 \frac{dz}{z}\frac{d\Delta R_{12}}{\Delta
R_{12}}\delta(x_J-z(1-z)\Delta R_{12}^2)\nonumber\\
&&= \frac{-\ln{\left(1-\sqrt{1-{4 x_J}/{D^2}}\right)}}{2x_J} \Theta\left[D^2/4 -
x_J\right] .
\label{eq:LL1}
\end{eqnarray}
Note we are integrating over the phase space of Fig.~\ref{fig:zRcontours1}, treating it as one-dimensional.  The resulting distribution is exhibited in Fig.~\ref{fig:LLmjet} for $D = 1.0$ where we have multiplied by a factor of $x_J$ to remove the explicit pole.  We observe both the cutoff at $x_J = D^2/4$ arising from the kinematics discussed in Section \ref{sec:qcd:jets:algorithm} and the $-\ln(x_J)/x_J$ small-$x_J$ behavior arising from the singular soft/collinear dynamics.  Even if the infrared singularity is regulated by virtual emissions and the distribution is resummed, we still expect QCD jet mass distributions (with fixed $p_{T_{J}}$) to be peaked at small mass values and be rapidly cutoff for $m_J > p_{T_{J}} D/2$.

\begin{figure}[htbp]
\begin{center}
\includegraphics[width = .5\columnwidth]{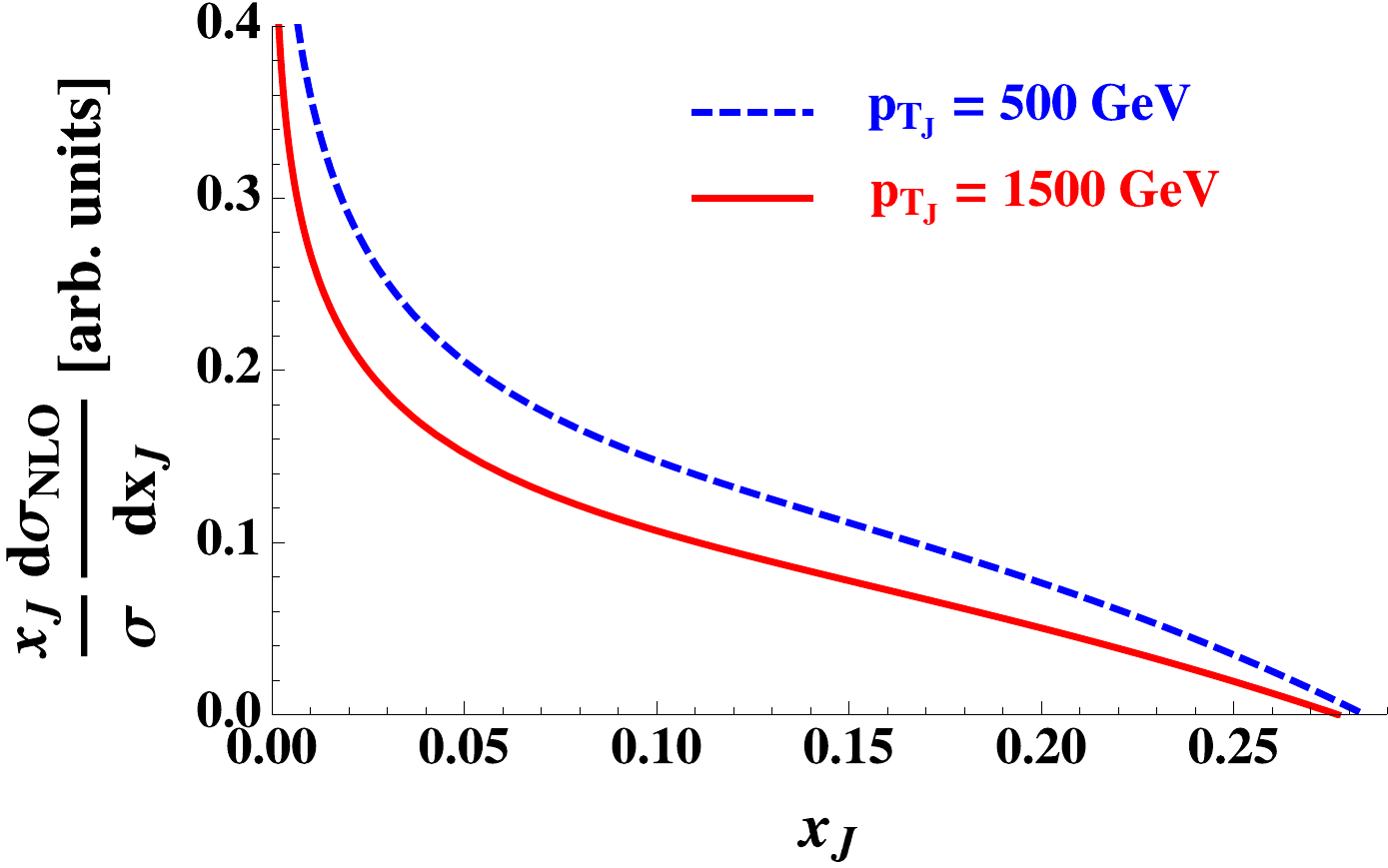}
\end{center}
\caption[NLO distribution in $x_J$ for $\kt$-style QCD jets]{NLO distribution in $x_J$ for $\kt$-style QCD jets with D = 1.0, $\sqrt{s} = 14$ TeV, and two values of $p_{T_{J}}$.}
\label{fig:NLOjetmass}
\end{figure}

We can improve this approximation somewhat by using the more quantitative perturbative analysis described in \cite{FamousJetReview}. In perturbation theory jet masses appear at next-to-leading order (NLO) in the overall jet process where two (massless) partons can be present in a single jet.  Strictly, the jet mass is then being evaluated at leading order (i.e., the jet mass vanishes with only one parton in a jet) and one would prefer a NNLO result to understand scale dependence (we take $\mu=p_{T_J}/2$).  Here we will simply use the available NLO tools \cite{Kunszt:92.1}. This approach leads to the very similar $x_J$ distribution displayed in Fig.~\ref{fig:NLOjetmass}, plotted for two values of $p_{T_J}$ (at the LHC, with $\sqrt{s} = 14$ TeV).
We are correctly including the full NLO matrix element (not simply the singular parts), the full kinematics of the jet mass (not just the small-angle approximation) and the effects of the parton distribution functions.  In this case the distribution is normalized by dividing by the Born jet cross section.  Again we see the dominant impact of the soft/collinear singularities for small jet masses.  Note also that there is little residual dependence on the value of the jet momentum and that again the distribution essentially vanishes for $x_{J} \gtrsim 0.25$, $m_{J}/p_{T_J} \gtrsim 0.5 = D/2$.  The average jet mass suggested by these results is $\langle m_{J}/p_{T_J} \rangle \approx 0.2 D$.  Because the jet only contains two partons at NLO, we are still ignoring the effects of the nonzero subjet masses and the effects of the ordering of mergings imposed by the algorithm itself.  For example, at this order there is no difference between the CA and $\kt$ algorithms.

\begin{figure}[htbp]
\begin{center}
\includegraphics[width = .5\columnwidth]{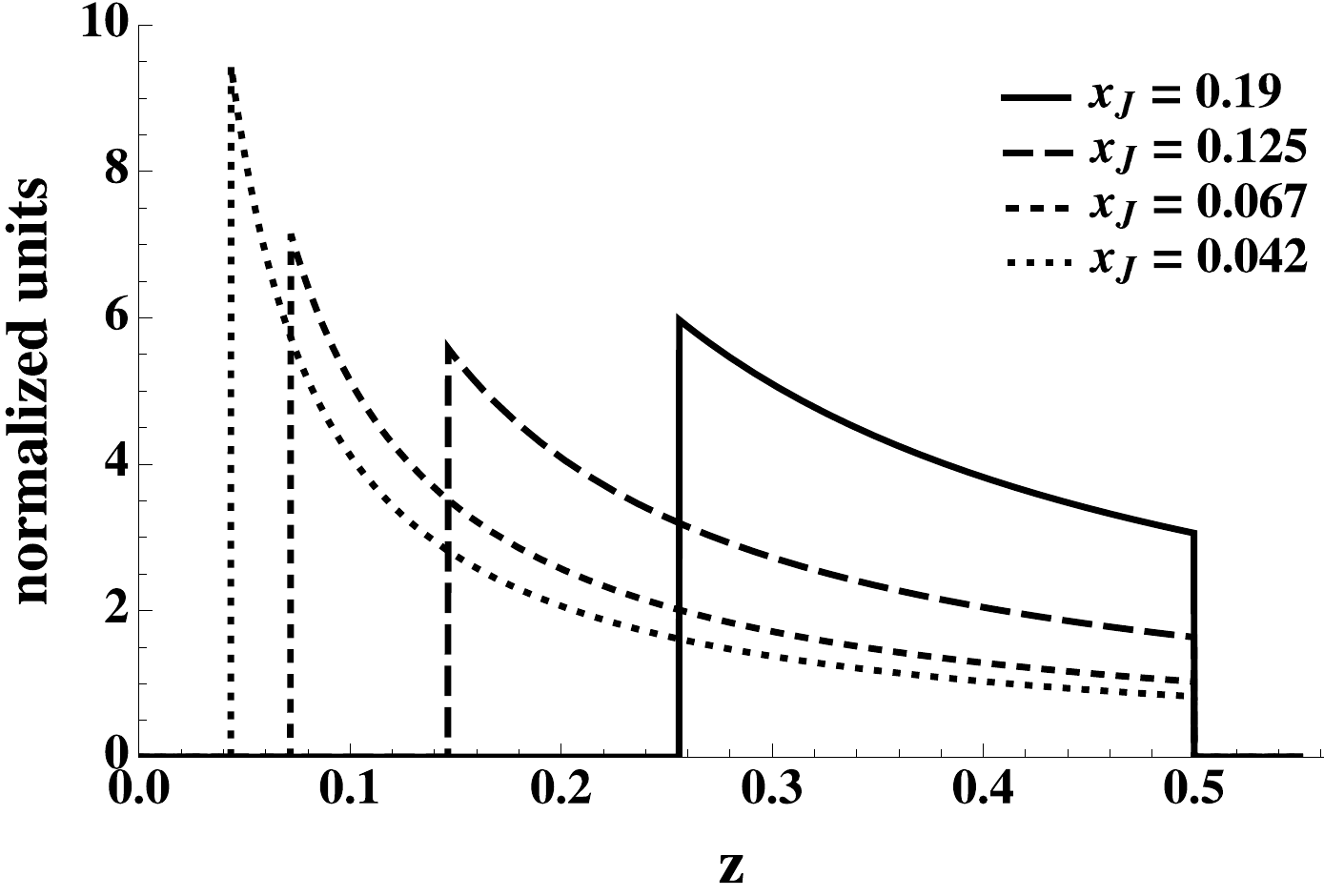}
\end{center}
\caption[Distribution in $z$ for LL QCD jets ]{Distribution in $z$ for LL QCD jets for $D = 1.0$ and various values of $x_J$.  The curves are normalized to have unit area.}
\label{fig:LLz}
\end{figure}

Next we consider the $z$ and $\Delta R_{12}$ distributions for the LL approximation where a single recombination of two (massless) partons is required to reconstruct as a jet of definite $p_{T_J}$ and mass (fixed $x_J$).  To that end we can ``undo'' one of the integrals in Eq.~(\ref{eq:LL1}) and consider the distributions for $z$ and $\Delta R_{12}$ .  We find for the $z$ distribution the form
\beq
\frac{1}{\sigma}\frac{d\sigma_{LL}}{dx_J dz} \sim \frac{1}{2 z x_J}\Theta{\left[z-\frac{1-\sqrt{1- 4 x_J/D^2}}{2}\right]} \Theta \left[ \frac{1}{2}-z \right].
\label{eq:LL2}
\eeq
As expected, we see the poles in $z$ and $x_J$ from the soft/collinear dynamics, but, as in Section \ref{sec:qcd:jets:algorithm} , the constraint of fixed $x_J$ yields a lower limit for $z$.  Recall that the upper limit for $z$ arises from its definition, again applied in the small-angle limit.  Thus the LL QCD distribution in $z$ is peaked at the lower limit but the characteristic turn-on point is fixed by the kinematics, requiring the branching at fixed $x_J$ to be in a jet of size $D$.  This behavior is illustrated in Fig.~\ref{fig:LLz} for various values of $x_J = 1/(\gamma^2-1)$ corresponding to those used in Section \ref{sec:qcd:jets:algorithm}.

\begin{figure}[htbp]
\begin{center}
\includegraphics[width = .5\columnwidth]{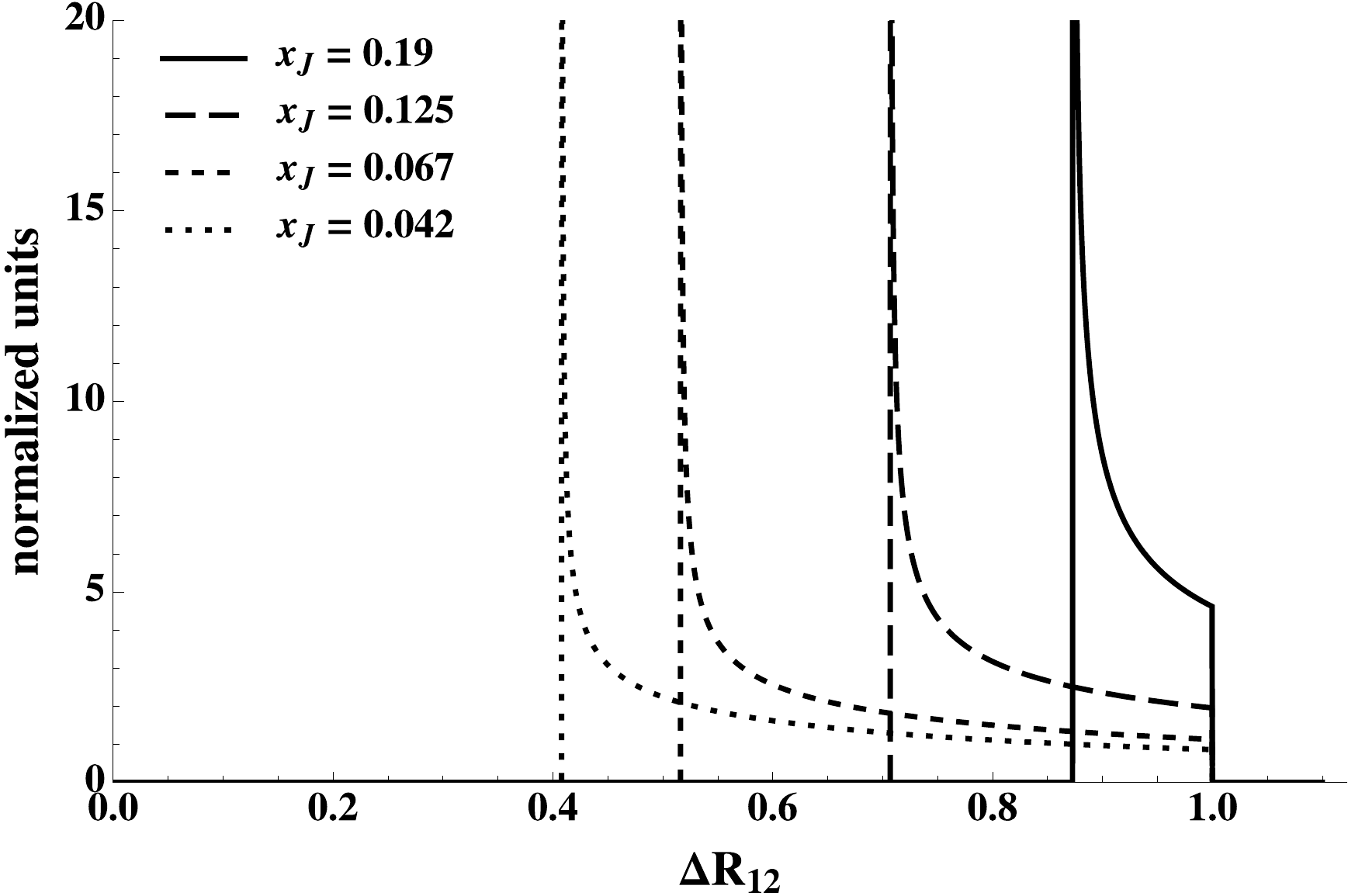}
\end{center}
\caption[Distribution in $\Delta R_{12}$ for LL QCD jets]{Distribution in $\Delta R_{12}$ for LL QCD jets for $D = 1.0$ and various values of $x_J$.  The curves are normalized to have unit area.}
\label{fig:LLDR}
\end{figure}

The expression for the $\Delta R_{12}$ dependence in the LL approximation is
\begin{eqnarray}
&&\frac{1}{\sigma}\frac{d\sigma_{LL}}{dx_J d\Delta R_{12}}\\
&&\sim \frac{2}{\Delta R_{12}^2}\frac{\Theta{\left[\Delta R_{12}-2 \sqrt{x_J}\right]}\Theta{ \left[ D - \Delta R_{12} \right]} }{\sqrt{\Delta R_{12}^2-{4 x_J}}\left(1-\sqrt{1-{4 x_J}/{\Delta R_{12}^2 }}\right)}. \nonumber
\label{eq:LL3}
\end{eqnarray}
This distribution is illustrated in Fig.~\ref{fig:LLDR} for the same values of $x_J$ as in Fig.~\ref{fig:LLz}.  As with the $z$ distribution the kinematic constraint of being a jet with a definite $x_J$ yields a lower limit, $\Delta R_{12} \gtrsim 2\sqrt{x_J}$, along with the expected upper limit, $\Delta R_{12} \leq D$. However, for $\Delta R_{12}$ the change of variables also introduces an (integrable) square root singularity at the lower limit. This square root factor tends to be numerically more important than the $1/\Delta R_{12}^2$ factor.\footnote{One factor of $\Delta R_{12}$ arises from the collinear QCD dynamics while the other comes from change of variables.  The soft QCD singularity is contained in the denominator factor $\left(1-\sqrt{1-{4 x_J}/{\Delta R_{12}^2 }}\right) \to 2z$ for $x_J\ll \Delta R^2$ (equivalently, $z \ll 1$).}  Since this square root singularity arises from the choice of variable (a kinematic effect), we will see that it is also present for heavy particle decays, suggesting that the $\Delta R_{12}$ variable will not be as useful as $z$ in distinguishing QCD jets from heavy particle decay jets.

Thus, in our toy QCD model with a single recombination, leading-logarithm dynamics and the small-angle jet mass definition, the constraints due to fixing $x_J$ tend to dominate the behavior of the $z$ and $\Delta R_{12}$ distributions, with limited dependence on the QCD dynamics and no distinction between the CA and $\kt$ algorithms.  However, this situation changes dramatically when we consider more realistic jets with full showering.  We will return to this subject after a brief interlude to consider the substructure of heavy particle decays.


\subsection[\textit{Substructure in heavy particle decays}]{Substructure in heavy particle decays}
\label{sec:sub:parton:decay}


Recombination algorithms have the potential to reconstruct the decay of a heavy particle.  Ideally, the substructure of a jet may be used to identify jets coming from a decay and reject the QCD background to those jets.  In this section, we investigate a pair of unpolarized parton-level decays, a heavy particle decaying into two massless quarks (a $1 \to 2$ decay) and a top quark decay into three massless quarks (a two-step decay).  For each decay, we study the available phase space in terms of the lab frame variables $\Delta R_{12}$ and $z$ and the shaping of kinematic distributions imposed by the requirement that the decay be reconstructed in a single jet.  We will determine the kinematic regime where decays are reconstructed, and contrast this with the kinematics for a $1\to2$ splitting in QCD.

\subsubsection{\texorpdfstring{$1\to2$ Decays}{1 -> 2 Decays}}
\label{sec:sub:parton:decay:onestep}

\begin{figure}[htbp]
\begin{center}
\includegraphics[width = .5\columnwidth]{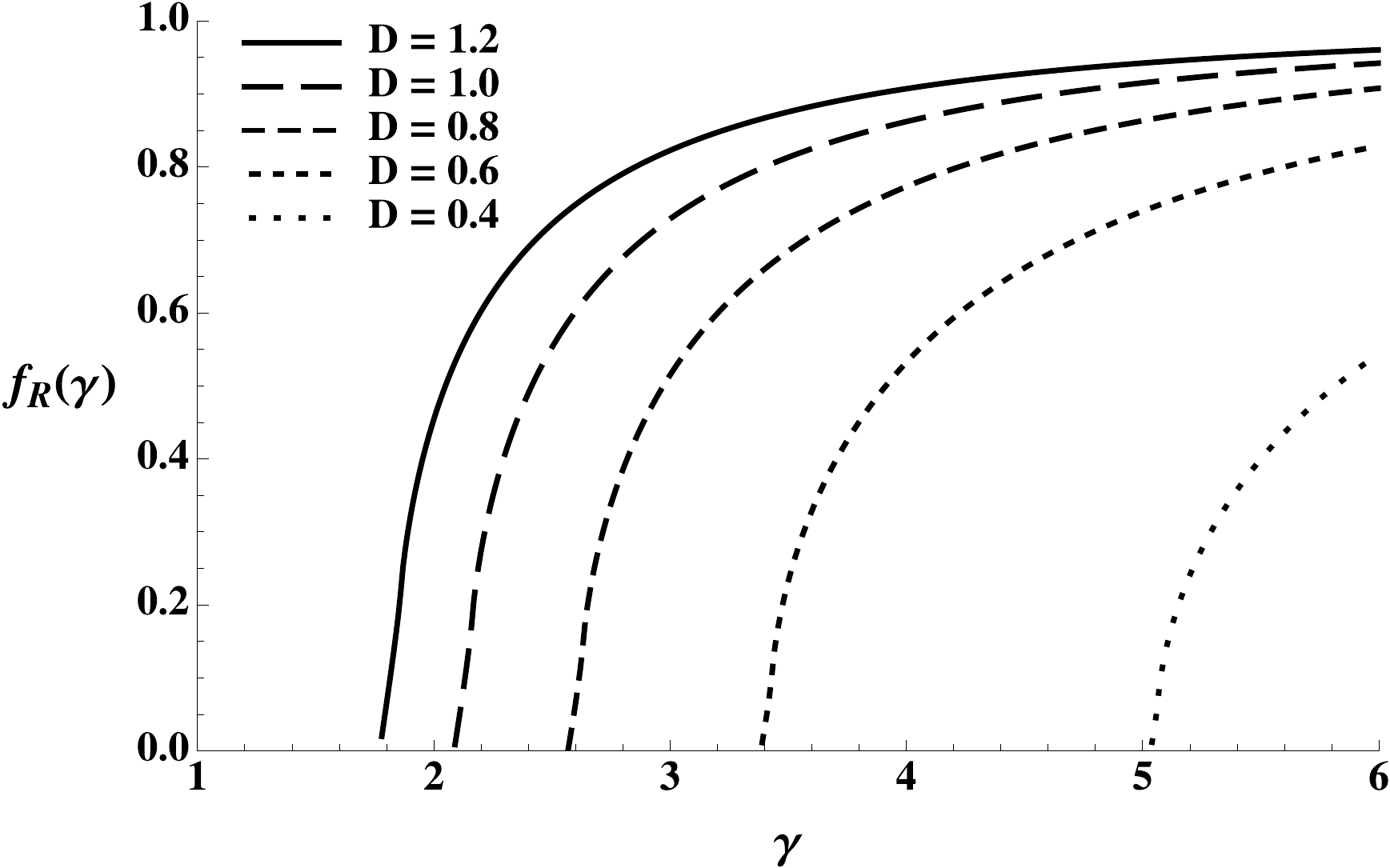}
\end{center}
\caption{Reconstruction fractions $f_R(\gamma)$ as a function of $\gamma$ for various $D$.}
\label{fig:fRdist}
\end{figure}

We begin by considering a $1\to2$ decay with massless daughters.  An unpolarized decay has a simple phase space in terms of the rest frame variables $\cos\theta_0$ and $\phi_0$:
\[
\frac{d^2N_0}{d\cos\theta_0 d\phi_0} = \frac{1}{4\pi} .
\label{eq:restframedist}
\]
Recall from Sec.~\ref{sec:qcd:jets:algorithm}  that $\cos\theta_0$ and $\phi_0$ are the polar and azimuthal angles of the heavier daughter particle in the parent particle rest frame relative to the direction of the boost to the lab frame.  In general, we will use $N_0$ to label the distribution of \emph{all} decays, while $N$ will label the distribution of decays \emph{reconstructed} inside a single jet. $N_0$ is normalized to unity, so that for any variable set $\Phi$,
\[
\int d\Phi\frac{dN_0}{d\Phi} = 1 .
\label{eq:normN0}
\]
The distribution $N$ is defined from $N_0$ by selecting those decays that fit in a single jet, so that generically
\[
\frac{dN}{d\Phi} \equiv \int d\Phi' \frac{dN_0}{d\Phi'}\delta(\Phi' - \Phi)\Theta(\text{single jet reconstruction}) .
\label{eq:dNdef}
\]
$N$ is naturally normalized to the total fraction of reconstructed decays.  The constraints of single jet reconstruction will depend on the decay and on the jet algorithm used, and abstractly take the form of a set of $\Theta$ functions.  For a $1\to2$ decay and a recombination-type algorithm, the only constraint is that the daughters must be separated by an angle less than $D$:
\[
\Delta R_{12} < D .
\label{eq:recon1to2}
\]
Since the kinematic limits imposed by reconstruction are sensitive to the boost $\gamma$ of the parent particle, we will want to consider the quantities of interest at a variety of $\gamma$ values.  To illustrate this $\gamma$ dependence, we first find the total fraction of all decays that are reconstructed in a single jet for a given value of the boost.  We call this fraction $f_R(\gamma)$:
\[
f_R(\gamma) \equiv \int d\cos\theta_0 d\phi_0 \frac{d^2N_0}{d\cos\theta_0d\phi_0}\Theta\left(D - \Delta R_{12}\right) .
\label{eq:fRgammadef}
\]
In Fig.~\ref{fig:fRdist}, we plot $f_R(\gamma)$ vs. $\gamma$ for several values of $D$.
The reconstruction fraction rises rapidly from no reconstruction to nearly complete reconstruction in a narrow range in $\gamma$.  This indicates that $\Delta R_{12}$ is strongly dependent on $\gamma$ for fixed $\cos\theta_0$ and $\phi_0$, which we will see below.  Conversely, the minimum boost necessary for a decay to fit in a jet depends strongly on $D$.  The turn-on for increasing $\gamma$ is the same effect as the ($z,\Delta R_{12}$) phase space moving into the allowed region below $\Delta R_{12}=D$ in Fig.~\ref{fig:zRcontours1} as $x_J$ is reduced.

To better understand the effect that reconstruction has on the phase space for decays, we would like to find the distribution of $1\to2$ decays in terms of lab frame variables,
\[
\frac{d^2N_0}{dzd\Delta R_{12}} .
\label{eq:decaylabvars}
\]
With two massless daughters, $\Delta R_{12}$ is given in terms of rest frame variables by
\begin{eqnarray}
\Delta R_{12}^2  =  \left[\tanh^{-1}\left(\frac{2\gamma\sin\theta_0\sin\phi_0}{ \sin^2\theta_0(\beta^2\gamma^2 + \sin^2\phi_0) + 1}\right)\right]^2 \nonumber\\
+ \left[\tan^{-1}\left(\frac{2\beta\gamma\sin\theta_0\cos\phi_0}{ \sin^2\theta_0(\beta^2\gamma^2 + \sin^2\phi_0) - 1}\right)\right]^2 .
\label{eq:dR}
\end{eqnarray}
with $\beta \equiv \sqrt{1 - \gamma^{-2}}$.  This relation is analytically
non-invertible, meaning we cannot write the Jacobian for the transformation
\[
\frac{d^2N_0}{d\cos\theta_0d\phi_0} \to \frac{d^2N_0}{dzd\Delta R_{12}}
\label{eq:disttransform}
\]
in closed form.  However, $\Delta R_{12}$ has some simple limits.  In particular, when the boost $\gamma$ is large, to leading order in $\gamma^{-1}$,
\[
\Delta R_{12} = \frac{2}{\gamma\sin\theta_0} + \mathcal{O}\left(\gamma^{-3}\right) .
\label{eq:dRlimit}
\]
This limit is only valid for $\sin\theta_0 \gtrsim \gamma^{-1}$, but as we will see this is the region of phase space where the decay will be reconstructed in a single jet.  The large-boost approximation describes the key features of the kinematics and is useful for a simple picture of kinematic distributions when particles are reconstructed in a single jet.

Since $\gamma = \sqrt{1+ 1/x_J}$, this limit is equivalent to the small-angle limit we took in Sec.~\ref{sec:sub:parton:qcd}.  (For $\Delta R^2 \ll 1$, $x_J \approx z(1-z)\Delta R^2 \ll 1$.)  We can see this in Eq.~(\ref{eq:dR}), where $\Delta R \approx 1/\gamma$.

The value of $z$ is also simple in the large-boost approximation.  In this limit,
\[
z = \frac{1 - \left|\cos\theta_0\right|}{2} +
\mathcal{O}\left(\gamma^{-2}\right) .
\label{eq:zlimit}
\]
With the large-boost approximation, $z$ and $\Delta R_{12}$ are both independent of $\phi_0$.  As noted earlier both $\Delta R_{12}$ and $z$ depend on $\phi_0$ only through terms that are suppressed by inverse powers of $\gamma$ (cf. Figs.~\ref{fig:thetaphicontours} and \ref{fig:zRcontours}).  In this limit we can integrate out $\phi_0$ and find the distributions in $z$ and $\Delta R_{12}$ for all decays.  For $z$ the distribution is simply flat:
\beq
\frac{dN_0}{dz} \approx 2\Theta\left(\frac12 - z\right)\Theta(z) .
\label{eq:dN0dz}
\eeq
We have included the limits for clarity.  For $\Delta R_{12}$, the distribution is
\beq
\frac{dN_0}{d\Delta R_{12}} \approx \frac{4}{\gamma^2\Delta R_{12}^2}\frac{\Theta\left(\Delta R_{12} - 2\gamma^{-1}\right)}{\sqrt{\Delta R_{12}^2 - 4\gamma^{-2}}} .
\label{eq:dN0ddR}
\eeq
This distribution has a lower cutoff requiring $\Delta R_{12} \ge 2\gamma^{-1}$.  This is close to the true lower limit on $\Delta R_{12}$, $\Delta R_{12} \ge 2\csc^{-1}\gamma$.  Note that in Eq.~(\ref{eq:dN0ddR}), there is a enhancement at the lower cutoff in $\Delta R_{12}$ due to the square root singularity arising from the change of variables, just as there was in the QCD result in Eq.~(\ref{eq:LL3}).

\begin{figure}[htbp]
\begin{center}
\includegraphics[width=.5\columnwidth]{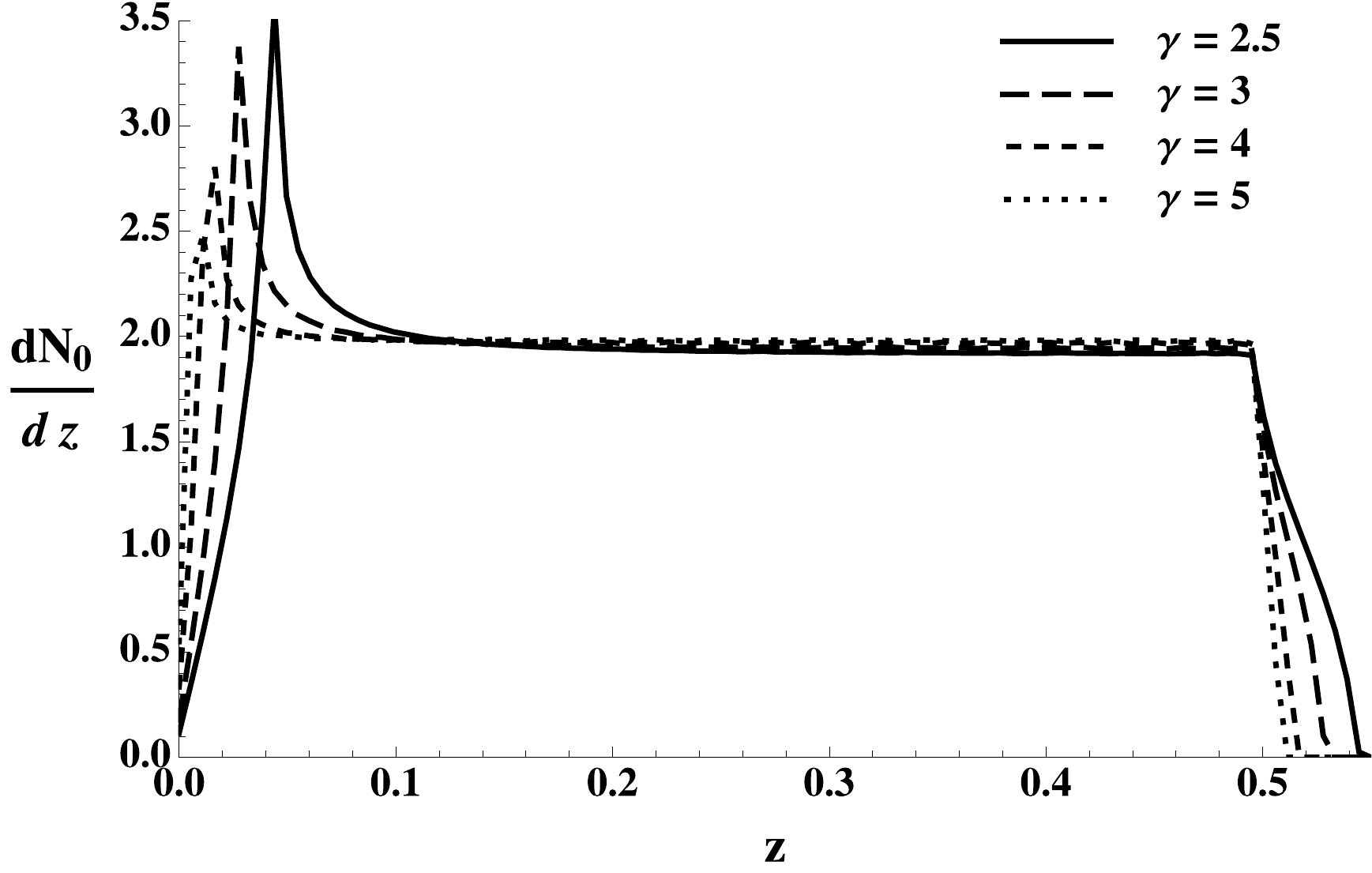}
\end{center}
\caption{The distribution of all decays in $z$ for several values of $\gamma$.}
\label{fig:zdistall}
\end{figure}

\begin{figure}[htbp]
\begin{center}
\includegraphics[width=.5\columnwidth]{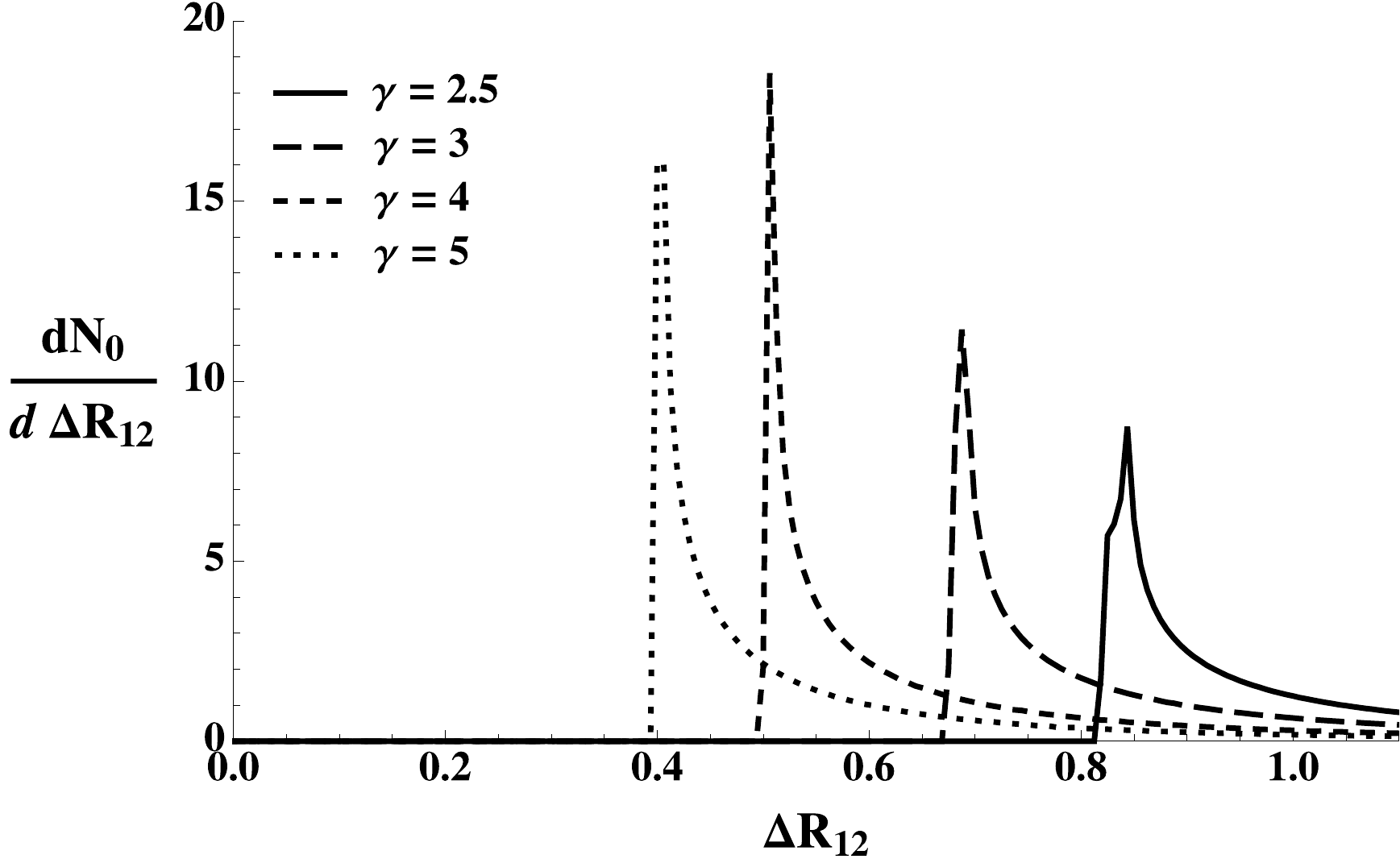}
\end{center}
\caption{The distribution of all decays in $\Delta R_{12}$ for several values of $\gamma$.}
\label{fig:DRdistall}
\end{figure}

In Fig.~\ref{fig:zdistall}, we plot the exact distribution $dN_0/dz$, found numerically, for several values of $\gamma$.
The true distribution is qualitatively similar to the approximate one in Eq.~(\ref{eq:dN0dz}), which is flat.  The peak in the distribution at small $z$ values comes from the reduced phase space as $z\to0$, and the peak is lower for larger boosts.  In Fig.~\ref{fig:DRdistall}, we plot the exact distribution $dN_0/d\Delta R_{12}$, which is again qualitatively similar to the large-boost result.
The distribution in $\Delta R_{12}$ is localized at the lower limit, especially for larger boosts.  This provides a useful rule: the opening angle of a decay is strongly correlated with the transverse boost of the parent particle.  Note that the relevant boost is the transverse one because the angular measure $\Delta R$ is invariant under longitudinal boosts (recall that in the example here, we have set the parent particle to be transverse).

\begin{figure}[htbp]
\begin{center}
\includegraphics[width = .5\columnwidth]{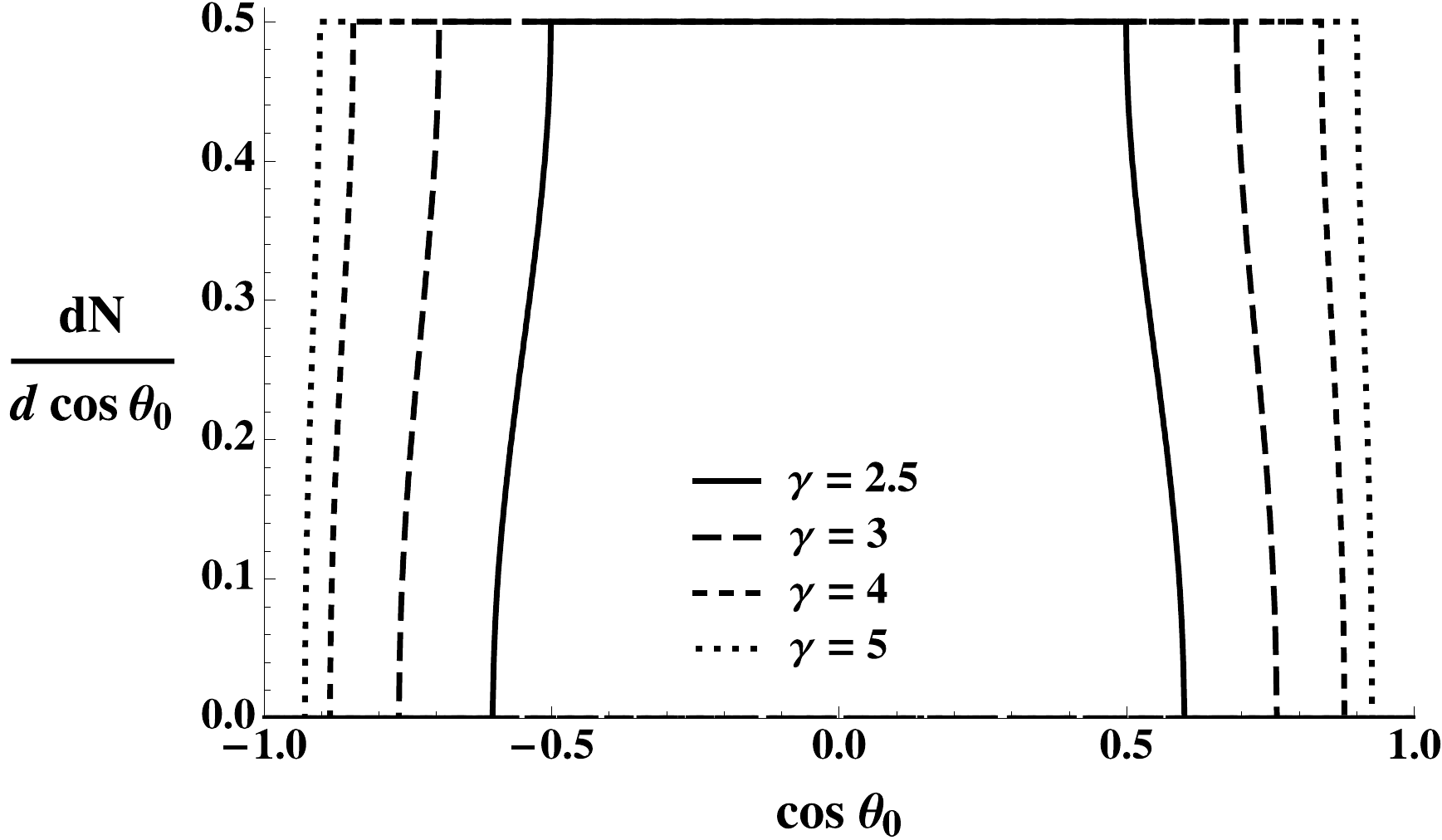}
\end{center}
\caption{The reconstructed distribution $dN/d\cos\theta_0$ with $D = 1.0$ for various values of $\gamma$.}
\label{fig:costhdist}
\end{figure}

The constraint imposed by reconstruction is simple in the large-boost approximation.  In terms of $\sin\theta_0$, the constraint $\Delta R_{12} < D$ requires $\sin\theta_0 > 2/\gamma D$, which excludes the region where the approximation breaks down.  Therefore the large-boost approximation is apt for describing the kinematics of a reconstructed decay.  In Fig.~\ref{fig:costhdist}, we plot the distribution, $dN/d\cos\theta_0$, where the implied sharp cutoff is apparent (and should be compared to what we observed in Fig.~\ref{fig:thetaphicontours1}).
This distribution is easy to understand in the rest frame of the decay.  When $|\cos\theta_0|$ is close to 1, one of the daughters is nearly collinear with the direction of the boost to the lab frame, and the other is nearly anti-collinear.  The anti-collinear daughter is not sufficiently boosted to have $\Delta R_{12} < D$ with the collinear daughter, and the parent particle is not reconstructed.  As $|\cos\theta_0|$ decreases, the two daughters can be recombined in the same jet; this transition is rapid because the $\phi_0$ dependence of the kinematics is small.  We now look at the distributions of $z$ and $\Delta R_{12}$ when we require reconstruction.

\begin{figure}[htbp]
\begin{center}
\includegraphics[width=.5\columnwidth]{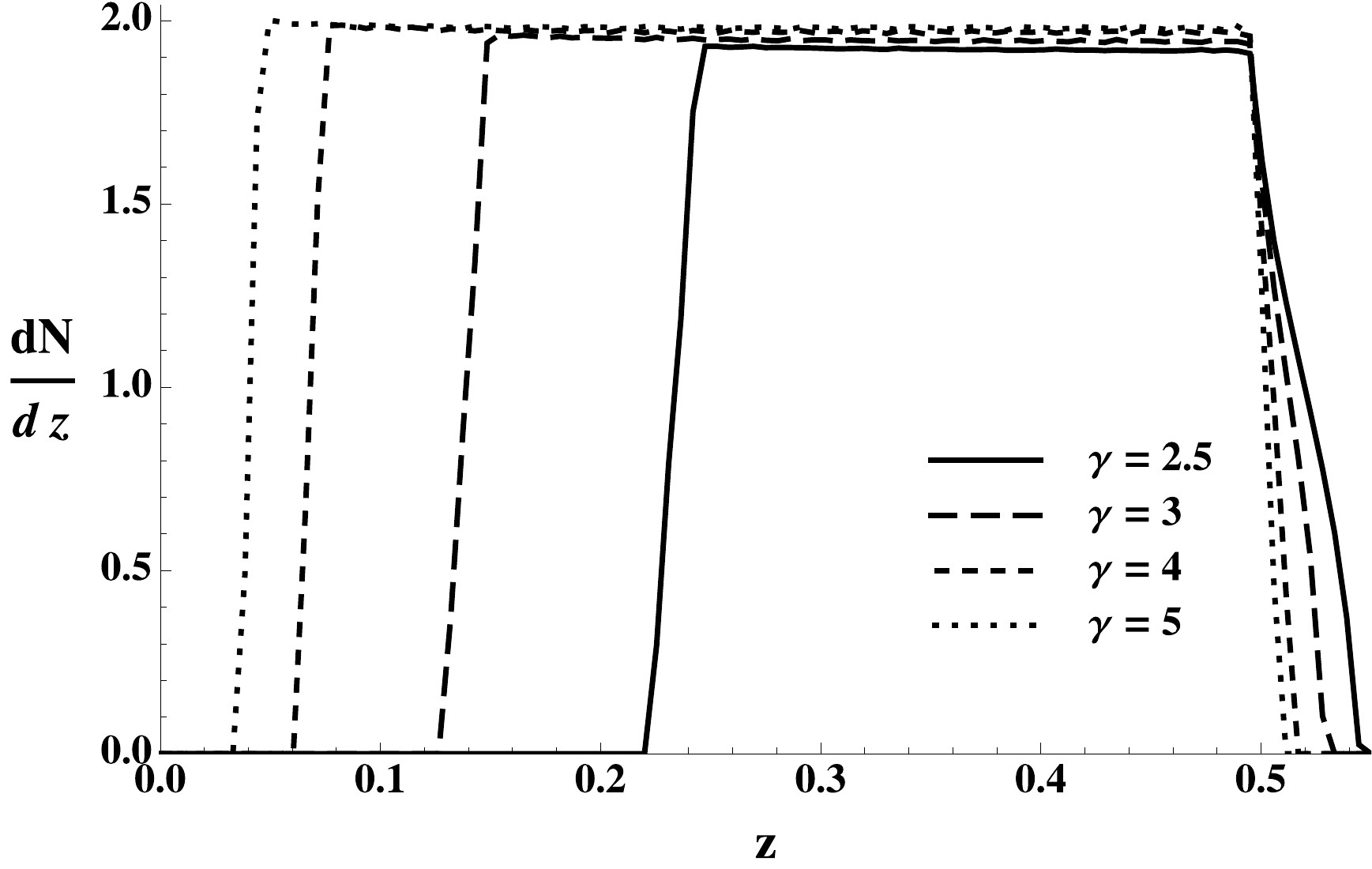}
\end{center}
\caption{The distribution of reconstructed decays in $z$ for several values of $\gamma$.}
\label{fig:zdistrecon}
\end{figure}

Because $z$ is linearly related to $\cos\theta_0$ at large boosts, the distribution in $z$ has a simple form:
\beq
\frac{dN}{dz} \approx 2 \Theta\left(z - \frac{1 - \sqrt{1 - 4/(\gamma^2D^2)}}{2}\right) \Theta\left(\frac12 - z\right) .
\label{eq:dNdz}
\eeq
Comparing to Eq.~(\ref{eq:dN0dz}), we see that requiring reconstruction simply cuts out the region of phase space at small $z$.  This is confirmed in the exact distribution $dN/dz$, shown in Fig.~\ref{fig:zdistrecon}.
The small-$z$ decays that are not reconstructed come from the regions of phase space with $|\cos\theta_0|$ near 1, just as in the previous discussion.  In these decays, the backwards-going (anti-collinear) daughter is boosted to have small $p_T$ in the lab frame.  Comparing to Fig.~\ref{fig:LLz}, the distribution in $z$ for QCD splittings, we see first that the cutoffs on the distributions are similar (they are not identical because of the LL approximation used in Fig.~\ref{fig:LLz}).  However, the QCD distribution has an enhancement at small $z$ values, due to the QCD soft singularity, that the distribution for reconstructed decays does not exhibit.

The distribution of reconstructed particles in the variable $\Delta R_{12}$ is related simply to the distribution of all decays in the same variable:
\beq
\frac{dN}{d\Delta R_{12}} = \frac{dN_0}{d\Delta R_{12}}\Theta\left(D - \Delta R_{12}\right) ,
\label{eq:dNddR}
\eeq
which means that the distribution $dN/d\Delta R_{12}$ is given by Fig.~\ref{fig:DRdistall} with a cutoff at $\Delta R_{12} = D$.  Note that this distribution is very close in shape to the distribution of QCD branchings versus $\Delta R_{12}$ displayed in Eq.~(\ref{eq:LL3}) and Fig.~\ref{fig:LLDR}.  This similarity arises from that the fact that the most important factor in the shape is the square root singularity, which arises from the change of variables in both cases and hides the underlying differences in dynamics.


\subsubsection{Two-step Decays}
\label{sec:sub:parton:decay:twostep}

We now turn our attention to two-step decays, which exhibit a more complex substructure.  Two-step decays offer new insights into the ordering effects of the $\kt$ and CA algorithms, highlight the shaping effects from the algorithm on the jet substructure, and offer a surrogate for the cascade decays that are often featured in new physics scenarios.  Even at the parton level the choice of jet algorithm matters in reconstructing a multi-step decay; different algorithms can give different substructure.  In studying this substructure we take the same approach as for the $1\to2$ decay, translating the simple kinematics of a parton-level decay into the lab frame variables $\Delta R_{12}$ and $z$.

The top quark is a good example of a two-step decay, and we focus on it in this section.  We will label the top quark decay $t\to Wb$, with $W\to qq'$.  In this discussion requiring that the top quark be reconstructed means that the $W$ must be recombined from $q$ and $q'$ first, then merged with the $b$.  The observed (3-parton) ``jet'' will then have the $W$ as one of its daughter subjets.

For the $\kt$ algorithm, reconstructing the top quark in a single jet imposes the following constraints on the partons:
\[
\begin{split}
\min(p_{Tq},p_{T{q'}})\Delta R_{qq'} &< \min(p_{Tq},p_{Tb})\Delta R_{bq}, \\
\min(p_{Tq},p_{T{q'}})\Delta R_{qq'} &< \min(p_{T{q'}},p_{Tb})\Delta R_{bq'}, \\
\Delta R_{qq'} &< D,~\text{and}\\
\Delta R_{bW} &< D.
\end{split}
\]
For the CA algorithm the relations are strictly in terms of the angle:
\[
\begin{split}
\Delta R_{qq'} &< \Delta R_{bq}, \\
\Delta R_{qq'} &< \Delta R_{bq'}, \\
\Delta R_{qq'} &< D,~\text{and}\\
\Delta R_{bW} &< D.
\end{split}
\]
The kinematic limits requiring the decay to be reconstructed in a single jet are the same for the two algorithms, but fixing the ordering of the two recombinations requires a different restriction for each algorithm, which in turn biases the distributions of kinematic variables.

The common requirements such that the top quark be reconstructed in a single jet, $\Delta R_{qq'} < D$ and $\Delta R_{Wb} < D$, are straightforward to understand in terms of the rest frame variable $\cos\theta_0$, which here is the polar angle in the top quark rest frame between the $W$ and the boost direction to the lab frame.  For $\cos\theta_0 \approx 1$, the $W$ has a large transverse boost in the lab frame, so $\Delta R_{qq'} < D$, but the angle between the $W$ and $b$ will be large (as was the case for the corresponding $1\to2$ decay in the previous section).  For $\cos\theta_0 \approx -1$, the $W$ transverse boost is small, and $\Delta R_{qq'}$ will be large.  Therefore, we only expect to reconstruct top quarks in a single jet when $|\cos\theta_0|$ is not near $1$.

If the CA algorithm correctly reconstructs the top quark, the two quarks from the $W$ decay must be the closest pair (in $\Delta R$) of the three final state particles.  This requirement strongly selects for decays where the $W$ opening angle, $\Delta R_{qq'}$, is smaller than the top quark opening angle, $\Delta R_{Wb}$.  Therefore, only decays with a large (transverse) $W$ boost will be reconstructed by the CA algorithm.  In terms of $\cos\theta_0$, the fraction of decays that are reconstructed will increase as we increase $\cos\theta_0$ towards the upper limit where $\Delta R_{Wb} \ge D$, and the reconstruction fraction will be small for lower values of $\cos\theta_0$.

The $\kt$ algorithm orders recombinations by $p_T$ as well as angle, and the set of reconstructed decays is understood most easily by contrasting with CA.  As the transverse boost of the $W$ decreases, on average the $p_T$ of the $q$ and $q'$ decrease while the $p_T$ of the $b$ increases.  Therefore, while $\Delta R_{qq'}$ is increasing, $\min(p_{Tq}, p_{Tq'})$ is decreasing, and these competing effects suggest that $\kt$ reconstructs decays with smaller values of $\cos\theta_0$ than CA, and that the dependence on $\cos\theta_0$ is not as strong.

\begin{figure}[htbp]
\begin{center}
\includegraphics[width = .5\columnwidth]{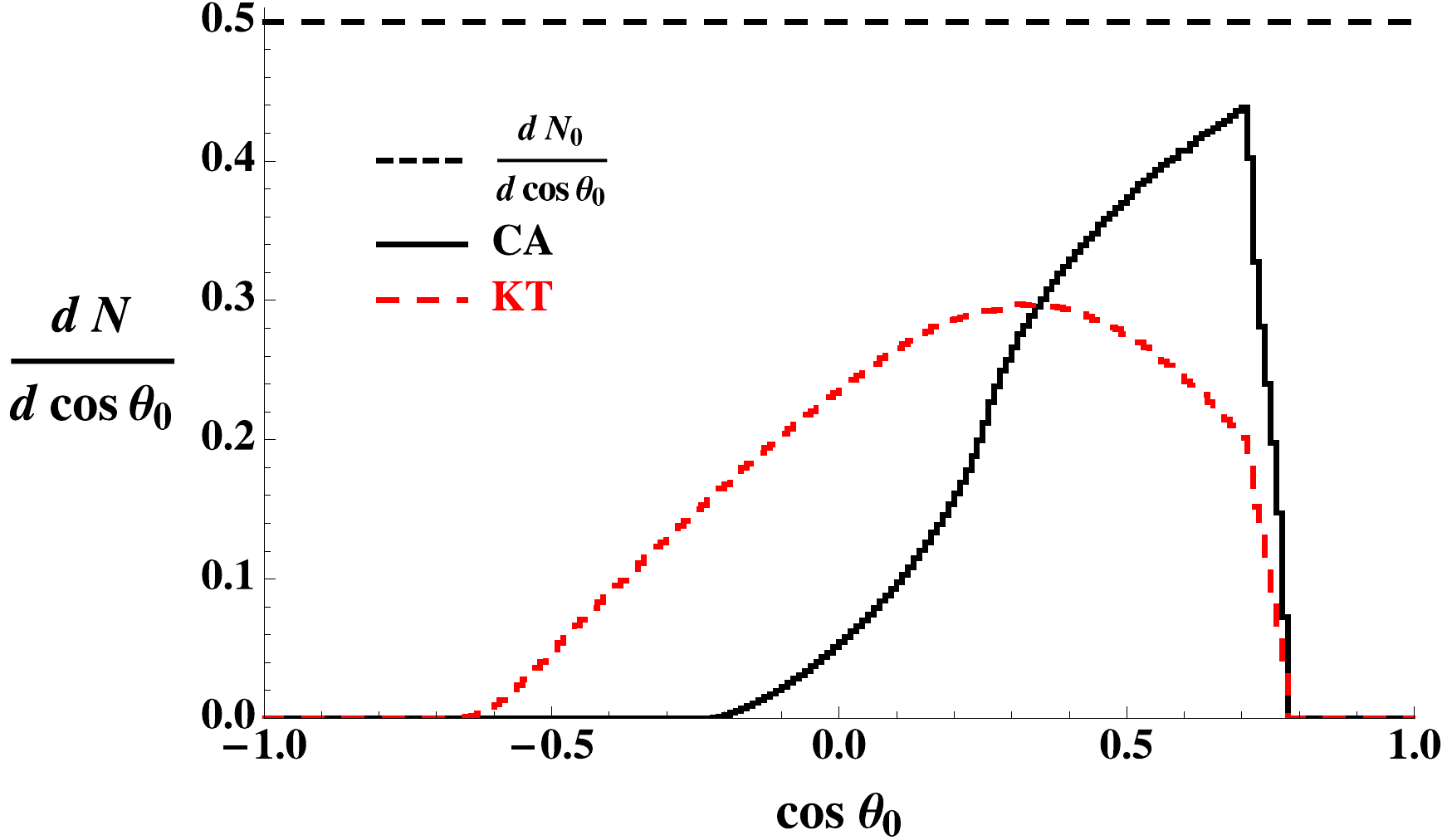}
\end{center}
\caption[$dN/d\cos\theta_0$ vs. $\cos\theta_0$, with $\gamma = 3$]{$dN/d\cos\theta_0$ vs. $\cos\theta_0$, with $\gamma = 3$, for both the $\kt$ and CA algorithms.  The underlying distribution $dN_0/d\cos\theta_0 = 1/2$ is plotted as the dotted line for reference.}
\label{fig:costhdist_top}
\end{figure}

The effect of the CA and $\kt$ algorithms on the observed distribution in $\cos\theta_0$ is shown in Fig.~\ref{fig:costhdist_top}, where we plot the distribution of $\cos\theta_0$ for reconstructed top quarks for both algorithms.  The top boost is fixed to $\gamma = 3$.
We observe the kinematic limit near $\cos\theta_0\approx 0.8$ is common between algorithms, and that $\cos\theta_0\approx-1$ is not accessed by either algorithm.  As expected, the distribution for the CA algorithm falls off more sharply than for $\kt$ at lower values of $\cos\theta_0$.

\begin{figure}[htbp] \begin{center}
\includegraphics[width = .5\columnwidth]{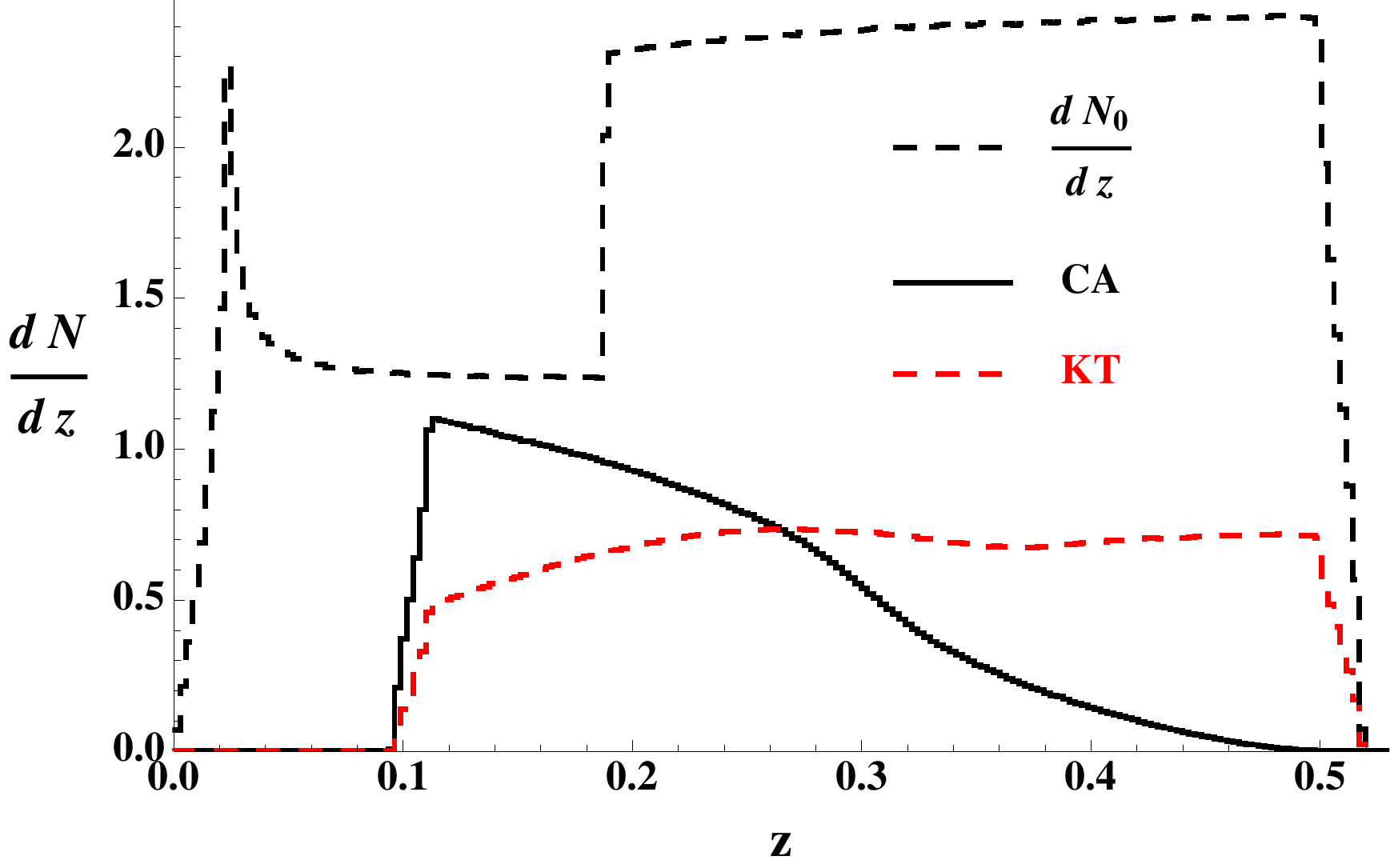}
\end{center}
\caption{$dN_0/dz$ (all decays) and $dN/dz$ (reconstructed decays), with $\gamma = 3$.}
\label{fig:topzdist}
\end{figure}

Next, we look at distributions in $z$ and $\Delta R_{Wb}$.  Just as in the $1\rightarrow2$ decay, we expect decays with small $z$ not to be correctly reconstructed.  Small values of $z$ will come when the $W$ or $b$ is soft, and therefore produced very backwards-going in the top rest frame.  This corresponds to $\cos\theta_0 \approx \pm 1$, and from Fig.~\ref{fig:costhdist_top} these decays are not reconstructed.  In Fig.~\ref{fig:topzdist}, we plot the distribution in $z$ for all decays, $dN_0/dz$, and the distribution for reconstructed decays, $dN/dz$, for a boost of $\gamma = 3$.

In $dN_0/dz$, the discontinuity at $z\approx0.2$ arises from the fact that the $W$ is sometimes softer than the $b$, but has a minimum $p_T$.  The extra weight in $dN_0/dz$ for $z$ above this value comes from the decays where the $W$ is softer than the $b$.  Note that these decays are rarely reconstructed, especially for CA: the distribution $dN/dz$ is smooth, and has little additional support in the region where the $W$ is softer.  This correlates with the fact that decays with negative $\cos\theta_0$ values are rarely reconstructed with CA, but more frequently with $\kt$.  The distribution $dN/dz$ has a lower cutoff that corresponds to the upper cutoff in Fig.~\ref{fig:costhdist_top}.  As the boost $\gamma$ of the top increases, the cutoff at small $z$ decreases, since the limit in $\cos\theta_0$ for which $\Delta R_{Wb} > D$ will increase towards 1.

\begin{figure}[htbp]
\begin{center}
\includegraphics[width = .5\columnwidth]{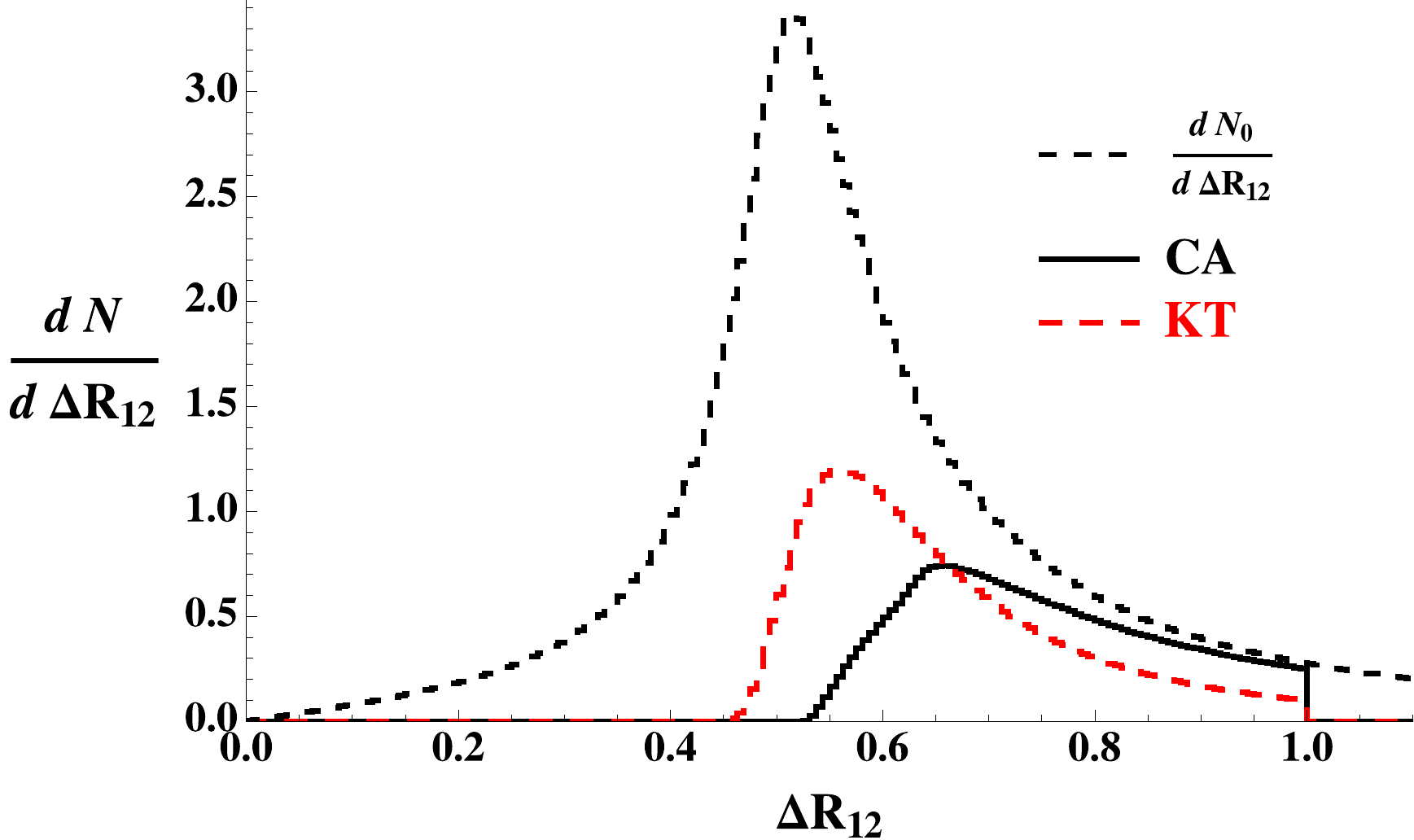}
\end{center}
\caption{$dN_0/d\Delta R_{Wb}$ (all decays) and $dN/d\Delta R_{12}$ (reconstructed decays), with $\gamma = 3$.}
\label{fig:topDRdist}
\end{figure}

The opening angle $\Delta R_{Wb}$ of the top quark decay also illustrates how strongly the kinematics are shaped by the jet algorithm.  When $\cos\theta_0 \approx -1$, for sufficient boosts $\Delta R_{Wb}$ is small because the $W$ is boosted forward in the lab frame, but these decays are not reconstructed because the ordering of recombinations will typically be incorrect and the $W$ decay may not have $\Delta R_{qq'} < D$.  For $\cos\theta_0 \approx 1$, $\Delta R_{Wb}$ will exceed $D$ and the top will not be reconstructed.  In Fig.~\ref{fig:topDRdist}, we plot the distribution $dN_0/d\Delta R_{Wb}$ of the angle between the $W$ and $b$ in all top decays for a top boost of $\gamma = 3$, as well as the distribution $dN/d\Delta R_{12}$ of the angle of the last recombination for reconstructed top quarks with the $\kt$ and CA algorithms.  Note that when the top quark is reconstructed at the parton level, $\Delta R_{12} = \Delta R_{Wb}$.
The difference in $dN/d\Delta R_{12}$ between the $\kt$ and CA algorithms reflects their different recombination orderings.  Because CA orders strictly by angle, the angle $\Delta R_{12}$ tends to be larger than for $\kt$ because CA requires $\Delta R_{12} = \Delta R_{Wb} > \Delta R_{qq'}$.

\subsubsection{Contrast with QCD}
\label{sec:sub:parton:decay:compare}

Contrasting the figures in the previous two subsections, we can see that at this level of approximation, QCD splittings and heavy particle decays have distinct kinematics.  In both cases the kinematical requirements of fixed mass and $p_T$ lead to cutoffs in phase space (recall \fig{fig:zRcontours}), but within these boundaries the differing dynamics shape the distributions.  For example, QCD splittings tend to have small $z$ (\fig{fig:LLz}), driven by the soft singularity of the QCD splitting function.  A one-step decay (\fig{fig:zdistrecon}) has a completely flat distribution in $z$, whereas a two-step decay (\fig{fig:topzdist}) has a more complicated shape once we require accurate reconstruction.  In the case of $\Delta R_{12}$, the differences are less dramatic.  In both cases kinematics drive $\Delta R_{12}$ to be as small as possible (compare Figs.~\ref{fig:LLDR} and \ref{fig:DRdistall}).  The two-step decay is more complicated, but the distributions (\fig{fig:topDRdist}) still have a peak at low values.

If we wish to jets representing heavy particle decays from their QCD background, after cutting on a jet mass we would presumably be interested in jet substructure.  We have seen that at the parton level, after fixing jet masses, the distributions in $z$ are still distinct enough to expect some additional discrimination.  $\Delta R_{12}$, on the other hand, does not appear useful.  To see if these kinematic differences can be exploited, we must first study how they appear in real jets, where we include the effects of showering and subsequent reconstruction.


\section{Algorithm systematics in \texorpdfstring{$\ee$}{ee} events}
\label{sec:sub:algeffects}

To obtain a more realistic understanding of jet substructure we must turn to simulated events.\footnote{Some of the discussion in this section and the next one are taken from Sections III--V of \cite{Pruning2}.  All of the figures, except Fig.~\ref{fig:topPartonVsReconZDR}, are new.}  Monte Carlo event generators replace our simple models with exact matrix element calculations, supplemented with parton shower algorithms that model the behavior of QCD showering.  Such generators produce events consisting of hundreds of outgoing hadrons, which we analyze via a jet algorithm.  If, after finding jets, we consider their final merging step (from the shower perspective, their first splitting), we might expect this branching to resemble the splittings of the previous section's models.  This would be the case if the jet algorithm could precisely undo the parton shower, but of course this can never be true.  In this section we will consider how the mirror processes of the QCD shower and the jet algorithm shape jet substructure.  

We begin by considering top-antitop and dijet events in $\ee$ collisions, since this excludes a variety of other effects we wish to postpone having to think about.  For both samples we consider events with a center-of-mass energy of 1200 GeV.  We will keep (incongruously) hadron collider language and analysis, since we're only using $\ee$ collisions as a proxy for ``clean'' events with no strongly interacting particles in the initial state.  The details of the event generation are given in Appendix~\ref{app:details}.

\subsection{QCD jets}
\label{sec:sub:algeffects:qcd}

We first consider simulated QCD jets.  As suggested earlier, we anticipate two important changes from the previous discussion.  First, the showering ensures that the daughter subjets at the last recombination have nonzero masses.  More importantly and as noted in Section \ref{sec:qcd:jets:algorithm}, the sequence of recombinations generated by the jet algorithm tends to force the final recombination into a particular region of phase space that depends on the recombination metric of the algorithm.  For the CA algorithm this means that the final recombination will tend to have a value of $\Delta R_{12}$ near the limit $D$, while the $\kt$ algorithm will have a large value of $z \Delta R_{12} p_{T{_J}}$.  This issue will play an important role in explaining the observed $z$ and $\Delta R_{12}$ distributions.

\begin{figure}[htbp]
\begin{center}
\includegraphics[width = .7\columnwidth]{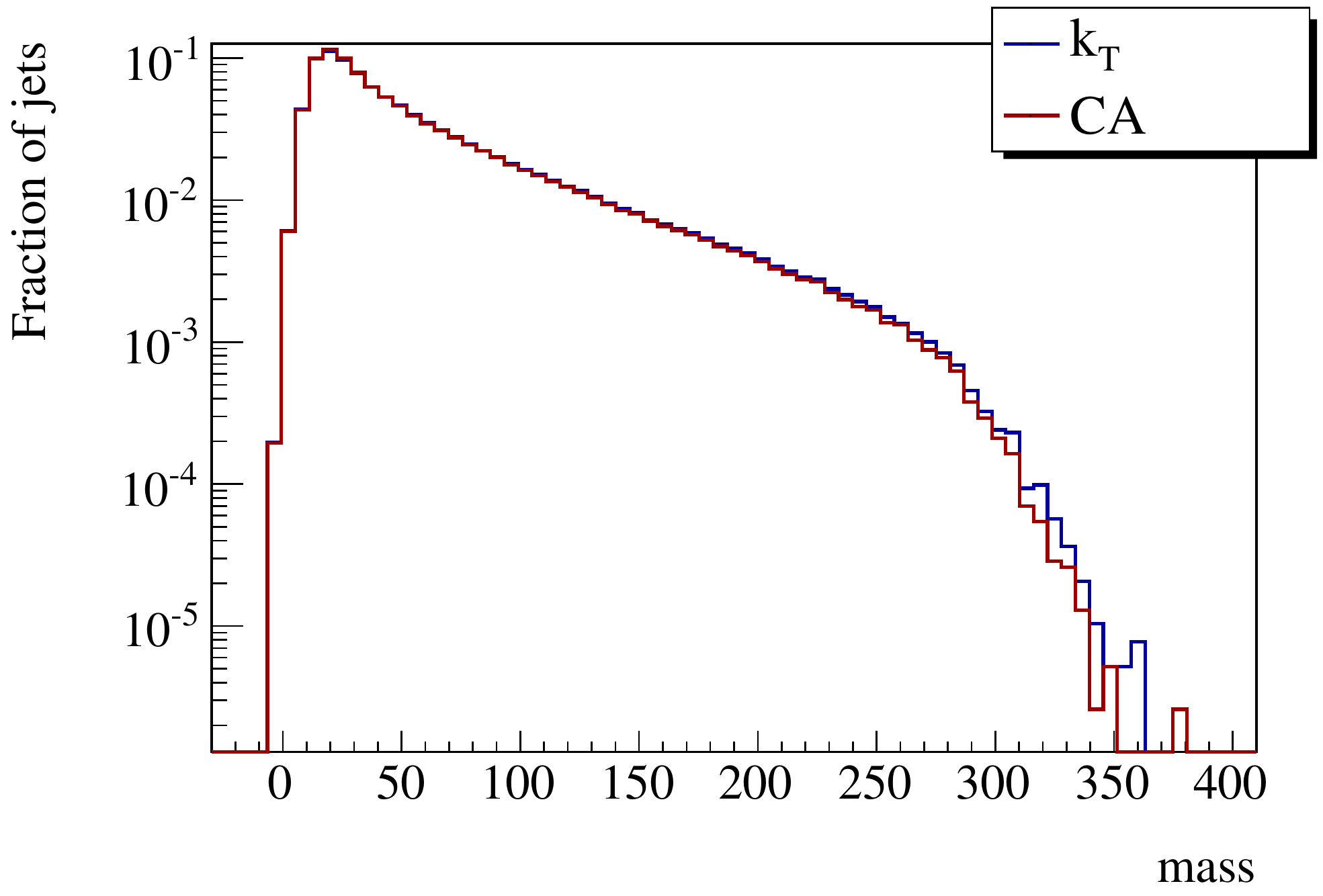}
\end{center}
\caption{Distribution in $m_J$ for QCD jets in $\ee \to q\bar q$ events, with D = 1.0.}
\label{fig:mjetplot}
\end{figure}

First, consider the jet mass distributions from the simulated event samples.  In Fig.~\ref{fig:mjetplot}, we plot the jet mass distributions for the $\kt$ and CA algorithms for all jets in the sample.
As expected, for both algorithms the QCD jet mass distribution smoothly falls from a peak only slightly displaced from zero (the remnant of the perturbative $-\ln(m^2)/m^2$ behavior).  There is a more rapid cutoff for $m_J > p_{T_J}D/2$, which corresponds to the expected kinematic cutoff from the LL approximation, but smeared by the spread in $p_T$, the nonzero subjet masses and the other small corrections to the LL approximation. The average jet mass, $\left\langle m_{J} \right\rangle \approx 100$ GeV, is in crude agreement with the perturbative expectation $\left\langle m_{J}/p_{T_J}\right\rangle \approx 0.2$.  Note that in these events the two algorithms give nearly identical distributions.

Other details of the QCD jet substructure are substantially more sensitive to the specific algorithm than the jet mass distribution.  To illustrate this point we will discuss the distributions of $z$, $\Delta R_{12}$, and the heavier subjet mass for the last recombination in the jet.  We can understand the observed behavior by combining a simple picture of the geometry of the jet with the constraints induced on the phase space for a recombination from the jet algorithm.  In particular, recall that the ordering of recombinations defined by the jet algorithm imposes relevant boundaries on the phase space available to the late recombinations (see Fig.~\ref{fig:algboundaries}).

While the details of how the $\kt$ and CA algorithms recombine protojets within a jet are different, the overall structure of a large-$p_T$ jet is set by the shower dynamics of QCD, i.e., the dominance of soft/collinear emissions.  Typically the jet has one (or a few) hard core(s), where a hard core is a localized region in $y$--$\phi$ with large energy deposition.  The core is surrounded by regions with substantially smaller energy depositions arising from the radiation emitted by the energetic particles in the core (i.e., the shower), which tend to dominate the area of the jet.  In particular, the periphery of the jet is occupied primarily by the particles from soft radiation, since even a wide-angle hard parton will radiate soft gluons in its vicinity.  This simple picture leads to very different recombinations with the $\kt$ and CA algorithms, especially the last recombinations.

The CA algorithm orders recombinations only by angle and ignores the $p_T$ of the protojets.  This implies that the protojets still available for the last recombination steps are those at large angle with respect to the core of the jet.  Because the core of the jet carries large $p_T$, as the recombinations proceed the directions of the protojets in the core do not change significantly.  Until the final steps, the recombinations involving the soft, peripheral protojets tend to occur only locally in $y$--$\phi$ and do not involve the large-$p_T$ protojets in the core of the jet. Therefore, the last recombinations defined by the CA algorithm are expected to involve two very different protojets.  Typically one has large $p_T$, carrying most of the four-momentum of the jet, while the other has small $p_T$ and is located at the periphery of the jet.  The last recombination will tend to exhibit large $\Delta R_{12}$, small $z$, large $a_1$ (near 1), and small $a_2$, where the last two points follow from the small $z$ and correspond to the $(z, \Delta R_{12})$ phase space of Fig.~\ref{fig:zRcontours3}.

In contrast, the $\kt$ algorithm orders recombinations according to both $p_T$ and angle.  Thus the $\kt$ algorithm tends to recombine the soft protojets on the periphery of the jet earlier than with the CA algorithm.  At the same time, the reduced dependence on the angle in the recombination metric implies the angle between protojets for the final recombinations will be lower for $\kt$ than CA.  While there is still a tendency for the last recombination in the $\kt$ algorithm to involve a soft protojet with the core protojet, the soft protojet tends to be not as soft as with the CA algorithm (i.e., the $z$ value is larger), while the angular separation is smaller.  Since this final soft protojet in the $\kt$ algorithm has participated in more previous recombinations than in the CA case, we expect the average $a_2$ value to be further from zero and the $a_1$ value to be further from 1.  Generally the $(z, \Delta R_{12})$ phase space for the final $\kt$ recombination is expected to be more like that illustrated in Figs.~\ref{fig:zRcontours2} and \ref{fig:zRcontours4} (coupled with the boundary in Fig.~\ref{fig:algboundariesKT}).

\begin{figure}[htbp] \begin{center}
\subfloat[$z$, CA] {\includegraphics[width = .48\columnwidth] {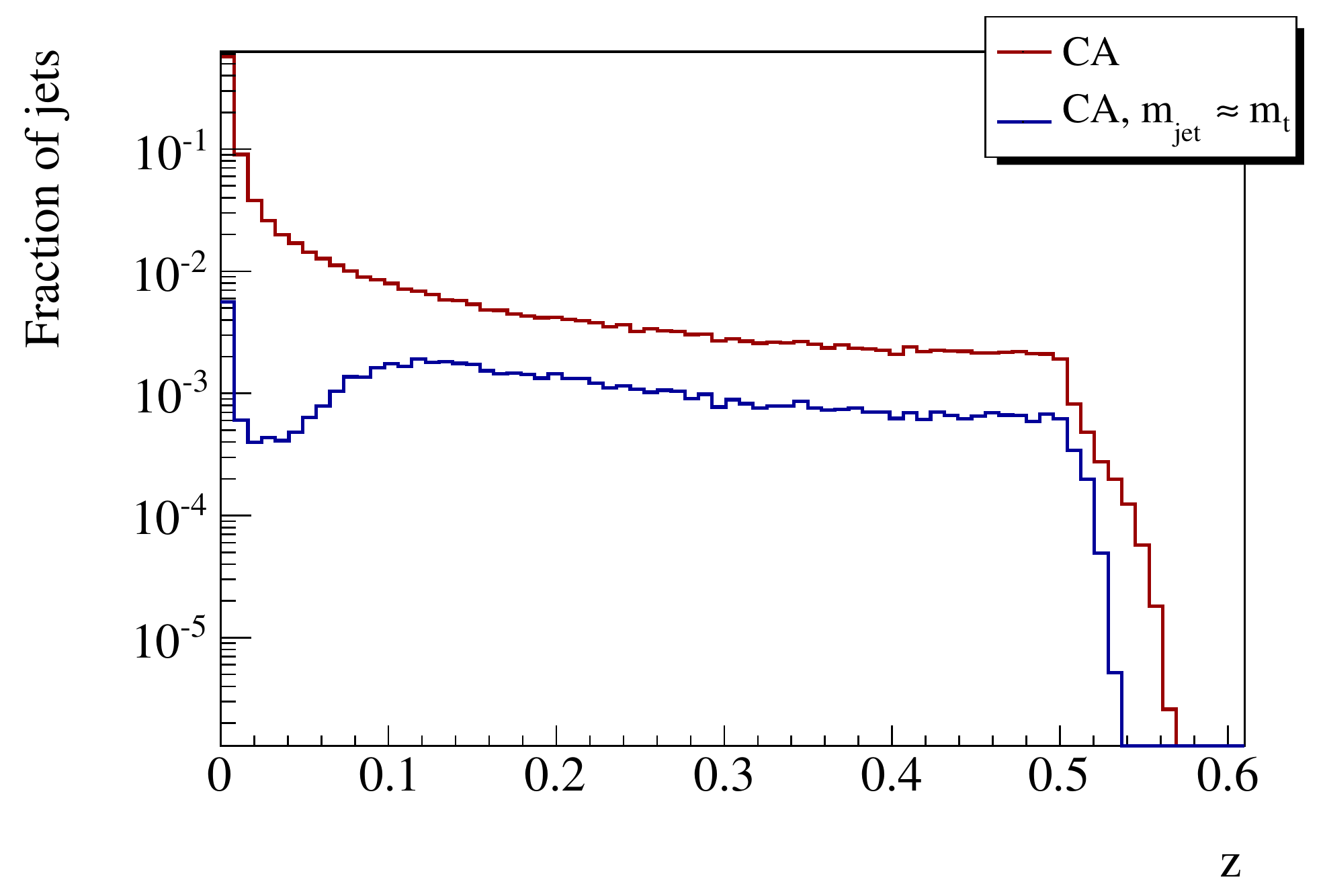} \label{fig:QCDkin:zCA}}
\subfloat[$z$, $\kt$]{\includegraphics[width = .48\columnwidth]{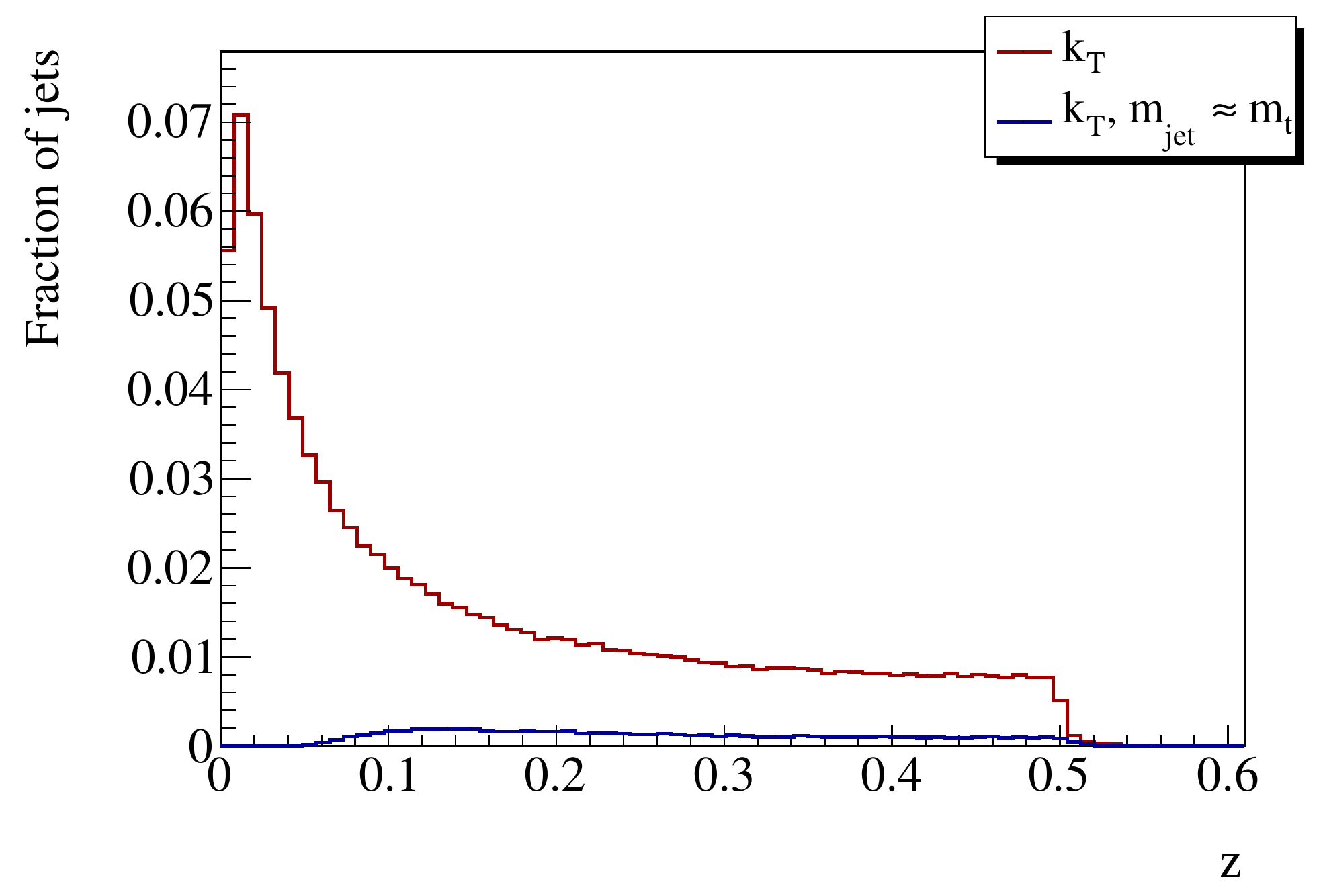} \label{fig:QCDkin:zKT}}

\subfloat[$\Delta R_{12}$, CA] {\includegraphics[width = .48\columnwidth] {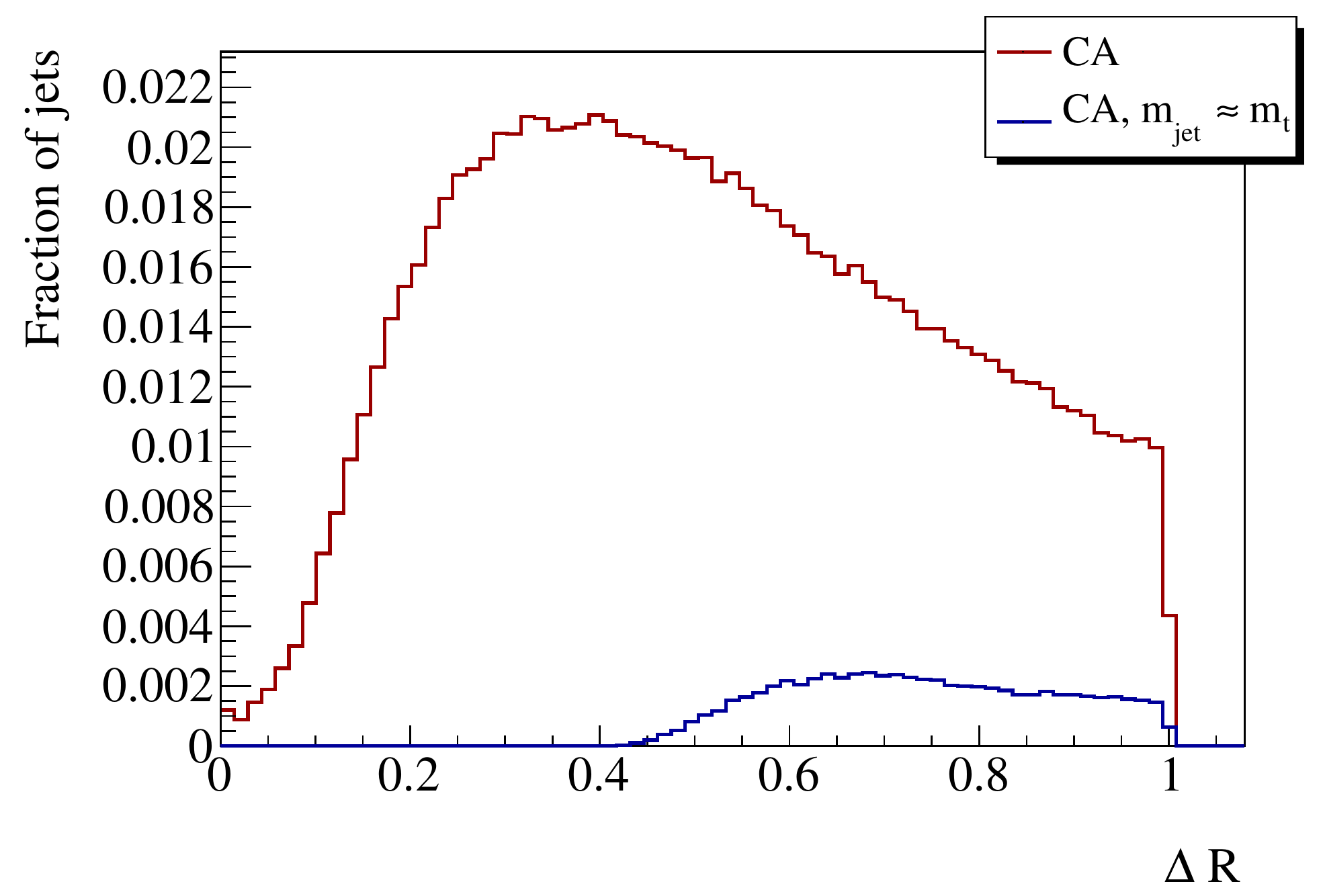}
\label{fig:QCDkin:DRCA}}
\subfloat[$\Delta R_{12}$, $\kt$] {\includegraphics[width = .48\columnwidth] {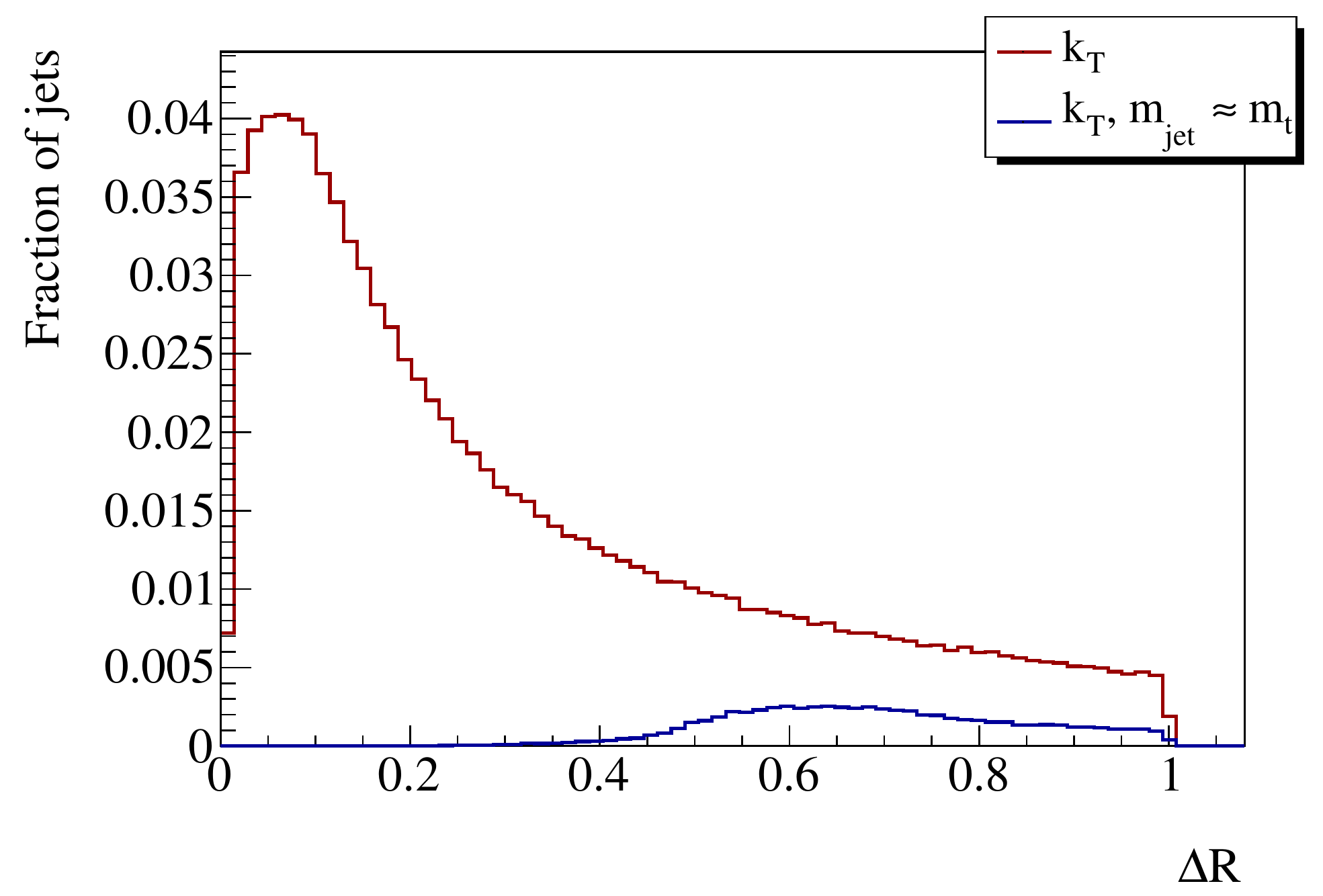} \label{fig:QCDkin:DRKT}}

\subfloat[$a_1$, CA] {\includegraphics[width = .48\columnwidth]{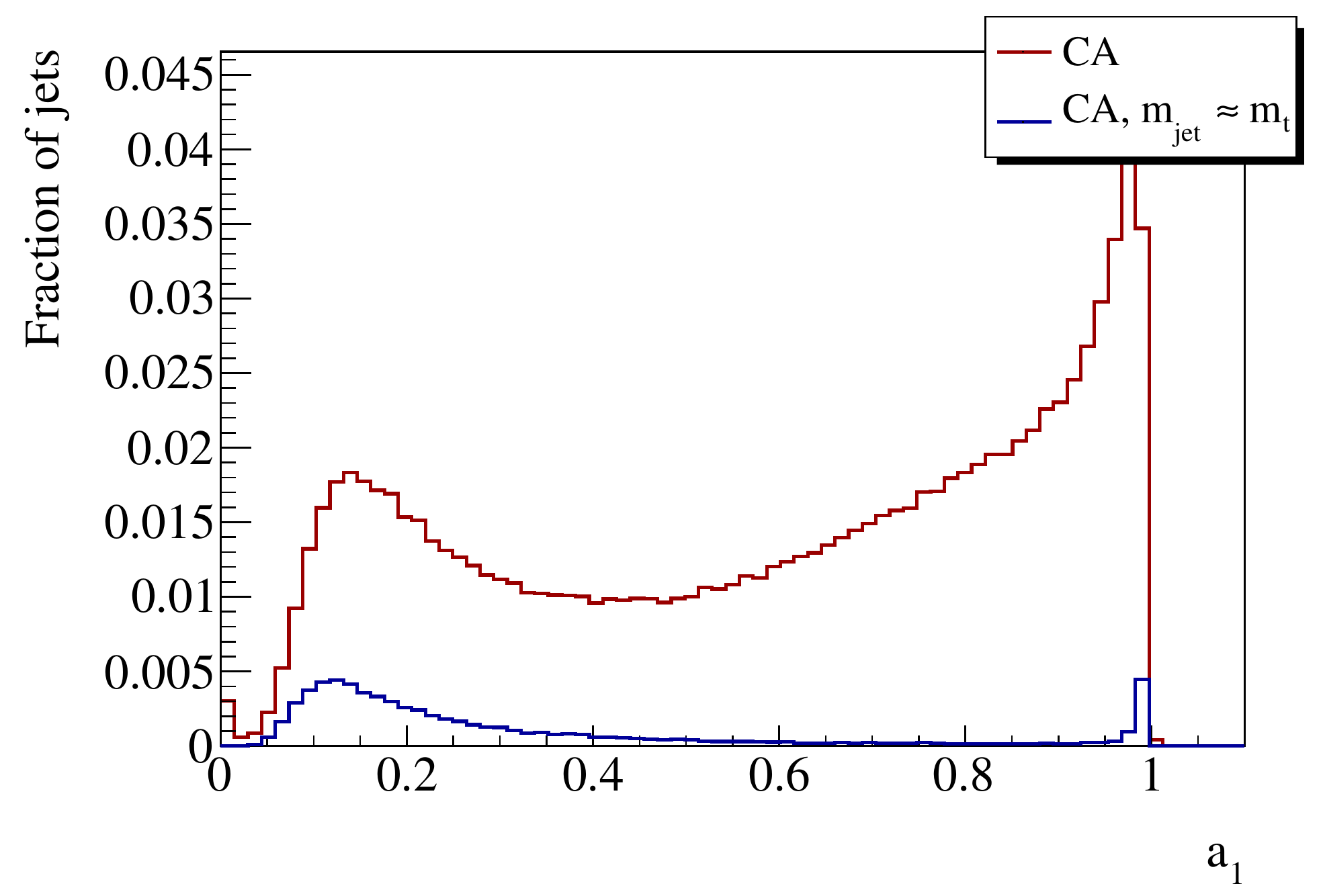} \label{fig:QCDkin:a1CA}}
\subfloat[$a_1$, $\kt$]{\includegraphics[width = .48\columnwidth]{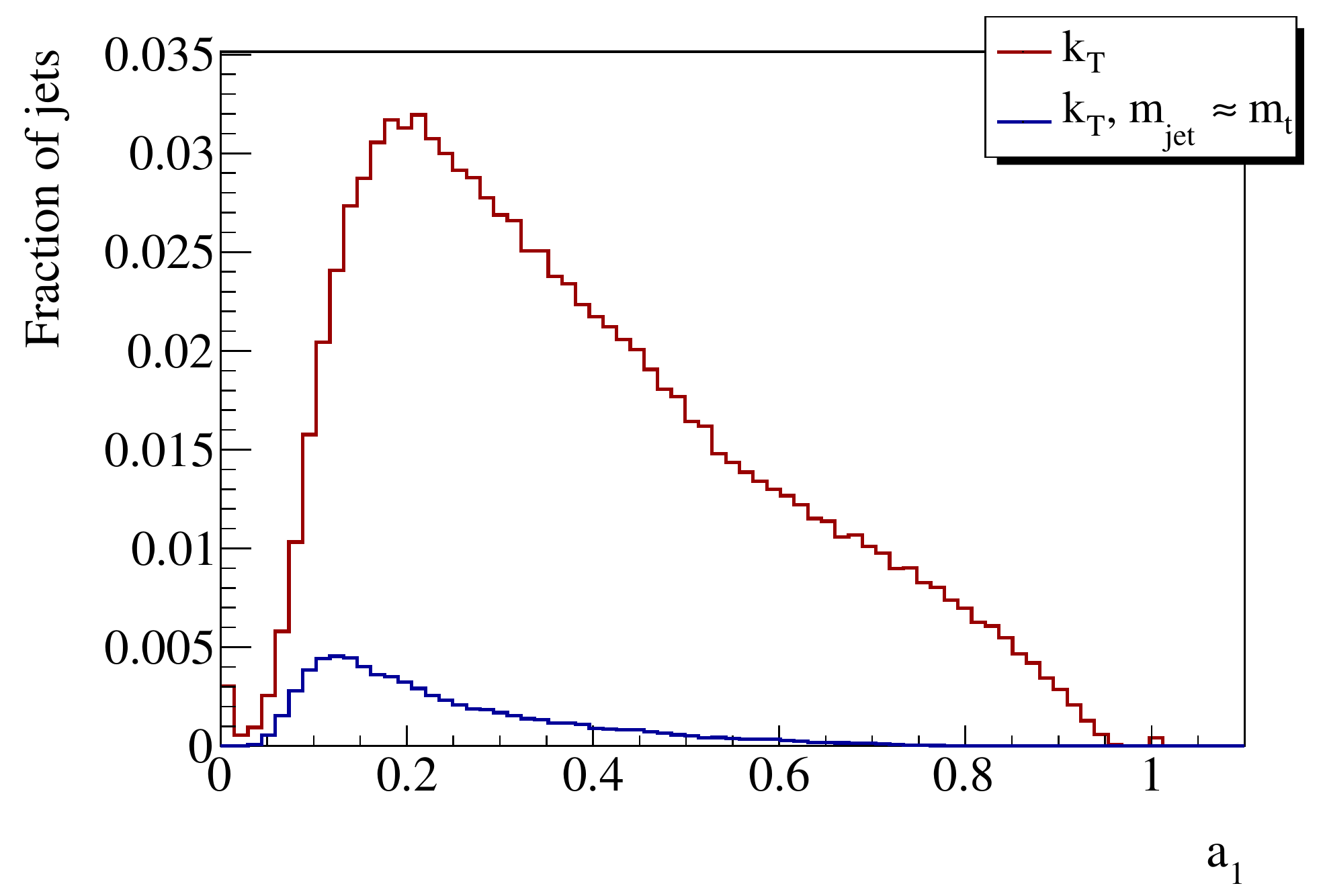}
\label{fig:QCDkin:a1KT}}
\end{center}
\caption[Distribution in $z$, $\Delta R_{12}$, and $a_1$ for QCD jets in $\ee \to q \bar q$ events]{Distribution in $z$, $\Delta R_{12}$, and the scaled (heavier) daughter mass $a_1$ for QCD jets in $\ee \to q \bar q$ events, using the CA and $\kt$ algorithms, with (red) and without (blue) a cut around the top quark mass.  D = 1.0.
}
\label{fig:QCDkin}
\end{figure}

To illustrate this discussion we have plotted distributions of $z$, $\Delta R_{12}$, and $a_1$ for the last recombination in a jet for the $\kt$ and CA algorithms in Fig.~\ref{fig:QCDkin}.  We plot distributions with and without a cut on the jet mass, where the cut is a narrow window ($\approx$ 15 GeV) around the top quark mass.  This cut selects heavy QCD jets: for the jets in this sample, with $p_T$ between 500--600 GeV, it corresponds to a cut on $x_J$ of 0.06--0.09.
These distributions reflect the combined influence of the QCD shower dynamics, the restricted kinematics from being in a jet, and the algorithm-dependent ordering effects discussed above.  Most importantly, note the very strong enhancement at the smallest values of $z$ for the CA algorithm in Fig.~\ref{fig:QCDkin:zCA}, which persists even after the heavy jet mass cut.  Note the log scale in Fig.~\ref{fig:QCDkin:zCA}!  While the $\kt$ result in Fig.~\ref{fig:QCDkin:zKT} is still peaked near zero when summed over all jet masses, the enhancement is not nearly as strong.  After the heavy jet mass cut is applied, the distribution shifts to larger values of $z$, with an enhancement remaining at small values.  Only in this last plot is there evidence of the lower limit on $z$ of order 0.1 expected from the earlier LL approximation results.

Fig.~\ref{fig:QCDkin:DRCA} illustrates the expected enhancement near $\Delta R_{12} = D = 1.0$ for CA.  Fig.~\ref{fig:QCDkin:DRKT} shows that $\kt$ exhibits a much broader distribution than CA with an enhancement for small $\Delta R_{12}$ values.  Once the heavy jet mass cut is applied, both algorithms exhibit the lower kinematic cutoff on $\Delta R_{12}$ suggested in the LL approximation results, as both distributions shift to larger values of the angle.  This shift serves to enhance the CA peak at the upper limit and moves the lower end enhancement in $\kt$ to substantially larger values of $\Delta R_{12}$.

\begin{figure}[htbp]
\begin{center}
\includegraphics[width = .46\columnwidth]{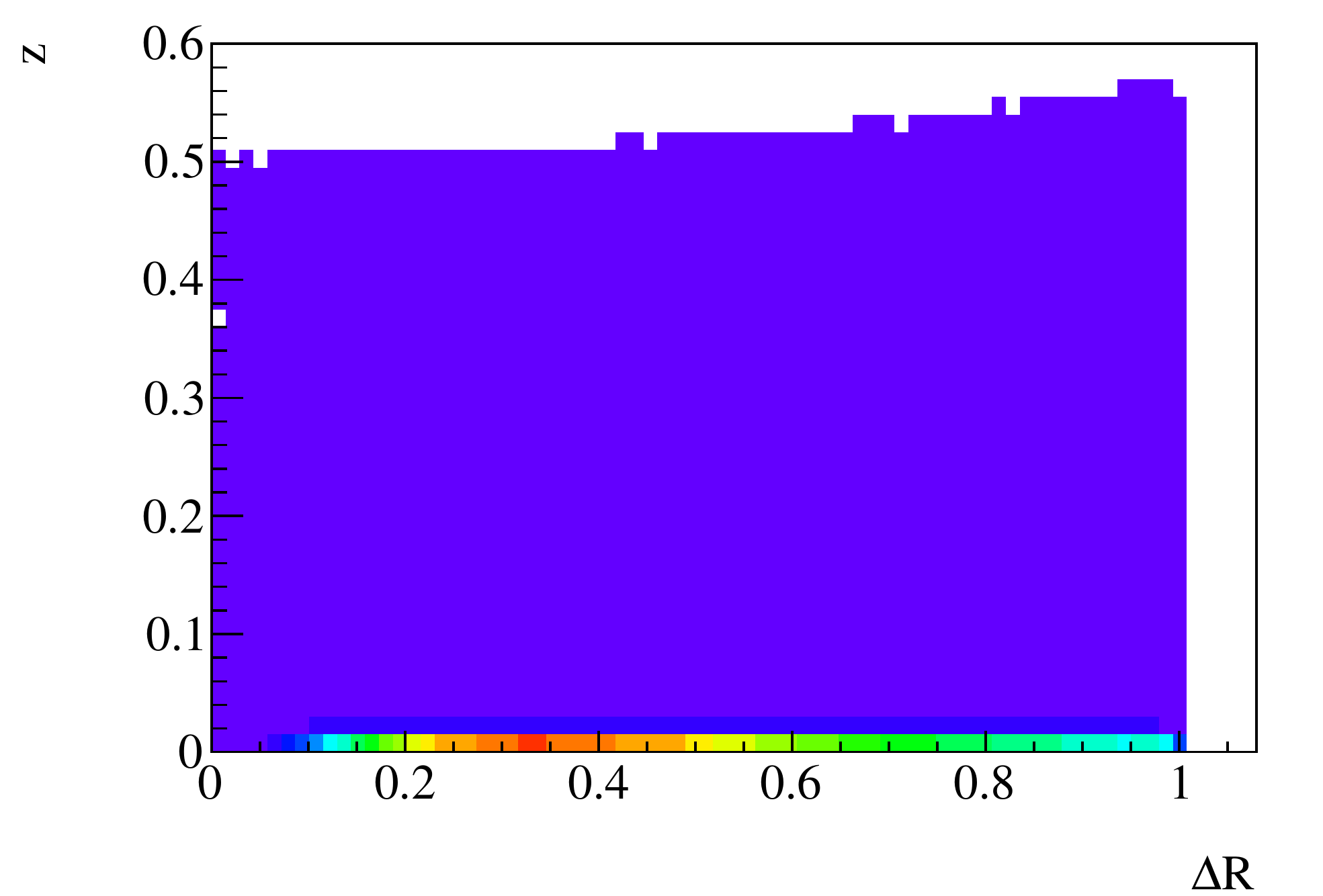}
\includegraphics[width = .52\columnwidth]{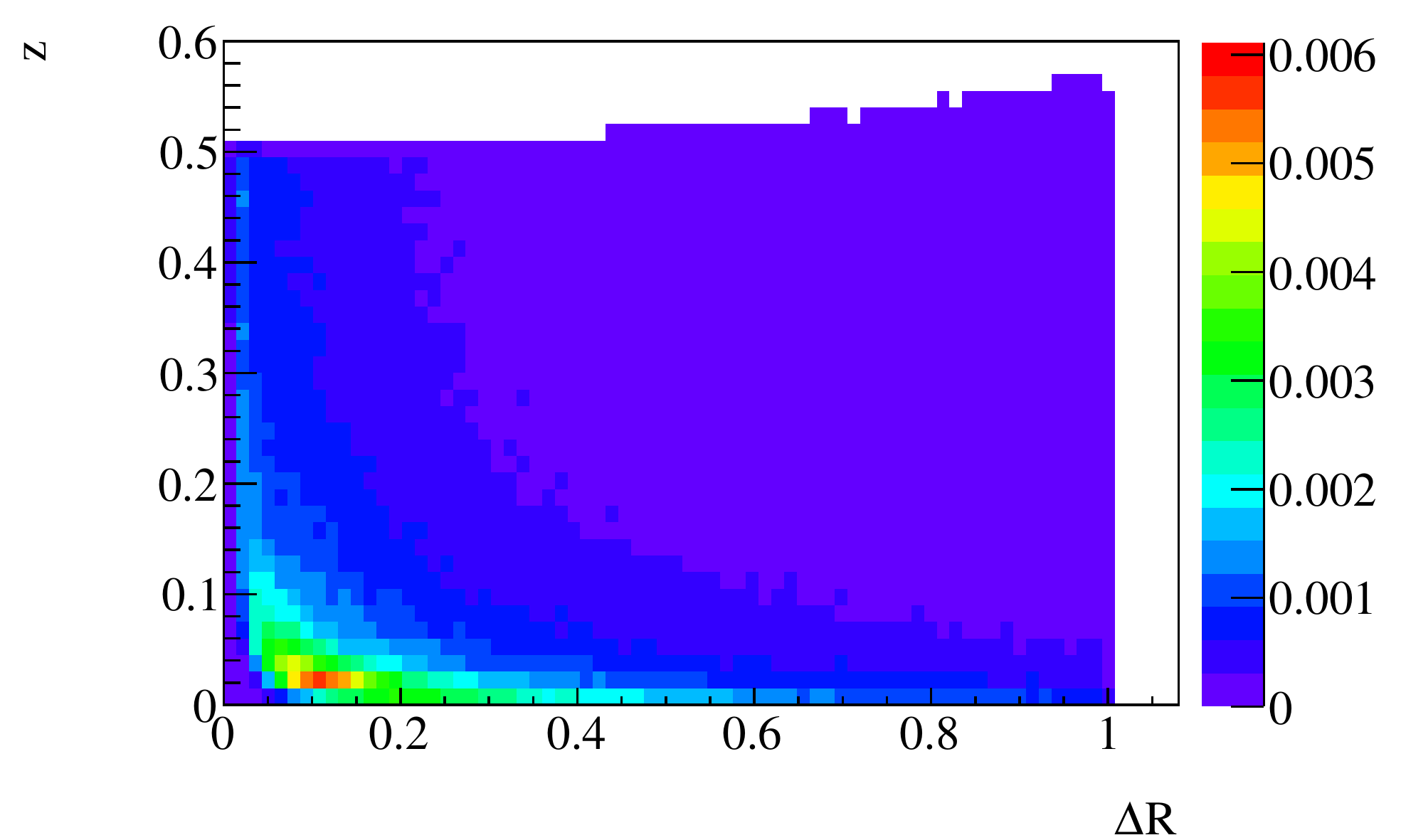}
\end{center}
\caption[Combined distribution in $z$ and $\Delta R_{12}$ for QCD jets in $\ee \to q \bar q$ events]{Combined distribution in $z$ and $\Delta R_{12}$ for QCD jets in $\ee \to q \bar q$ events, using the CA (left) and $\kt$ (right) algorithms, for jets with $p_T >$ 500 and D = 1.0.
}
\label{fig:QCDzDR}
\end{figure}

The CA algorithm bias toward large $a_1$ is demonstrated in Fig.~\ref{fig:QCDkin:a1CA}.  We can see that requiring a heavy jet enhances the large-$a_1$ peak.  The $\kt$ distribution in $a_1$, shown in Fig.~\ref{fig:QCDkin:a1KT}, exhibits a broad enhancement around $a_1 \approx 0.4$.  This distribution is relatively unchanged after the jet mass cut.  To give some insight into the correlations between $z$ and $\Delta R_{12}$, in Fig.~\ref{fig:QCDzDR} we plot the distribution of both variables simultaneously for both algorithms, with no jet mass cut applied.
The very strong enhancement at small $z$ and large $\Delta R_{12}$ for CA is evident in this plot.  For $\kt$, there is still an enhancement at small $z$, but there is support over the whole range in $z$ and $\Delta R_{12}$ with the impact of the shaping due to the $z \times \Delta R_{12}$ dependence in the metric clearly evident.  Note that the $\kt$ distribution is closer to what one would expect from QCD alone, with enhancements at \emph{both} small $z$ and small $\Delta R_{12}$, while the CA distribution is asymmetrically shaped away from the QCD-like result.  Finally we should recall, as indicated by Fig.~\ref{fig:mjetplot}, that the jets found by the two algorithms tend to be slightly different, with the $\kt$ algorithm recombining slightly more of the original (typically soft) protojets at the periphery and leading to slightly larger jet masses.

Because the QCD shower is present in all jets, and is responsible for the complexity in the jet substructure, the systematic effects discussed above will be present in all jets.  While the kinematics of a heavy particle decay is distinct from QCD in certain respects, we will find in the next subsection that these effects still present themselves in jets containing the decay of a heavy particle.

\subsection{Jets from heavy particle decays}

For an example of a heavy particle decay, we know consider the systematic effects of showering and the jet algorithm on top quark jets.  We consider $\ee \to t \bar t$ events with $Q = 1200$ GeV, so each top quark will have $E = 600$ GeV, and will tend to appear within a single jet (we use $D$ = 1.0).  Details of the event generation are given in Appendix \ref{app:details}.  Note that even in the relatively clean $\ee \to t\bar t$ events, the top quarks are not themselves color singlets, so hadronization connects the jets --- fortunately this is a small effect since the top quarks' energy and mass are both much larger than the hadronization scale.  After reconstructing ``top jets'', we expect that the kinematics of the last few mergings/splittings will differ in important ways from our parton-level predictions.  For instance, with the CA algorithm we expect that soft recombinations will occur at the last recombination step, even for jets that contain the decay products of a top quark.  This can make the substructure look more like a heavy QCD jet than a top quark decay, and subsequently the jet may not be properly identified.

\begin{figure}[htbp] \begin{center}
\includegraphics[width = .7\columnwidth]{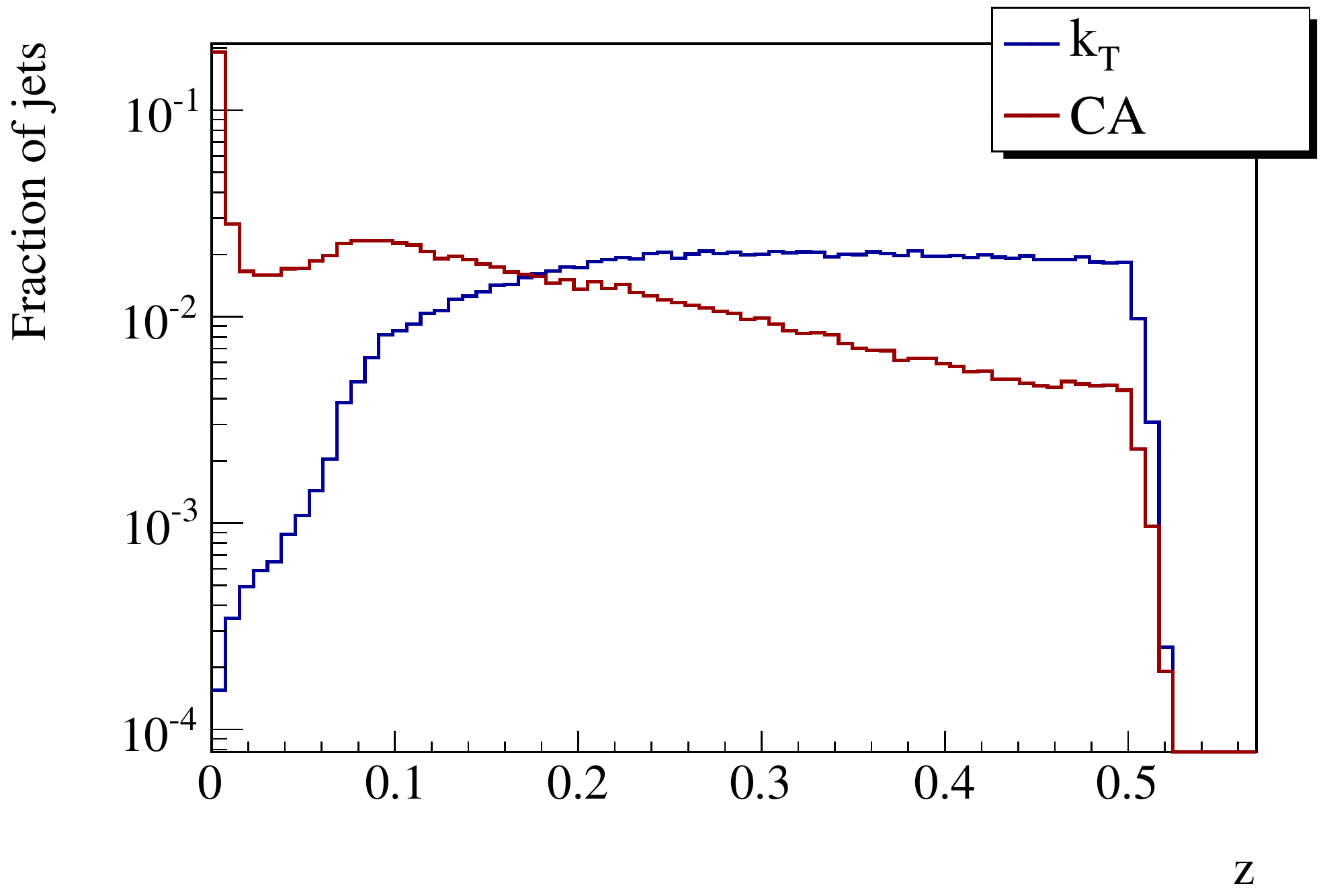}
\end{center}
\caption[Distribution in $z$ for jets with the top mass in $\ee \to t \bar t$ events]{Distribution in $z$ for jets with the top mass in $\ee \to t \bar t$ events.  D = 1.0.}
\label{fig:topz}
\end{figure}

To demonstrate this point, in Fig.~\ref{fig:topz} we plot the distribution in $z$ for jets with mass within a window around the top quark mass.  The distribution for CA jets is very different from the parton-level distribution (Fig.~\ref{fig:topzdist}).  The excess at small values of $z$ arises from soft recombinations in the CA algorithm, which make the distribution similar to that for QCD jets (Figs.~\ref{fig:QCDkin:zCA} and \ref{fig:QCDkin:zKT}).  For the $\kt$ algorithm, there are rarely soft recombinations late in the algorithm, because the metric orders according to $z$ as well as $\Delta R$.

\begin{figure}[htbp] \begin{center}
\includegraphics[width = .7\columnwidth]{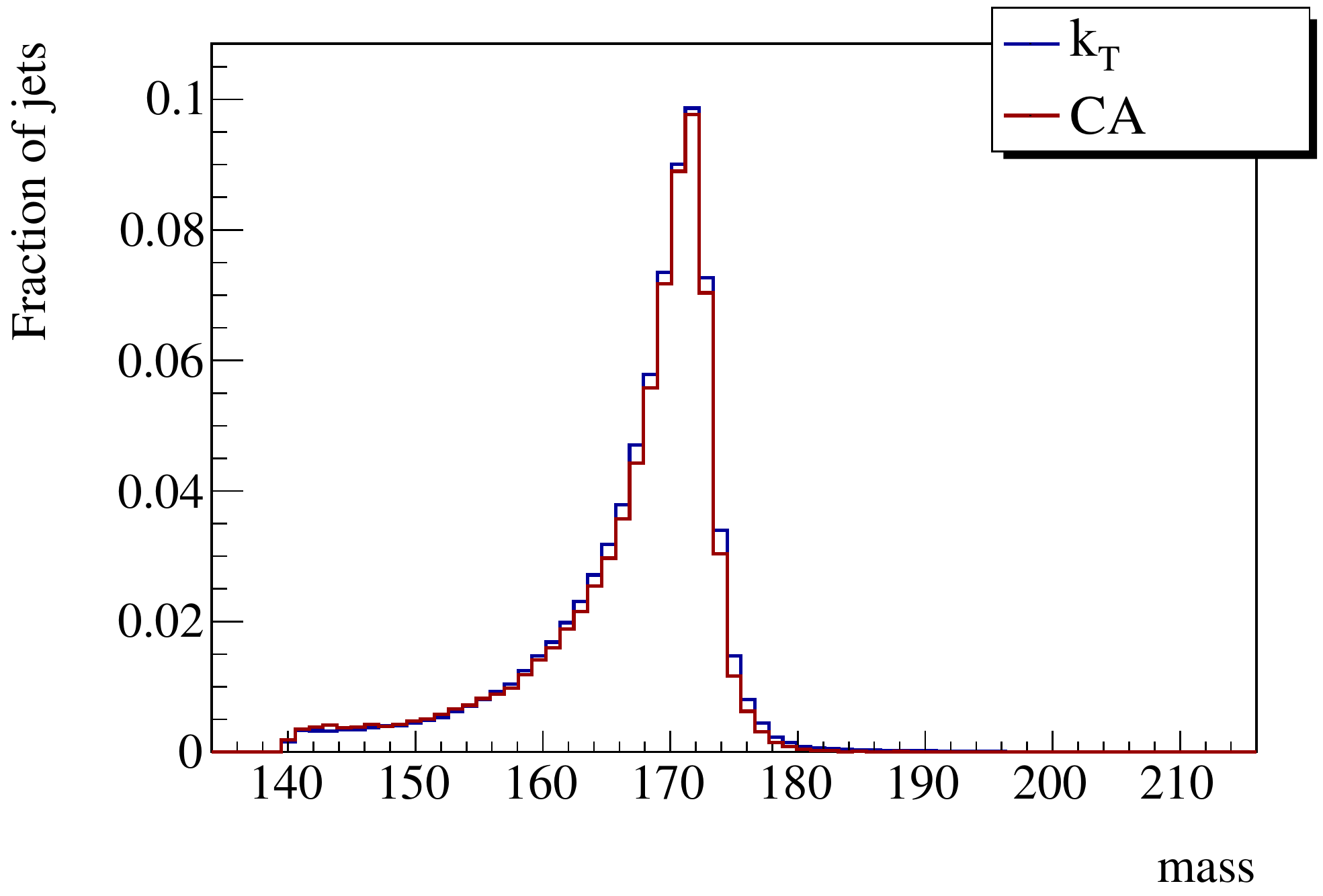}
\end{center}
\caption[Distribution in jet mass for jets in the neighborhood of the top mass in $\ee \to t\bar{t}$ events]{Distribution in jet mass for jets in the neighborhood of the top mass in $\ee \to t\bar{t}$ events for the CA (black) and $\kt$ (red) algorithms.  D = 1.0.}
\label{fig:topmass_EE}
\end{figure}

In these relatively clean events, the $\kt$ and CA algorithms find very nearly the same jets.  This can be seen in Fig.~\ref{fig:topmass_EE}, where we plot the jet mass distribution for both algorithms.  Thus the effects we have seen stem from different \emph{ordering} in the algorithms, not differences in the particles that get included in the jet.  We will see in the next section that differences in what is included in the jet play a bigger role at hadron colliders.

\section{Event effects on jet substructure in hadron collisions}
\label{sec:sub:eventeffects}

At a hadron collider like the LHC, there are additional systematic effects on jet substructure.  We need to account for the combined effect of splash-in from several sources: initial state radiation (ISR,  the radiation from the incoming partons in the hard scattering), the underlying event (UE, the rest of the $pp$ interaction), and pile-up (other $pp$ collisions that occur in the same time bin).  All of these sources add particles to jets that are typically soft and approximately uncorrelated.  Splash-in particles will mostly be located at large angle to the jet core, simply because there is more area there.  How these particles affect jet substructure depends on the algorithm used.  We expect them to contribute similarly to soft radiation from the QCD shower, discussed in the previous section.  In this section we will consider the effects of adding ISR and UE.  We expect the effects of pile-up will be of a similar nature, although possibly of a much greater magnitude depending on the collider luminosity.

We should note that such a clean separation of different effects is artificial.\footnote{Except in the case of pile-up, where the separation is perfectly well defined.}  Whether outgoing gluons were radiated from initial- or final-state partons is not quantum-mechanically meaningful, so the amplitudes for initial- and final-state radiation must interfere.  The same is true for the underlying event.  In addition to interference at the perturbative level, hadronization in general will, and often must, link these different processes together.  The particles seen by the detector are of course color singlets, so quarks and gluons in the ``final state'' must connect with each other or the rest of the event to form hadrons.  This makes the question of whether a hadron belongs to final-state radiation, initial-state radiation, or underlying event inherently ambiguous.  In this section as we progressively include more of the event in our Monte Carlo samples, we should think of this as building a progressively more realistic model of QCD into the Monte Carlo, not as simply adding another source of final-state hadrons.

\subsection{Mass effects}
\label{sec:sub:eventeffects:mass}

\begin{figure}[htbp] \begin{center}
\subfloat[CA] {\includegraphics[width = .48\columnwidth] {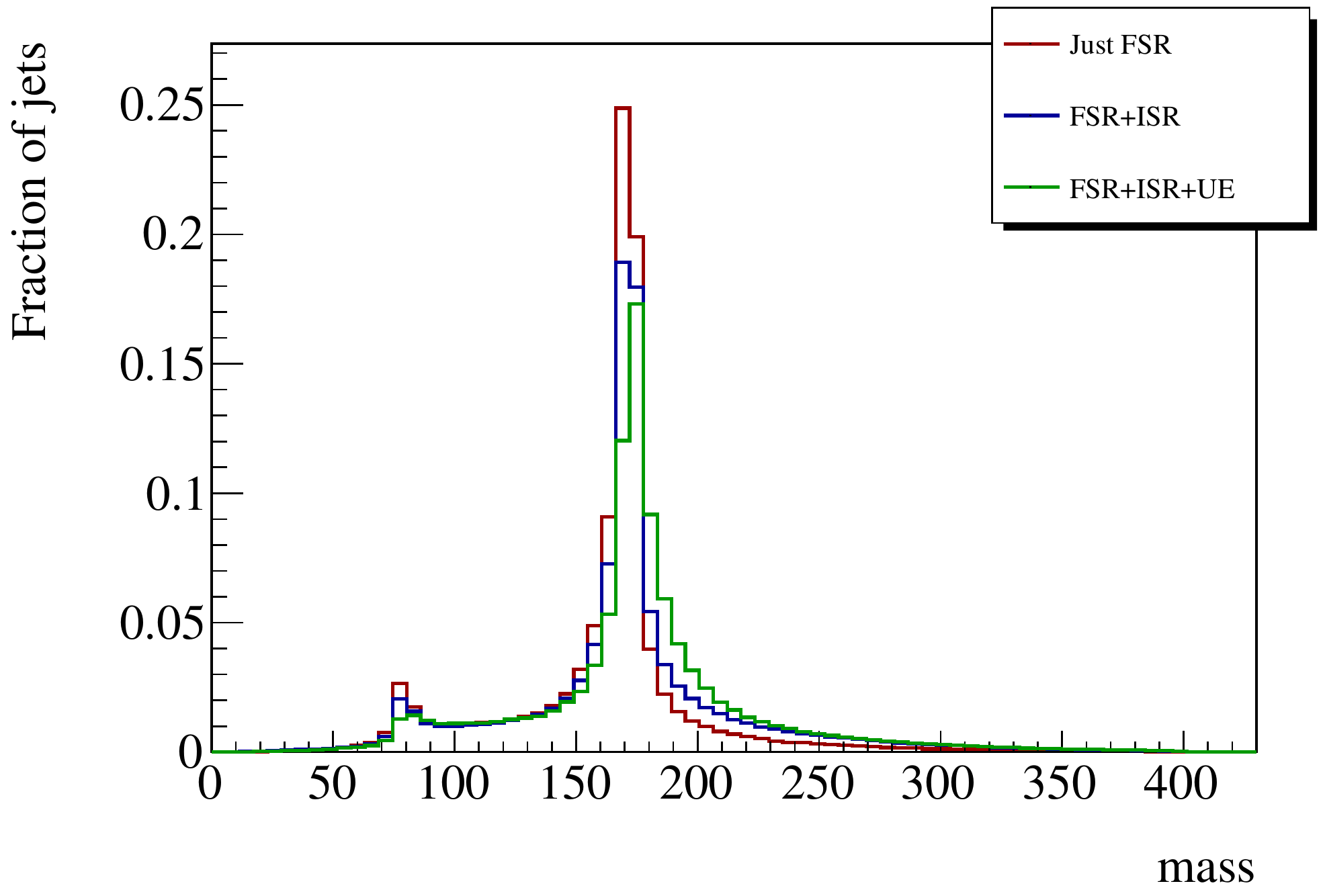} \label{fig:ttbarPPMassMerged:CA}}
\subfloat[CA (zoomed in)] {\includegraphics[width = .48\columnwidth] {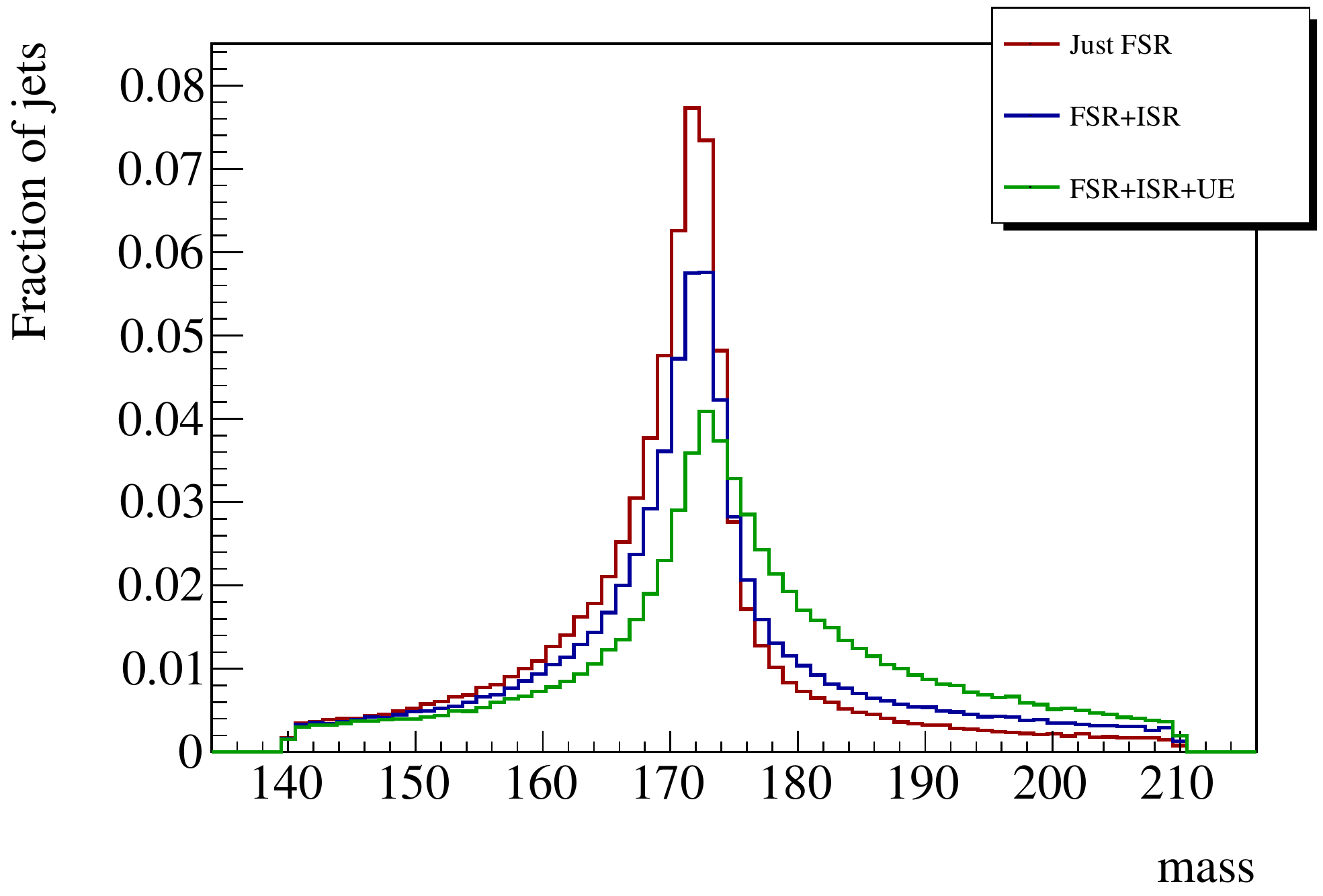}
\label{fig:ttbarPPMassMerged:CAnarrow}}

\subfloat[$\kt$]{\includegraphics[width = .48\columnwidth]{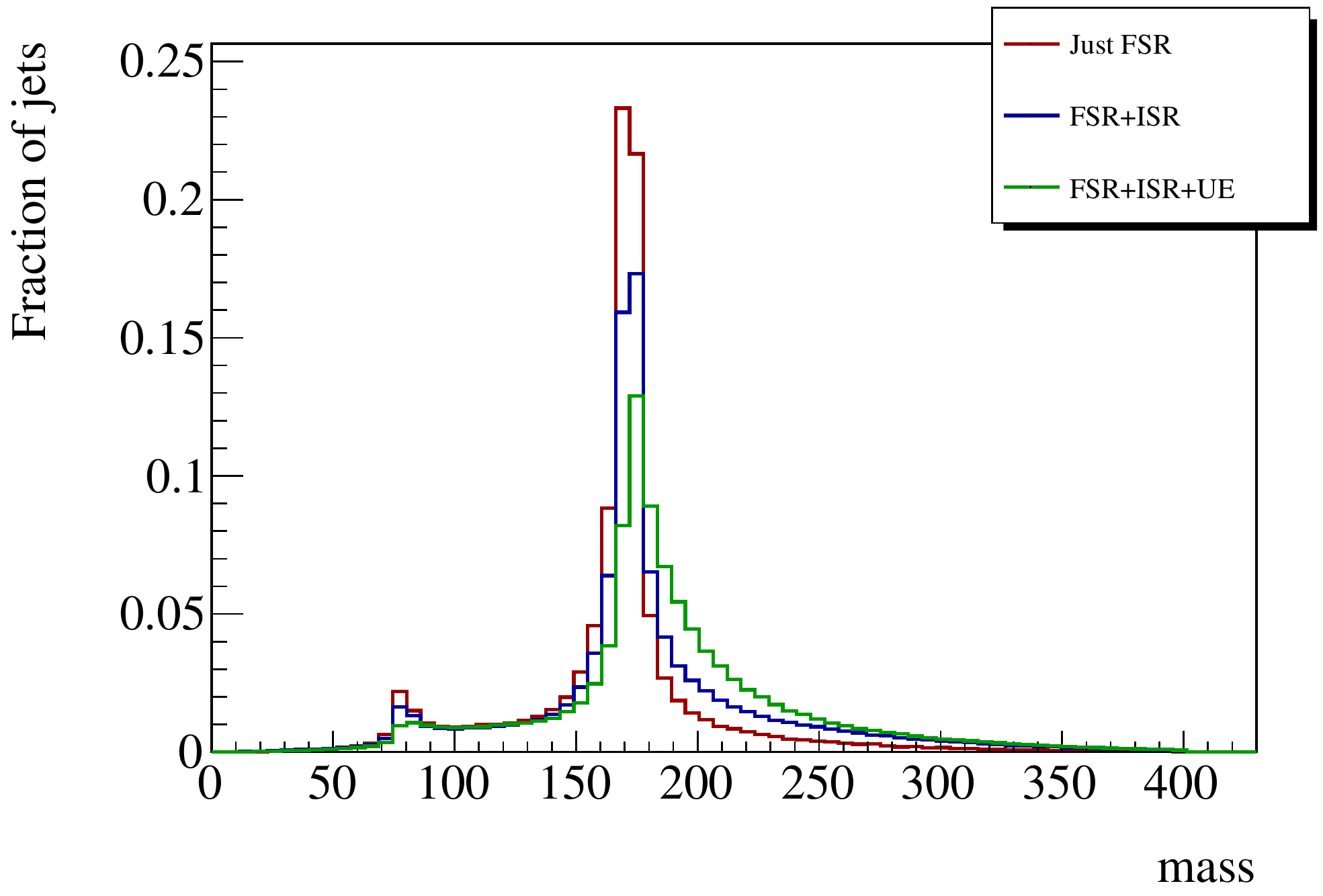} \label{fig:ttbarPPMassMerged:KT}}
\subfloat[$\kt$ (zoomed in)] {\includegraphics[width = .48\columnwidth] {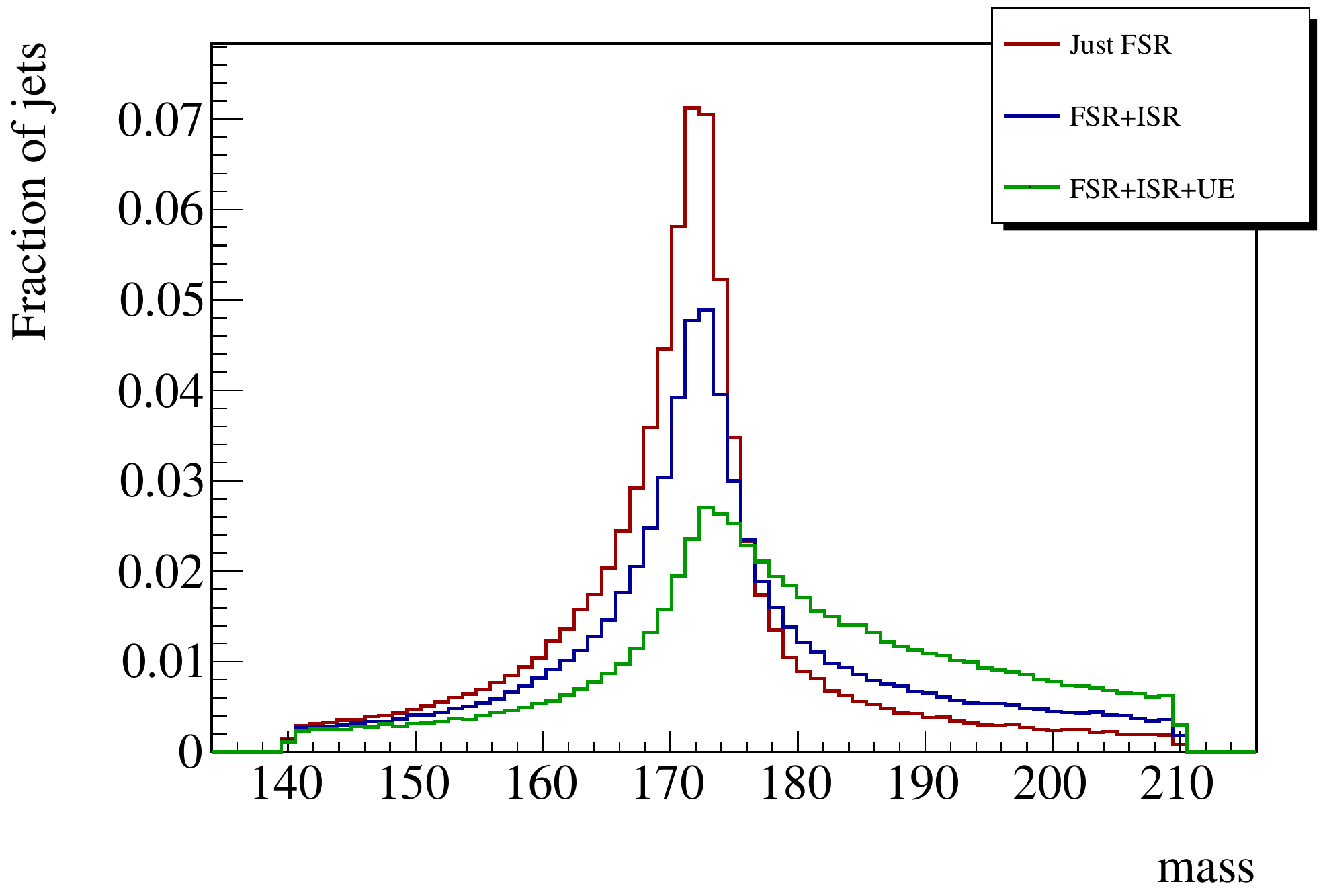} \label{fig:ttbarPPMassMerged:KTnarrow}}
\end{center}
\caption[Distribution in $m_J$ for ttbar jets in $pp \to t \bar t$ events]{Distribution in $m_J$ for ttbar jets in $pp \to t \bar t$ events, using the CA and $\kt$ algorithms, with only FSR (blue), including ISR (red), and including ISR and UE (green).  Jets have $p_T > 500$ GeV and D = 1.0.}
\label{fig:ttbarPPMassMerged}
\end{figure}

As we consider jets in hadron collisions, the natural place to begin is jet masses.  In Fig.~\ref{fig:ttbarPPMassMerged} we plot the mass distribution for jets in $t\bar t$ events with the CA and $\kt$ algorithms.  In each plot we show the distribution for three kinds of Monte Carlo samples from \prog{Pythia}: events where we only include radiation from final-state partons, events including initial-state radiation, and events including both initial-state radiation and underlying event activity.  The precise details of these samples are given in Appendix \ref{app:details}.

As we include more of the full event's activity, the jet mass distribution is broadened, with a peak that shifts upward.  We can contrast these results with Fig.~\ref{fig:topmass_EE} where we found the equivalent distribution for $\ee \to t\bar t$ events.  In that case, the jet mass distribution had a clean upper bound at the top mass, with a tail for lower masses.  This has a simple interpretation: in $\ee \to t\bar t$ events, essentially all final-state hadrons come from one top decay or the other, and for high-$Q^2$ events these are well separated.  We reconstruct jets that can encompass the entire top quark decay, but there is nothing else to pick up so the mass distribution cuts off at $m_t$.  Some amount of radiation will in general be emitted outside the jet radius, leading to the tail at lower masses.

This lower tail shows up again in $pp$ events, but now a high-mass tail is present as well.  For the FSR sample, the tail is slightly larger than for $\ee$ events: even without full ISR and UE simulation, \prog{Pythia} must arrange color connections to produce outgoing hadrons and deal with the beam remnants, so even these events are not as clean as in $\ee$ collisions.  Note that whereas the process $\ee \to t\bar t$ always occurs through an electroweak boson, at a $pp$ collider the dominant process is gluon fusion ($gg \to g^* \to t\bar t$), so the final state is not a color singlet.

As we include the full effects of ISR and UE, there is more radiation in the event that can be included in the top quark jets.  This naturally leads to broader mass distributions and a higher mass peak.  We can see that adding UE has a much bigger effect than adding ISR.  In understanding Monte Carlo simulations as well as data, UE will be the more important consideration.

In $\ee$ collisions, the CA and $\kt$ algorithms found essentially the same jets despite their different substructure ordering.  We can see by comparing the upper and lower figures in Fig.~\ref{fig:ttbarPPMassMerged} that this is not the case in $pp$ events.  $\kt$ jets, while similar to CA jets in the FSR sample, are substantially more susceptible to the mass broadening induced by the addition of ISR and UE.  This effect is a manifestation of the $\kt$ algorithm's larger and more irregular ``jet area'' \cite{JetAreas}.

\begin{figure}[htbp] \begin{center}
\subfloat[CA] {\includegraphics[width = .48\columnwidth] {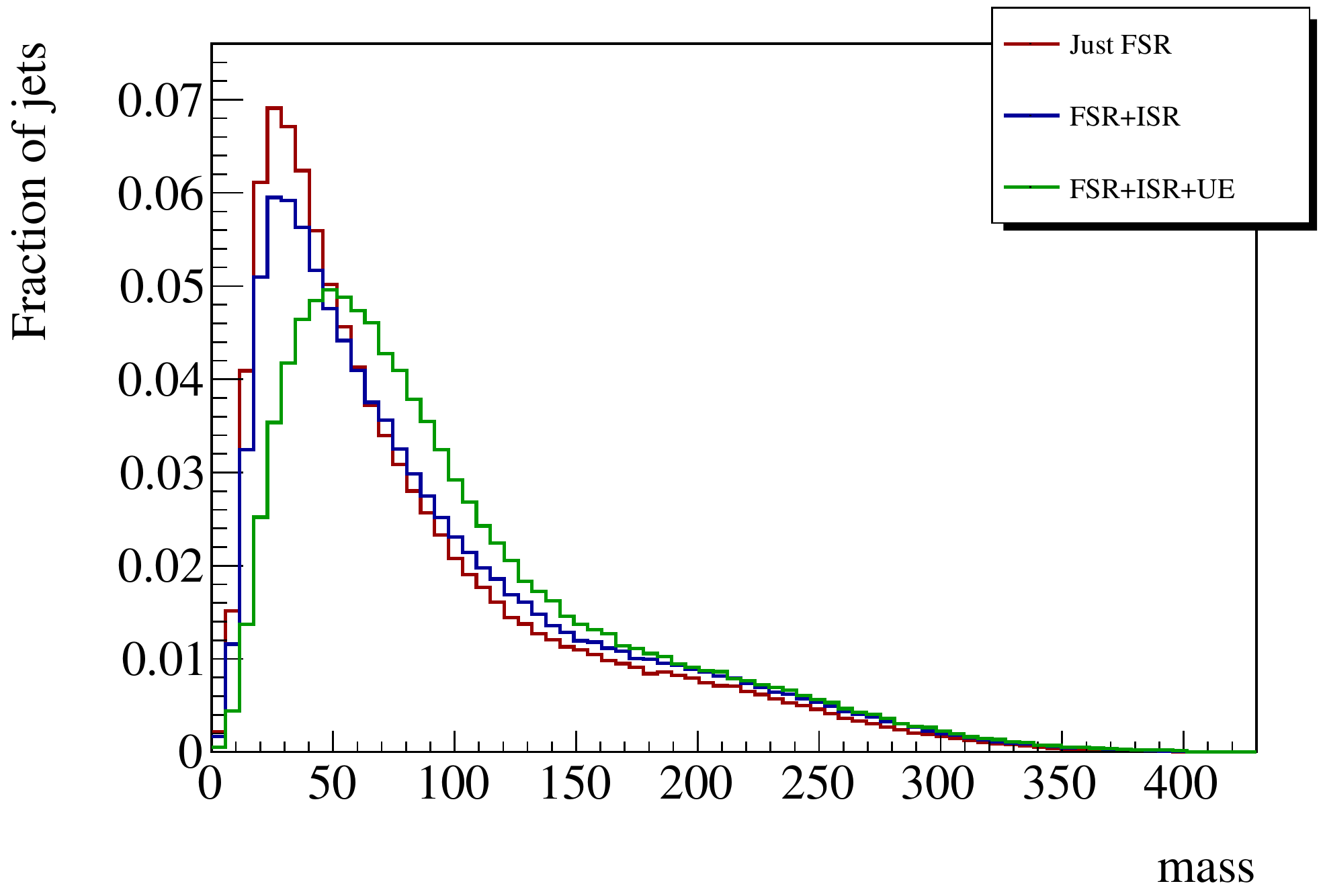} \label{fig:mQCDPPMassMerged:CA}}
\subfloat[CA (zoomed in)] {\includegraphics[width = .48\columnwidth]{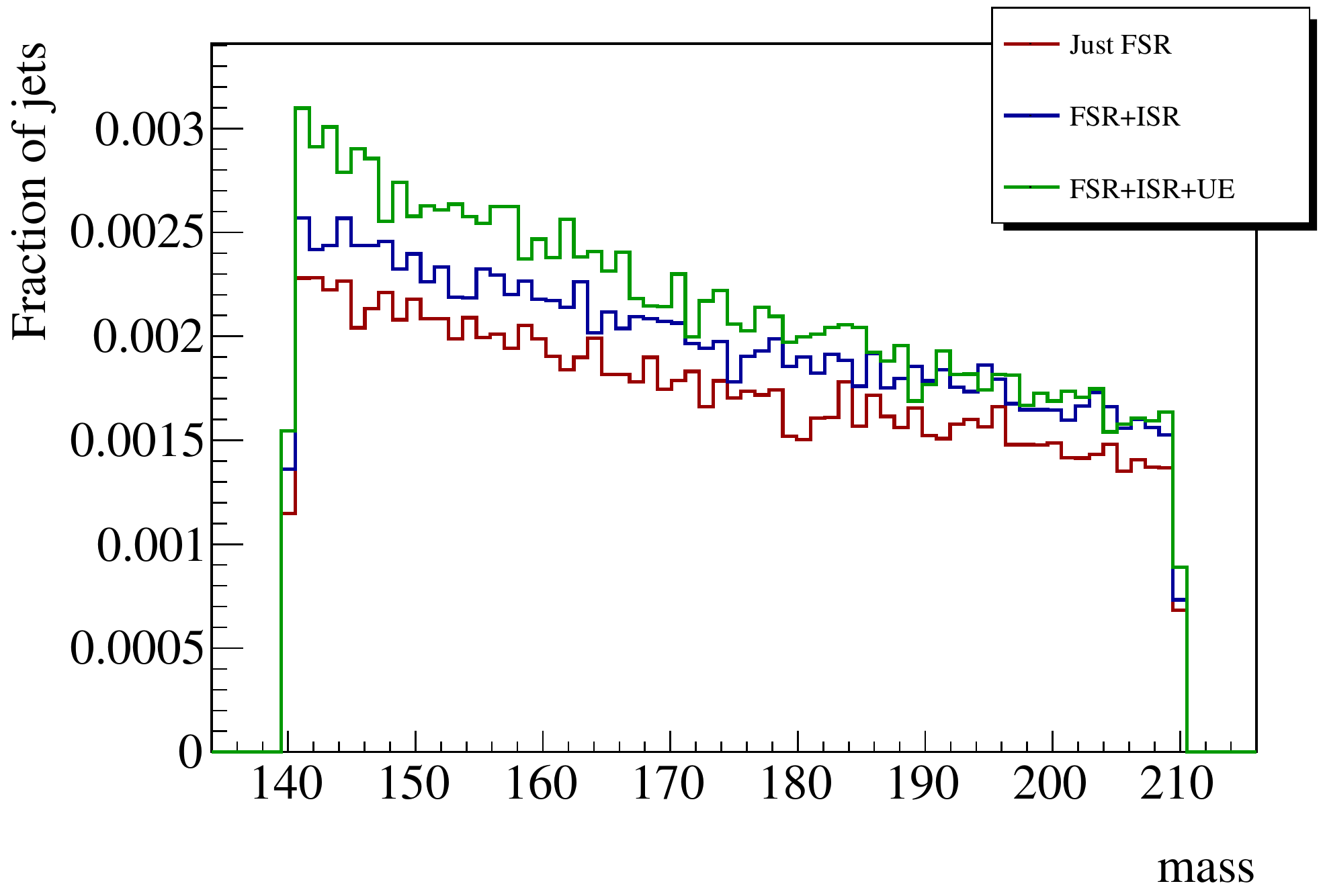}
\label{fig:mQCDPPMassMerged:CAnarrow}}

\subfloat[$\kt$]{\includegraphics[width = .48\columnwidth]{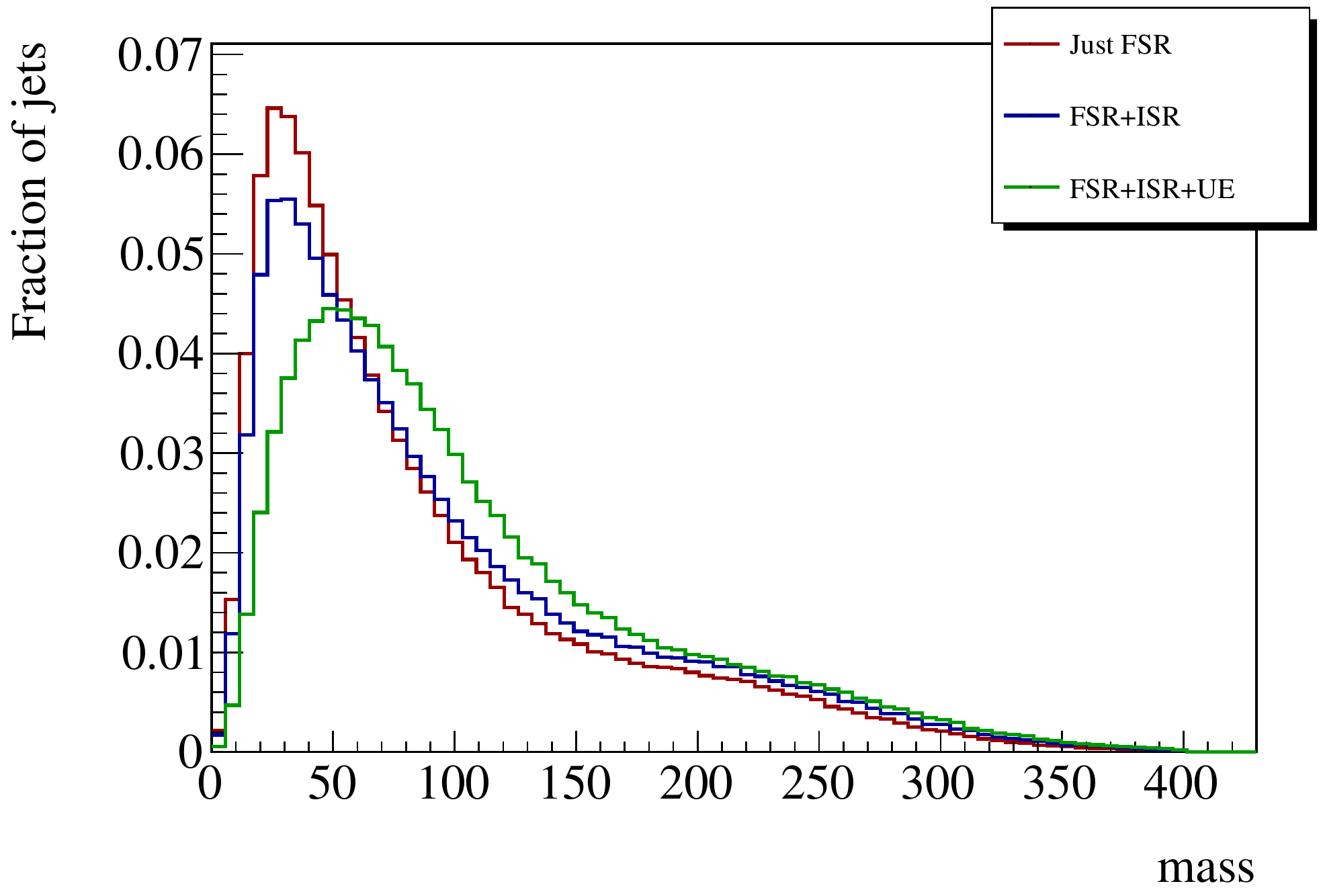} \label{fig:mQCDPPMassMerged:KT}}
\subfloat[$\kt$ (zoomed in)] {\includegraphics[width = .48\columnwidth] {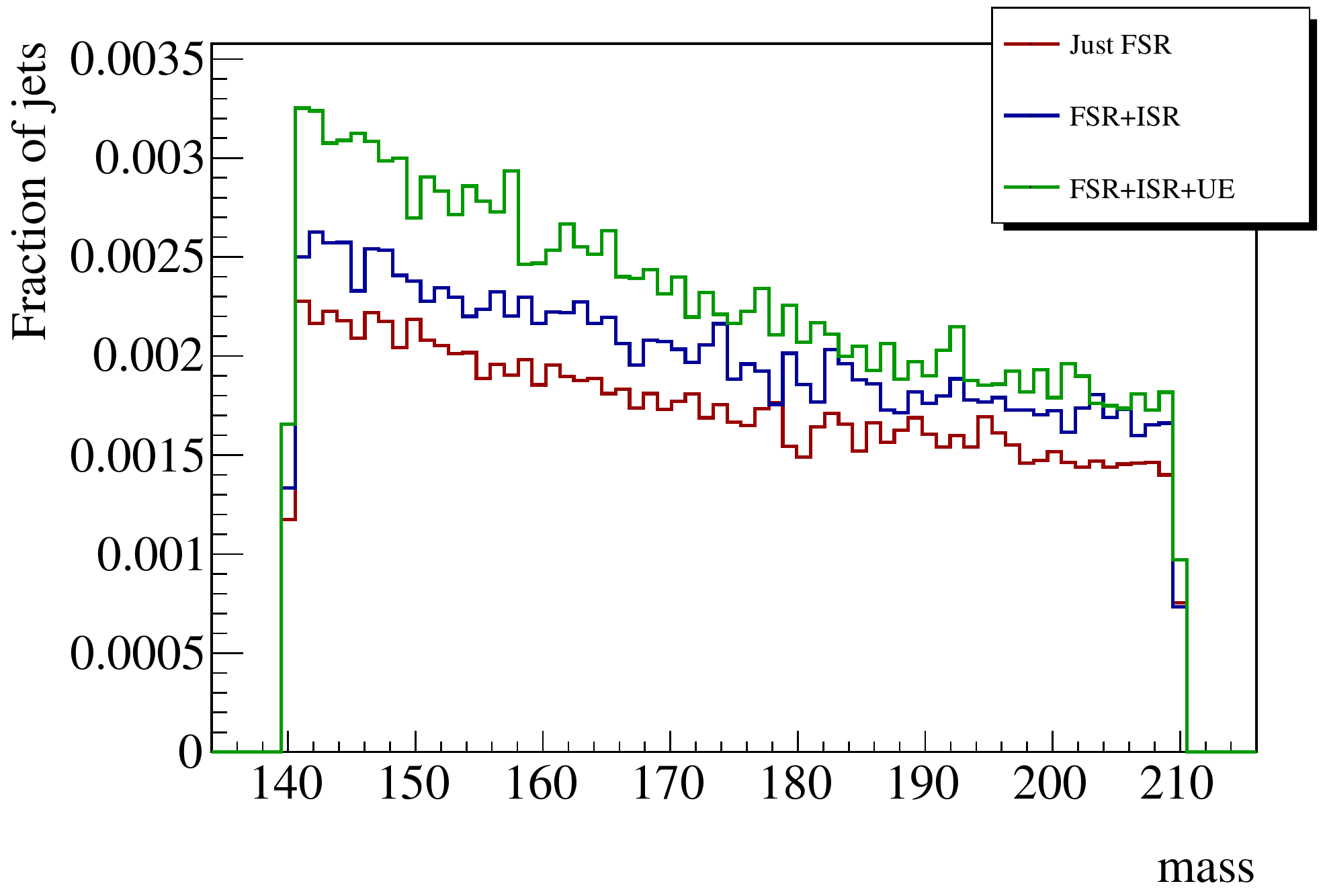} \label{fig:mQCDPPMassMerged:KTnarrow}}
\end{center}
\caption[Distribution in $m_J$ for QCD jets in matched $pp \to \text{jets}$ events]{Distribution in $m_J$ for QCD jets in matched $pp \to \text{jets}$ events, using the CA and $\kt$ algorithms, with only FSR (blue), including ISR (red), and including ISR and UE (green).  Jets have $p_T > 500$ GeV and D = 1.0.}
\label{fig:mQCDPPMassMerged}
\end{figure}

In Fig~\ref{fig:mQCDPPMassMerged}, we show the analogous plots for jet masses in QCD multijet events.  Broadly, the same effects are visible as in $t\bar t$ events.  Adding ISR and UE shifts the jet mass distribution upward, significantly increasing the number of jets falling inside the top quark's mass window.  The $\kt$ algorithm is again more susceptible to the extra radiation although the effect is less pronounced.

If our goal is to search for top quarks by looking for jets with a mass near $m_t$, we can see that the hadronic environment has two pernicious effects.  First, the mass distribution for top quark jets --- the signal distribution --- is broadened and has a lower peak.  Second, the multijet mass distribution --- the background distribution --- is shifted upwards so that the number of background events is larger in the region we're interested in.  For the reasons discussed at the beginning of this section, completely removing the effects of ISR and UE is not possible even in principle.  But to the extent  we can remove them we will improve our ability to identify heavy particles in jets.

\subsection{Substructure effects}
\label{sec:sub:eventeffects:sub}

To further explore the effects of the hadronic environment on heavy particle jets, we now turn to jet substructure.  An understanding of how event effects appear in jet substructure will help us see how to mitigate them.

\begin{figure}[htbp] \begin{center}
\subfloat[$z$, CA] {\includegraphics[width = .48\columnwidth] {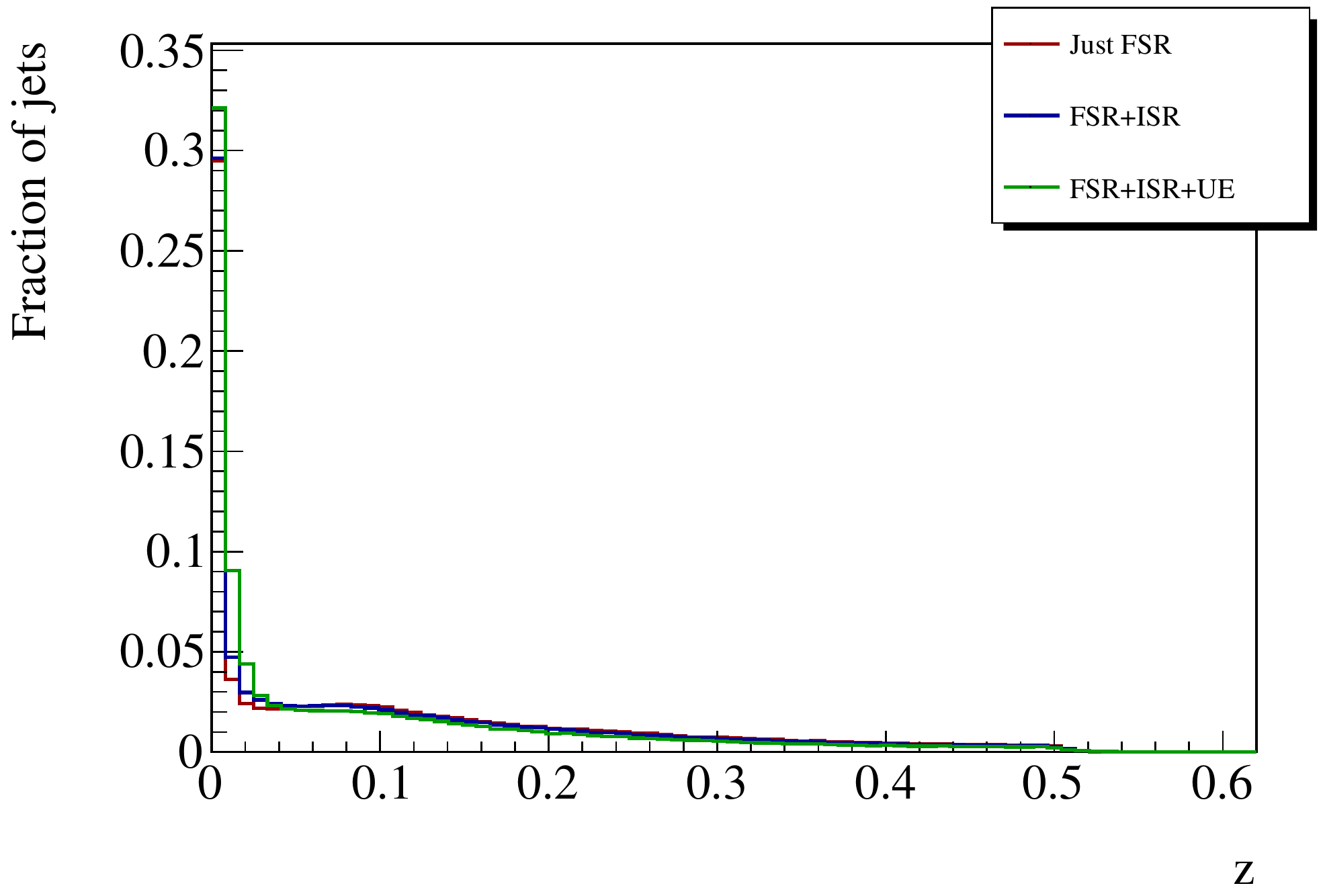} \label{fig:ttbarPPMerged:zCA}}
\subfloat[$z$, $\kt$]{\includegraphics[width = .48\columnwidth]{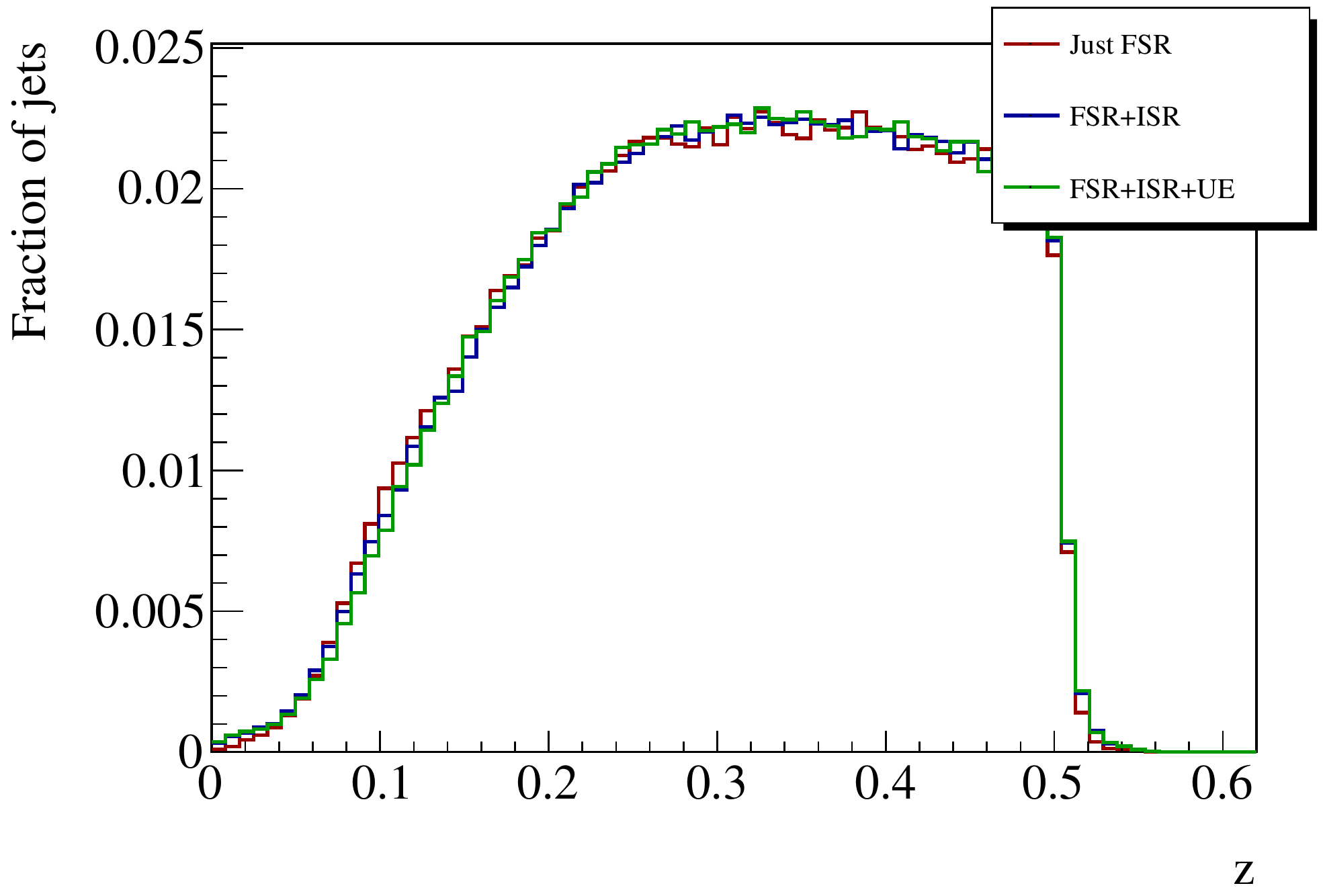} \label{fig:ttbarPPMerged:zKT}}

\subfloat[$\Delta R_{12}$, CA] {\includegraphics[width = .48\columnwidth] {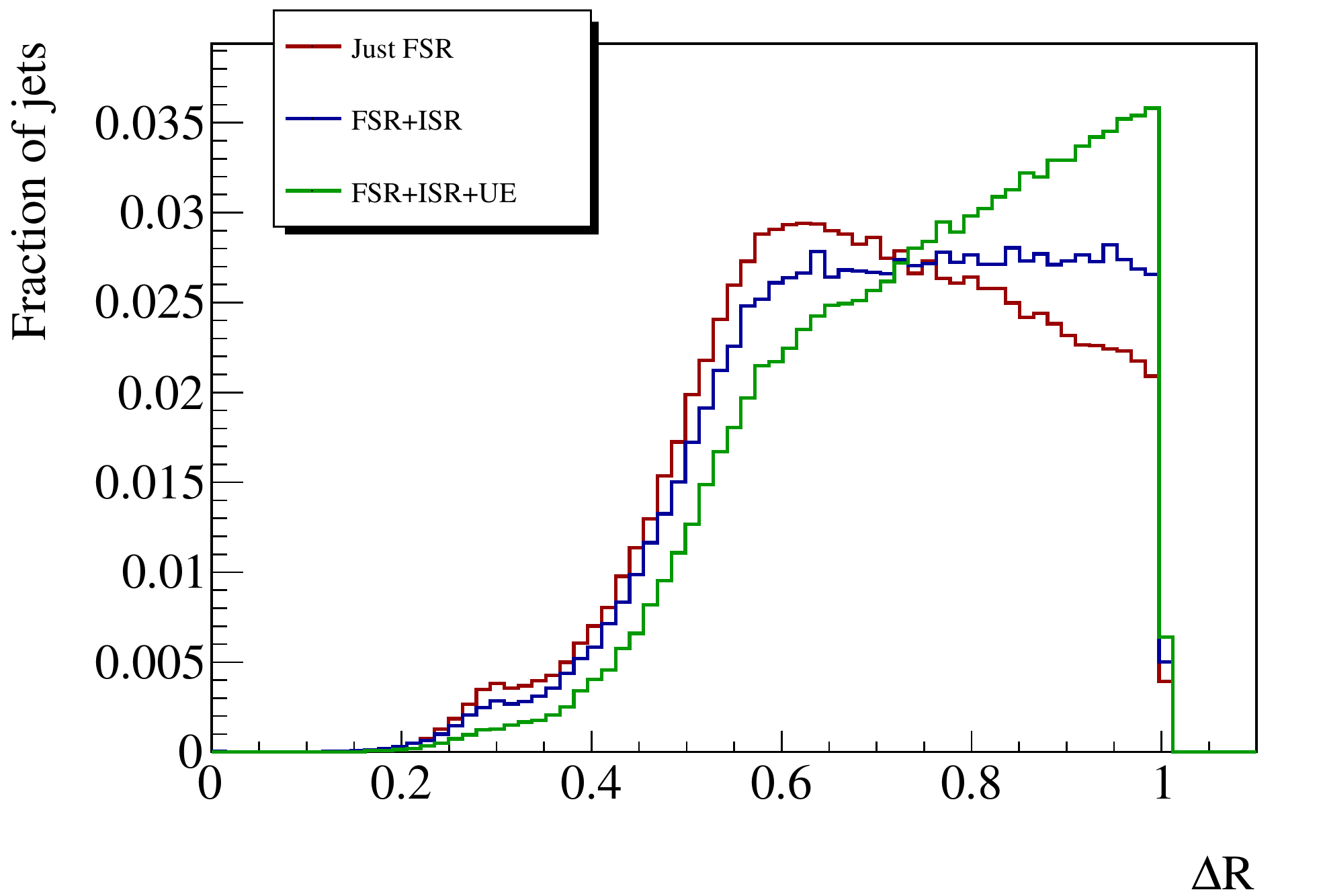}
\label{fig:ttbarPPMerged:DRCA}}
\subfloat[$\Delta R_{12}$, $\kt$] {\includegraphics[width = .48\columnwidth] {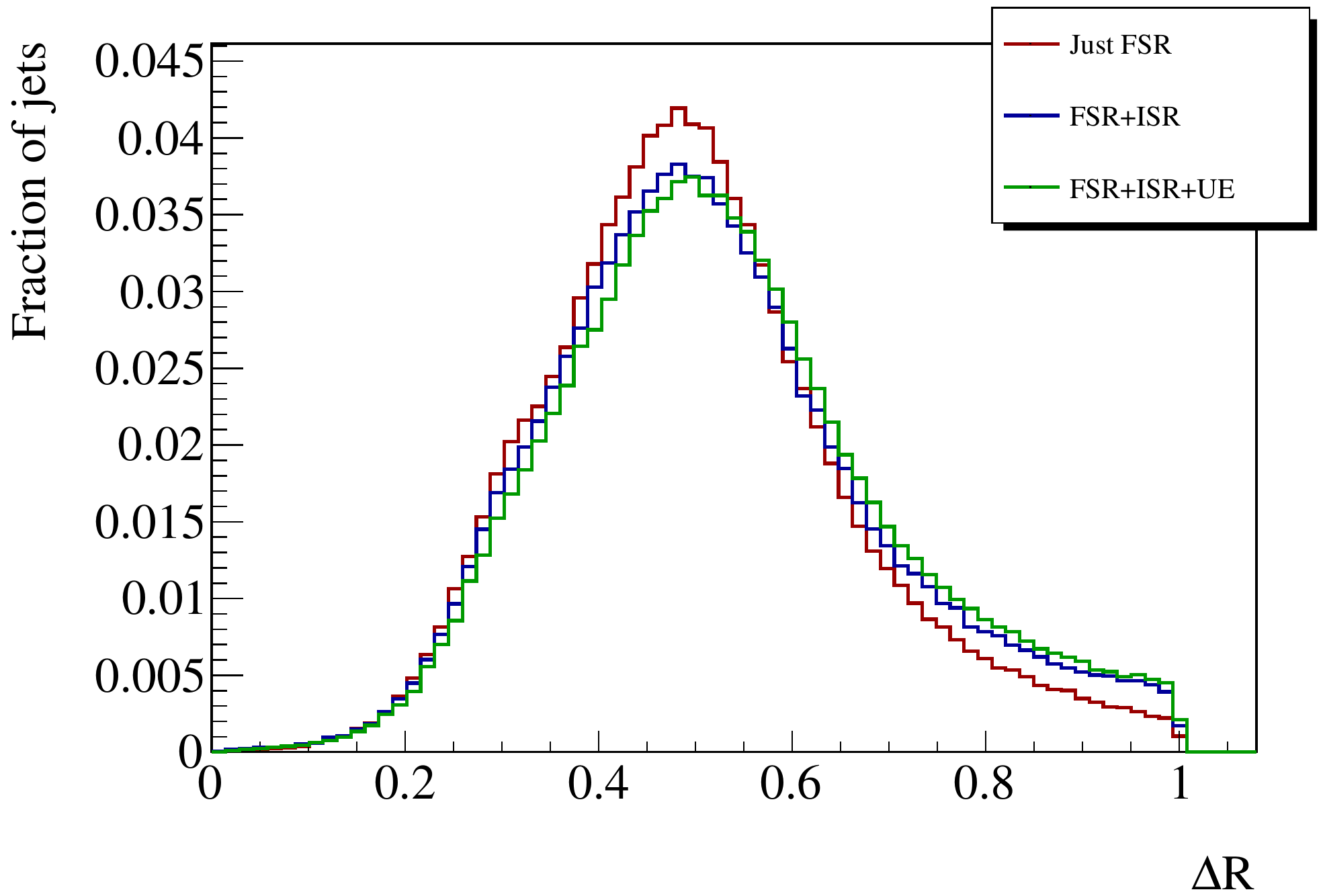} \label{fig:ttbarPPMerged:DRKT}}

\subfloat[$a_1$, CA] {\includegraphics[width = .48\columnwidth]{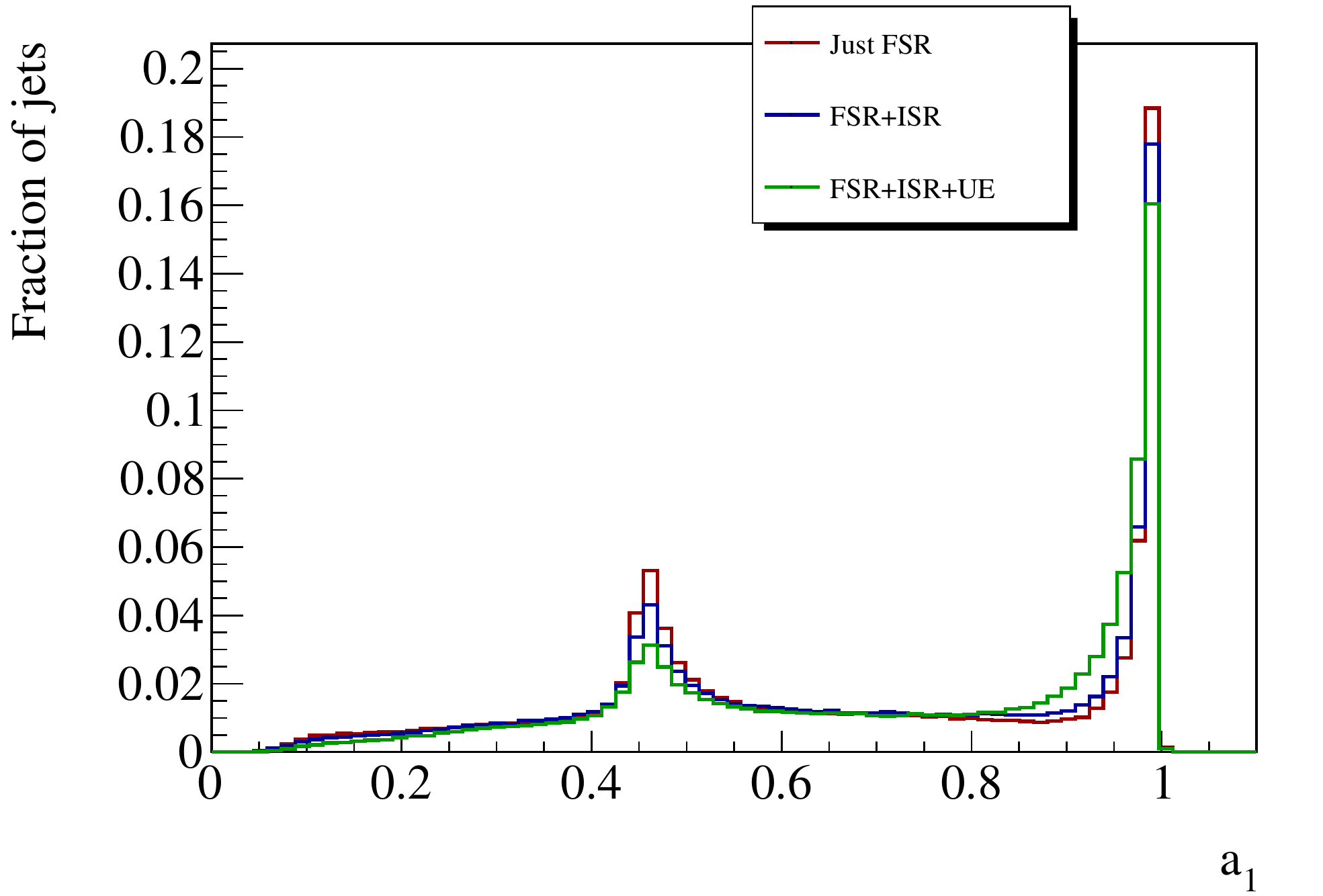} \label{fig:ttbarPPMerged:a1CA}}
\subfloat[$a_1$, $\kt$]{\includegraphics[width = .48\columnwidth]{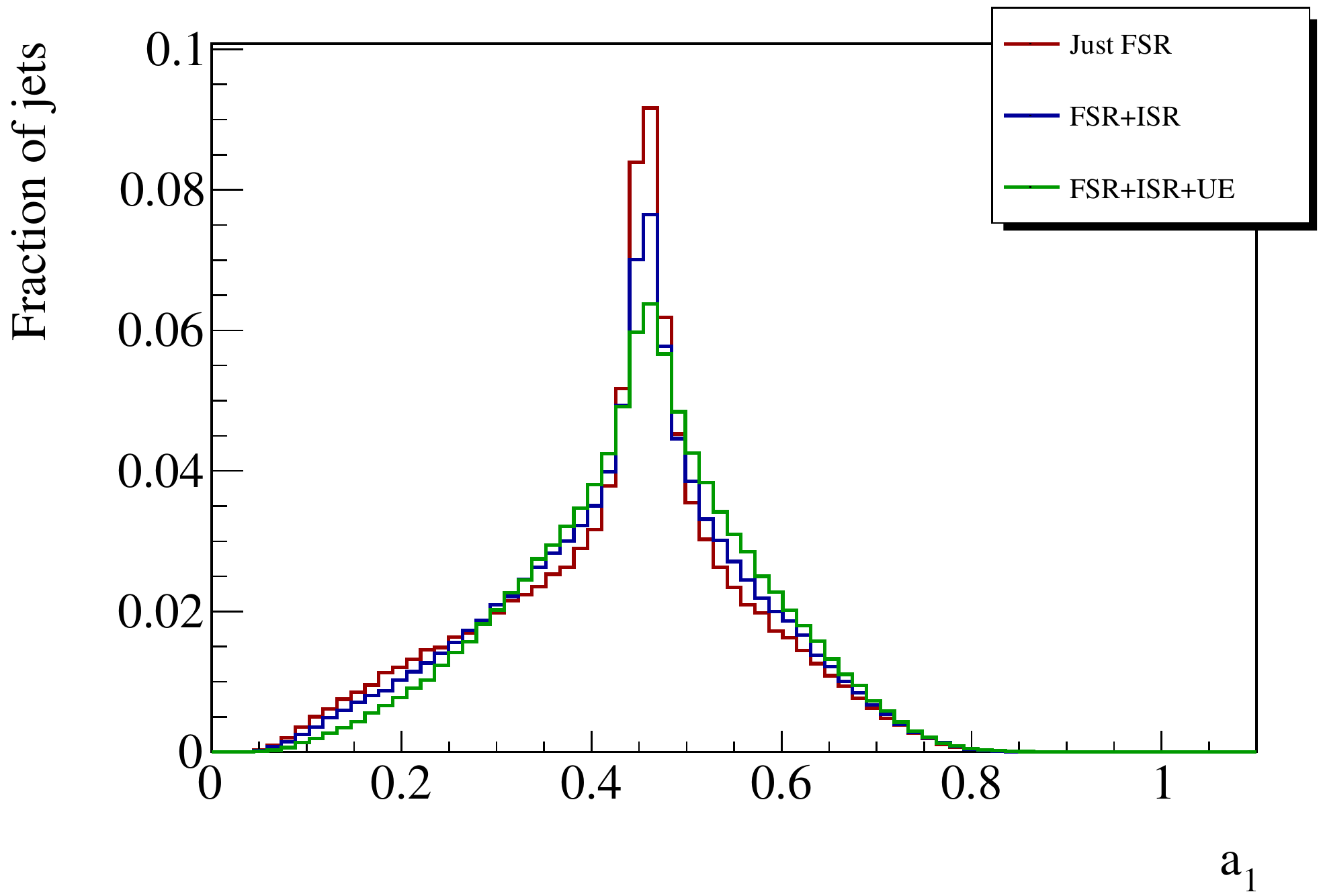}
\label{fig:ttbarPPMerged:a1KT}}
\end{center}
\caption[Distribution in $z$, $\Delta R_{12}$, and $a_1$ for ttbar jets in $pp \to t \bar t$ events]{Distribution in $z$, $\Delta R_{12}$, and the scaled (heavier) daughter mass $a_1$ for ttbar jets in $pp \to t \bar t$ events, using the CA and $\kt$ algorithms, with only FSR (blue), including ISR (red), and including ISR and UE (green).  Jets have $p_T > 500$ GeV and D = 1.0.}
\label{fig:ttbarPPMerged}
\end{figure}

In Fig.~\ref{fig:ttbarPPMerged} we plot the substructure variables $z$, $\Delta R_{12}$, and $a_1$ for $pp \to t\bar t$ events, the same three samples as in the previous subsection.  For CA jets the changes are clearest in the $\Delta R_{12}$ distribution (Fig.~\ref{fig:ttbarPPMerged:DRCA}).  ISR and UE push upward the final angle of recombination $\Delta R_{12}$.  The CA algorithm recombines protojets in order of angular separation, so the final mergings already tend to be at large angles.  To push the final angle even larger, ISR and UE must be adding radiation at the periphery of the jet which can be merged in late in the jet algorithm.  In Fig.~\ref{fig:ttbarPPMerged:DRKT} we can see that this effect does not occur in $\kt$ jets.  $\kt$ orders by $p_T$ as well as angle, so soft radiation at the periphery is merged into the jet early on, leaving as the final merging the combination of moderately-separated hard protojets --- perhaps representing the $W$ and $b$ in the case of a top quark jet.  We can conclude that the effects seen in the CA distributions are due to soft, large-angle radiation, and not to a more fundamental shift in the hard subjet dynamics, because this would show up in the $\kt$ distribution.

Moreover, if we consider the distributions in $a_1$, the scaled heavier subjet mass, CA jets are pushed more toward $a_1 = 1$.  This corresponds to one subjet having the same mass as the jet, with the other subjet having close to zero: the heavier subjet should presumably be associated with the top quark whereas the light subjet is likely to be soft radiation.  This radiation is quite possibly from ISR or UE, but in any case is not contributing significantly to the jet mass.

For $\kt$ there is no obvious systematic effect on $z$ or $\Delta R_{12}$, but we see that the distribution in $a_1$, peaked at $m_w/m_t$ is broadened just like the jet mass distribution.

\begin{figure}[htbp] \begin{center}
\subfloat[$z$, CA] {\includegraphics[width = .48\columnwidth] {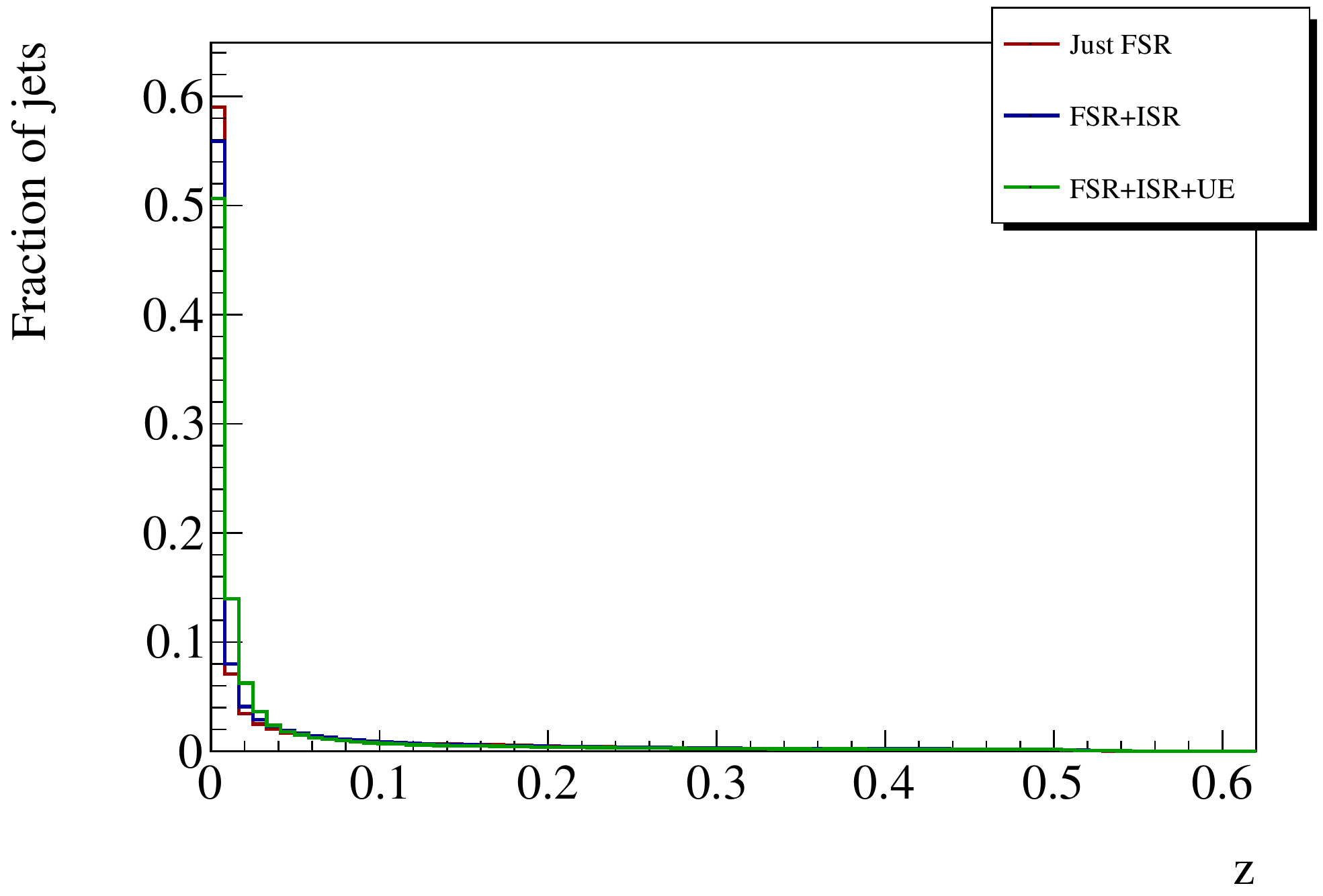} \label{fig:mQCDPPMerged:zCA}}
\subfloat[$z$, $\kt$]{\includegraphics[width = .48\columnwidth]{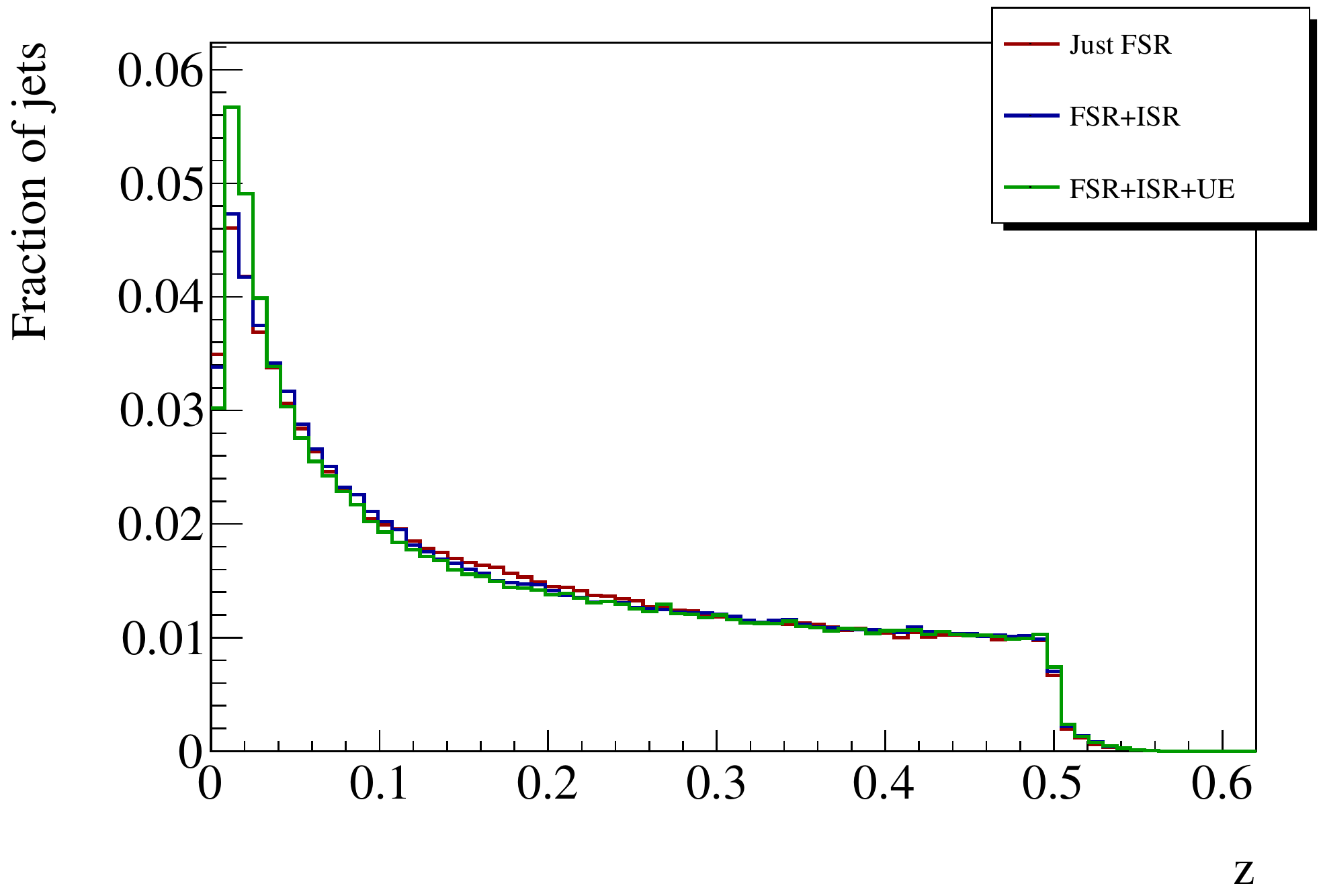} \label{fig:mQCDPPMerged:zKT}}

\subfloat[$\Delta R_{12}$, CA] {\includegraphics[width = .48\columnwidth] {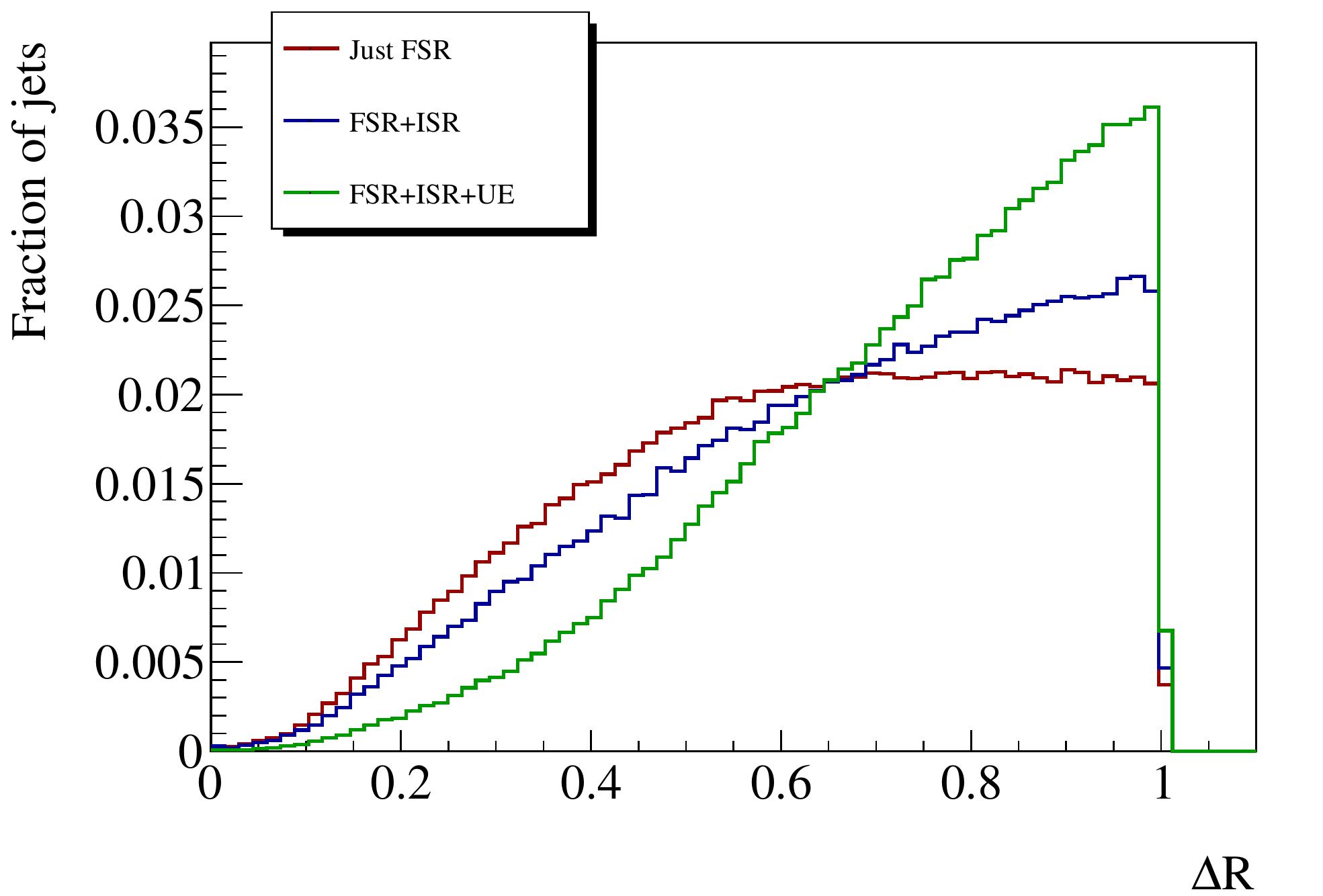}
\label{fig:mQCDPPMerged:DRCA}}
\subfloat[$\Delta R_{12}$, $\kt$] {\includegraphics[width = .48\columnwidth] {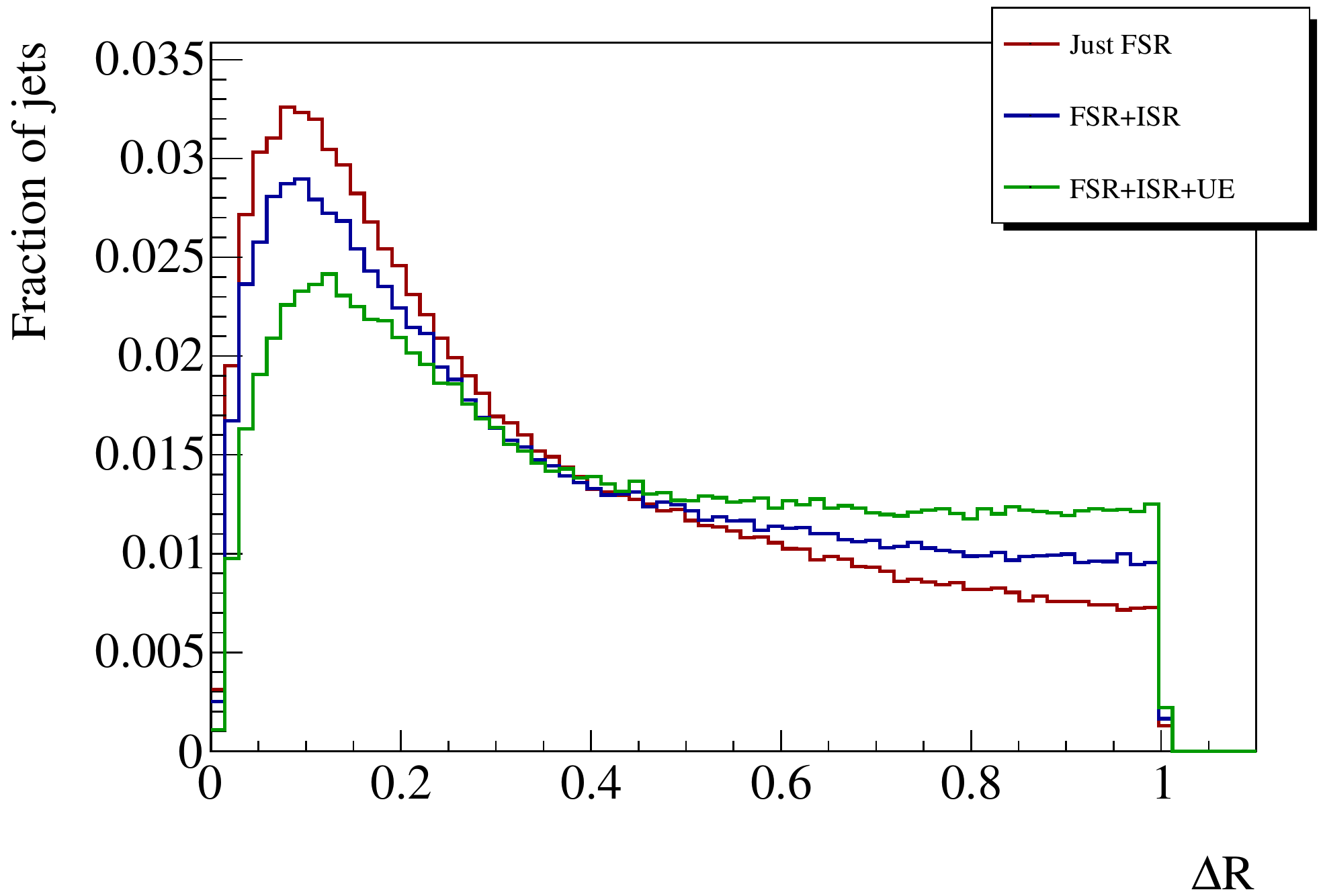} \label{fig:mQCDPPMerged:DRKT}}

\subfloat[$a_1$, CA] {\includegraphics[width = .48\columnwidth]{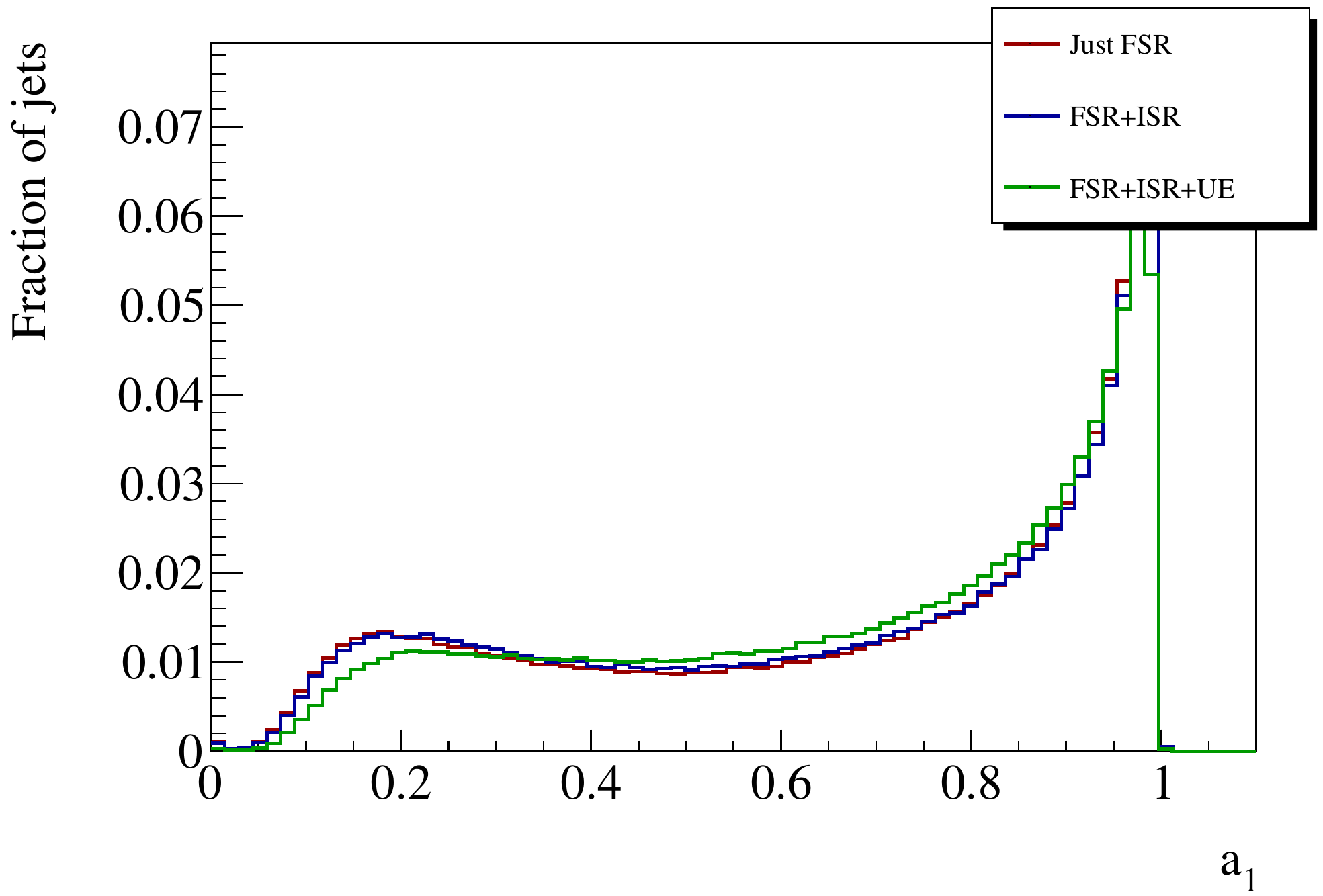} \label{fig:mQCDPPMerged:a1CA}}
\subfloat[$a_1$, $\kt$]{\includegraphics[width = .48\columnwidth]{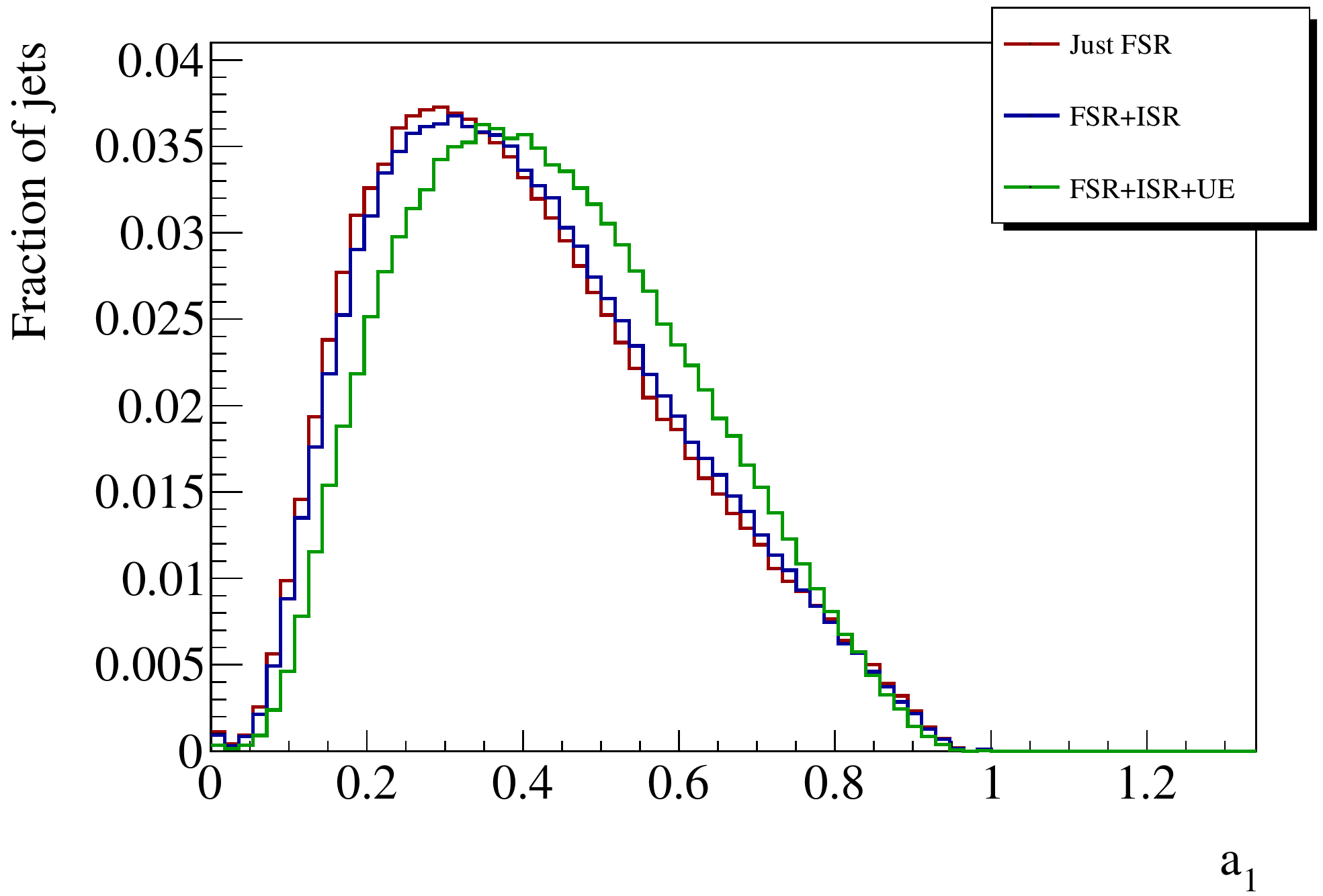}
\label{fig:mQCDPPMerged:a1KT}}
\end{center}
\caption[Distribution in $z$, $\Delta R_{12}$, and $a_1$ for QCD jets in matched $pp \to \text{jets}$ events]{Distribution in $z$, $\Delta R_{12}$, and the scaled (heavier) daughter mass $a_1$ for QCD jets in matched $pp \to \text{jets}$ events, using the CA and $\kt$ algorithms, with only FSR (blue), including ISR (red), and including ISR and UE (green).  Jets have $p_T > 500$ GeV and D = 1.0.}
\label{fig:mQCDPPMerged}
\end{figure}

In Fig.~\ref{fig:mQCDPPMerged} we show the same plots for the QCD multijet samples.  We can again see that ISR and UE add additional soft radiation at large angle, pushing up the distribution in $\Delta R_{12}$.  This is true even for $\kt$ jets: for QCD events $\kt$ jets have a smaller typical opening angle than $t\bar t$ events so the scope for contamination from ISR and UE is greater.  Other than the shift in $\Delta R_{12}$ the substructure variables are not strongly affected.

The substructure distributions for signal and background suggest that a large part of the effect of ISR and UE consists of the addition of soft, large-angle radiation.  The effects are less pronounced for $\kt$ jets, especially in the signal sample, but $\kt$ jets have their own disadvantage.  Whereas CA jets tend to have ISR and UE radiation included toward the end of the algorithm, shifting the kinematics of the final substructure, $\kt$ jets include \emph{more} extra radiation earlier.  This can be seen in the mass distributions in Figs.~\ref{fig:ttbarPPMassMerged} and \ref{fig:mQCDPPMassMerged}.  The distortions in CA substructure are in fact an advantage: they will give us a tool for removing (some of) the contributions of ISR and UE.

\section{Summary}
\label{sec:sub:summary}

We have seen numerous examples that the kinematics of the jet substructure in the last recombination for CA is a poor indicator for the physics of the jet.  However, we can characterize the aberrant substructure very simply.  For the CA algorithm, late recombinations (necessarily at large $\Delta R$) with small $z$ are more likely to arise from systematics effects of the algorithm than from the dynamics of the underlying physics in the jet.  For the $\kt$ algorithm, the poor mass resolution of the jet arises from earlier recombinations of soft protojets.  The last recombination for $\kt$ is representative of the physics of the jet, but the degraded mass resolution makes it difficult to efficiently discriminate between jets reconstructing heavy particle decays and QCD.  While small-$z$, large-$\Delta R$ recombinations are not as frequent late in the $\kt$ algorithm as in CA, they do contribute the most to the poor mass resolution of $\kt$.

As a simple example of the sensitivity of the mass to small-$z$, large-$\Delta R$ recombinations, consider the recombination $i,j\to p$ of two massless objects in the small-angle approximation.  The mass of the parent $p$ is given by $m_p^2 = p_{T_p}^2z(1-z)\Delta R_{ij}^2$, as in Eq.~(\ref{eq:simplejmass}).  Suppose the value of the $\kt$ recombination metric, $\rho_{ij}(\kt) = p_{T_p}z\Delta R_{12}$ is bounded below by a value $\rho_0$ (say by previous recombinations), and the recombination $i,j\to p$ occurs at $\rho_{ij}(\kt) = \rho_0$.  Then the mass of the parent is $m_p^2 = \rho_0^2(1-z)/z$, which is maximized for small $z$.  Therefore, at a given stage of the algorithm, small-$z$ recombinations have a large effect on the mass of the jet.

\begin{figure}[htbp]
\begin{center}
\subfloat[$m_J$ cut, $z$] {\label{fig:topPartonVsReconZDR:zmjet}
 \includegraphics[width = 0.45\columnwidth] {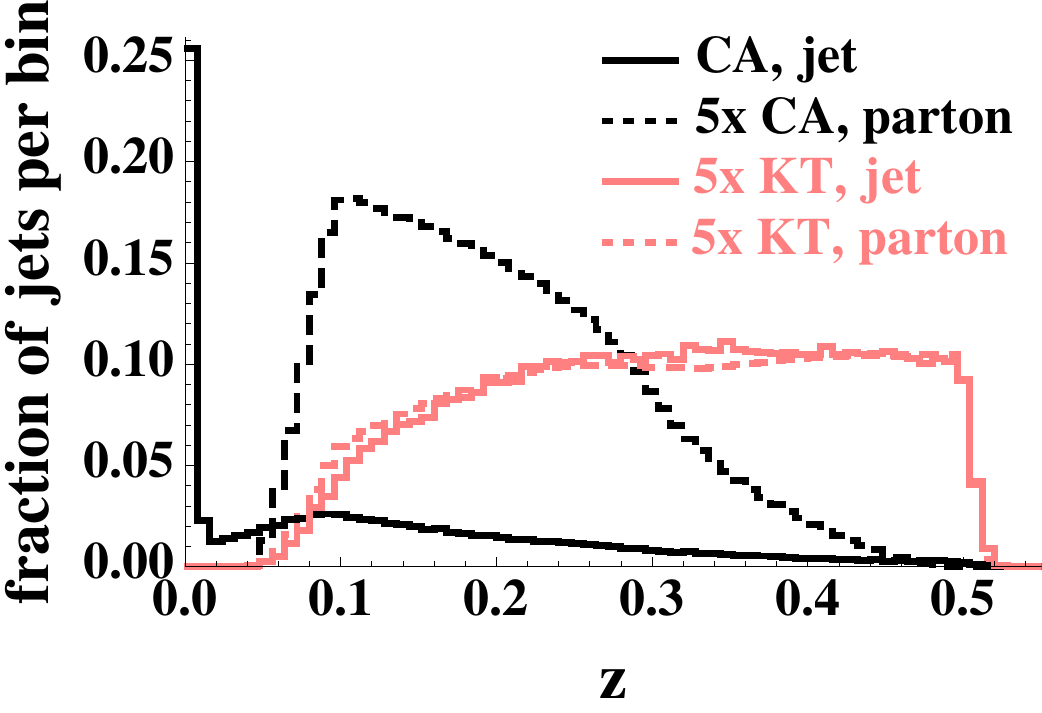}}
\subfloat[$m_J$ cut, $\Delta R_{12}$] {\includegraphics[width = 0.45\columnwidth]{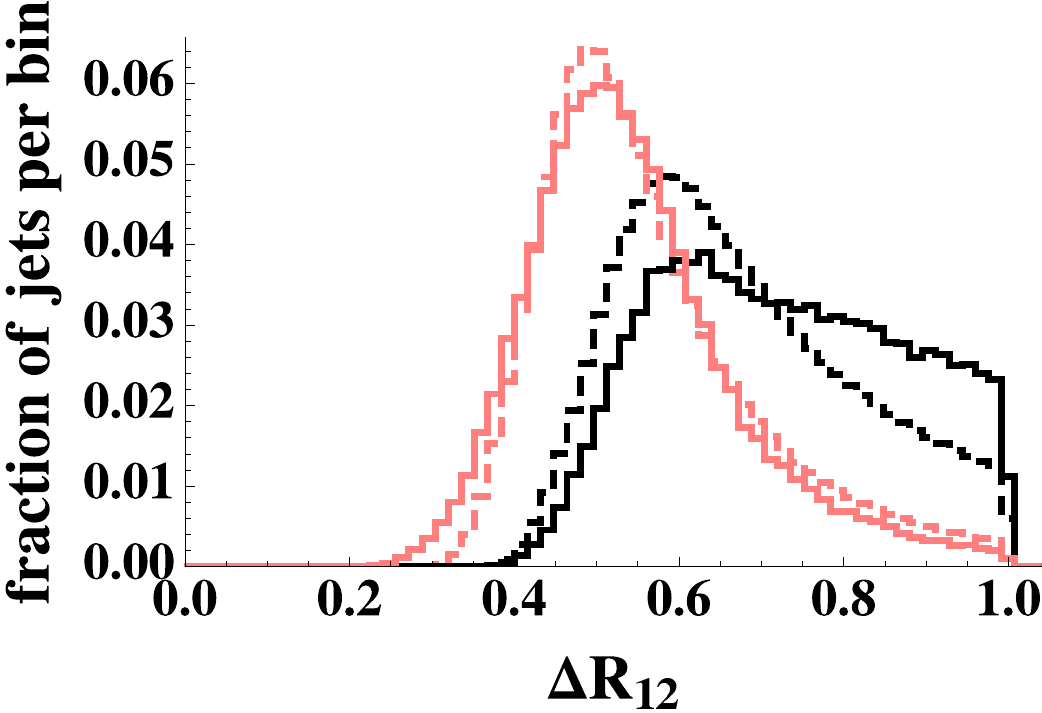}}

\subfloat[$m_J$ and $m_{\text{Sub}J}$ cuts, $z$] {\includegraphics[width = 0.45\columnwidth] {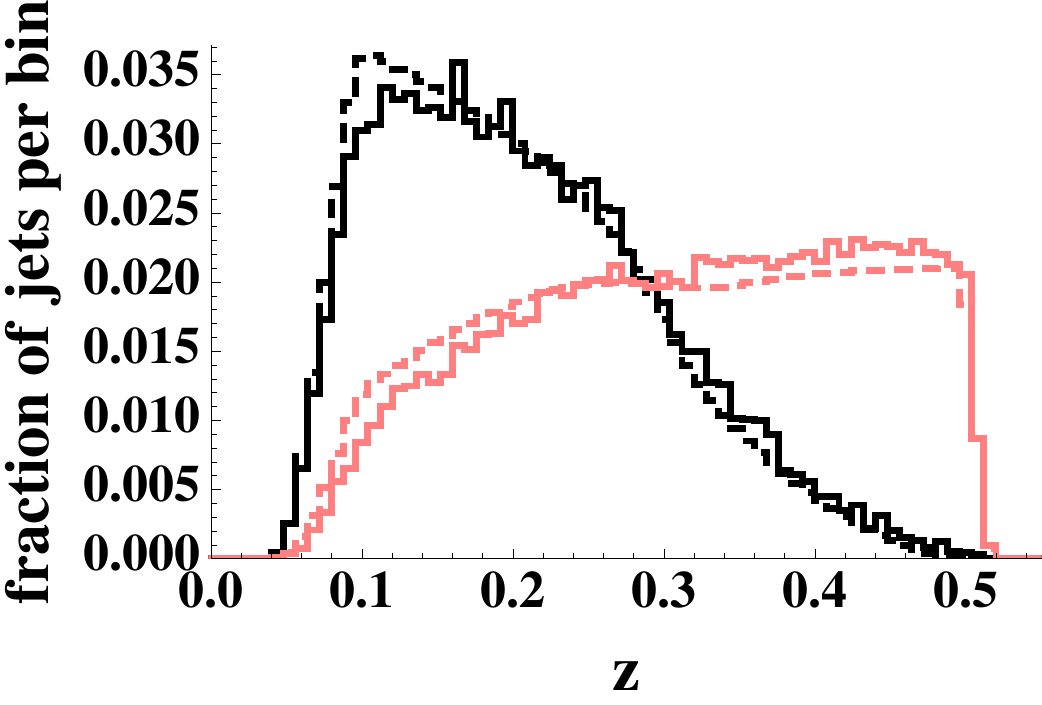}}
\subfloat[$m_J$ and $m_{\text{Sub}J}$ cuts, $\Delta R_{12}$] {\includegraphics[width =
0.45\columnwidth] {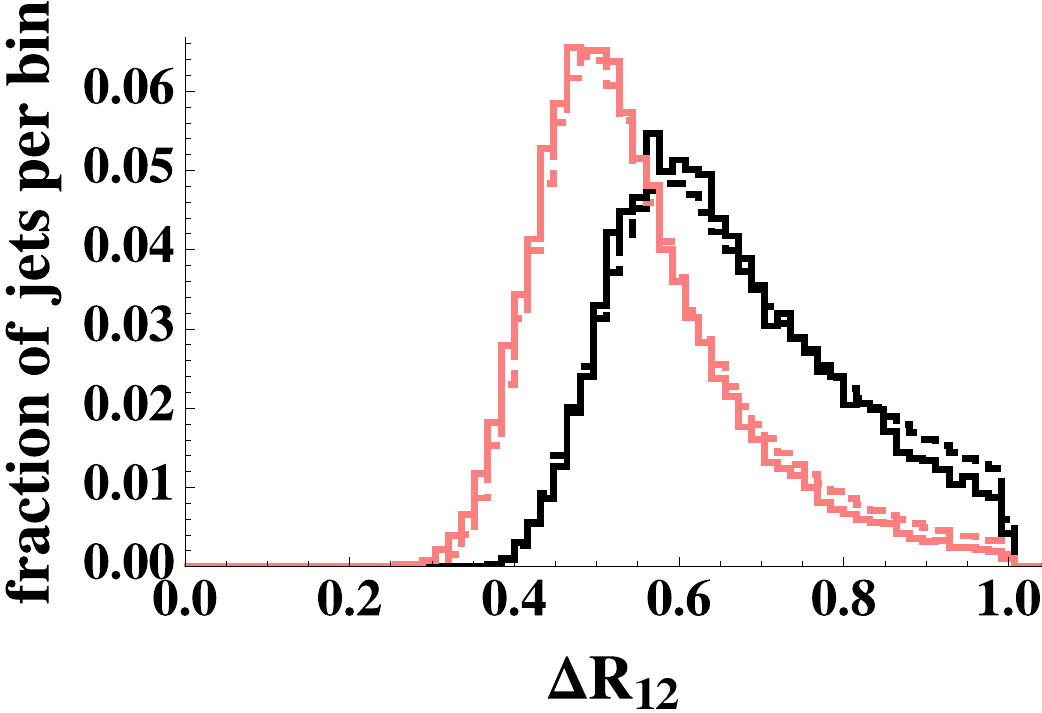}}
\end{center}

\caption[Distributions in $z$ and $\Delta R_{12}$ comparing for top quark decays at the parton level and from Monte Carlo events]{Distributions in $z$ and $\Delta R_{12}$ comparing for top quark decays at the parton level and from Monte Carlo events.  The jets have $p_T$ between 500 and 700 GeV, and have D = 1.0.  The parton-level top decays have the same distribution of boosts as the Monte Carlo top jets.  Jets in the upper plots have a mass cut on the jet; the lower plots include a subjet mass cut.  The details of these cuts are described in Sec.~\ref{sec:prune:study}.}
\label{fig:topPartonVsReconZDR}
\end{figure}

When we can resolve the mass scales of a decay in a jet, the distribution of kinematic variables matches closely what we expect from the parton-level kinematics of the decay.  For the example of the top quark decay, if we select jets with the top mass that have a daughter subjet with the $W$ mass, the kinematic distributions of $z$ and $\Delta R_{12}$ closely match the distributions from the parton-level decay of the top quark.  We show this in Fig.~\ref{fig:topPartonVsReconZDR}, where we make a top quark ``hadron-parton'' comparison for $z$ and $\Delta R_{12}$.  The specifics of the mass cuts are described in Sec.~\ref{sec:prune:study}.  In the parton-level events, we simply require that the top quark decay to three partons be fully reconstructed by the algorithm in a single jet, namely that the $W$ is correctly recombined first from its decay products before recombination with the $b$ quark to make the top.  The parton-level events have the same distribution of top quark boosts as the top jets in the hadron-level events.
It is clear that simply requiring the hadron-level jet to have the top mass, which makes no cut on the substructure, leads to kinematic distributions in $z$ and $\Delta R_{12}$ for CA that do not match the parton-level distributions, although the distributions do match quite well for the $\kt$ algorithm.  The excess of small-$z$ recombinations for CA in the hadron-level jet with only a jet mass cut arises from jet algorithm effects discussed previously.  After the subjet mass cut, these are removed and the distribution of $z$ in the jet matches the reconstructed parton-level decay very well.

Therefore, when we can accurately reconstruct the mass scales of a decay in a jet, the kinematics of the jet substructure tend to reproduce the parton-level kinematics of the decay.  This suggests that if we can reduce systematic effects that generate misleading substructure, we can improve heavy particle identification and separation from background.  Reducing these systematic effects can also improve the mass resolution of the jet, which will aid in identifying a heavy particle decay reconstructed in a jet and in rejecting the QCD background.



\graphicspath{{chapter5/graphics/}}

\chapter{Improving Heavy Particle Searches with Jet Substructure: ``Jet Pruning''}
\label{sec:prune}

\section[\textit{Pruning: ``Cleaning up'' jet substructure}]{Pruning: ``Cleaning up'' jet substructure}
\label{sec:prune:define}

We now define a technique that modifies the jet substructure to reduce the systematic effects that obscure heavy particle reconstruction.\footnote{This section is taken, with small modifications, from Sec.~VI of \cite{Pruning2}.}  In general, we will think of a \emph{pruning procedure} as using a criterion on kinematic variables to determine whether or not a branching is likely to represent accurate reconstruction of a heavy particle decay.  This takes the form of a cut: if a branching does not pass a set of cuts on kinematic variables, that recombination is vetoed.  This means that one of the two branches to be combined (determined by some test on the kinematics) is discarded and the recombination does not occur.

In Sec.~\ref{sec:sub:summary}, we identified recombinations that are unlikely to represent the reconstruction of a heavy particle.  These can be characterized in terms of the variables $z$ and $\Delta R$: recombinations with large $\Delta R$ and small $z$ are much more likely to arise from systematic effects of the jet algorithm and in QCD jets rather than heavy particle reconstruction (compare the upper and lower figures in Fig.~\ref{fig:topPartonVsReconZDR}).  We expect that removing (\emph{pruning}) these recombinations will tend to improve our ability to measure jet substructure, including subjet masses.  We also expect that this procedure will systematically shift the QCD mass distribution lower, reducing the background in the signal mass window.  Finally this procedure is expected to reduce the impact of uncorrelated soft radiation from the underlying event and pile-up. We therefore define the following pruning procedure:

\begin{itemize}
\item[0.]  Start with a jet found by any jet algorithm, and collect the objects (such as calorimeter towers) in the jet into a list $L$.  Define parameters $D_{\cut}$ and $z_{\cut}$ for the pruning procedure.
\item[1.]  Rerun a jet algorithm on the list $L$, checking for the following condition in each recombination $i,j\to p$:
\beq
z = \frac{\min(p_{Ti},p_{Tj})}{p_{Tp}} < z_{\cut} \quad \text{and} \quad \Delta R_{ij} > D_{\cut}.
\label{eq:pruneTest}
\eeq
This algorithm must be a recombination algorithm such as the CA or $\kt$ algorithms, and should give a ``useful'' jet substructure (one where we can meaningfully interpret recombinations in terms of the physics of the jet).
\item[2.]  If the conditions in 1. are met, do not merge the two branches $1$ and $2$ into $p$.  Instead, discard the softer branch, i.e., veto on the merging.  Proceed with the algorithm.
\item[3.]  The resulting jet is the \emph{pruned jet}, and can be compared with the jet found in Step 0.
\end{itemize}

This technique is intended to be generically applicable in heavy particle searches.  It generalizes analysis techniques suggested by other authors, including ``filtering'' \cite{FilteringHiggs} and ``top-tagging'' \cite{TopTagging}, in that these methods also modify the jet substructure to assist separate a particular signal from backgrounds.  In particular, the use of the variables $z$ and $\Delta R_{ij}$ follows the use of $\delta_p$ and $\delta_r$ in \cite{TopTagging}, with the significant difference that $\delta_p$ measures softness relative to the total jet, and we define $z$ to be a ``local'' variable that only depends on the two protojets being recombined.  A more important distinction is that filtering and top-tagging are designed to find a specific number of subjets to map onto a specific decay, whereas pruning is intended to be applied to an entire jet with no bias toward a specific substructure configuration.  While we think this generality is novel, we emphasize that pruning is an evolution from earlier methods and relies on the same physical effects.  We have endeavored to justify our claim for generality with the discussions in Chapter~\ref{sec:sub}, which demonstrate that the interpretation of jet substructure is subject to generic systematic effects that can be well characterized.  Pruning is not the only option, but offers some advantages which we explore in further studies below.

In the analysis of pruning, we will explore the dependence of the pruned jets on the value of $D$ from the jet algorithm.  When reconstructing a boosted heavy particle in a single jet, without pruning the reconstruction is optimized if the value of $D$ is fit to the expected opening angle of the decay.  However, this angle depends on the mass of the particle (which is not known in a search) and its $p_T$.  We will show that pruning reduces the sensitivity to $D$ and allows one to use large-$D$ jets over a broad range in $p_T$ to search for heavy particles.

Values for the two parameters of the pruning procedure, $z_{\cut}$ and $D_{\cut}$, can be well motivated.  In the following studies, we will show that the results of pruning are rather insensitive to the parameters, and that the optimal parameters are similar for different searches.  That is, it is not necessary to tune the pruning procedure for individual searches.

The parameter $z_{\cut}$ can be chosen based on the analysis of single-step and multi-step decays in Sec.~\ref{sec:sub:parton:decay}.  Near the limit in boost where decays are reconstructed in a single jet, the value of $z$ is typically large.  It is only at large boosts, where the production rate of heavy particles is much smaller, that small values of $z$ are allowed for reconstructed decays (see Fig.~\ref{fig:zdistrecon}).  Therefore, we can choose a value of $z_{\cut}$ that will keep all reconstructed parton-level decays at small boost, and only remove a small fraction of decays at larger boosts.  We expect that a $z_\cut \sim 0.10$ will be a reasonable compromise.  Note that Fig.~\ref{fig:topPartonVsReconZDR:zmjet} indicates that much of the soft radiation distorting the substructure for CA jets has $z \lesssim 0.02$, so that at least for CA a $z_\cut$ not much bigger than this should be effective.

The parameter $D_{\cut}$ can be determined on a jet-by-jet basis, allowing pruning to be more adaptive than a fixed-parameter procedure.  $D_{\cut}$ determines how much of the jet substructure can be pruned, with smaller values allowing for more pruning.  $D_\cut$ should be sufficiently small so that if a decay is ``hidden'' inside the jet substructure by late recombinations of, say, UE particles, the substructure can be pruned and the decay can be found.  A value that is too small, however, will result in over-pruning.  A natural scale for $D_{\cut}$ is the opening angle of the jet.  However, this is an infrared unsafe quantity, as soft radiation can change the opening angle.  Instead, the dimensionless ratio $m_J/p_{T_J}$ for the jet is related to the opening angle: typically, $\Delta R_{12} \approx 2m_J/p_{T_J}$.  Therefore, we choose $D_{\cut}$ to scale with $2m_J/p_{T_J}$.  $D_{\cut} = m_J/p_{T_J}$ is a reasonable starting value.


\section{Effects of pruning in \texorpdfstring{$\ee$}{ee} collisions}
 \label{sec:prune:ee}

Having defined the pruning procedure, we now wish to study its effects.  In this study, we use the parameters $D_{\cut} = m_J/p_{T_J}$ for both algorithms, and $z_{\cut} = 0.10$ for the CA algorithm and 0.15 for the $\kt$ algorithm.  We will motivate these parameters in Sec.~\ref{sec:prune:results:params}.

\begin{figure}[htbp]
\subfloat[$z$, CA] {\includegraphics[width = .48\columnwidth] {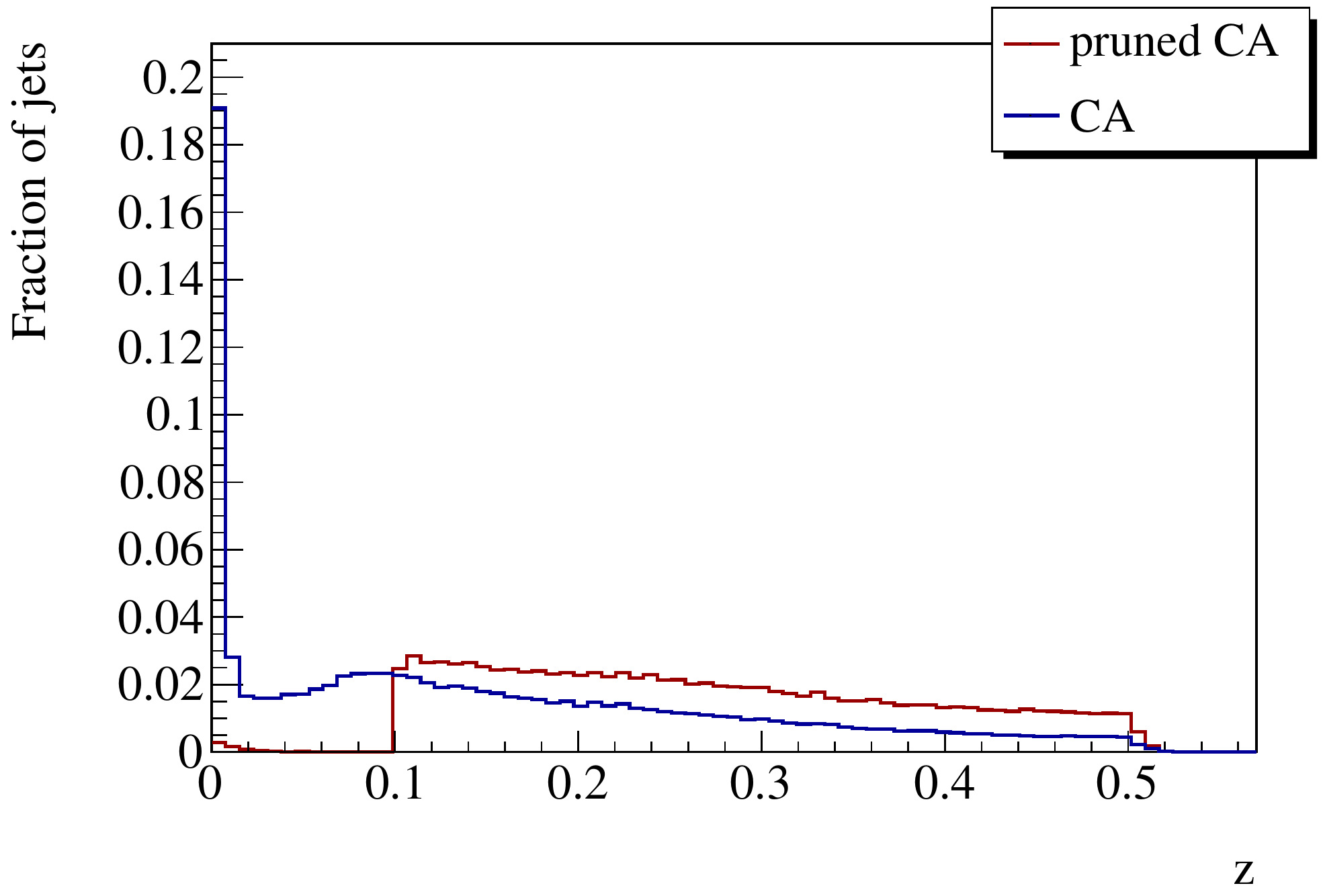} \label{fig:epem_ttbar_pCompare:zCA}}
\subfloat[$z$, $\kt$]{\includegraphics[width = .48\columnwidth]{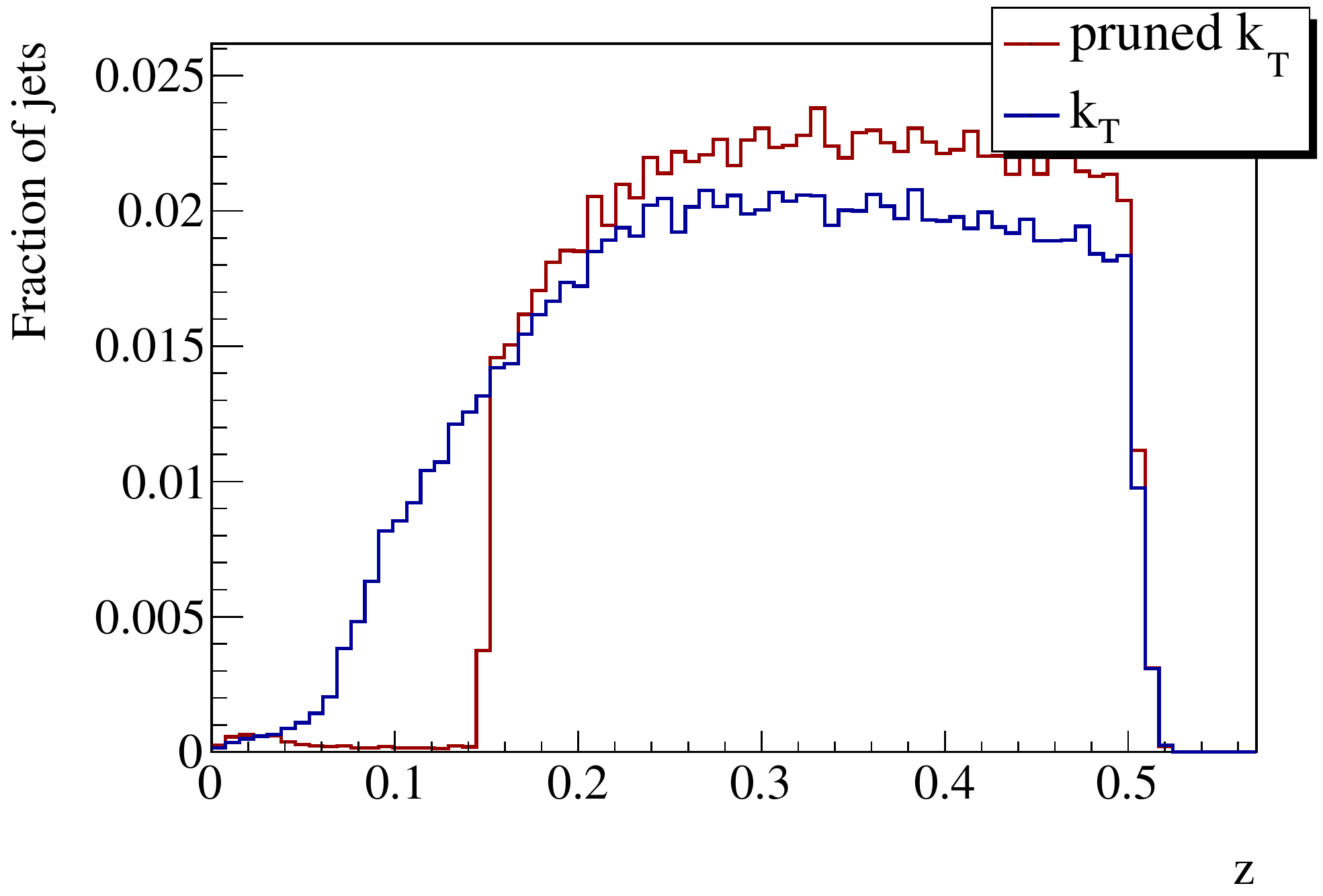} \label{fig:epem_ttbar_pCompared:zKT}}

\subfloat[$\Delta R_{12}$, CA] {\includegraphics[width = .48\columnwidth] {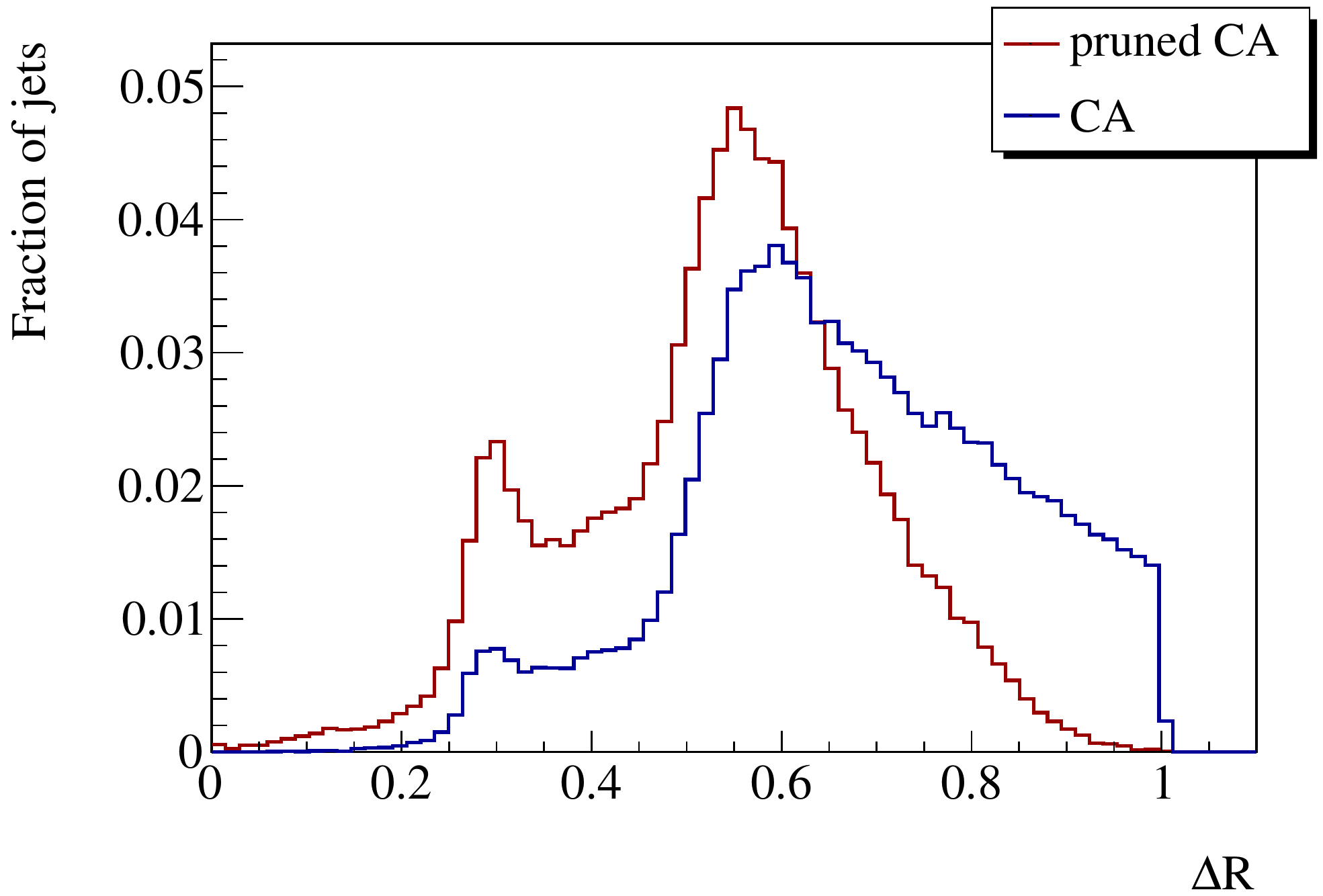}
\label{fig:epem_ttbar_pCompare:DRCA}}
\subfloat[$\Delta R_{12}$, $\kt$] {\includegraphics[width = .48\columnwidth] {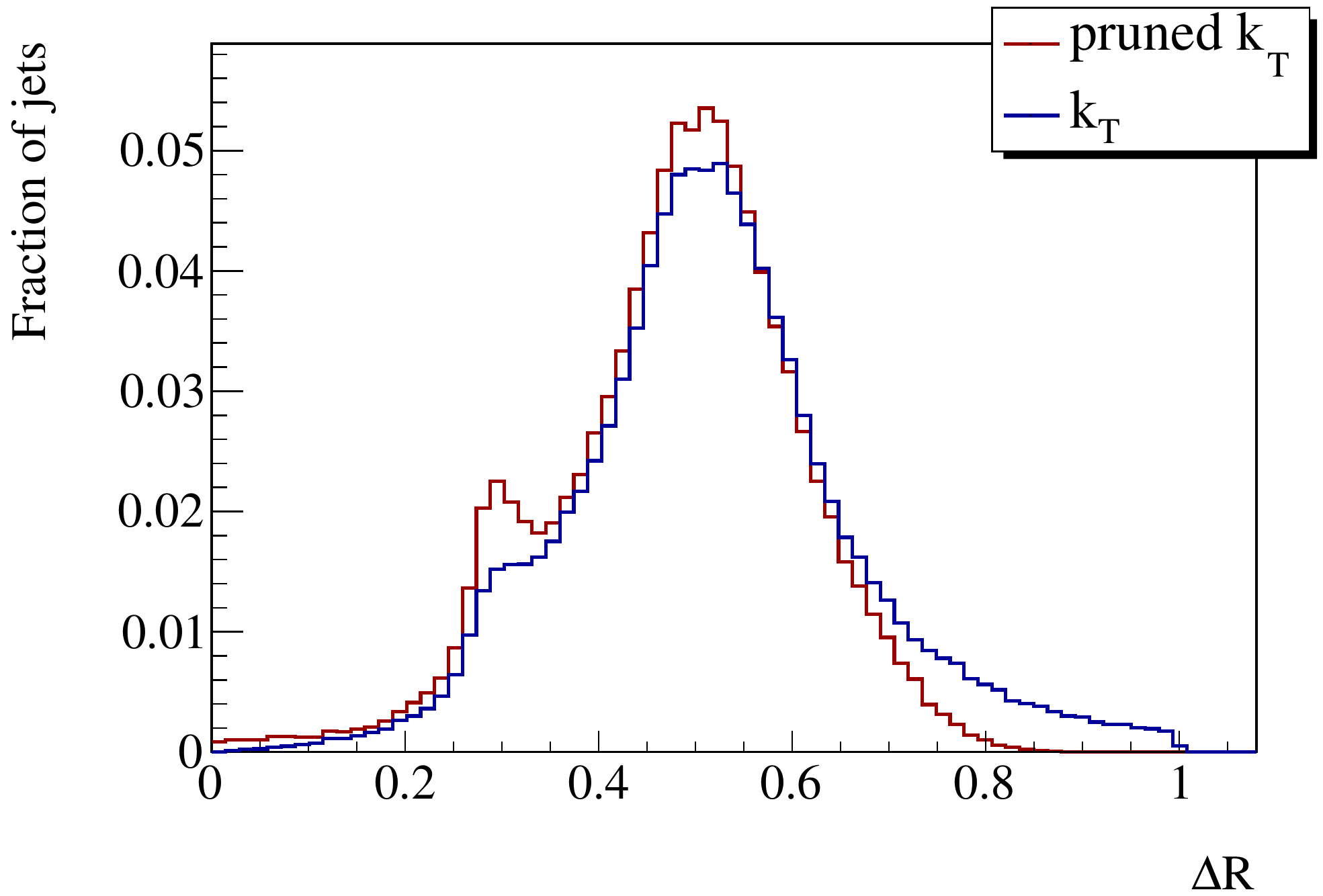} \label{fig:epem_ttbar_pCompare:DRKT}}

\subfloat[$a_1$, CA] {\includegraphics[width = .48\columnwidth]{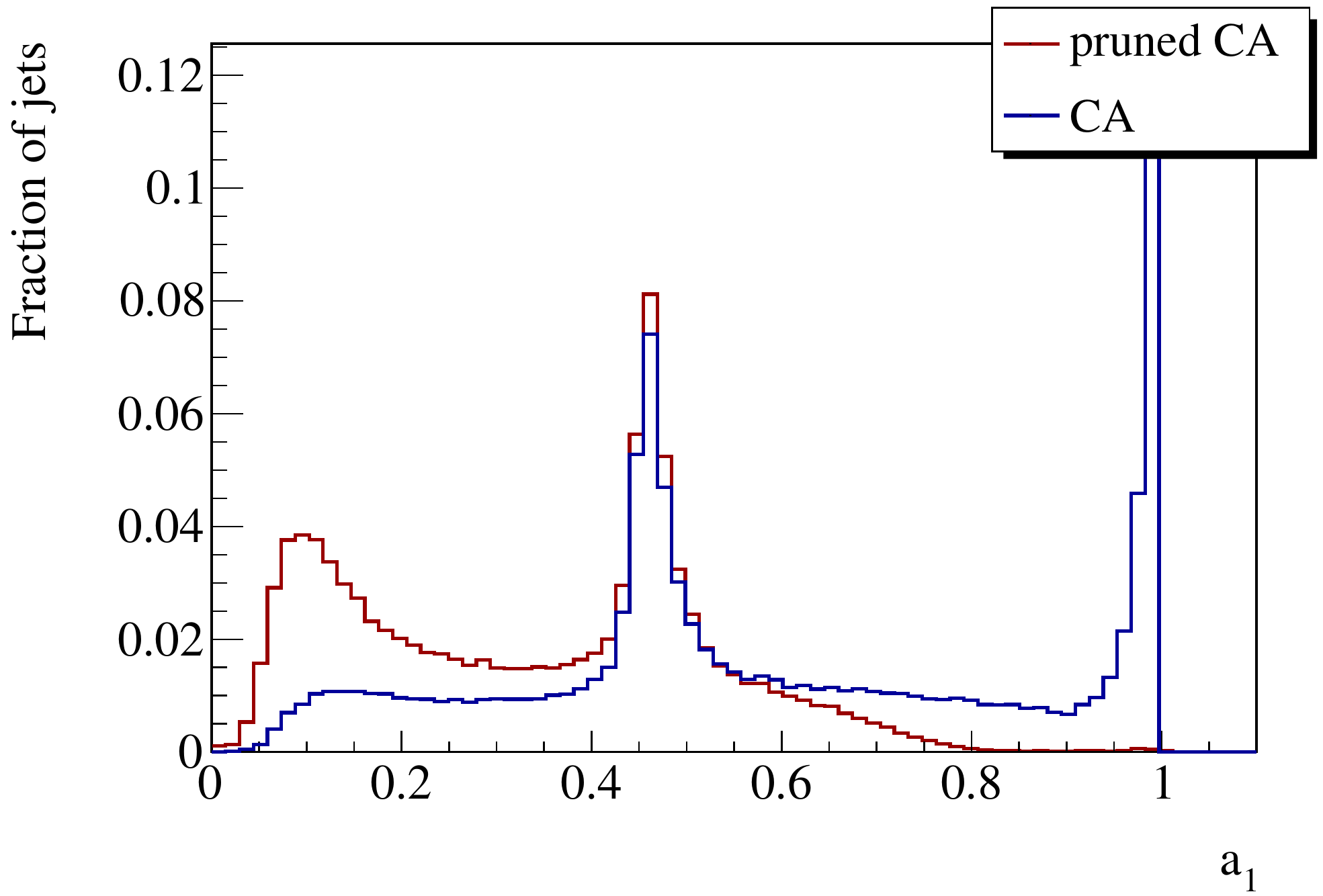} \label{fig:epem_ttbar_pCompare:a1CA}}
\subfloat[$a_1$, $\kt$]{\includegraphics[width = .48\columnwidth]{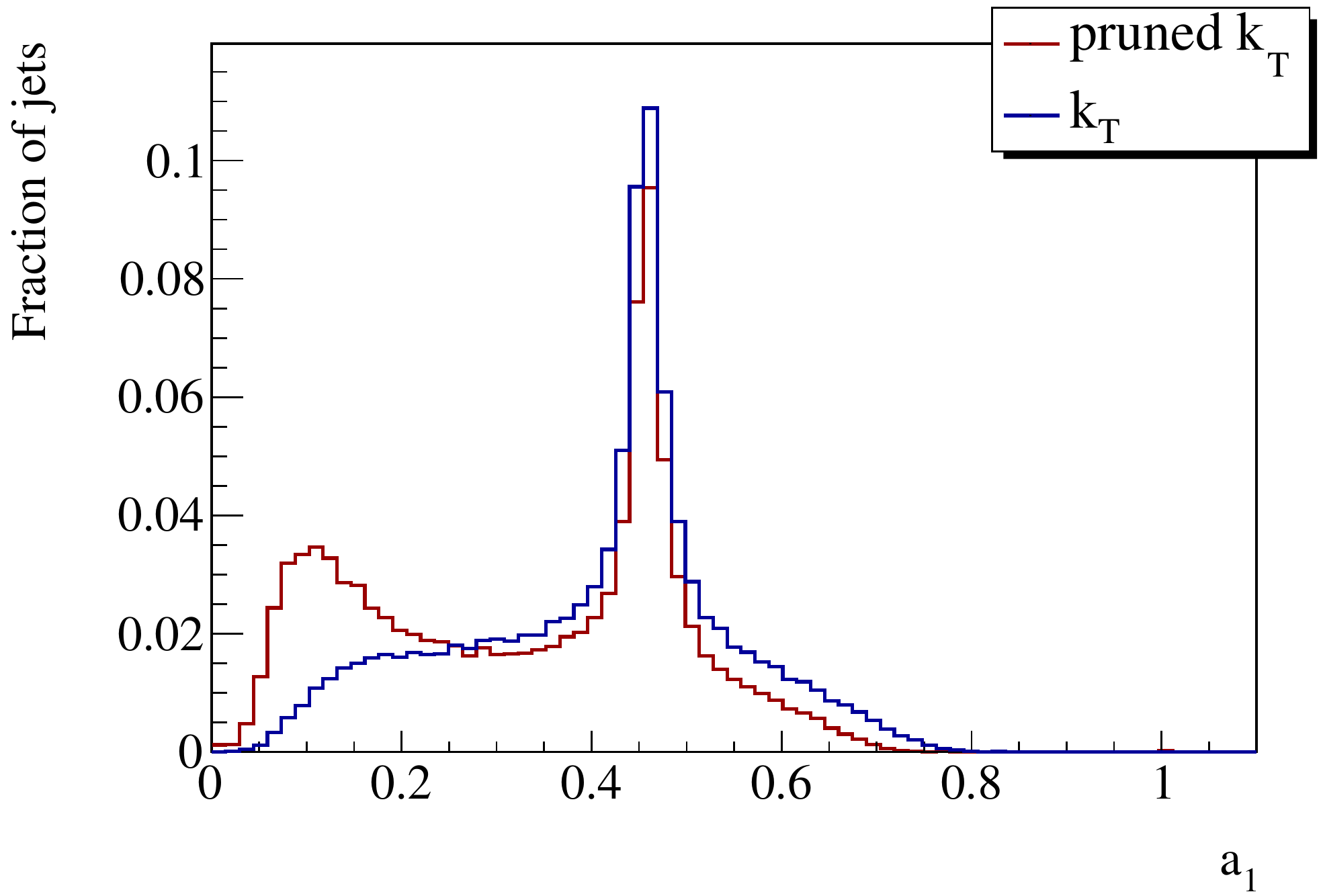}
\label{fig:epem_ttbar_pCompare:a1KT}}

\caption{Distributions in $z$, $\Delta R_{12}$, and $a_1$ for pruned and unpruned jets in $\ee \to t\bar t$ events.}
\label{fig:epem_ttbar_pCompare}
\end{figure}

We begin with jets in $\ee$ collisions as a baseline.  Although we expect pruning will be most useful at hadron colliders, it is instructive to consider how it affects jets in a simpler environment.  In Fig.~\ref{fig:epem_ttbar_pCompare} we show the distribution in substructure kinematics for $\ee \to t\bar t$ events.

For the $\kt$ algorithm, pruning does not significantly affect the kinematics of the final branching.  Pruning only removes soft, wide-angle mergings, which rarely occur as the last merging in a $\kt$ jet.  The small reduction in the $a_1$ peak corresponds to occasionally identifying the $W+b$ merging correctly but discarding the $b$ for being too soft.  That pruning does occasionally happen can also be seen in the depletion of jets with $z < 0.15$, the softness cutoff used in these plots.

For CA on the other hand pruning has a large effect.  Nearly 20\% of unpruned jets had $z < 0.01$; these mergings have nearly all been eliminated.  (The requirement that only mergings with $\Delta R_{12} > D_\text{cut}$ can be pruned means that some jets survive with $z < z_\text{cut}$.)  Since the final merging(s) of CA jets are often pruned, we see that the distributions in $\Delta R_{12}$ and $a_1$ are shifted.  Jets with $z \approx 0$ have $a_1 \approx 1$, so this peak has disappeared.  The typical final opening angle has also been shifted downward.  The double peaks correspond to the kinematically typical opening angles for top quark and $W$ boson decays at this $p_T$.

\begin{figure}[htbp]
\subfloat[$z$, CA] {\includegraphics[width = .48\columnwidth] {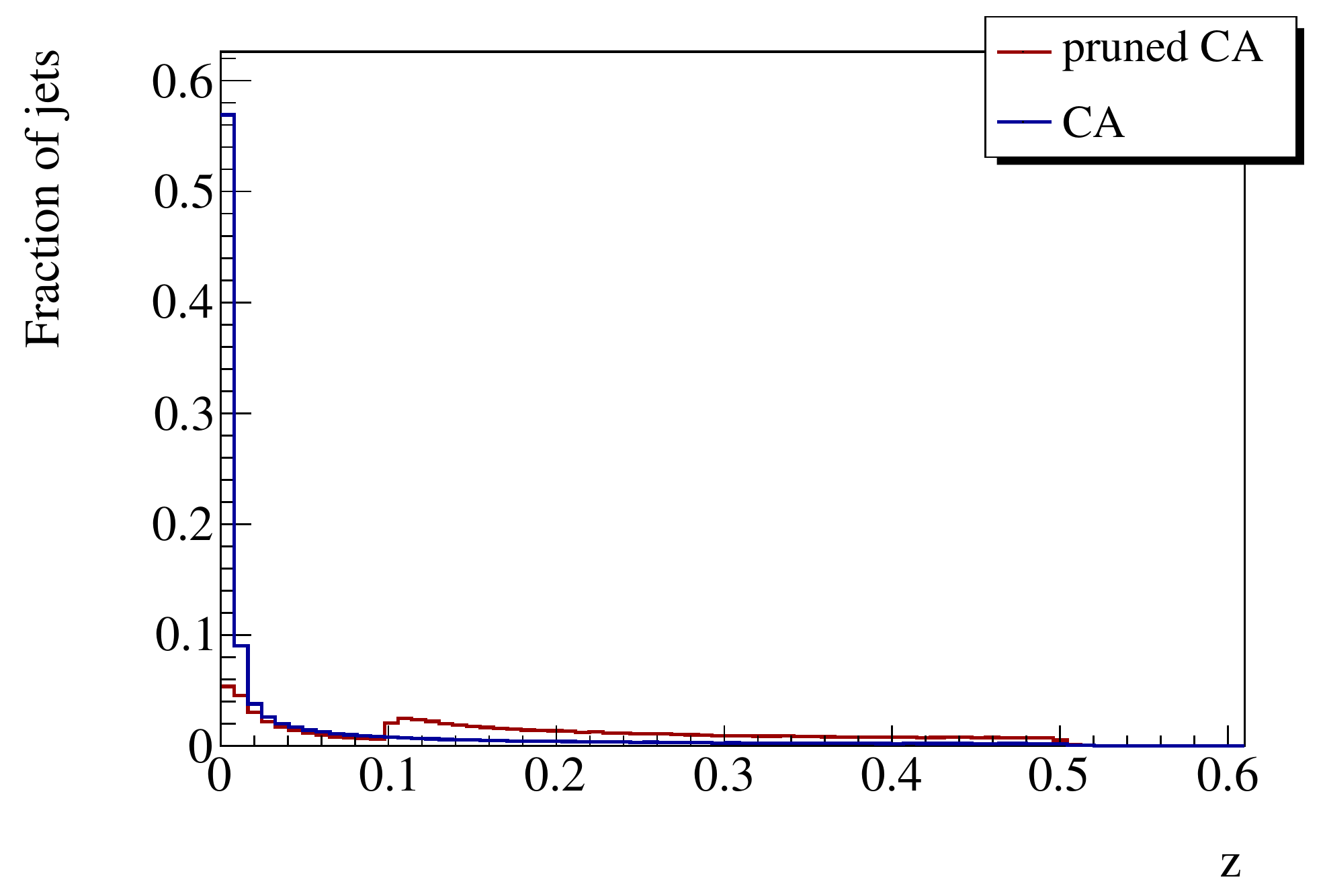} \label{fig:epem_QCD_pCompare:zCA}}
\subfloat[$z$, $\kt$]{\includegraphics[width = .48\columnwidth]{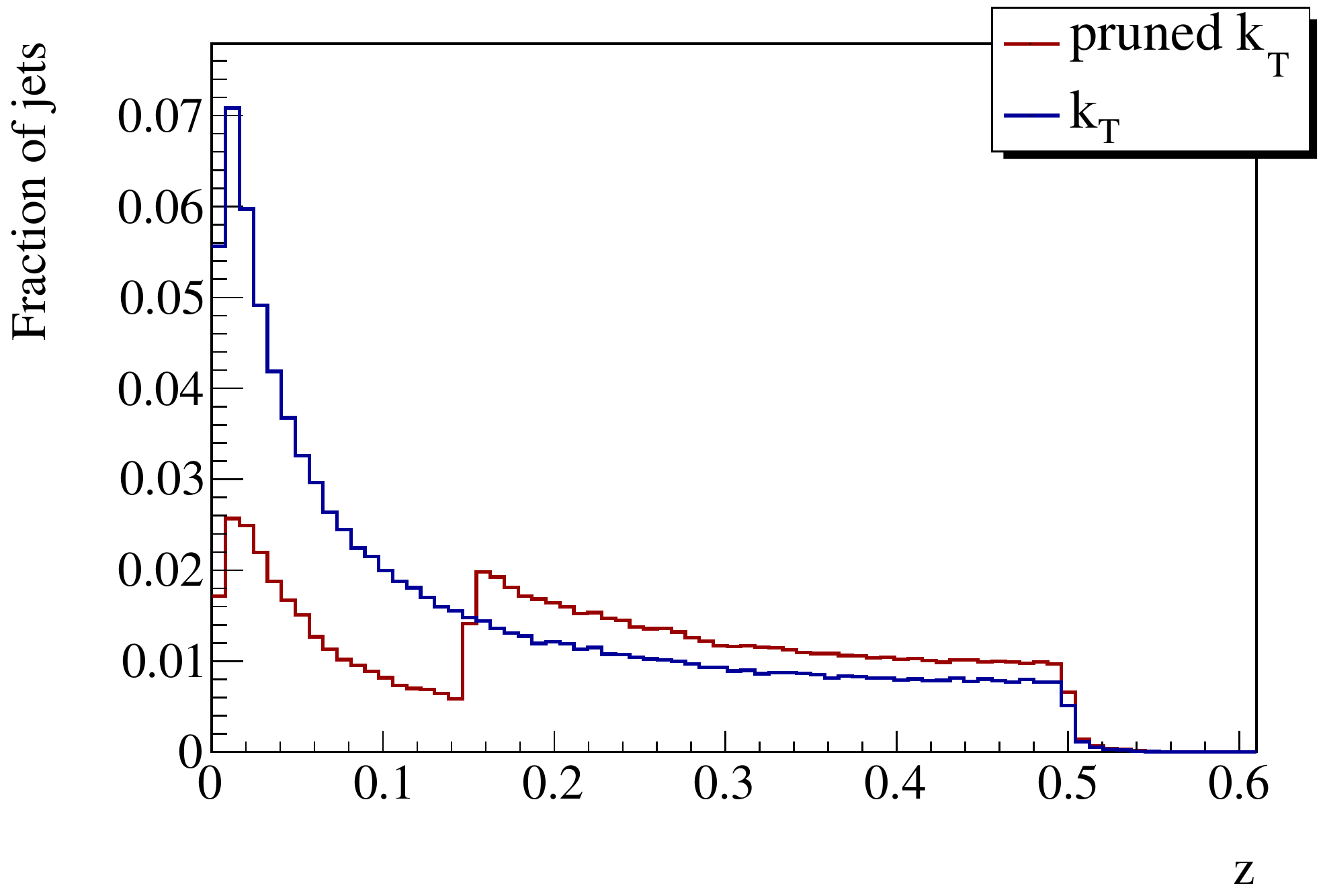} \label{fig:epem_QCD_pCompared:zKT}}

\subfloat[$\Delta R_{12}$, CA] {\includegraphics[width = .48\columnwidth] {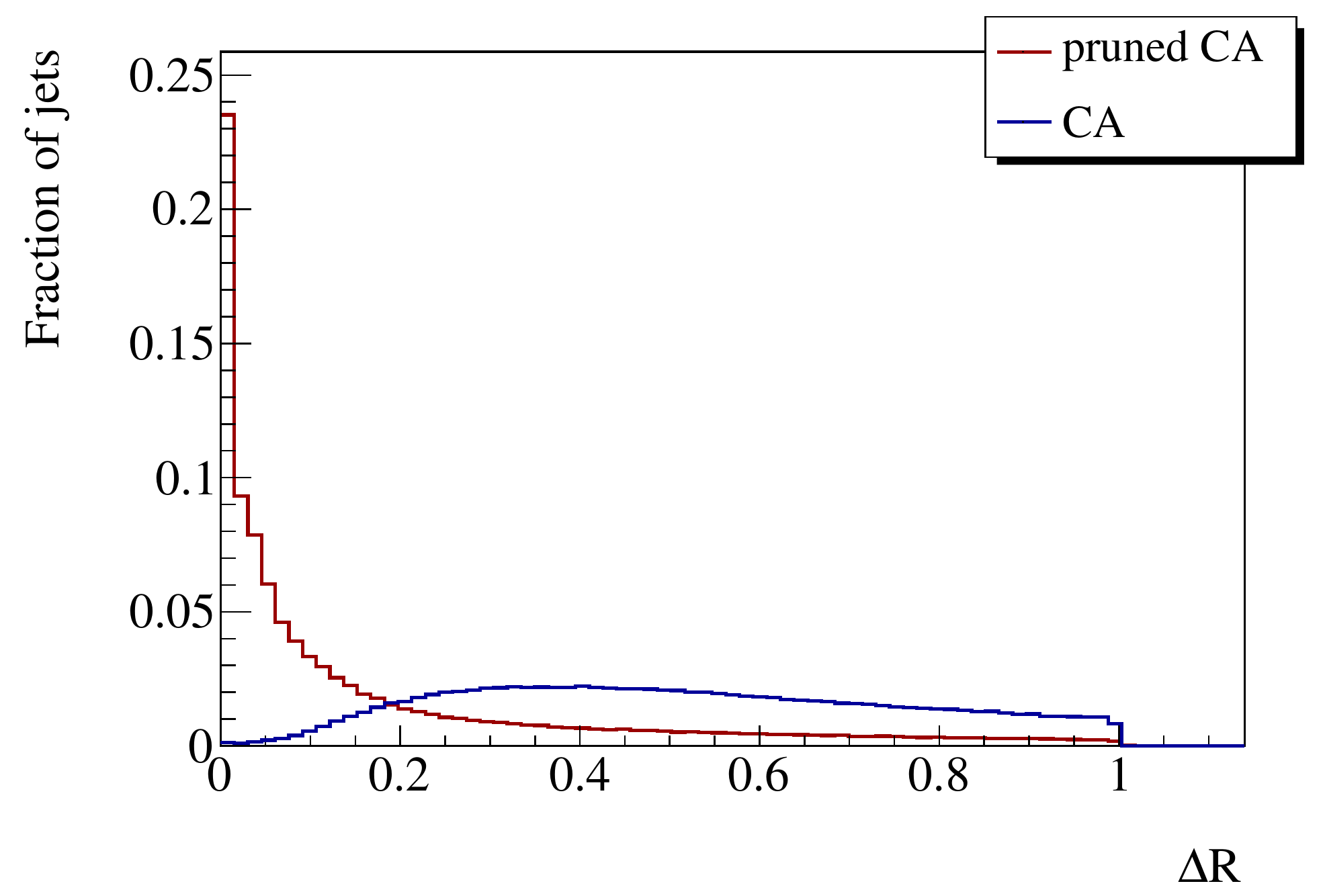}
\label{fig:epem_QCD_pCompare:DRCA}}
\subfloat[$\Delta R_{12}$, $\kt$] {\includegraphics[width = .48\columnwidth] {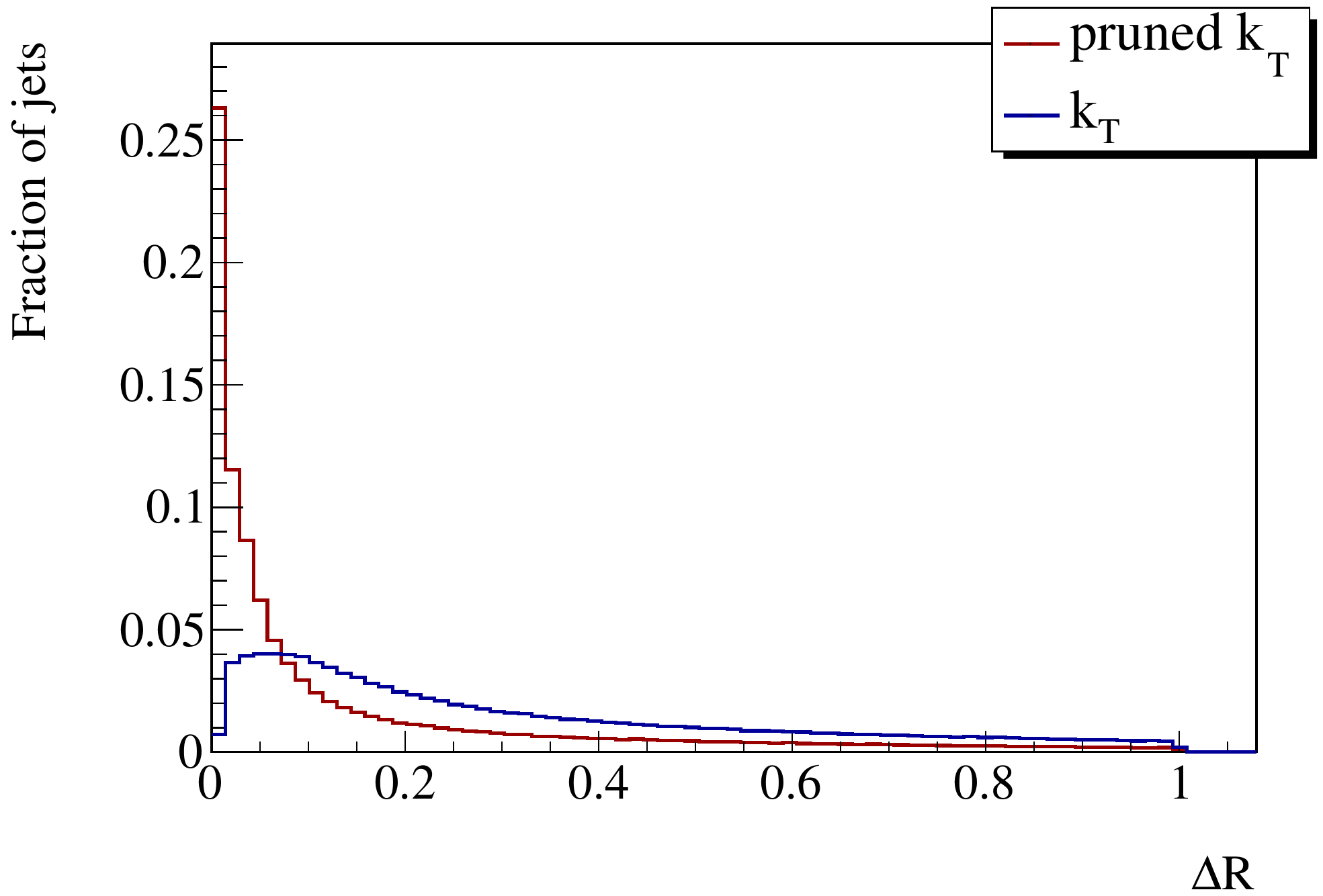} \label{fig:epem_QCD_pCompare:DRKT}}

\subfloat[$a_1$, CA] {\includegraphics[width = .48\columnwidth]{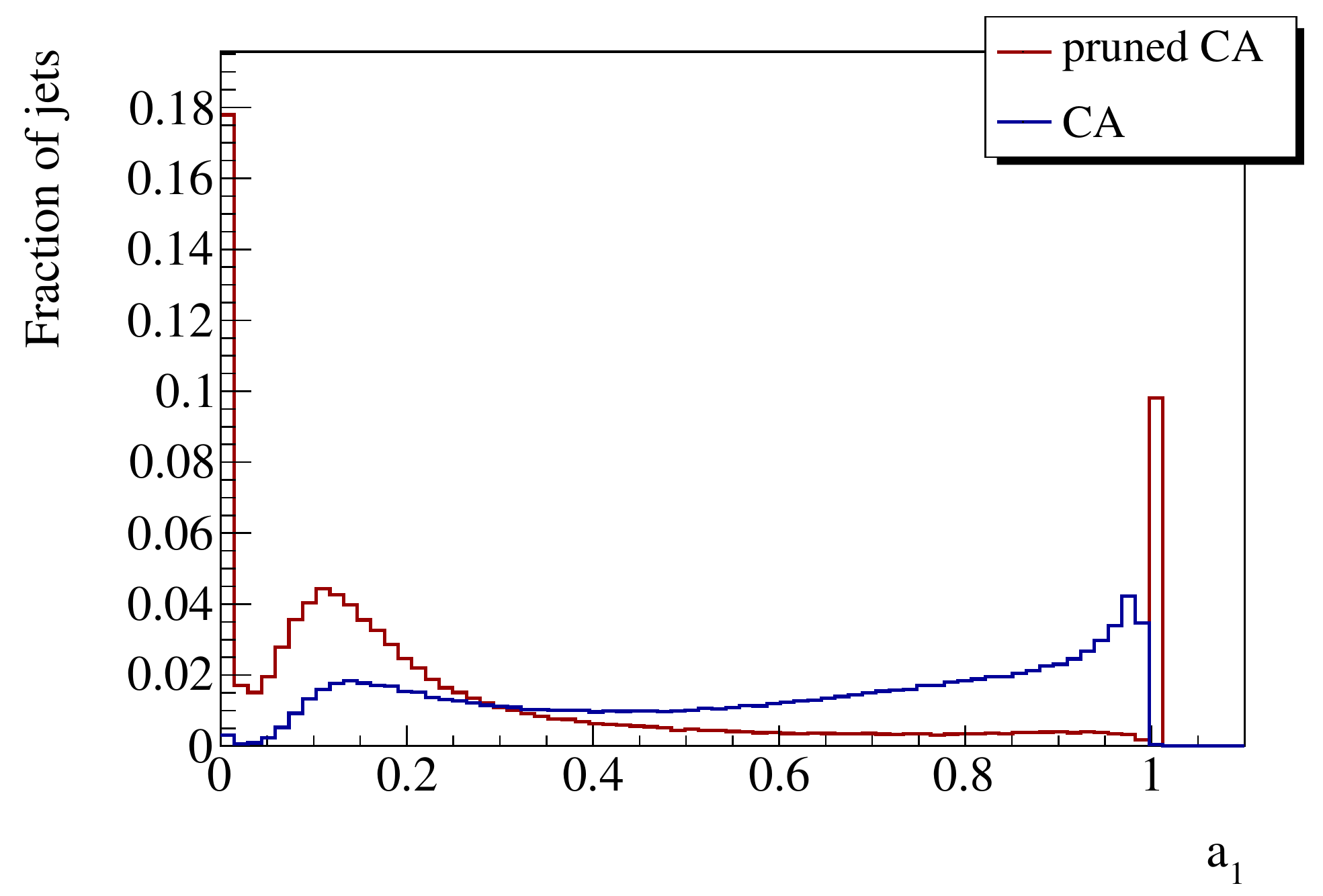} \label{fig:epem_QCD_pCompare:a1CA}}
\subfloat[$a_1$, $\kt$]{\includegraphics[width = .48\columnwidth]{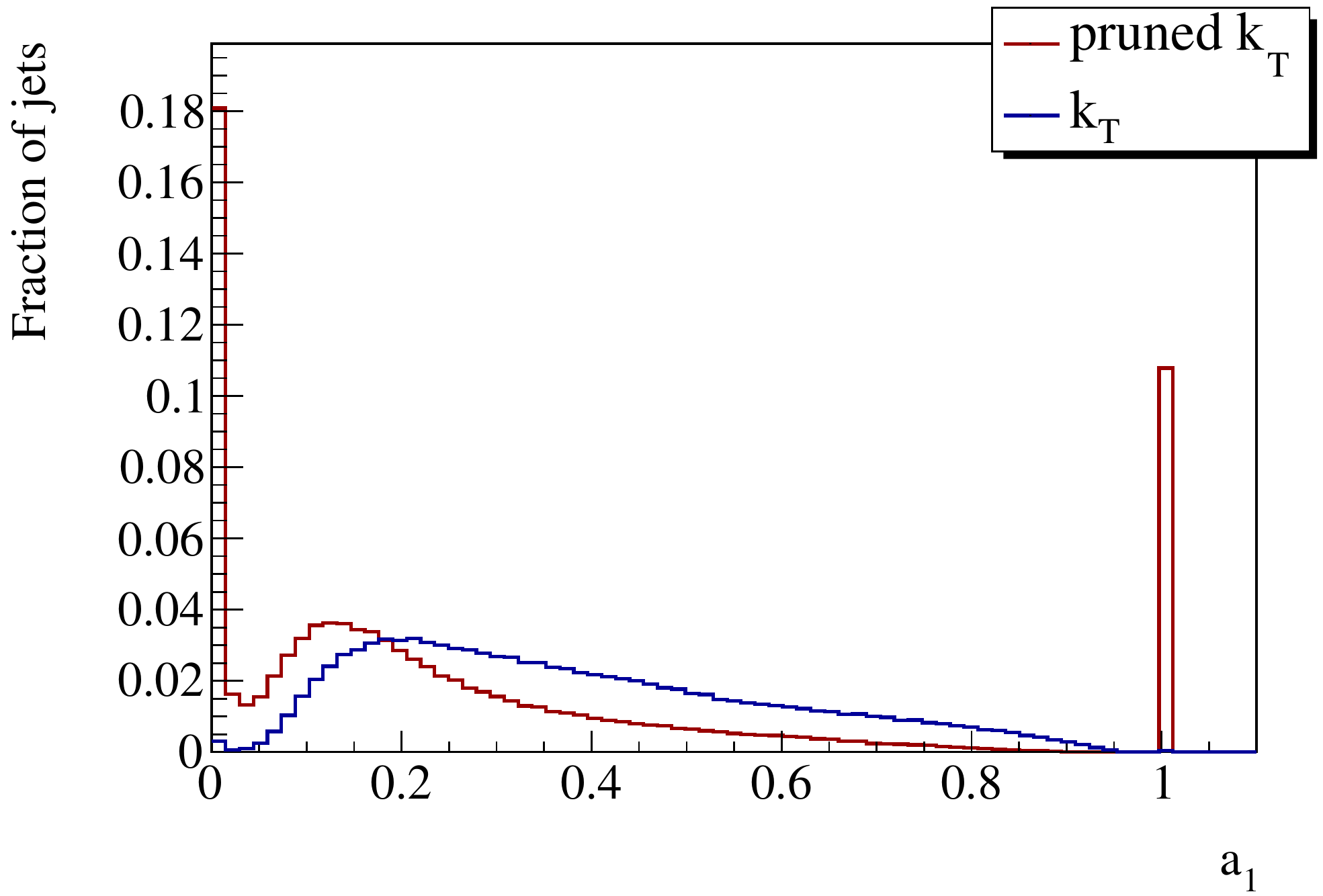}
\label{fig:epem_QCD_pCompare:a1KT}}

\caption{Distributions in $z$, $\Delta R_{12}$, and $a_1$ for pruned and unpruned jets in $\ee \to q\bar q$ events.}
\label{fig:epem_QCD_pCompare}
\end{figure}

In Fig.~\ref{fig:epem_QCD_pCompare} we show the same plots for the $\ee \to q\bar q$ sample.  For both algorithms pruning has a significant effect.  $z$ and $\Delta R_{12}$ are pushed toward zero, indicating that all but a very narrow hard core of the QCD jets are being pruned away.  For each variable, jets with only one constituent are included in the zero bin, which explains the excess at $z \approx 0$ for $\kt$.  The distributions in $a_1$ suggest that asymmetric mergings are pruned away from jets until all that remains are a few reasonably symmetric low-mass protojets.

\begin{figure}[htbp] \begin{center}
\subfloat[CA] {\includegraphics[width = .48\columnwidth] {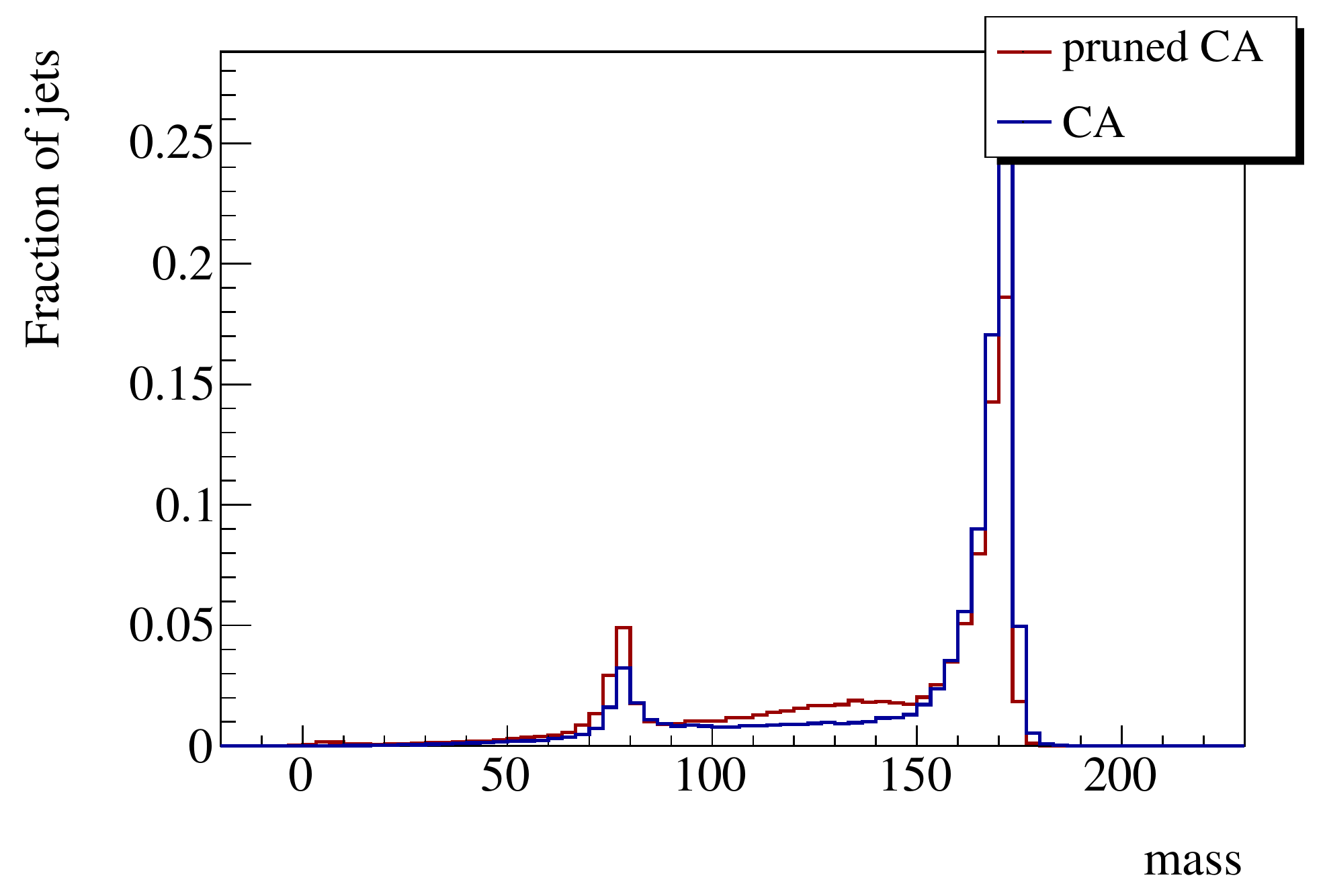} \label{fig:epem_ttbar_pCompare_mass:CA}}
\subfloat[CA (zoomed in)] {\includegraphics[width = .48\columnwidth]{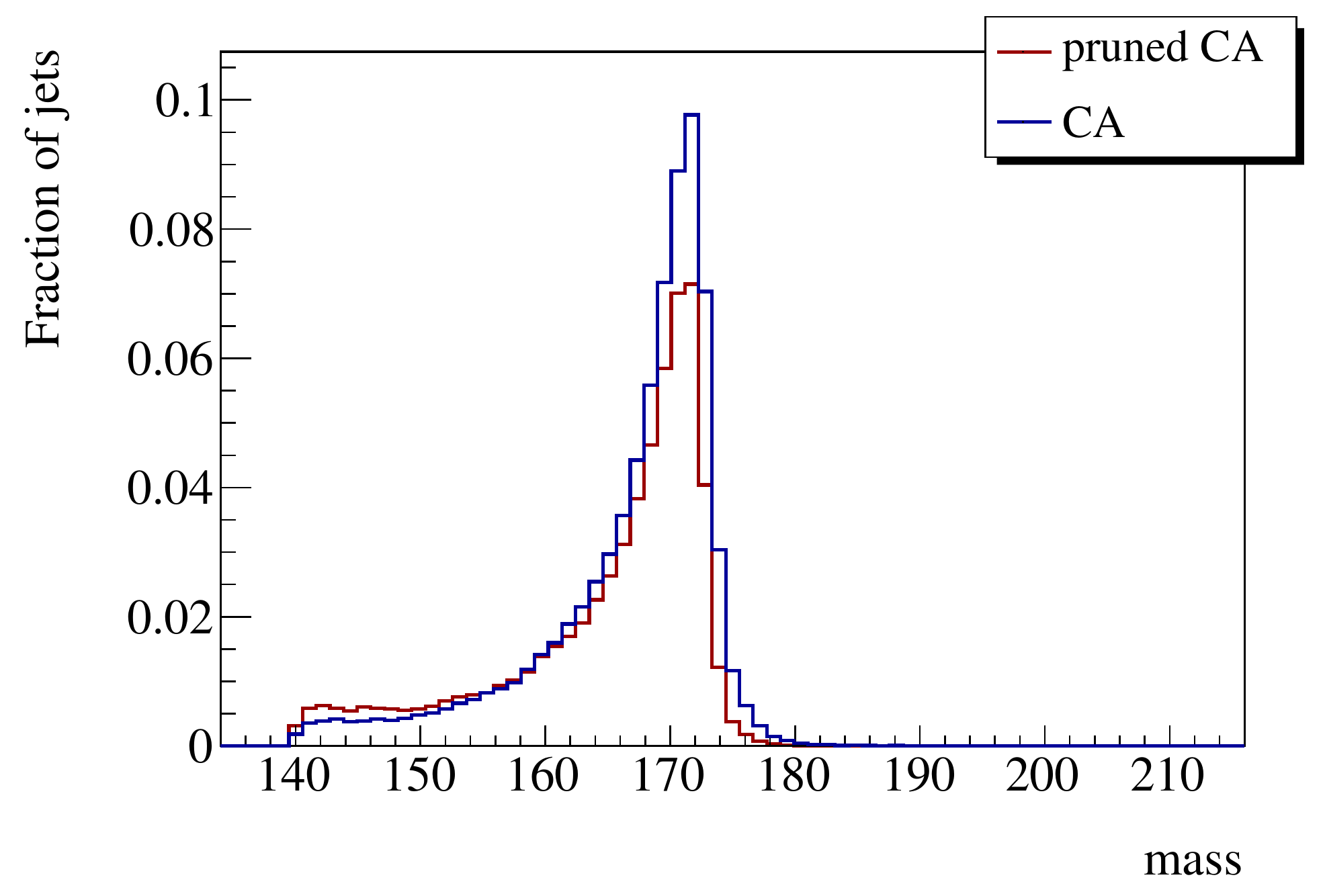}
\label{fig:epem_ttbar_pCompare_mass:CAnarrow}}

\subfloat[$\kt$]{\includegraphics[width = .48\columnwidth]{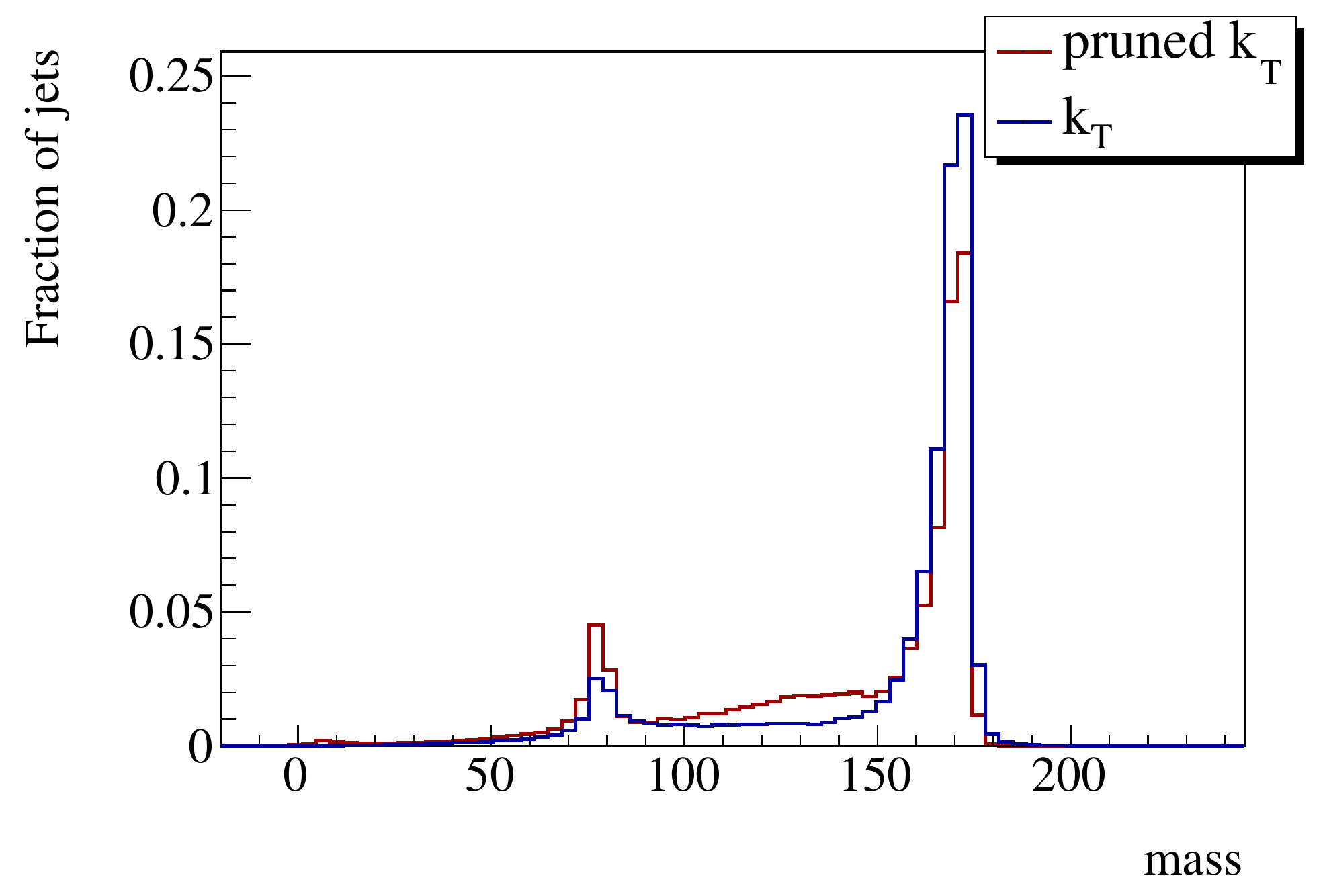} \label{fig:epem_ttbar_pCompare_mass:KT}}
\subfloat[$\kt$ (zoomed in)] {\includegraphics[width = .48\columnwidth] {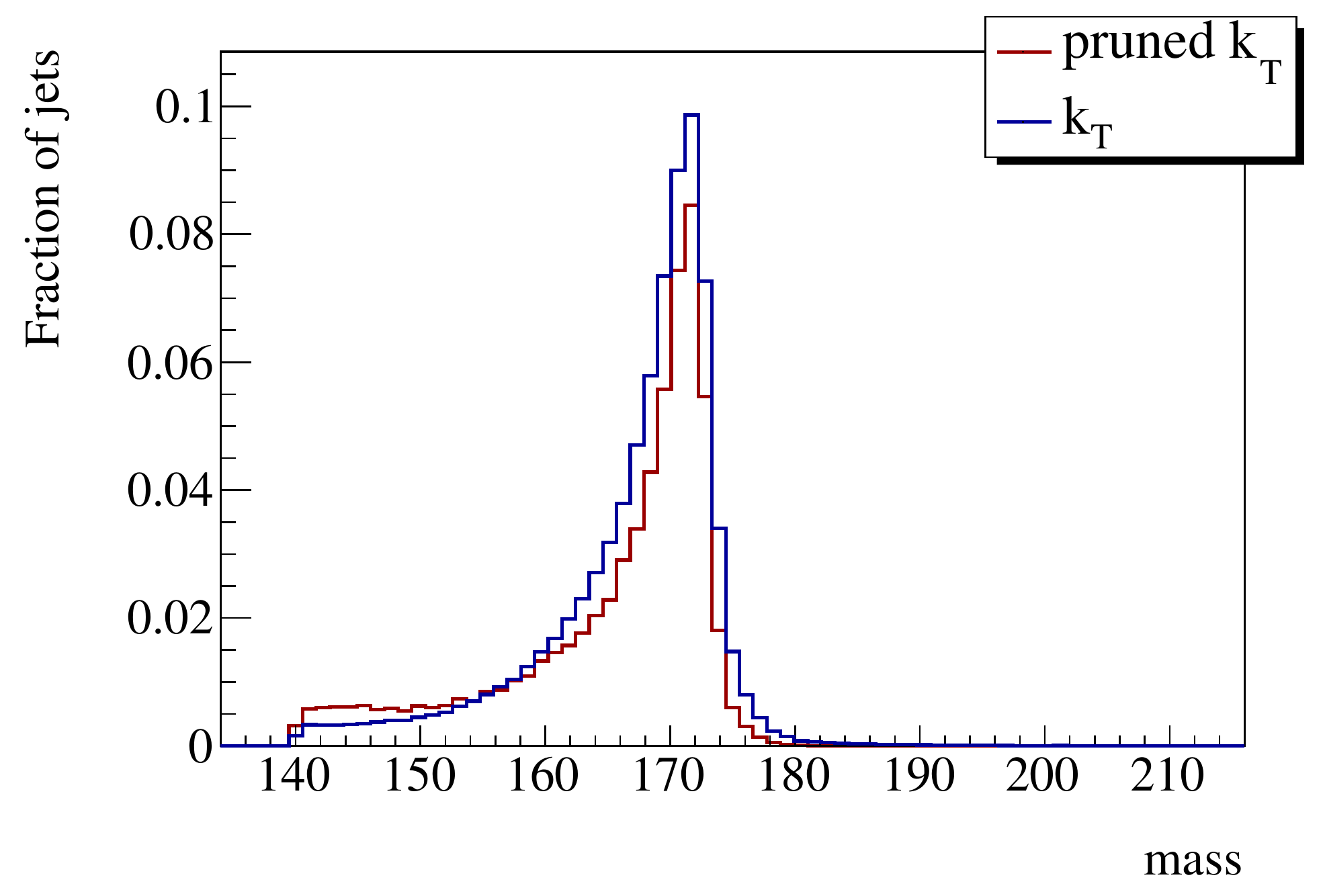} \label{fig:epem_ttbar_pCompare_mass:KTnarrow}}
\end{center}
\caption[Distribution in $m_J$ for pruned and unpruned jets in $\ee \to t\bar t$ events]{Distribution in $m_J$ for pruned and unpruned jets in $\ee \to t\bar t$ events, using the CA and $\kt$ algorithms.  Jets have $p_T > 500$ GeV and D = 1.0.}
\label{fig:epem_ttbar_pCompare_mass}
\end{figure}

\begin{figure}[htbp] \begin{center}
\subfloat[CA] {\includegraphics[width = .48\columnwidth] {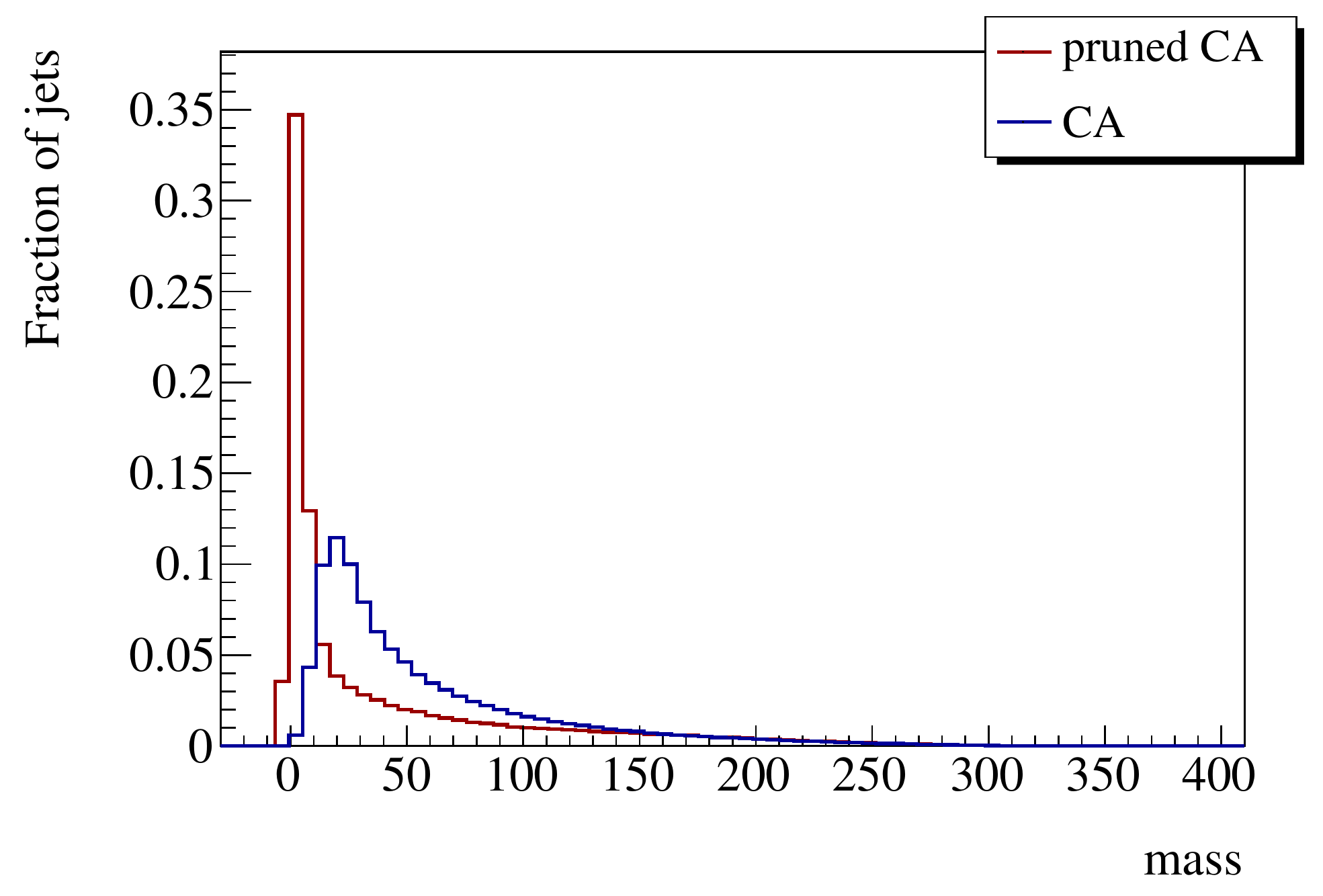} \label{fig:epem_QCD_pCompare_mass:CA}}
\subfloat[CA (zoomed in)] {\includegraphics[width = .48\columnwidth]{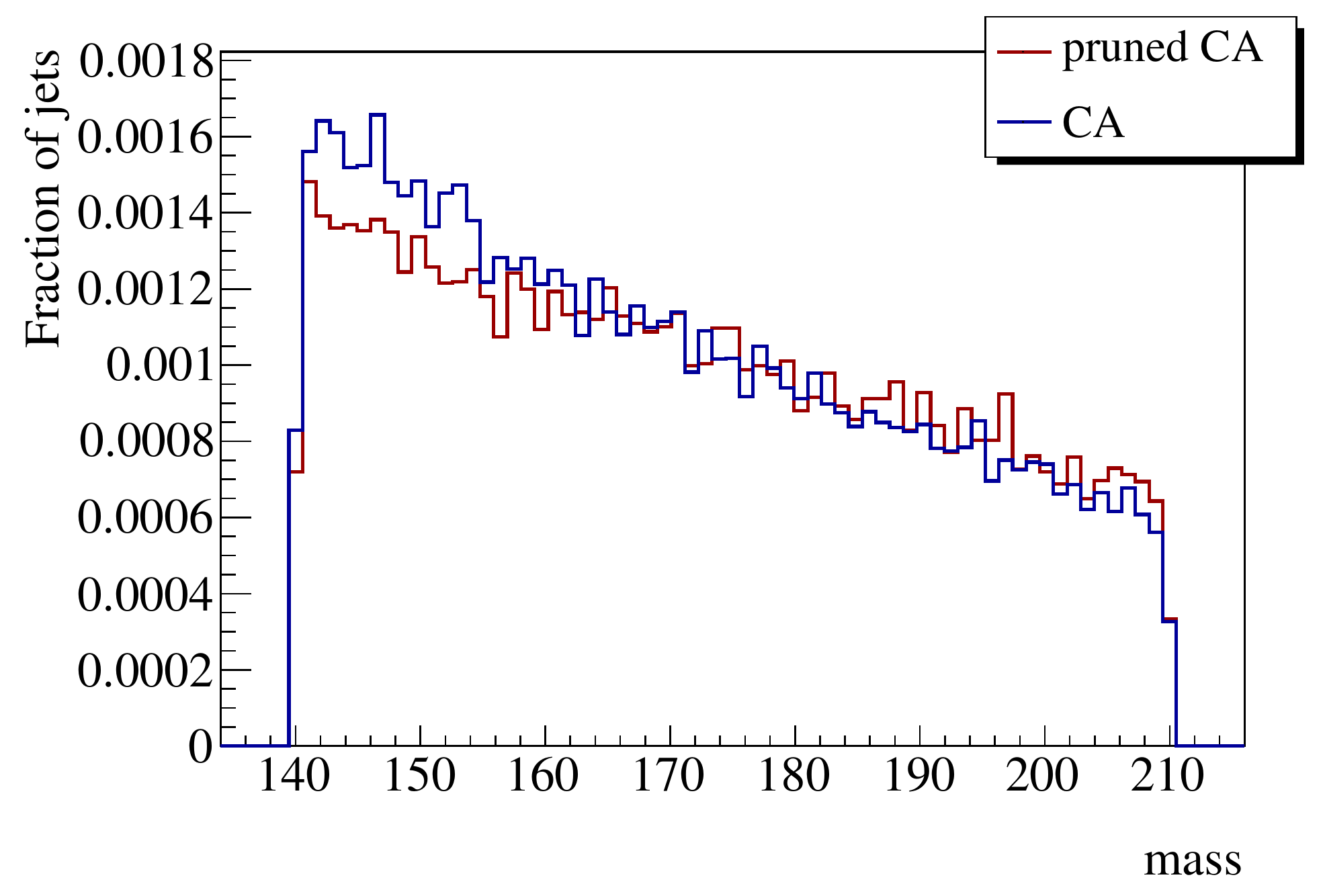}
\label{fig:epem_QCD_pCompare_mass:CAnarrow}}

\subfloat[$\kt$]{\includegraphics[width = .48\columnwidth]{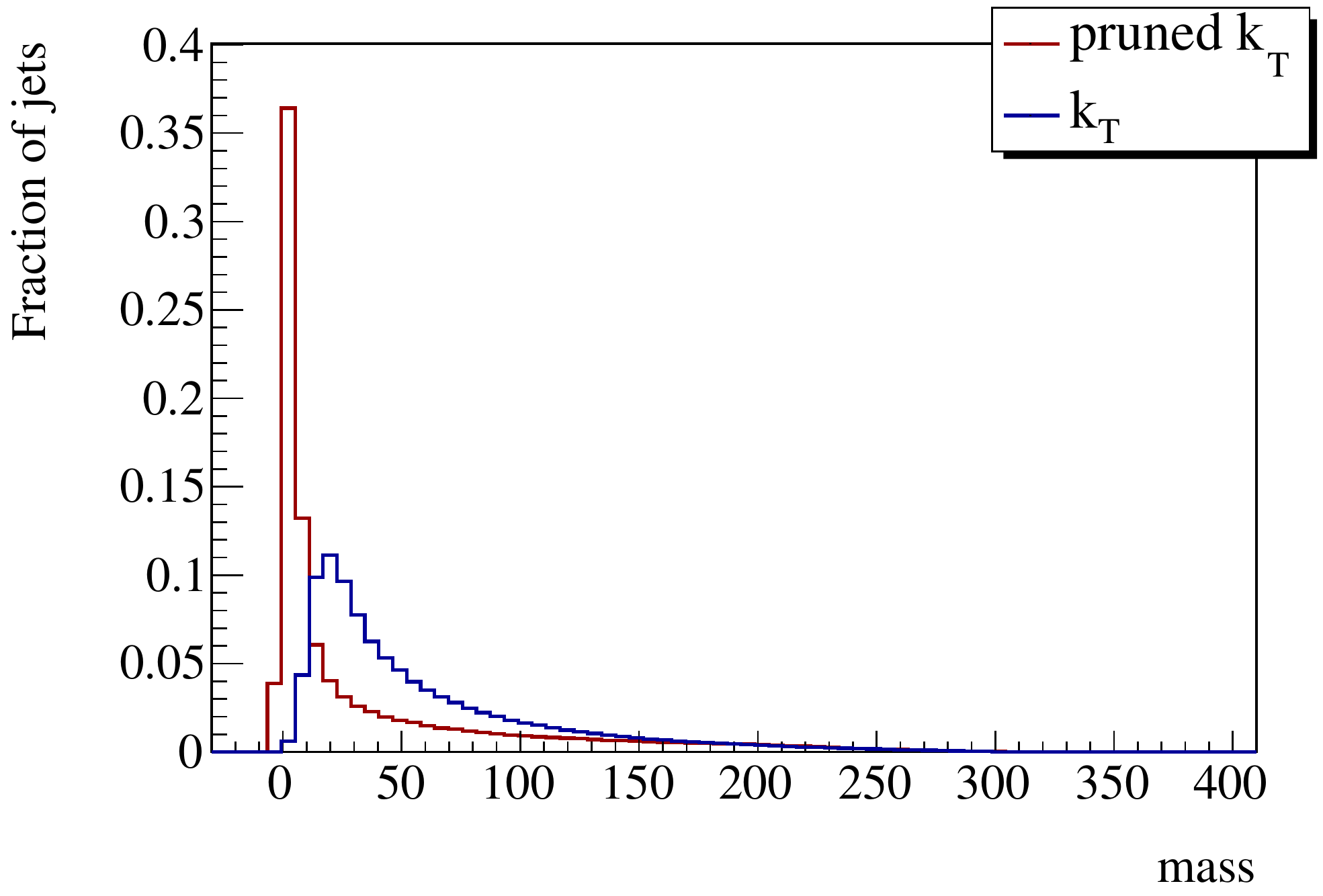} \label{fig:epem_QCD_pCompare_mass:KT}}
\subfloat[$\kt$ (zoomed in)] {\includegraphics[width = .48\columnwidth] {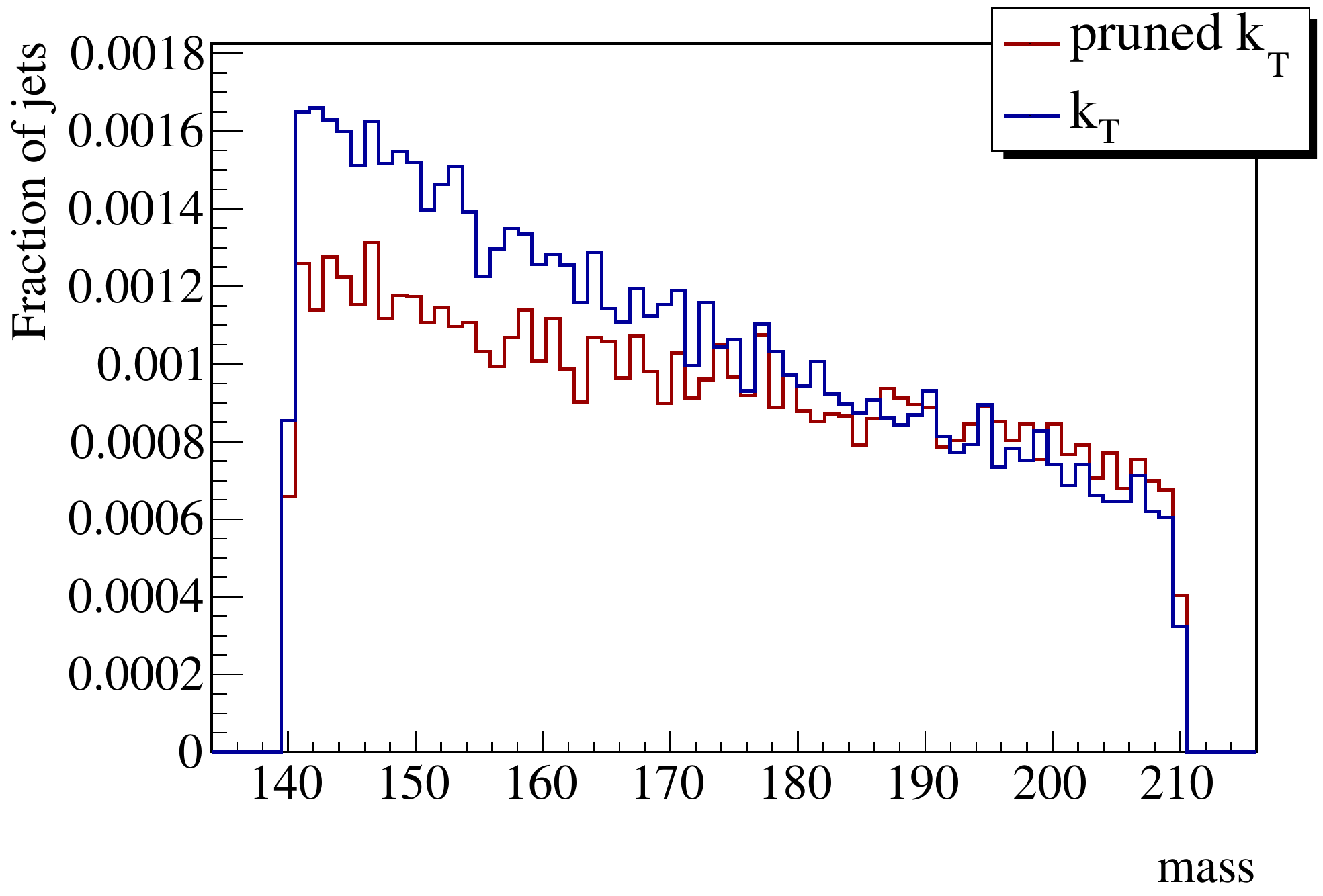} \label{fig:epem_QCD_pCompare_mass:KTnarrow}}
\end{center}
\caption[Distribution in $m_J$ for pruned and unpruned jets in $\ee \to q\bar q$ events]{Distribution in $m_J$ for pruned and unpruned jets in $\ee \to q\bar q$ events, using the CA and $\kt$ algorithms.  Jets have $p_T > 500$ GeV and D = 1.0.}
\label{fig:epem_QCD_pCompare_mass}
\end{figure}

Of course the most important effect of pruning is on the jet mass distribution, which we plot in Fig.~\ref{fig:epem_ttbar_pCompare_mass} for the $t\bar t$ sample and in Fig.~\ref{fig:epem_QCD_pCompare_mass} for the dijet sample.  In the signal sample, the unpruned algorithms already performed quite well at finding tops, and pruning degrades this somewhat.  Note that the $W$ peak increases for both algorithms, indicating we sometimes prune a top down to a $W$.  In the background sample, pruning shifts the mass distribution down considerably, but the effect is negligible in the top mass window --- the jets with masses this large are not affected by pruning.  We can conclude that pruning is probably not very useful in a search for top quarks in this case, although it might be useful in a search for decays with a smaller $m/p_T$.

\section{Effects of pruning in \texorpdfstring{$pp$}{pp} collisions}
 \label{sec:prune:pp}

We saw in Sec.~\ref{sec:sub:eventeffects} that hadron collisions are noisier than electron collisions, with radiation coming from initial partons as well as multiple interactions of beam remnants.  Pruning is intended to remove as much of this ``extra'' radiation as possible, so we now repeat the analysis of the previous section for $pp$ event samples to see its effects.  We compare pruned and unpruned jets, acting on the ``FSR'' (just final-state radiation) and ``FSR+ISR+UE'' (full simulation) samples from Sec.~\ref{sec:sub:eventeffects}.  We might hope that pruning, acting on jets in the latter sample, would yield results similar to the former.  In fact pruning systematically shifts the kinematics of both samples, but in such a way that the end results are similar: pruned FSR jets are remarkably similar to pruned FSR+ISR+UE jets.

\begin{figure}[htbp]
\subfloat[$z$, CA] {\includegraphics[width = .48\columnwidth] {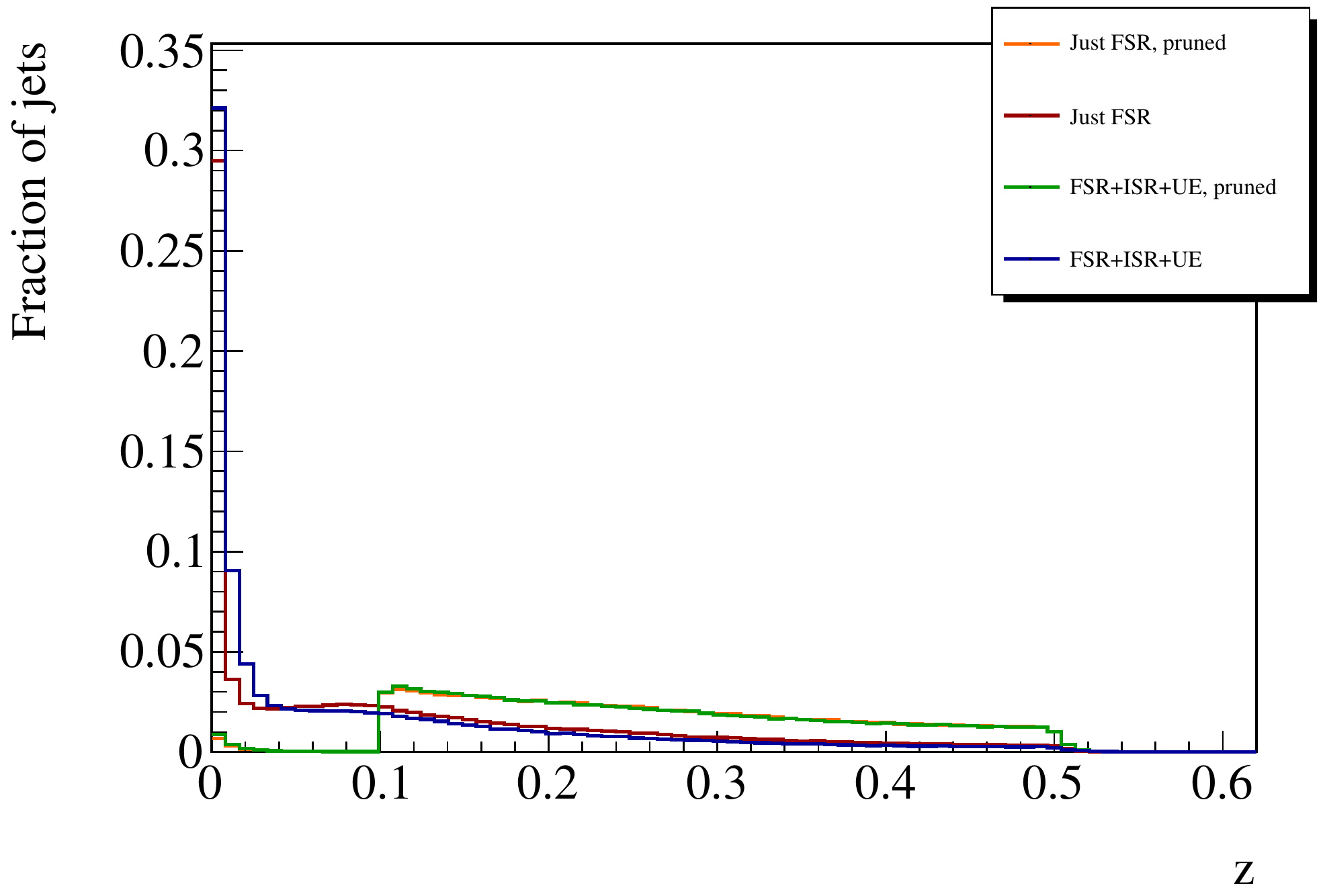} \label{fig:ttbar_pCompare:zCA}}
\subfloat[$z$, $\kt$]{\includegraphics[width = .48\columnwidth]{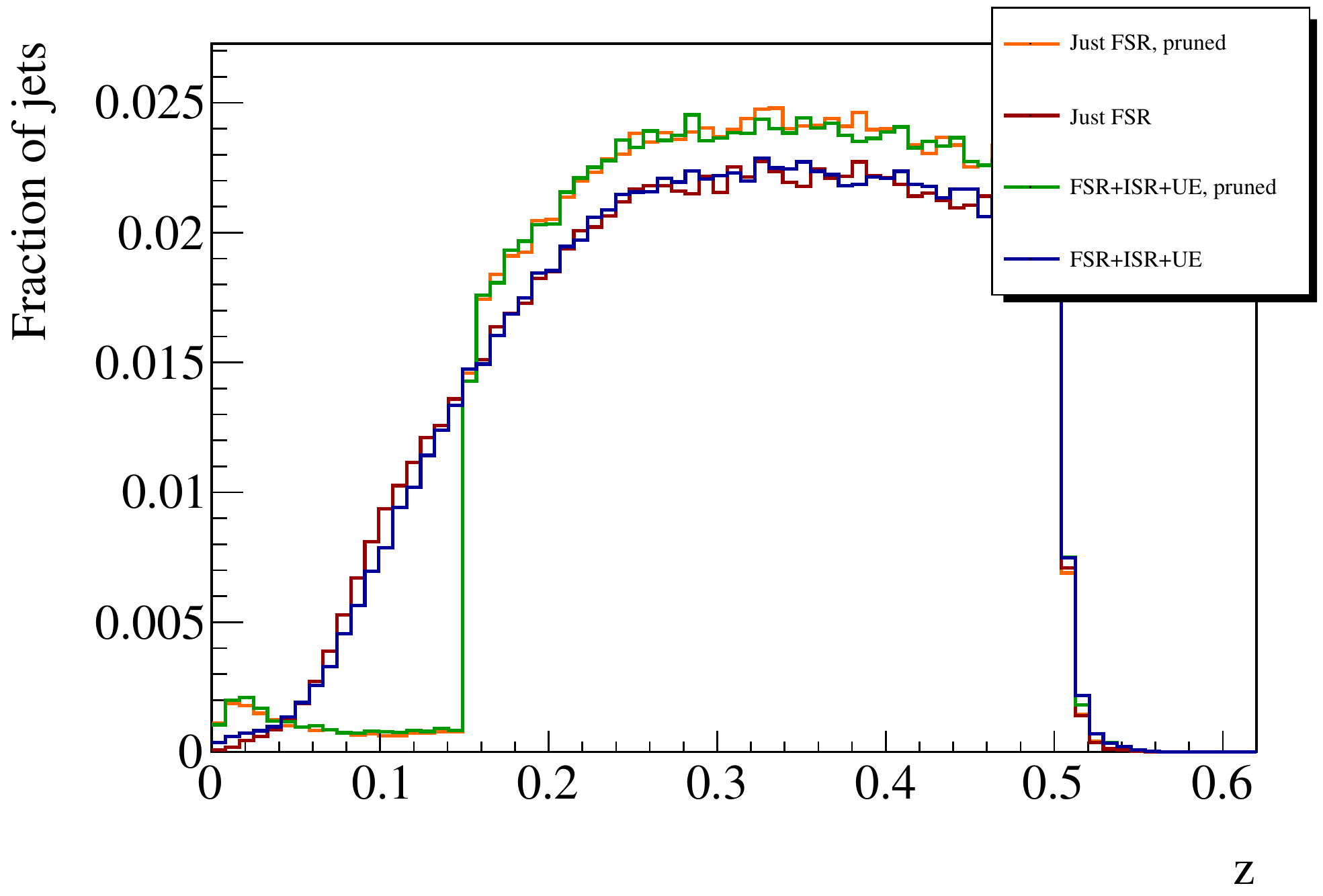} \label{fig:ttbar_pCompared:zKT}}

\subfloat[$\Delta R_{12}$, CA] {\includegraphics[width = .48\columnwidth] {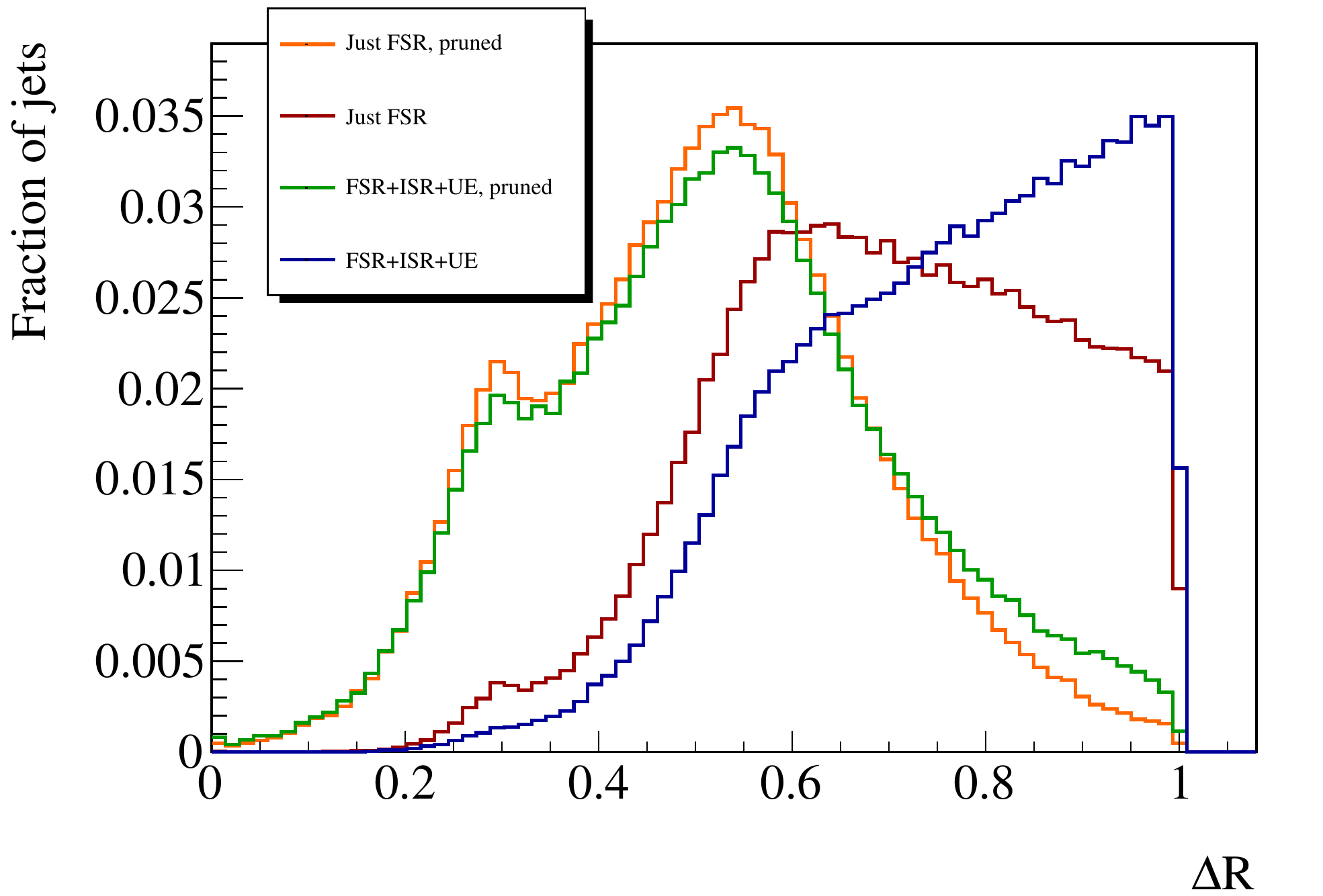}
\label{fig:ttbar_pCompare:DRCA}}
\subfloat[$\Delta R_{12}$, $\kt$] {\includegraphics[width = .48\columnwidth] {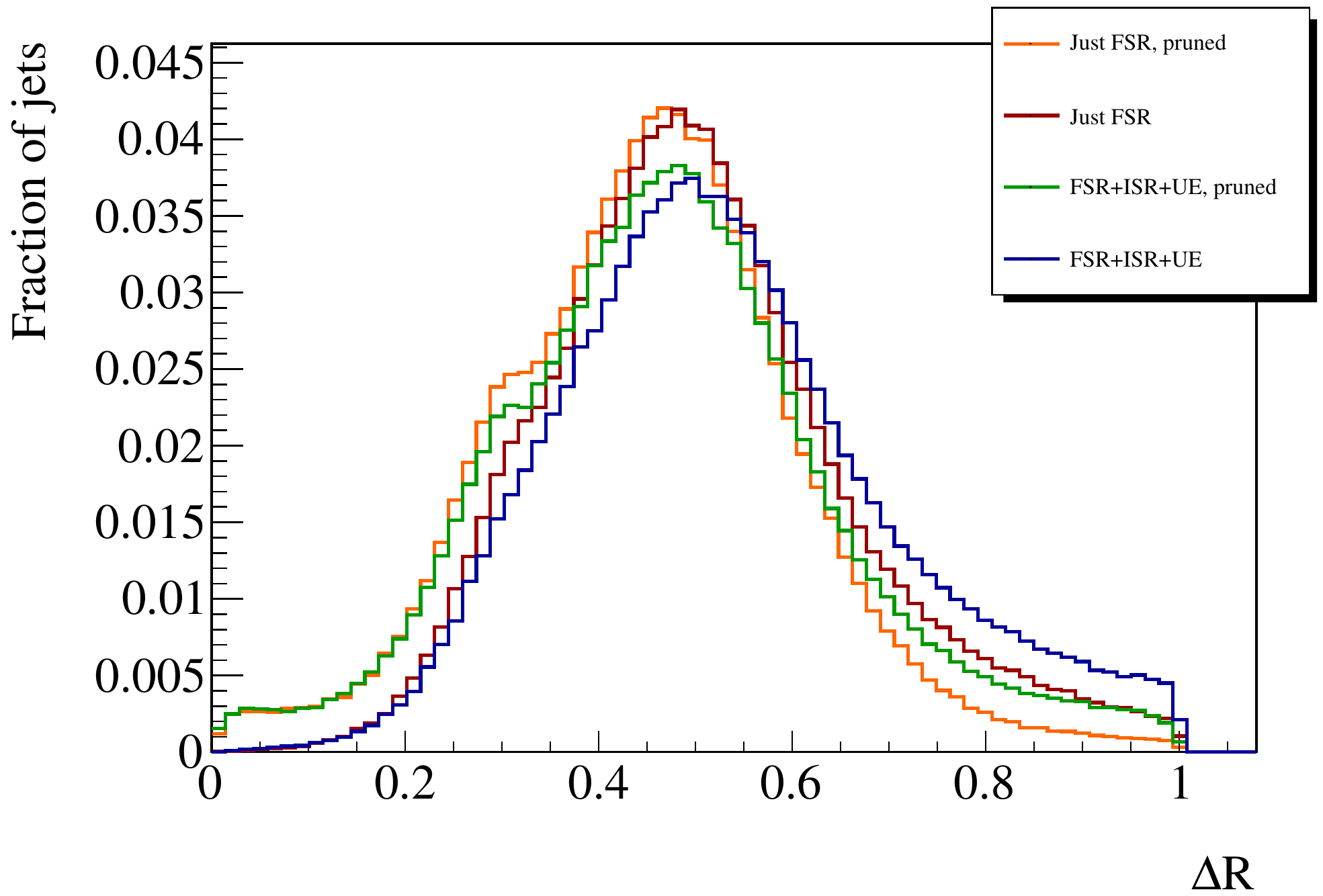} \label{fig:ttbar_pCompare:DRKT}}

\subfloat[$a_1$, CA] {\includegraphics[width = .48\columnwidth]{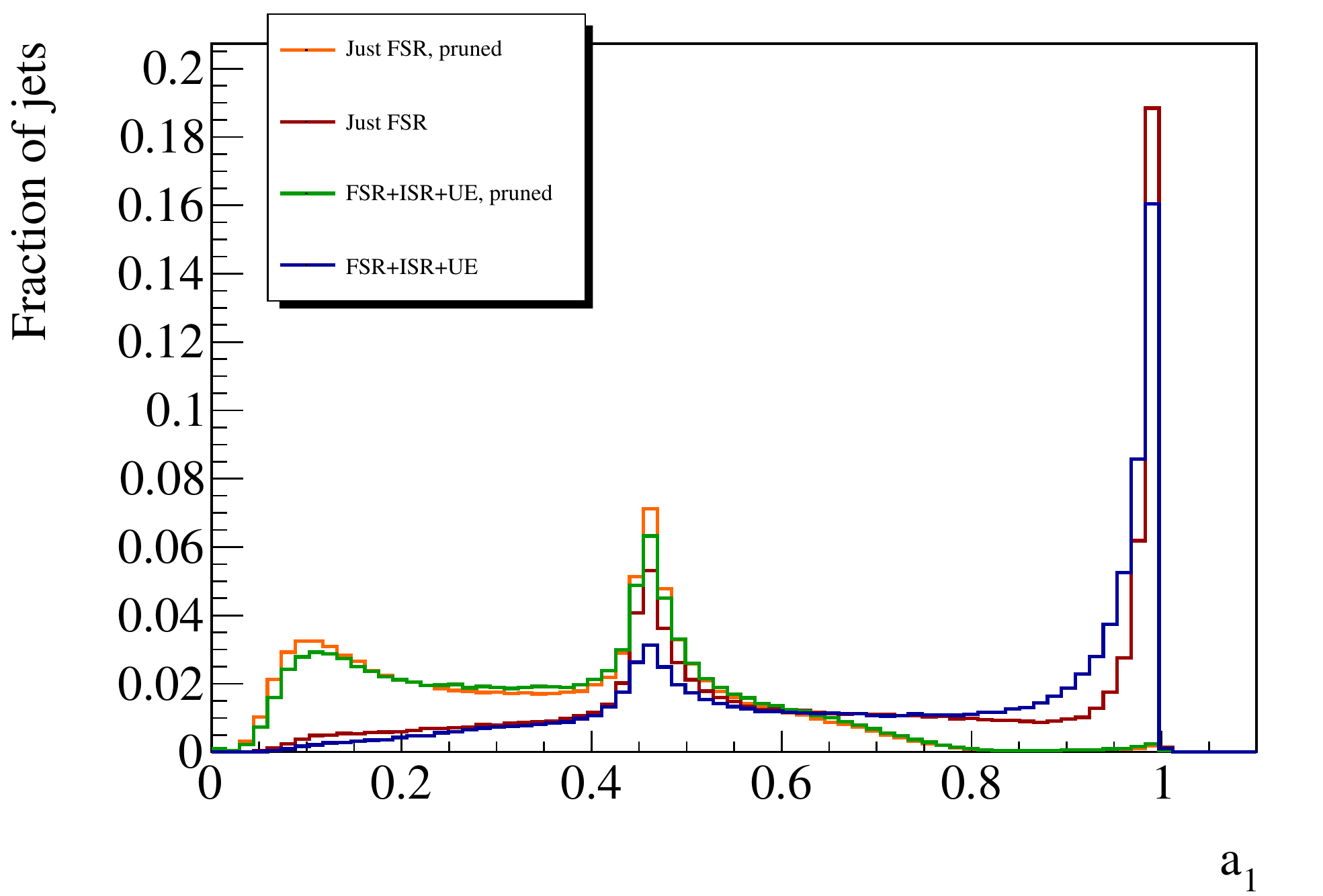} \label{fig:ttbar_pCompare:a1CA}}
\subfloat[$a_1$, $\kt$]{\includegraphics[width = .48\columnwidth]{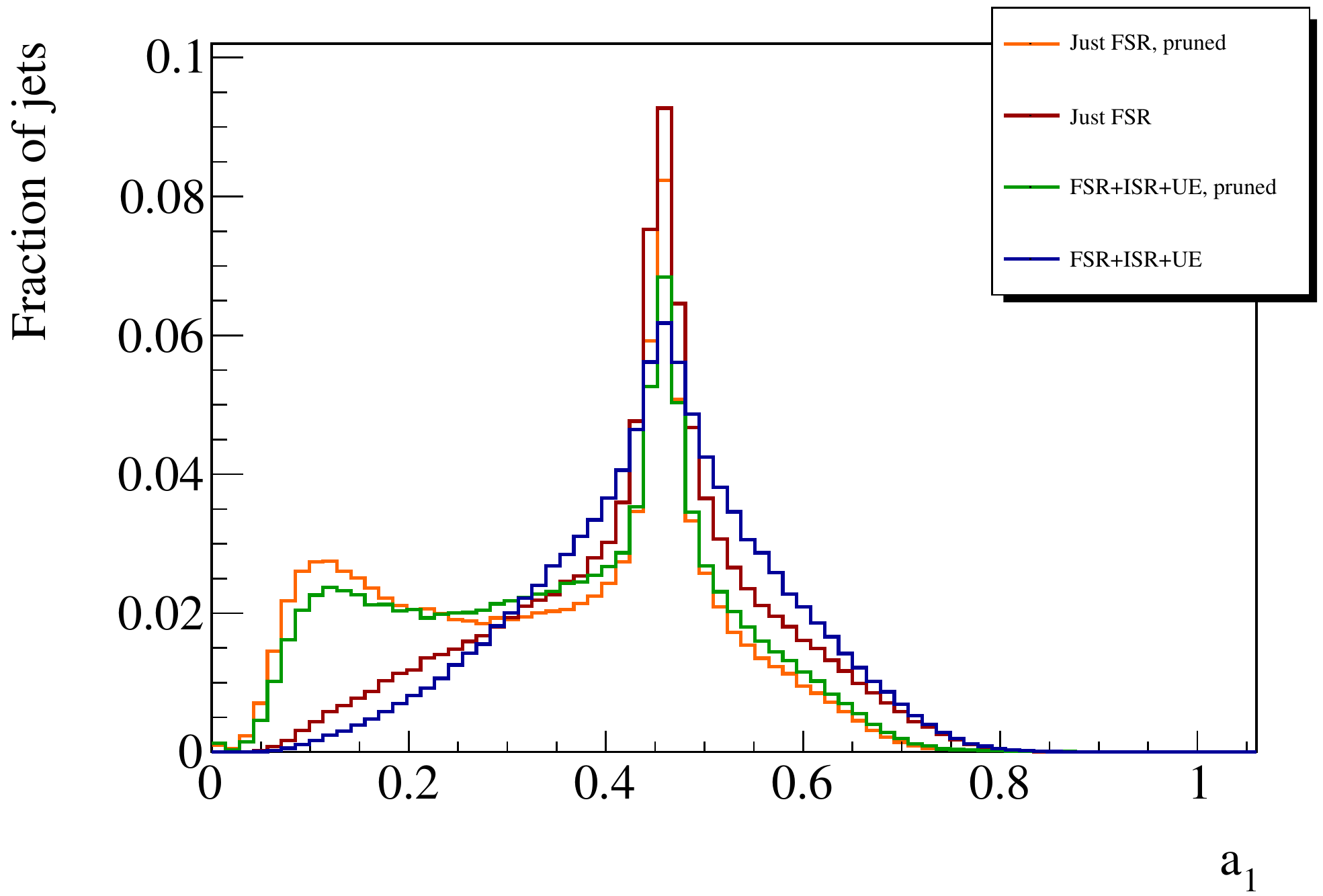}
\label{fig:ttbar_pCompare:a1KT}}

\caption{Distributions in $z$, $\Delta R_{12}$, and $a_1$ for pruned and unpruned jets in $pp \to t\bar t$ events.}
\label{fig:ttbar_pCompare}
\end{figure}

In Fig.~\ref{fig:ttbar_pCompare} we plot substructure kinematic distributions for pruned and unpruned jets in the $pp \to t\bar t$ samples.  As in the $\ee$ events, pruning removes soft, large-angle radiation and hence depletes the small-$z$, large-$\Delta R_{12}$ regions of phase space.  The $z \approx 0$ and $a_1 \approx 1$ peaks for CA disappear, while for $\kt$ the substructure is largely unaffected.  Unlike in $\ee$ events, the $a_1$ peak is strongly enhanced for both algorithms: pruning improves our ability to resolve a $W$ subjet.  For CA, it is notable that while including ISR and UE shifts the substructure distributions --- particularly $\Delta R_{12}$ --- this difference is greatly reduced after pruning.  Pruning is largely removing the effect of extra radiation.

\begin{figure}[htbp]
\subfloat[$z$, CA] {\includegraphics[width = .48\columnwidth] {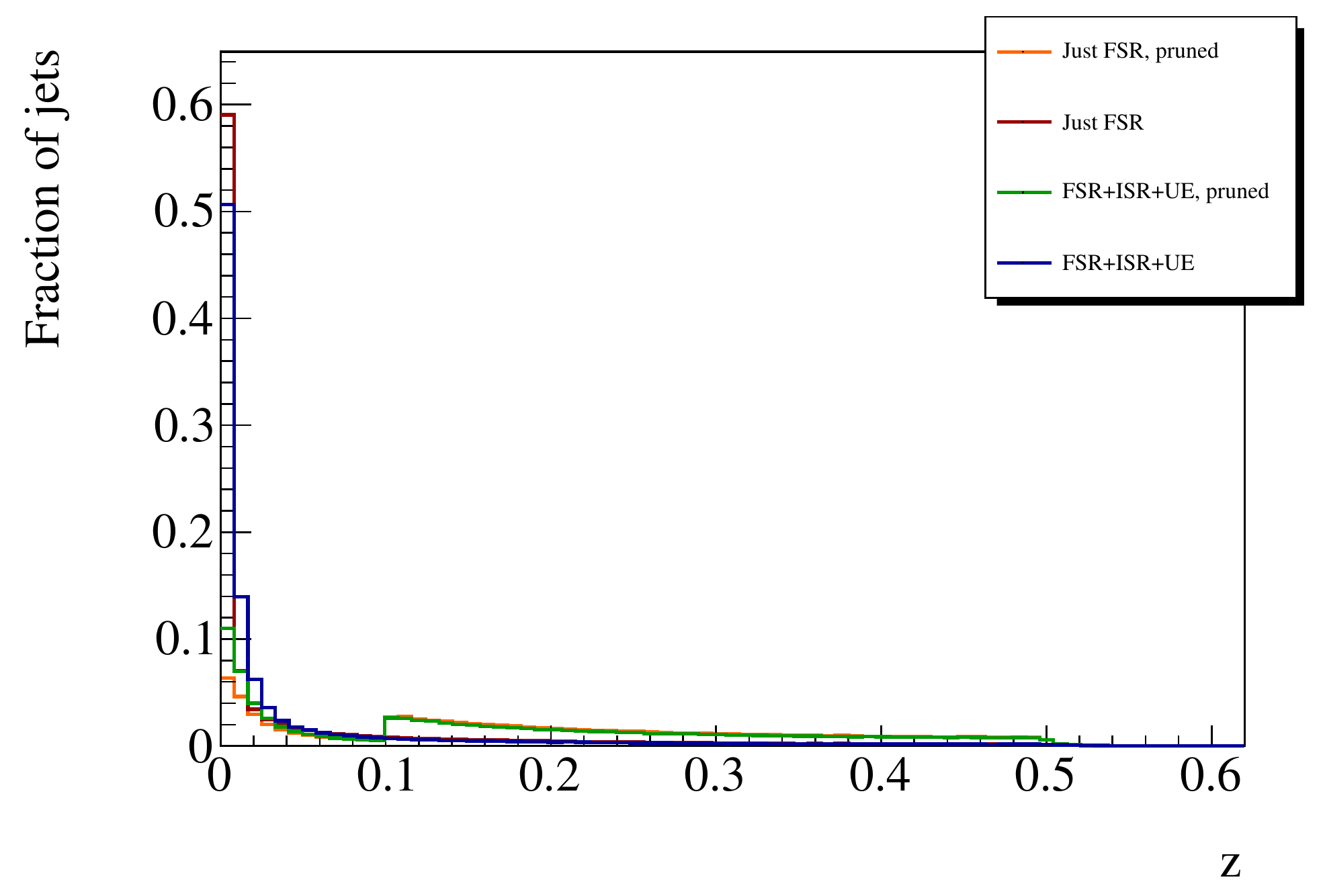} \label{fig:mQCD_pCompare:zCA}}
\subfloat[$z$, $\kt$]{\includegraphics[width = .48\columnwidth]{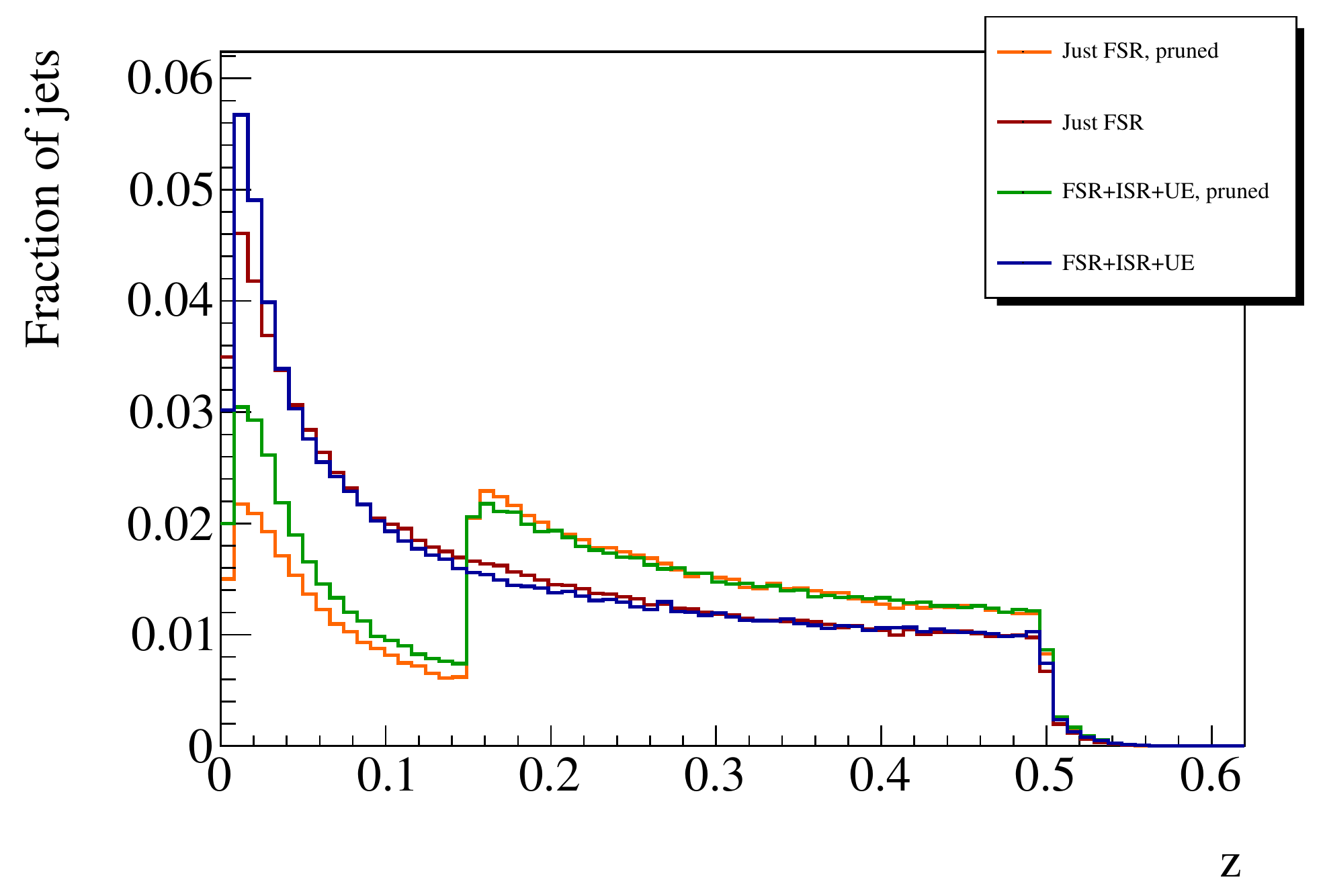} \label{fig:mQCD_pCompared:zKT}}

\subfloat[$\Delta R_{12}$, CA] {\includegraphics[width = .48\columnwidth] {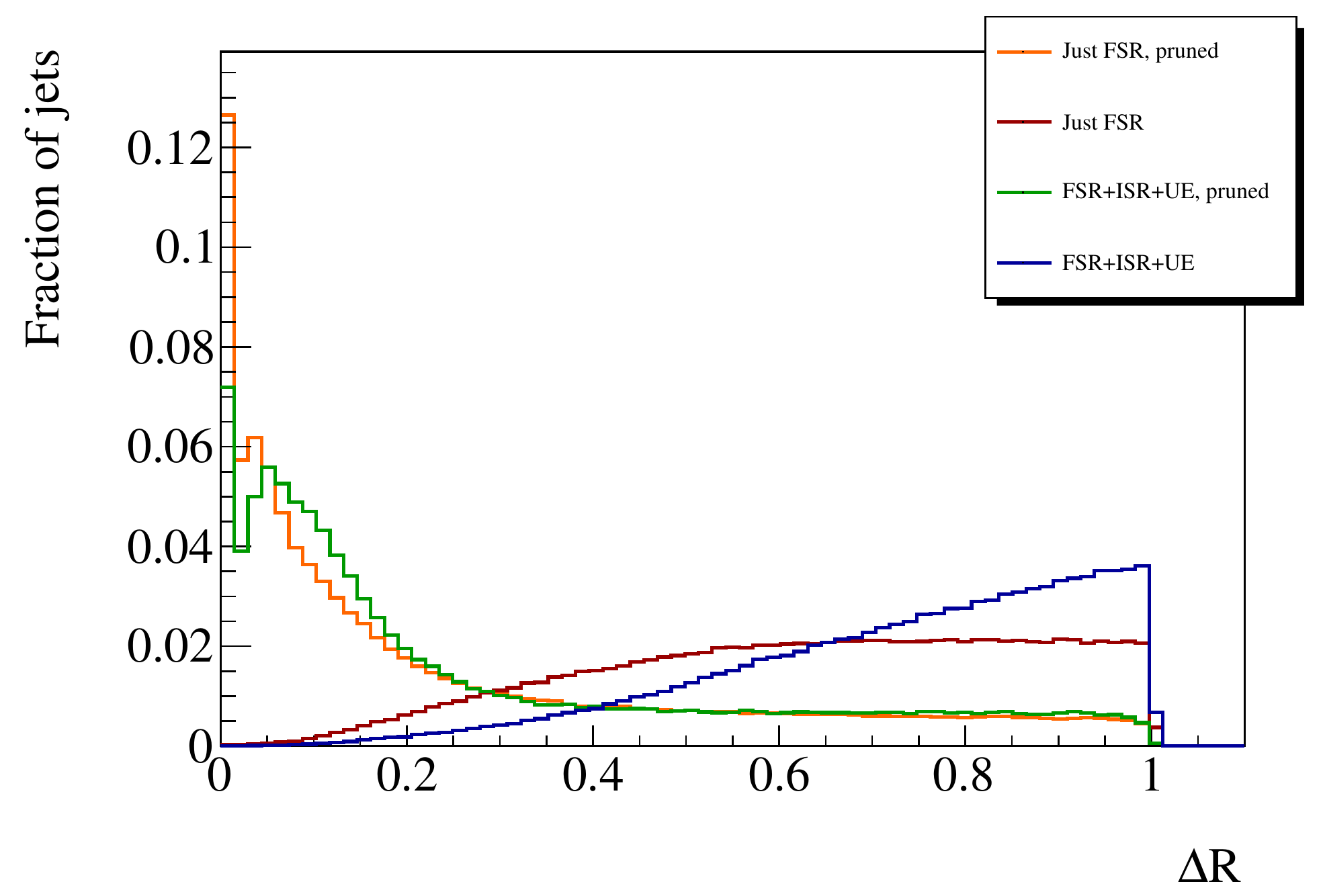}
\label{fig:mQCD_pCompare:DRCA}}
\subfloat[$\Delta R_{12}$, $\kt$] {\includegraphics[width = .48\columnwidth] {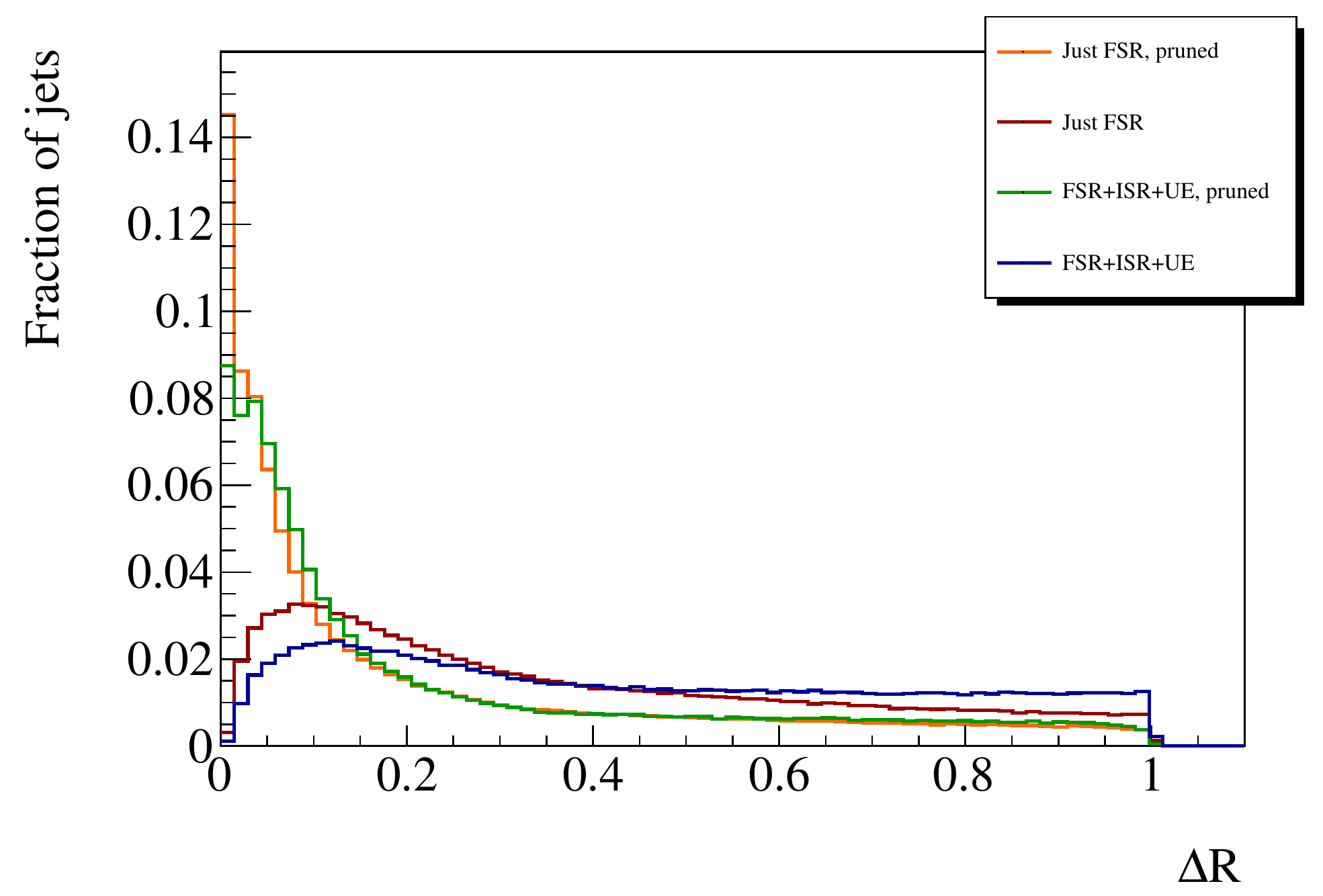} \label{fig:mQCD_pCompare:DRKT}}

\subfloat[$a_1$, CA] {\includegraphics[width = .48\columnwidth]{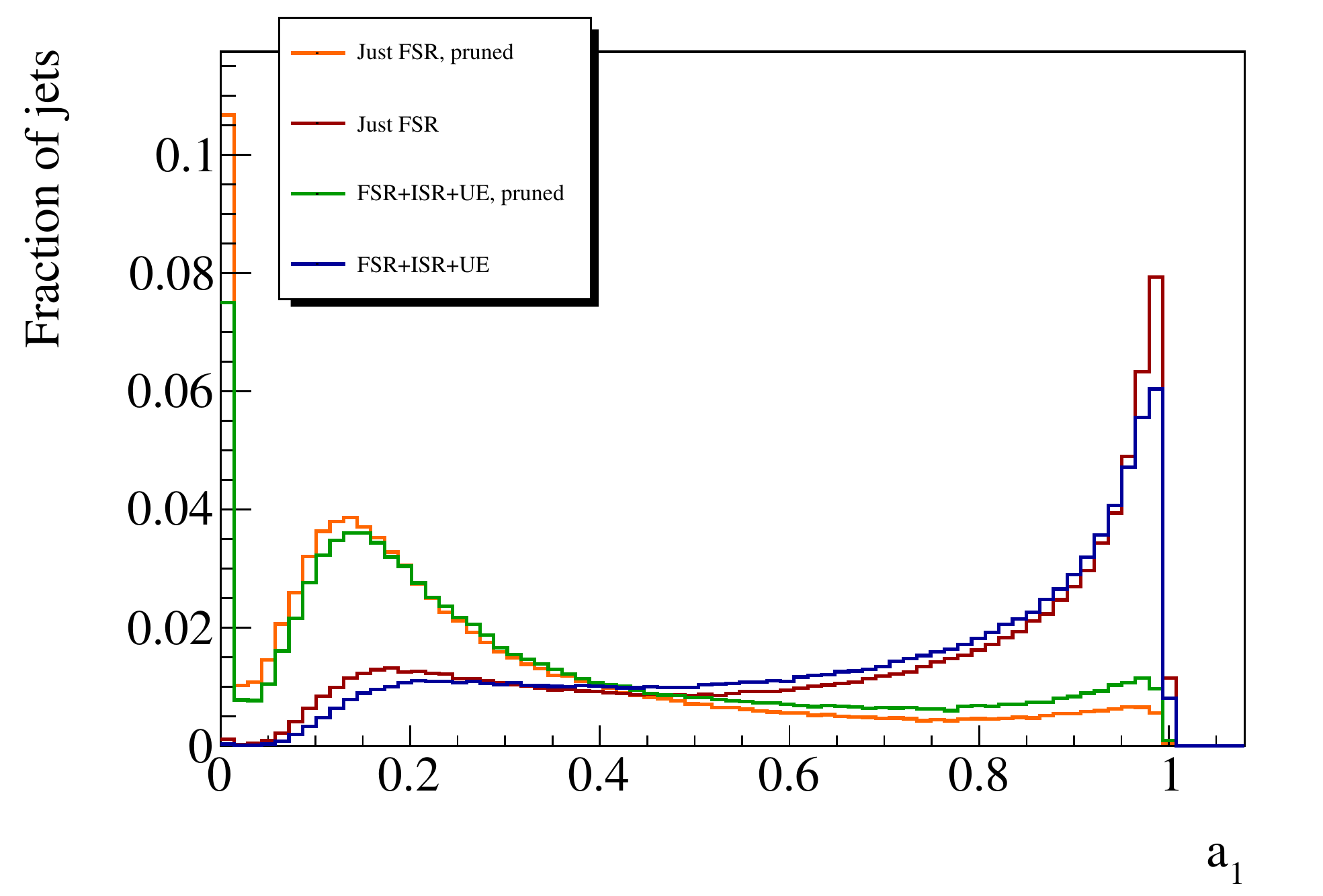} \label{fig:mQCD_pCompare:a1CA}}
\subfloat[$a_1$, $\kt$]{\includegraphics[width = .48\columnwidth]{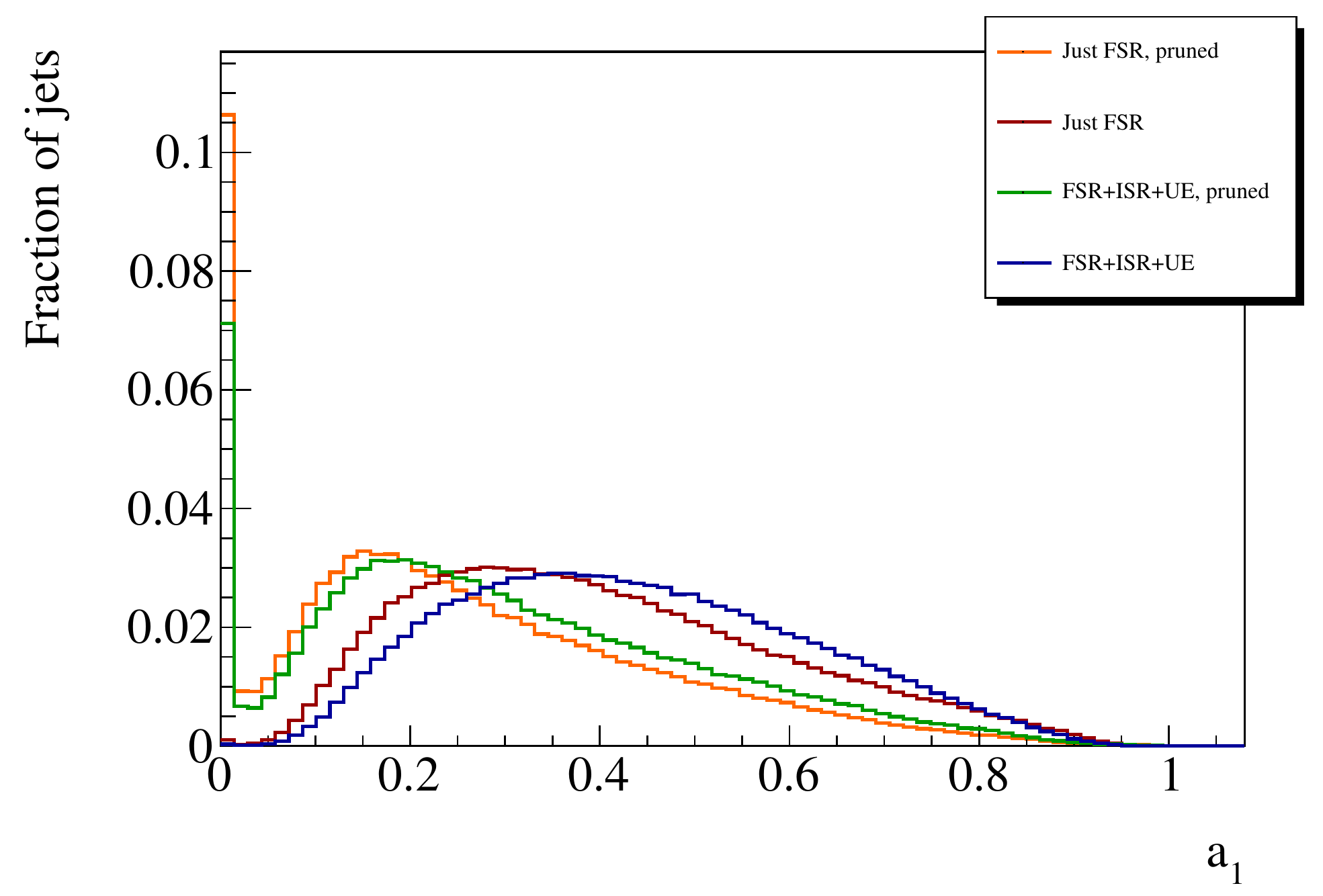}
\label{fig:mQCD_pCompare:a1KT}}

\caption{Distributions in $z$, $\Delta R_{12}$, and $a_1$ for pruned and unpruned jets in matched $pp \to \text{jets}$ events.}
\label{fig:mQCD_pCompare}
\end{figure}

In Fig.~\ref{fig:mQCD_pCompare} we show the same plots for the matched $pp \to \text{jets}$ samples.  As in the $\ee$ events, we can see that jets are being ``pruned back'' to have small $\Delta R_{12}$ and $a_1$, with a spike at $\Delta R_{12} = 0$ representing jets with only one constituent left.  As in the $t\bar t$ sample, the FSR and FSR+ISR+UE distributions are more similar after pruning than before.

\begin{figure}[htbp] \begin{center}
\subfloat[CA] {\includegraphics[width = .48\columnwidth] {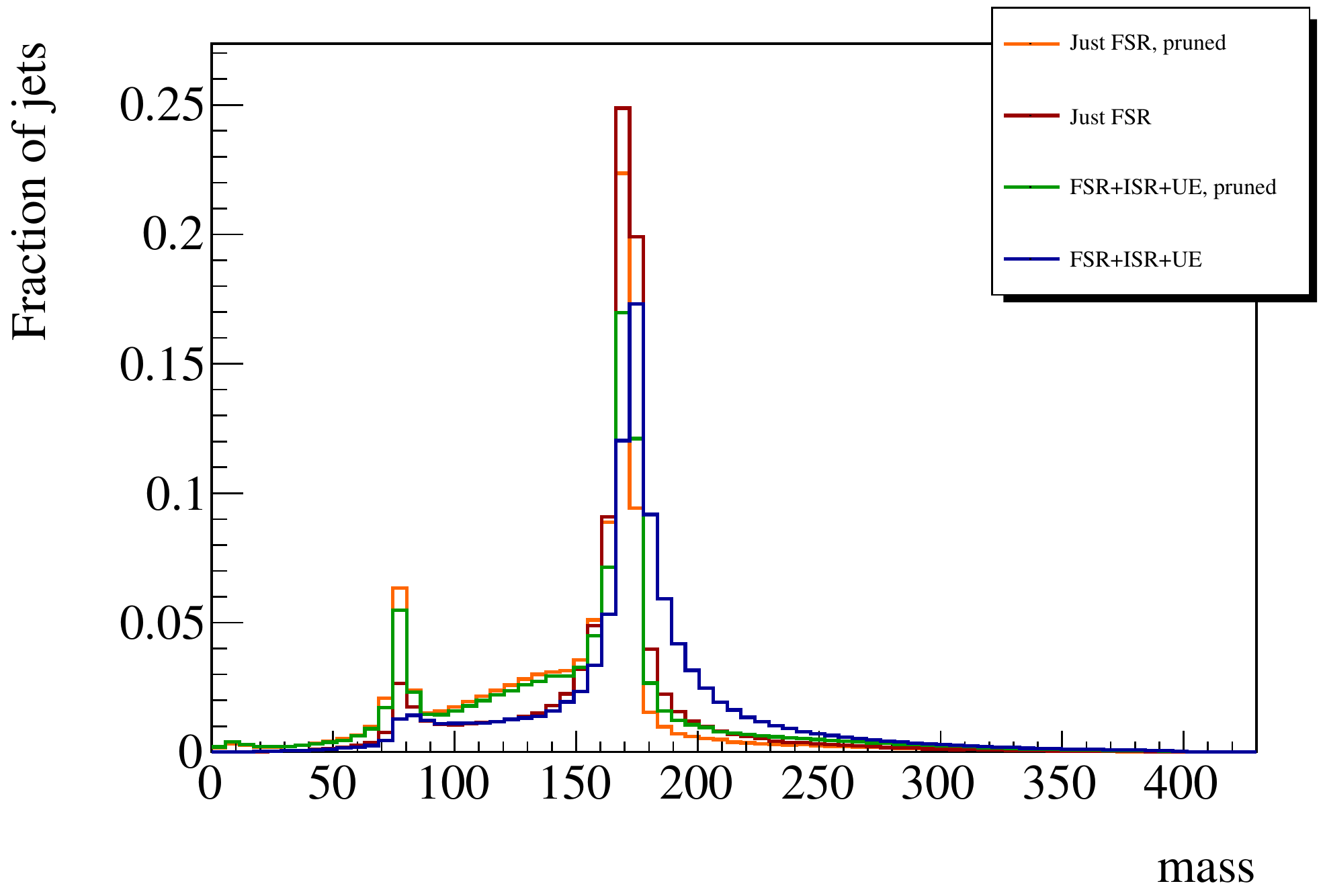} \label{fig:ttbar_pCompare_mass:CA}}
\subfloat[CA (zoomed in)] {\includegraphics[width = .48\columnwidth]{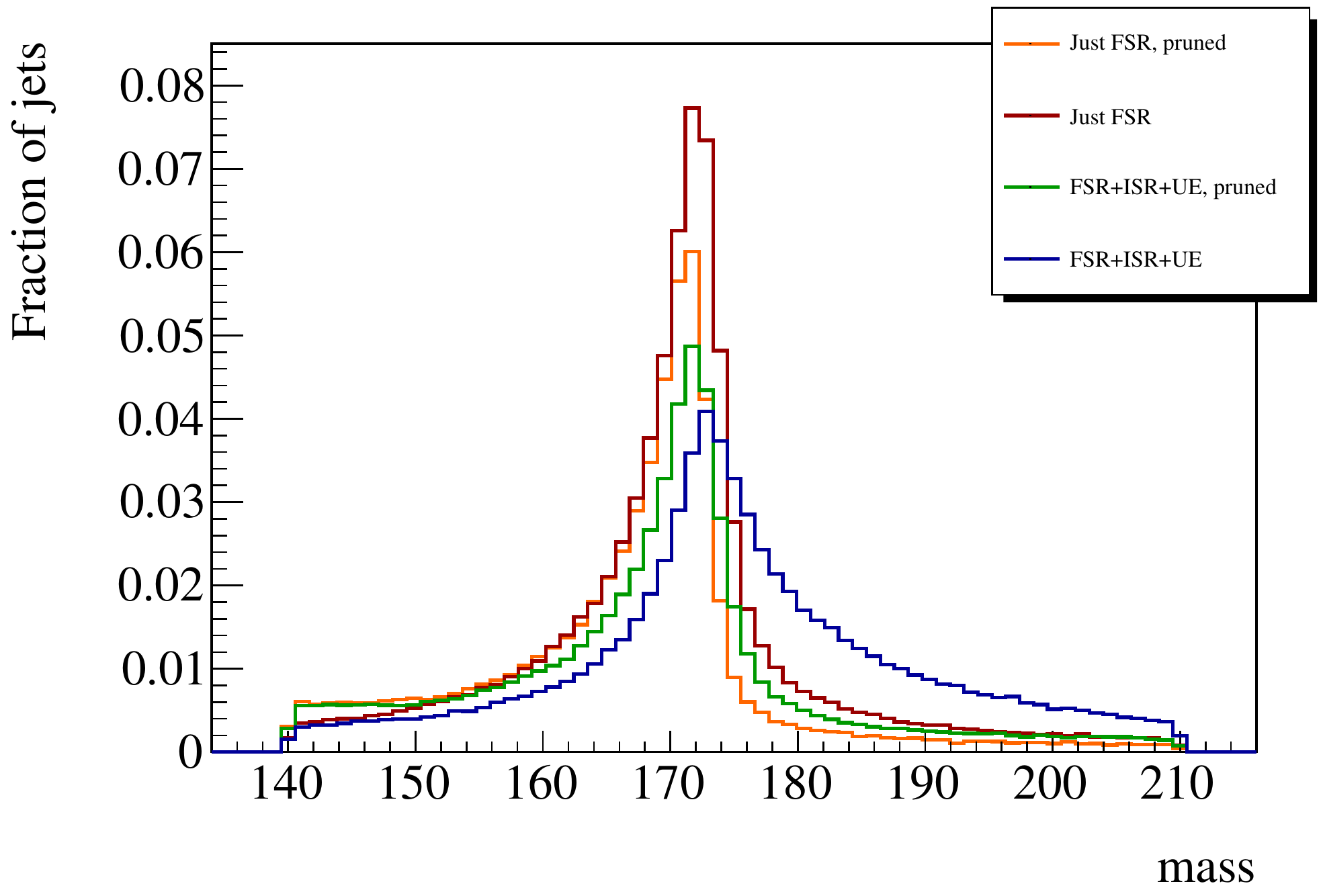}
\label{fig:ttbar_pCompare_mass:CAnarrow}}

\subfloat[$\kt$]{\includegraphics[width = .48\columnwidth]{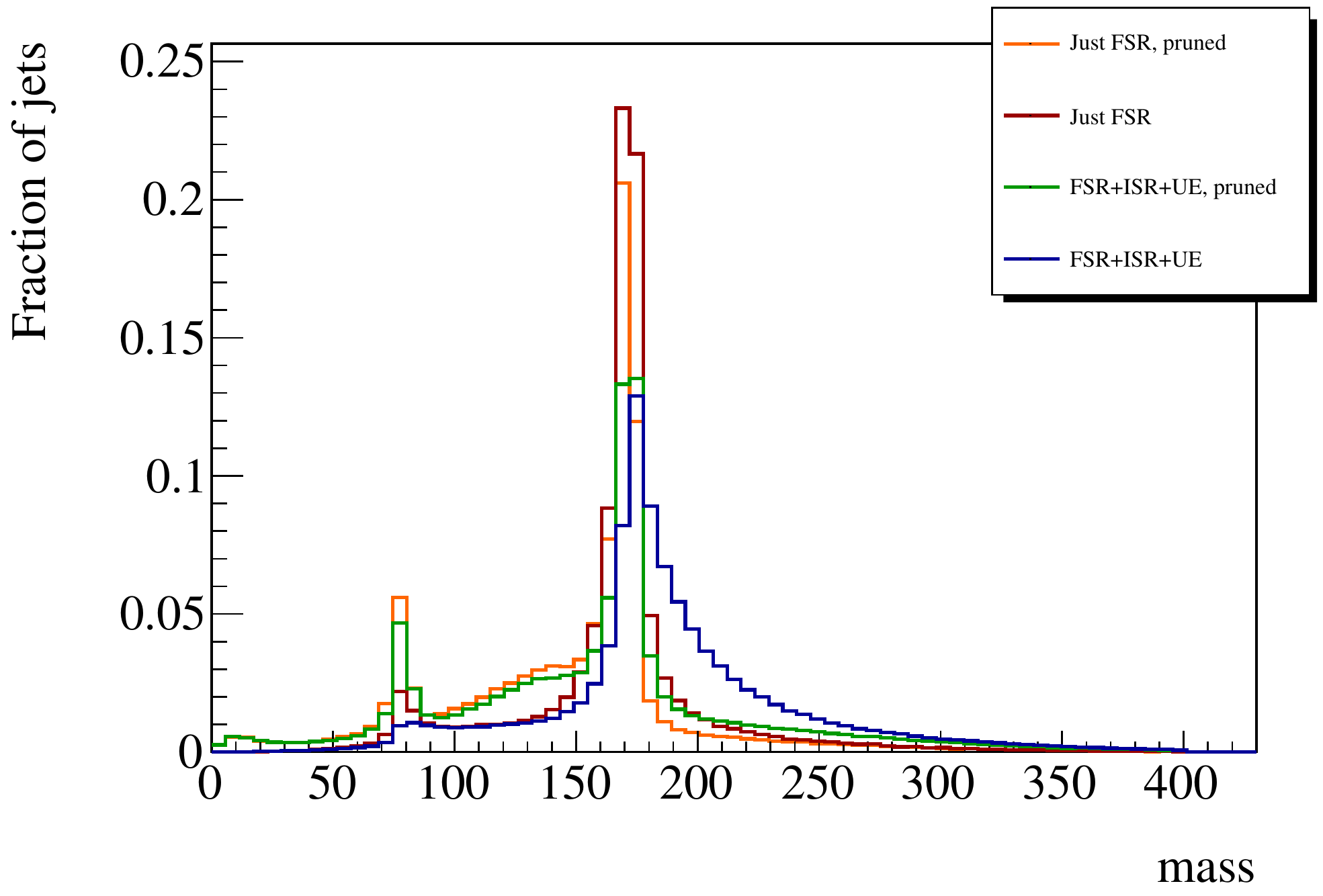} \label{fig:ttbar_pCompare_mass:KT}}
\subfloat[$\kt$ (zoomed in)] {\includegraphics[width = .48\columnwidth] {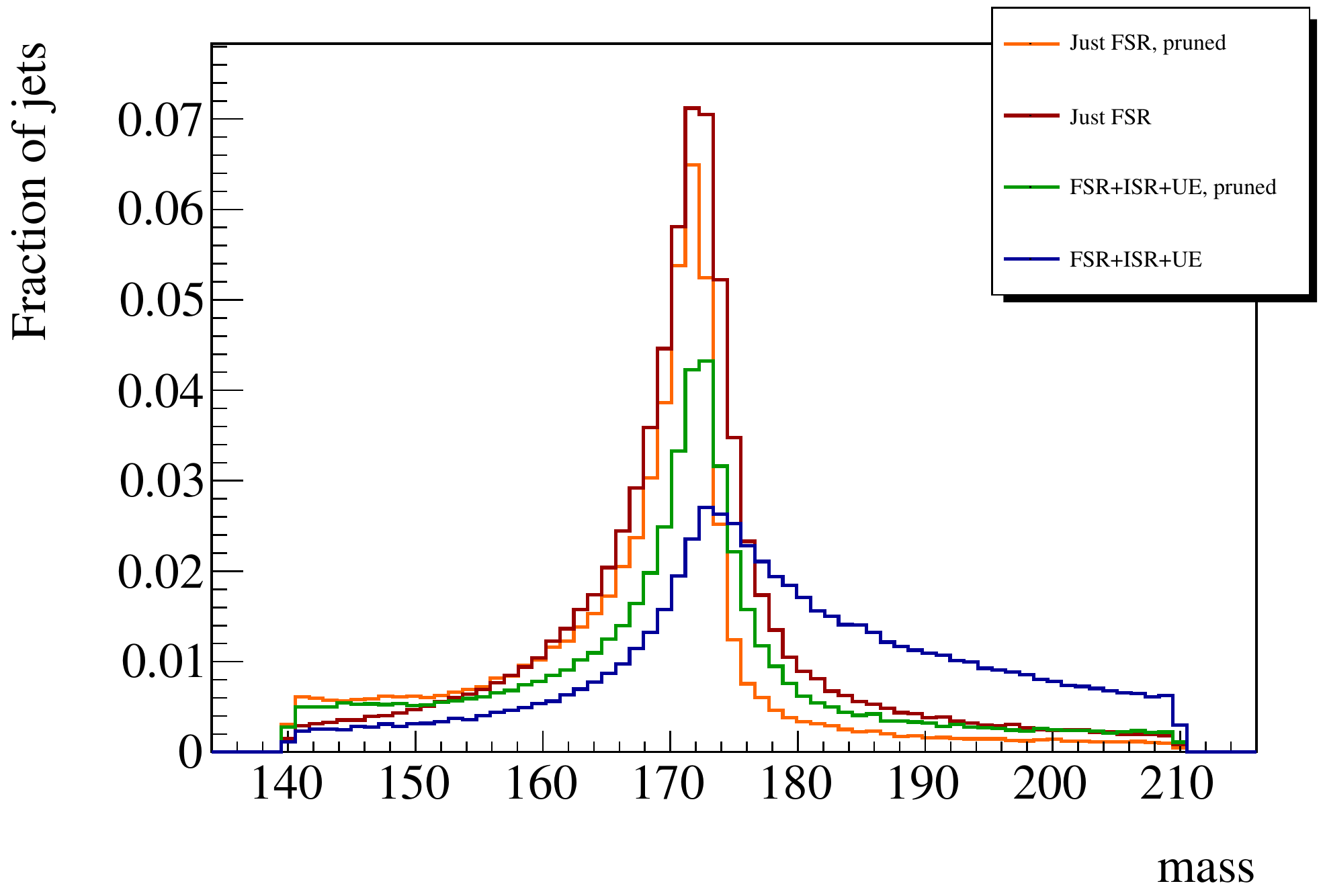} \label{fig:ttbar_pCompare_mass:KTnarrow}}
\end{center}
\caption[Distribution in $m_J$ for pruned and unpruned jets in $pp \to t\bar t$ events]{Distribution in $m_J$ for pruned and unpruned jets in $pp \to t\bar t$ events, using the CA and $\kt$ algorithms.  Jets have $p_T > 500$ GeV and D = 1.0.}
\label{fig:ttbar_pCompare_mass}
\end{figure}

\begin{figure}[htbp] \begin{center}
\subfloat[CA] {\includegraphics[width = .48\columnwidth] {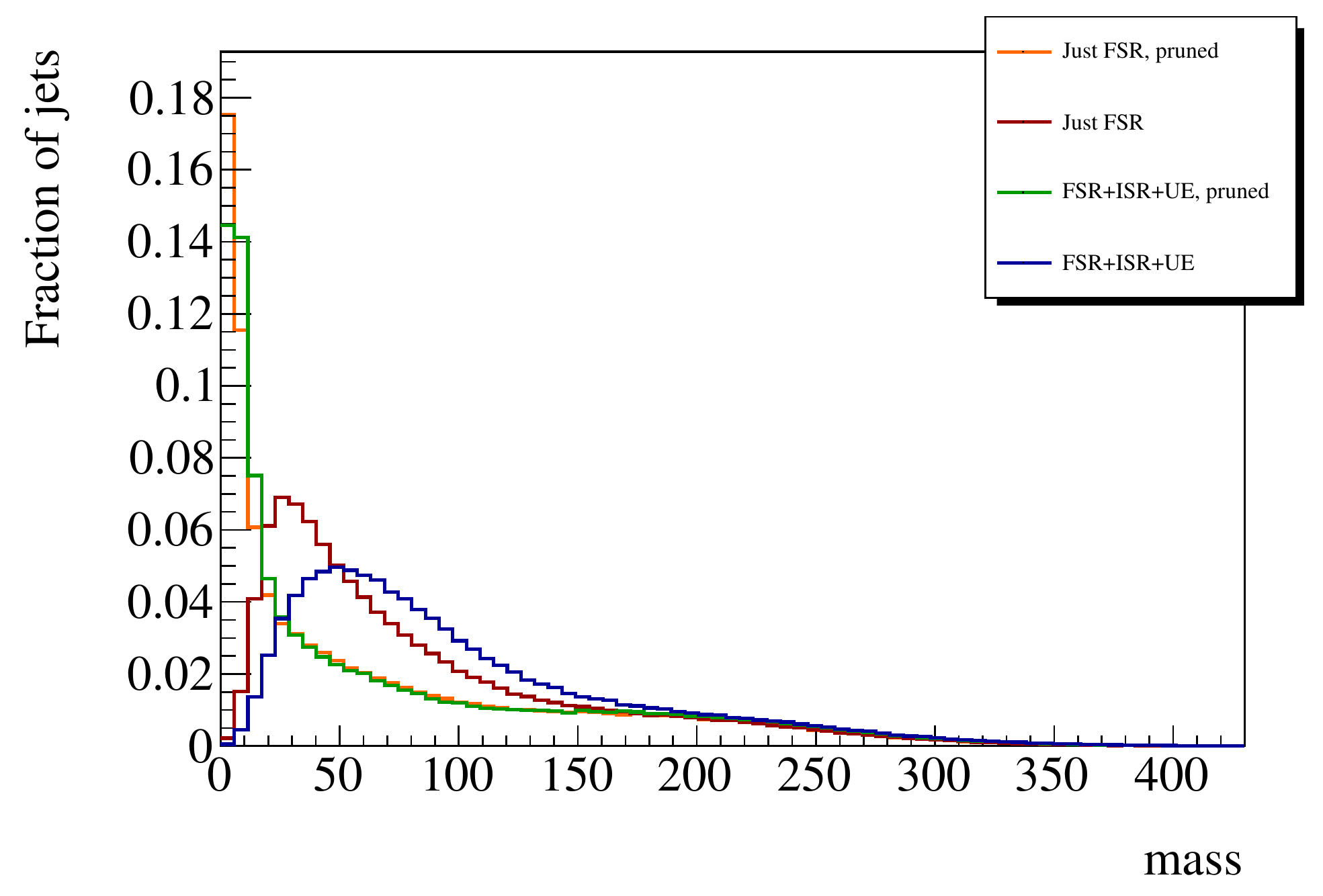} \label{fig:mQCD_pCompare_mass:CA}}
\subfloat[CA (zoomed in)] {\includegraphics[width = .48\columnwidth]{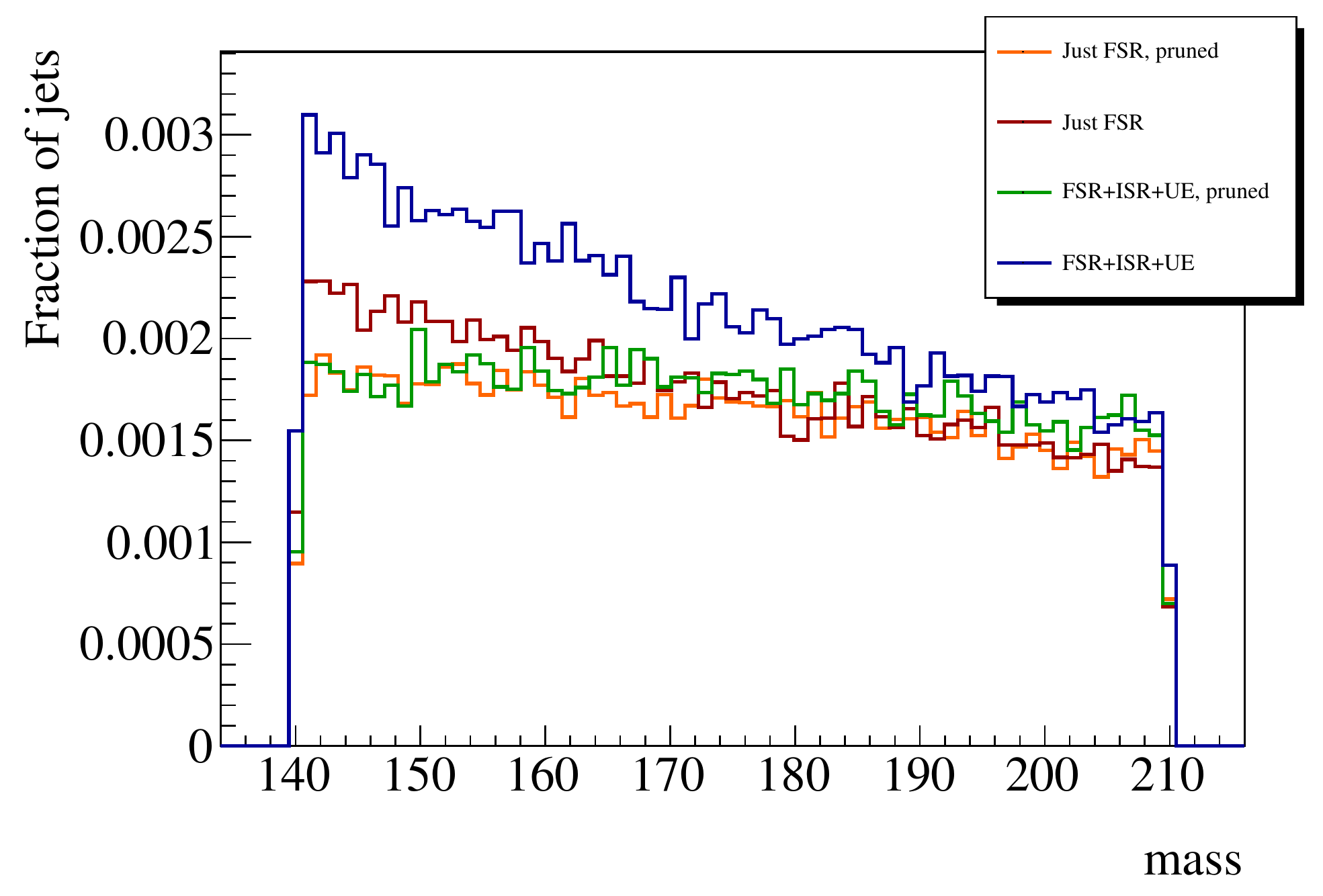}
\label{fig:mQCD_pCompare_mass:CAnarrow}}

\subfloat[$\kt$]{\includegraphics[width = .48\columnwidth]{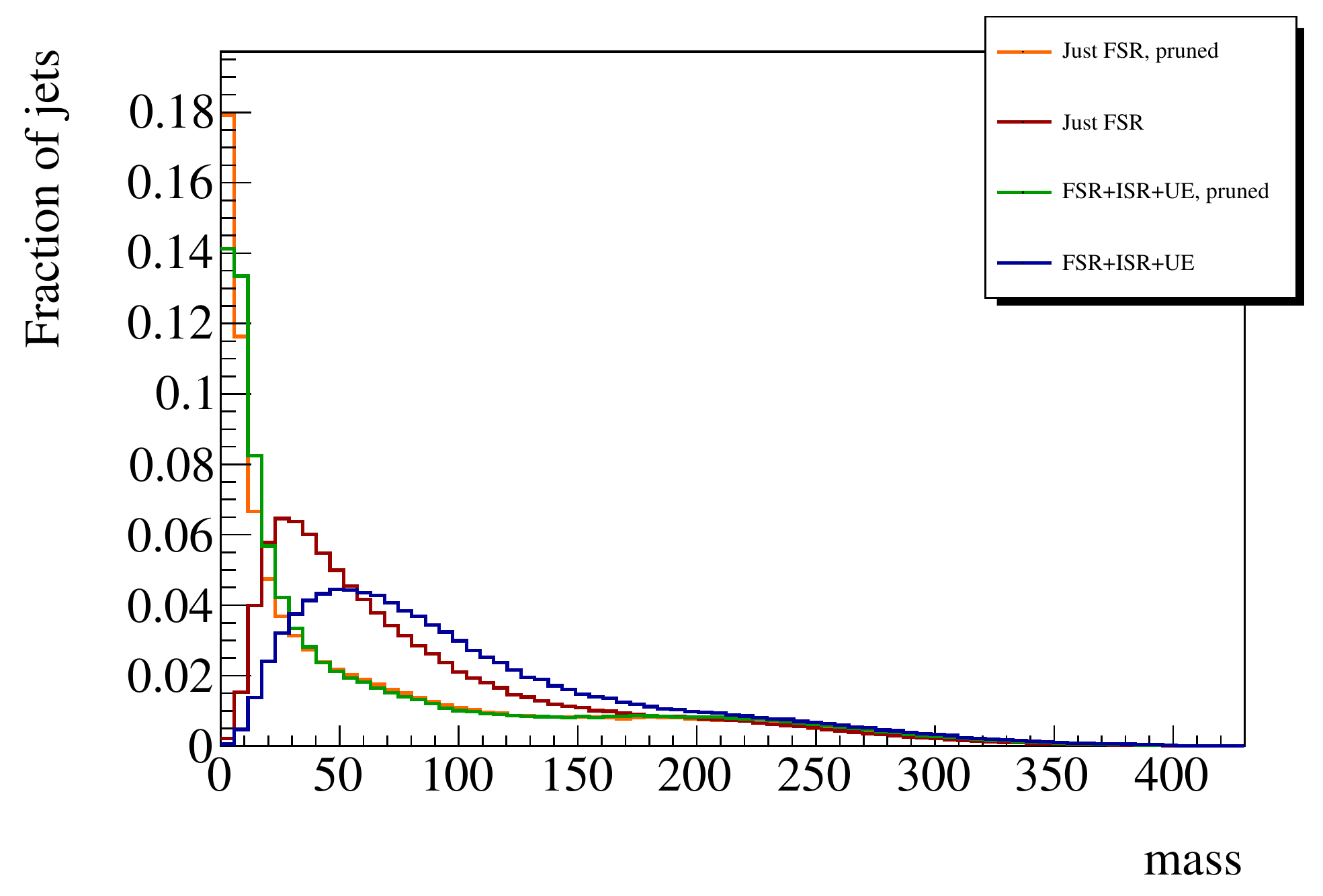} \label{fig:mQCD_pCompare_mass:KT}}
\subfloat[$\kt$ (zoomed in)] {\includegraphics[width = .48\columnwidth] {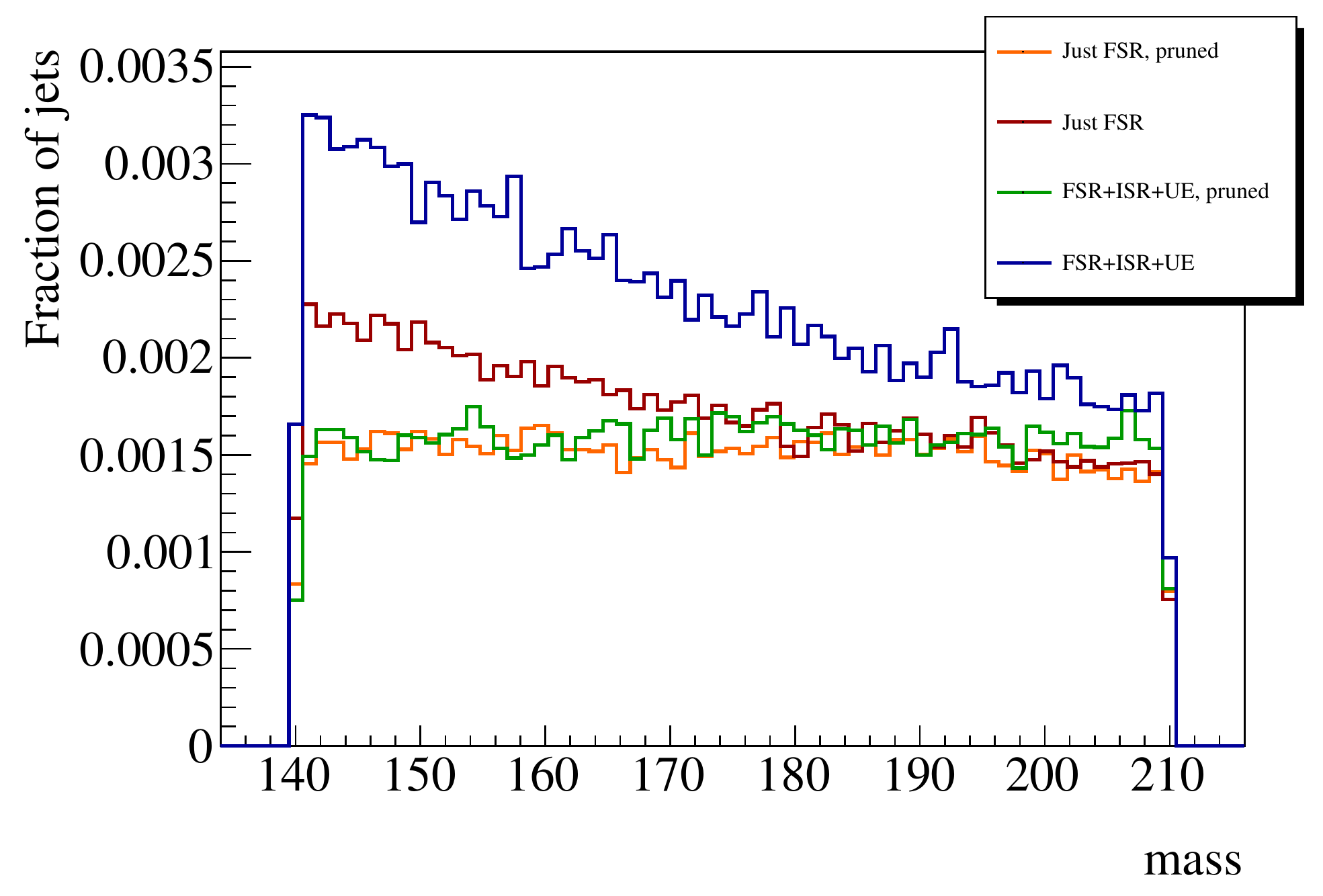} \label{fig:mQCD_pCompare_mass:KTnarrow}}
\end{center}
\caption[Distribution in $m_J$ for pruned and unpruned jets in matched $pp \to \text{jets}$ events]{Distribution in $m_J$ for pruned and unpruned jets in matched $pp \to \text{jets}$ events, using the CA and $\kt$ algorithms.  Jets have $p_T > 500$ GeV and D = 1.0.}
\label{fig:mQCD_pCompare_mass}
\end{figure}

We arrive at last at the key metric for pruning: jet masses in $pp$ events.  In Fig.~\ref{fig:ttbar_pCompare_mass} we plot jet masses before and after pruning for the $t\bar t$ samples; in Fig.~\ref{fig:mQCD_pCompare_mass} we show the same plots for the multijet background samples.  In the signal sample we see that pruning narrows the peak near the top mass, especially for $\kt$.  The peak for pruned FSR+ISR+UE jets is not as sharp as for FSR jets, but pruning provides a clear improvement.  Recall from Sec.~\ref{sec:sub:eventeffects} that the separation in FSR/ISR/UE is to some extent artificial, and we should not expect any method on fully simulated events to reproduce the simplicity of the FSR sample.

In the background mass plots, we can see that unlike in the $\ee$ case, here pruning lowers the number of jets in the top mass window.  The distinction between $\ee$ and $pp$ jets is related to the contrast between FSR and FSR+ISR+UE jets.  As for $\ee$ events, pruning has little effect on high-mass jets in the FSR sample.  Here large jet masses are presumably coming from hard, large-angle radiation that pruning cannot remove.  Recall that the $pp$ sample is a matched sample with 2, 3, or 4 final state partons.  By contrast, in the FSR+ISR+UE sample, moderately heavy jets have their masses increased by the inclusion of additional radiation from the rest of the event, pushing them into the top mass window.  Pruning can remove this radiation, moving these jets back out of the top window and reducing the background to the top quark signal.

\subsection[\textit{Parton-hadron comparison}]{Parton-hadron comparison}

\begin{figure}[htbp]
\begin{center}
\subfloat[$m_J$ cut, $z$]{\includegraphics[width = 0.45\columnwidth]{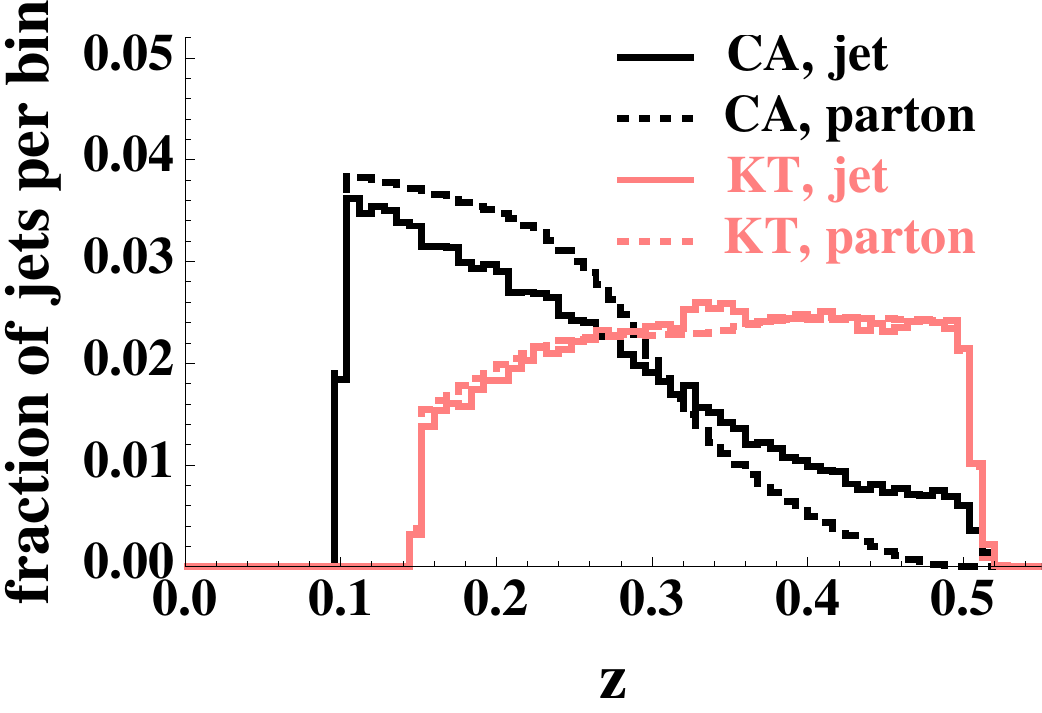}}
\subfloat[$m_J$ cut, $\Delta R_{12}$]{\includegraphics[width = 0.45\columnwidth]{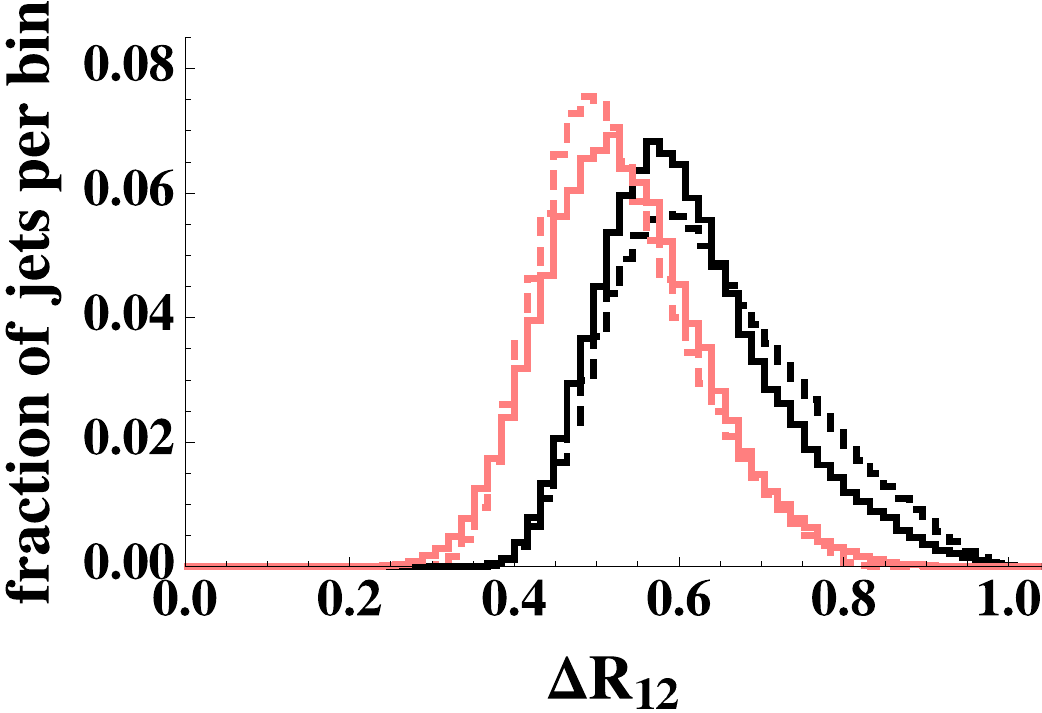}}

\subfloat[$m_J$ and $m_{\text{Sub}J}$ cuts, $z$] {\includegraphics[width = 0.45\columnwidth] {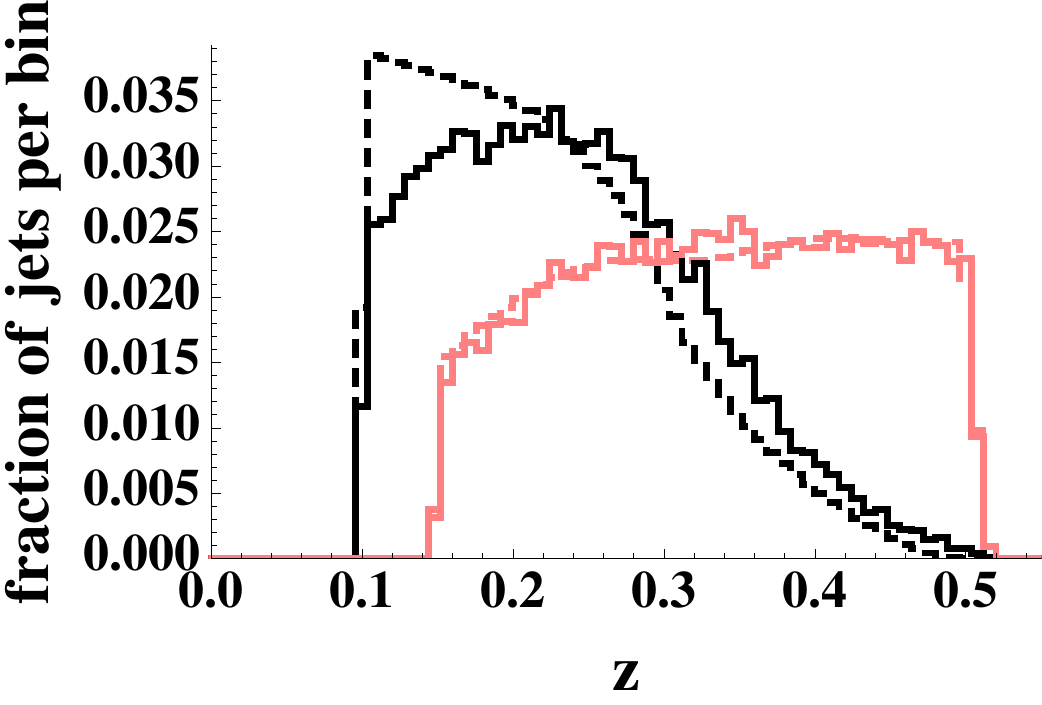}}
\subfloat[$m_J$ and $m_{\text{Sub}J}$ cuts, $\Delta R_{12}$] {\includegraphics[width =
0.45\columnwidth] {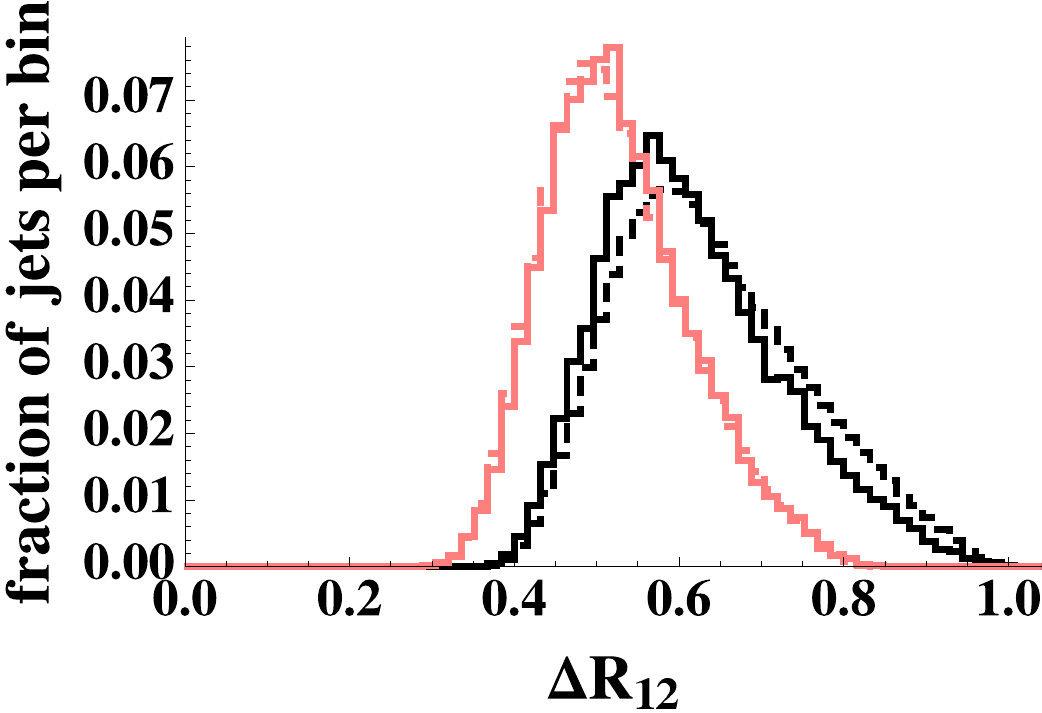}}
\end{center}

\caption[Distributions in $z$ and $\Delta R_{12}$ comparing top quark decays at the parton-level and from Monte Carlo events, after pruning]{Distributions in $z$ and $\Delta R_{12}$ comparing top quark decays at the parton-level and from Monte Carlo events, after implementing pruning.  This figure uses the same samples and cuts as Fig.~\ref{fig:topPartonVsReconZDR}.}
\label{fig:topPartonVsReconZDRpruned}
\end{figure}

Finally, it is instructive to revisit the ``parton-hadron'' comparison from Sec.~\ref{sec:sub:summary}.\footnote{This subsection is taken, with small modifications, from Sec.~VIA of \cite{Pruning2}.}  In Fig.~\ref{fig:topPartonVsReconZDRpruned}, we reproduce Fig.~\ref{fig:topPartonVsReconZDR}, using pruning at both the hadron and parton level.  The parton-level pruning is implemented in the same way as defined above, treating the three partons of the reconstructed top quark as the jet.

By comparing Figs.~\ref{fig:topPartonVsReconZDR} and \ref{fig:topPartonVsReconZDRpruned}, we again can see that pruning has removed much of the systematic effects in the CA algorithm; when only a jet mass cut is made, the distribution in $z$ and $\Delta R_{12}$ for pruned jets match the parton-level distribution much better than unpruned jets.  When both mass and subjet mass cuts are made, pruning shows a slightly poorer agreement to the parton-level kinematics than the unpruned case.  Note however that for pruned jets, the efficiency of the subjet mass cut is considerably greater since we more often identify one of the daughter subjets as a $W$ (see the discussion of Fig.~\ref{fig:VaryZcut} in Sec.~\ref{sec:prune:results:params}).

We move on to examine pruning through a set of studies using Monte Carlo simulated events.  We will investigate the parameter dependence of pruning, motivating the parameters used above.  We will extensively study both top and $W$ reconstruction with pruning, and quantify the improvements from pruning in terms of basic statistical measures.  These studies will provide evidence of the insensitivity of pruning to the value of $D$ in the jet algorithm.


\section[\textit{Study overview}]{Study overview}
 \label{sec:prune:study}

The parameter space for questions about pruning procedures is very large.\footnote{This section is taken from Sec.~VII of \cite{Pruning2}.}  We will not be able to answer all possible questions, but we will attempt to answer the most important.  We use Monte Carlo samples to study $W$ reconstruction and the rejection of $W$ + jets backgrounds, as well as top quark reconstruction and the rejection of QCD multijet backgrounds.  To test the usefulness of pruning across a range of jet $m/p_T$, and hence the heavy particle boost, we study both signals in four $p_T$ bins.  We will also be able to compare a signal with a single mass scale (the $W$) to one with two (the top).  The details of the Monte Carlo samples and their generation are described in Appendix~\ref{app:details}.

In the following sections, we define a particular method to identify the heavy particles using jet substructure, and examine pruning in this context.  We are more concerned with the \emph{improvements} provided by pruning than its absolute performance.  Therefore, we compare pruning to an analysis procedure where the jets are left unpruned.  This comparison removes dependence on quantities that have large uncertainties, such as signal and background cross sections, or are not specified, such as the integrated luminosity.  Instead, the performance of pruning is quantified in terms of how much \emph{better} pruning resolves the physically relevant substructure of the jet and separates signal and background processes versus using the substructure from unpruned jets.

Additionally, we test the performance of pruning as parameters of the jet algorithm and the pruning procedure are varied, including $D$.  We expect the $D$ dependence to be closely correlated with the jet $p_T$, as it is a direct measure of the boost of the heavy particle.  We aim to draw some basic conclusions about how pruning should be applied in a search.

\subsection[\textit{Measures used to quantify pruning}]{Measures used to quantify pruning}
\label{sec:prune:study:metrics}

Mass variables are by far the strongest discriminator between QCD jets and jets reconstructing heavy particle decays.  QCD jets have a smooth mass distribution set by the jet $p_T$ (see Sec.~\ref{sec:sub:parton:qcd}), while a decaying particle can have multiple intrinsic mass scales.  We define simple criteria to identify a jet as coming from a top quark: if the jet mass is in the top mass window and one of the two subjets has a mass in the $W$ mass window, then we tag the jet as a \emph{top jet}.  The top and $W$ mass windows are defined by fitting the relevant mass peaks of the signal sample, which we describe in detail below.  The $W$ study proceeds analogously with only a jet mass cut.  In a real search for a particle of unknown mass, one obviously cannot fit a ``signal sample''.  However, we employ this method to demonstrate two effects of pruning: sharpening the signal mass peak and reducing the QCD background in this region.  These two effects will determine how well pruning improves our ability to find bumps in jet mass distributions.

We use a common set of variables to measure the difference between a jet algorithm and its pruned version.  Let $N_{\text{\tiny{S}}}(A)$ be the number of jets in the signal sample identified as a reconstructed heavy particle for algorithm $A$, and $N_{\text{\tiny{B}}}(A)$ the analogous number of jets in the background sample.  Use $pA$ to denote the pruning procedure run on jets found with algorithm $A$.  Then the variables we use are:
\[
\begin{split}
\epsilon &= \frac{N_{\text{\tiny{S}}}(pA)}{N_{\text{\tiny{S}}}(A)} , \\
R &= \frac{N_{\text{\tiny{S}}}(pA) / N_{\text{\tiny{B}}}(pA)}{N_{\text{\tiny{S}}}(A) / N_{\text{\tiny{B}}}(A)}, ~\text{and}\\
S &= \frac{N_{\text{\tiny{S}}}(pA) / \sqrt{N_{\text{\tiny{B}}}(pA)}}{N_{\text{\tiny{S}}}(A) / \sqrt{N_{\text{\tiny{B}}}(A)}} .
\end{split}
\]
$\epsilon$ is the relative efficiency of pruning in identifying heavy particles in the signal sample, while $R$ and $S$ are the relative signal-to-background and signal-to-noise ratios for the pruned and unpruned algorithms.  We also evaluate the relative mass window widths, which we label $w_\text{rel}$.  For the $W$ study, this is the ratio of the $W$ mass window width for pruning relative to not pruning; for the top study it is the ratio in the top mass window width.  Note that in the top study, a $W$ subjet mass cut is also used.  A value of $w_\text{rel} < 1$ means pruning has improved the mass resolution of the jets.  These ratios are independent of the integrated luminosity and the total cross sections, and are representative of the improvements that pruning would provide in an analysis.

\begin{figure}[htbp]
\begin{center}
\includegraphics[width=.5\columnwidth]{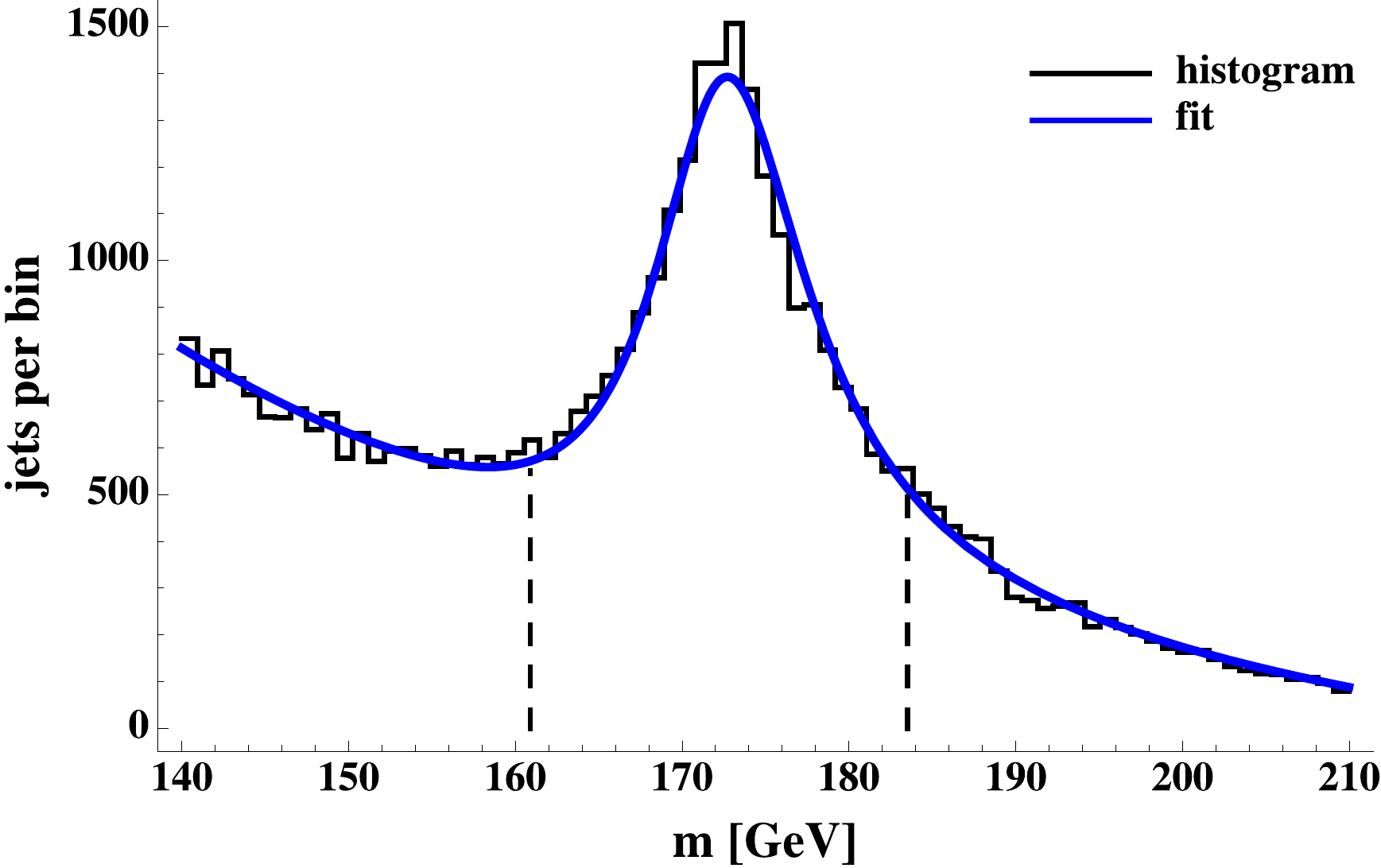}
\end{center}
\caption[A sample fit showing the jet mass distribution and sample fit for CA jets from $t\bar{t}$ events]{A sample fit showing the jet mass distribution (black histogram) and sample fit (blue curve) for CA jets from $t\bar{t}$ events.}
\label{fig:SampleBWfit}
\end{figure}

To determine the mass window for a particular signal sample, we fit the mass peak to determine the window width.  In these studies, a skewed Breit-Wigner is sufficient to fit the peak, with a power law continuum background.  These functions used to fit mass peaks are:
\[
\begin{split}
\text{peak: } f(m) &= \frac{M^2\Gamma^2}{(m^2-M^2)^2 + M^2\Gamma^2}\left(a + b(m-M)\right) ; \\
\text{continuum: } g(m) &= \frac{c}{m} + \frac{d}{m^2} .
\end{split}
\]
$M$ is the location of the mass peak; $\Gamma$ is the width of the peak.  A sample fit it shown in Fig.~\ref{fig:SampleBWfit}.
The mass window $[M-\Gamma,M+\Gamma]$ is found to be nearly optimal, given this functional form, in measures similar to $\epsilon$, $R$, and $S$: the area in the window ($\sim\epsilon$), the ratio of area to the window width ($\sim R$), and the ratio of area to the square root of the width ($\sim S$).


\section[\textit{Results}]{Results}
 \label{sec:prune:results}
 
In this section we present results comparing analyses with pruned jets to unpruned jets.\footnote{This section is taken from Sec.~VIII of \cite{Pruning2}.}  We demonstrate two main points: first, pruning is useful and broadly applicable, and second, its parameters do not need fine tuning for it to provide significant improvement.

The natural starting point is to investigate the parameters particular to the pruning procedure, $D_{\cut}$ and $z_{\cut}$.  The most important question is whether these need to be tuned to the signal.  To answer this, in Sec.~\ref{sec:prune:results:params} we study the performance of pruning as we vary its parameters for two different signals across the full $p_T$ range for the samples.  We find that optimal choices of $z_\cut$ and $D_\cut$ vary slowly with $m/p_T$, but that our choice of parameters is not far from optimal in all cases.

After fixing $z_\cut$ and $D_\cut$, we consider the effect of varying $D$ in the jet algorithm.  In Sec.~\ref{sec:prune:results:fixedD} we study pruning with $D$ fixed at 1.0 over all $p_T$ bins.  This type of analysis is like a search where the mass (and hence $m/p_T$) of the new heavy particle is not known.  For comparison, in Sec.~\ref{sec:prune:results:varD} we redo the analysis, but with $D$ adjusted for each bin to fit the expected angular size of the decay in that bin.  In this case, the unpruned jet algorithm performs better than with a constant $D$, as expected, but pruning still shows improvements in finding $W$'s and tops.  In all cases, pruned jets are a better way to identify heavy particles than unpruned.  In Sec.~\ref{sec:prune:results:compD} we compare the results of Secs.~\ref{sec:prune:results:fixedD} and \ref{sec:prune:results:varD}.  Significantly, if jets are pruned, we find that it does not make much difference what the initial $D$ value was, indicating that searches with large fixed $D$ do not suffer in power compared to searches with $D$ tuned to known or suspected $m/p_T$.

In Sec.~\ref{sec:prune:results:absolute} we give some absolute measures of top-finding with pruned jets for comparison to other methods.  In Sec.~\ref{sec:prune:results:algComparison} we directly compare the CA and $\kt$ algorithms, before and after pruning.  Finally, in Sec.~\ref{sec:prune:results:smearing} we consider the effect of a crude detector model where we smear the energies of all particles in the calorimeter.  We find that the performance of the pruned and unpruned algorithms are degraded, but that pruning still provides significant improvement.

\subsection[\textit{Dependence on Pruning Parameters}]{Dependence on Pruning Parameters}
\label{sec:prune:results:params}

The pruning procedure we have defined has two free parameters (in addition to those of the jet algorithms themselves).  In introducing the procedure, we argued that $z_{\cut} = 0.10$ and $D_{\cut} = m_J/p_{T_J}$ were sensible choices.  We now investigate how pruning performs when each of these parameters is varied while the other is held fixed, for both ($W$ and top) signals and across the four $p_T$ bins for each signal.

We will look at the values of the metrics $w_\text{rel}$, $\epsilon$, $R$, and $S$ defined in Sec.~\ref{sec:prune:study:metrics}.  The priority in choosing particular values for $z_{\cut}$ and $D_{\cut}$ should be in optimizing $S$, as it is the criterion for discovery.  That being said, $\epsilon$ and $R$ are still important measures as they determine the total size of the signal and remaining fraction relative to the background.  We also evaluate $w_\text{rel}$ because the mass window width drives the other three metrics.  As the relative width decreases, in general the measures $R$ and $S$ will increase because the heavy particle is better resolved and more of the background is rejected, but $\epsilon$ will tend to decrease simply because the narrower width selects fewer signal jets.  $\epsilon$ can, however, increase with decreasing mass window width if enough high-mass signal jets are being pruned into the mass window.

\begin{figure*}[htbp]
\begin{center}
\subfloat[$W$'s, CA jets]{\label{fig:VaryZcut:WCA}\includegraphics[width=0.22\textwidth]{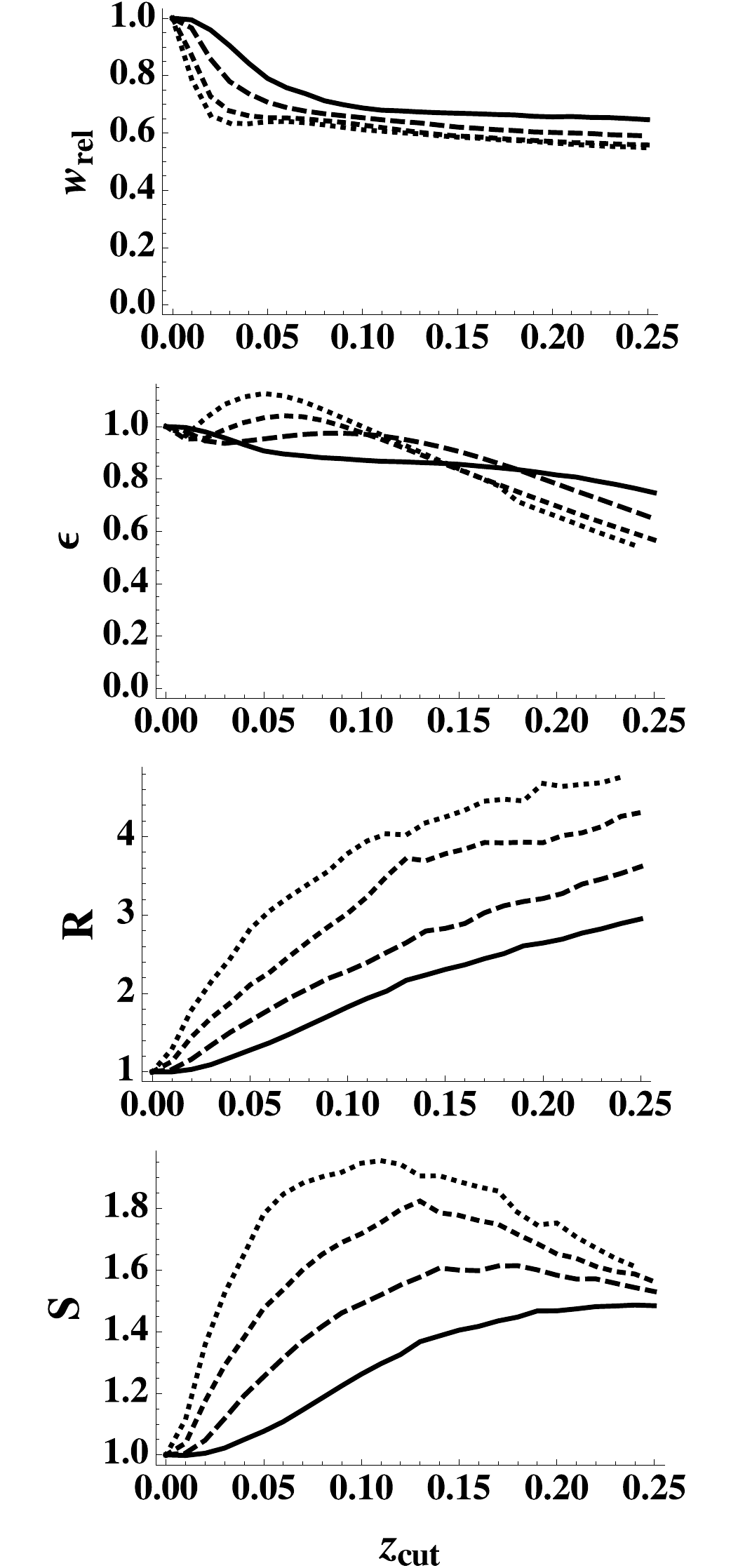}}
\subfloat[tops, CA jets]{\label{fig:VaryZcut:tCA}\includegraphics[width=0.22\textwidth]{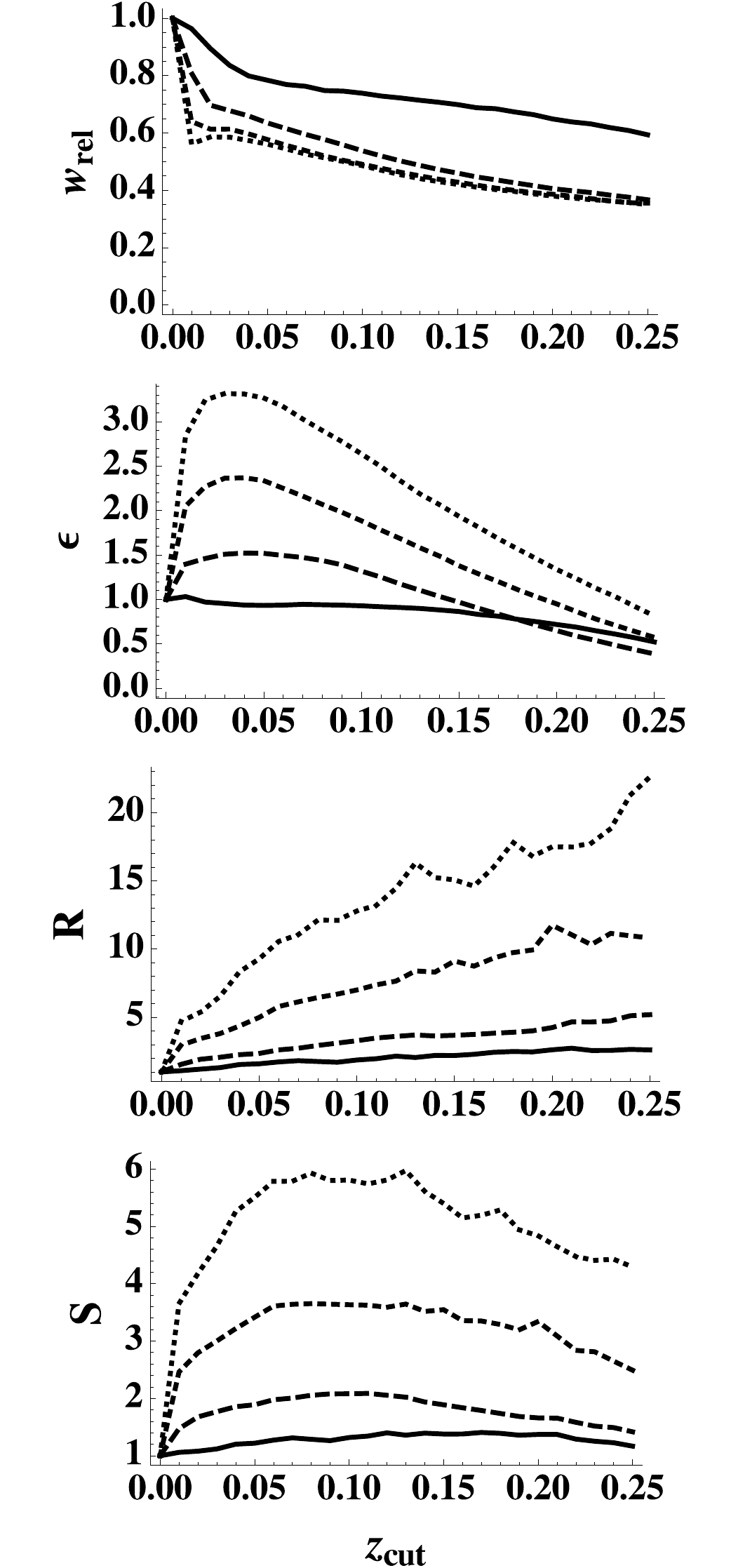}}
\subfloat[$W$'s, $\kt$ jets]{\label{fig:VaryZcut:WkT}\includegraphics[width=0.22\textwidth]{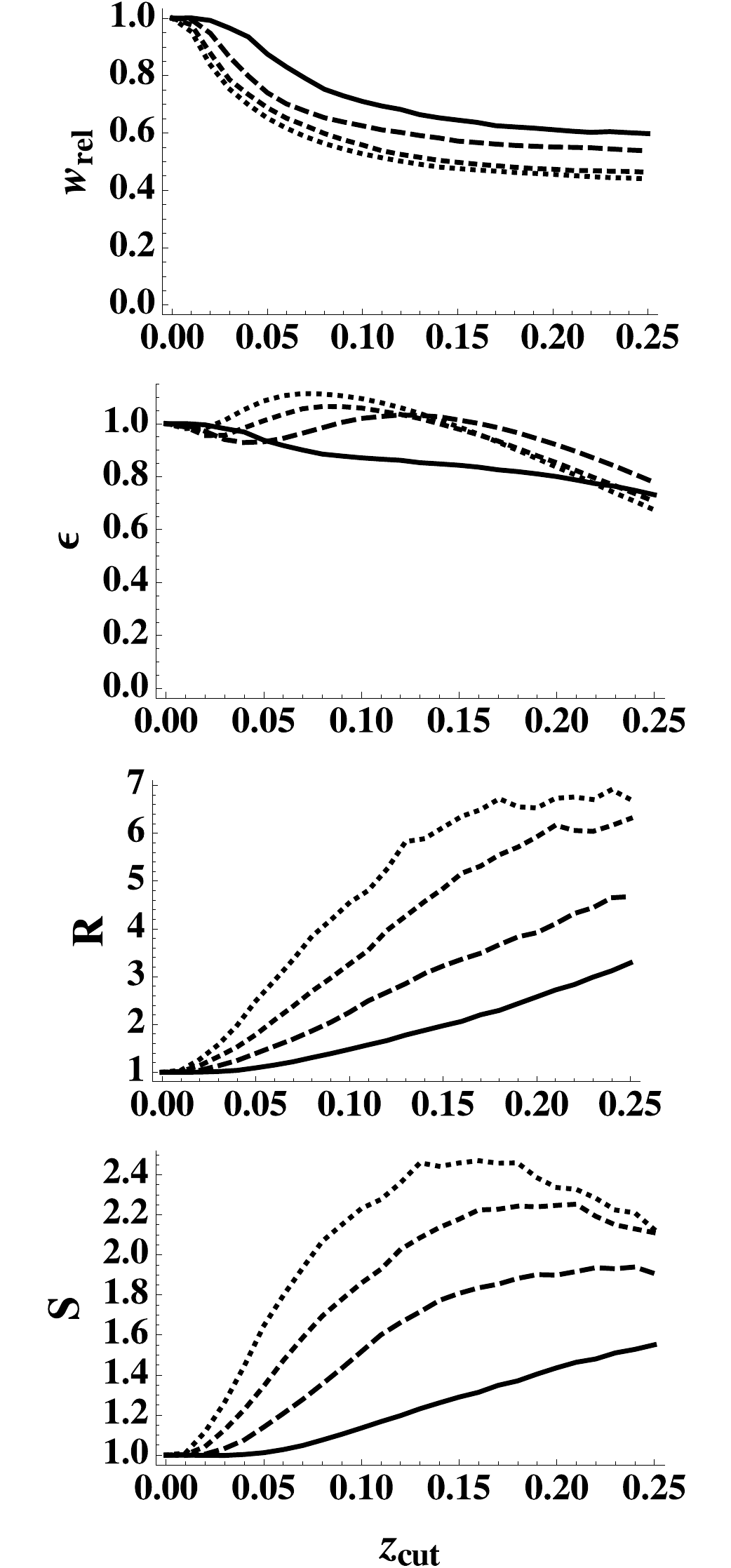}}
\subfloat[tops, $\kt$ jets]{\label{fig:VaryZcut:tkT}\includegraphics[width=0.22\textwidth]{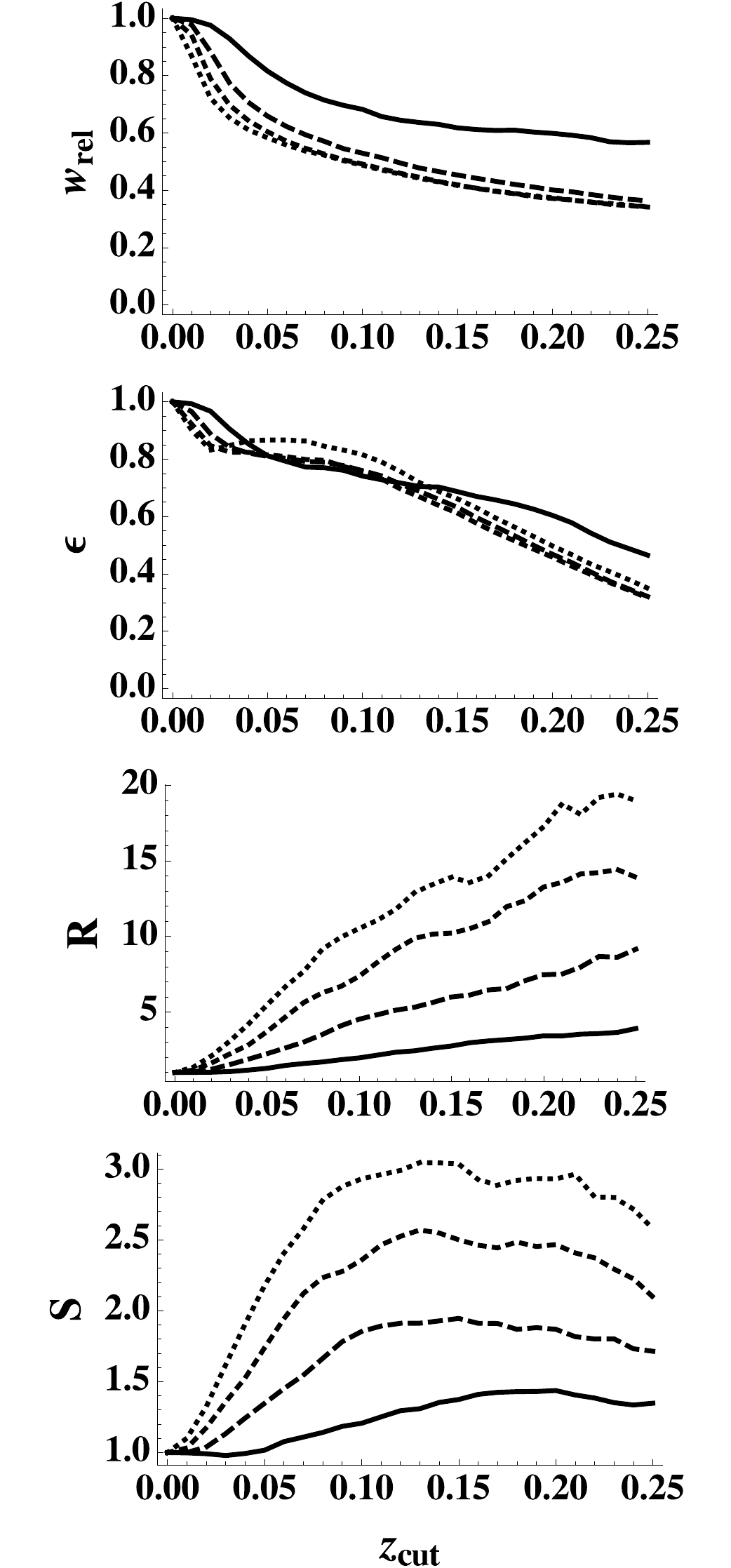}}
\includegraphics[width=0.10\textwidth]{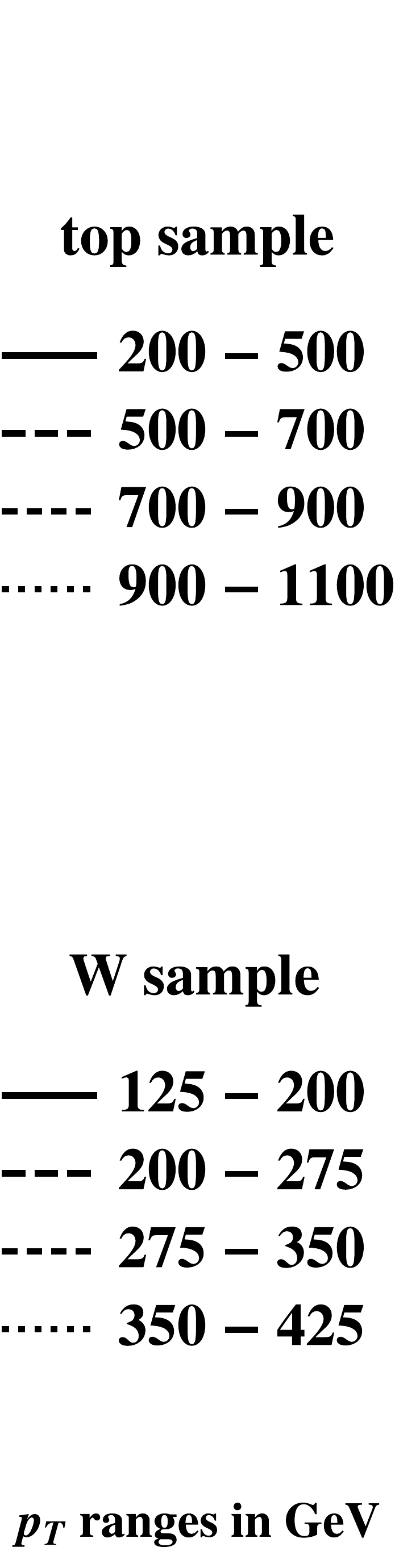}
\end{center}

\caption[Relative statistical measures $w_\text{rel}$, $\epsilon$, $R$, and $S$ vs. $z_{\cut}$ for $W$'s and tops, for four $p_T$ bins]{Relative statistical measures $w_\text{rel}$, $\epsilon$, $R$, and $S$ vs. $z_{\cut}$ for $W$'s and tops, using CA and $\kt$ jets.  Four $p_T$ bins are shown for each sample.  Statistical errors (not shown) are $\mathcal{O}(1\%)$ for $w_\text{rel}$ and $\epsilon$, and $\mathcal{O}(10\%)$ for $R$ and $S$.}
\label{fig:VaryZcut}
\end{figure*}

In Fig.~\ref{fig:VaryZcut}, we show all four metrics for top and $W$ jets, for both CA and $\kt$ jets.  $D_\cut$ is set to $m_J/p_{T_J}$ throughout, and $z_\cut$ is varied in [0, 0.25].  $z_\cut = 0$ represents no pruning and we can see that all metrics are 1 here.  With increasing pruning, the mass window width initially decreases rapidly, then levels out.  In all but the smallest $p_T$ bin, the relative signal efficiency $\epsilon$ increases as the width narrows, suggesting that signal jets that had ``vacuumed up'' too much UE or soft radiation are being pruned back into the mass window.  Note that for the top quark sample with the $\kt$ algorithm, $\epsilon$ merely flattens out for a range in $z_\cut$, and does not increase as it does for the other samples.  Once the window stops shrinking significantly (around $z_\cut = 0.05$), the relative signal efficiency starts decreasing; now the dominant effect is over-pruning signal jets out of the mass window.  Note, however, that even though the relative signal efficiency is \emph{decreasing}, the relative signal-to-background ratio $R$ is \emph{increasing} over the full range.  So even as signal jets are being removed from the mass window, background jets are being removed even faster.  If we look at signal-to-noise, $S$, there appears to be a broad optimal range in $z_{\cut}$ that depends somewhat on the signal, on the $p_T$ bin and on the jet algorithm.

There are two important lessons to be learned from these plots.  First, more pruning is required for $\kt$ jets than for CA to achieve similar results.  The right two columns ($\kt$) are similar to the left two (CA) except that features are shifted out in $z_\cut$.  Second, the peak in $S$ does not depend strongly on the signal or the $p_T$, in the three largest $p_T$ bins.  The dependence on $S$ in the smallest $p_T$ bin, however, is different from the others due to threshold effects of the heavy particle being reconstructed in a single jet.  In this bin, the boosts of the $W$'s or tops are small enough that many decays are just at the threshold for being reconstructed.  Decays at the reconstruction threshold typically have poor mass resolution, and cutting more aggressively on $z$ reduces these threshold effects and significantly decreases the background, leading to an increase in $S$ over the whole range in $z_{\cut}$.  For CA, our ``reasonable choice'' of $z_\cut$ of 0.10 looks close to optimal for the upper three bins, and not far off for the smallest.  For $\kt$, a larger $z_\cut$ is needed; 0.15 is close to optimal.

\begin{figure}[htbp]
\begin{center}
\subfloat[$p_T$ bin 1, 200--500 GeV]{\includegraphics[width = 0.45\columnwidth]{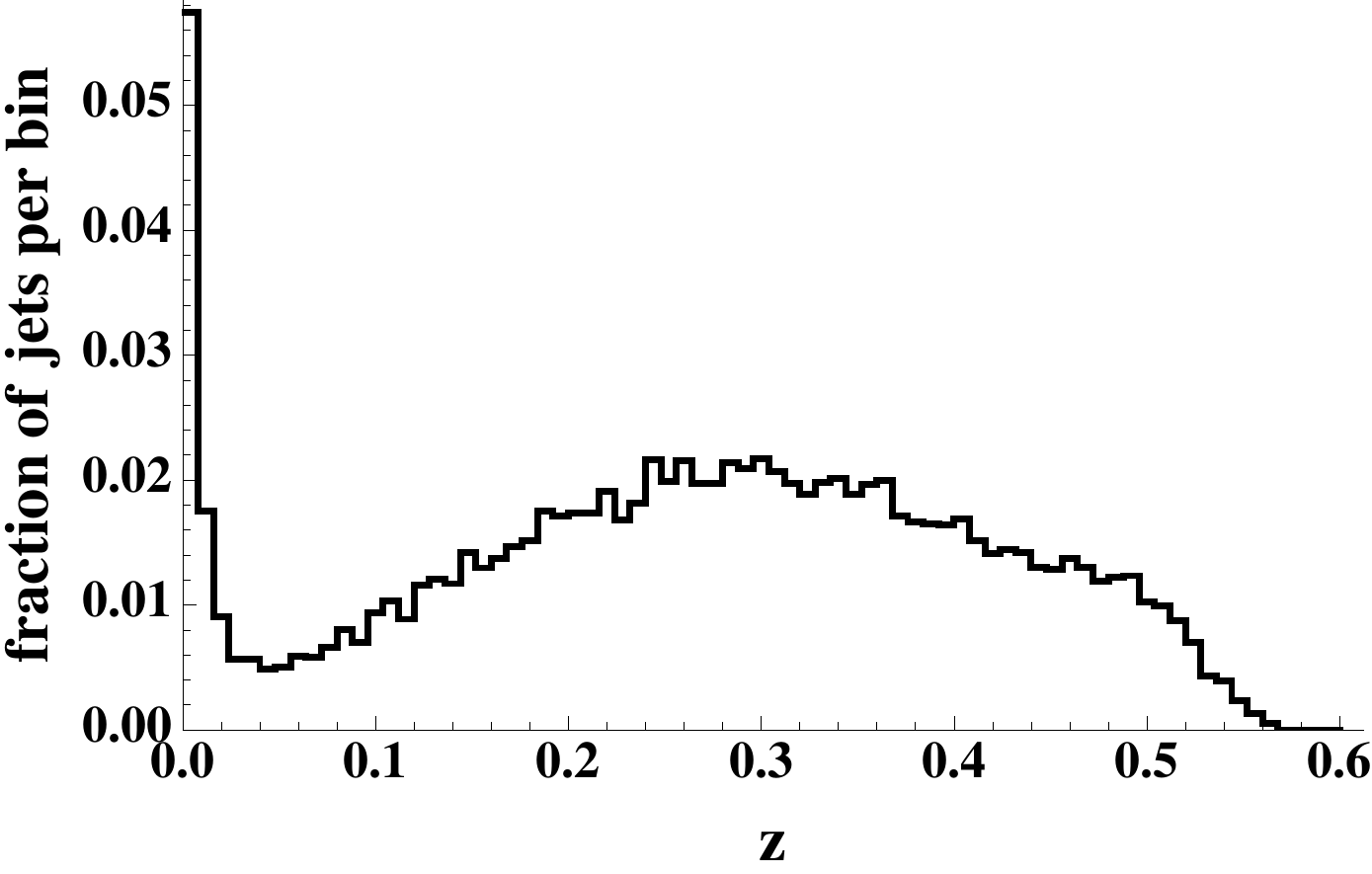}}
\subfloat[$p_T$ bin 4, 900--1100 GeV]{\includegraphics[width = 0.45\columnwidth]{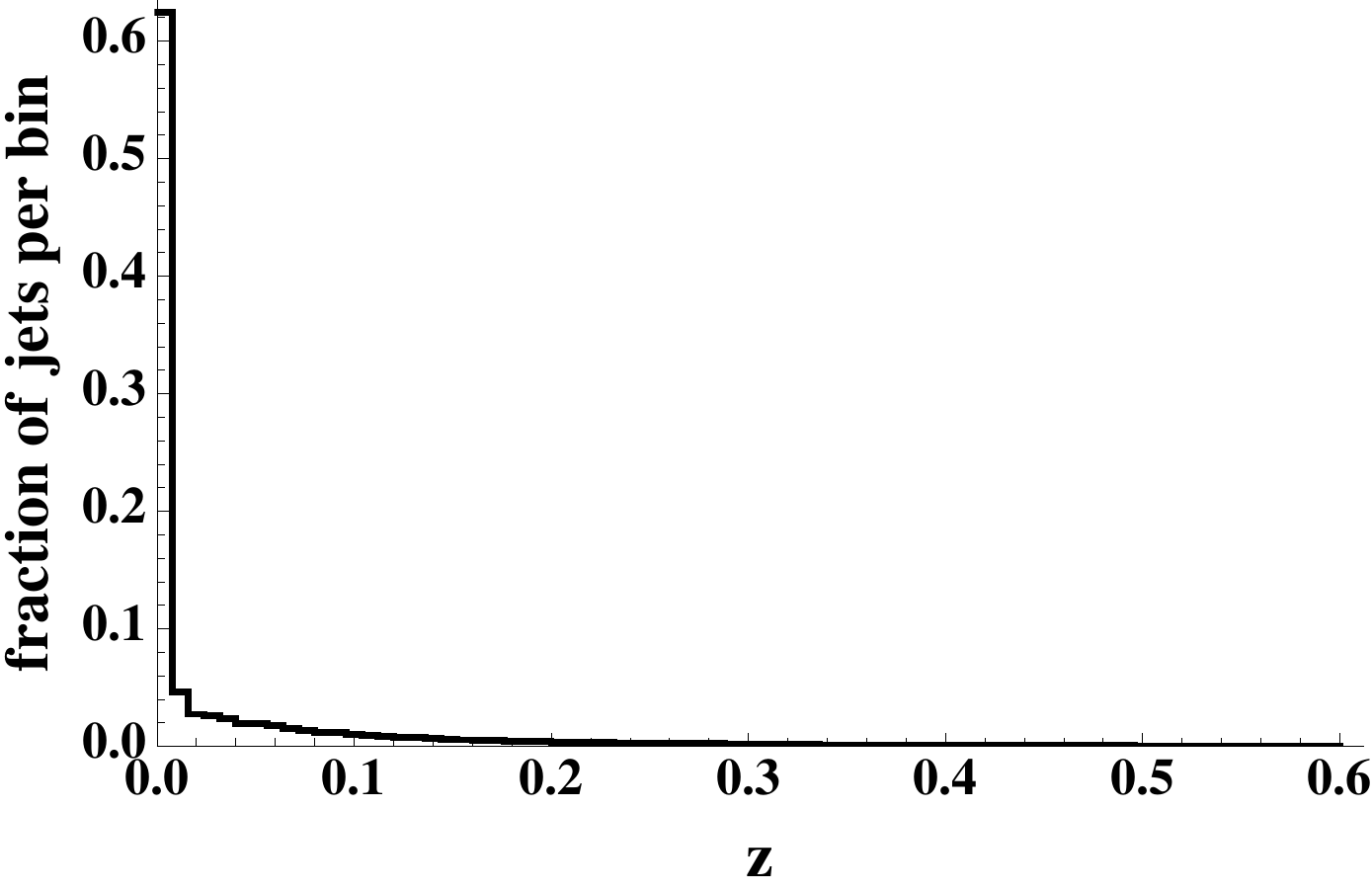}}
\end{center}

\caption[Distribution in $z$ for unpruned CA jets in the top mass window for two $p_T$ bins]{Distribution in $z$ for unpruned CA jets in the top mass window for two $p_T$ bins.  The small $p_T$ bin distribution (left) has only a small enhancement of entries at small $z$, while the large $p_T$ bin distribution (right) is dominated by small $z$.}
\label{fig:zCAmtopPT1&4}
\end{figure}

\begin{figure}[htbp]
\begin{center}
\includegraphics[width=.5\columnwidth]{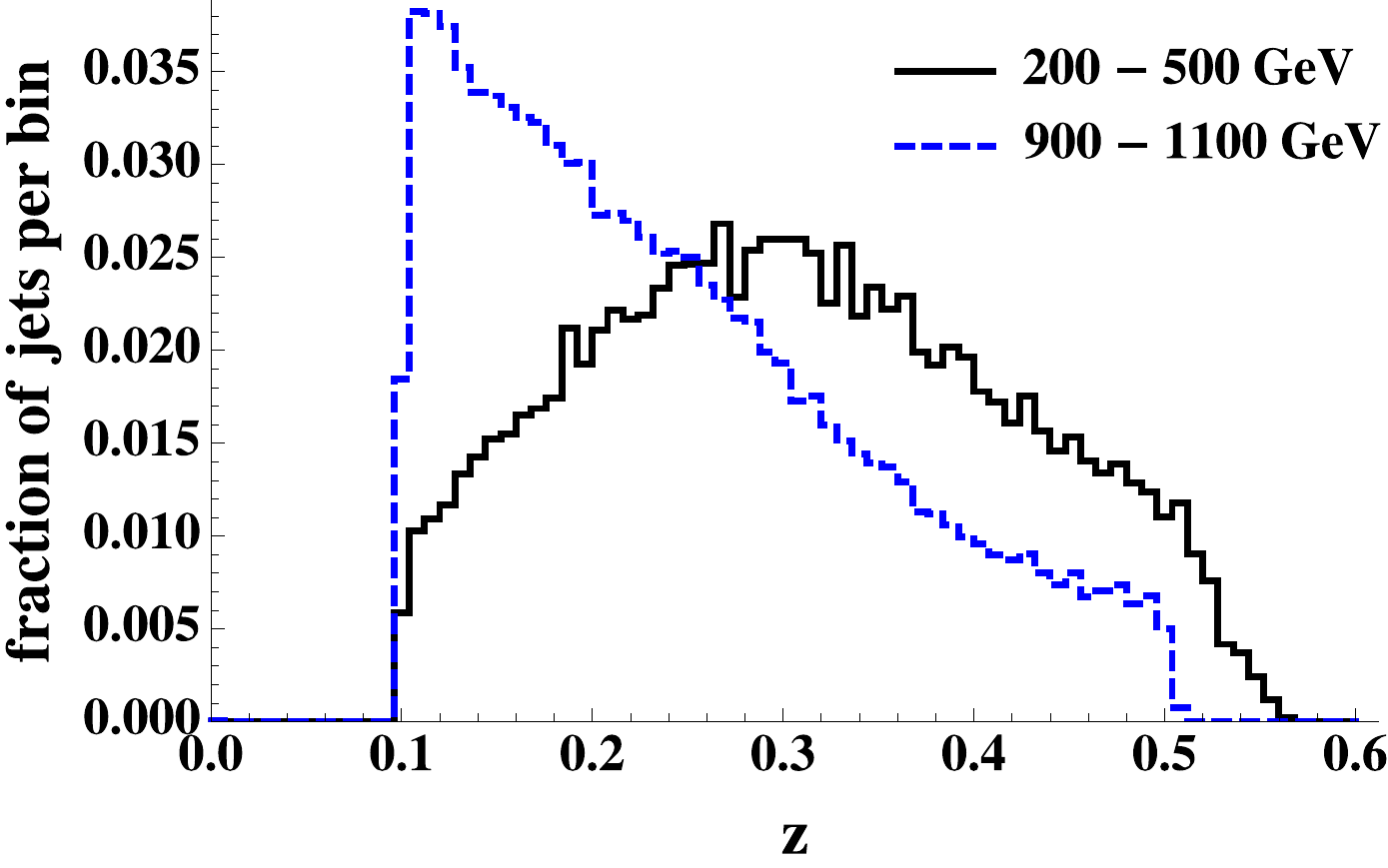}
\end{center}
\caption[Distribution in $z$ for pruned CA jets in the top mass window for two $p_T$ bins]{Distribution in $z$ for pruned CA jets in the top mass window for two $p_T$ bins, using $z_\text{cut}$ = 0.10.}
\label{fig:zpCAmtopPT1&4}
\end{figure}

Additionally, these plots offer an interesting perspective on the role of $z$ in jet substructure.  The $t\bar{t}$ sample for the CA algorithm is the most instructive.  In this case, small values of $z_{\cut}$ lead to dramatically increased efficiency for finding top jets in the larger $p_T$ bins.  This is due to the improved ability after pruning to find the $W$ as a subjet of the top.  At large $p_T$ with a fixed $D = 1.0$, the opening angle of the top quark decay is much smaller than $D$.  This means that the top quark decay is very localized in the jet, and much of the jet area includes soft radiation.  For the CA algorithm, which recombines solely by the angle between protojets, this tends to delay recombining the soft peripheral radiation until the end of the algorithm.  The result is substructure with small $z$ at the last recombination that is not representative of the top quark decay --- neither daughter protojet of the top has the $W$ mass.  As an illustration of this point, in Fig.~\ref{fig:zCAmtopPT1&4} we plot the distribution of $z$ for unpruned jets in the top mass range for the CA algorithm in the largest and smallest $p_T$ bins.  Note that in the largest $p_T$ bin, where the top quark decay is highly localized in the jet and the decay angle is much less than $D$, there is a substantially increased fraction of jets with a small value of $z$.  This does not occur in the smallest $p_T$ bin, where most of the reconstructed tops are at threshold for being just inside the jet.
When pruning is implemented, however, much of this soft radiation is removed.  In Fig.~\ref{fig:zpCAmtopPT1&4}, we plot the same distributions as in Fig.~\ref{fig:zCAmtopPT1&4}, but for pruned jets.  In this case, no jets with the top mass have small $z$, since pruning has removed those recombinations.  This leads to a highly enhanced efficiency to resolve the $W$ subjet and identify the jet and a top jet.  In Sec.~\ref{sec:prune:results:fixedD}, we will study pruning when the value of $D$ is matched to the average angle of the heavy particle decay, and we will see that the performance of the unpruned CA algorithm improves.

By contrast, this situation does not occur for the $\kt$ algorithm.  Even when the value of $D$ is mismatched with the top quark decay angle, the soft radiation on the periphery of the jet is recombined early in the $\kt$ algorithm because of the $p_T$ weighting in the recombination metric.  Therefore, there is no increase in efficiency with increasing $z_\cut$ for large $p_T$, and the decrease in $\epsilon$ comes from the narrower width of the top and $W$ mass distributions.  The small variation in the measures $R$ and $S$ for the $\kt$ algorithm at small $z_{\cut}$ is evidence of the fact that $\kt$ tends to have many fewer small-$z$ recombinations at the end of the algorithm, and supports the larger value of $z_{\cut} = 0.15$ for the $\kt$ algorithm that we will use in the remainder of the study.

\begin{figure*}[htbp]
\begin{center}
\subfloat[$W$'s, CA jets]{\label{fig:VaryDcut:WCA}\includegraphics[width=0.22\textwidth]{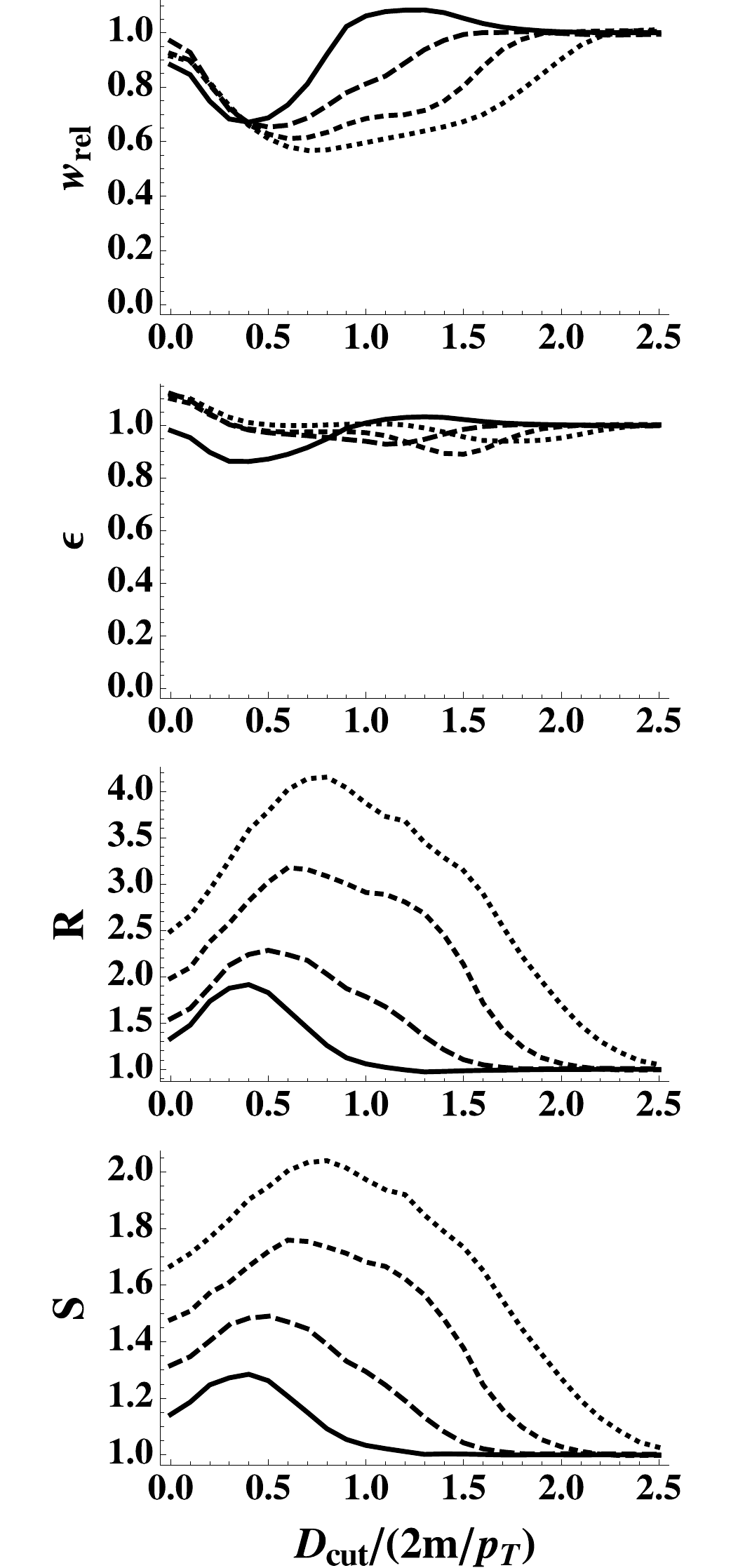}}
\subfloat[tops, CA jets]{\label{fig:VaryDcut:tCA}\includegraphics[width=0.22\textwidth]{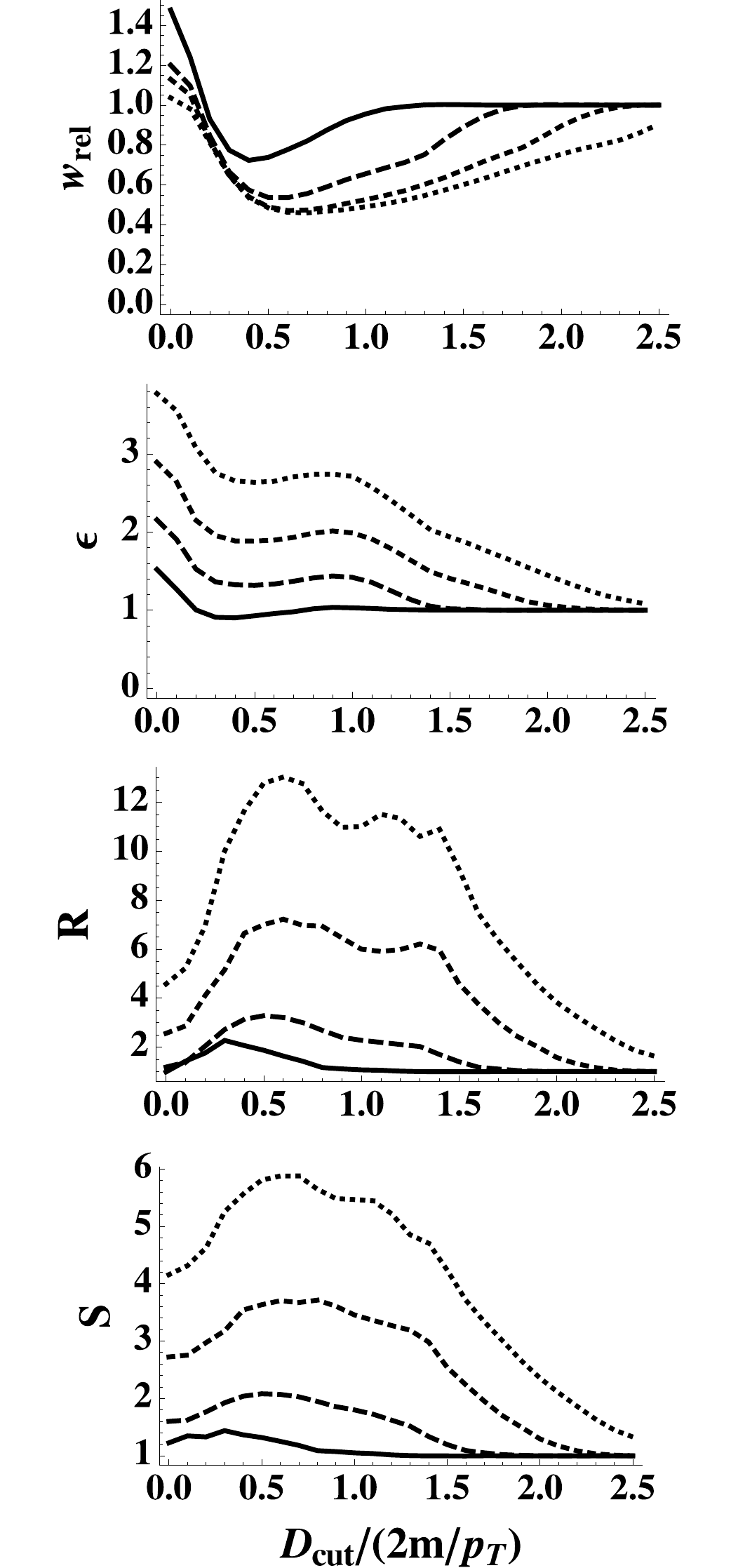}}
\subfloat[$W$'s, $\kt$ jets]{\label{fig:VaryDcut:WkT}\includegraphics[width=0.22\textwidth]{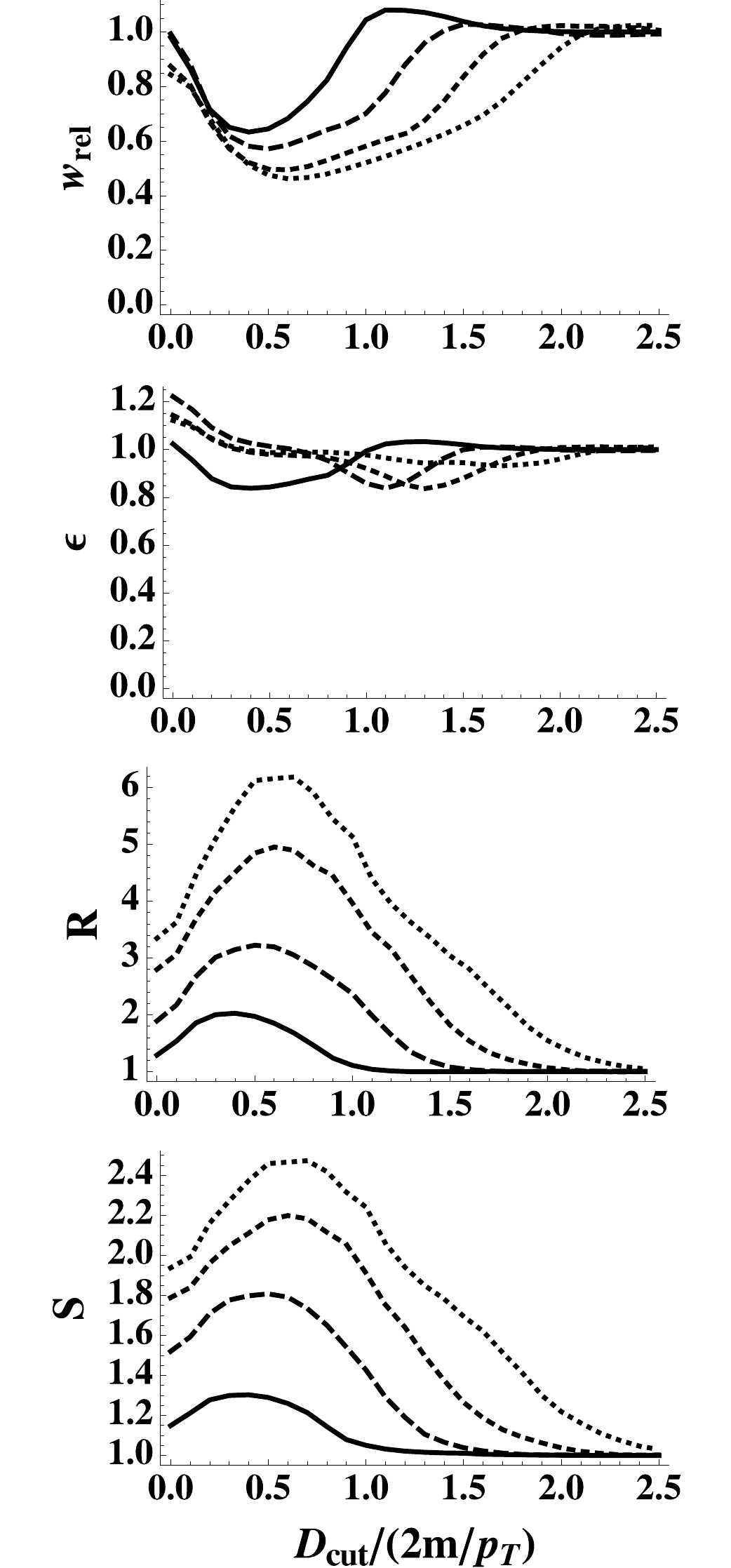}}
\subfloat[tops, $\kt$ jets]{\label{fig:VaryDcut:tkT}\includegraphics[width=0.22\textwidth]{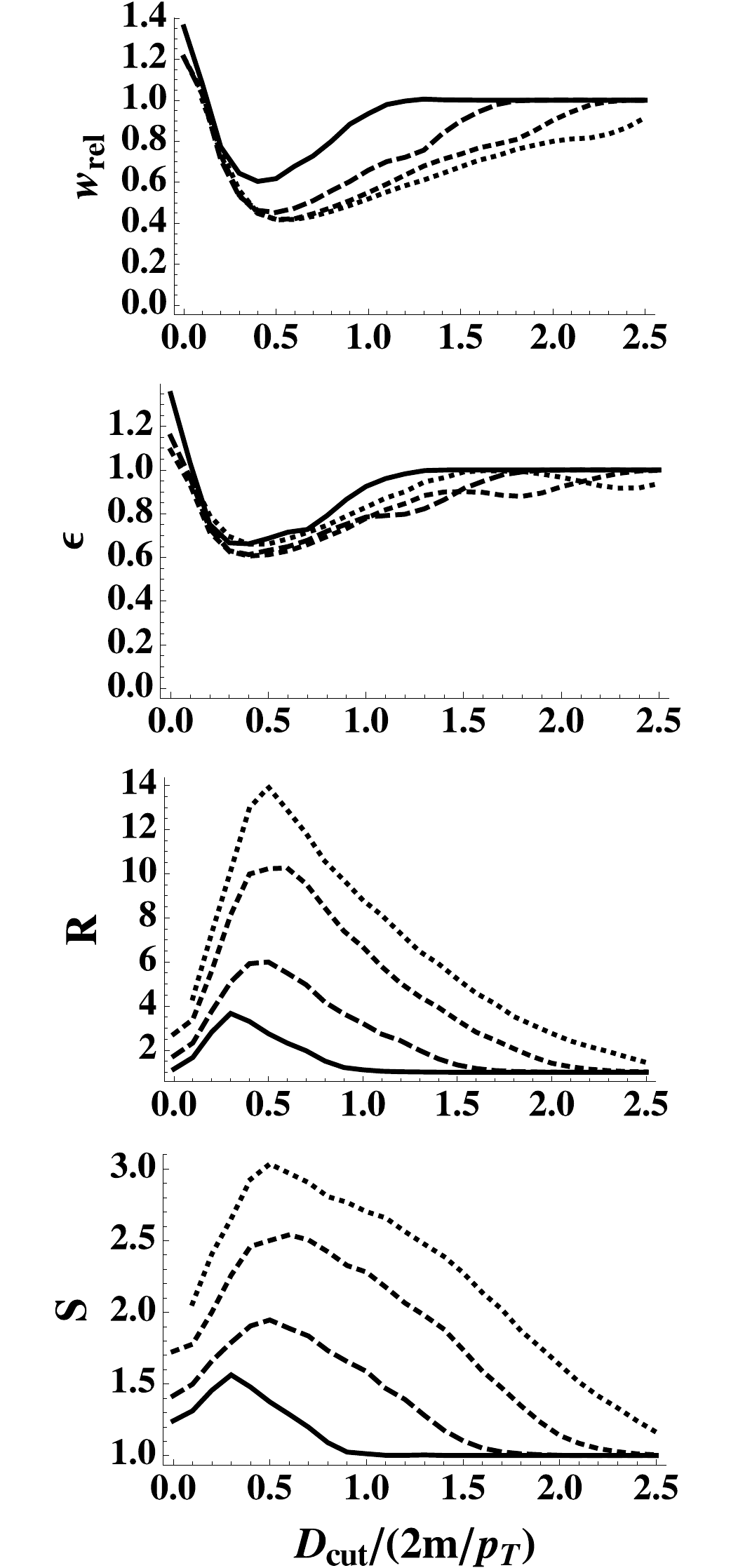}}
\includegraphics[width=0.10\textwidth]{pTbinlegend.pdf}
\end{center}

\caption[Relative statistical measures $w_\text{rel}$, $\epsilon$, $R$, and $S$ vs. $D_{\cut}/\frac{2 m_J}{p_{T_J}}$ for $W$'s and tops]{Relative statistical measures $w_\text{rel}$, $\epsilon$, $R$, and $S$ vs. $D_{\cut}/\frac{2 m_J}{p_{T_J}}$ for $W$'s and tops, using CA and $\kt$ jets.  Four $p_T$ bins are shown for each sample.  Statistical errors (not shown) are $\mathcal{O}(1\%)$ for $w_\text{rel}$ and $\epsilon$, and $\mathcal{O}(10\%)$ for $R$ and $S$.}
\label{fig:VaryDcut}
\end{figure*}

We now fix $z_{\cut}$ to study the dependence on $D_{\cut}$.  For the CA algorithm we choose $z_{\cut} = 0.1$, and for $\kt$ we choose 0.15.  In Fig.~\ref{fig:VaryDcut}, we plot $w_\text{rel}$, $\epsilon$, $R$, and $S$ as $D_{\cut}$ is varied in [0, $5 m_J/p_{T_J}$].  While $z_\cut$ sets the minimum $p_T$ asymmetry that recombinations can have, $D_\cut$ sets the minimum opening angle for recombinations that can be pruned.  We can think of $D_\cut$ as determining which recombinations can be pruned, and $z_\cut$ as determining whether or not that pruning takes place.  This difference is clearer when we consider two limiting values of $D_\cut$ and their impact on the pruned jet substructure.

As $D_\cut$ grows past $2m_J/p_{T_J}$, any recombination must have a large opening angle between the daughters to be pruned.  Note that the limit $D_\cut \to \infty$ is the limit of no pruning.  For both the CA and $\kt$ algorithms, in this limit only very late recombinations in the algorithm can be pruned (if the jet can be pruned at all).  In this limit, we expect the statistical measures to tend to one as the amount of pruning decreases.

The second limit is $D_\cut \to 0$.  In this limit any recombination can be pruned, since the minimum opening angle needed is very small.  As $D_\cut$ decreases towards zero, more of the jet substructure can be pruned.  In particular, earlier recombinations --- those with smaller opening angle on average --- can be pruned as $D_\cut$ decreases.  In general, these early recombinations are associated with the QCD shower, and pruning them can degrade the mass resolution of the jet because too much radiation is being removed.  Therefore, we expect the performance of pruning to be poor in this region.

Both of these limits are present in Fig.~\ref{fig:VaryDcut}, and our expectations about these limits are correct.  It is in the intermediate region, where $D_\cut \approx m_J/p_{T_J}$, that the performance of pruning is optimal, with a maximum in $S$ that is not very sensitive to the $p_T$ bin, sample, or algorithm.  This value of $D_\cut = m_J/p_{T_J}$ is sensible when we recognize that the average opening angle of the jet is approximately $2m_J/p_{T_J}$, and half this value allows for pruning of late recombinations but not the soft, small-angle recombinations associated with the QCD shower.

For the remainder of the study, we fix the pruning parameters $z_{\cut} = 0.1$ for the CA algorithm and $z_{\cut} = 0.15$ for the $\kt$ algorithm, as well as $D_{\cut} = m_J/p_{T_J}$ for both algorithms.  With these parameters fixed, we move on to discuss more interesting tests of the pruning procedure.

\subsection[\textit{Top and \texorpdfstring{$W$}{W} Identification with Constant \texorpdfstring{$D$}{D}}]{Top and \texorpdfstring{$W$}{W} Identification with Constant \texorpdfstring{$D$}{D}}
\label{sec:prune:results:fixedD}

\begin{figure*}[htbp!]
\subfloat[$W$'s, CA jets]{\label{fig:VarypT:WCA}\includegraphics[width=0.24\textwidth]{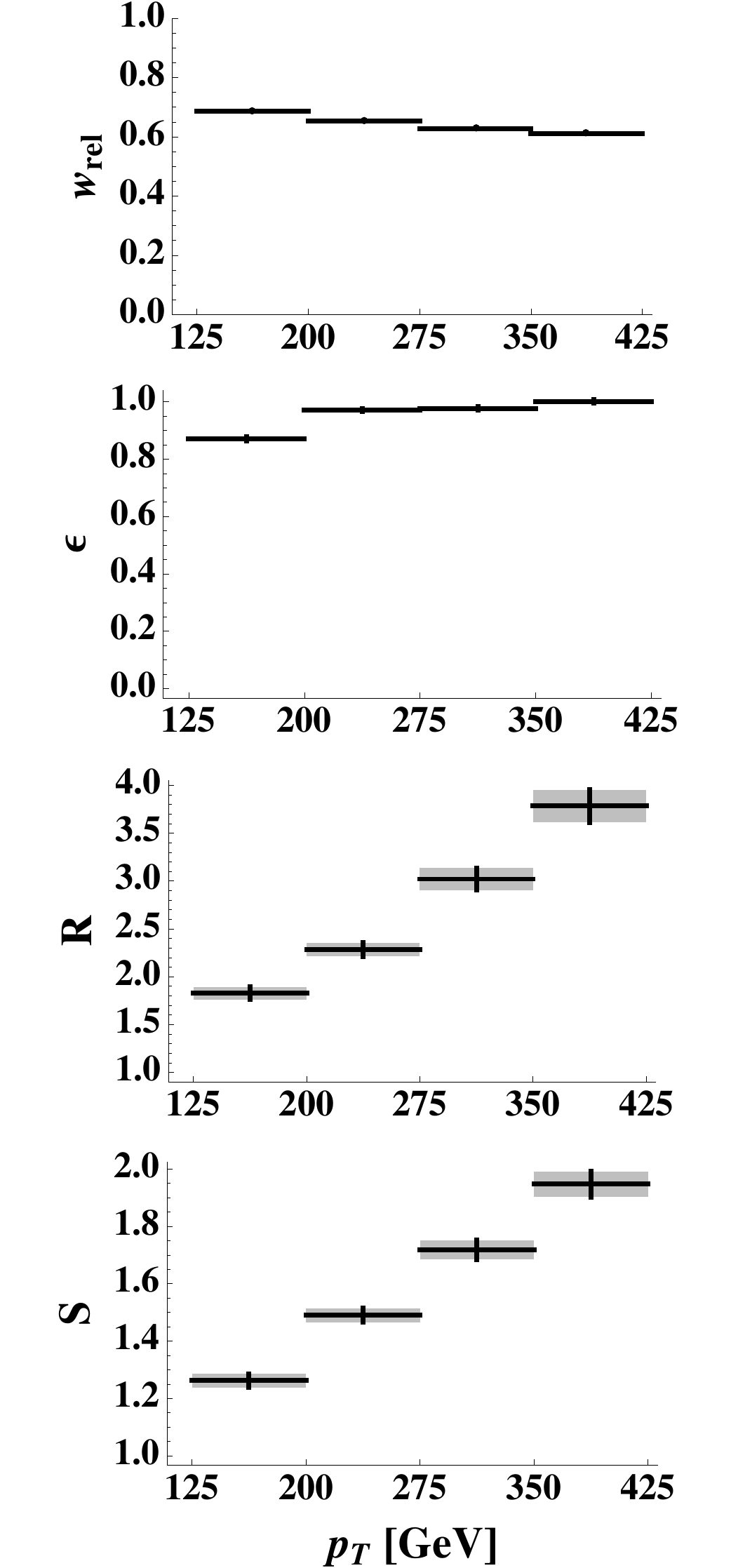}}
\subfloat[tops, CA jets]{\label{fig:VarypT:tCA}\includegraphics[width=0.24\textwidth]{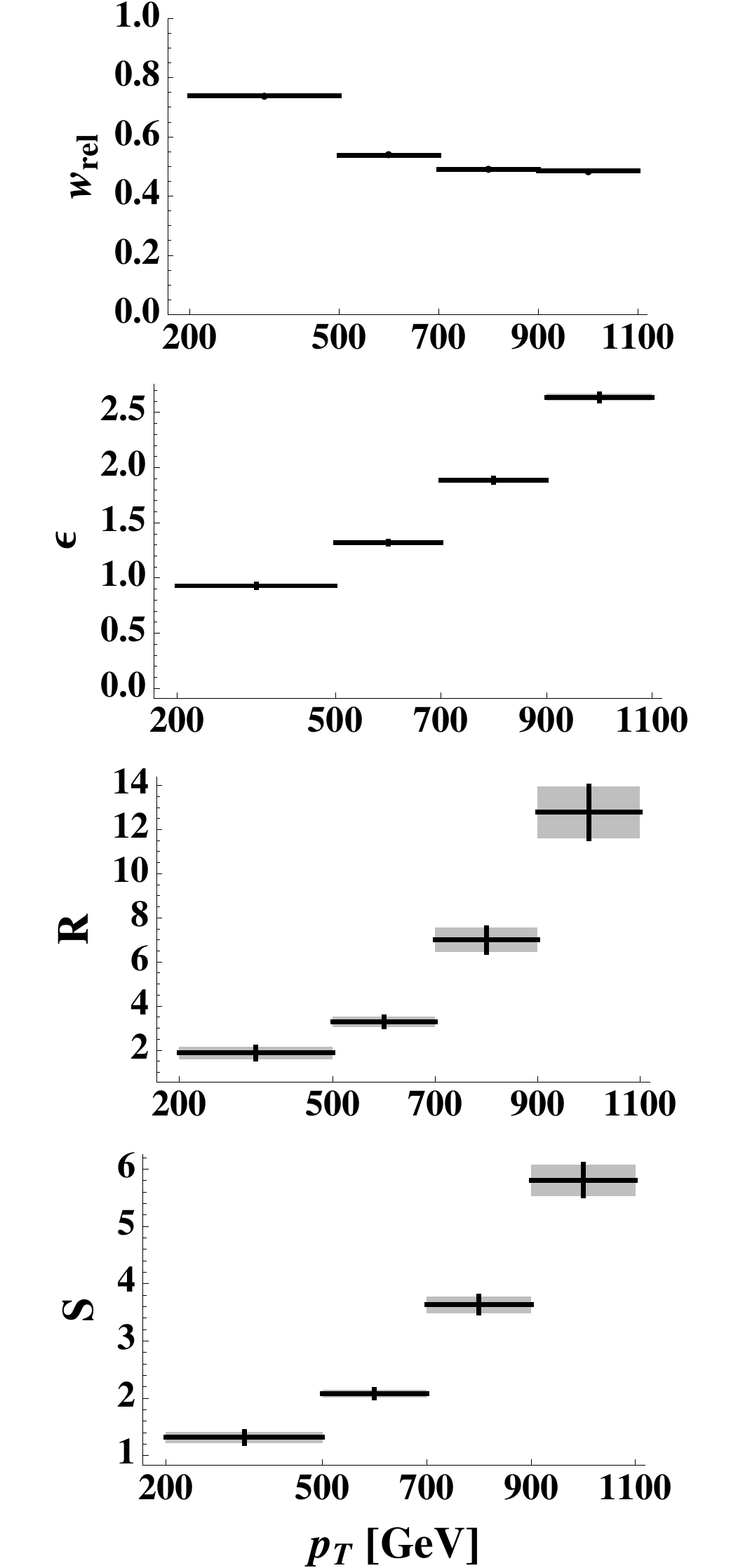}}
\subfloat[$W$'s, $\kt$ jets]{\label{fig:VarypT:WkT}\includegraphics[width=0.24\textwidth]{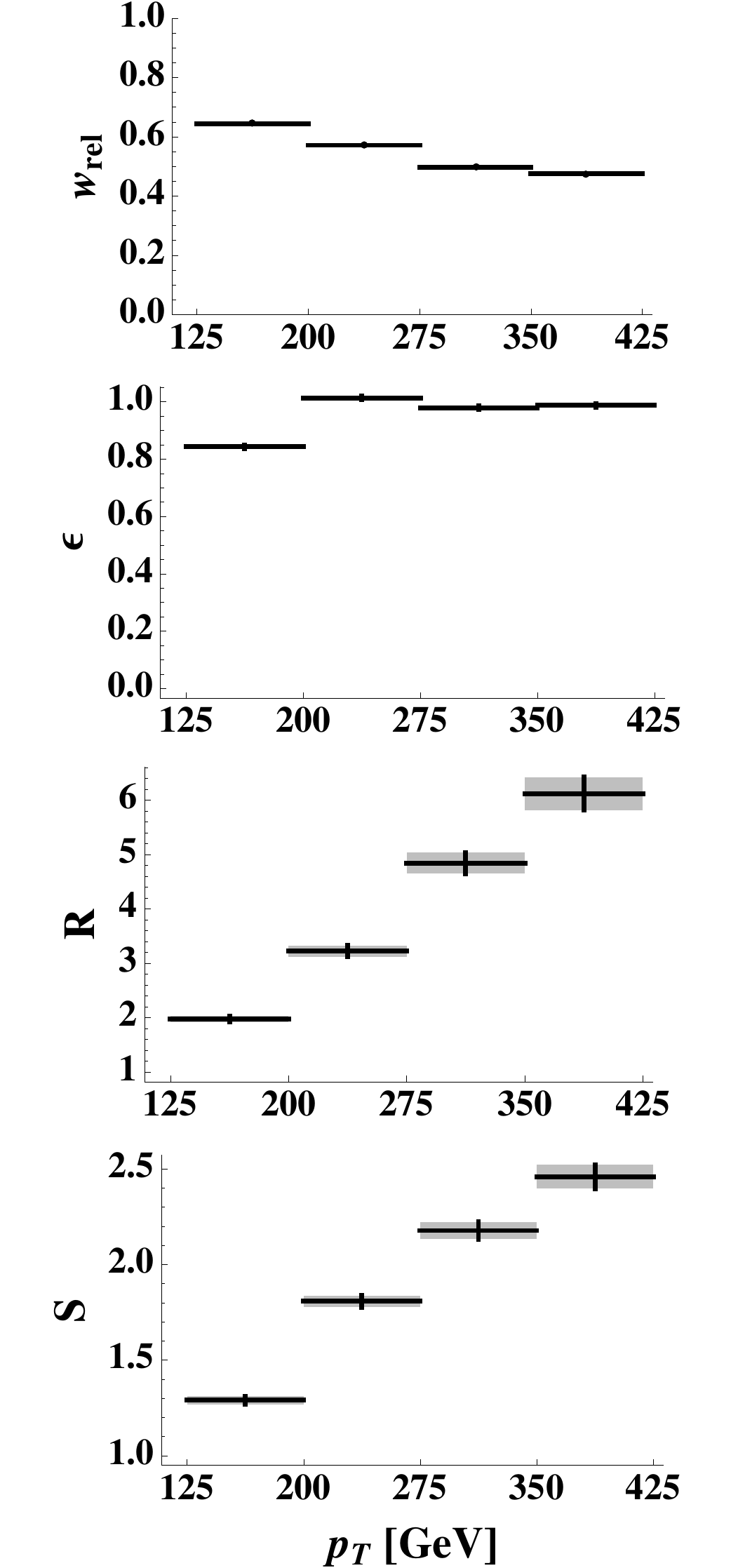}}
\subfloat[tops, $\kt$ jets]{\label{fig:VarypT:tkT}\includegraphics[width=0.24\textwidth]{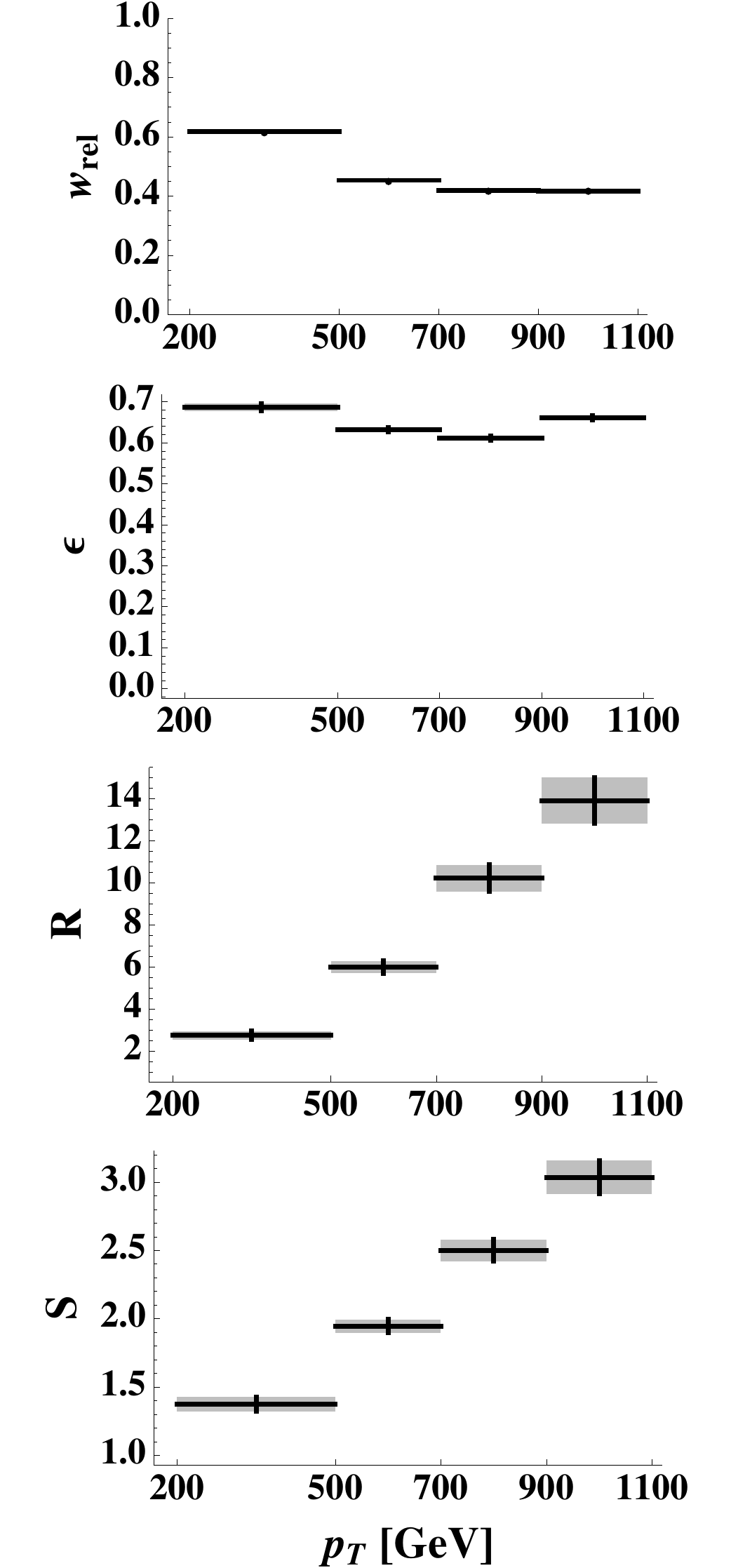}}
\caption[Relative statistical measures $w_\text{rel}$, $\epsilon$, $R$, and $S$ vs. $p_T$ for $W$'s and tops]{Relative statistical measures $w_\text{rel}$, $\epsilon$, $R$, and $S$ vs. $p_T$ for $W$'s and tops, using CA and $\kt$ jets with D = 1.0.  Statistical errors are shown.}
\label{fig:VarypT}
\end{figure*}

In a search for heavy particles decaying into jets, it may be unfeasible to divide a sample into $p_T$ bins and use a tailored jet algorithm to look for local excesses in the jet mass distribution in each $p_T$ bin.  (A ``variable-$R$'' method for avoiding $p_T$-binning, which we do not consider here, has recently been suggested \cite{VariableR}.  This still requires knowing or guessing the mass of the new particle, since it is $m/p_T$ that determines the relevant angular size.)  For instance, the appropriate angular scale may be unknown because the mass of the heavy particle is not known or the production mechanism is not well understood (so that the spectrum of heavy particle boosts is not known).  In this case, a large-$D$ jet algorithm may be used to search for heavy particles reconstructed in single jets.  To mimic such an analysis, and provide a reference point for further tests of pruning, we find our statistical measures for $W$ and top quark jets with a fixed $D$ of 1.0.

In Fig.~\ref{fig:VarypT} we plot the values for $w_\text{rel}$, $\epsilon$, $R$, and $S$ versus $p_T$ bin for $W$'s and tops, using the CA and $\kt$ algorithms.\footnote{The statistical error bars shown are primarily due to the limited number of events in the background sample after pruning.}  Pruning improves $W$ and top finding for both algorithms, with substantial improvements for large $p_T$.  The measure $S$ in the smallest $p_T$ bins ranges from 30--40\%, growing to values between 100--600\% in the largest $p_T$ bins.  At large $p_T$ in the top quark study, the improvement in signal-to-noise for the CA algorithm is larger than for the $\kt$ algorithm, as is the relative efficiency to identify tops.  This arises because the CA algorithm is poor at reconstructing the $W$ as a subjet of the top jet at large $p_T$ when the value of $D$ is not matched to the opening angle of the decay.  We will investigate this case further in the rest of the analysis.

\subsection[\textit{Top Identification with Variable \texorpdfstring{$D$}{D}}]{Top Identification with Variable \texorpdfstring{$D$}{D}}
\label{sec:prune:results:varD}

\begin{figure*}[htbp]
\subfloat[$W$'s, CA jets]{\label{fig:VarypToptD:WCA}\includegraphics[width=0.24\textwidth]{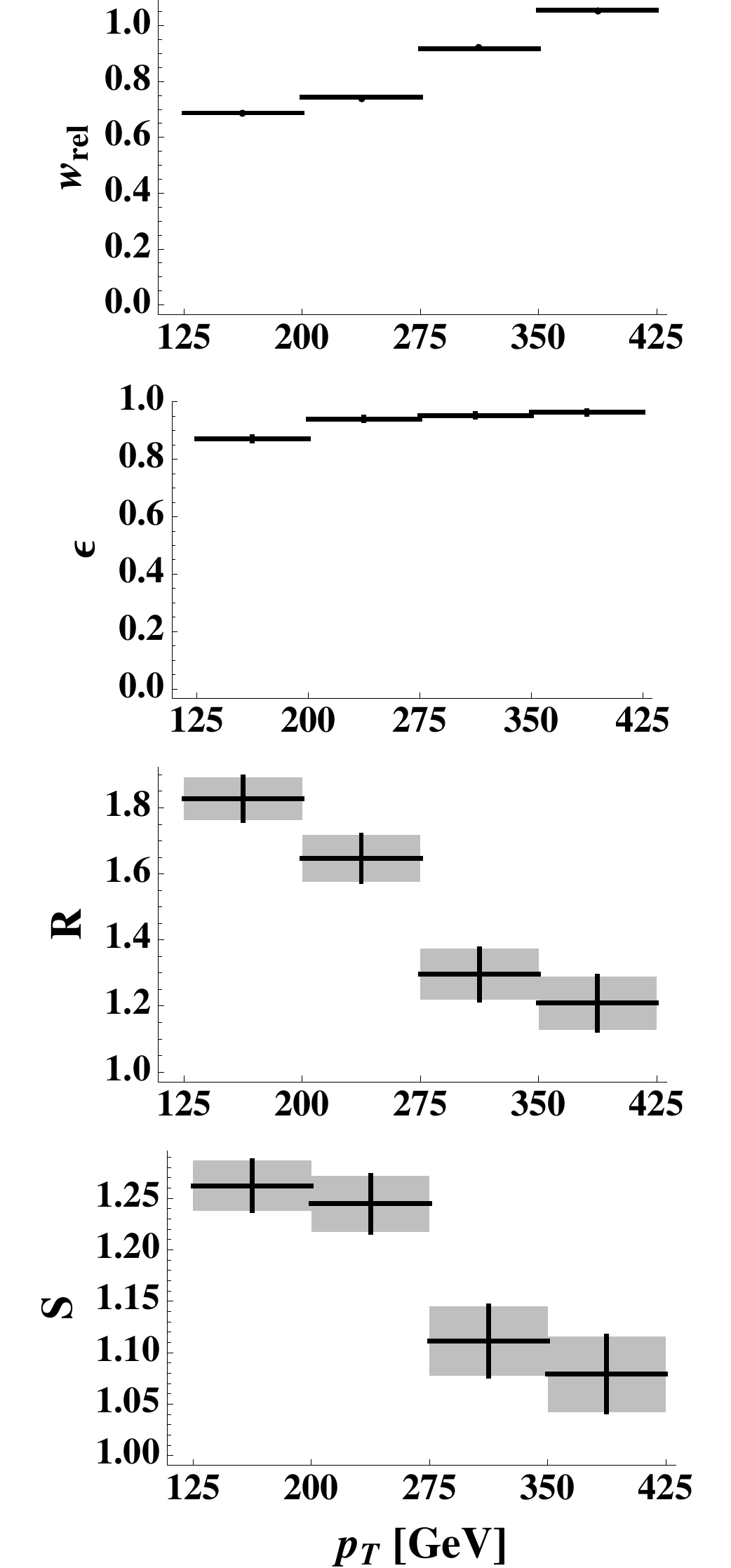}}
\subfloat[tops, CA jets]{\label{fig:VarypToptD:tCA}\includegraphics[width=0.24\textwidth]{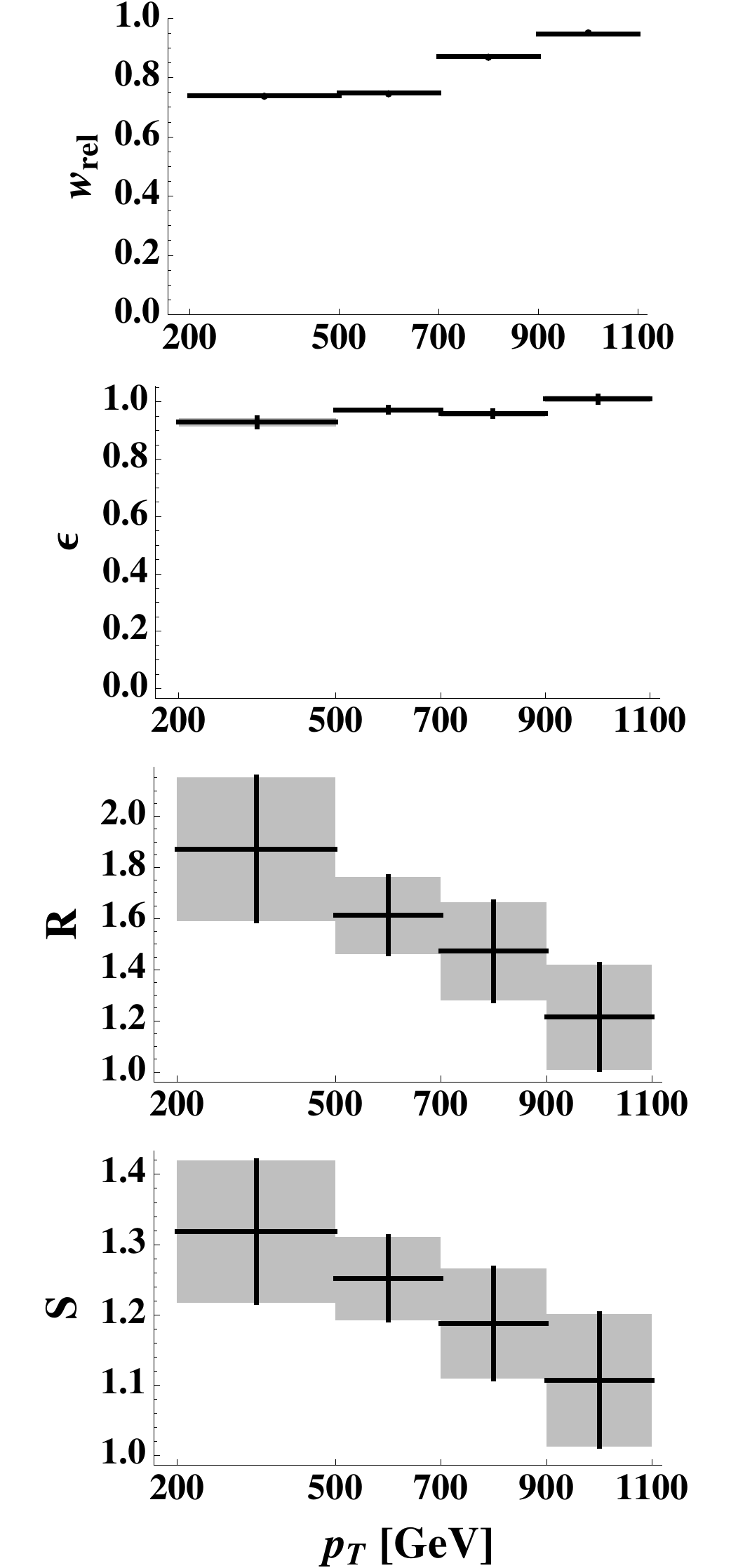}}
\subfloat[$W$'s, $\kt$ jets]{\label{fig:VarypToptD:WkT}\includegraphics[width=0.24\textwidth]{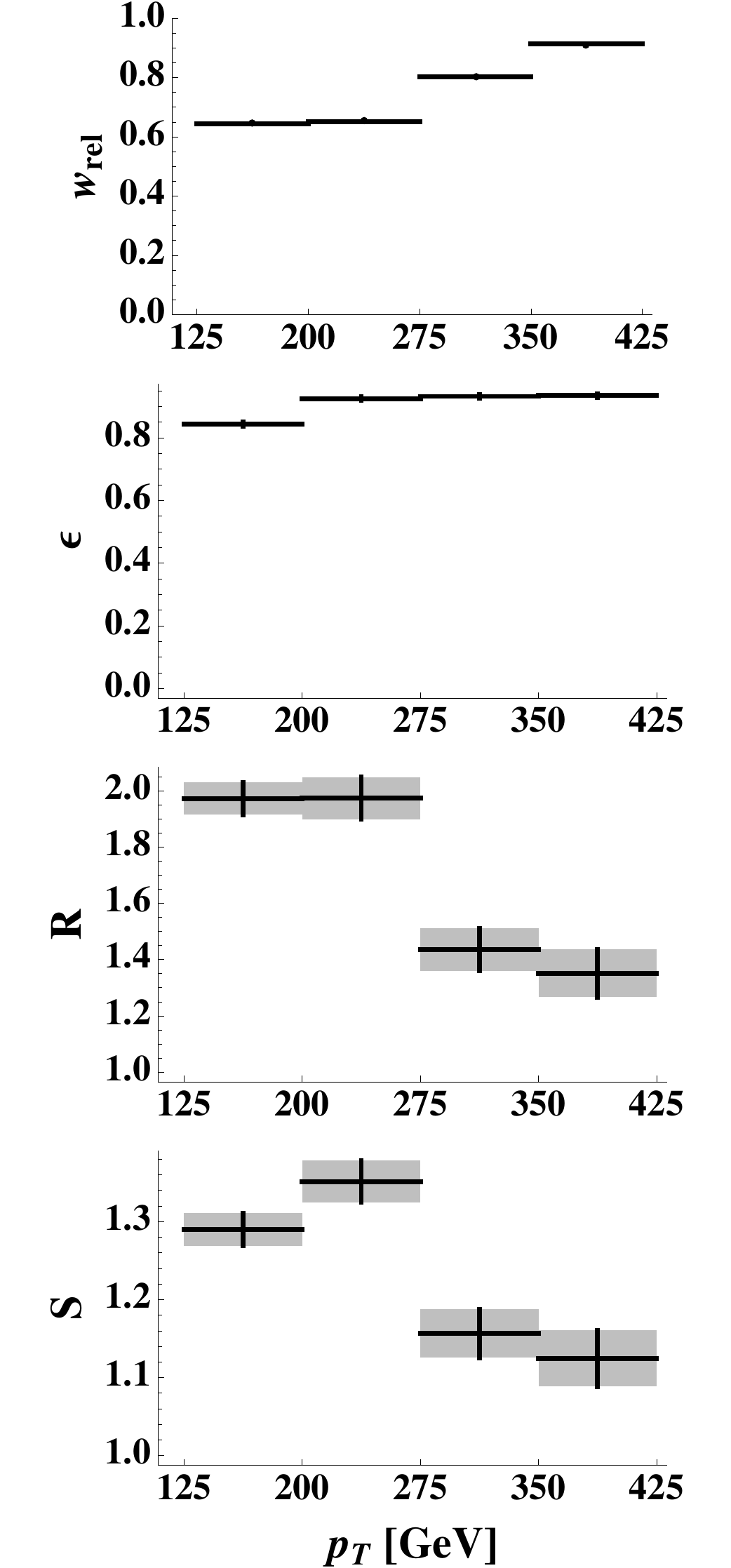}}
\subfloat[tops, $\kt$ jets]{\label{fig:VarypToptD:tkT}\includegraphics[width=0.24\textwidth]{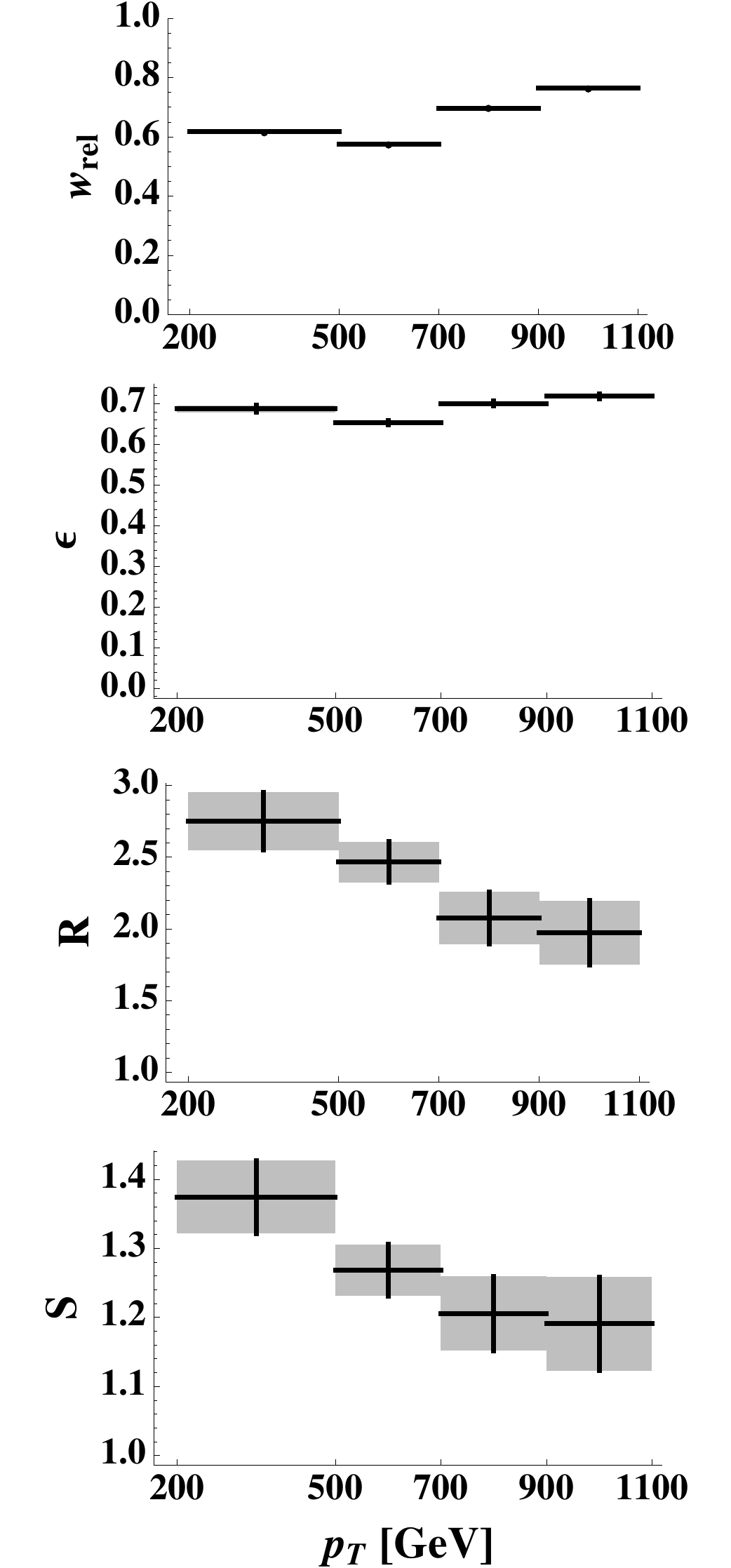}}
\caption[Relative statistical measures $w_\text{rel}$, $\epsilon$, $R$, and $S$ vs. $p_T$ for $W$'s and tops]{Relative statistical measures $w_\text{rel}$, $\epsilon$, $R$, and $S$ vs. $p_T$ for $W$'s and tops, using CA and $\kt$ jets.  Instead of a fixed $D = 1.0$, a tuned $D$ is used for each $p_T$ bin (see Table \ref{table:D}).  Statistical errors are shown.}
\label{fig:VarypToptD}
\end{figure*}

\begin{table}[htbp]
\begin{center}
\begin{tabular}{|r||c|c|c|c|}
\hline
 & \multicolumn{4}{c|}{$W$} \\
 \hline
$p_T$ (GeV) & 125--200 & 200--275 & 275--350 & 350-425 \\
\hline
``tuned'' $D$ & 1.0 & 0.8 & 0.6 & 0.4 \\
\hline
\hline
 & \multicolumn{4}{c|}{top} \\
\hline
$p_T$ (GeV) &  200--500 & 500--700 & 700--900 & 900--1100 \\
\hline
``tuned'' $D$ & 1.0 & 0.7 & 0.5 & 0.4 \\
\hline
\end{tabular}
\end{center}

\caption[``Tuned'' D values for $W$ and top $p_T$ bins]{``Tuned'' D values for $W$ and top $p_T$ bins.  The fixed-$D$ analysis used $D = 1.0$, so the smallest bin does not change.}
\label{table:D}
\end{table}

For an analysis where the heavy particle mass is known, the jet algorithm can be tailored to the jet $p_T$.  The $D$ value can be chosen using the relation
\[
D = \min\left(1.0, 2\frac{m}{p_T}\right) .
\label{varD}
\]
where $m$ is the heavy particle mass and $p_T$ is the transverse momentum of the jet.  We take 1.0 to be the maximum allowed value of $D$.  The $D$ values we use are given in Table~\ref{table:D}.  In Fig.~\ref{fig:VarypToptD}, we plot $w_\text{rel}$, $\epsilon$, $R$, and $S$ for jets with these $D$ values used for each $p_T$ bin.  Note that Eq.~(\ref{varD}) neglects the differences between algorithms, which depend on the particular decay.  As an example of the fidelity of this relation for $D$, recall Fig.~\ref{fig:topDRdist}, which plotted the distribution in $\Delta R$ for reconstructed parton-level top quark decays with a top boost of $\gamma = 3$.  Eq.~(\ref{varD}) suggests the value $D = 0.7$, while the means of the CA and $\kt$ distributions for the reconstructed parton-level decay are 0.75 and 0.65 respectively.  Because the distribution in opening angles of the reconstructed decay is broad, by using a smaller, fixed $D$ some decays will not be reconstructed by the jet algorithm.

The difference between the case of constant $D = 1.0$ and variable $D$ is readily apparent.  When the $D$ value is matched to the expected opening angle of the decay, the improvements in pruning are flatter over the whole range in $p_T$, and generally decreasing towards high $p_T$.  The decreased efficiency for pruning, especially for the $\kt$ algorithm, is outweighed by the increases in $R$ and $S$ over the whole range in $p_T$.

\begin{figure*}[htbp!]
\subfloat[$W$'s, CA jets]{\label{fig:VarypTcompareD:WCA}\includegraphics[width=0.24\textwidth]{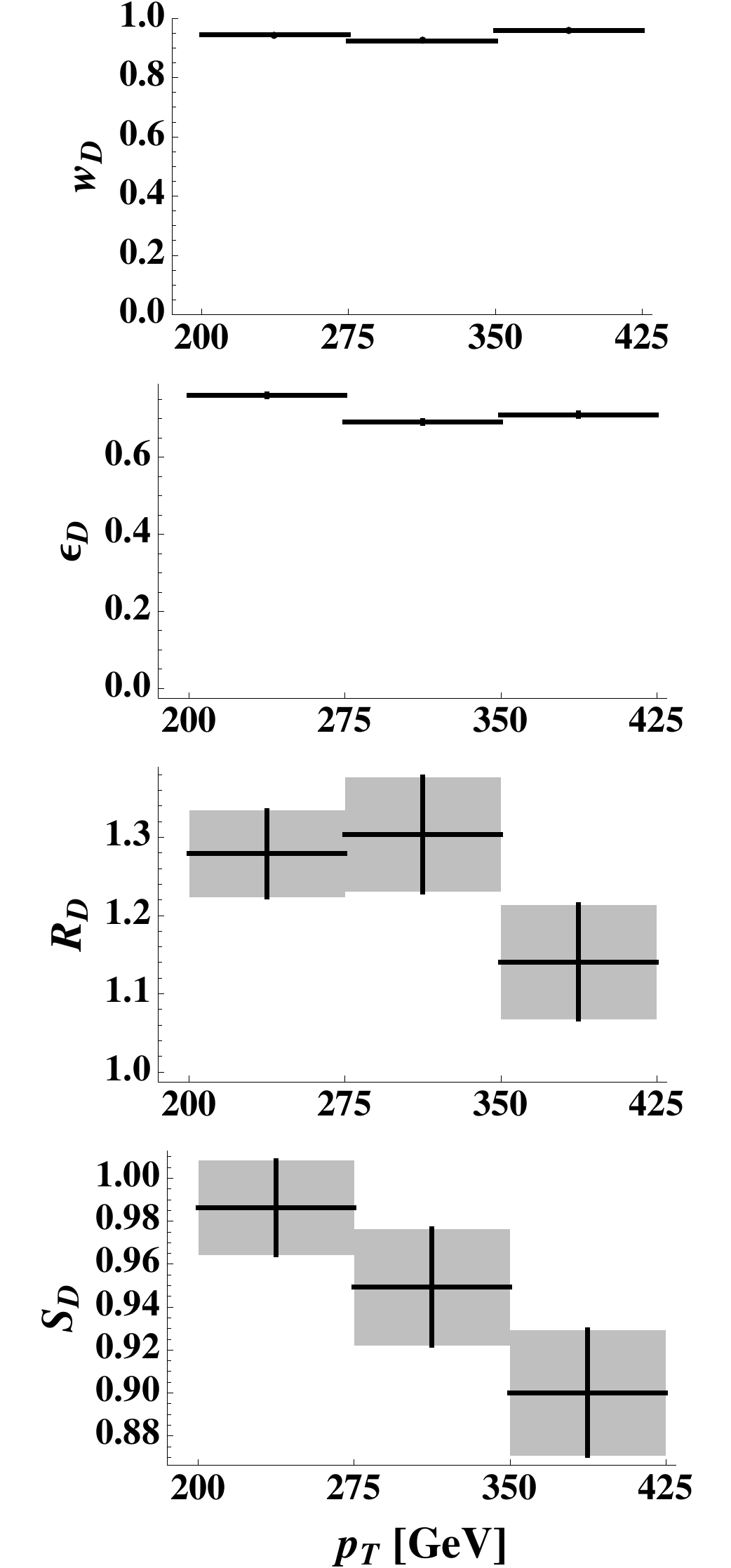}}
\subfloat[tops, CA jets]{\label{fig:VarypTcompareD:tCA}\includegraphics[width=0.24\textwidth]{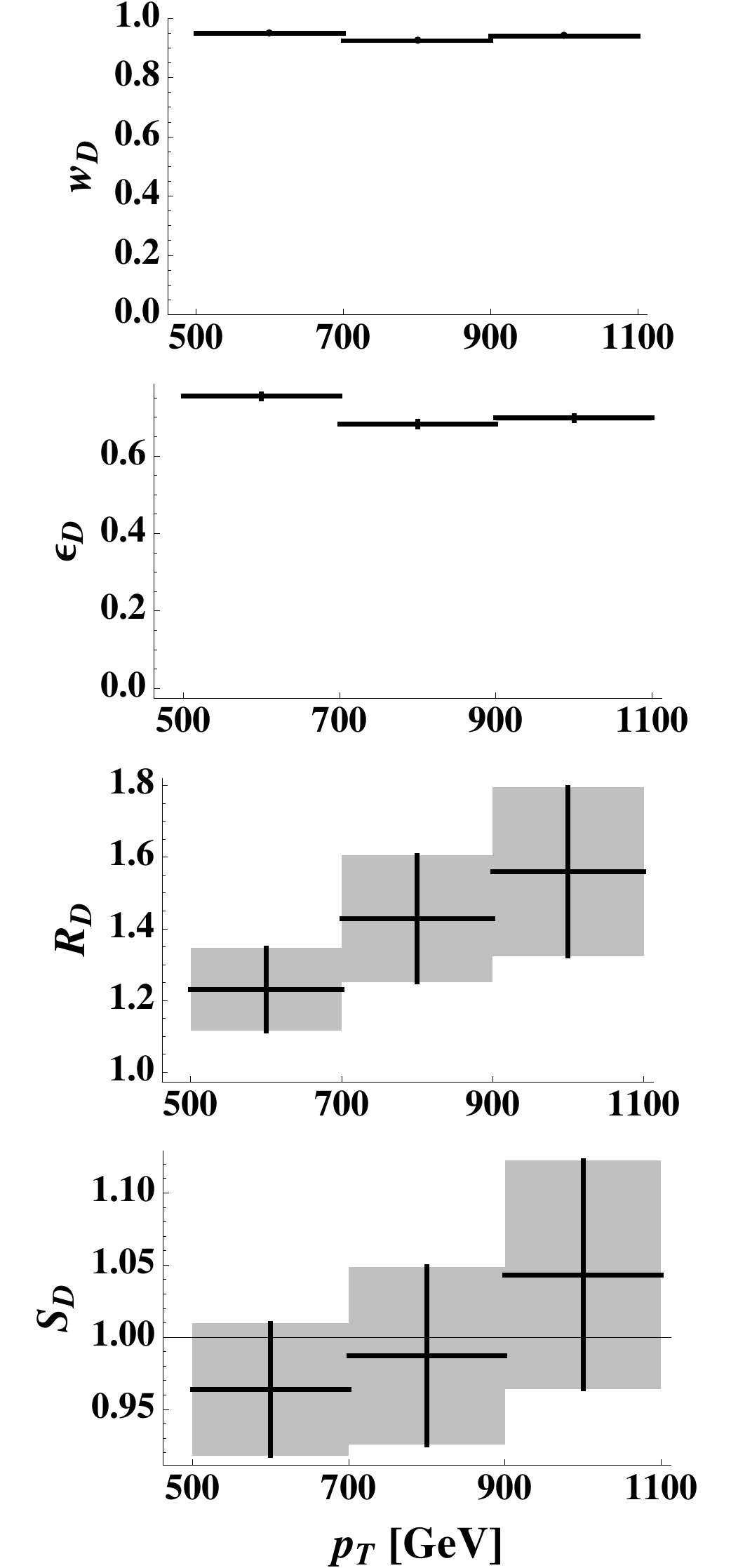}}
\subfloat[$W$'s, $\kt$ jets]{\label{fig:VarypTcompareD:WkT}\includegraphics[width=0.24\textwidth]{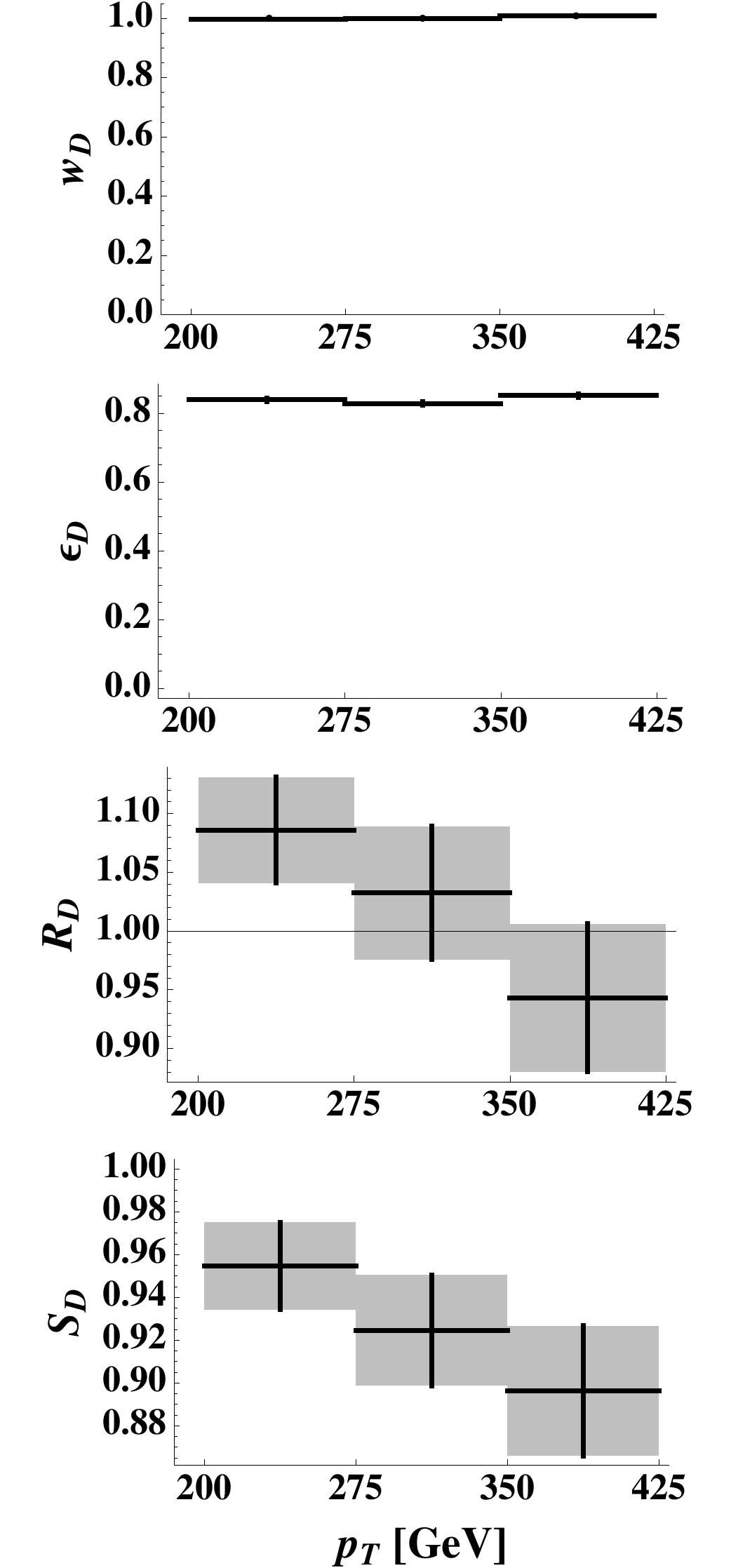}}
\subfloat[tops, $\kt$ jets]{\label{fig:VarypTcompareD:tkT}\includegraphics[width=0.24\textwidth]{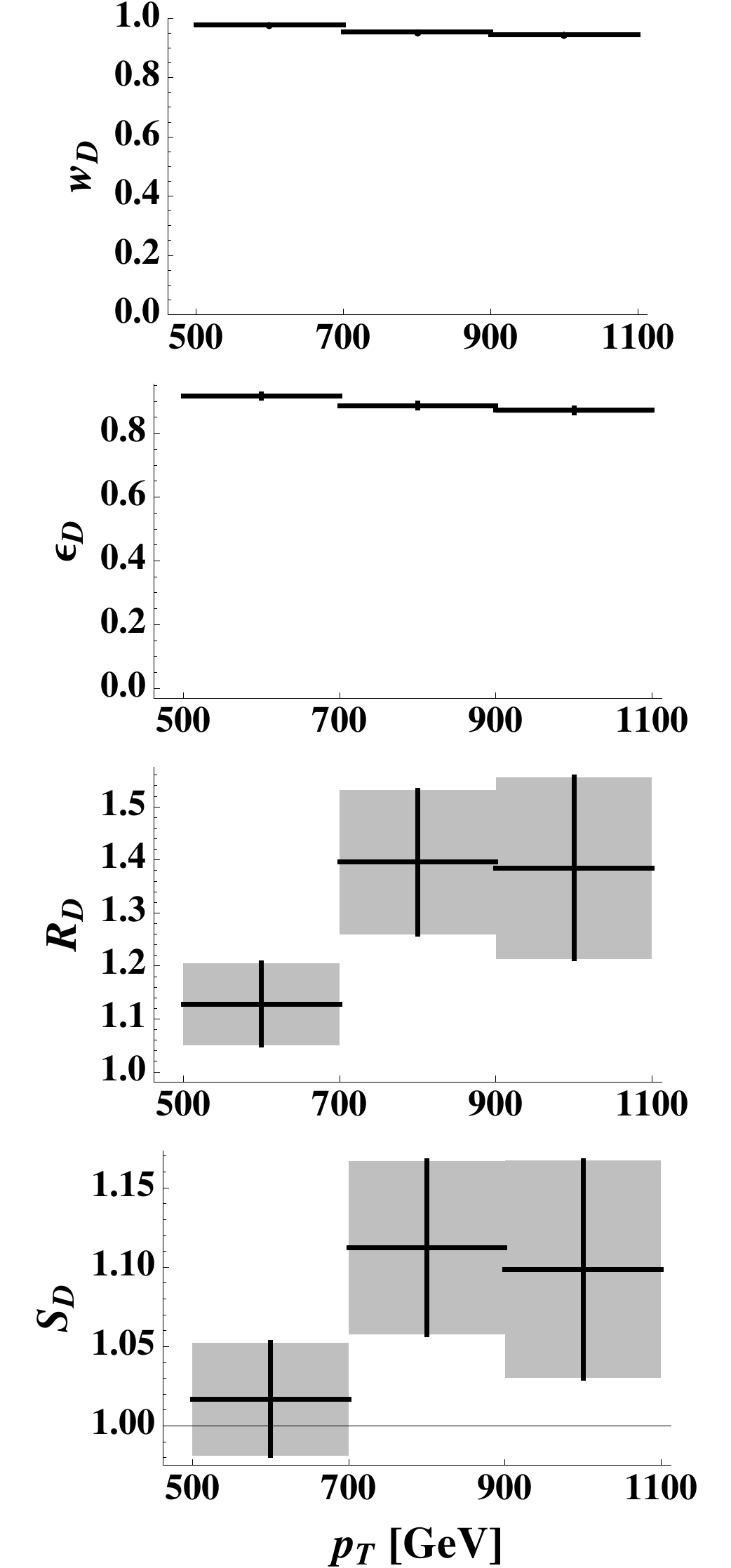}}
\caption[Relative statistical measures $w_D$, $\epsilon_D$, $R_D$, and $S_D$ vs. $p_T$ for $W$'s and tops]{Relative statistical measures $w_D$, $\epsilon_D$, $R_D$, and $S_D$ vs. $p_T$ for $W$'s and tops, using CA and $\kt$ jets.  The measures now compare pruning with a tuned $D$ value in each $p_T$ bin to pruning with a fixed $D$.  Statistical errors are shown.}
\label{fig:VarypTcompareD}
\end{figure*}

\subsection[\textit{Comparing Pruning with Different \texorpdfstring{$D$}{D} Values}]{Comparing Pruning with Different \texorpdfstring{$D$}{D} Values}
\label{sec:prune:results:compD}

In the previous two subsections we saw that an unpruned analysis performs much better when $D$ is tuned to the $m/p_T$ of the signal.  We now consider whether this is true of a pruned analysis.

In each $p_T$ bin, we can compare the results of pruned jets with $D = 1.0$ with pruned jets using value of $D$ fit to the expected size of the decay.  Because the naive expectation is that the tuned value of $D$ will yield better separation from background, we find the improvements in pruning when $D$ is tuned, relative to pruning with a fixed $D$ of 1.0.  Analogous metrics, $w_D$, $\epsilon_D$, $R_D$, and $S_D$, are used, but now they compare the results from pruning with the tuned $D$ value to the results from pruning with $D = 1.0$.  For instance,
\[
R_D \equiv \frac{S/B\text{ from pruning with tuned }D}{S/B\text{ from pruning with } D = 1.0}.
\]
Note that $x_D > 1$ indicates that tuning $D$ yields an improvement.  The values of these four measures are shown in Fig.~\ref{fig:VarypTcompareD} over the range of $p_T$.\footnote{The statistical errors now have significant contributions from both pruned background samples.  Each ``measurement'' compares the results of two methods, where each method has an associated uncertainty (the error bars in Figures \ref{fig:VarypT} and \ref{fig:VarypToptD}).  These errors are not independent because the same initial background sample is used in each case.  The combined uncertainties in this figure assume that the individual errors are independent, so should be viewed as an upper bound and at best a rough estimate of the statistical uncertainty.}  Note that since the tuned value of $D$ in the smallest $p_T$ bin is 1.0, the comparison there is trivial and so is not shown.

These results show only small improvements in $S_D$, with the statistical error bars at most data points including the value $S_D = 1$.  They indicate that the results after pruning are roughly independent of the value of $D$ used in the jet algorithm, as long as that $D$ is large enough to fit the expected size of the decay in a single jet.  From the point of view of heavy particle searches, we can conclude that pruning removes much of the $D$ dependence of the jet algorithm in the search.

\subsection[\textit{Absolute Measures of Pruning}]{Absolute Measures of Pruning}
\label{sec:prune:results:absolute}

So far, we have only considered measures of pruning relative to a similar analysis without pruning, because this factors out much of the dependence on details of the samples.  However, several recent studies report absolute performance metrics for heavy particle identification, so we examine similar measures here for completeness.  In addition, we directly compare the CA and $\kt$ algorithms, with and without pruning.

\begin{figure}[htbp!]
\begin{center}
\subfloat[top mass window width] {\label{fig:MassWidths:top}\includegraphics[width=0.45\columnwidth]{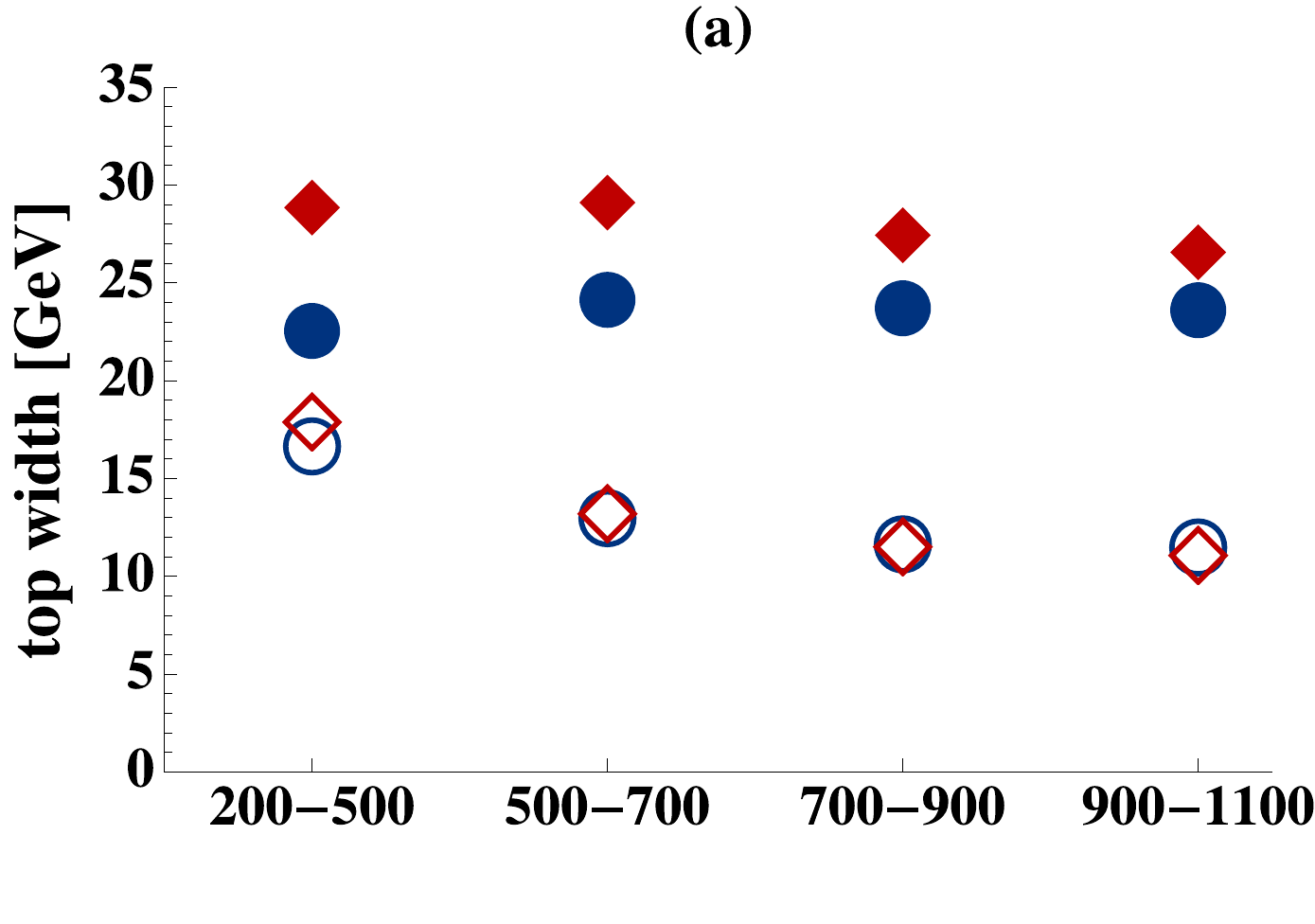}}
{\label{fig:MassWidths:legend}\includegraphics[width=0.45\columnwidth]{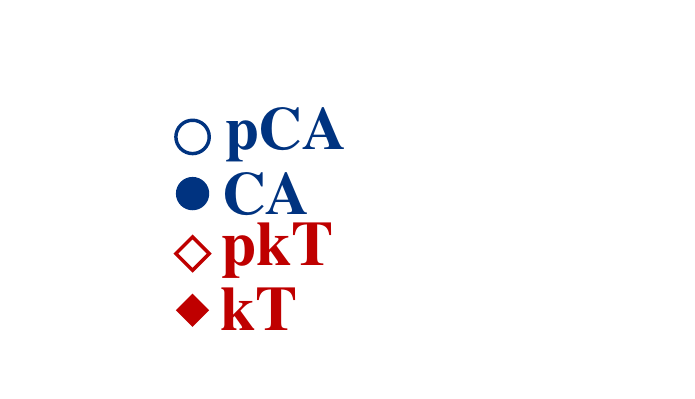}}

\subfloat[$W$ mass window width, top sample] {\label{fig:MassWidths:topW}\includegraphics[width=0.45\columnwidth]{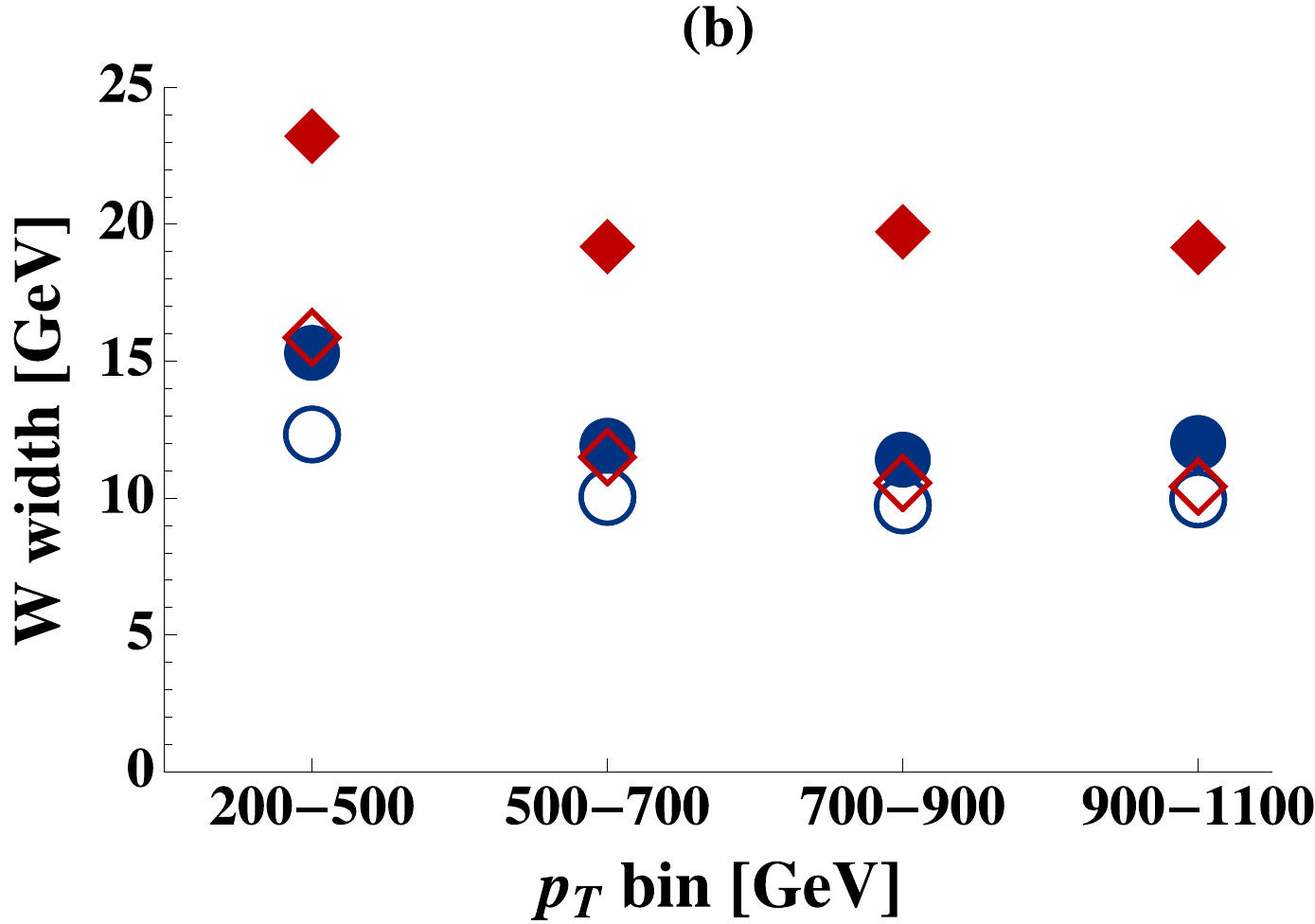}}
\subfloat[$W$ mass window width, $W$ sample] {\label{fig:MassWidths:W}\includegraphics[width=0.45\columnwidth]{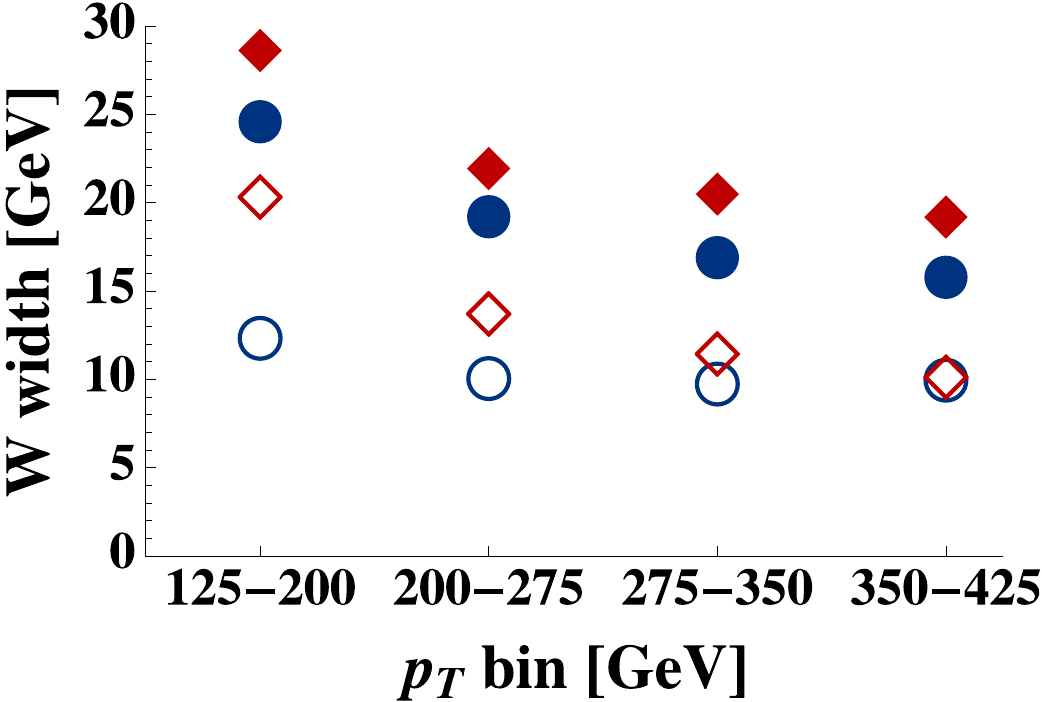}}
\end{center}

\caption[Widths of the top jet, $W$ subjet of the top jet, and $W$ jet mass windows]{Widths of the top jet (a), $W$ subjet of the top jet (b), and $W$ jet (c) mass windows for the top and $W$ signal samples.}
\label{fig:MassWidths}
\end{figure}

As can be seen from the plots of $w_\text{rel}$ in previous sections, pruning reduces the width of the mass distribution for heavy particles.  In Fig.~\ref{fig:MassWidths}, we plot the absolute widths of the fitted mass distributions for both the top and $W$ in the $t\bar{t}$ sample and the $W$ in the $WW$ sample, over all $p_T$ bins.  We plot this width for the pruned and unpruned version of the CA and $\kt$ algorithms.

Note that the heavy particle identification method we use in this work selects jets within a range of width $2\Gamma$, with $\Gamma$ coming from a fit to the signal sample.  This gives rise to a mass range cut that is typically much narrower than fixed width ranges used in other studies, and hence the absolute efficiency to identify heavy particles is lower.

\begin{figure}[htbp!]
\begin{center}
\subfloat[$\epsilon_{\text{abs}}$, tops]{\label{fig:AbsoluteEfficiencies:topabs}\includegraphics[width=0.45\columnwidth]{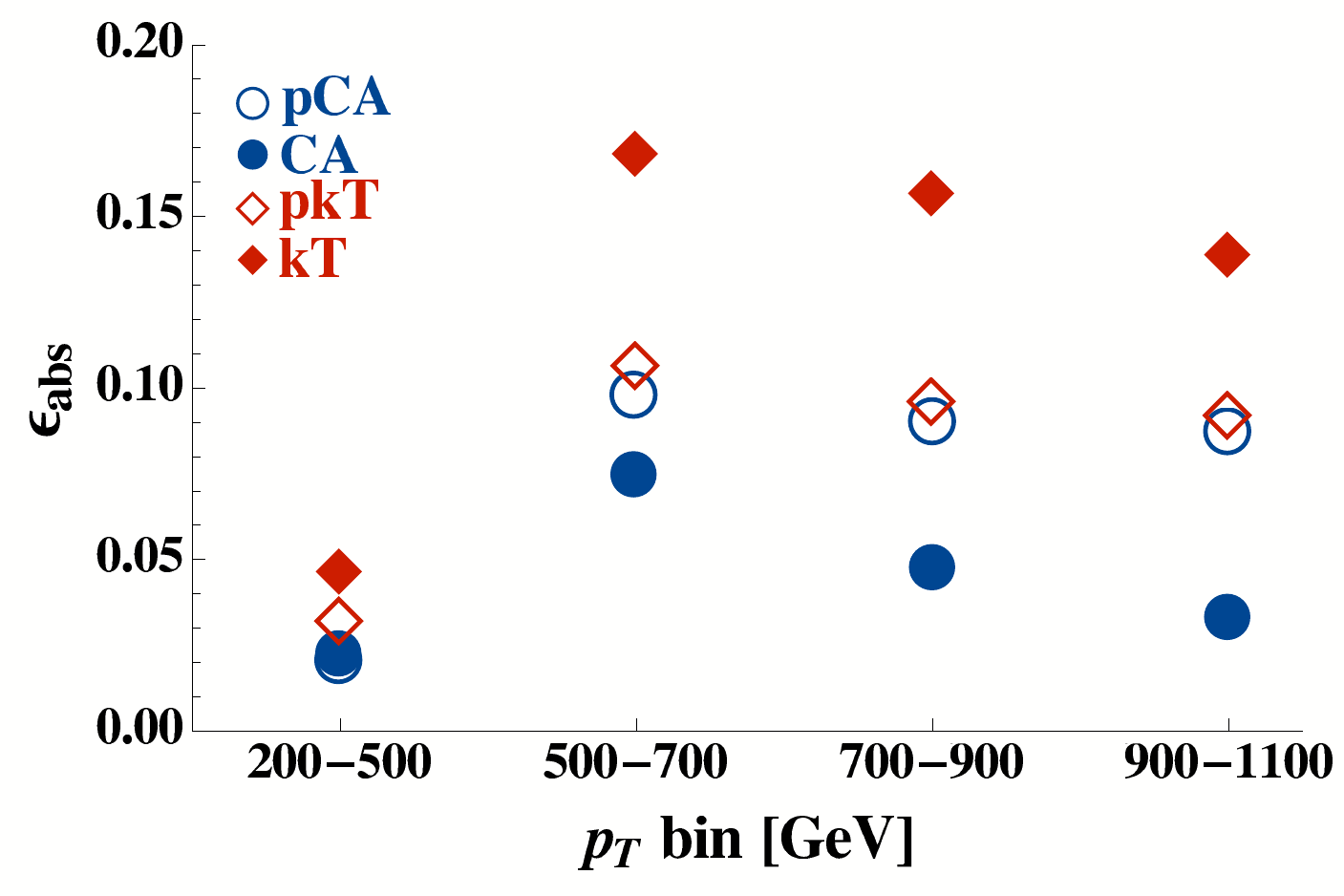}}
\subfloat[$\epsilon_{\text{abs}}$, $W$s]{\label{fig:AbsoluteEfficiencies:Wabs}\includegraphics[width=0.45\columnwidth]{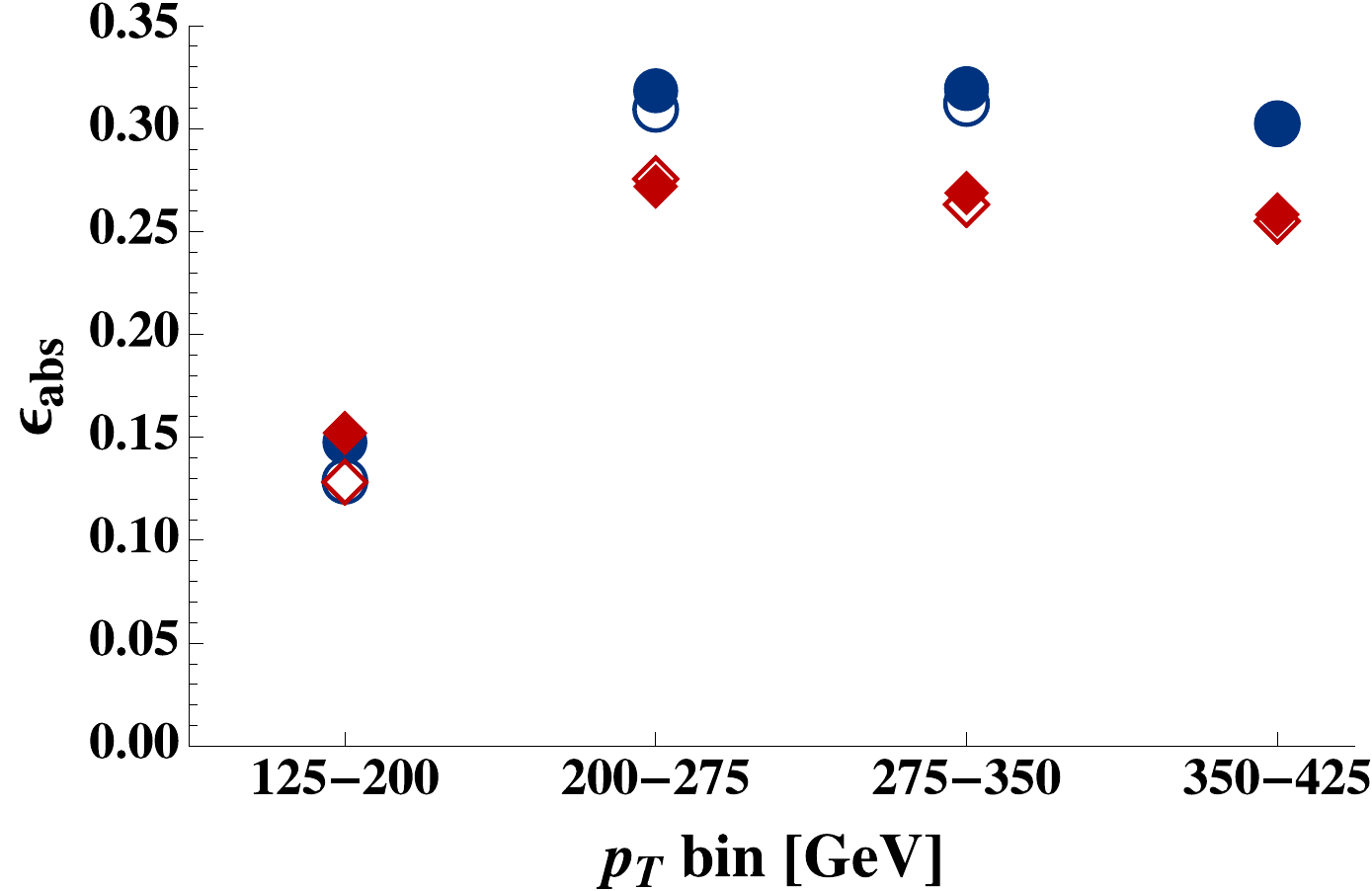}}

\subfloat[$\epsilon_{\text{fake}}$, tops]{\label{fig:AbsoluteEfficiencies:topfake}\includegraphics[width=0.45\columnwidth]{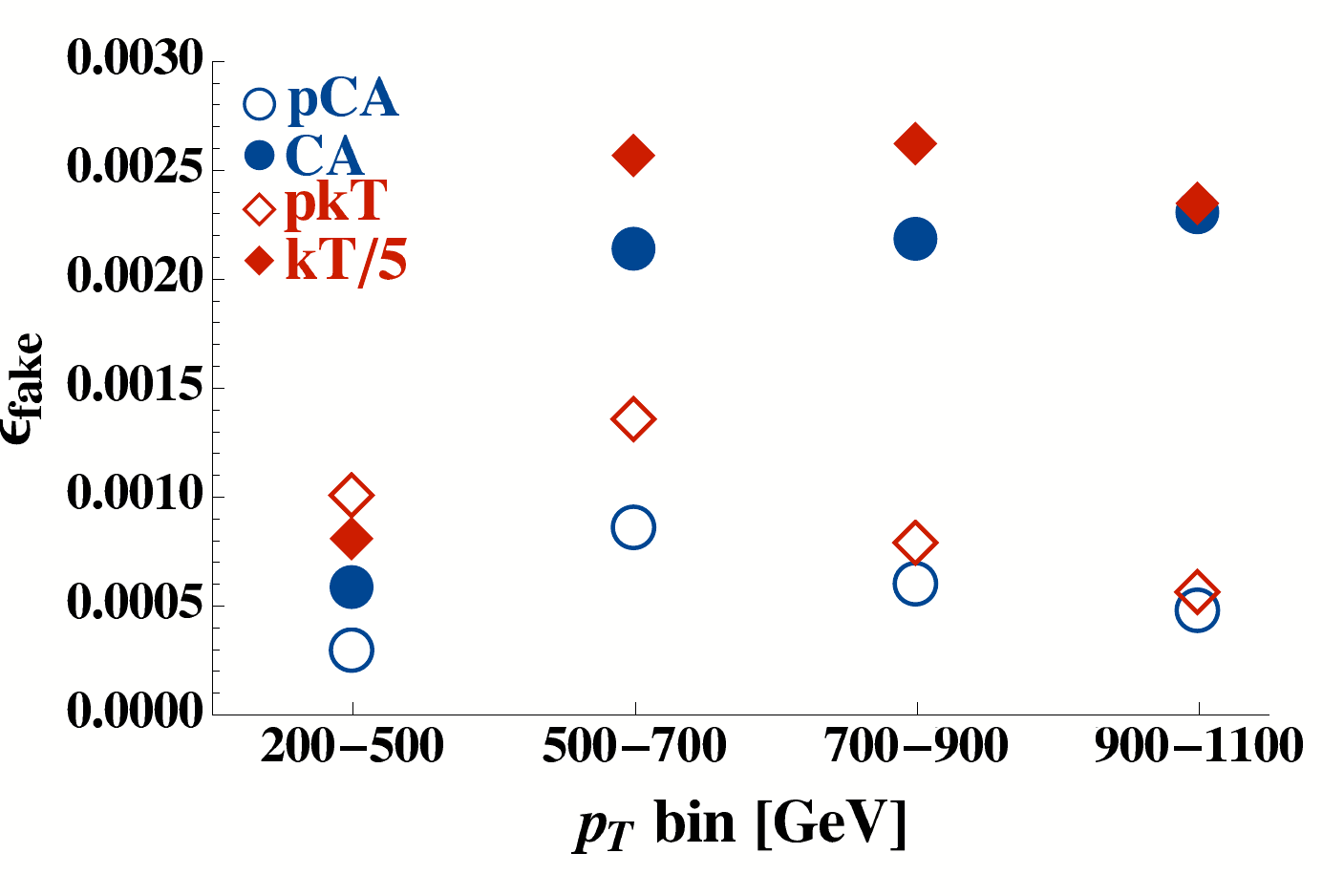}}
\subfloat[$\epsilon_{\text{fake}}$, $W$s]{\label{fig:AbsoluteEfficiencies:Wfake}\includegraphics[width=0.45\columnwidth]{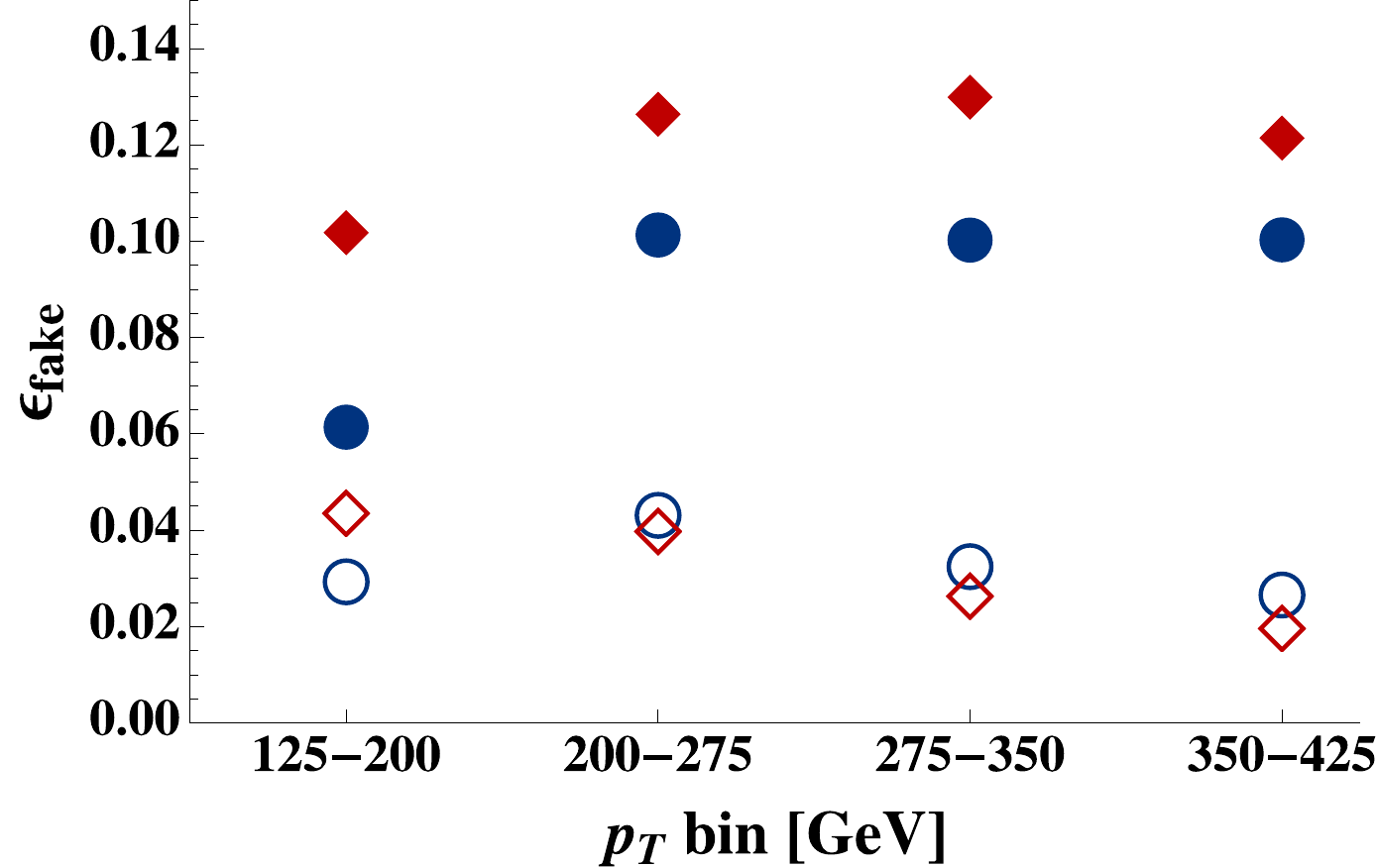}}
\end{center}

\caption[$\epsilon_\text{abs}$ and $\epsilon_\text{fake}$ vs. $p_T$ bin, for the CA and $\kt$ algorithms with and without pruning]{$\epsilon_\text{abs}$ and $\epsilon_\text{fake}$ vs. $p_T$ bin, for the CA and $\kt$ algorithms with and without pruning, using $D = 1.0$.  A ``p'' before the algorithm name denotes the pruned version.  The legend for figure (a) applies to figures (b) and (d) --- note the scale difference for $\kt$ jets in (c).}
\label{fig:AbsoluteEfficiencies}
\end{figure}

In Figs.~\ref{fig:AbsoluteEfficiencies:topabs} and \ref{fig:AbsoluteEfficiencies:Wabs}, we plot the absolute efficiency to identify tops and $W$s in the two signal samples for both algorithms, with and without pruning.  For the top sample, this efficiency $\epsilon_\text{abs}$ is the ratio
\[
\epsilon_\text{abs} \equiv \frac{\text{\# of top jets in the signal sample}}{\text{\# of parton-level tops in the $p_T$ range}}
\]
for each $p_T$ bin, with $\epsilon_\text{abs}$ defined analogously for the $W$ sample.  Because the substructure of the $W$ decay is much simpler than the top decay, with no secondary mass cut, the absolute identification efficiencies are similar between all algorithms.

The efficiency to find top quarks is only meaningful when compared to the fake rate for QCD jets to be misidentified as a top quark.  We define this fake rate as
\[
\epsilon_\text{fake} \equiv \frac{\text{\# of fake top jets in the background sample}}{\text{\# of unpruned jets in the $p_T$ range}}
\]
for each $p_T$ bin, and analogously for the $W$ sample.  In Figs.~\ref{fig:AbsoluteEfficiencies:topfake} and \ref{fig:AbsoluteEfficiencies:Wfake}, we plot $\epsilon_\text{fake}$ for tops and $W$s in the two background samples for both algorithms, with and without pruning.  The fake rate is significantly reduced for pruned jets compared to unpruned jets, for both the top and $W$ studies.  The decrease in absolute efficiency arising from using a narrow mass window is compensated by a correspondingly small fake rate for QCD jets.

For top quarks, the efficiencies shown in Fig.~\ref{fig:AbsoluteEfficiencies} can be compared with those given in Table~5 of \cite{Jetography} for several other top-finding methods.  Our highest $p_T$ bin is relevant for the comparison.  More than a few words of caution are in order, however.  Unlike the pruning-to-not-pruning comparisons we have presented so far, comparisons between methods using absolute efficiencies will depend on the details of the signal and background samples, as well as the details of the various cuts included in each analysis.  For example, the cuts we have used in this analysis are narrower than fixed mass window cuts used in other top-finding algorithms, and hence our top identification efficiency and background fake rate are both lower than described in other methods.  We intend to perform a more thorough comparison between different substructure approaches in a future work.

\begin{figure*}[htbp!]
\subfloat[$W$'s, CA vs. $\kt$]{\label{fig:CompareCAKT:W}\includegraphics[width=0.24\textwidth]{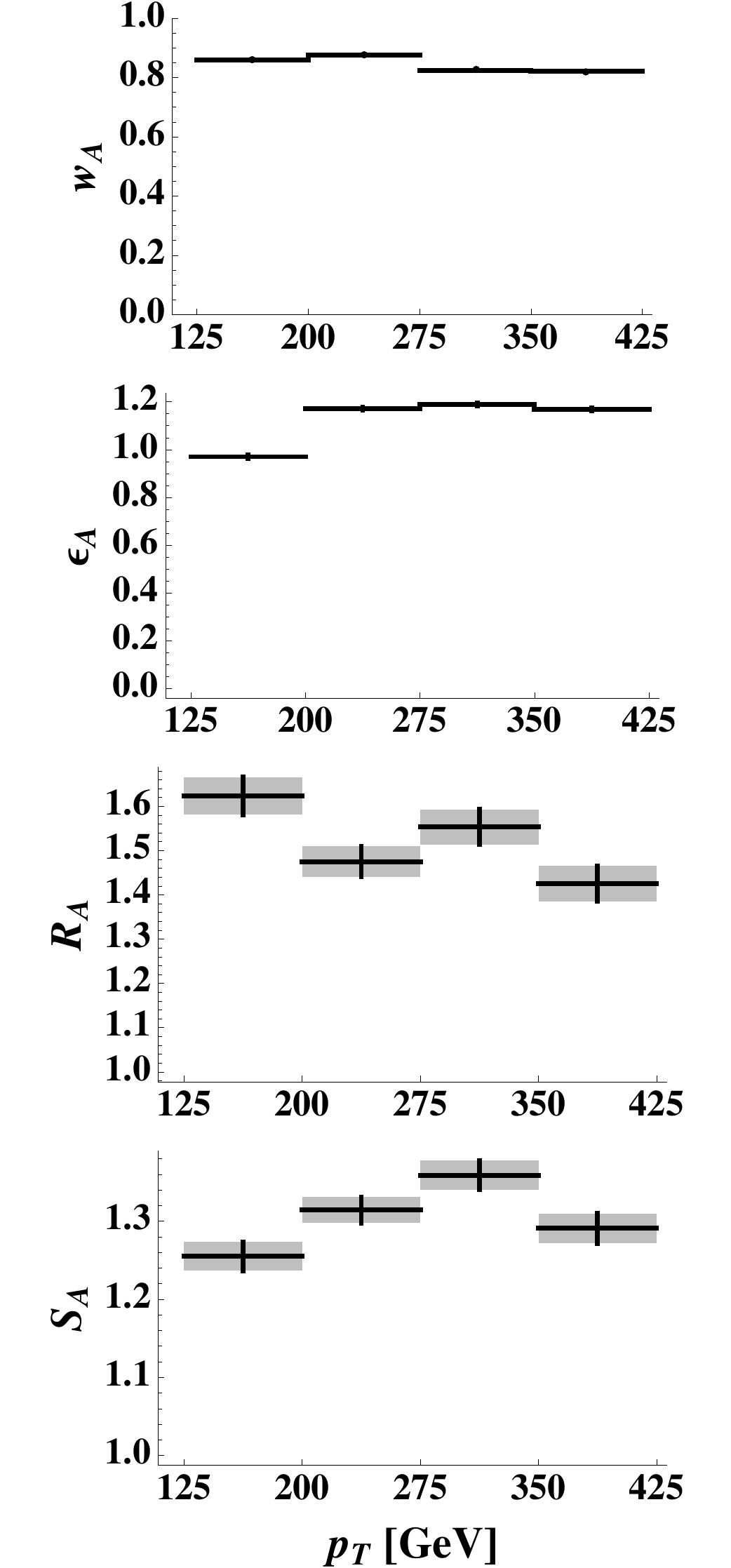}}
\subfloat[tops, CA vs. $\kt$]{\label{fig:CompareCAKT:t}\includegraphics[width=0.24\textwidth]{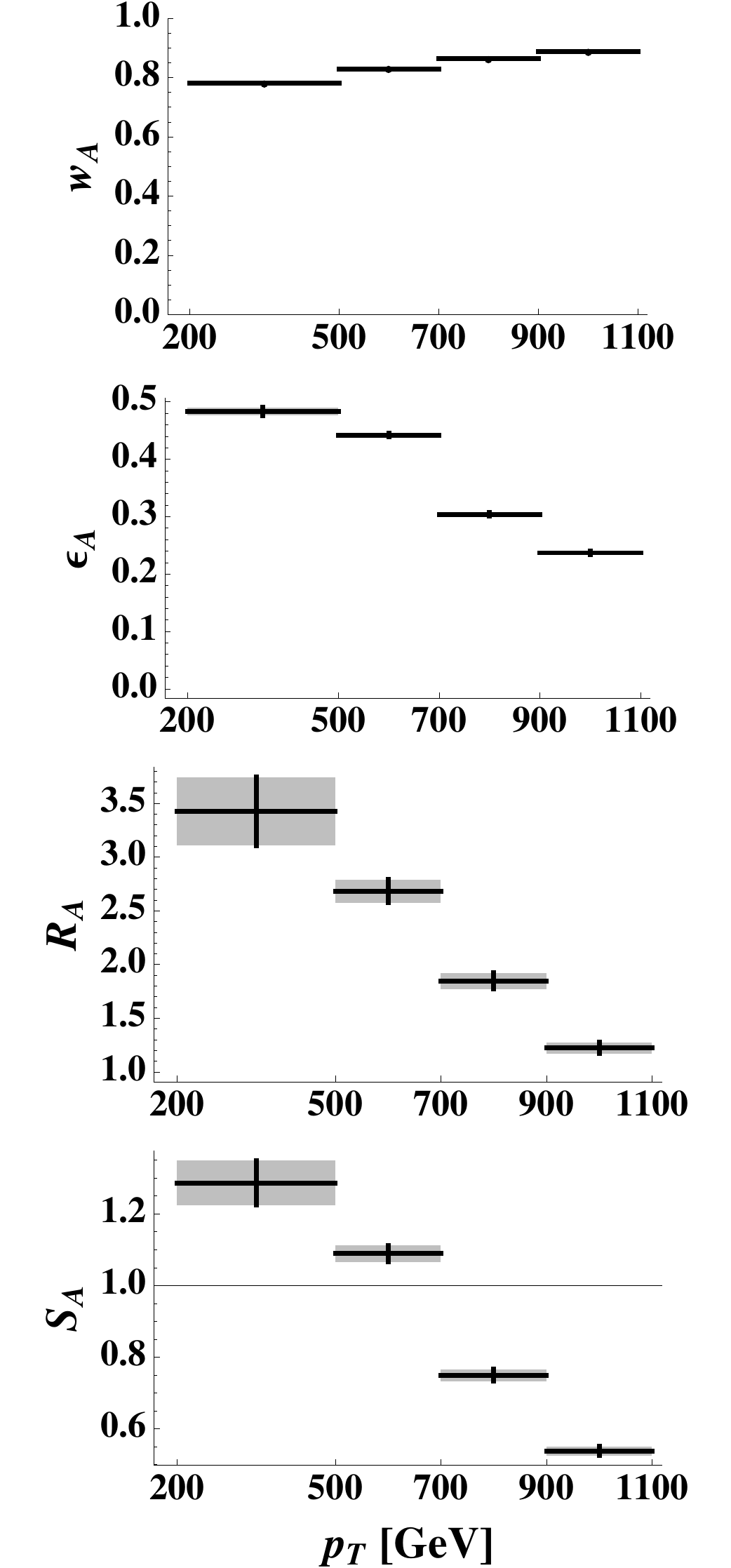}}
\subfloat[$W$'s, pCA vs. p$\kt$]{\label{ComparepCApKT:W}\includegraphics[width=0.24\textwidth]{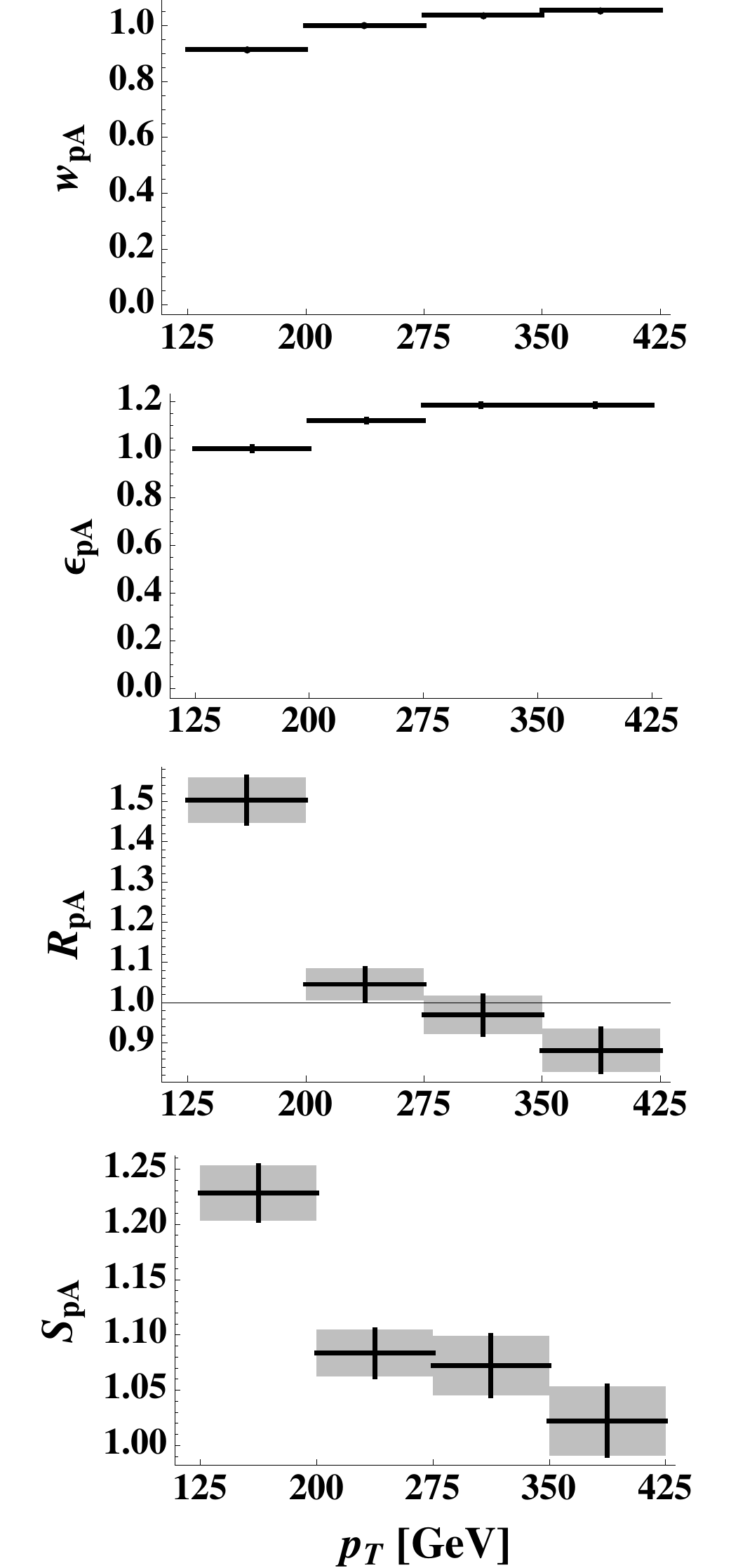}}
\subfloat[tops, pCA vs. p$\kt$]{\label{ComparepCApKT:t}\includegraphics[width=0.24\textwidth]{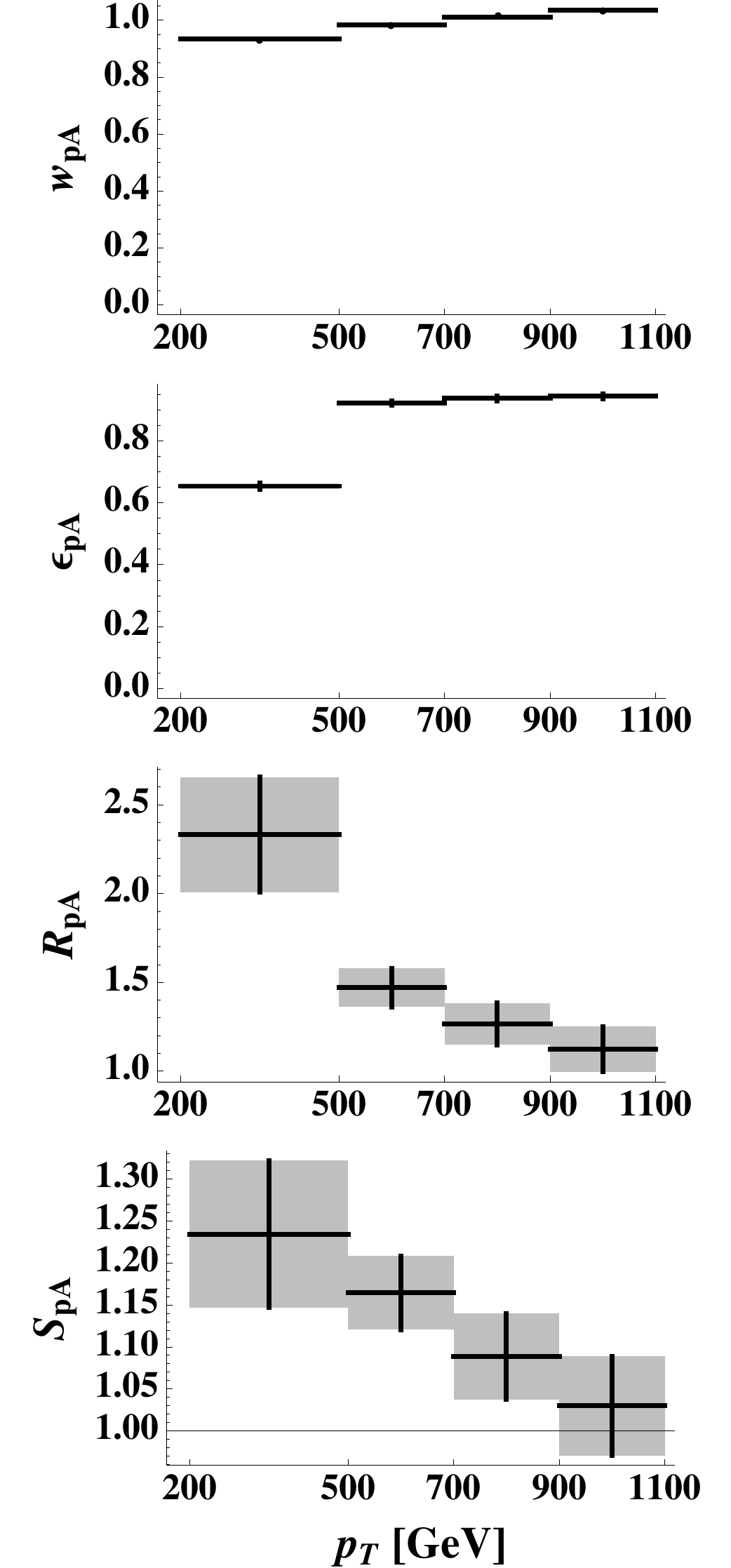}}
\caption[Relative statistical measures comparing CA to $\kt$ jets and pruned CA to pruned $\kt$ jets vs. $p_T$ for $W$'s and tops]{Relative statistical measures comparing CA to $\kt$ jets and pruned CA to pruned $\kt$ jets vs. $p_T$ for $W$'s and tops, using D = 1.0.  Statistical errors are shown.}
\label{fig:CompareCAKT}
\end{figure*}

\subsection[\textit{Algorithm Comparison}]{Algorithm Comparison}
\label{sec:prune:results:algComparison}

Throughout this paper, we have studied how pruning compares to not pruning for the CA and $\kt$ algorithms.  However, it is also of interest to study how the CA and $\kt$ algorithms compare, with and without pruning.  To do this, we use statistical measures $w_A$, $\epsilon_A$, $R_A$, and $S_A$ analogous to $w_\text{rel}$, $\epsilon$, $R$, and $S$.  For instance,
\[
R_A \equiv \frac{S/B \text{ from the CA algorithm with D = 1.0}}{S/B \text{ from the $\kt$ algorithm with D = 1.0}} .
\]
We will change the subscript to $pA$ to compare the pruned versions of the algorithms, e.g.,
\[
R_{pA} \equiv \frac{S/B \text{ from pruned CA with D = 1.0}}{S/B \text{ from pruned $\kt$ with D = 1.0}} .
\]
In Fig.~\ref{fig:CompareCAKT}, we plot the measures comparing CA to $\kt$ and pruned CA to pruned $\kt$ for both the $WW$ and $t\bar{t}$ samples.

These comparisons illustrate many of the effects that we have observed throughout this paper.  For the unpruned algorithm comparison, CA tends to have a much lower efficiency to identify tops than $\kt$.  As $p_T$ increases, CA performs more poorly relative to $\kt$, with the efficiency decreasing significantly.  This arises because the CA has a decreasing efficiency to identify the $W$ at high $p_T$, when the top quark becomes more localized in the fixed $D$ jet.  Pruning corrects for this, though the performance of CA relative to $\kt$ still decreases at high $p_T$.

The $WW$ sample is instructive because it lets us compare the effectiveness of pruning between CA and $\kt$ across a wide range in $p_T$.  For the unpruned algorithms, the performance of CA relative to $\kt$ is fairly consistent over all $p_T$, reflecting the fact that $W$ identification is simpler than top identification, with accurate mass reconstruction the only requirement.  However, when the jets are pruned, the performance of pruned CA relative to pruned $\kt$ improves in the smallest $p_T$ bin and worsens in the largest $p_T$ bin, as compared to the performance of CA versus $\kt$ for unpruned jets.  This skewing indicates that pruning is more effective for CA than $\kt$ at small $p_T$, where threshold effects are important, and more effective for $\kt$ than CA at large $p_T$.

\subsection[\textit{Detector Effects}]{Detector Effects}
\label{sec:prune:results:smearing}

\begin{figure}[htbp]
\subfloat[tops, CA jets]{\label{fig:CompareTopMassSmeared:CA}\includegraphics[width=0.5\columnwidth]{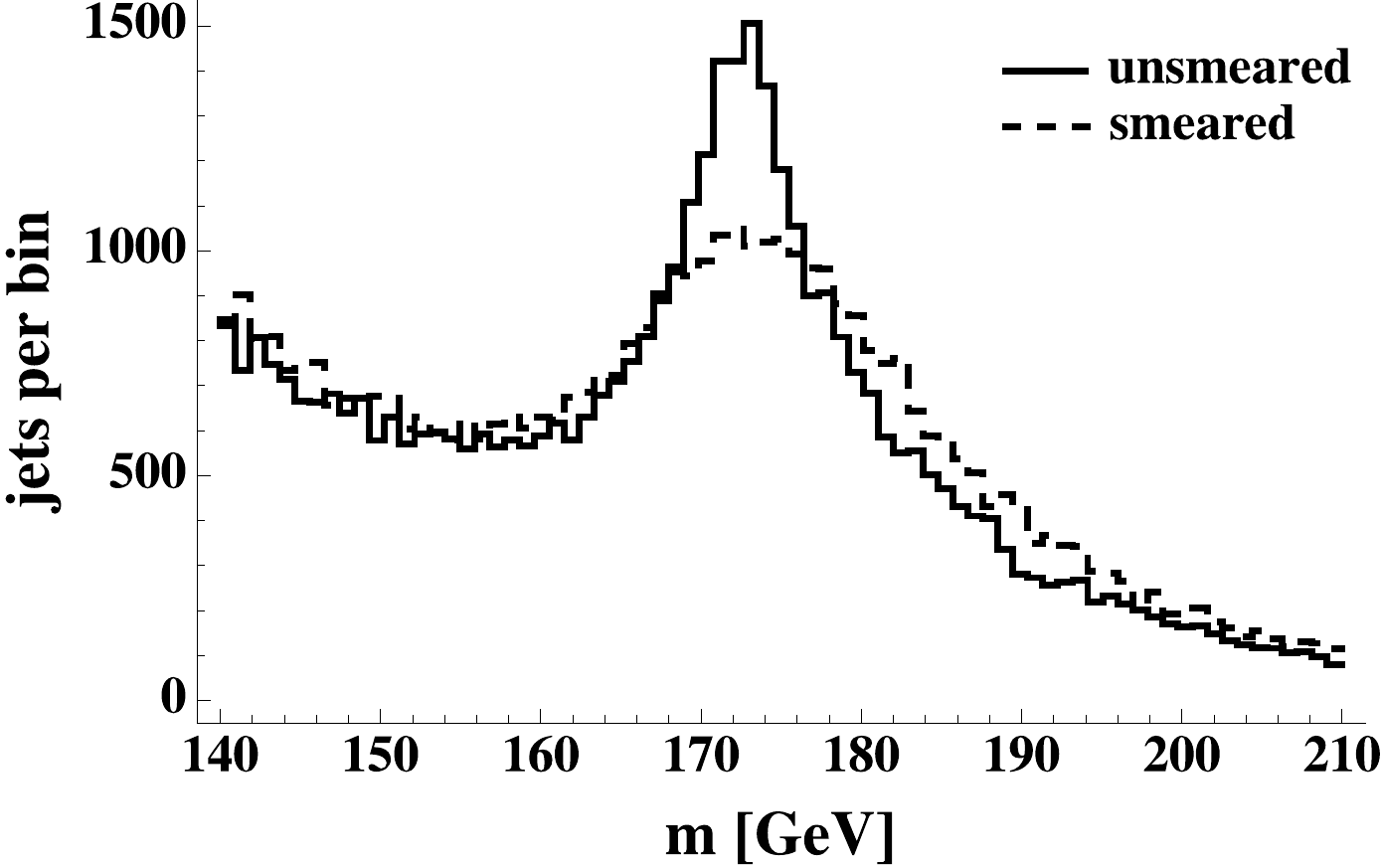}}
\subfloat[tops, $\kt$ jets]{\label{fig:CompareTopMassSmeared:KT}\includegraphics[width=0.5\columnwidth]{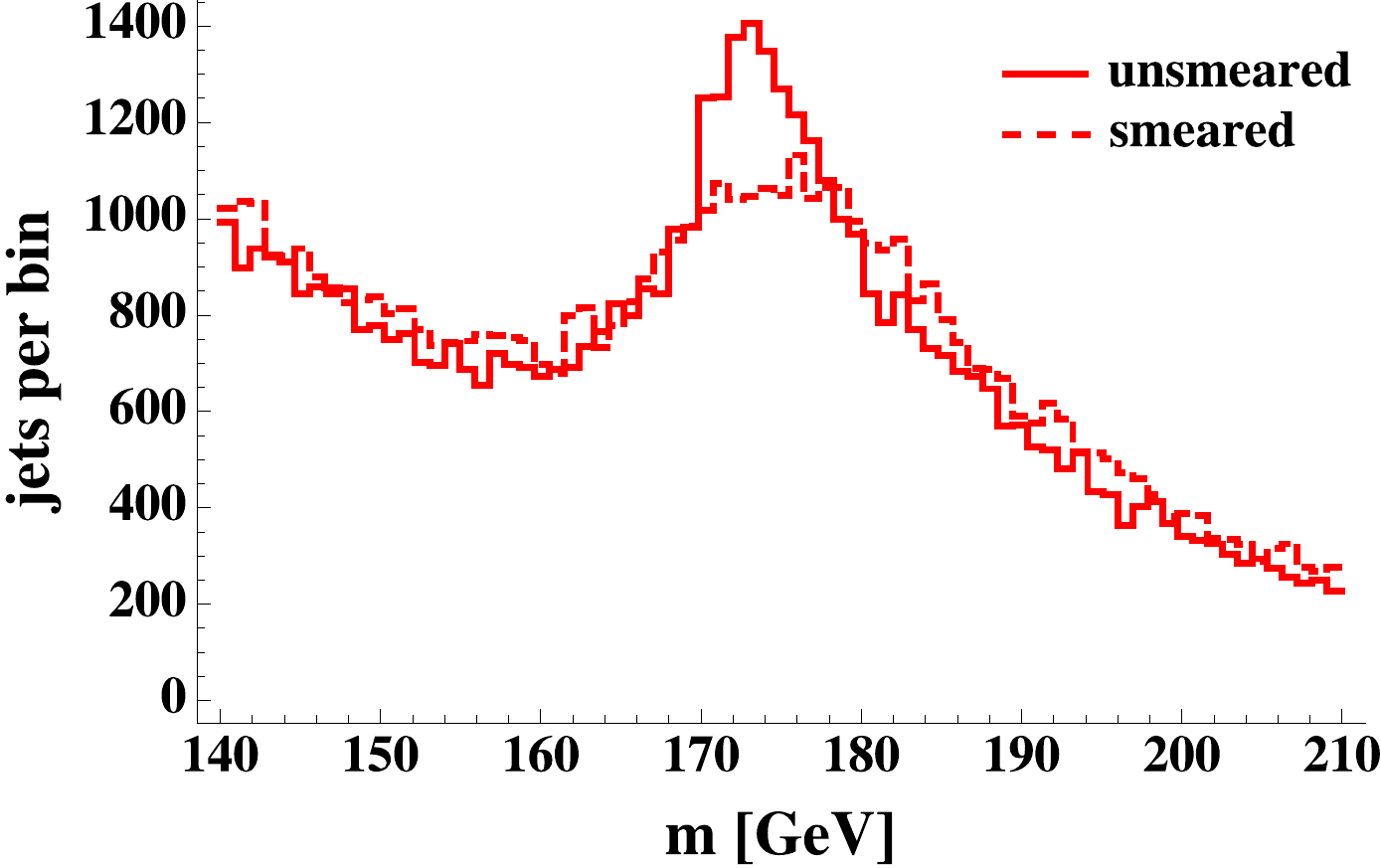}}
\caption{Distribution in jet mass for $t\bar{t}$ events, with (dashed) and without (solid) energy smearing.  The jets have $p_T$ of 200--500 GeV and $D = 1.0$, and there is no pruning.}
\label{fig:CompareTopMassSmeared}
\end{figure}

\begin{figure}[htbp]
\subfloat[tops, pruned CA jets]{\label{fig:CompareTopMassSmearedPruned:CA}\includegraphics[width=0.5\columnwidth]{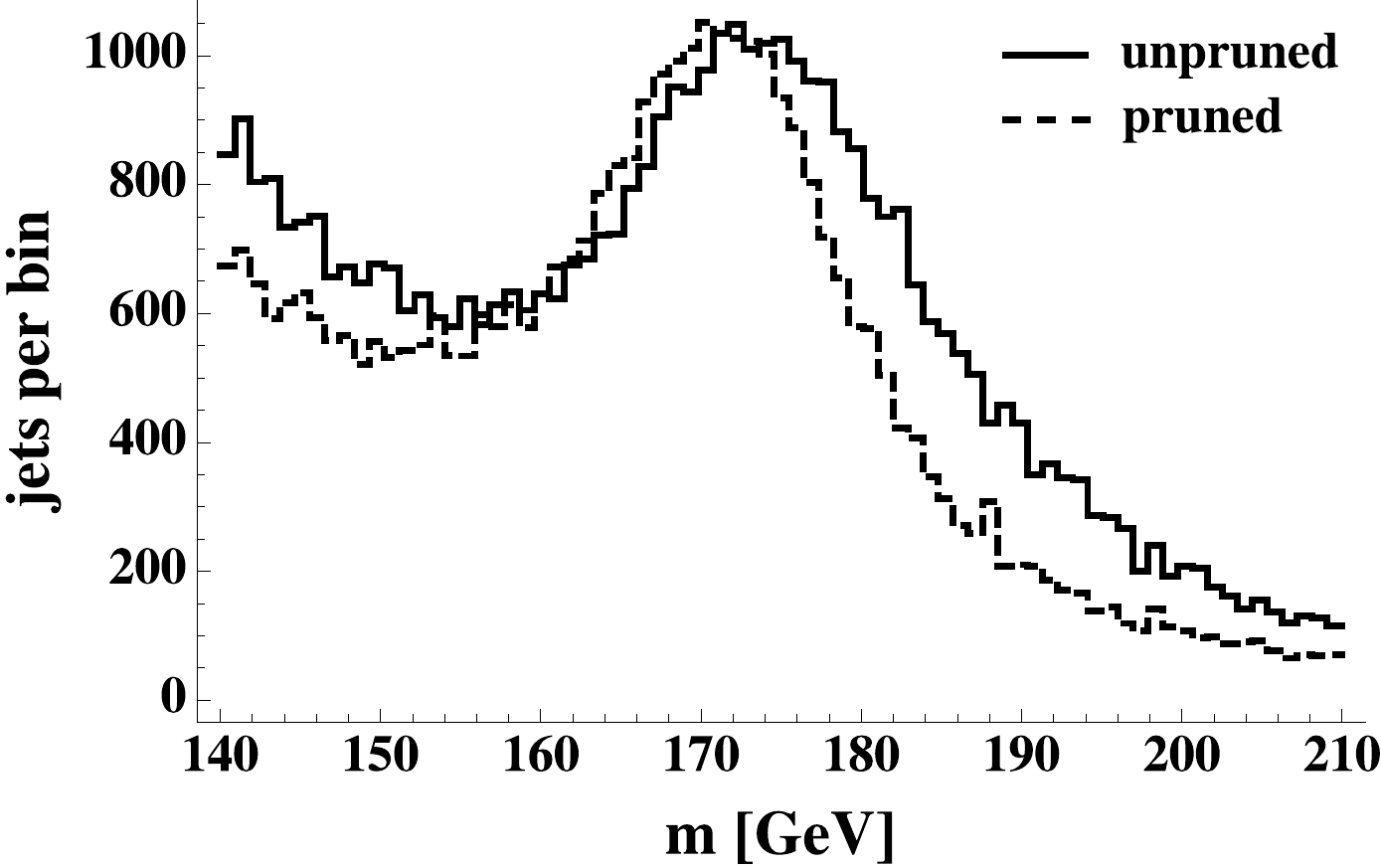}}
\subfloat[tops, pruned $\kt$ jets]{\label{fig:CompareTopMassSmearedPruned:KT}\includegraphics[width=0.5\columnwidth]{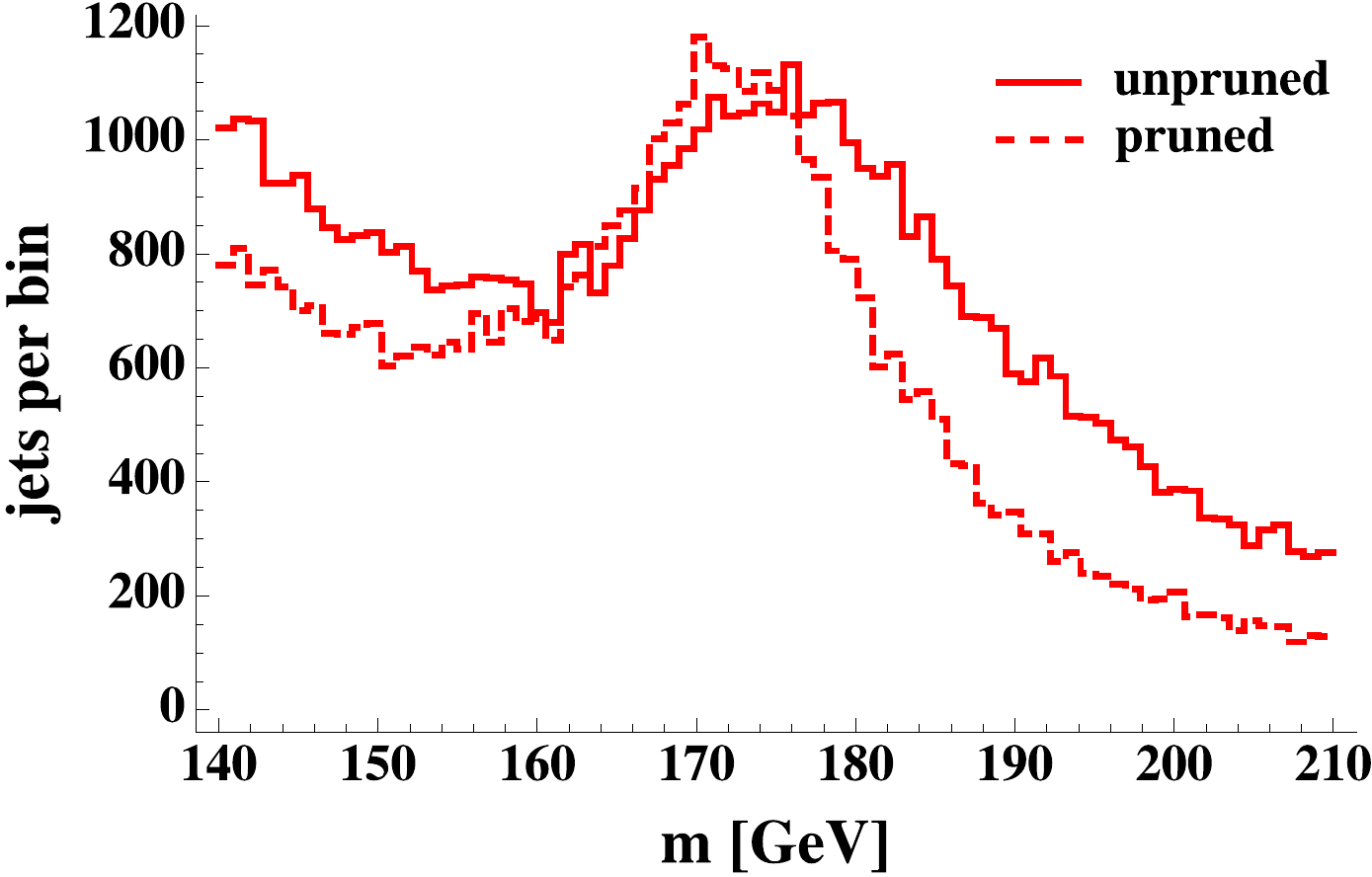}}
\caption[Distribution in jet mass for pruned and unpruned jets, for $t\bar{t}$ events with energy smearing]{Distribution in jet mass for pruned (dashed) and unpruned (solid) jets, for $t\bar{t}$ events with energy smearing.  The jets have $p_T$ of 200--500 GeV and $D = 1.0$.}
\label{fig:TopMassSmearedPruned}
\end{figure}

So far, no detector simulation has been applied to our events aside from clustering particles into massless calorimeter cells.  We now consider a technique that approximates the impact that detector resolution has on the effectiveness of pruning.  We modify our top and $W$ jet analyses by smearing the energy $E$ of each calorimeter cell with a factor sampled from a Gaussian distribution with mean $E$ and standard deviation $\sigma$ given by
\[
\sigma(E) = \sqrt{a^2 E + b^2 + c^2 E^2} .
\]
We consider a parameter set motivated by the expected ATLAS hadronic calorimeter resolution \cite{ATLAS:08.1}, $\{a,b,c\} = \{0.65, 0.5, 0.03\}$.  One obvious effect of the detector smearing is degraded mass resolution.  In Fig.~\ref{fig:CompareTopMassSmeared}, we show this effect by plotting the jet mass distribution for the $t\bar{t}$ sample in the first $p_T$ bin.  Even after smearing, however, pruning improves the jet mass resolution.  In Fig.~\ref{fig:TopMassSmearedPruned}, we plot the pruned and unpruned jet mass distribution for the $t\bar{t}$ sample in the first $p_T$ bin.  Note that because the QCD jet mass distribution is smooth, only the overall size of the sample in the mass window changes, so we do not plot these distributions.

\begin{figure*}[htbp]
\subfloat[$W$'s, CA jets]{\label{fig:VarypTSmeared:WCA}\includegraphics[width=0.24\textwidth]{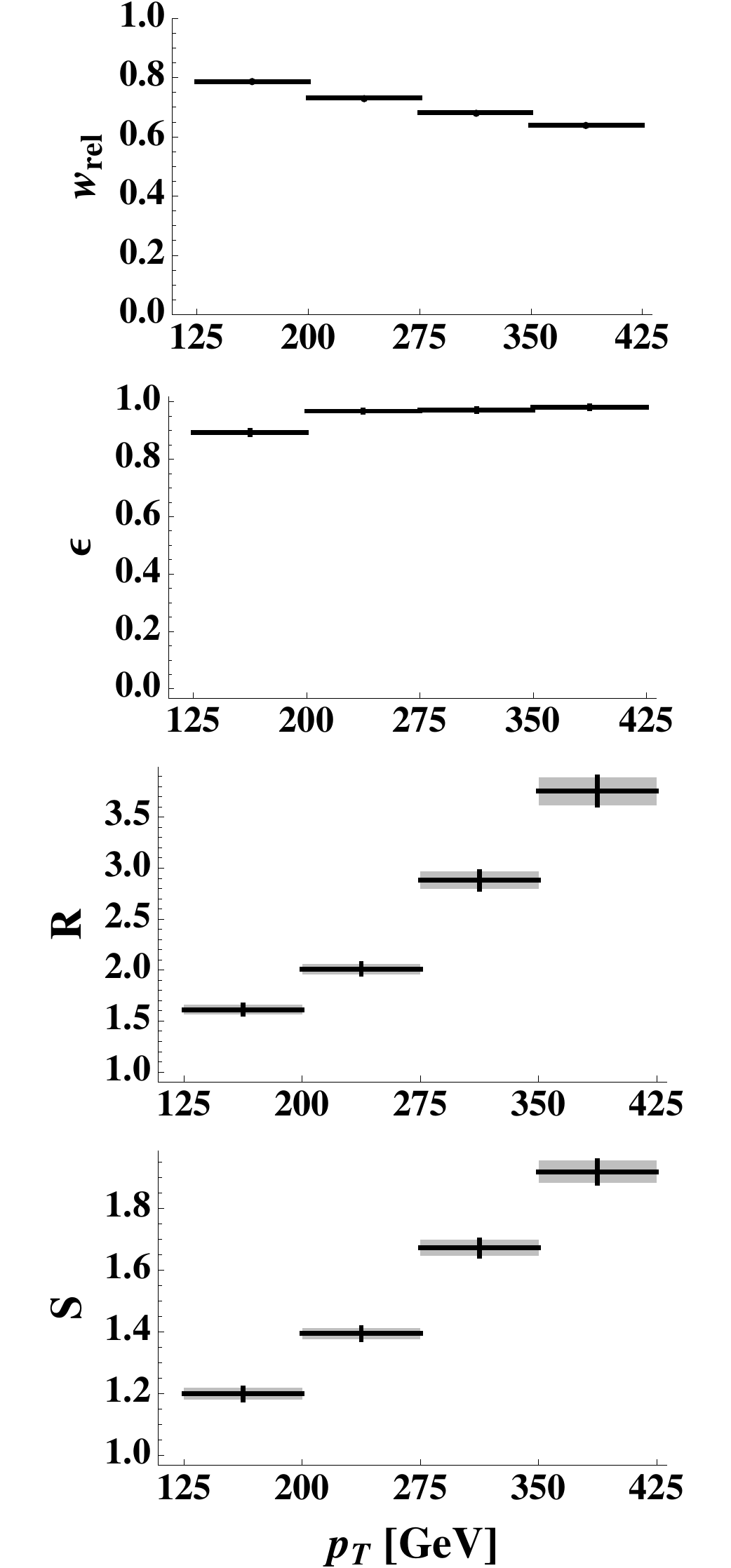}}
\subfloat[tops, CA jets]{\label{fig:VarypTSmeared:tCA}\includegraphics[width=0.24\textwidth]{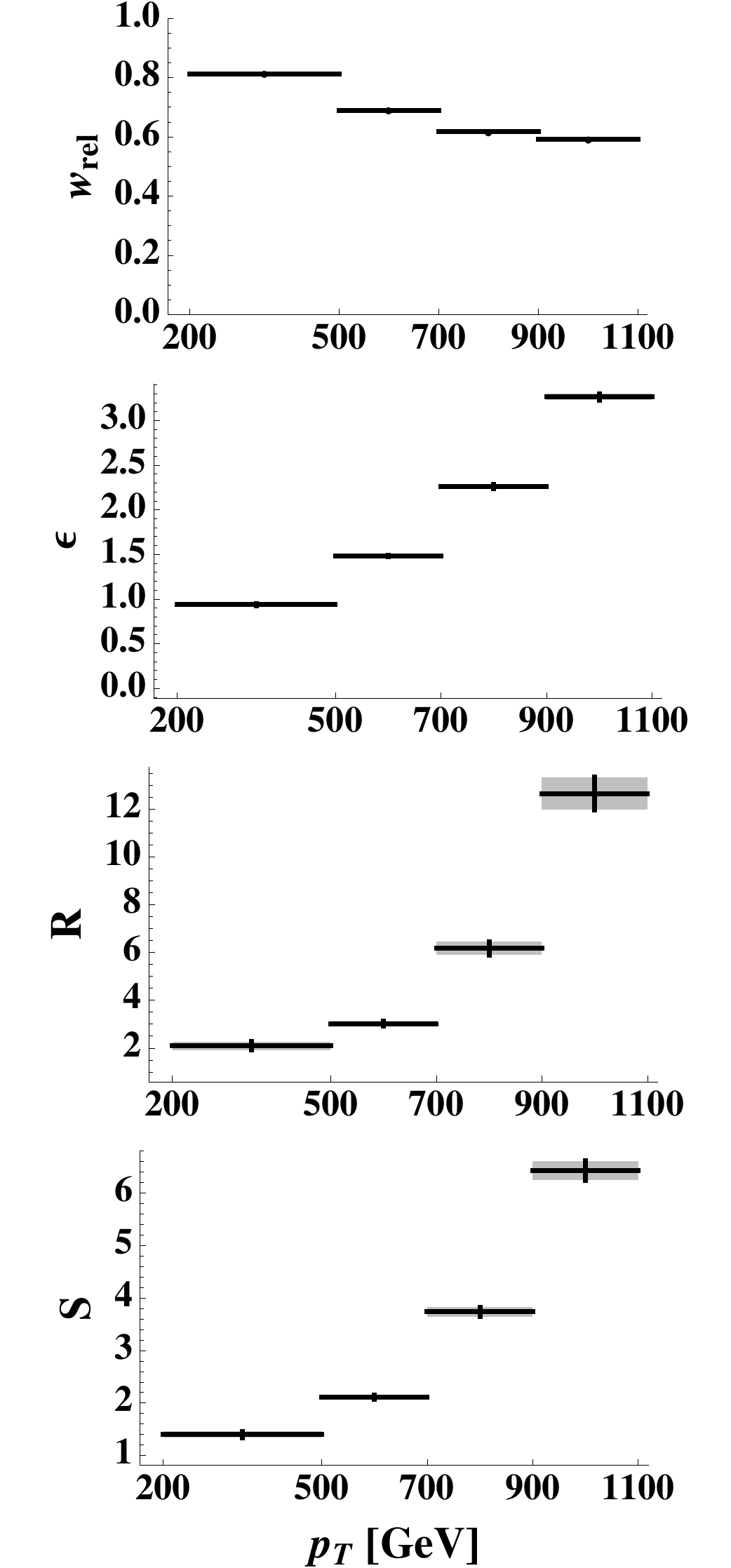}}
\subfloat[$W$'s, $\kt$ jets]{\label{fig:VarypTSmeared:WkT}\includegraphics[width=0.24\textwidth]{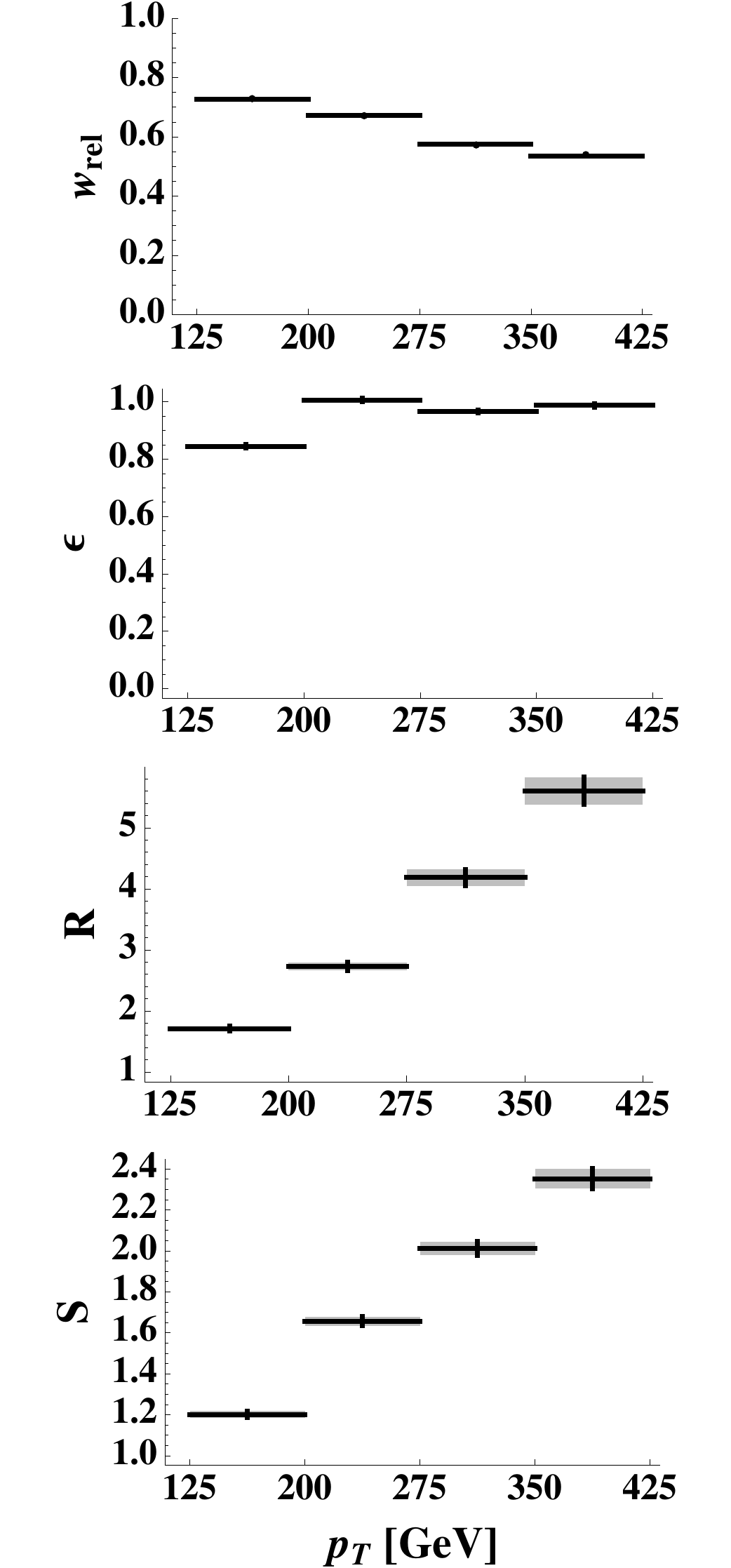}}
\subfloat[tops, $\kt$ jets]{\label{fig:VarypTSmeared:tkT}\includegraphics[width=0.24\textwidth]{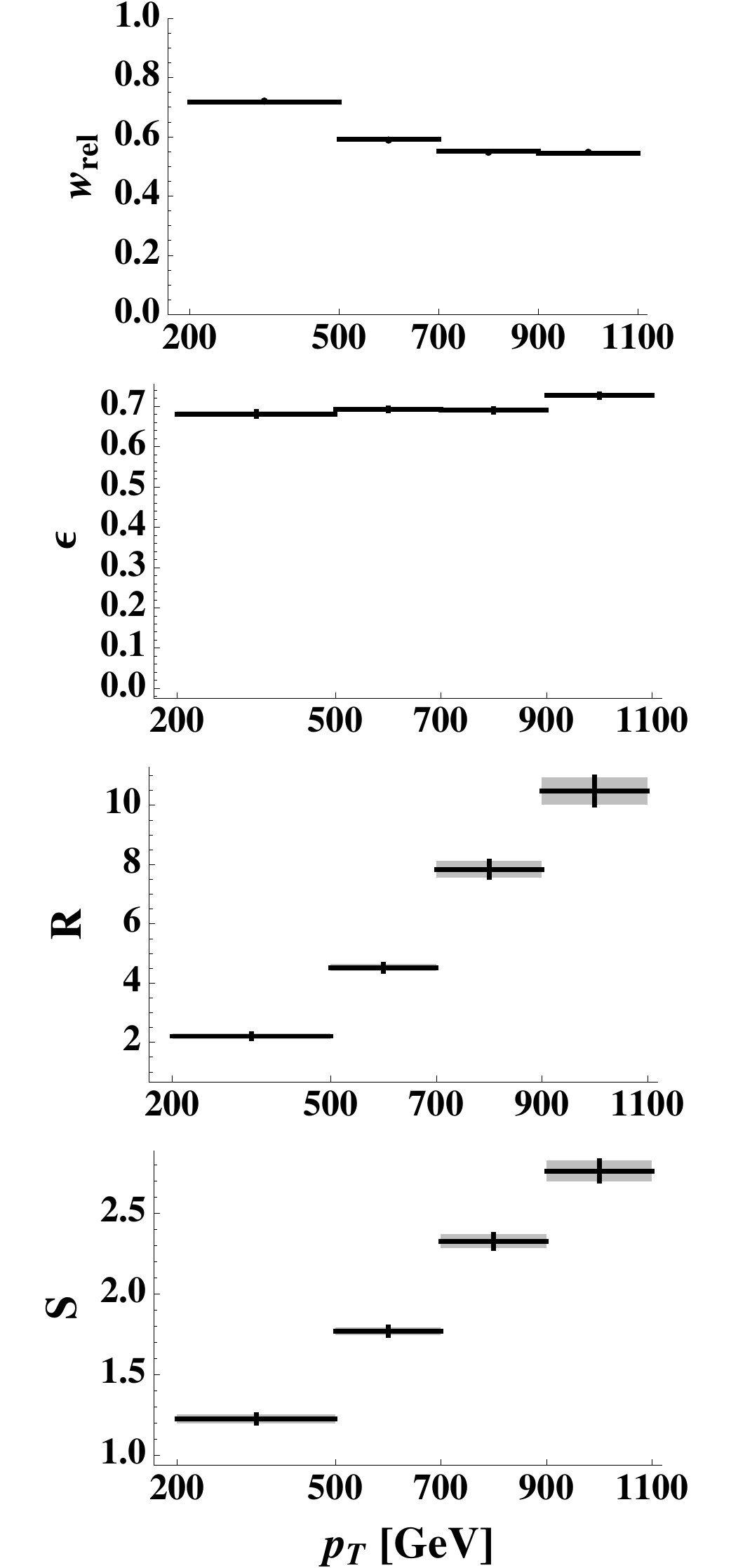}}
\caption[Relative statistical measures $w_\text{rel}$, $\epsilon$, $R$, and $S$ vs. $p_T$ for $W$'s and tops]{Relative statistical measures $w_\text{rel}$, $\epsilon$, $R$, and $S$ vs. $p_T$ for $W$'s and tops, using CA and $\kt$ jets.  Calorimeter cell energies are smeared as described in the text.  Statistical errors are shown.}
\label{fig:VarypTSmeared}
\end{figure*}

If Fig.~\ref{fig:VarypTSmeared}, we repeat the basic analysis of Sec.~\ref{sec:prune:results:fixedD}, applying the detector smearing described above.  This figure can be compared to Fig.~\ref{fig:VarypT} from the previous analysis, which plots the same measures when no energy smearing is used.  The improvements are very similar to those for unsmeared jets, good evidence that pruning may retain its utility in a more realistic detector simulation or in real data.

\section{Relation to other methods}
 \label{sec:prune:others}

To the best of this author's knowledge, the earliest paper addressing heavy particle identification with jet substructure is a 1994(!) paper by Michael Seymour \cite{Seymour94:1}, which considers $W$ finding in the context of a Higgs search.  In addition to a mass cut on the $W$ jet, cuts are applied on $\Delta R_{jj}$ and $\Delta R_{jJ}$, the angles from each subjet to the other and to the jet axis.  To reduce the effect of the underlying event, a reclustering --- \emph{filtering} --- procedure is applied.  Germinal forms of the concepts of jet areas and variable $R$ parameters are also discussed.  \textit{Sub sole nihil novi est.}

Interest in substructure perked up again several years later with two papers by Butterworth, et al. \cite{WW, SparticleSpectra}, which proposed using the variable $y p_{T\text{, jet}}^2 \equiv d^2_{ij}$.  This is the merging distance for the last step in the $\kt$ algorithm, expected to be $\cO(M^2)$ for the decay of a heavy particle in a single jet.  The restrictions on the branching kinematics that appear in subsequent substructure methods are all variations of this idea.

The ``mass-drop filter'' method proposed in \cite{FilteringHiggs} contained a novel feature:  instead of using the $\kt$ algorithm to construct substructure and cut on the final $d_{ij}$, the ``mass-drop'' step involved discarding elements of the substructure from the top down.  After first clustering with Cambridge-Aachen, the top-level merging is checked for a large mass drop (indicating that the mass of the merged jet is coming from the kinematics of a decay, not just a heavy subjet).  If this cut fails, instead of rejecting the jet, the lighter subjet is discarded and the search continues on the heavier subjet.  After discarding extraneous subjets in this way, the remaining jet is ``filtered'' in a method similar to \cite{Seymour94:1}: the constituents of the jet are reclustered with a smaller $R$ and the hardest three jets are kept.

The ``top-tagging'' method proposed in \cite{TopTagging} implements a variant of the mass-drop scheme for identifying the relevant substructure in a heavy particle jet, and by repeating the subjet-splitting procedure twice also achieves some of the success of filtering.  The top-tagging procedure also involves finding jets with CA and then looking backwards through the merging history for a large-scale splitting.  Branchings where one subjet is very soft are discarded; branchings where both subjets are soft or the subjets are too close together are ``irreducible'', and branchings were neither of these is true are valid splittings.  After looking for a valid top-level splitting, the procedure is repeated once on each subjet, resulting in up to four subjets.  The tagger requires that at least three be found.

Jet pruning can be thought of as a generalization of the subjet-identification step of top-tagging, but with two important distinctions.  First, pruning is run from the bottom up, with any merging failing a kinematic cut being discarded as a jet is built up.  Second, because the procedure is bottom-up, the kinematic comparisons are both local --- in top tagging, a subjet is too soft if $z \equiv p_T^j/p_T^\text{jet}$ is too small; in pruning the relevant value is $z \equiv p_T^{j}/p_T^{i+j}$, where $(i+j)$ represents the merger of subjets $i$ and $j$.

The discussions in this chapter have demonstrated that pruning is a \emph{generic} tool: it is successful on a variety of signals over a wide range in $m/p_T$, and does not require fore-knowledge of the number of subjets expected or any particle masses.  The precisely optimal parameters will depend on these details, but as we have seen (see Figs.~\ref{fig:VaryZcut} and \ref{fig:VaryDcut}) this dependence is not strong.

Another ``grooming'' method, ``jet trimming'' was proposed in \cite{Trimming}.  Trimming is similar to filtering, but instead of keeping some fixed number of subjets, subjets which contain at least some fraction of the jet's $p_T$ are kept.

Instead of using jet and subjet masses that have been improved by grooming, several studies have proposed other substructure variables to distinguish decays from heavy particles \cite{BoostedTops, TopYSplitter}.  These are generally based on the kinematics of the last few mergings in $\kt$ jets.

An alternative to considering the properties of subjets is to use jet shape or energy flow variables, as in \cite{Almeida:08.1, Almeida:2008tp, Chekanov:2010vc, Chekanov:2010gv, TemplateOverlap}.  An interesting idea, ``N-Subjettiness'', was described in \cite{NSubjettiness} that interpolates the number of subjets as a smooth jet shape.

Finally, one more difference between decays and QCD was exploited in \cite{Pull}: color flow.  The variable ``pull'' was shown to characterize the fact that a color singlet's decay products are color connected, whereas partons in a QCD jet are often color connected to other parts of the event.

While all of these methods rely on similar physics, it turns out that combinations of them can be even more useful \cite{SoperSpannowsky, MultivariateHiggs, WTagging}.  In fact, \cite{WTagging} found that it took 25 substructure variables to saturate the improvement in $W$ finding.

Which of the various substructure methods is ``best'' is a largely open --- and largely unanswerable --- question, with the answer presumably depending on the signal in question and potentially on details of the detector, luminosity, event topology, etc.  The bewildered and justifiably irritated experimentalist is perhaps to be consoled only with the assurance that tools such as the \FJ plugin mechanism and the \SJ package will make comparisons simple to perform.  In the example \SJ analysis in Appendix \ref{app:example}, I will show some comparisons between pruning and its relatives in top and $W$ finding.

\section{Using pruning}
 \label{sec:prune:using}
 
For readers interested in using pruning in their own analyses, the author has released a software package, \FP \cite{FastPrune}, to make this simple.  The \FP package includes two simple means of including pruning in a jet analysis.  A pruning \FJ plugin allows the user to find pruned jets (specifying the finding and pruning jet algorithms, as well as the $z_\text{cut}$ and $D_\text{cut}$ parameters) in precisely the same manner as for any other jet algorithm.  The latest version also includes a pruning tool for use with \SJ.  This tool takes as input jets found with some other algorithm, and returns the pruned versions.  In an analysis that compares pruned with unpruned algorithms, this saves the step of finding the unpruned jets twice (once for the unpruned analysis and once as the first step of the pruning \FJ plugin).  The use of these tools is described more fully in Chapter \ref{sec:tools}.


\section[\textit{Summary}]{Summary}
\label{sec:prune:conclusions}

In Chapter~\ref{sec:sub}, we demonstrated that a variety of systematic effects shape the substructure of heavy particles reconstructed in single jets.\footnote{This section, with small modifications, is taken from Sec.~IX of \cite{Pruning2}.}  We have identified regions in the variables $z$ and $\Delta R$ where individual recombinations are unlikely to represent the kinematics of a reconstructed heavy particle.  Specifically, soft, large-angle recombinations are unlikely to arise from the accurate reconstruction of a heavy particle decay, and are likely to come from QCD jets, uncorrelated radiation, or systematic effects of the jet algorithm.  For the CA algorithm, we have demonstrated that these soft, large-angle recombinations are a key systematic effect that shapes the substructure of the jet, in particular the final recombinations.

In this chapter we have presented a procedure, called pruning, that eliminates soft, large-angle recombinations from the substructure of the jet.  Using hadronically decaying top quarks and $W$ bosons as test cases, we have demonstrated that the pruning procedure improves the separation between heavy particles decays and a QCD multijet background.  We have motivated the parameters of the pruning procedure and demonstrated that they roughly optimize the improvements from pruning in our study for both top quarks and $W$ bosons.

Our studies on pruning have demonstrated many positive results of the procedure.  In a heavy particle search, the jet is sensitive to the parameter $D$, and if the value of $D$ is not well matched to the decay of a heavy particle then the ability to identify that particle in single jets is greatly reduced.  Our results indicate that pruning removes much of the jet algorithm's dependence on $D$.  Pruning shows improvements even when $D$ is adjusted to fit the expected decay of the heavy particle.  We have demonstrated that pruning largely removes the effects of the underlying event, as the underlying event mainly contributes soft, uncorrelated radiation that can be pruned away.  Additionally, we have shown that the results of pruning are robust to a basic energy-smearing applied to the calorimeter cells used to seed the jet algorithm.  Finally, we have quantified absolute measures of the pruning procedure that can be used to compare to other jet substructure methods.

It should be reiterated that pruning systematizes methods that have been proposed by other authors for specific searches.  Pruning should be applicable to a wide range of searches, and is intended to be a generic jet analysis tool.  We have detailed the ideas behind why pruning works and why it should be used, and presented an in-depth discussion of many of the physics issues arising when studying jet substructure.

\subsection[\textit{Future Prospects}]{Future Prospects}

The conclusions in this chapter, like those for any analysis technique not demonstrated on real data, must be taken cautiously.  This is especially true for studies like this one on jet substructure, where a majority of the work has been in exploring techniques that may --- or may not --- actually be useful in an experiment.  However, new techniques like jet substructure offer great promise.  All studies thus far indicate that jet substructure, and in general a more innovative approach to jets, will be a useful tool for understanding the physics in events with jets at collider experiments.

The most obvious and immediate application of pruning, and jet substructure tools in general, is in rediscovery of the Standard Model at the LHC.  As the LHC collects data from high-energy collisions, there will be an abundant sample of high-$p_T$ top quarks, and $W$ and $Z$ bosons with fully hadronic decays.  As these channels are observed using standard analyses, jet substructure techniques can be applied and tested.  These channels can also serve as key calibration tools for jet substructure methods applied in the search for new physics.

From the theoretical side, improvements in jet-based analyses can come from a variety of sources.  As calculations in perturbative QCD progress, they can be used to improve predictions for jet-based observables in QCD.  Improved Monte Carlo tools, such as the continued implementation of next-to-leading order matrix elements and better parton showers, will lead to more accurate studies and a better understanding of jet physics.  Additionally, the SCET framework will improve our understanding of QCD jets.  As SCET is adapted to describe a wider variety of event topologies and realistic jet algorithms are implemented in the effective theory, it can be used to calculate resummed predictions \cite{Bauer:2006qp, Stewart:2009yx, Cheung:2009sg} for jet-based observables and accurately describe processes that are difficult to access with fixed-order perturbative QCD.  Jets will likely play a central role in new physics searches at the LHC, and a better understanding of jets and jet substructure can aid in the discovery process.




\graphicspath{{chapter6/graphics/}}
 
\chapter{Tools to Study Jet Substructure}
\label{sec:tools}
 
An ``event'' at a hadron collider typically consists of very many\footnote{Actually, without adding some kinematic restrictions, ``how many particles are observed'' is not a well-defined quantity!  The $t\bar t$ Monte Carlo events used in Chapters \ref{sec:sub} and \ref{sec:prune} typically have $\sim 500$ outgoing particles from Pythia, $\sim 250$ particles with $|\eta| < 5$ and $p_T > 0.5$ GeV, and $\sim 150$ calorimeter cells with $p_T > 1$ GeV.  This includes the effects of the underlying event, but not pile-up.} outgoing particles, mostly electrons, photons, and hadrons.  Given this multiplicity, calculating cross sections differential in the momenta of all outgoing particles is clearly intractable.  We can perform analytic calculations for suitably inclusive quantities, such as the total cross section for a specific process, but we certainly cannot make fine-grained predictions of, say, jet substructure.  In addition, if we want to simulate the effects of the detector, we need a way to produce realistic, high-multiplicity events, either with an appropriate distribution or with known weight factors.  This is the task of a Monte Carlo event generator.

The output of an event generator is a list of particles and their four-momenta.  The next step in a realistic analysis is to simulate the output of a particle detector such as ATLAS or CMS, given a specific particle-level event.  This can involve detailed simulation of the interaction of particles passing through the various materials of the detector as well as instrumental response, or much cruder approximations where particles are grouped together into ``calorimeter cells'' and assumed to be measured with some resolution.

If the final state involves jets, detector outputs such as calorimeter cells must be clustered into jets.  As we have seen this can be done in a variety of ways, and in general an analysis will involve multiple jet algorithms and ``jet manipulations'' such as filtering or pruning.  Being able to test and compare multiple jet tools at this step is essential.  Finally, having found jets, as well as other final state objects such as isolated leptons, a specific physics analysis can be performed.

In the following sections I review the individual steps in performing a physics study using Monte-Carlo-simulated data, noting at each step the various software packages available for that task.  In discussing jet finding and analysis, I will pay particular attention to the \FJ and \SJ packages; I have made significant contributions to the development of the latter.


\section{Analysis chain overview}
\label{sec:tools:chain}

\subsection{Event generation: ME/PS/Matching}

A ``complete'' Monte Carlo event generator can be broken into three parts, typically performed by separate computer programs.  First, a low-multiplicity ``parton-level'' event is generated, and given a weight corresponding to the exact matrix element squared for that process.  Processes with hadrons in the initial or final state ($\ee \to$ jets or $pp \to ZH$, for example), are treated as involving some fixed number of quarks or gluons ($\ee \to q\bar q$ or $gg \to ZH$, for example).  The matrix elements are calculated to some fixed order in $\alpha_s$, often just to tree level; logarithmic resummation can also be included at this step.  To ensure that events have finite weights, kinematic cuts on the outgoing particles are typically required.

To produce the multiplicity of particles seen in the detector, the matrix-element-level generator must be combined with a ``parton shower Monte Carlo", which takes outgoing partons (quarks and gluons) and iteratively radiates gluons and splits gluons into $q\bar q$ pairs until all particles have energy (or some other scale such as virtuality) below some fixed lower scale.  This process typically assumes that emissions are independent of each other, and approximates the matrix elements for gluon radiation and splitting, making sure to be accurate in the limit that a splitting is soft or collinear (the singular limits of the matrix element).  This obviously does not reproduce the QCD shower exactly, but is correct to leading logarithmic precision.

Without some care, the matrix-element-level generation and the parton shower will not cover all of phase space exactly once.  Consider an event sample that is represented at the parton level as $\ee \to q \bar q g$.  Each of the quarks will radiate gluons as part of the parton shower; one of these gluons could end up with the same momentum as the gluon produced by the matrix element generator unless we impose some sort of restriction one or both Monte Carlos.  One solution is to require that partons in a matrix-element-level event be well-separated by some criterion ($\kt$ distance, say), and that parton shower emissions can never be separated by this much.  Generically, a method for combining matrix-element generators with parton shower generators is called a ``matching procedure''.

One final step is necessary before we can send our events to a detector simulator.  A parton shower produces a multiplicity of quarks and gluons, but of course these are not the particles we observe.  QCD is confining, and the outgoing quarks and gluons, after showering down to some low energy of order $\Lambda_\text{QCD}$, will re-arrange into bound states --- hadrons.  This is a fundamentally non-perturbative process, and the best we can do is model it and fit the model to data.  Such a hadronization procedure is typically included at the end of a parton shower Monte Carlo.

In hadron collisions, we must also consider the initial state.  First, rather than generate events with incoming partons of fixed energy, we must include partons with arbitrary fractions of the incoming hadrons' momenta and convolute with the probability that, at the energy scale involved, we find two partons with those two momentum fractions.  These probabilities are known as ``parton density functions''.  They cannot be calculated perturbatively, although their renormalization group flow can, so after measuring their form at some energy scale we can predict them at any other scale.  We must also, in analogy with the showering of outgoing partons, consider radiation from the incoming partons (``initial state radiation'').  Finally, the ``beam remnants'' --- the valence and sea quarks from the incoming hadrons that did participate in the main interactions --- can themselves interact.  The output of these interactions is known as the ``underlying event''.  To the extent that these ``multiple interactions'' are independent of the rest of the event, they will typically involve low (transverse) energy scales, since in the absence of analysis cuts (i.e., ``minimum bias'') all events typically involve low scales, so having two high-energy interactions in a single collision is rare.  However, note that if the final state is not a color singlet ($q\bar q \to g^* \to t\bar t$, e.g.) the underlying event \emph{cannot} be completely independent of the primary interaction due to color connections.  In fact, it is observed that the underlying event is independent to a good approximation \cite{UE}.  All of these effects can either be incorporated into the parton shower Monte Carlo or generated independently.  Note that the outgoing quarks and gluons from initial state radiation and the underlying event must themselves shower and hadronize.

A fairly complete database of Monte Carlo event generators is available at the CEDAR HepCode page \cite{HepCode}.



\subsubsection{Monte Carlo programs used in this work}

In the studies discussed in this thesis, we use the \prog{MadGraph/MadEvent} package \cite{MadGraph} to generate matrix-element-level events.  For the $pp$ studies, MLM matching is used.  Both MLM \cite{MLM} and CKKW \cite{CKKW} matching are included in the \prog{MG/ME}-\prog{Pythia} interface included with the \prog{MG/ME} package.  We use \prog{MG/ME}'s included \prog{Pythia} package (version 6.4 \cite{Pythia6}) to shower incoming and outgoing partons, as well as generate multiple interactions (the underlying event).  \prog{Pythia }also models the hadronization of partons.


\subsection{Detector simulation}

After generating particle-level events, sets of output particles should be passed to some kind of detector simulator.  Very detailed simulators of the detectors for all major particle physics experiments exist (see, e.g. \cite{Geant4}), but these are typically overkill for speculative theoretical studies.  For these, a general purpose simulator that captures the broad features of calorimetry is sufficient: \prog{PGS} \cite{PGS} and \prog{Delphes} \cite{Delphes} are two examples.

In the studies described in this thesis, we have used our own crude detector simulation, which rejects invisible and outside-of-detector particles, clusters particles into calorimeter cells, isolates leptons, and imposes a minimum $p_T$ cut on calorimeter cells.  We have also incorporated Gaussian smearing of calorimeter cell energies to roughly model detector resolution effects.


\subsection{Jet finding and analysis}

To study events with jets, a jet algorithm must be applied to the outputs of the detector simulation, typically calorimeter cells.  An enormous variety of such algorithms exist (see \cite{Jetography} for a survey), all of which have been implemented in software.  Historically this was done individually by experimental groups and theorists, occasionally in subtly different ways (see, e.g., the discussion of seeded cone algorithms in \cite{Seeds}).
Now the \FJ package \cite{FastJet}, is fast becoming standard among jet practitioners.  \FJ implements most, if not all, commonly used jet algorithms, and through a plugin mechanism can be extended to implement other algorithms as well.  Many \FJ algorithms incorporate insights from computational geometry, making them faster than previous implementations.  More details on \FJ are given in Sec.~\ref{sec:tools:fj}.

Another tool for studying jets has recently emerged: \SJ (\cite{FamousJetReview}, \cite{SpartyJetWebsite}).  \SJ incorporates jet finding with \FJ and adds several useful layers of input, analysis, and output, partially based on \prog{ROOT} \cite{ROOT}.  Many analysis components can be glued together with simple Python scripts.  More details on \SJ are given in Sec.~\ref{sec:tools:sj}.

In the studies described in this thesis, we have used \SJ for jet finding and analysis; the jet algorithms were implemented in \FJ via the \SJ wrapper.  The plots new to this thesis were all generated from the SpartyJet GUI.


\section{FastJet}
\label{sec:tools:fj}

\FJ is the new standard in jet finding.  This section gives a brief overview of its capabilities.  A more detailed description of \FJ's features, use, and implementation is given in the official \FJ manual \cite{FastJetManual}.  At the end of this section I will also discuss the \FJ plugin I have written to implement jet pruning in a simple and standard way.

\subsection{Overview}

The achievement of \FJ is two-fold:  First, to standardize the implementation of jet algorithms between and among experimentalists and theorists, eliminating the possibility of subtle and hidden discrepancies.  Second, to bring together in one place advances in jet finding technology, for example introducing the technique of Voronoi diagrams (see the discussion and references in \cite{FastJet}) for efficient distance finding for very large numbers of particles.  \FJ also includes several implementations of ``jet area'' finding \cite{JetAreas} for arbitrary jet algorithms, which I will not discuss.

\FJ is a package of \prog{C++} libraries that implement jet finding and related tools.  The primary classes are:

\begin{lstlisting}
class fastjet::PseudoJet;
class fastjet::JetDefinition;
class fastjet::ClusterSequence;
\end{lstlisting}

\code{PseudoJet} is basic four-vector class, adding a pair of indices: one for cluster ordering and one left to the user.  \code{JetDefinition} collects the full specification of a jet definition, including an algorithm like $\kt$, $R$ and any other parameters necessary, and a recombination scheme.\footnote{A recombination scheme specifies how to make \code{PseudoJet p} from merged \codes{PseudoJet} \code{p1} and \code{p2}.  To combine four momenta, by far the most common is the ``E-scheme'', where $p = p_1 + p_2$.  The indices on a \code{PseudoJet} allow expanded schemes where, for example, the user index tracks the parton flavor which the recombination scheme can be designed to propagate.}  The actual business of jet finding is done by the \code{ClusterSequence} class.  Given a list of \codes{PseudoJet} and a \code{JetDefinition}, a \code{ClusterSequence} constructs the set of final jets.  For recombination algorithms, a merging history is also constructed.\footnote{Actually, \FJ stores a merging history for all algorithms, including cone-type algorithms, but for the latter the history is not meaningful.}  Both the jets and the clustering history can be accessed with a variety of methods:

\begin{lstlisting}
// Set up input particles
vector<PseudoJet> inputs;
// ...   fill this vector somehow

// Set up a jet definition
JetAlgorithm algorithm = kt_algorithm;
double R = 1.0;
RecombinationScheme recomb_scheme = E_scheme;
Strategy strategy = Best;
JetDefinition jet_def(algorithm, R, recomb_scheme, strategy);

// Get jets and merging history
ClusterSequence cluster_seq(inputs, jet_def);

// *****  Access methods **************

// inclusive jets
vector<PseudoJet> inc_jets = cluster_seq.inclusive_jets (pt_min);
// exclusive jets, with a dcut
vector<PseudoJet> exc_dcut_jets = cluster_seq.exclusive_jets (dcut);
// exclusive jets, stop at N jets
vector<PseudoJet> exc_jets = cluster_seq.exclusive_jets (Njets);

// get constituents of a given jet
PseudoJet jet = inc_jets[0];
vector<PseudoJet> consts = cluster_seq.constituents(jet);

// look at substructure
PseudoJet child, parent1, parent2;
child = jet;
while (cluster_seq.has_parents(child, parent1, parent2)) {
  child = parent1;
}
// child is now an input particle from jet
//   child = parent1(parent1( ... parent1(jet) ... ))

// Can also go the other way:
PseudoJet pj = child;
PseudoJet new_child;
if (cluster_seq.has_child(pj, new_child)) {
  // ...
}
// new_child is set to pj's child

\end{lstlisting}

Note that in \FJ language, two ``parent'' pseudojets merge into a ``child'' pseudojet, in contrast to the parent/daughter language used in Sections~\ref{sec:sub} and \ref{sec:prune}.


\subsection{Built-in versus plugin algorithms}

The set of algorithms that run natively in \FJ are shown in Table~\ref{table:FJalgs}.  Note that all of the native algorithms are specific cases of the generalized $\kt$ algorithm for either hadron or $\ee$ collisions.

\begin{table}[htbp]
\begin{center}
\begin{tabular}{|r||c|c|c|}
\hline
Algorithm & Name & $d_{ij}$ & $d_i$ \\
\hline
\multicolumn{4}{|c|}{$pp$} \\
 \hline
$\kt$ & \code{kt\_algorithm} & $\min (p_{Ti}^2, p_{Tj}^2) \Delta R_{ij}^2/R^2$ & $p_{Ti}^2$ \\
\hline
Cambridge/Aachen & \code{cambridge\_algorithm} & $ \Delta R_{ij}^2/R^2$ & $1$ \\
\hline
anti-$\kt$ & \code{antikt\_algorithm} & $\min (p_{Ti}^{-2}, p_{Tj}^{-2}) \Delta R_{ij}^2/R^2$ & $p_{Ti}^{-2}$ \\
\hline
Generalized $\kt$ & \code{genkt\_algorithm} & $\min (p_{Ti}^{2p}, p_{Tj}^{2p}) \Delta R_{ij}^2/R^2$ & $p_{Ti}^{2p}$ \\
\hline
\hline
\multicolumn{4}{|c|}{$\ee$} \\
\hline
$\kt$ & \code{ee\_kt\_algorithm} &  $\min (E_{i}^2, E_{j}^2) \frac{(1 - \cos\theta_{ij})}{(1 - \cos R)}$  & $E_i^2$ \\
\hline
Generalized $\kt$ & \code{ee\_genkt\_algorithm} &  $\min (E_{i}^{2p}, E_{j}^{2p}) \frac{(1 - \cos\theta_{ij})}{(1 - \cos R)}$  & $E_i^{2p}$ \\
\hline
\end{tabular}
\end{center}

\caption{Native \texorpdfstring{\FJ}{FastJet} algorithms}
\label{table:FJalgs}
\end{table}

This set of algorithms is implemented internally in \FJ, but a much broader (and growing) class of jet algorithms is accessible through the plugin mechanism.  A \FJ plugin is derived from the abstract base class \code{fastjet::JetDefinition::Plugin}.  A plugin defines the \code{run\_clustering(ClusterSequence \&)} function, using an internal interface to the passed \code{ClusterSequence}.  Many algorithms beyond $\kt$ variants are shipped with \FJ as plugins; here is an example of their use from the \FJ manual \cite{FastJetManual}:

\begin{lstlisting}
// have some plugin class derived from the Plugin base class
class CDFMidPointPlugin : public fastjet::JetDefinition::Plugin {...};

// create an instance of the CDFMidPointPlugin class
CDFMidPointPlugin cdf_midpoint( [... options ...] );

//create the jet definition
fastjet::JetDefinition jet_def = fastjet::JetDefinition( & cdf_midpoint);

// then create ClusterSequence with the input particles and jet_def,
// and use it to extract jets as usual
\end{lstlisting}

For a better idea of how a plugin is actually implemented, see the description of the \FP plugin in the next subsection.

A list of plugins available in \FJ is given in Table~\ref{table:FJPlugins}.  In addition, several recent proposals for new jet finding techniques have been accompanied by the release of \FJ plugins (e.g., \cite{VariableR}, \cite{Trimming}, and \cite{Pruning2}).  \FJ's capabilities continue to grow.  The nature of the plugin mechanism allows arbitrary new jet methods to be plugged directly into old analyses with minimal effort.

\begin{sidewaystable}[p]
\caption{Plugin algorithms shipped with \FJ}
\label{table:FJPlugins}
\end{sidewaystable}

\begin{sidewaystable}[p]
\begin{tabular}{|l|l|l|}
\hline
Algorithm & Class name & Description \\
\hline
\multicolumn{3}{|c|}{$pp$} \\
 \hline
\prog{SISCone} & \code{SISConePlugin} & Seedless Infrared Safe Cone algorithm \\
\hline
CDF Midpoint & \code{CDFMidPointPlugin} & Midpoint-type iterative cone used at CDF Run II \\
\hline
\prog{JetClu} & \code{CDFJetCluPlugin} & CDF's main Run I alg., other Run II alg. \\
\hline
D0 Run II cone & \code{D0RunIIConePlugin} & D0's main Run II algorithm \\
\hline
ATLAS iterative cone & \code{ATLASConePlugin} & A deprecated cone alg. for ATLAS \\
\hline
CMS iterative cone & \code{CMSIterativeConePlugin} & A deprecated cone alg. for CMS \\
\hline
\prog{PxCone} & \code{PxConePlugin} & Abandonware, in Fortran \\
\hline
\prog{TrackJet} & \code{TrackJetPlugin} & Used for track-based jets at the Tevatron \\
\hline
\hline
\multicolumn{3}{|c|}{$\ee$} \\
\hline
``Original'' Cambridge & \code{EECambridgePlugin} & $\ee$ version of Cambridge/Aachen, with a different promotion criterion \\
\hline
\prog{JADE} & \code{JadePlugin} & The immortal and beloved \prog{JADE} algorithm \\
\hline
\end{tabular}
\end{sidewaystable}


\subsection{The \FP plugin}

Having read Sec.~\ref{sec:prune}, the reader is no doubt eager to try jet pruning at home.  Rest assured, gentle reader: nothing could be easier.  I have written \FP, a \FJ plugin implementing pruning, for just this purpose.  The plugin is available online \cite{FastPrune}.  This subsection gives an overview of the plugin's features and use; all code is taken from version \verb+0.4.1+.

Like any \FJ plugin, \FP is implemented as a class deriving from \code{fastjet::JetDefinition::Plugin}.  The following constructors are available:

\begin{lstlisting}

// Basic constructor
FastPrunePlugin (const JetDefinition & find_definition,
                 const JetDefinition & prune_definition,
                 const double & zcut = 0.1,
                 const double & Rcut_factor = 0.5);

// Lets the user specify a Recombiner class
FastPrunePlugin (const JetDefinition & find_definition,
                 const JetDefinition & prune_definition,
                 const JetDefinition::Recombiner* recomb,
                 const double & zcut = 0.1,
                 const double & Rcut_factor = 0.5);

// Two new constructors that allow you to pass your own CutSetter.
//  This lets you define zcut and Rcut on a jet-by-jet basis.
FastPrunePlugin (const JetDefinition & find_definition,
                 const JetDefinition & prune_definition,
                 CutSetter* const cut_setter);

FastPrunePlugin (const JetDefinition & find_definition,
                 const JetDefinition & prune_definition,
                 CutSetter* const cut_setter,
                 const JetDefinition::Recombiner* recomb);

\end{lstlisting}

The parameters \code{zcut} and \code{Rcut\_factor} correspond the the parameters $z_\text{cut}$ and $D_\cut$ in Sec.~\ref{sec:prune:define}, where the actual $D_\cut$ used for a given jet is \code{Rcut\_factor} $\times 2 m_J/p_{T_J}$.  Two jet definitions need to be passed.  The first is used to find initial jets (Step 0 in Sec.~\ref{sec:prune:define}).  The second is used in the pruning procedure (Step 1), so should be a recombination algorithm like CA or $\kt$.  The user can specify their own \code{Recombiner}, for example to preserve flavor information in the merging.  Setting the \code{Recombiner} for the pruning jet definition will have the same effect.  The user can also specify a CutSetter class, which stores values for \code{zcut} and \code{Rcut} and implements the function \code{SetCuts(const PseudoJet \&, const ClusterSequence \&)}.  \code{CutSetter}, as well as an example \code{DefaultCutSetter} are defined in \code{FastPrunePlugin.hh}.

\FP works in three stages.  First, unpruned jets are found with the \code{JetDefinition} \code{find\_definition}.  Second, each individual jet and its constituents are then passed to a second \code{ClusterSequence} using the \code{prune\_definition}.  The \code{Recombiner} for the pruned \code{JetDefinition} is set to be a \code{PrunedRecombiner}, a helper class that implements the pruning test.  It wraps the \code{Recombiner} in \code{prune\_definition}, checking for the pruning test given in Eq.~\ref{eq:pruneTest}.  If the test fails (i.e., the softer branch should be pruned), the recombination does not happen and the index of the pruned \code{PseudoJet} is stored.  Finally, the \code{ClusterSequence} built up by this process is transferred to the output via the standard plugin interface.\footnote{In the final \code{ClusterSequence}, pruned \codes{PseudoJet} appear in the merging history as steps with \code{Invalid} children  --- they are never merged with other \codes{PseudoJet} or the beam.}

The most important step is the running of the pruned \code{JetDefinition}, with its \code{PrunedRecombiner}.  A few notes are in order.  Since the jet definition is supplied by the user, any algorithm that \FJ knows about can be pruned.  Moreover, \FP doesn't need to implement any actual jet finding since this is outsourced to existing \FJ code.  Since the only difference between a pruned algorithm and the unpruned sort is that some recombinations are vetoed, the same \code{JetDefinition} can be used ---just with a new \code{Recombiner}.  If the user supplies their own \code{Recombiner}, this is passed to the plugin's \code{PrunedRecombiner}.  \code{PrunedRecombiner} first checks if a recombination should be pruned, then if not does the recombination with the user's \code{Recombiner}.  If no \code{Recombiner} is passed, then \FJ's \code{DefaultRecombiner} is used.  Finally, \FP preserves the user indices for input \codes{PseudoJet}, and these can be used, for example, by the user's \code{Recombiner} class.

Here is a shortened version of the example program indicating how the plugin is used:

\pagebreak

\begin{lstlisting}
// setup
JetDefinition jet_def(cambridge_algorithm, 1.0, E_scheme, Best);
JetDefinition jet_def_bigR(cambridge_algorithm(), 0.5*pi, E_scheme, Best);
FastPrunePlugin *PRplugin = new FastPrunePlugin(jet_def, jet_def_bigR, 0.1, 0.5);
JetDefinition pruned_def(PRplugin);

vector<PseudoJet> inputs;
/* ... fill inputs somehow ... */
// find jets
ClusterSequence pruned_seq(inputs, pruned_def);
// access jets
vector<PseudoJet> pruned_jets = pruned_seq.inclusive_jets(20.0);
/* ... do stuff with jets ... */
// can also see which subjets were pruned
// pruned_subjets[0] are subjets pruned from highest-pT jet, pruned_subjets[1] are next-highest, etc.
vector<vector<PseudoJet> > pruned_subjets = PRplugin->pruned_subjets();



\end{lstlisting}

\section{SpartyJet}
\label{sec:tools:sj}

\SJ is a jet analysis package that complements and extends jet finding with \FJ.  \SJ provides a framework for jet finding and analysis that includes support for a variety of input and output formats and easy combination of many jet manipulation and measurement tools.  \FJ is a tool for \emph{finding} jets; \SJ is a tool for \emph{studying} jets.  This section gives an overview of \SJ's capabilities, and is intended to complement the manual, available at \cite{SpartyJetWebsite}. 

\subsection{Input and output}

\SJ can take input particle in put from a variety of sources; the user only needs to specify the location of an input file and its format.  A full list of possible input formats is given in Table \ref{table:SJinput}.  Configuring input is simple: just create an instance of the appropriate input class, typically passing it a file name:
\begin{lstlisting}
SpartyJet::InputMaker *input = new SpartyJet::StdHepInput("events.hep");
\end{lstlisting}
All input classes derive from \code{SpartyJet::InputMaker}; an object of this type is passed to jet analysis.  Several add-ons to input reading are available, including checking for bad input (e.g., four-momenta with negative energy) and storing PDG ID codes.

\begin{table}[htbp]
\begin{center}
\begin{tabular}{|l||l|l|}
\hline
Format & Class name & Description \\
\hline
\hline
\prog{ROOT} NTuple & \code{NtupleInputMaker} & Reads 4-vectors from a \code{TTree} \\
\hline
ASCII text & \code{StdTextInput} & Reads lines of \code{"E px py pz"} text \\
\hline
StdHEP & \code{StdHepInput} & Reads StdHEP XDR files \\
\hline
CALCHEP & \code{CalchepPartonTextInput} & Reads CALCHEP files \\
\hline
HepMC & \code{HepMCInput} & Reads HepMC ASCII output \\
\hline
\end{tabular}
\end{center}
\caption{Available \texorpdfstring{\SJ}{SpartyJet} input formats}
\label{table:SJinput}
\end{table}

\SJ output is stored in a \code{TTree} in a \prog{ROOT} file.  Four-momenta for all jets (for an arbitrary set of jet finders) are stored, as well as four-momenta for all input particles and indices to keep track of which input particles ended up in which jet.  Complete merging history (as in \FJ's \code{ClusterSequence}) storage is stored internally, but not written to the output file.  Persistency for the clustering history is in development.  As described below, an arbitrary set of jet ``moments'' can be added to any or all jet finders; the values of these moments are also stored as \code{TTree} branches.


\subsection{Jet algorithms}

Previous versions of \SJ offered a large number of native jet algorithms, as well as access to a subset of native \FJ algorithms.  Most of the native algorithms are collaboration-specific implementations of cone and $\kt$-type algorithms, for example CDF's \prog{JetClu}.  As experiments move to standardized algorithms, and non-standard algorithms are implemented in \FJ, built-in \SJ jet algorithms have become deprecated.   Currently, the only \SJ native algorithm not available through \FJ is an implementation of \prog{Pythia}'s \prog{CellJet}.
With version 3.4, \SJ can now use any \FJ \,\code{JetDefinition}, including native algorithms like $\kt$, included plugins like \prog{SISCone}, or user-supplied plugins like \FP.  Any jet algorithm that can be implemented as a \FJ plugin can be used with \SJ and this is now the preferred method of adding a new jet algorithm to \SJ.

Here are some examples of creating jet finders in \SJ.  Jet finder classes derive from the more general \code{JetTool} class, about which more will be said in the next subsection.

\begin{lstlisting}
// ***  Old-style jet finders  (see examples_C/multiAlgExample.cc)  ***
// Add a Midpoint alg
cdf::MidPointFinder * tool1 = new cdf::MidPointFinder();  
tool1->set_coneRadius(0.4);   // can set all parameters like this
tool1->set_name("MidPoint4");
builder.add_default_alg(tool1);
  
// Add a Jet Clu alg
builder.add_default_alg( new cdf::JetClustFinder("myJetClu"));
  
// Add a CellJet alg --- second parameter turns off constituent storage,
//   which does not work in CellJet
builder.add_default_alg( new pythia::CellJetFinder("myCellJet"),false);

// ***  New-style (FastJet) finders (see examples_C/FJExample.cc) ***
// Add an algorithm (AntiKt) - uses the fastjet::JetDefinition::JetAlgorithm enum
FastJetFinder *anti4 = new FastJetFinder("AntiKt4",antikt_algorithm,0.4,false);
builder.add_default_alg(anti4);

// Same algorithm, uses your own JetDefinition
JetDefinition jet_def(antikt_algorithm, 0.4);
FastJetFinder *anti4_2 = new FastJetFinder(&jet_def,"AntiKt4_2",false);
builder.add_default_alg(anti4_2);

// More interesting example: FastJet Plugin
// Note that SISCone is included in FastJet, but is implemented as a plugin
// To use your own plugin, you will need to link against the relevant library
double coneRadius = 0.4, overlapThreshold = 0.75;
SISConePlugin plugin(coneRadius,overlapThreshold);
JetDefinition plugin_jet_def(&plugin);
FastJetFinder *siscone4 = new FastJetFinder(&plugin_jet_def,"SISCone4",false);
builder.add_default_alg(siscone4);
\end{lstlisting}


\subsection{\texorpdfstring{\codes{JetCollection}}{JetCollections} and \texorpdfstring{\codes{JetTool}}{JetTools}: Constructing an analysis}

The basic \emph{object} of a \SJ analysis is a \code{JetCollection}; the basic \emph{action} of an analysis is described by a sequence of \codes{JetTool}.  A \code{JetCollection} is just a set of \codes{Jet} together with with a map of jet and event ``moments'', which can represent any measurement on a jet or an event --- these are discussed further below.  A \code{JetCollection} also stores the clustering history of the event it represents.

A \code{JetTool} is an abstract base class that operates on a \code{JetCollection}: a \code{JetTool} must define the method \code{JetTool::execute(JetCollection \&)}.  The most important \codes{JetTool} are jet finders like those seen in the previous subsection.  A jet finder takes a \code{JetCollection} representing a set of input particles and replaces it with a \code{JetCollection} containing a set of found jets together with their clustering history.  Other examples include \code{JetPtSelectorTool}, which removes all jets failing a $p_T$ cut, \code{JetMomentTool}, an abstract class for tools that calculate and store jet moments for the input \code{JetCollection}, and \code{MinBiasInserterTool}, which adds particles representing pile-up events to an input \code{JetCollection}.

A \code{JetAlgorithm} in \SJ is a sequence of \codes{JetTool}; a complete analysis consists of a set of \codes{JetAlgorithm}.  The key ability of \SJ is to provide a very simple way to construct and compare multiple analyses, represented as chains of \codes{JetTool}.  An interesting example of a complete \SJ analysis is given in Appendix \ref{app:example}, where I compare pruning to top-tagging and mass-drop filtering.


\subsection{\texorpdfstring{\SJ}{SpartyJet}/\texorpdfstring{\FJ}{FastJet} interoperability}

Recent developments in \SJ, in addition to streamlining the use of \FJ jet finders, have added the ability to convert back and forth between the main analysis objects in each framework: \code{fastjet::ClusterSequence} and \code{SpartyJet::JetCollection}, including transfer of clustering history.  In practical terms, this means that with minimal wrapping, \SJ\,\codes{JetTool} can be used in a \FJ-based analysis and likewise \FJ-based tools can easily be inserted into \SJ-based analyses.

Wrapping of \FJ tools is done via the \code{FastJetTool} class, which converts a \code{JetCollection} to a \code{ClusterSequence}, calls \code{execute(ClusterSequence \&)}, and finally converts the modified \code{ClusterSequence} back to a \code{JetCollection}.  Derived tools then implement some function on a \code{ClusterSequence}. 

Tools that use features already implemented in \FJ, e.g. the \FP tool described in the next section, are naturally written as \codes{FastJetTool}.  Other tools, such as the \code{TopDownPruneTool}, which prunes away asymmetric branchings (used in several \SJ implementations of jet substructure tools), are simpler to implement in terms of \codes{JetCollection}, which are easier to modify in place than \codes{ClusterSequence}.


\subsubsection{\texorpdfstring{\code{FastPruneTool}}{FastPruneTool}: an example \texorpdfstring{\FJ}{FastJet}-based tool}

\code{FastPruneTool} is a variant of the \FP plugin, now included in the \FP package, that is intended to be inserted into a \SJ analysis.  Instead of acting as a \FJ plugin, \code{FastPruneTool} inherits from \code{SpartyJet::JetTool}.  Given a \code{JetCollection} representing jets found with some jet finder, it returns a \code{JetCollection} representing the pruned versions of those jets.  This simplifies the insertion of pruning into an existing analysis.  If the analysis compares pruned jets to unpruned jets, the pruning tool eliminates the computational effort of finding jets twice (relative to using the \FJ plugin, which finds unpruned jets itself).

\subsection{Jet moments}

In addition to storing a set of jets (and their substructure) at each point in an analysis chain, \SJ stores jet ``moments'' --- arbitrary pieces of additional information about each jet.  Examples include a PDG ID code, stored as a jet moment for an input ``jet'', or a jet area, which is calculated by a \FJ jet finder and then stored as a jet moment.  Moments are implemented via the \code{Moment} and \code{JetMomentMap} classes.  Every \code{JetCollection} holds a \code{JetMomentMap}, which stores a set of moments for each jet in the collection.  Moments can be saved and retrieved by name, and there can be any number of jet moments.  Event moments, which do not correspond to any particular jet, can be created, stored, and retrieved in a similar manner.  Every jet or event moment is stored as a branch in the output \code{TTree}.

Jet and event moments are implemented via the \code{JetMoment<T>} and \code{EventMoment<T>} classes, which both inherit from \code{Moment}.  \code{T} can be any basic type or class that \prog{ROOT} has a dictionary for (so it can be stored in the output file).  The \code{JetMomentTool} tool stores a user-supplied \code{JetMoment<T>}-derived object that calculates the given moment for each jet in a \code{JetCollection}; the tool then stores this in the \code{JetMomentMap} for that collection.  See \verb+JetTools/JetMomentTool.hh+ for examples.  Once a moment has been stored, it can be accessed by subsequent tools, e.g. \code{JetMomentSelectorTool}, which selects jets based on whether a given moment falls within a given range.  See \verb+examples_py/TopTaggerExample.py+ for an example.

\subsection{Substructure tools}

A number of jet substructure tools have recently been introduced to \SJ.  These include tools for jet filtering, ``top-down pruning'' as in the mass-drop step of \cite{FilteringHiggs} or the subjet-finding step in top-tagging \cite{TopTagging}, and subjet manipulation.  Some of these tools simply wrap existing \FJ tools (the wrapper is necessary so that the tool behaves like a \code{JetTool}, modifying a \code{JetCollection} in place); others are natively implemented in \SJ.  See the substructure section of the \SJ user manual, and the scripts in \verb+examples_py/+ for more examples and details.

\subsection{Graphical interface}

\begin{figure}[htbp]
\begin{center}
\includegraphics[width=\textwidth]{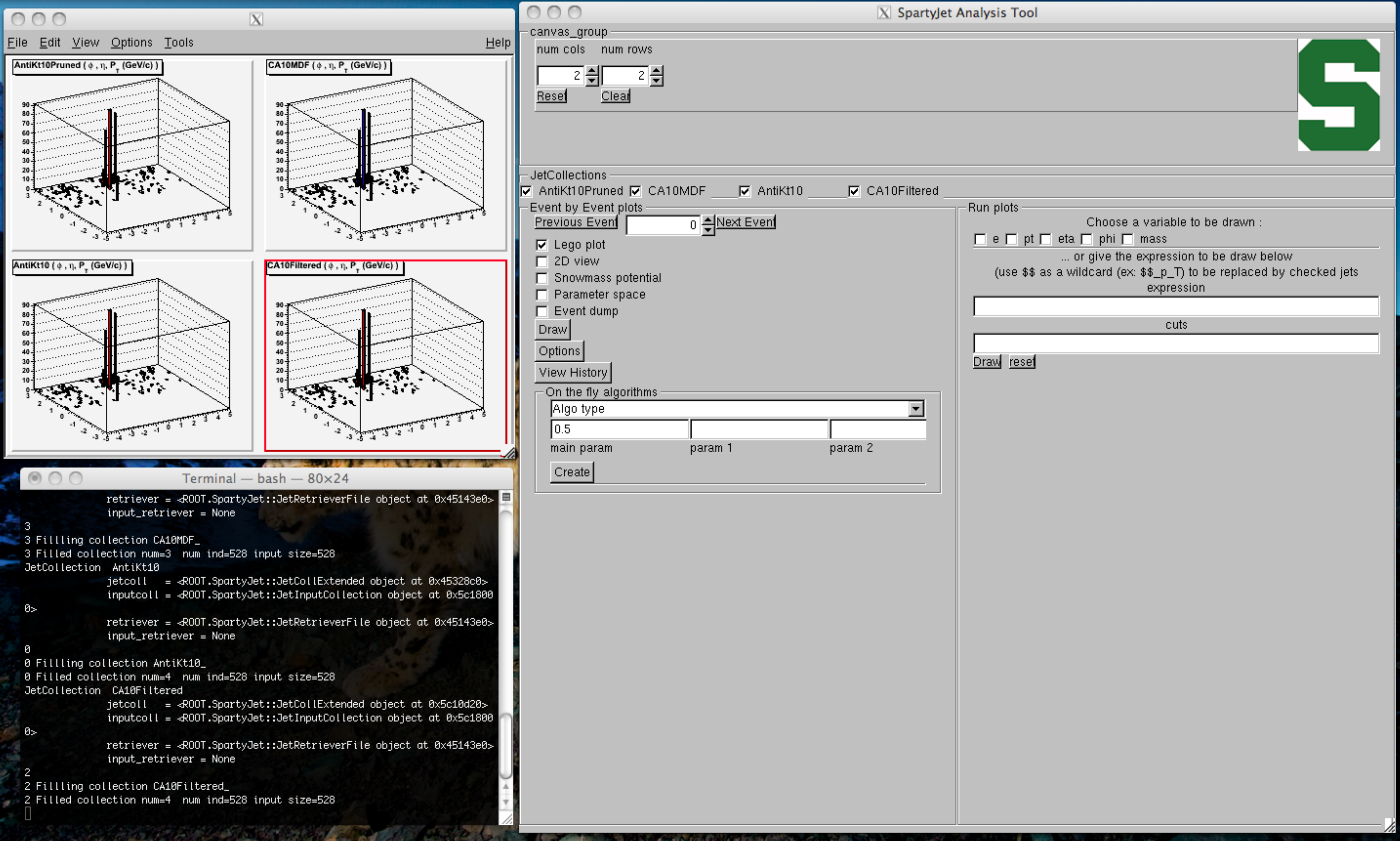}
\end{center}
\caption{A screenshot of the \texorpdfstring{\SJ}{SpartyJet} GUI.}
\label{fig:SJscreenshot}
\end{figure}

\SJ contains an (in development) graphical user interface (GUI) for comparing results for found jets.  The developers of \SJ hope that in the near future this will become a powerful and easy to use tool for visually comparing the results of different analyses.  An example screenshot is shown in Fig.~\ref{fig:SJscreenshot}.  The GUI loads a specified output \prog{ROOT} file and the user can display a variety of plots for one or more of the saved \codes{JetCollection}.  For example, a user could quickly plot the jet area and a jet shape variable, both calculated and stored as jet moments, for two different \codes{JetAlgorithm}.  Both event displays and full-run plots are available, and more types of display are planned.

\chapter{Conclusions}
\label{sec:conc}
 
The last few years have seen a proliferation of new theoretical and experimental techniques to search for new physics at the Large Hadron Collider.  No silver bullets have been discovered, and none will be.  Many complementary advances will no doubt contribute to the most significant results at the LHC.

\subsubsection{SCET}

Physics at the LHC inescapably involves jets.  The best possible theoretical description of jet physics is therefore indispensable.  Soft/collinear effective theory is proving to be a powerful tool in this regard.  As shown in Chapter \ref{sec:scet}, SCET provides a simple framework for factorization, and hence resummation of the logarithms arising in each separate piece of a calculation.  SCET captures of the dominant physics of QCD while allowing for systematic improvements to the approximations used.

For SCET calculations to be useful at the LHC, however, several advances have been necessary.  First, the effects of strongly-interacting particles in the initial state must be taken into account through ``beam functions'' \cite{Stewart:2009yx, QuarkBeamFunction} --- essentially the application of a jet function to the ``beam jet''.  (see, e.g., \cite{BeamJet}).  Second, a useful calculation at the LHC must be in terms of \emph{jets}.  Whereas event shapes were interesting and useful measures of hadronic activity in $\ee$ collisions, the environment of a high-luminosity hadron collider is less well suited to event measures and it is useful to think instead in terms of ``jet shapes''.  Our work on jet angularity measurements (Chapter \ref{sec:scet}, \cite{SCETLetter, Ellis:2010rw}) is a step in this direction although it does not yet incorporate the additional complications of a hadron collider.  Other groups have also made progress in incorporating jet algorithms--- and jets --- into SCET calculations \cite{Cheung:2009sg, Jouttenus:2009ns}.  An intriguing alternative involving an event-shape like measure instead of traditional jets was presented in \cite{NJettiness}.

What remains is to apply these improved theoretical predictions to specific applications.  One goal claimed in \cite{Ellis:2010rw} is the use of angularities in distinguishing quark and gluon jets.  An obvious extension would be to use jet shapes to distinguish jets involved in new physics (top jets, for example) from their QCD backgrounds --- as in the template overlap method of \cite{TemplateOverlap}.  As theoretical predictions converge with experimental methods, another challenge is incorporating the effects of jet modifications such as filtering-type techniques and pile-up subtraction.  One step in this direction has been the calculation of non-global logarithms in filtered jets in \cite{FilteredLogs}.

\subsubsection{Pruning}
 
While one approach to better LHC studies is better QCD predictions, another is to simply discard the parts of the event that are hardest to understand.  This is the essential goal of grooming methods such as jet pruning.  Of course this can only be done on average, but to some extent this approach allows us to focus on the high-energy, perturbative physics we understand well and pull out the signals we are interested in.  As we saw in Chapters \ref{sec:sub} and \ref{sec:prune}, pruning significantly reduced the new-physics-obscuring effects of splash-in from many sources.  Pruning also greatly reduced the mass of pure QCD jets --- typically moving background jets out of the signal region.  We explored in Chapter \ref{sec:sub} the reasons for these improvements.  The branchings removed by pruning almost never represent the substructure of a heavy particle decay but are instead characteristic of QCD radiation or splash-in.  Removing such branchings tends to clean up the signal and prune back the background.

As methods for modifying jet substructure have proliferated, it has become clear that while they all exploit the same underlying physics, there can be subtle differences between methods that will moreover vary between analyses (see, e.g., the comparison of top-tagging methods in \cite{SpannowskyTalk}).  The field awaits a synthesis of such techniques that explains these differences.  A full theory of jet substructure and filtering methods will require integrating our understanding of the QCD parton shower with the effects of initial state radiation, the underlying event, and pile-up.
 
\subsubsection{Software tools}
 
In the mean time, the experimentalist or phenomenologist is confronted with a surfeit of choice in designing a new physics search.  Fortunately tools exist for penetrating this thicket --- pruning it back, as it were.  In addition to the variety of jet \emph{algorithms} available within the \FJ package, there is a growing number of jet \emph{tools} implemented in software.  In the author's estimation the simplest use of these tools exists within the \SJ package, which provides a framework for assembling a jet analysis from a large --- and rapidly increasing --- number of jet filtering, measuring, and selecting tools.  The goal of the \SJ package, thus far partially attained, is to simplify to the greatest extent possible the design and comparison of jet analyses.  Improvements planned for the near future include greater inclusion of proposed jet tools and a more powerful, easier to use graphical interface for studying and comparing the final results.

\subsubsection*{}

The Large Hadron Collider, run by some of the most highly funded and technologically sophisticated experimental collaborations in the history of science, will nonetheless require the advances in prediction, technique, and software that will be provided by the theory community.  It is humbly hoped that the tools described in this thesis are a step in the right direction.

\printendnotes

%
%

\bibliographystyle{JHEP}
\bibliography{thesis}
%
%
\appendix
\raggedbottom\sloppy



\setlength{\arraycolsep}{2pt}

\chapter{\texorpdfstring{$\ee \to $}{ee ->} hadrons: an example QCD calculation in depth}
\label{app:eeNLO}

\section{Introduction}

In this appendix I explain in detail how to calculate $\sigma(e^+e^- \to \text{hadrons})$ to $\mathcal{O}(\alpha_s)$, which is a nice example of a one loop calculation, requiring non-trivial regularization.  Throughout I will take all masses to be zero, and use dimensional regularization to regulate infrared divergences.  The NLO diagrams have soft-collinear divergences which show up as $1/\epsilon$ poles, which cancel in the final inclusive cross section.

Some notes on the calculation.  I will use Peskin and Schroeder \cite{Peskin} conventions throughout, notably a $(+---)$ metric.  I take $s \ll m_Z^2$, so I can neglect contributions from the $Z$ propagator.  These are irrelevant to the consideration of the NLO strong correction.  The Feynman diagrams at $\mathcal{O}(\alpha_s)$ are given in Fig.~\ref{fig:eeFeynman}.

\begin{figure}[htbp]
\begin{center}

\subfloat{
\addtocounter{subfigure}{-1}
\begin{fmffile}{eeqqa}
	{\fmfreuse{eeqq}}
\end{fmffile}
\qquad \qquad
\begin{fmffile}{eeqqNLO1a}
	{\fmfreuse{eeqqNLO1}}
\end{fmffile}
}

\vspace{30pt}

\subfloat[]{
\begin{fmffile}{eeqqNLO2a}
	{\fmfreuse{eeqqNLO2}}
\end{fmffile}
\qquad \qquad
\begin{fmffile}{eeqqNLO3a}
	{\fmfreuse{eeqqNLO3}}
\end{fmffile}
}

\vspace{20pt}

\subfloat[]{
\begin{fmffile}{eeqqg1a}
	{\fmfreuse{eeqqg1}}
\end{fmffile}
\qquad\qquad
\begin{fmffile}{eeqqg2a}
	{\fmfreuse{eeqqg2}}
\end{fmffile}
}

\end{center}
\caption[Feynman diagrams for $\ee \to q\bar q$ and $\ee \to q \bar q g$]{Feynman diagrams for (a) $\ee \to q\bar q$ and (b) $\ee \to q \bar q g$.}
\label{fig:eeFeynman}
\end{figure}

In the following, I take the initial momenta to be $p_1$ and $p_2$, the photon momentum to be $q$ (with $q^2 \equiv s$), and the final quark momenta to be $k_1$ and $k_2$.  For the real emission diagram, I label the gluon momentum $k_3$.  For this diagram, the non-trivial phase space dependence makes it useful to define the scalars $x_i \equiv 2 k_i \cdot q/s$.  Note that $x_1 + x_2 + x_3 = 2$.  After summing over spins and gluon polarization, the cross section only depends on $x_1$ and $x_2$.  I work in the center of momentum frame ($(p_1+p_2)^{\mu} = q^\mu = (\sqrt{s},0,0,0)$) throughout.

\section{Some general results}

\subsection{Factorization of cross section}

For all contributing diagrams, the amplitude is composed of a leptonic part ($e^+e^-\to\gamma^*$) and a hadronic part ($\gamma^*\to q \bar{q} (g)$), with the general form
\beq \begin{split}
i \mathcal{M} & = L^\mu \left(\frac{-i g_{\mu \nu}}{q^2}\right)H^\nu   \\
\Rightarrow \mathcal{M} & = -\frac{1}{s}L^\mu H_\mu.  
\end{split} \eeq
At tree level, only one diagram contributes, so
\beq
|\mathcal{M}|^2 = \frac{1}{s^2}L^{\mu *}L^{\nu} H_{\mu}^* H_\nu.  
\eeq
It is convenient to split the $1/s^2$ factor between the leptonic and hadronic parts to make them dimensionless (for $2\to2$):
\beq
\label{eq:lmunu}
L^{\mu \nu} \equiv \frac{1}{s}L^{\mu *}L^{\nu}, \quad H^{\mu \nu} \equiv \frac{1}{s}H^{\mu *}H^{\nu}. 
\eeq
In this notation,
\beq
\label{eq:mesimple}
|\mathcal{M}|^2 = L^{\mu \nu}H_{\mu \nu}.
\eeq
$|\mathcal{M}|^2$ can in general be written in this form, up to electroweak corrections that connect the incoming and outgoing particles.  The Ward identity (or gauge invariance, or current conservation, etc.) guarantees that
\beq
q_\mu L^{\mu \nu} = q_\nu L^{\mu \nu} = q_\mu H^{\mu \nu} = q_\nu H^{\mu \nu} = 0.  
\eeq
We're interested in calculating the total cross section, so we will, in the end, integrate over the final phase space.  This means that after this integration, there are no vectors $H^{\mu \nu}$ can depend on other than $q^\mu$, so we can write:
\beq
\int d\Pi_2 H^{\mu \nu} \to \int d\Pi_2 \left(g^{\mu \nu} - \frac{q^\mu q^\nu}{q^2}\right) H'.  
\eeq
This means that we can re-express the phase space integral of $|\mathcal{M}|^2$:
\beq \begin{split}
\int d\Pi_n |\mathcal{M}|^2 &= L^{\mu \nu} \int d\Pi_n H_{\mu \nu}  \\
           &= L^{\mu \nu} \int d\Pi_n \left(g^{\mu \nu} - \frac{q^\mu q^\nu}{q^2}\right) H'  \\
           &= \left(g_{\mu \nu}L^{\mu \nu}\right) \int d\Pi_n H'.  
\end{split} \eeq

The Ward identity has been used to discard the $q^\mu q^\nu$ term going from the second to the third line.  $n=2$ for the tree-level diagram and the virtual correction; $n=3$ for the real emission.  Noting that in $d$ dimensions, $g^{\mu \nu}g_{\mu \nu} = d$, 
\beq \begin{split}
g_{\mu \nu} H^{\mu \nu} &= g_{\mu \nu}  \left(g^{\mu \nu} - \frac{q^\mu q^\nu}{q^2}\right) H'   \\
   &= (d-1)H'.  
\end{split} \eeq
Defining $L \equiv g_{\mu \nu}L^{\mu \nu}$ and $H \equiv g_{\mu \nu}H^{\mu \nu}$, we can write
\beq
\label{eq:contracted}
\int d\Pi_n |\mathcal{M}|^2 = \frac{1}{d-1}L \int d\Pi_n H.
\eeq
Generically, a $2 \to N$ cross section has the form
\beq
\sigma(2 \to N) = \frac{1}{2s}\int d\Pi_N |\mathcal{M}|^2,
\eeq
so for the contributions we consider, we can write
\beq
\label{eq:xsection}
\sigma_{2(3)} = \frac{1}{2s}\frac{1}{d-1}L\int d\Pi_{2(3)} H.
\eeq
%

\subsection{Phase space in \texorpdfstring{$d$}{d} dimensions}

Since we are regularizing the calculation by performing it in $d$ dimensions, we must work out the phase space factors in arbitrary dimension.

\subsubsection{Two-body}

The two-body final states have trivial dependence on the phase space variables, so we just need to calculate the total integral:

\beq \begin{split}
\int d\Pi_2 &= \mu^{4-d} \int \frac{d^{d-1} k_1}{(2 \pi)^{d-1} 2 E_1} \frac{d^{d-1} k_2}{(2 \pi)^{d-1} 2 E_2}(2 \pi)^d \delta^d (p_1 + p_2 - k_1 - k_2)  \\
   &= \mu^{4-d} \int \frac{d^{d-1} k_1}{(2 \pi)^{d-2} 4 E_1^2} \delta (\sqrt{s} - 2 E_1)  \\
   &= \mu^{4-d} \int \frac{E_1^{d-2} d E_1 d^{d-2}\Omega}{(2 \pi)^{d-2} 8 E_1^2} \delta (\sqrt{s}/2 - E_1)  \\
&= \frac{1}{8} \left(\frac{2 \mu}{\sqrt{s}}\right)^{4-d} \int \frac{d^{d-2}\Omega}{(2 \pi)^{d-2}}  \\
&= \frac{1}{8} \left(\frac{2 \mu}{\sqrt{s}}\right)^{4-d} \frac{1}{(2 \pi)^{d-2}} \frac{2 \pi^{(d-1)/2}} {\Gamma((d-1)/2)}.  
\end{split} \eeq

The $\mu^{4-d}$ factor is inserted to keep the overall dimension correct.  The last line uses a standard result for the surface area of an $n$-sphere.\footnote{The Wikipedia page for "Spherical coordinates" has a number of nice results relating to this.}  Writing $d = 4 - 2 \epsilon$,
\beq \begin{split}
\label{eq:PS2}
\int d\Pi_2 &= \frac{1}{8 \pi} \left(\frac{16 \pi \mu^2}{s} \right)^\epsilon \frac{\sqrt{\pi}}{2 \Gamma (\frac{3}{2} - \epsilon)}  \\
  &= \frac{1}{8 \pi}  \left(\frac{4 \pi \mu^2}{s} \right)^\epsilon \frac{\Gamma(1-\epsilon)}{\Gamma (2 - 2 \epsilon)}.
\end{split} \eeq
In the second line, we used the relation $\Gamma(z) \Gamma(z+\frac{1}{2}) = \frac{\sqrt{4 \pi}}{2^{2 z}}\Gamma(2 z)$.

\subsubsection{Three-body}

In this case the final state contains three vectors $|\mathcal{M}|^2$ can depend on, but one can show (it's pretty easy) that all possible scalar products between them can be expressed as a function of $s$ and the energy fractions $x_1$ and $x_2$.  We need to integrate out the other $3(d-1)-2$ variables and save the $x_1$ and $x_2$ integrals until we know the integrand.  The key trick is using the energy-conserving delta function to integrate over an \emph{angle}, not an energy.  We start with the trivial integral over the three-momentum delta function:

\beq \begin{split}
\int d\Pi_3 &= \mu^{2(4-d)} \int \frac{d^{d-1} k_1}{(2 \pi)^{d-1} 2 E_1} \frac{d^{d-1} k_2}{(2 \pi)^{d-1} 2 E_2} \frac{d^{d-1} k_3}{(2 \pi)^{d-1} 2 E_3} (2 \pi)^d \delta^d (p_1 + p_2 - k_1 - k_2 - k_3)  \\
   &= \mu^{2(4-d)} \int \frac{d^{d-1} k_1}{(2 \pi)^{d-1} 2 E_1} \frac{d^{d-1} k_2}{(2 \pi)^{d-1} 2 E_2} \frac{(2 \pi)}{2 E_3} \delta (\sqrt{s} - E_1 - E_2 - |\vec{k_1} + \vec{k_2}|).  
\end{split} \eeq

We can split the remaining two integrals into energy (magnitude) and angular parts.  One angular integral is trivial, but the other will include integrating over the delta function, which we write:

\beq \begin{split}
\delta (\sqrt{s} - \Sigma E_i) &=  \delta (\sqrt{s} - E_1 - E_2 - |\vec{k_1} + \vec{k_2}|)  \\
&=  \delta \left(\sqrt{s} - E_1 - E_2 - \sqrt{E_1^2 + E_2^2 + 2 E_1 E_2 \cos\theta}\right)  \\
&=  \frac{E_3}{E_1 E_2} \delta (\cos\theta - \cos\theta_0),  
\end{split} \eeq
where $\cos\theta_0$ is defined by
\beq \begin{split}
\cos\theta_0 &\equiv \frac{(\sqrt{s} - E_1 - E_2)^2 - E_1^2 - E_2^2}{2 E_1 E_2}  \\
&= \frac{(1-x_1)(1-x_2)}{x_1x_2}.  
\end{split} \eeq

Note that $E_3$ is fixed by $E_1$ and $E_2$.  Integrating over the delta function gives a theta function that limits us to the physical region in the energy integrals.  Returning to the phase space integral:

\beq \begin{split}
\int d\Pi_3 &= \frac{\mu^{2(4-d)}}{(2 \pi)^{2 d-1}} \int \frac{d^{d-1} k_1}{ 2 E_1} \frac{d^{d-1} k_2}{2 E_2} \frac{1}{2 E_3} \delta (\sqrt{s} - E_1 - E_2 - |\vec{k_1} + \vec{k_2}|)  \\
&= \frac{ \mu^{2(4-d)}}{(2 \pi)^{2 d-1}} \int \frac{E_1^{d-2} d E_1 d^{d-2}\Omega_1}{ 2 E_1} \frac{E_2^{d-2} d E_2 d^{d-2}\Omega_2}{2 E_2} \frac{1}{2 E_3} \delta (\sqrt{s} - E_1 - E_2 - |\vec{k_1} + \vec{k_2}|)  \\
&= \frac{ \mu^{2(4-d)}}{8 (2 \pi)^{2 d-1}} \int \frac{E_1^{d-2} d E_1 d^{d-2}\Omega_1}{ E_1} \frac{E_2^{d-2} d E_2 d^{d-2}\Omega_2}{ E_2}\frac{1}{E_3} \frac{E_3}{E_1 E_2} \delta (\cos\theta - \cos\theta_0) \\
&= \frac{\mu^{2(4-d)}}{8 (2 \pi)^{2 d-1}} \left(\int \frac{d E_1 d E_2}{(E_1 E_2)^{4-d}}\right) \left(\int d^{d-2}\Omega_1\right) \left(\int d^{d-2}\Omega_2  \delta (\cos\theta - \cos\theta_0)\right)  \\
&= \frac{\mu^{2(4-d)}}{8 (2 \pi)^{2 d-1}} \left(\int \frac{d E_1 d E_2}{(E_1 E_2)^{4-d}}\right) \frac{2 \pi^{(d-1)/2}} {\Gamma((d-1)/2)} \left(\int d^{d-2}\Omega_2  \delta (\cos\theta - \cos\theta_0)\right)  \\
&= \frac{ \mu^{2(4-d)}}{8 (2 \pi)^{2 d-1}}\left( \frac{s}{4} \right)^{d-3} \left(\int \frac{d x_1 d x_2}{(x_1 x_2)^{4-d}}\right) \frac{2 \pi^{(d-1)/2}} {\Gamma((d-1)/2)} \left(\int d^{d-2}\Omega_2  \delta (\cos\theta - \cos\theta_0)\right).  
\end{split} \eeq

For the last angular integral, we need to break a $(d-2)$-dimensional angular space into one azimuthal angle and the rest:
\beq
\int d^{d-2}\Omega = \int d^{d-3}\Omega \int d \theta \sin^{d-3}\theta,  
\eeq
so:
\beq \begin{split}
\int d^{d-2}\Omega  \delta (\cos\theta - \cos\theta_0) &= \left(\int d^{d-3}\Omega\right) \int d \theta \sin^{d-3}\theta  \delta (\cos\theta - \cos\theta_0)  \\
&=\frac{2 \pi^{(d-2)/2}}{\Gamma((d-2)/2)}  \int d \cos\theta \sin^{d-4}\theta  \delta (\cos\theta - \cos\theta_0)  \\
&=\frac{2 \pi^{(d-2)/2}}{\Gamma((d-2)/2)} \sin^{d-4}\theta_0  \\
&=\frac{2 \pi^{(d-2)/2}}{\Gamma((d-2)/2)} \left(1 - \left(\frac{(1-x_1)(1-x_2)}{x_1x_2}\right)^2 \right)^{(d-4)/2}  \\
&=\frac{2 \pi^{(d-2)/2}}{\Gamma((d-2)/2)} \left( \frac{4(1-x_1)(1-x_2)(1-x_3)}{x_1^2 x_2^2} \right)^{(d-4)/2}.  
\end{split} \eeq
Putting it all together,
\beq \begin{split}
\int d\Pi_3 &= \mu^{2(4-d)} \frac{1}{8 (2 \pi)^{2 d-1}} \left( \frac{s}{4} \right)^{d-3} \frac{2 \pi^{(d-1)/2}} {\Gamma((d-1)/2)} \frac{2 \pi^{(d-2)/2}}{\Gamma((d-2)/2)}  \\
&\qquad \times \left(\int \frac{d x_1 d x_2}{(x_1 x_2)^{4-d}}\right) \left( \frac{4(1-x_1)(1-x_2)(1-x_3)}{x_1^2 x_2^2} \right)^{(d-4)/2}  \\
&= \frac{\mu^{2(4-d)} s^{d-3}}{2^{2d-5/2}(2\pi)^{d-3/2}} \frac{1} {\Gamma((d-1)/2)\Gamma((d-2)/2)} \\
&\qquad \times \int d x_1 d x_2 \left[(1-x_1)(1-x_2)(1-x_3)\right]^{(d-4)/2}.  
\end{split} \eeq

Again writing $d = 4 - 2 \epsilon$,
\beq \begin{split}
\int d\Pi_3 &= \frac{s(\mu^2/s)^{2\epsilon}}{2^{11/2-4\epsilon}(2\pi)^{5/2-2\epsilon}} \frac{1} {\Gamma(\frac{3}{2}-\epsilon)\Gamma(1-\epsilon)} \int d x_1 d x_2 \left[(1-x_1)(1-x_2)(1-x_3)\right]^{-\epsilon}  \\
&= \frac{s}{16 (2\pi)^3} \frac{(4\pi)^{2\epsilon}(\mu^2/s)^{2\epsilon}} {\Gamma(2-2\epsilon)} \int d x_1 d x_2 \left[(1-x_1)(1-x_2)(1-x_3)\right]^{-\epsilon}. 
\end{split} \eeq
In the last line we have used the relation $\Gamma(z)\Gamma(z+1/2) = 2^{1/2-2z}\sqrt{2\pi}\Gamma(2z)$.  Our final result:
\beq
\label{eq:PS3}
\int d\Pi_3 = \frac{s}{128 \pi^3} \frac{(4\pi \mu^2 /s)^{2\epsilon}} {\Gamma(2-2\epsilon)} \int d x_1 d x_2 \left[(1-x_1)(1-x_2)(1-x_3)\right]^{-\epsilon}.
\eeq

Whew!  Now we can actually start calculating diagrams.

\section{Tree-level cross section}
We first calculate the tree-level cross section $\sigma(e^+e^- \to q\bar{q})$.  The calculation is identical to $e^+e^- \to \mu^+\mu^-$, up to an overall color factor.  There are no divergences to worry about, so we will go ahead and set $d=4$ for this part of the calculation.  The matrix element is:
\beq
i\mathcal{M} = \left[\bar{v}(p_2)(i e \gamma^\mu) u(p_1) \right] \frac{-i g_{\mu \nu}}{q^2} \left[ \bar{u}(k_1)(i e \gamma^\nu) v(k_2) \right].
\eeq
In the notation of Eqs.~\ref{eq:lmunu} and \ref{eq:mesimple}, we can write $|\mathcal{M}|^2 = L_{\mu \nu} M^{\mu \nu}$, with:
\begin{align}
\label{eq:treelmunu}
L^{\mu \nu} &= \frac{e^2}{s}\left[\bar{v}(p_2) \gamma^\mu u(p_1) \right] \left[\bar{u}(p_1)\gamma^\nu v(p_2) \right]; \\
\label{eq:treehmunu}
H^{\mu \nu} &= \frac{e^2 Q_f^2}{s}\left[\bar{u}(k_2) \gamma^\mu v(k_1) \right] \left[\bar{v}(k_1)\gamma^\nu u(k_2) \right].
\end{align}
As shown in Eq.~\ref{eq:contracted}, we only need $g_{\mu \nu}L^{\mu \nu}$ and $g_{\mu \nu}H^{\mu \nu}$.  We now calculate these, summing over final spins and averaging over initial spins.
\begin{align}
L \equiv g_{\mu \nu}\bar{L}^{\mu \nu} &= \frac{e^2}{4 s} \text{Tr} \left[ \slashed{p}_2 \gamma^\mu \slashed{p}_1 \gamma_\mu \right]  \nonumber \\
&=  \frac{-2 e^2}{ s} p_1 \cdot p_2 \nonumber  \\
\label{eq:L}
&= -e^2. \\
H \equiv g_{\mu \nu}\bar{H}^{\mu \nu} &= \frac{e^2 Q_f^2}{s} \text{Tr} \left[ \slashed k_1 \gamma^\mu \slashed k_2 \gamma_\mu \right] \nonumber  \\
&= -4 e^2 Q_f^2. \label{eq:H}
\end{align}
Plugging into Eq.~\ref{eq:xsection}, and using Eq.~\ref{eq:PS2}, we can find the cross section:
\beq \begin{split}
\sigma_\text{tree} &= \frac{1}{2s}\frac{1}{3}(-e^2)\frac{1}{8 \pi} (-4 e^2 Q_f^2)  \\
&= \frac{4 \pi \alpha^2}{3 s} Q_f^2. 
\end{split} \eeq
Summing over quark charges and colors, we get our final expression:
\beq \begin{split}
\label{eq:xsec_tree}
\sigma_\text{tree} &= \frac{4 \pi \alpha^2}{3 s} \times N_c \sum_f Q_f^2  \\
&\equiv \sigma_0 \times N_c \sum_f Q_f^2.
\end{split} \eeq
The astute reader will note that $\sigma_0$ is the total (tree-level, massless) cross section for $e^+e^-\to\mu^+\mu^-$.  In the massless limit, the only difference for quarks is the color and charge factors.

\section{Virtual correction}
\label{virtual}

We'll start with the virtual corrections.  The two leg corrections involve scaleless integrals (there is no dimensionful quantity that the integral over the loop momentum could depend on).  In dimensional regularization, the integrals have non-zero dimension and therefore must be equal to zero. The relevant diagram is the tree-level diagram with a gluon connecting the quark lines.  The $\mathcal{O}(\alpha_s)$ contribution to the total cross section comes from the interference term $2\ \mathfrak{Re}(\mathcal{M}^*_\text{tree}\mathcal{M}_\text{virtual})$.  The matrix elements share the same structure on the leptonic side, so $L^{\mu\nu}$ (Eq.~\ref{eq:treelmunu}) is unchanged.  Meanwhile, $H^{\mu\nu}$ shares one factor with the tree-level calculation (Eq.~\ref{eq:treehmunu}).  We only need to calculate the other half.  As in Eq.~\ref{eq:lmunu}:
\beq \begin{split}
H^{\mu \nu} &\equiv \frac{1}{s}H_\text{tree}^{\mu *}H_\text{virt}^{\nu} ; \\
H_\text{tree}^{\mu *} &= \bar{v}(k_2)(-i e Q_f \gamma^\mu) u(k_1) ; \\
H_\text{virt}^{\nu} &= \mu^{4-d} \int \frac{d^d p}{(2 \pi)^d} \bar{u}(k_1)(i g \gamma^\alpha t^A) \frac{i (\slashed k_1 + \slashed p)}{(k_1+p)^2}(i e Q_f \gamma^\nu )\frac{i (\slashed p - \slashed k_2)}{(p - k_2)^2}(i g \gamma^\beta t^B) v(k_2) \frac{-i g_{\alpha \beta} \delta^{A B}}{p^2}  \\
 &= g^2 e Q_f \mu^{4-d} \int \frac{d^d p}{(2 \pi)^d} \frac{\bar{u}(k_1) (t^A t^A) \gamma^\alpha (\slashed k_1 + \slashed p) \gamma^\nu (\slashed p - \slashed k_2)\gamma_\alpha v(k_2)}{(k_1+p)^2 (p - k_2)^2 p^2}.
\end{split} \eeq
We work in $d$ dimensions from the start (hence the $\mu$ factor).  Summing over final-state spins, but not color (we leave an implicit $\delta$ function in color space, as in the tree-level calculation), and using $t^A t^A = C_F \mathcal{I}$:
\beq \begin{split}
\label{eq:virt_gH_bef_tr}
g_{\mu \nu} H^{\mu \nu} &= \frac{-i}{s} g^2 e^2 Q_f^2 C_F \mu^{4-d} \int \frac{d^d p}{(2 \pi)^d} \frac{\tr \left[ \gamma_\mu \slashed k_1 \gamma^\alpha (\slashed k_1 + \slashed p) \gamma^\mu (\slashed p - \slashed k_2)\gamma_\alpha \slashed k_2 \right]}{(k_1+p)^2 (p - k_2)^2 p^2}.
\end{split} \eeq
Performing the trace requires $\gamma$ matrix contractions in $d$ dimensions:
\begin{align}
\gamma^a\gamma_a &= d \label{eq:con0}\\
\gamma^a \gamma^b \gamma_a &= -(d-2)\gamma^b \label{eq:con1}\\
\gamma^a \gamma^b \gamma^c \gamma_a &= 4 g^{cd} -(4-d)\gamma^b \gamma^c \label{eq:con2}\\
\gamma^a \gamma^b \gamma^c \gamma^d \gamma_a &= -2 \gamma^d \gamma^c \gamma^b +(4-d)\gamma^b \gamma^c \gamma^d \label{eq:con3}.
\end{align}
Now, we can do the trace, as usual setting $d = 4 - 2\epsilon$ for simplicity:
\beq
\begin{split}
\tr [ \ldots ]& = -2 \tr \left[(\slashed k_1 + \slashed p)  \gamma^\alpha \slashed k_1 (\slashed p - \slashed k_2)\gamma_\alpha \slashed k_2  \right]  + 2 \epsilon \tr \left[ \slashed k_1 \gamma^\alpha (\slashed k_1 + \slashed p) (\slashed p - \slashed k_2)\gamma_\alpha \slashed k_2 \right]  \\
 &=  -2 \left\{ 16 k_1 \cdot (p - k_2) k_2 \cdot (k_1+p)  - 2 \epsilon \tr \left[\slashed p \slashed k_1 \slashed p \slashed k_2  \right] \right\}  \\
 & \quad + 2 \epsilon \left\{ 8 s (k_1+p) \cdot (p-k_2) - 4 \epsilon s p^2 \right\}   \\
 &=  -2 \left\{ 16 k_1 \cdot (p - k_2) k_2 \cdot (k_1+p)  - 8 \epsilon \left[ 2 p \cdot k_1 p \cdot k_2 - \frac{s}{2} p^2 \right] \right\}  \\
 & \quad + 2 \epsilon \left\{ 8 s (k_1+p) \cdot (p-k_2) - 4 \epsilon s p^2 \right\}   \\
 &=  -32 k_1 \cdot (p - k_2) k_2 \cdot (k_1+p) - 8 \epsilon^2 s p^2  \\
 & \quad +16 \epsilon \left[ 2 p \cdot k_1 p \cdot k_2 - \frac{s}{2} p^2 + s (k_1+p) \cdot (p-k_2)\right]  \\
 &=  32 k_1 \cdot (k_2-p) k_2 \cdot (k_1+p) - 8 \epsilon^2 s p^2  \\
 & \quad +32 \epsilon \left[p \cdot k_1 p \cdot k_2 + \frac{s}{2} (k_1-k_2) \cdot p - \frac{s^2}{4} \right] + 8\epsilon s p^2  \\
 &=  32 k_1 \cdot (k_2-p) k_2 \cdot (k_1+p) + 8 (1- \epsilon) \epsilon s p^2  \\
 & \quad -32 \epsilon \left[ (\frac{s}{2} - p \cdot k_1) (\frac{s}{2} + p \cdot k_2) \right]  \\
 &=  32 k_1 \cdot (k_2-p) k_2 \cdot (k_1+p) + 8 (1- \epsilon) \epsilon s p^2  \\
 & \quad -32 \epsilon \left[ k_1 \cdot (k_2-p) k_2 \cdot (k_1+p) \right]  \\
 &=  8 (1-\epsilon) \left[ 4 k_1 \cdot (k_2-p) k_2 \cdot (k_1+p) + \epsilon s p^2 \right].
\end{split}
\eeq
In the second line, we have used $\slashed a \slashed a = a^2$ and $k_1^2 = k_2^2 = 0$.  We have also used $k_1 \cdot k_2 = \frac{1}{2} (k_1 + k_2)^2 = \frac{s}{2}$.  Plugging into Eq.~\ref{eq:virt_gH_bef_tr}:
\beq
\label{eq:virt_gH_aft_tr}
g_{\mu \nu} H^{\mu \nu} = \frac{-8i}{s} g^2 e^2 Q_f^2 C_F (1-\epsilon) \mu^{2 \epsilon} \int \frac{d^d p}{(2 \pi)^d} \frac{ 4 k_1 \cdot (k_2-p) k_2 \cdot (k_1+p) + \epsilon s p^2}{(k_1+p)^2 (p - k_2)^2 p^2}.
\eeq

We now introduce Feynman parameters in the standard way:
\beq
\frac{1}{(k_1+p)^2 (p - k_2)^2 p^2} = \int_0^1 dx \int_0^{1-x} dy \frac{2}{\left[ x(k_1+p)^2  + y(p - k_2)^2 +(1-x-y)p^2 \right]^3}. 
\eeq
We can rewrite the denominator and shift variables to write it in the form $[l-\Delta]^3$:
\beq
\begin{split}
D &\equiv p^2 + 2x p \cdot k_1 - 2 y p \cdot k_2 \\
 &= p^2 + 2 p \cdot (x k_1 - y k_2) \\
l &\equiv p + (x k_1 - y k_2) \\
D &= l^2 - (x k_1 - y k_2)^2 \\
 &= l^2 + x y s 
 \end{split}
\eeq
We can change the integration variable to $l$ because we're integrating over all $p$ and $l$ is just an constant additive shift.  Now we need to rewrite the numerator in terms of $l$ instead of $p$:
\beq
\begin{split}
N &\equiv 4 k_1 \cdot (k_2-p) k_2 \cdot (k_1+p) + \epsilon s p^2 \\
&= 4 k_1 \cdot \left((1-y)k_2- l + x k_1\right) k_2 \cdot \left((1-x) k_1+ l -y k_2 \right) + \epsilon s \left( l^2 - 2 l \cdot (xk_1 - yk_2) - x y s \right).
\end{split}
\eeq
The integral over an odd number of $l^\mu$ factors will vanish by parity, so we can drop terms linear in $l$:
\begin{equation}
\begin{split}
N &= 4 k_1 \cdot \left((1-y)k_2- l \right) k_2 \cdot \left((1-x) k_1+ l \right) + \epsilon s \left( l^2 - x y s \right) \\
&= 4 \left[ (1-x)(1-y)\frac{s^2}{4} - (k_1\cdot l) (k_2 \cdot l) \right] + \epsilon s \left( l^2 - x y s \right).
\end{split}
\end{equation}
Again using symmetry, we can replace $l^\mu l^\nu \to \frac{1}{d} l^2 g^{\mu \nu}$ inside the integral (after integrating, the tensor structure of the integral can only come from $g^{\mu\nu}$; contracting with $g_{\mu\nu}$ fixes the coefficient):
\begin{equation}
\begin{split}
N &= 4 \left[ (1-x)(1-y)\frac{s^2}{4} - k_1^\mu k_2^\nu l_\mu l_\nu \right] + \epsilon s \left( l^2 - x y s \right) \\
&= 4 \left[ (1-x)(1-y)\frac{s^2}{4} - \frac{1}{2(2-\epsilon)} \frac{s}{2} l^2 \right] + \epsilon s \left( l^2 - x y s \right) \\
&=  \left[ (1-x)(1-y) -\epsilon x y \right] s^2 + \left[ -\frac{1}{2-\epsilon} + \epsilon \right] s l^2.
\end{split}
\end{equation}
Plugging everything into Eq.~\ref{eq:virt_gH_aft_tr}:
\beq
\label{eq:virt_gH_with_ints}
\begin{split}
g_{\mu \nu} H^{\mu \nu} =& \frac{-16i}{s} g^2 e^2 Q_f^2 C_F (1-\epsilon)  \mu^{2 \epsilon} \\
 &\times \int_0^1 dx \int_0^{1-x} dy \int \frac{d^d l}{(2 \pi)^d} \frac{ \left[ (1-x)(1-y) -\epsilon x y \right] s^2 + \left[ -\frac{1}{2-2\epsilon} + \epsilon \right] s l^2}{(l^2 + x y s)^3} \\
\equiv& \frac{-16i}{s} g^2 e^2 Q_f^2 C_F (1-\epsilon)  \mu^{2\epsilon} \int dx\,dy \left[ C_0 I_0(-x y s) + C_2 I_2(-x y s) \right] .
\end{split}
\eeq
In the second line we've separated the simple integrals from their coefficients.  I'll just pull the standard forms out of Peskin and Schroeder:
\beq
\label{eq:ints}
\begin{split}
I_0(\Delta) & \equiv \int \frac{d^d l}{(2 \pi)^d} \frac{1}{(l^2 - \Delta)^3} \\
 &= \frac{-i}{(4 \pi)^{2-\epsilon}} \frac{\Gamma (1+\epsilon)}{2} \left(\frac{1}{\Delta} \right)^{1+\epsilon}; \\
I_2(\Delta) & \equiv \int \frac{d^d l}{(2 \pi)^d} \frac{l^2}{(l^2 - \Delta)^3} \\
 &= \frac{i}{(4 \pi)^{2-\epsilon}} \frac{(2-\epsilon)}{2} \Gamma(\epsilon) \left(\frac{1}{\Delta} \right)^\epsilon \\
 &= \frac{i}{(4 \pi)^{2-\epsilon}} \frac{(2-\epsilon)}{2} \frac{\Gamma(1+\epsilon)}{\epsilon} \left(\frac{1}{\Delta} \right)^\epsilon \\ 
 &= -\frac{(2-\epsilon)}{\epsilon}\Delta I_0(\Delta).
 \end{split}
\eeq
Combining the two terms:
\beq
\label{eq:virt_gH_ints}
\begin{split}
&C_0 I_0(-x y s) + C_2 I_2(-x y s) \\
&= \left[ (1-x)(1-y) -\epsilon x y \right] s^2 I_0(-x y s) + \left[ -\frac{1}{2-\epsilon} + \epsilon \right] s I_2(-x y s) \\
&= \left\{ \left[ (1-x)(1-y) -\epsilon x y \right] s^2 + \left[ \frac{1}{2-\epsilon} - \epsilon \right] \frac{(2-\epsilon)}{\epsilon} (-x y s) s \right\} I_0(-x y s) \\
&= \left\{ \left[ (1-x-y +x y) -\epsilon x y \right] -x y \left[ \frac{1}{\epsilon} - (2-\epsilon) \right] \right\} s^2 I_0(-x y s) \\
&= \left\{ (1-x-y) + x y \left[ 3-2\epsilon - \frac{1}{\epsilon} \right]  \right\} s^2 I_0(-x y s) \\
&= \left\{ (1-x-y) - \frac{x y}{\epsilon} (1-\epsilon)(1-2\epsilon) \right\} s^2 I_0(-x y s). \\
\end{split}
\eeq
We can write out $I_0(-x y s)$:
\beq
\begin{split}
I_0(-x y s) &=  \frac{-i}{(4 \pi)^{2-\epsilon}} \frac{\Gamma (1+\epsilon)}{2} \left(\frac{1}{-x y s} \right)^{1+\epsilon} \\
&=  \frac{i}{32 \pi^2 s} \left(\frac{-4 \pi }{s}\right)^\epsilon \Gamma (1+\epsilon) \left(\frac{1}{x y } \right)^{1+\epsilon}.
\end{split}
\eeq
Plugging Eq.~\ref{eq:virt_gH_ints} back into Eq.~\ref{eq:virt_gH_with_ints}:
\beq
\begin{split}
g_{\mu \nu} H^{\mu \nu} &= \frac{-16i}{s} g^2 e^2 Q_f^2 C_F (1-\epsilon)  \mu^{2\epsilon} \int dx\,dy \left[ C_0 I_0(-x y s) + C_2 I_2(-x y s) \right] \\
&= \frac{-16i}{s} g^2 e^2 Q_f^2 C_F (1-\epsilon)  \mu^{2\epsilon} \\
&\qquad \times \int dx\,dy \left\{ (1-x-y) - \frac{x y}{\epsilon} (1-\epsilon)(1-2\epsilon) \right\} s^2 I_0(-x y s)\\
&= \frac{1}{2\pi^2} g^2 e^2 Q_f^2 C_F (1-\epsilon) \left(\frac{-4 \pi \mu^2 }{s}\right)^\epsilon \Gamma (1+\epsilon) \\
&\qquad \times \int dx\,dy \left\{ \frac{(1-x-y)}{(x y)^{1+\epsilon}} + \frac{(1-\epsilon)(1-2\epsilon)}{\epsilon(xy)^\epsilon}  \right\} \\
&\equiv \frac{1}{2\pi^2} g^2 e^2 Q_f^2 C_F (1-\epsilon) \left(\frac{-4 \pi \mu^2 }{s}\right)^\epsilon \Gamma (1+\epsilon) I_\text{virt}(\epsilon). \\
\end{split}
\eeq
The first integral has a $1/\epsilon^2$ pole; the second is finite but multiplies $1/\epsilon$ so we must keep the integral to $\mathcal{O}(\epsilon)$.  The integrals  can be performed using Beta functions:
\beq
\begin{split}
I_\text{virt}(\epsilon) &\equiv \int_0^1 dx \int_0^{1-x} dy \left\{ \frac{(1-x-y)}{(x y)^{1+\epsilon}} + \frac{(1-\epsilon)(1-2\epsilon)}{\epsilon(xy)^\epsilon}  \right\} \\
&= \frac{1}{\epsilon^2} + \frac{3}{2 \epsilon} + 4 - \frac{\pi^2}{6} + \mathcal{O}(\epsilon).
\end{split}
\eeq
Before plugging in this form, let's collect all the factors in $\sigma_\text{virt}$, using Eqs.~\ref{eq:xsection}, \ref{eq:PS2}, \ref{eq:L}.  Note that while $L^{\mu\nu}$ has not changed, the contraction in Eq.~\ref{eq:L} has to be modified in $d$ dimensions, giving an extra factor of $(1-\epsilon)$.
\beq
\begin{split}
\sigma_\text{virt} =& \frac{1}{2s}\frac{1}{d-1}L\int d\Pi_{2} H \\
=& \frac{1}{2s}\frac{1}{3-2\epsilon} \left(-(1-\epsilon) e^2\right) \frac{1}{8 \pi}  \left(\frac{4 \pi \mu^2}{s} \right)^\epsilon \frac{\Gamma(1-\epsilon)}{\Gamma (2 - 2 \epsilon)} \\
&\times 2 \mathfrak{Re} \left[ \frac{1}{2\pi^2} g^2 e^2 Q_f^2 C_F (1-\epsilon) \left(\frac{-4 \pi \mu^2 }{s}\right)^\epsilon \Gamma (1+\epsilon) I_\text{virt}(\epsilon) \right]\\
=& \frac{-g^2 e^4 Q_f^2 C_F}{48 \pi^3 s}\frac{3}{3-2\epsilon} \frac{\Gamma(1-\epsilon) \Gamma(1+\epsilon)}{\Gamma (2 - 2 \epsilon)} (1-\epsilon)^2 \left(\frac{4 \pi \mu^2 }{s}\right)^{2\epsilon} I_\text{virt}(\epsilon) \mathfrak{Re}\left[(-1)^\epsilon \right] \\
=& \frac{-g^2 e^4 Q_f^2 C_F}{48 \pi^3 s} \Gamma(1-\epsilon) \Gamma(1+\epsilon)I_\text{virt}(\epsilon) \mathfrak{Re}\left[(-1)^\epsilon \right] H(\epsilon)\\
\end{split}
\eeq
We have pulled out the strange-looking term
\beq
\label{eq:Heps}
H(\epsilon) \equiv \frac{3}{3-2\epsilon} \frac{1}{\Gamma (2 - 2 \epsilon)} (1-\epsilon)^2 \left(\frac{4 \pi \mu^2 }{s}\right)^{2\epsilon} = 1 + \mathcal{O}(\epsilon)
\eeq
because this factor will appear in $\sigma_\text{real}$, too.  Now we can expand the rest of $\sigma_\text{virt}$ in $\epsilon$.  The only tricky bit is:
\beq
\mathfrak{Re}\left[(-1)^\epsilon \right] = \mathfrak{Re}\left[e^{\pm i \pi \epsilon} \right] = \mathfrak{Re}\left[ 1 \pm i \pi \epsilon - \epsilon^2 \frac{\pi^2}{2} + \cdots \right] = 1 - \epsilon^2 \frac{\pi^2}{2} + \cdots. 
\eeq
The $\pm$ comes from choosing which side of the branch cut to pick, and hence the sign of the $i\varepsilon$ in the propagators we ignored in Eq.~\ref{eq:virt_gH_bef_tr}; in the end taking the real part lets us ignore this subtlety.  With the expansion of the integral and the $\Gamma$ functions, we have (dropping terms $\mathcal{O}(\epsilon)$):
\beq
\begin{split}
\label{eq:xsec_virt}
\sigma_\text{virt} &= \frac{g^2 e^4 Q_f^2 C_F}{96 \pi^3 s} H(\epsilon)\left[ -\frac{2}{\epsilon^2} - \frac{3}{\epsilon} - 8 + \pi^2 \right] \\
&= \frac{4 \pi \alpha^2}{3 s} \frac{Q_f^2 \alpha_s C_F}{2 \pi} H(\epsilon)\left[ -\frac{2}{\epsilon^2} - \frac{3}{\epsilon} - 8 + \pi^2 \right] \\
&\Rightarrow \sigma_0 \left(\sum_f Q_f^2\right) N_c C_F \frac{\alpha_s}{2 \pi} H(\epsilon)\left[ -\frac{2}{\epsilon^2} - \frac{3}{\epsilon} - 8 + \pi^2 \right].
\end{split}
\eeq
In the last line we have summed over flavors and colors; recall the implicit $\delta$ function in color space.

\section{Real emission}

Now that we have calculated the correction to $\sigma$ from a virtual gluon, we need to consider the emission of a gluon.  There are two diagrams that contribute to the real correction: emission of a gluon from either of the quarks.  Since the final state is distinct from the tree-level and virtual diagrams, there is no interference.  So what we want to calculate is the sum of the two real diagrams.  We follow the same procedure as above and break the calculation into leptonic and hadronic parts.  $L$ will be the same.  Adding the two diagrams, we find $H$:
\beq
\begin{split}
i \mathcal{M}^\mu_\text{had} &= \frac{i}{2} g e Q_f \varepsilon_\alpha^*(k_3) \bar{u}(k_1) t^A \left[ \gamma^\mu \frac{(\slashed k_2 + \slashed k_3)}{k_2 \cdot k_3} \gamma^\alpha - \gamma^\alpha \frac{(\slashed k_1 + \slashed k_3)}{k_1 \cdot k_3} \gamma^\mu \right] v(k_2) \\
\Rightarrow H^\mu &= \frac{1}{2} g e Q_f \varepsilon_\alpha^*(k_3) \bar{u}(k_1) t^A \left[ \gamma^\mu \frac{(\slashed k_2 + \slashed k_3)}{k_2 \cdot k_3} \gamma^\alpha - \gamma^\alpha \frac{(\slashed k_1 + \slashed k_3)}{k_1 \cdot k_3} \gamma^\mu \right] v(k_2); \\
H &= \frac{1}{s} g_{\mu\nu} H^\mu H^{\nu *} \\
&= \frac{1}{4} g^2 e^2 Q_f^2 \tr(t^A t^A) \varepsilon_\alpha^* (k_3) \varepsilon_\beta (k_3) \bar{v}(k_2) \left[ \gamma^\alpha \frac{(\slashed k_2 + \slashed k_3)}{k_2 \cdot k_3} \gamma^\mu - \gamma^\mu \frac{(\slashed k_1 + \slashed k_3)}{k_1 \cdot k_3} \gamma^\alpha \right] \\
& \quad \times u(k_1)\bar{u}(k_1) \left[ \gamma_\mu \frac{(\slashed k_2 + \slashed k_3)}{k_2 \cdot k_3} \gamma^\beta - \gamma^\beta \frac{(\slashed k_1 + \slashed k_3)}{k_1 \cdot k_3} \gamma_\mu \right] v(k_2).
\end{split}
\eeq 
Doing the spin and polarization sums (the last allows the replacement $\varepsilon_\alpha^* (k_3) \varepsilon_\beta (k_3) \Rightarrow -g_{\alpha \beta}$):
\beq
\begin{split}
H &= -\frac{g^2 e^2 Q_f^2 C_F}{4 s} \tr \Bigg\{ \slashed k_2 \left[ \gamma^\alpha \frac{(\slashed k_2 + \slashed k_3)}{k_2 \cdot k_3} \gamma^\mu - \gamma^\mu \frac{(\slashed k_1 + \slashed k_3)}{k_1 \cdot k_3} \gamma^\alpha \right] \\
& \quad \times \slashed k_1 \left[ \gamma_\mu \frac{(\slashed k_2 + \slashed k_3)}{k_2 \cdot k_3} \gamma_\alpha - \gamma_\alpha \frac{(\slashed k_1 + \slashed k_3)}{k_1 \cdot k_3} \gamma_\mu \right] \Bigg\}.
\end{split}
\eeq
When the dust settles (no tricks here, just use the contraction formulae), we have:
\beq
H = -\frac{8 g^2 e^2 Q_f^2 C_F}{s} (1-\epsilon) \left[ \frac{(1-\epsilon) (x_1^2 + x_2^2) + 2\epsilon (1-x_3)}{(1-x_1)(1-x_2)} - 2\epsilon \right].
\eeq
Recall the definitions $x_i \equiv 2 k_i \cdot q / s$, where $q$ is the photon momentum; $x_3$ is fixed by the other two.  The only dependence on the final-state phase space is on $x_1$ and $x_2$.  Let's collect the factors in $\sigma_\text{real}$, using Eqs.~\ref{eq:xsection}, \ref{eq:PS3}, and \ref{eq:L}:
\beq
\begin{split}
\sigma_\text{real} &= \frac{1}{2s}\frac{1}{d-1}L\int d\Pi_3 H \\
&= \frac{1}{2s}\frac{\left(-(1-\epsilon) e^2\right)}{3-2\epsilon} \frac{s}{128 \pi^3} \frac{(4\pi \mu^2 /s)^{2\epsilon}} {\Gamma(2-2\epsilon)} \int d x_1 d x_2 \left((1-x_1)(1-x_2)(1-x_3)\right)^{-\epsilon} H \\
&= \frac{g^2 e^4 Q_f^2 C_F}{96 \pi^3 s} \frac{3 (1-\epsilon)^2}{3-2\epsilon }\frac{(4\pi \mu^2 /s)^{2\epsilon}} {\Gamma(2-2\epsilon)} \\
&\qquad \times \int d x_1 d x_2 \left[ \frac{(1-\epsilon) (x_1^2 + x_2^2) + 2\epsilon (1-x_3)}{(1-x_1)(1-x_2)} - 2\epsilon \right] \frac{1}{P(x_1,x_2)} \\
&= \frac{4 \pi \alpha^2}{3 s}\frac{\alpha_s Q_f^2 C_F}{2 \pi} H(\epsilon) \int d x_1 d x_2 \left[ \frac{(1-\epsilon) (x_1^2 + x_2^2) + 2\epsilon (1-x_3)}{(1-x_1)(1-x_2)} - 2\epsilon \right] \frac{1}{P(x_1,x_2)} \\
&= \sigma_0 \frac{\alpha_s Q_f^2 C_F}{2 \pi} H(\epsilon) I_\text{real}(\epsilon).
\end{split}
\eeq
where $P(x_1,x_2) \equiv \left[(1-x_1)(1-x_2)(1-x_3)\right]^\epsilon$.  The integral is
\beq
I_\text{real}(\epsilon) = \frac{2}{\epsilon^2} + \frac{3}{\epsilon} + 19/2 - \pi^2 + \mathcal{O}(\epsilon).
\eeq
All together, this yields (to $\mathcal{O}(1)$)
\beq
\sigma_\text{real}  = \sigma_0 \frac{\alpha_s Q_f^2 C_F}{2 \pi} H(\epsilon) \left[ \frac{2}{\epsilon^2} + \frac{3}{\epsilon} + 19/2 - \pi^2 \right].
\eeq
Finally, adding sums over flavor and color, we get
\beq
\label{eq:xsec_real}
\sigma_\text{real}  = \sigma_0 \left(\sum_f Q_f^2 \right) C_F N_c \frac{\alpha_s}{2 \pi} H(\epsilon) \left[ \frac{2}{\epsilon^2} + \frac{3}{\epsilon} + 19/2 - \pi^2 \right].
\eeq
%

\section{Final result}
\label{conclusions}

We now have our final result.  Combining Eqs.~\ref{eq:xsec_tree}, \ref{eq:xsec_virt}, and \ref{eq:xsec_real}, we find:
\beq
\label{eq:final}
\begin{split}
\sigma(e+e- \to \text{hadrons}) &= \sigma_0 \left(\sum_f Q_f^2 \right) N_c \left[1 + \frac{\alpha_s}{\pi} \frac{3 C_F}{4} \right] \\
&= \sigma_0 \left(\sum_f 3 Q_f^2 \right) \left[1 + \frac{\alpha_s}{\pi} \right].
\end{split}
\eeq
In the last line we have inserted the appropriate color factors for $SU(3)$.

\section{References}

The ``pink book'' \cite{PinkBook} is a good reference for the general ideas here.  For the calculational details any field text book should suffice; I've made extensive reference to Peskin and Schroeder \cite{Peskin}.  The ``Handbook of Perturbative QCD'' \cite{pQCD} is also a useful reference.  The CTEQ collaboration maintains a website with many useful and interesting QCD links \cite{CTEQ}.  Of particular note is a similar one-loop calculation of the Drell-Yann process by Bj\"orn P\"otter \cite{OneLoop}.


\graphicspath{{appendixB/graphics/}}

 
\chapter{The quark jet function in SCET}
\label{app:jetfunc}
 
In this appendix I give an example calculation in soft-collinear effective theory: the quark jet function (\eq{quark}) at next-to-leading order.\footnote{This appendix is taken from Sections 5.1, 5.2, and A.1 from \cite{Ellis:2010rw}.  Additional steps and explanations have been added.}  I repeat \eq{quark} here for reference (changing notation slightly):
\begin{align}
J^q_{n, \omega}(\tau_a) =&  \frac{1}{2 N_C} \Tr\sum_{X_n}\int\frac{dn\mcdot l}{2\pi} \int d^4 x \, e^{-il\cdot x} \frac{\bnslash}{2}  \delta_{N(\mathcal{J}(X_n)) - 1}  \nn \\
&\times\bra{0} \chi_{n,\omega}(x)\ket{X_n}\bra{X_n}\bar\chi_{n,\omega}(0)\ket{0} \delta(\tau_J - \tau_a(J(X_n))).
\label{Jdef}
\end{align}
From here on I will drop the ``$n$'' subscript on the jet function; the collinear direction will always be $n$.
 
The jet functions can be divided into two categories: those for measured jets, which are fixed to have a specific angularity $\tau_a$, and those for unmeasured jets, which are not.  I will denote the quark jet function by $J_{\omega}^q$, where $\omega$ is the label momentum, and the jet function $J_{\omega}^q(\tau_a)$ with an argument of $\tau_a$ denotes a measured jet.  I will calculate the jet function for the two classes of jet algorithms, $\kt$-type and cone-type algorithms.

\section{Phase Space Cuts}

To calculate the jet functions for a particular algorithm, we must impose phase space restrictions in the matrix element.  From the jet function definitions, these cuts take two forms.  One kind, imposed by the operator $\delta_{N(\mathcal{\hat{J}}(X_n)) - 1}$ in \eq{Jdef}, is common to every jet function.  It is the set of phase space restrictions related to the jet algorithm, and requires exactly one jet to arise from each collinear sector of SCET.  The other, imposed by the operator $\delta({\tau}_a - \hat \tau_a)$, is implemented only on measured jets and restricts the kinematics of the cut final states to produce a fixed value of the jet shape.  In this section we describe these phase space cuts in detail.

\begin{figure}[htbp]
\begin{center}
\includegraphics[width=0.4\textwidth]{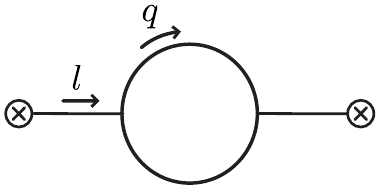}
\end{center}
\caption[A representative diagram for the NLO quark and gluon jet functions]{A representative diagram for the NLO quark and gluon jet functions. The incoming momentum is $l = \frac{n}{2}\omega + \frac{\bar n}{2} l^+$ and particles in the loop carry momentum $q$ (``particle 1'') and $l-q$ (``particle 2'').}
\label{fig:generic-jet}
\end{figure}

The typical form of the NLO diagrams in the jet functions is shown in \fig{fig:generic-jet}.  As shown in the figure, the momentum flowing through the graph has label momentum $l^- \equiv \bn\cdot l = \omega$ and residual momentum $l^+ \equiv n\cdot l$, and the loop momentum is $q$.  We will label ``particle 1'' as the particle in the loop with momentum $q$ and ``particle 2'' as the particle in the loop with momentum $l - q$.  For the quark jet, we take particle 1 as the emitted gluon and particle 2 as the quark.

As usual, the total forward scattering matrix element can be written as a sum over all cuts. Cutting through the loops corresponds to the interference of two real emission diagrams, each with two final state particles, whereas cutting through a lone propagator that is connected to a current corresponds to the interference between a tree-level diagram and a virtual diagram, each with a single final state particle. Thus, the phase space restrictions and measurements we impose act differently depending on where the diagrams are cut. In addition, since we will be working in dimensional regularization (with $d = 4-2\epsilon$), which sets scaleless integrals to zero, the only diagrams that contribute are the cuts through the loops. This means that we only need to focus on the form of phase-space restrictions and angularities in the case of final states with two particles.

The regions of phase space for two particles created by cutting through a loop in the jet function diagrams can be divided into three contributions:
\begin{enumerate}
\item Both particles are inside the jet.
\item Particle 1 exits the jet with energy $E_1<\Lambda$.
\item Particle 2 exits the jet with energy $E_2<\Lambda$.
\end{enumerate}
In contributions (2) and (3), the jet has only one particle, which is the remaining particle with $E > \Lambda$.  In principle, an exiting particle could have $E_i > \Lambda$ if it entered another jet.  As long as the jets are all well separated, this contribution is power suppressed, since it requires a collinear particle to be at large angle to the collinear direction.  If it were not power suppressed it would break factorization, since a given jet function does not know about the directions of other jets --- this is one reason we require $t_{ij} \gg 1$ (\eq{tijdef}).

It is well known\footnote{To those who know it well, of course --- e.g., \cite{Manohar:2006nz}.} that collinear integrations of jet functions can be allowed to extend over all values of loop momenta so long as a ``zero-bin subtraction'' is taken from the result to avoid double counting the soft region already accounted for in the soft function.  We will demonstrate that contributions (2) and (3) are power suppressed by $\cO(\Lambda/\omega)$, which scales as $\lambda^2$, after the zero-bin subtraction.

The phase space cuts that enforce both particles to be in the jet depend on the jet algorithm.  There are two classes of jet algorithm that we consider, cone-type algorithms and (inclusive) $\kt$-type algorithms, and all the algorithms in each class yield the same phase space cuts.  We label the phase space restrictions as $\Theta_\text{cone}$ and $\Theta_{\kt}$, generically $\Theta_\text{alg}$.  For the cone-type algorithms,
\[
\Theta_\text{cone} \equiv \Theta_\text{cone}(q, l^+) = \Theta \left( \tan^2{\frac{R}{2}} >  \frac{q^+}{q^- }\right) \Theta \left( \tan^2{\frac{R}{2}} > \frac{l^+ - q^+}{\omega - q^-} \right) .
\]
These $\Theta$ functions demand that both particles are within $R$ of the label direction.  For the $\kt$-type algorithms, the only restriction is that the relative angle of the particles be less than $R$:
\begin{align}
\Theta_{\kt} \equiv \Theta_{\kt}(q, l^+)  &= \Theta \left( \cos{R} < \frac{\vec{q} \cdot \vec{l} - q^2}{q \sqrt{l^2 + q^2 - 2 \vec{q} \cdot \vec{l}}} \right) \nn\\
&= \Theta \left( \tan^2{\frac{R}{2}} >  \frac{q^+\omega^2}{q^- \left(\omega - q^-\right)^2} \right) .
\end{align}
In the second line we took the collinear scaling of $q$ ($q^+ \ll q^-$). While this is not strictly needed, it makes the calculations significantly simpler.

For the phase space restrictions of zero-bin subtractions, we take the soft limit of the above restrictions (all components of $q$ scale like $\lambda^2$). The zero-bin subtractions are the same for all the algorithms we consider. For the case of particle 1, which has momentum $q$, the zero-bin phase space cuts are given by
\begin{align}
\Theta_\text{alg}^{(0)} = \Theta_\text{cone}^{(0)} =
\Theta_{\kt}^{(0)} =  \Theta \left( \tan^2{\frac{R}{2}} >  \frac{q^+}{q^- }\right)
.\end{align}
For the quark jet function, we don't need a zero bin for particle 2, since the quark is never soft.

For all the jet algorithms we consider, the zero-bin subtractions of the unmeasured jet functions are scaleless integrals.\footnote{Note that algorithms do exist that give nonzero zero-bin contributions to unmeasured jet functions \cite{Cheung:2009sg}.} However, for the measured jet functions, the zero-bin subtractions give nonzero contributions that are needed for the consistency of the effective theory.

In the case of a measured jet, in addition to the phase space restrictions we also demand that the jet contributes to the angularity by an amount $\tau_a$ with the use of the delta function $\delta_R = \delta(\tau_a - \hat \tau_a) $, which is given in terms of $q$ and $l$ by
\begin{equation}
\delta_R \equiv \delta_R(q, l^+) = \delta\left(\tau_a - \frac{1}{\omega}(\omega - q^-)^{a/2}(l^+ - q^+)^{1-a/2} - \frac{1}{\omega}(q^-)^{a/2}(q^+)^{1-a/2}\right)
.\end{equation}
In the zero-bin subtraction of particle 1, the on-shell conditions can be used to write the corresponding zero-bin $\delta$-function as
\begin{equation}
\delta^{(0)}_R = \delta\left(\tau_a - \frac{1}{\omega}(q^-)^{a/2}(q^+)^{1-a/2}\right).
\end{equation}

\section{Quark Jet Function}
\begin{figure}[htbp]
\includegraphics[width=\textwidth]{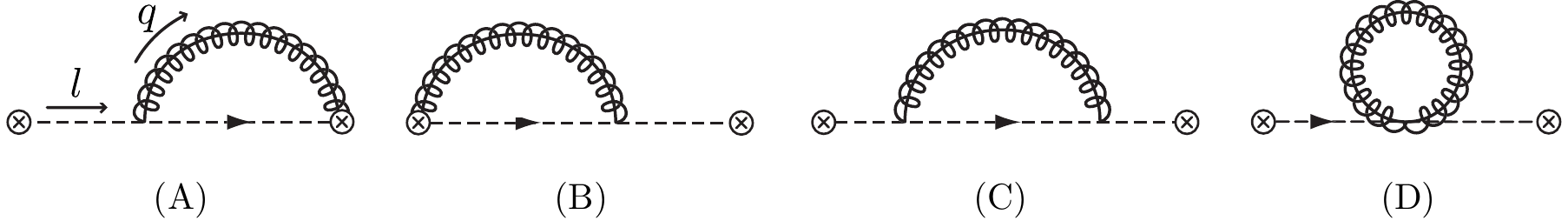}

\caption[Diagrams contributing to the quark jet function]{Diagrams contributing to the quark jet function. (A) and (B) Wilson line emission diagrams; (C) and (D) QCD-like diagrams.}
\label{fig:quark-jet-function}
\end{figure}

The diagrams corresponding to the quark jet function are shown in Fig.~\ref{fig:quark-jet-function}.  The fully inclusive quark jet function is defined as
\[
\int\!d^4 x \, e^{i l\cdot x} \bra{0}\chi_{n, \omega}^{a \alpha}(x) \bar\chi_{n, \omega}^{b \beta} (0)\ket{0} \equiv \delta^{a b}\left(\frac{\nslash}{2}\right)^{\alpha\beta} J^q_\omega(l^+) ,
\]
and has been computed to NLO (see, e.g., \cite{Bauer:2003pi, Bosch:2004th}) and to NNLO \cite{Becher:2006qw}.  Below we compute the quark jet function at NLO with phase space cuts for the jet algorithm for both the measured jet, $J^q_{\omega}(\tau_a)$, and the unmeasured jet, $J^q_{\omega}$.  As discussed above, the only nonzero contributions come from cuts through the loop when both particles are inside the jet.

\subsection{Measured Quark Jet}
\label{ssec:quark-meas}

The measured quark jet function includes contributions from naive Wilson line graphs (A) and (B) and QCD-like graphs (C) and (D) in \fig{fig:quark-jet-function}.  Using the SCET Feynman rules \cite{Bauer:2000yr}, the matrix element for graph (A), cut through the loop, is:
\begin{alignat}{2}
\Disc[\mathcal{M}_\text{A}] &=& \mu^{2\epsilon}\int\frac{d^d q}{(2\pi)^d} & \left(-\frac{g}{q^-}T^A \bar n^\mu\right)\left(i \frac{\nslash}{2} \frac{\omega - q^-}{(l-q)^2}\right)\left(i g T^B n_\nu \frac{\bnslash}{2}\right) \left(i \frac{\nslash}{2} \frac{\omega}{l^2} \right) \left(-i \frac{g^{\mu\nu} \delta^{AB}}{q^2}\right) \nn \\
 &&\times &  \left(-2 \pi i q^2 \delta(q^2) \Theta(q^0) \right) \left(-2 \pi i (l-q)^2 \delta\left((l-q)^2 \Theta(l^0 - q^0) \right) \right) \Theta_\text{alg} \delta_R\nn \\
 &=&  \mu^{2\epsilon} \int\frac{d^d q}{(2\pi)^d} & (4 \pi^2 g^2) (n\cdot \bn) \left(\frac{\nslash}{2}\frac{\bnslash}{2}\frac{\nslash}{2}\right) \left(T^AT^B\delta^{AB}\right) \left(\frac{\omega(\omega - q^-)}{q^- \, l^2}\right) \nn\\
 &&\times & \, \delta(q^2) \Theta(q^0) \delta\left((l-q)^2\right) \Theta(l^0 - q^0) \Theta_\text{alg} \delta_R \nn \\
 &=&  \mu^{2\epsilon} \int\frac{d^d q}{(2\pi)^d} & (8 \pi^2 g^2) \frac{\nslash}{2} C_F \mathbf{1} \left(\frac{\omega(\omega - q^-)}{q^- \, l^2}\right) \nn \\
 && \times & \delta(q^2)\Theta(q^0) \delta\left((l-q)^2\right)\Theta(l^0 - q^0) \Theta_\text{alg} \delta_R.
\end{alignat}
In the third line we have used the $SU(N)$ identity $T^AT^A = C_F \mathbf{1}$, where $\mathbf{1}$ is the identity matrix in color space, $\tr(\mathbf{1}) = N_C$.  The last two parentheticals in the (continued) first line represent the cut across the two propagators in the loop.  The factor of $\mu^{2\epsilon}$ is there to ensure the whole expression has the correct dimension.

Graph (B) is just the reflection of (A) and therefore has the same value.  As noted in \cite{Hornig:2009vb}, the sum of graphs (C) and (D) is equivalent to the plain QCD diagram, bracketed by projections onto the collinear propagator: $\mathcal{M}_\text{C} + \mathcal{M}_\text{D} = P_n \mathcal{M}_\text{QCD} P_\bn$.  This is because we can freely boost to a frame where the momenta in the QCD diagram have collinear scaling.  The projected and cut matrix element is thus:
\begin{alignat}{2}
\label{eq:QCDnaive1}
\Disc[\mathcal{M}_\text{C+D}] &=&  \mu^{2\epsilon}\int\frac{d^d q}{(2\pi)^d} & P_n \frac{i \slashed l}{l^2} \left( i g \gamma^\mu T^A \right) \left(\frac{i (\slashed l - \slashed q)}{(l-q)^2}\right)\left(i g \gamma^\nu T^B \right) \left(i \frac{\slashed l}{l^2} \right) P_\bn  \left(-i \frac{g^{\mu\nu} \delta^{AB}}{q^2}\right) \nn \\
 &&\times & \left(-2 \pi i q^2 \delta(q^2)\Theta(q^0)\right) \left(-2 \pi i (l-q)^2 \delta\left((l-q)^2\right) \right) \nn \\
 &&\times& \Theta(l^0 - q^0) \Theta_\text{alg} \delta_R\nn \\
 &=&  \mu^{2\epsilon} \int\frac{d^d q}{(2\pi)^d} & (-4 \pi^2 g^2) C_F \mathbf{1} \frac{1}{\omega^2 (l^+)^2} \omega \frac{\nslash}{2} \gamma^\mu \left(\slashed l - \slashed q\right) \gamma_\mu \omega \frac{\nslash}{2} \nn \\  
&& \times & \, \delta(q^2)\Theta(q^0) \delta\left((l-q)^2\right)\Theta(l^0-q^0) \Theta_\text{alg} \delta_R.
\end{alignat}
In the second line we have used several identities involving the collinear projection operators: $P_n \slashed l = P_n \left(\omega \frac{\nslash}{2} + l^+ \frac{\bnslash}{2}\right) = P_n \omega \frac{\nslash}{2}$ and likewise $\slashed l P_\bn = \omega \frac{\nslash}{2} P_\bn$; and $P_n \frac{\nslash}{2} = \frac{\nslash}{2}$, $\frac{\nslash}{2} P_\bn = \frac{\nslash}{2}$.  The Dirac structure can be simplified as follows:
\begin{align*}
\frac{\nslash}{2} \gamma^\mu \left(\slashed l - \slashed q\right) \gamma_\mu \frac{\nslash}{2} &= -(d-2) \frac{\nslash}{2} \left(\slashed l - \slashed q\right) \frac{\nslash}{2} \\
&=  -(d-2) \frac{\nslash}{2} \left((\omega - q^-) \frac{\nslash}{2} + (l^+ - q^+)\frac{\bnslash}{2} - \slashed{q}_{\perp} \right) \frac{\nslash}{2} \\
&=  -(d-2) \frac{\nslash}{2} (l^+ - q^+)\frac{\bnslash}{2} \frac{\nslash}{2} \\
&=  -2(1-\epsilon) (l^+ - q^+) \frac{\nslash}{2}.
\end{align*}
In the first line we have used a $\gamma$-matrix contraction in $d$ dimensions.  In the second we have used the facts that $(\nslash)^2 = 0$ and that $\nslash$ anticommutes with $\slashed{p}_\perp$.  Putting this back into \eq{eq:QCDnaive1} we have:
\begin{align}
\Disc[\mathcal{M}_\text{C+D}] =  \mu^{2\epsilon} \int\frac{d^d q}{(2\pi)^d} &(8 \pi^2 g^2) C_F \mathbf{1} \frac{1}{(l^+)^2} (1-\epsilon) (l^+ - q^+) \frac{\nslash}{2}  \nn \\
\times & \, \delta(q^2)\Theta(q^0) \delta\left((l-q)^2\right)\Theta(l^0-q^0) \Theta_\text{alg} \delta_R.
\end{align}

The total cut matrix element is:
\begin{align}
\Disc[\mathcal{M}] = (8 \pi^2 g^2)  C_F \mathbf{1} \frac{\nslash}{2} \mu^{2\epsilon} \int\frac{d^d q}{(2\pi)^d} & \left(\frac{2 (\omega - q^-)}{q^- \, l^+} + (1-\epsilon) \frac{l^+ - q^+}{(l^+)^2} \right) \nn \\
\times & \, \delta(q^2)\Theta(q^0) \delta\left((l-q)^2\right)\Theta(l^0-q^0) \Theta_\text{alg} \delta_R.
\end{align}

We can now plug this into \eq{quark} to find the full naive quark jet function:
\begin{align}
\label{Jnaivein}
\qjetnaive^q_{\omega}(\tau_a) =&  \frac{1}{2\CA} \Tr\sum_{X_n}\int\frac{dn\mcdot l}{2\pi} \int d^4 x \, e^{-il\cdot x} \frac{\bnslash}{2}  \delta_{n(\mathcal{J}(X_n)) - 1} \nn \\
&\times\bra{0} \chi_{n,\omega}(x)\ket{X_n}\bra{X_n}\bar\chi_{n,\omega}(0)\ket{0} \delta(\tau_J - \tau_a(J(X_n))) \nn \\
=& \frac{1}{2\CA} \Tr \int\frac{d l^+}{2\pi} \frac{\bnslash}{2} \Disc_{\tau_a, \text{alg}}[\mathcal{M}] \nn \\
=& \Tr\left(\frac{\nslash}{2} \frac{\bnslash}{2}\right) (4 \pi^2 g^2) C_F  \mu^{2\epsilon} \int\frac{d l^+}{2\pi} \int\frac{d^d q}{(2\pi)^d} \left(\frac{2 (\omega - q^-)}{q^- \, l^+} + (1-\epsilon) \frac{l^+ - q^+}{(l^+)^2} \right) \nn \\
&\times \, \delta(q^2)\Theta(q^0) \delta\left((l-q)^2\right)\Theta(l^0-q^0) \Theta_\text{alg} \delta_R \nn \\
=& g^2 C_F  \mu^{2\epsilon} \int\frac{d l^+}{2\pi} \frac{1}{(l^+)^2} \int\frac{d^d q}{(2\pi)^d} \left(\frac{4 l^+}{q^-} + 2 (1-\epsilon) \frac{l^+ - q^+}{\omega - q^-} \right) \nn \\
& \times \, 2 \pi \delta(q^+ q^- - q^2_\perp)\Theta(q^+)\Theta(q^-) 2 \pi \delta\left(l^+ - q^+ - \frac{q^2_\perp}{\omega - q^-} \right)\nn \\
& \times \Theta(\omega - q^-) \Theta(l^+ - q^+) \Theta_\text{alg} \delta_R.
\end{align}
The trace in the second line is over Dirac and color indices.  The contribution proportional to $1-\epsilon$ comes from the QCD-like graphs (C) and (D) in \fig{fig:quark-jet-function}.  Only the Wilson line graphs have a nonzero zero-bin limit, which comes from taking the scaling limit $q\sim\lambda^2$ of the naive contribution:
\begin{equation}
\label{Jzeroin}
\begin{split}
J^{q(0)}_{\omega}(\tau_a) &= 4g^2\mu^{2\epsilon} C_F \int\frac{dl^+}{2\pi}\frac{1}{l^+}\int \frac{d^d q}{(2\pi)^d}\frac{1}{q^-} 2\pi\delta(q^- q^+ - q_\perp^2)\Theta(q^-)\Theta(q^+) \\
&\quad \times 2\pi\delta\left(l^+ - q^+\right)\Theta(l^+ - q^+) \,  \Theta_\text{alg}^{(0)} \delta_R^{(0)}
.\end{split}
\end{equation}
All jet algorithms that we use yield the same zero-bin contribution, since the phase space cuts are the same.

To evaluate these integrals, we can start with the trivial $l^+$ integral over the $\delta$ function (note that the factor $\delta(q^2)$ enforces $q^2_\perp = q^+q^-$):
\begin{align*}
\qjetnaive^q_{\omega}(\tau_a) =& g^2 C_F  \mu^{2\epsilon} \int\frac{d l^+}{2\pi} \frac{1}{(l^+)^2} \int\frac{d^d q}{(2\pi)^d} \left(\frac{4 l^+}{q^-} + 2 (1-\epsilon) \frac{l^+ - q^+}{\omega - q^-} \right) \\
& \times \, 2 \pi \delta(q^+ q^- - q^2_\perp)\Theta(q^+)\Theta(q^-) 2 \pi \delta\left(l^+ - q^+ - \frac{q^2_\perp}{\omega - q^-} \right)\Theta(\omega - q^-) \Theta(l^+ - q^+) \Theta_\text{alg} \delta_R \\
=& g^2 C_F  \mu^{2\epsilon} \int\frac{d^d q}{(2\pi)^d}  \left(\frac{\omega-q^-}{\omega q^+}\right)^2 \left(\frac{4 \omega q^+}{q^-(\omega - q^-)} + 2 (1-\epsilon) \frac{q^- q^+}{(\omega - q^-)^2} \right) \\
& \times \, 2 \pi \delta(q^+ q^- - q^2_\perp)\Theta(q^+)\Theta(q^-) \Theta(\omega - q^-) \Theta_\text{alg} \delta_R.
\end{align*}

It is easiest to split the $q$ integral into light-cone components:
\begin{align*}
\frac{d^dq}{(2\pi)^d} &= \frac{1}{2} dq^+ dq^- d^{d-2}q_\perp \\
&=  \frac{1}{2} dq^+ dq^- \Omega^{d-3} q^{d-3}_\perp dq_\perp \\
&= \frac{1}{2} dq^+ dq^- \Omega^{1-2\epsilon} q^{1-2\epsilon}_\perp dq_\perp \\
&= \frac{1}{4} dq^+ dq^- \Omega^{1-2\epsilon} q^{-2\epsilon}_\perp dq^2_\perp \\
&= \frac{1}{2} dq^+ dq^- \frac{\pi^{1-\epsilon}}{\Gamma(1-\epsilon)} \frac{dq^2_\perp}{q^{2\epsilon}_\perp}.
\end{align*}
In the second line we have integrated out the $d-3$ angles of the $\vect{q}_\perp$ subspace, which do not appear in the integrand.  Returning to the full integral:
\begin{align*}
\qjetnaive^q_{\omega}(\tau_a) =& g^2 C_F  \mu^{2\epsilon} \frac{\pi^{1-\epsilon}}{\Gamma(1-\epsilon)} \frac{1}{(2\pi)^{3-2\epsilon}} \int\frac{dq^+ dq^-}{(q^+ q^-)^\epsilon}  \left(\frac{\omega-q^-}{\omega q^+}\right)^2 \left(\frac{2 \omega q^+}{q^-(\omega - q^-)} + (1-\epsilon) \frac{q^- q^+}{(\omega - q^-)^2} \right) \\
& \times \, \Theta(q^+)\Theta(q^-) \Theta(\omega - q^-) \Theta_\text{alg} \delta_R \\
=& \frac{g^2 C_F}{16 \pi^2} \frac{\left(4 \pi \mu^2\right)^\epsilon}{\Gamma(1-\epsilon)} \frac{1}{\omega^2} \int_0^\omega \frac{dq^-}{(q^-)^{1+\epsilon}} \int_0^\infty \frac{dq^+}{(q^+)^{1+\epsilon}} \left(4 \omega (\omega - q^-) + 2 (1-\epsilon) (q^-)^2 \right) \Theta_\text{alg} \delta_R \\
=& \frac{\as C_F}{2 \pi} \left( \frac{4 \pi \mu^2}{\omega^2}\right)^\epsilon \frac{1}{\Gamma(1-\epsilon)} \int_0^1 \frac{dx}{x^{1+\epsilon}} \int_0^\infty \frac{dy}{y^{1+\epsilon}} \left(2(1 - x) + (1-\epsilon) x^2 \right) \Theta_\text{alg} \delta_R.
\end{align*}
In the last line we have introduced scaled variables $x \equiv q^-/\omega$ and $y\equiv q^+/\omega$.  To go further we must plug in explicit forms for $\Theta_\text{alg}$ and $\delta_R$.  Note that all we have needed to know so far is that they are both independent of the direction of $\vect{q}_\perp$.  For now we will consider the case of a cone-type algorithm:
\begin{align*}
\Theta_\text{cone} &= \Theta \left( \tan^2{\frac{R}{2}} >  \frac{q^+}{q^- }\right) \Theta \left( \tan^2{\frac{R}{2}} > \frac{l^+ - q^+}{\omega - q^-} \right) \\
&= \Theta \left( \tan^2{\frac{R}{2}} >  \frac{q^+}{q^- }\right) \Theta \left( \tan^2{\frac{R}{2}} > \frac{q^+q^-}{(\omega - q^-)^2} \right) \\
&= \Theta \left( \tan^2{\frac{R}{2}} >  \frac{y}{x}\right) \Theta \left( \tan^2{\frac{R}{2}} > \frac{xy}{(1-x)^2} \right).
\end{align*}
Meanwhile, the $\tau_a$-enforcing $\delta$ function is:
\begin{align*}
\delta_R &= \delta\left(\tau_a - \frac{1}{\omega}(\omega - q^-)^{a/2}(l^+ - q^+)^{1-a/2} - \frac{1}{\omega}(q^-)^{a/2}(q^+)^{1-a/2}\right) \\
&= \delta\left(\tau_a - \frac{1}{\omega}(\omega - q^-)^{a-1}(q^-q^+)^{1-a/2} - \frac{1}{\omega}(q^-)^{a/2}(q^+)^{1-a/2}\right) \\
&= \delta\left(\tau_a - (1 - x)^{a-1}(xy)^{1-a/2} -(x)^{a/2}(y)^{1-a/2}\right) \\
&= \delta\left(\tau_a - (xy)^{1-a/2} \left( (1 - x)^{a-1} - (x)^{a-1} \right) \right).
\end{align*}

Putting this all together, we have:
\begin{align*}
\qjetnaive^{q}_\text{cone}(\tau_a) =& \frac{\as C_F}{2 \pi} \left( \frac{4 \pi \mu^2}{\omega^2}\right)^\epsilon \frac{1}{\Gamma(1-\epsilon)} \int_0^1 \frac{dx}{x^{1+\epsilon}} \int_0^\infty \frac{dy}{y^{1+\epsilon}} \left(2(1 - x) + (1-\epsilon) x^2 \right) \\
&\times \Theta \left( \tan^2{\frac{R}{2}} >  \frac{y}{x}\right) \Theta \left( \tan^2{\frac{R}{2}} > \frac{xy}{(1-x)^2} \right) \\
&\times \delta\left(\tau_a - (xy)^{1-a/2} \left((1 - x)^{a-1} - (x)^{a-1} \right) \right) \\
=& \frac{\as C_F}{2 \pi} \left( \frac{4 \pi \mu^2}{\omega^2}\right)^\epsilon \frac{1}{\Gamma(1-\epsilon)} \int_0^1 \frac{dx}{x^{1+\epsilon}} \int_0^{rx} \frac{dy}{y^{1+\epsilon}} \left( 2(1 - x) + (1-\epsilon) x^2 \right) \\
&\times \Theta \left(r > \frac{xy}{(1-x)^2} \right) \delta\left(\tau_a - (xy)^{1-a/2} \left( (1 - x)^{a-1} - (x)^{a-1} \right) \right),
\end{align*}
using the abbreviation $r \equiv \tan^2(R/2)$.  Doing the $y$ integral over the $\delta$ function:
\begin{align*}
\qjetnaive^{q}_\text{cone}(\tau_a) =& \frac{\as C_F}{2 \pi} \left( \frac{4 \pi \mu^2}{\omega^2}\right)^\epsilon \frac{1}{\Gamma(1-\epsilon)} \int_0^1 dx \left( \frac{1}{1-a/2} \right) \frac{1}{\tau^{\frac{\epsilon}{1-a/2}}} \left( x^{a-1} + (1-x)^{a-1} \right)^{\frac{\epsilon}{1-a/2}} \\
&\times \left(2\frac{1 - x}{x} + (1-\epsilon) x \right) \Theta \left(f_\text{cone}(x) > \frac{\tau_a}{r^{1-a/2}} \right),
\end{align*}
where $f_\text{cone}(x)$ is defined as
\[
f_\text{cone} = 
\begin{cases}
x^{2-a} \left( x^{a-1} + (1-x)^{a-1} \right) & \qquad x < 1/2 \\
(1-x)^{2-a} \left( x^{a-1} + (1-x)^{a-1} \right) & \qquad x > 1/2.
\end{cases}
\]

The integration region is plotted in \fig{fig:jetfunctions-integrations}.  We can exploit the symmetry of the $\Theta$ function around $x = 1/2$ and rewrite the $x$ integral as being from 0 to $1/2$:
\begin{align}
\qjetnaive^{q}_\text{cone}(\tau_a) =& \frac{\as C_F}{2 \pi} \left( \frac{4 \pi \mu^2}{\omega^2}\right)^\epsilon \frac{1}{\Gamma(1-\epsilon)} \left( \frac{1}{1-a/2} \right) \int_0^{1/2} dx  \frac{1}{\tau^{\frac{\epsilon}{1-a/2}}} \left( x^{a-1} + (1-x)^{a-1} \right)^{\frac{\epsilon}{1-a/2}} \nn \\
&\times \left(2\frac{1 - x}{x} + 2\frac{x}{1-x} + (1-\epsilon) \right) \Theta \left(f_\text{cone}(x) > \frac{\tau_a}{r^{1-a/2}} \right).
\label{eq:Jnaive1}
\end{align}

To evaluate the remaining integral, we can analytically extract the coefficient of $\delta(\tau_a)$ by integrating over $\tau_a$ and using the fact that the remainder is a plus distribution.  We define plus distributions as \cite{Stewart:2009yx}:
\begin{equation}
\label{mitplus}
[\Theta(x)g(x)]_+ = \lim_{\epsilon\to 0}\frac{d}{dx}[\Theta(x-\epsilon) G(x)],\qquad \text{with} \qquad G(x) = \int_1^x dx' g(x'),
\end{equation}
defined so as to satisfy the boundary condition $\int_0^1 dx[\Theta(x)g(x)]_+ = 0$.  If we write $\qjetnaive^{q}_\text{cone}(\tau_a) = A \delta(\tau_a) + \left[B \frac{\Theta(\tau_a)}{\tau_a}\right]_+$,
\begin{align*}
A =& \frac{\as C_F}{2 \pi} \left( \frac{4 \pi \mu^2}{\omega^2}\right)^\epsilon \frac{1}{\Gamma(1-\epsilon)} \left( \frac{1}{1-a/2} \right) \\
&\times \int_0^{1/2} dx  \left( x^{a-1} + (1-x)^{a-1} \right)^{\frac{\epsilon}{1-a/2}} \left(2\frac{1 - x}{x} + 2\frac{x}{1-x} + (1-\epsilon) \right) \\
&\times \int_0^\infty d\tau_a \frac{1}{\tau^{\frac{\epsilon}{1-a/2}}} \Theta \left(f_\text{cone}(x) > \frac{\tau_a}{r^{1-a/2}} \right).
\end{align*}
The $\tau_a$ integral is then simple:
\begin{align*}
\int_0^\infty d\tau_a \frac{1}{\tau^{\frac{\epsilon}{1-a/2}}} \Theta \left(f_\text{cone}(x) > \frac{\tau_a}{r^{1-a/2}} \right) =& \int_0^{\tau_a^\text{max}(x)} d\tau_a \frac{1}{\tau^{\frac{\epsilon}{1-a/2}}} \\
=& -\frac{1-a/2}{\epsilon}(\tau_a^\text{max}(x))^\frac{-\epsilon}{1-a/2},
\end{align*}
where $\tau_a^\text{max}(x) = r^{1-a/2} f_\text{cone}(x)$.  This leaves
\begin{align*}
A =& - \frac{\as C_F}{2 \pi} \left( \frac{4 \pi \mu^2}{\omega^2}\right)^\epsilon \frac{1}{\Gamma(1-\epsilon)} \frac{1}{\epsilon} \\
&\times \int_0^{1/2} dx  \left( x^{a-1} + (1-x)^{a-1} \right)^{\frac{\epsilon}{1-a/2}} \left(2\frac{1 - x}{x} + 2\frac{x}{1-x} + (1-\epsilon) \right) \\
& \qquad \times \left[ r^{1-a/2} x^{2-a} \left( x^{a-1} + (1-x)^{a-1} \right) \right]^\frac{-\epsilon}{1-a/2} \\
=& - \frac{\as C_F}{2 \pi} \left( \frac{4 \pi \mu^2}{r \omega^2}\right)^\epsilon \frac{1}{\Gamma(1-\epsilon)} \frac{1}{\epsilon} \int_0^{1/2} \frac{dx}{x^{2\epsilon}} \left(2\frac{1 - x}{x} + 2\frac{x}{1-x} + (1-\epsilon) \right).
\end{align*}
In the $x$ integral, only the $2/x^{1+2\epsilon}$ term diverges as $\epsilon \to 0$, and this term can be easily integrated exactly.  The rest of the terms can be expanded to $\mathcal{O}(\epsilon)$ and then integrated.  The result is
\beq
A = \frac{\as C_F}{2\pi} \left( \frac{4 \pi \mu^2}{r \omega^2}\right)^\epsilon \frac{1}{\Gamma(1-\epsilon)} \left( \frac{1}{\epsilon^2} + \frac{3}{2\epsilon} + \frac{7}{2} - \frac{\pi^2}{3} + 3 \ln 2 \right).
\eeq

We can find the rest of $\qjetnaive^{q}_\text{cone}(\tau_a)$ by taking $\tau_a > 0$ in \eq{eq:Jnaive1}, which enforces a lower cutoff in the $x$ integral.  This renders the whole integration finite and we can take $\epsilon \to 0$.  This yields
\begin{align}
&\left[B \frac{\Theta(\tau_a)}{\tau_a}\right]_+ \nn \\
&=  \frac{\as C_F}{2 \pi} \left( \frac{1}{1-a/2} \right) \left[ \int_{x_\text{cone}}^{1/2} dx \left(2\frac{1 - x}{x} + 2\frac{x}{1-x} + 1 \right) \frac{\Theta(\tau_a) \Theta(\tau_a^\text{max} - \tau_a)}{\tau_a} \right]_+ \nn \\
&=  \frac{\as C_F}{2 \pi} \left( \frac{1}{1-a/2} \right) \left[ \left( 2 \ln \left(\frac{1-x_\text{cone}}{x_\text{cone}}\right) - \frac{3}{2} (1-2 x_\text{cone}) \right) \frac{\Theta(\tau_a) \Theta(\tau_a^\text{max} - \tau_a)}{\tau_a} \right]_+,
\label{eq:Bcone}
\end{align}
where $f_\text{cone}(x_\text{cone}) = \frac{\tau_a}{r^{1-a/2}}$.  The upper cutoff $\Theta(\tau_a^\text{max} - \tau_a)$ appears because this equation has no solution for $\tau_a > \tau_a^\text{max} = r^{1-a/2}$.

All together the naive contribution is
\beq
\label{Jnaiveinresult}
\qjetnaive^q_{\omega}(\tau_a) = \frac{\alpha_s C_F}{2\pi} \frac{1}{\Gamma(1-\epsilon)} \left(\frac{4\pi\mu^2}{\omega^2\tan^2\frac{R}{2}}\right)^\epsilon\left( \frac{1}{\epsilon^2}  + \frac{3}{2\epsilon} \right) \delta(\tau_a)
 +  \frac{\alpha_s}{2\pi} \qjetnaive^q_\text{alg} (\tau_a), 
\eeq
where for cone-type algorithms we have found
\begin{align}
\label{Jnaivecone}
\qjetnaive^q_\text{cone} (\tau_a) =& \CF \left( \frac{7}{2} - \frac{\pi^2}{3} + 3 \ln 2 \right) \delta(\tau_a) \nn \\
&+ \left( \frac{\CF}{1-a/2} \right) \left[ \left( 2 \ln \left(\frac{1-x_\text{cone}}{x_\text{cone}}\right) - \frac{3}{2} (1-2 x_\text{cone}) \right)  \frac{\Theta(\tau_a) \Theta(\tau_a^\text{max} - \tau_a)}{\tau_a} \right]_+.
\end{align}

The only difference between the jet algorithms that we consider resides in the finite distribution $\qjetnaive^q_\text{alg}(\tau_a)$.  We have calculated this piece explicitly for cone-type algorithms, and give the result for $\kt$-type algorithms below.  Note that the divergent part of the naive contribution is proportional to $\delta(\tau_a)$. This is due to the fact that the jet algorithm regulates the distribution for $\tau_a >0$. The divergent plus distributions come entirely from the zero-bin subtraction, to which we now turn.

The zero-bin subtraction for the quark jet function is given by \eq{Jzeroin}, which we can evaluate similarly to the naive result:
\begin{align}
J^{q(0)}_{\omega}(\tau_a) =& 4g^2\mu^{2\epsilon} \CF \int\frac{dl^+}{2\pi}\frac{1}{l^+}\int \frac{d^d q}{(2\pi)^d}\frac{1}{q^-} 2\pi\delta(q^- q^+ - q_\perp^2)\Theta(q^-)\Theta(q^+) \nn \\
& \times 2\pi\delta\left(l^+ - q^+\right)\Theta(l^+ - q^+) \,  \Theta_\text{alg}^{(0)} \delta_R^{(0)} \nn \\
=& \frac{\as \CF }{\pi} \left(4\pi\mu^2 \right)^{\epsilon}  \frac{1}{\Gamma(1-\epsilon)}  \int_0 \frac{d q^+ d q^-}{(q^-q^+)^{1+\epsilon}} \Theta\left( r - q^+/q^- \right) \delta \left( \tau_a - \frac{1}{\omega} (q^+)^{1-a/2} (q^-)^{a/2} \right) \nn \\
=& \frac{\as \CF }{\pi} \left(\frac{4\pi\mu^2}{\omega^2} \right)^{\epsilon}  \frac{1}{\Gamma(1-\epsilon)}  \int_0^\infty \frac{dx}{x^{1+\epsilon}}  \int_0^{rx} \frac{dy}{y^{1+\epsilon}} \delta \left( \tau_a - y^{1-a/2} x^{a/2} \right) \nn \\
=& \frac{\as \CF }{\pi} \left(\frac{4\pi\mu^2}{\omega^2} \right)^{\epsilon}  \frac{1}{\Gamma(1-\epsilon)} \frac{1}{1-a/2} \tau_a^{-\left(1+\frac{\epsilon}{1-a/2}\right)} \int_0^\infty \frac{dx}{x^{1+2\epsilon}} \Theta(x r^{1-a/2} - \tau_a) \nn \\
=& \frac{\alpha_s C_F}{\pi} \left(\frac{4\pi\mu^2\tan^{2(1-a)}\frac{R}{2}}{\omega^2}\right)^\epsilon  \frac{1}{\Gamma(1-\epsilon)} \frac{1}{(1-a)} \frac{1}{\epsilon} \frac{1}{\tau_a^{1+2\epsilon}}.
\label{Jzeroinresult}
\end{align}
This can be broken into $\delta(\tau_a)$ and plus distribution pieces using the relation
\[
\frac{\Theta(x)}{x^{1+2\epsilon}} = -\frac{\delta(x)}{2\epsilon} + \left[ \frac{\Theta(x)}{x} \right]_+ - 2 \epsilon \left[ \frac{\Theta(x) \ln(x)}{x} \right]_+ + \mathcal{O}(\epsilon^2),
\]
valid for $\epsilon < 0$.

Adding the leading-order contribution to all of the NLO graphs and expanding in powers of $\epsilon$, adopting the $\overline{\text{MS}}$ scheme (i.e., taking $\mu^2 \to \frac{\mu^2}{4\pi} e^{\gamma_E}$), we find the total quark jet function
\begin{align}
\label{Jtotal}
J^q_\omega(\tau_a) = \delta (\tau_a) + \qjetnaive^q_{\omega}(\tau_a) - J^{q(0)}_{\omega}(\tau_a)  &= \Biggl\{ 1 + \frac{\alpha_s C_F}{\pi}\Biggl[ \frac{1-\frac{a}{2}}{1-a}\frac{1}{\epsilon^2} + \frac{1-\frac{a}{2}}{1-a}\frac{1}{\epsilon}\ln\frac{\mu^2}{\omega^2} + \frac{3}{4\epsilon}
 \Biggr]\Biggr\}\delta(\tau_a) \nn \\
 & \qquad - \frac{\alpha_s C_F}{\pi}\Biggl[ \frac{1}{\epsilon}\frac{1}{1-a} \frac{\Theta(\tau_a)}{\tau_a} \Biggr]_+  + \frac{\as} {2\pi} J^q_\text{alg}(\tau_a).
\end{align}
This agrees with the standard jet function $J(k^+)$ given in  \cite{Bauer:2003pi, Bosch:2004th} by setting $a=0$ and $k^+ = \omega\tau_a$.  We have shown the divergent terms explicitly, and collect the finite pieces in $J_\text{alg}^q(\tau_a)$, given below.  Note that there is no jet algorithm dependence in the divergent parts of the jet function at this order in perturbation theory.

\subsubsection{Finite Parts of the Measured Quark Jet Function}

\begin{figure}[htbp]
\includegraphics[width = \textwidth]{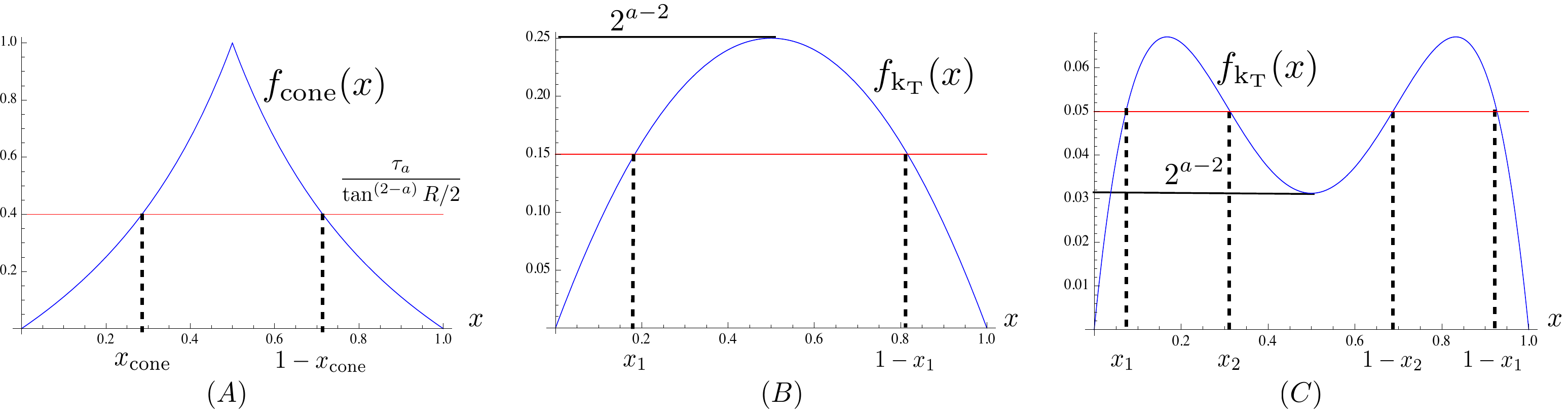}
\vspace{-.3in}
\caption[Regions of integration for finite parts of the quark jet function]{Regions of integration for the (A) cone and $\kt$-type algorithms for (B) $a>-1$ and (C) $a<-1$. The allowed region of $x$ is when the (blue) functions $f_{{\rm cone}, \, \kt}(x)$ lie above the (red) lines of constant $\tau_a/\tan^{(2-a)}{R/2}$. When $a<-1$ for the $\kt$ algorithm, there are two regions of integration when $\tau_a> 2^{a-2} \tan^{(2-a)}{R/2}$.}
\label{fig:jetfunctions-integrations}
\end{figure}

Having found $\qjetnaive^q_\text{cone}(\tau_a)$ explicitly, we merely quote the result for $\qjetnaive^q_{\kt}(\tau_a)$, which can be found similarly:
\begin{align}
\qjetnaive^{q}_{\kt}(\tau_a) &= \CF \left(\frac{13}{2} - 2\frac{\pi^2}{3} \right) \delta (\tau_a) + \frac{\CF}{1-\frac{a}{2}} \left[\mathcal{I}^{q}_{\kt}\,  \frac{\Theta(\tau_a)\Theta(\tau_a^{\max} - \tau_a)}{\tau_a}\right]_\dotplus  . 
\end{align}
$\mathcal{I}^{q}_{\kt}$ is given by
\[
\mathcal{I}^{q}_{\kt} = \int_{\mathcal{R}}dx\,\frac{2(1-x) + x^2}{x},
\]
where $\mathcal{R}$ is the region in $x$ where the constraint
\[
\label{kTtauconstraint}
f_{\kt}(x) \equiv x^{2-a} (1-x)^{2-a} [x^{-1+a} + (1-x)^{-1+a}] \ge \frac{ \tau_a}{\tan^{2-a}\frac{R}{2}}
\]
is satisfied. We plot this region in \fig{fig:jetfunctions-integrations}B and C for the cases $a>-1$ and $a<-1$, repsectively.  The boundaries of this region are the points $x_{1,2}$ illustrated in the figure, and are given by the equation
\begin{equation}
f_{\kt}(x_{1,2}) = \frac{ \tau_a}{\tan^{2-a}\frac{R}{2}} ,
\end{equation}
where we take $x_2>x_1$ if $x_2$ exists.
The upper limit $\tau_a^{\text{max}}$ is given by the maximum value over $x$ of the right-hand side of \eq{kTtauconstraint}.  In general, the constraint \eq{kTtauconstraint} is symmetric about $x=\frac12$, and so the region $\mathcal{R}$ is symmetric about the same point.  In general, if $a > -1$ or $\tau_a <  2^{a-2}\tan^{(2-a)}\frac{R}{2}$, then $\mathcal{R}$ is a single range in $x$.  Otherwise, $\mathcal{R}$ is two disjoint ranges in $x$. Since  $\tau_a \ge 2^{a-2}\tan^{(2-a)}\frac{R}{2}$ can only occur for $a<-1$, we can write $\mathcal{I}^{q}_{\kt}$ as
\begin{align}
\mathcal{I}^{q}_{\kt} &= \int_{x_1}^{1-x_1}dx\,\frac{2(1-x) + x^2}{x} - \Theta \left(\tau_a > 2^{a-2}\tan^{(2-a)}\frac{R}{2} \right)  \int_{x_2}^{1-x_2}dx\,\frac{2(1-x) + x^2}{x}.
\end{align}

Note that $\mathcal{I}^{q}_{\kt}$ involves the same integrand as in \eq{eq:Bcone}, but for $\kt$-type algorithms the integral is over a different range. In addition, both $\xcone$ and $x_{1}$ approach the same limiting value for small $\tau_a$,
\[
x \xrightarrow{\tau_a \to 0} \frac{\tau_a}{\tan^{(2-a)}\frac{R}{2}}
.\]
Thus, we can extract the small $\tau_a$ behavior of both distributions by writing
\[
\label{eq:tau-extract}
\left[ \frac{1}{\tau_a} \ln \left( \frac{1-x}{x}\right) \right]_\dotplus  =  \left[\frac{1}{\tau_a} \ln \left( \frac{\tau_a}{\tan^{(2-a)} \frac{R}{2}} \frac{1-x}{x} \right)\right]_\dotplus - \left[ \frac{1}{\tau_a} \ln \left( \frac{\tau_a}{\tan^{(2-a)} \frac{R}{2}}\right)\right]_+
,\]
where $x = \xcone$ or $x_{1}$ for the cone and $\kt$ algorithms, respectively.
Defining
\[
\label{eq:rq}
r_q(x) = 3 x + 2 \ln \frac{1-x}{x}
,\]
using \eq{eq:tau-extract}, and including the zero-bin subtraction in \eq{Jzeroinresult}, we find that the finite distributions of the full measured quark jet functions are
\begin{subequations}
\label{eq:quark_meas_finite}
\begin{align}
\label{eq:quark_meas_finite_cone}
J^{q}_\text{cone}(\tau_a) &=   \CF\Biggl[  \frac{3}{2} \ln\frac{\mu^2}{\omega^2\tan^2\frac{R}{2}} + \frac{1-\frac{a}{2}}{1-a} \ln^2\frac{\mu^2}{\omega^2} + \left(1-\frac{a}{2}\right) \ln^2\tan^2\frac{R}{2} + \frac{7}{2} + 3\ln 2  \nn\\
&\quad  - \frac{\pi^2}{6} \left(2 + \frac{1-\frac{a}{2}}{1-a}\right) \Biggr] \delta(\tau_a) -  \CF\Biggl[\Biggl( \frac{4}{1-a}\ln\frac{\mu\tan^{1-a}\frac{R}{2}}{\omega \tau_a }   \Biggr)\frac{\Theta(\tau_a - \tau_a^{\text{max}})}{\tau_a} \Biggr]_+ \nn\\
&\quad -  \frac{\CF}{1-\frac{a}{2}}\Biggl[\frac{\Theta(\tau_a)\Theta(\tau_a^{\max} - \tau_a)}{\tau_a} \biggl( \frac{3}{2} + \frac{2-a}{1-a}\ln\frac{\mu^2}{\omega^2\tau_a^{\frac{1}{1-a/2}}} \nn \\
&\qquad \qquad \qquad \qquad\qquad\qquad\qquad   -  r_q(\xcone) - 2\ln\frac{\tau_a}{\tan^{2-a}\frac{R}{2}}\biggr) \Biggr]_\dotplus
\end{align}
and
\begin{align}
\label{eq:quark_meas_finite_kt}
J^{q}_{\kt}(\tau_a) &=   \CF\Biggl[ \frac{3}{2}\ln\frac{\mu^2}{\omega^2\tan^2\frac{R}{2}} + \frac{1-\frac{a}{2}}{1-a} \ln^2\frac{\mu^2}{\omega^2} + \left(1-\frac{a}{2}\right) \ln^2\tan^2\frac{R}{2}  + \frac{13}{2}  \nn\\
& \quad - \frac{\pi^2}{6} \left(4 + \frac{1-\frac{a}{2}}{1-a}\right) \Biggr] \delta(\tau_a) -  \CF\Biggl[ \Biggl(  \frac{4}{1-a}\ln\frac{\mu\tan^{1-a}\frac{R}{2}}{\omega \tau_a }  \Biggr) \frac{\Theta(\tau_a -\tau_a^{\text{max}})}{\tau_a} \Biggr]_+ \nn\\
&\quad -  \frac{\CF}{1-\frac{a}{2}} \bigg\{ \frac{\Theta(\tau_a)\Theta(\tau_a^{\max} - \tau_a)}{\tau_a} \bigg[ \frac{3}{2} + \frac{2-a}{1-a}\ln\frac{\mu^2}{\omega^2\tau^{\frac{1}{1-a/2}}}  \nn\\
&\qquad\qquad\qquad  -r_q( x_1)  - 2\ln\frac{\tau_a}{\tan^{2-a}\frac{R}{2}} +\Theta \left(\tau_a^{\frac{1}{2-a}} > 2\tan\frac{R}{2} \right)\biggl( r_q(x_2) - \frac{3}{2} \biggr)\bigg]\bigg\}_\dotplus
 .
\end{align}
\end{subequations}
For $a=0$, these expressions for the jet functions can be simplified further to give 
\begin{subequations}
\label{eq:quark_meas_finite_a0}
\begin{align}
\label{eq:quark_meas_finite_cone_a0}
J^{q}_\text{cone}(\tau_0) &=  J^{q}_{\text{incl}}(\tau_0)  + C_F\left[3 \frac{\Theta(\tau_0)\Theta\left(\tan^2\frac{R}{2}\!-\!\tau_0\right)}{\tau_0 + \tan^2\frac{R}{2}} + \frac{\Theta\left(\tau_0 \!-\! \tan^2\frac{R}{2}\right)}{\tau_0}\left(2\ln\frac{\tau_0}{\tan^{2}\frac{R}{2}} + \frac{3}{2}\right) \right] ,
\end{align}
for the cone jet function, and
\begin{align}
\label{eq:quark_meas_finite_kt_a0}
J^{q}_{\kt}(\tau_0) &=  J^{q}_{\text{incl}}(\tau_0) +  C_F\Biggl\{\frac{\Theta(\tau_0)\Theta\left(\frac{1}{4}\tan^2\frac{R}{2} - \tau_0\right)}{\tau_0}\left[3x_1 + 2\ln\left(\frac{1-x_1}{x_1}\frac{\tau_0}{\tan^2\frac{R}{2}}\right)\right] \nn \\
&\quad \qquad\qquad\qquad + \frac{\Theta\left(\tau_0  -  \frac{1}{4}\tan^2\frac{R}{2}\right)}{\tau_0}\left(2\ln\frac{\tau_0}{\tan^{2}\frac{R}{2}} + \frac{3}{2}\right) \Biggr\},
\end{align}
\end{subequations}
for the $\kt$ jet function. In \eq{eq:quark_meas_finite_kt_a0}, $x_1$ is given by its value for $a=0$,
\begin{equation}
\label{x1_a0}
x_1 = \frac{1}{2}\left(1-  \sqrt{1-\frac{4\tau_0}{\tan^2\frac{R}{2}}}\right).
\end{equation}
In \eq{eq:quark_meas_finite_a0}, we have divided the cone and $\kt$ jet functions into the  contribution $J^{q}_{\text{incl}}(\tau_0)$ to the inclusive jet function \cite{Bauer:2003pi, Bosch:2004th}, given by
\begin{equation}
J^{q}_{\text{incl}}(\tau_0) =  C_F\left\{ \delta(\tau_0) \left( \frac{3}{2} \ln\frac{\mu^2}{\omega^2} + \ln^2\frac{\mu^2}{\omega^2}+ \frac{7}{2} - \frac{\pi^2}{2}\right)   - \left[\frac{\Theta(\tau_0)}{\tau_0} \left(\frac{3}{2} + 2\ln\frac{\mu^2}{\omega^2\tau}\right)\right]_+ \right\} ,
\end{equation}
and algorithm-dependent parts. The algorithm-dependent part of the $a=0$ cone jet function \eq{eq:quark_meas_finite_cone_a0} agrees with \cite{Jouttenus:2009ns}.  Note that if one takes $R$ to be parametrically larger than $\tau_0$ (cf. Sec.~\ref{sec:resum} and \eq{scalechoices}), the algorithm-dependent parts of \eq{eq:quark_meas_finite_a0} are power suppressed, and the cone and $\kt$ jet functions reduce to the inclusive jet function.

\subsection{Gluon Outside Measured Quark Jet}
\label{ssec:quark-out}

In this section we calculate the contribution to the quark jet function from the region of phase space in which the gluon exits the jet carrying an energy $E_g<\Lambda$.  This cut causes the contribution to be power suppressed by $\Lambda/\omega$, which scales as $\lambda^2$.  However, we elect to evaluate this case explicitly as it provides a clear example of the zero-bin subtraction giving the proper scaling to the total contribution.  We only evaluate this contribution for the cone algorithm; the details of the $\kt$ algorithm calculation are similar.  Note that the contribution when the quark is out of the jet is power suppressed at the level of the Lagrangian given in \ref{sec:qcd:eft:scet}, in which soft quarks do not couple to collinear partons at leading order in $\lambda$.

For the cone algorithm, the gluon exits the jet when the angle between the jet axis, $\vect{n}$, and the gluon is greater than $R$.  When the gluon is not in the jet, the cone axis is the quark direction, and so it makes no contribution to the angularity.  Therefore, this region of phase space contributes only to the $\delta(\tau_a)$ part of the angularity distribution.

For the naive contributions, requiring the gluon to be outside the jet and have energy less than $\Lambda$, we have the integral
\begin{align}
\label{Jnaiveout}
\qjetnaive^{q, \rm out}_{\omega}(\tau_a) &= g^2\mu^{2\epsilon} C_F \int\frac{d l^+}{2\pi}\frac{1}{(l^+)^2}\int \frac{d^d q}{(2\pi)^d}\left(4\frac{l^+}{q^-} + (d-2)\frac{l^+ - q^+}{\omega - q^-}\right) 2\pi\delta(q^- q^+ - q_\perp^2) \nn \\
&\quad \times\Theta(q^-)\Theta(q^+) 2\pi\delta\left(l^+ - q^+ - \frac{q_\perp^2}{\omega - q^-}\right)\Theta(\omega - q^-)\Theta(l^+ - q^+) \nn \\
&\quad\times \Theta\left(\frac{q^+}{q^-} - \tan^2\frac{R}{2}\right)\Theta\left(2\Lambda - q^-\right)\delta(\tau_a) .
\end{align}
This is simply \eq{Jnaivein} with different phase space $\Theta$ functions and $\delta_R$ replaced by $\delta(\tau_a)$.  Note that the theta function requiring $q^- < 2\Lambda$ is more restrictive than $q^- < \omega$.  Evaluating \eq{Jnaiveout} yields a contribution that scales with $\Lambda$ only below the leading term in $1/\epsilon$:
\begin{equation}
\label{Jnaiveoutresult}
\qjetnaive^{q, \rm out}_{\omega}(\tau_a) = -\frac{\alpha_s C_F}{2\pi} \frac{1}{\Gamma(1-\epsilon)} \left(\frac{4\pi\mu^2}{(2\Lambda\tan\frac{R}{2})^2}\right)^{\epsilon} \delta(\tau_a) \left(\frac{1}{\epsilon^2} + \frac{1}{\epsilon}\left(\frac{4\Lambda}{\omega} - \frac{2\Lambda^2}{\omega^2}\right) + \frac{8\Lambda}{\omega}\right).
\end{equation}
The zero-bin subtraction of Eq.~\ref{Jnaiveout} is
\begin{align}
\label{Jzeroout}
\qjetnaive^{q, {\rm out}(0)}_{\omega}(\tau_a) &= g^2\mu^{2\epsilon} C_F \int\frac{d l^+}{2\pi}\frac{1}{(l^+)^2}\int \frac{d^d q}{(2\pi)^d}\left(4\frac{l^+}{q^-} + (d-2)\frac{l^+ - q^+}{\omega - q^-}\right) 2\pi\delta(q^- q^+ - q_\perp^2)\nn \\
&\quad \times\Theta(q^-)\Theta(q^+) 2\pi\delta\left(l^+ - q^+\right) \Theta\left(\frac{q^+}{q^-} - \tan^2\frac{R}{2}\right)\Theta\left(2\Lambda - q^-\right)\delta(\tau_a) .
\end{align}
Evaluating \eq{Jzeroout}, we find the zero bin will exactly remove the leading term in $1/\epsilon$:
\begin{equation}
\label{Jzerooutresult}
\qjetnaive^{q, {\rm out}(0)}_{\omega}(\tau_a) = -\frac{\alpha_s C_F}{2\pi} \frac{1}{\Gamma(1-\epsilon)} \left(\frac{4\pi\mu^2}{(2\Lambda\tan\frac{R}{2})^2}\right)^{\epsilon} \delta(\tau_a) \frac{1}{\epsilon^2}.
\end{equation}
Therefore, the difference is power suppressed only after the zero bin is included.  Because other contributions when one particle is outside of the jet are similarly power suppressed, we will drop them in our remaining discussion of the jet functions.

\subsection{Unmeasured Quark Jet}
\label{ssec:quark-unmeas}

When the angularity of a jet is not measured, the jet function has no $\tau_a$ dependence.  The naive and zero-bin contributions are the same as \eqs{Jnaivein}{Jzeroin} except for the factor of $\delta_R$.  The zero-bin contribution is
\begin{equation}
\label{J2zeroin}
\begin{split}
J^{q(0)}_{\omega} &= 4 g^2\mu^{2\epsilon} \CF \int\frac{dl^+}{2\pi}\frac{1}{l^+}\int \frac{d^d q}{(2\pi)^d}\frac{1}{q^-} 2\pi\delta(q^- q^+ - q_\perp^2)\Theta(q^-)\Theta(q^+) \\
&\quad \times 2\pi\delta\left(l^+ - q^+\right)\Theta(l^+ - q^+) \,  \Theta_\text{alg}^{(0)}
.\end{split}
\end{equation}
This integral is scaleless and therefore equal to 0 in dimensional regularization.  This implies that the NLO part of the quark jet function for an unmeasured jet is just the naive result.
We find, making the divergent part explicit, in the $\overline{\text{MS}}$ scheme,
\begin{equation}
\label{Jq}
J^q_\omega = 1+ \qjetnaive^q_\omega = 1 + \frac{\alpha_s C_F}{2 \pi} \Biggl\{ \frac{1}{\epsilon^2} + \frac{3}{2\epsilon} + \frac{1}{ \epsilon} \ln \left(\frac{\mu^2}{\omega^2\tan^2\frac{R}{2}}\right) \Biggr\} + \frac{\alpha_s}{2 \pi} J^q_\text{alg}  ,
\end{equation}
where the finite parts $J^q_\text{alg}$ are\footnote{The unmeasured jet function \eq{Jq} is not simply obtained by integrating the measured jet function \eq{Jtotal} over $\tau_a$. This is due to the different relative scaling of $R$ with the SCET expansion parameter $\lambda_i$ in a measured and unmeasured jet sector, as noted earlier. Namely, $R\sim \lambda_i^0$ in a measured jet sector (where $\lambda\sim\sqrt{\tau_a}$) while $\lambda_k\sim\tan(R/2)$ in an unmeasured jet sector.}
\[
\label{eq:quark_unmeas_finite}
J^q_\text{alg} =  \frac{3 \CF}{2} \ln \left(\frac{\mu^2}{\omega^2 \tan^2\frac{R}{2}} \right) + \frac{\CF}{2}\ln^2 \left(\frac{\mu^2}{\omega^2 \tan^2\frac{R}{2}} \right) +d^{q, \, \text{alg}}_J,
\]
with the constant terms
\begin{equation}
\label{eq:dJq}
d^{q, \, \text{cone}}_J = \CF\left( \frac{7}{2} + 3\ln 2 - \frac{5\pi^2}{12}   \right) \ , \quad
d^{q, \, \kt}_J =  \CF\left( \frac{13}{2} - \frac{3 \pi^2}{4}  \right).
\end{equation}



 \graphicspath{{appendixD/graphics/}}

\chapter{Our analysis in detail}
\label{app:details}

I here give a brief summary of the computational tools employed to do the studies in this thesis.\footnote{Parts of this appendix are taken from Appendix A of \cite{Pruning2}.}  We simulate high-energy collisions using {\tt MadGraph/MadEvent} v4.4.21 \cite{MadGraph} interfaced with {\tt Pythia} v6.4 \cite{Pythia6}.  From the hadron-level output of {\tt Pythia}, we group final-state particles into ``cells'' based on the segmentation of the ATLAS hadronic calorimeter ($\Delta \eta=0.1$, $\Delta\phi=0.1$ in the central region). We sum the four-momenta of all particles in each cell and rescale the resulting three-momentum to make the cell massless. After a threshold cut on the cell energy of 1 GeV, cells become the inputs to the jet algorithm. Our implementation of recombination algorithms uses \FJ \cite{FastJet} interfaced with \SJ.

Several of the plots in Sections~\ref{sec:sub} and \ref{sec:prune} involve mass cuts on jets.  The details of these cuts are provided in Sec.~\ref{sec:prune:study}.

\section{\texorpdfstring{$\ee$ events}{e+e- events}}

\begin{figure}[htbp]
\begin{center}
\includegraphics[width = .7\columnwidth]{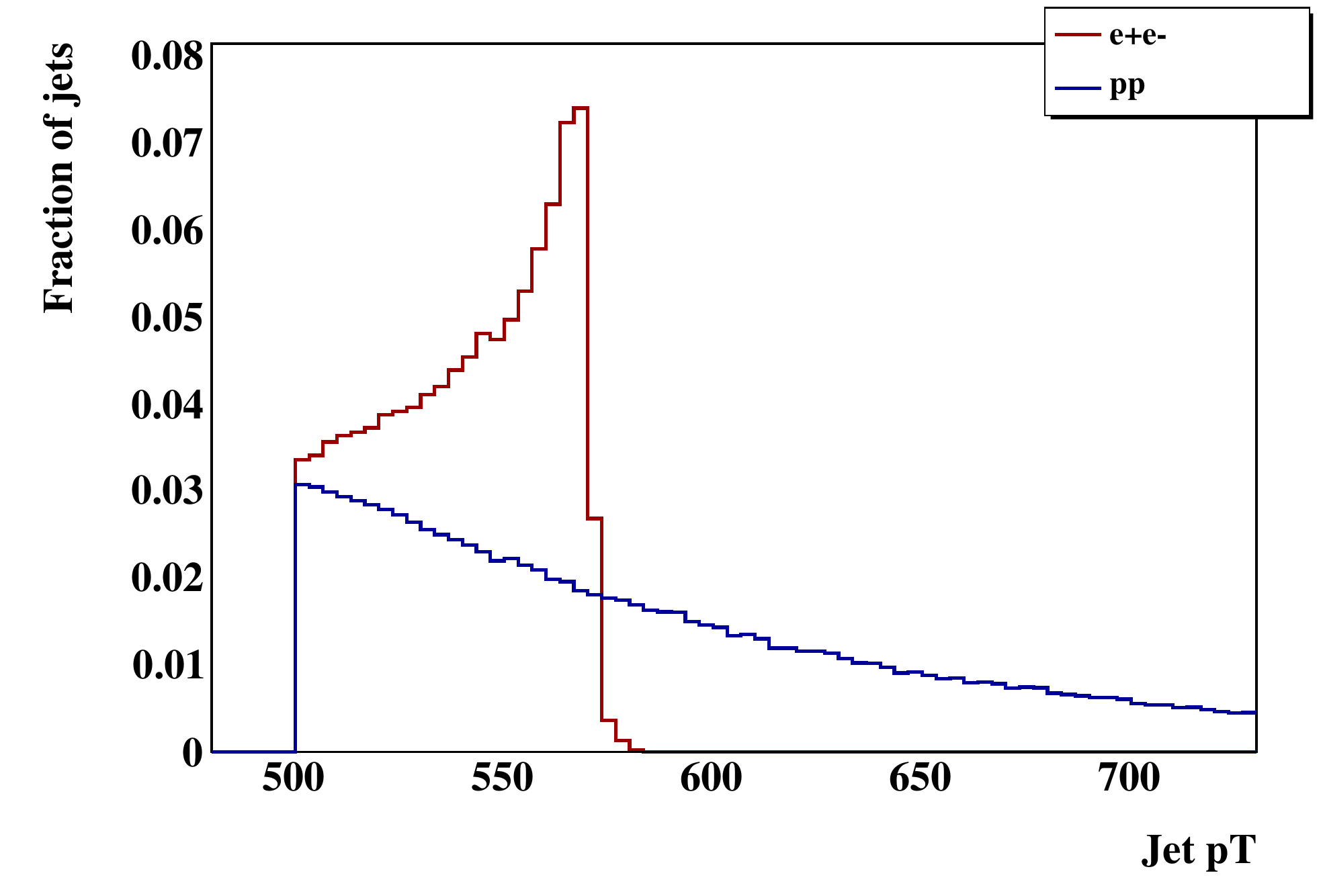}
\end{center}
\caption[Distribution in $p_T$ for top quark jets in the $\ee$ sample and the $pp$ sample]{Distribution in $p_T$ for top quark jets in the $\ee$ sample (red) and the $pp$ sample (blue).}
\label{fig:pT_ttbar_eeVSpp}
\end{figure}

For the $\ee$ studies in Sec.~\ref{sec:sub:algeffects}, we generate $\ee \to q\bar q$ and $\ee \to t\bar t$ events with center of mass energy $Q = 1200$ GeV.  In the $t\bar t$ events, the top quarks are required to decay hadronically.  We then apply the same minimal detector simulation and analysis as for our simulated LHC events --- we are only considering $\ee$ collisions as a way to study jets without the effect of initial state radiation, multiple interactions, pile-up, etc., although of course $\ee$ collisions are interesting in their own right.  The center of mass energy has been chosen so that the $p_T$ distribution of the jets is similar to that for our second $p_T$ bin $pp \to t\bar t$ sample below.  The two distributions are shown in Fig.~\ref{fig:pT_ttbar_eeVSpp}.  Note that whereas jets in the $pp$ sample have a falling $p_T$ distribution with a lower cutoff, jets in the $\ee$ sample have a natural upper cutoff, along with the same imposed lower cutoff.

\section{\texorpdfstring{$pp$}{pp} events}

We also study jets in $pp$ collisions.  We employ MLM-style matching, implemented in {\tt MadGraph} (see, e.g., \cite{MatchingExample}), on the backgrounds.  We have checked that our matching parameters are reasonable using the tool {\tt MatchChecker} \cite{MatchChecker}.  We use the DWT tune \cite{PythiaTunes} in Pythia to give a ``noisy'' underlying event (UE).  For the hadron-level studies in Sec.~\ref{sec:sub:algeffects}, we exclude (include) the underlying event by setting the {\tt Pythia} parameter {\tt MSTP(81)} to zero (one), turning off (on) multiple interactions.  To exclude (include) initial state radiation, we set \code{MSTP(61)} to zero (one).  Both ISR and UE are on unless otherwise noted.

We perform no detector simulation, other than the calorimeter clustering noted above, so we can isolate the ``best case'' effects of our method.  In Sec.~\ref{sec:prune:results:smearing}, we examine the effects of Gaussian smearing on the energies of final state particles from {\tt Pythia} to get a sense for how much the results may change with a detector.

For the $W$ study, the signal sample is $W^+W^-$ pair production, with exactly one $W$ required to decay leptonically.  The background is a matched sample of a leptonically decaying $W$ and one or two light partons (gluons and the four lightest quarks) before showering.  These partons must be in the central region, $|\eta| < 2.5$.  Signal and background samples are divided into four $p_T$ bins: [125, 200], [200, 275], [275, 350], and [350, 425] (all in GeV).  Each bin is defined by a $p_T$ cut that is applied to single jets in the analysis.  These bins confine the $W$ boost to a narrow range and allow us to study the performance of pruning as the jet $p_{T}$ (or $W$ boost) varies.

For each $p_T$ bin $[p_T^\text{min}, p_T^\text{max}]$, both samples are generated with a $p_T$ cut on the leptonic $W$ of $p_T^\text{min} - 25$ GeV.  For the background, we set the matching scales $(Q^\text{ME}_\text{cut}, Q_\text{match})$ to be (10, 15) GeV in all four bins.

For the top quark reconstruction study, the signal sample is $t\bar{t}$ production with fully hadronic decays.  The background is a matched sample of QCD multijet production with two, three, or four light partons, with the same cut on parton centrality as in the $W$ study.  Samples are again divided into four $p_T$ bins: [200, 500], [500, 700], [700, 900], and [900, 1100] (all in GeV).

We generate signal and background samples with a parton-level $h_{T}$ cut for generation efficiency, where $h_{T}$ is the scalar sum of all $p_{T}$ in the event.  For each $p_T$ bin $[p_T^\text{min}, p_T^\text{max}]$, the parton-level $h_T$ cut is $p_T^\text{min} - 25\ \text{GeV} \le h_T/2 \le  p_T^\text{max} + 100\ \text{GeV}$.  For the background, we use matching scales (20, 30) GeV for the smallest $p_T$ bin and (50, 70) GeV in the other three bins.

\subsection{Matched vs. unmatched samples}

We use matched samples for our QCD backgrounds --- that is, samples where the full matrix element weighting is used for additional partons in the hard process.  This gives background samples with somewhat heavier mass distributions and ``harder'' substructure.  Large jet masses and significant substructure are perturbative effects, and are enhanced by including the full matrix elements.  We expect that substructure predictions made with matched backgrounds will be more reliable.

\begin{figure}[htbp]
\begin{center}
\subfloat[$m_\text{jet}$] {\includegraphics[width = .48\columnwidth]{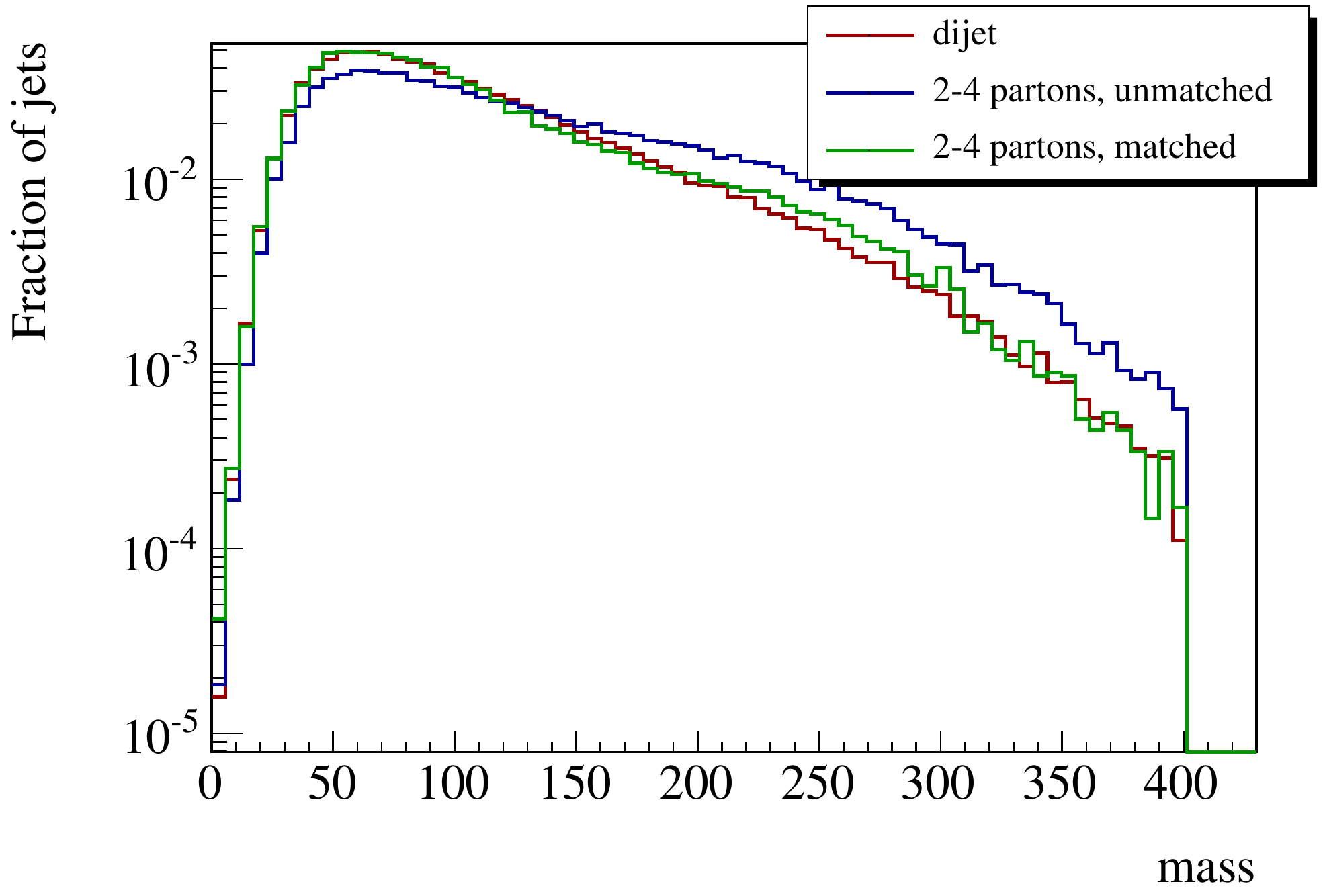}}
\subfloat[$m_\text{jet}/p^T_\text{jet}$] {\includegraphics[width = .48\columnwidth]{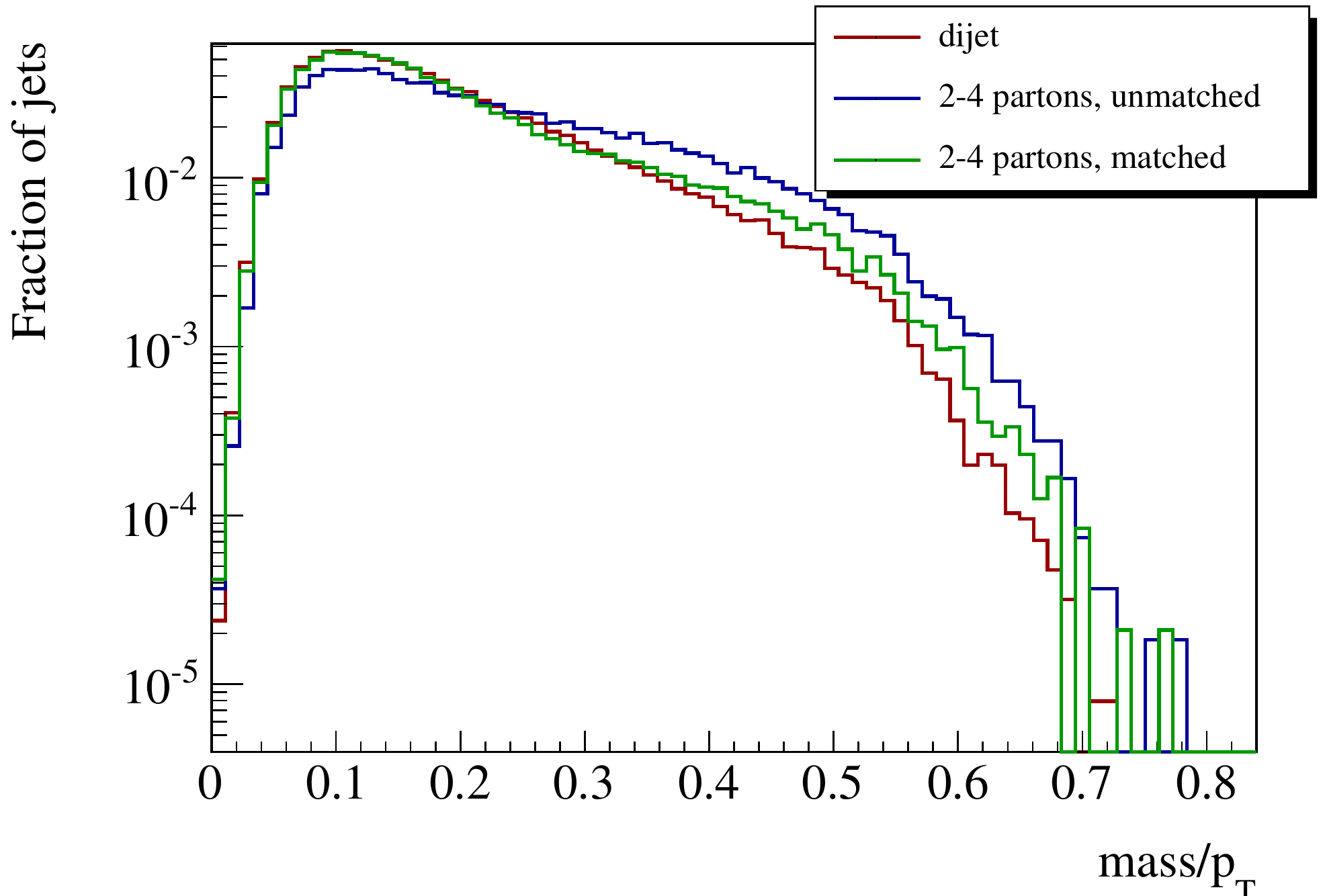}}

\subfloat[$a_1$] {\includegraphics[width = .48\columnwidth]{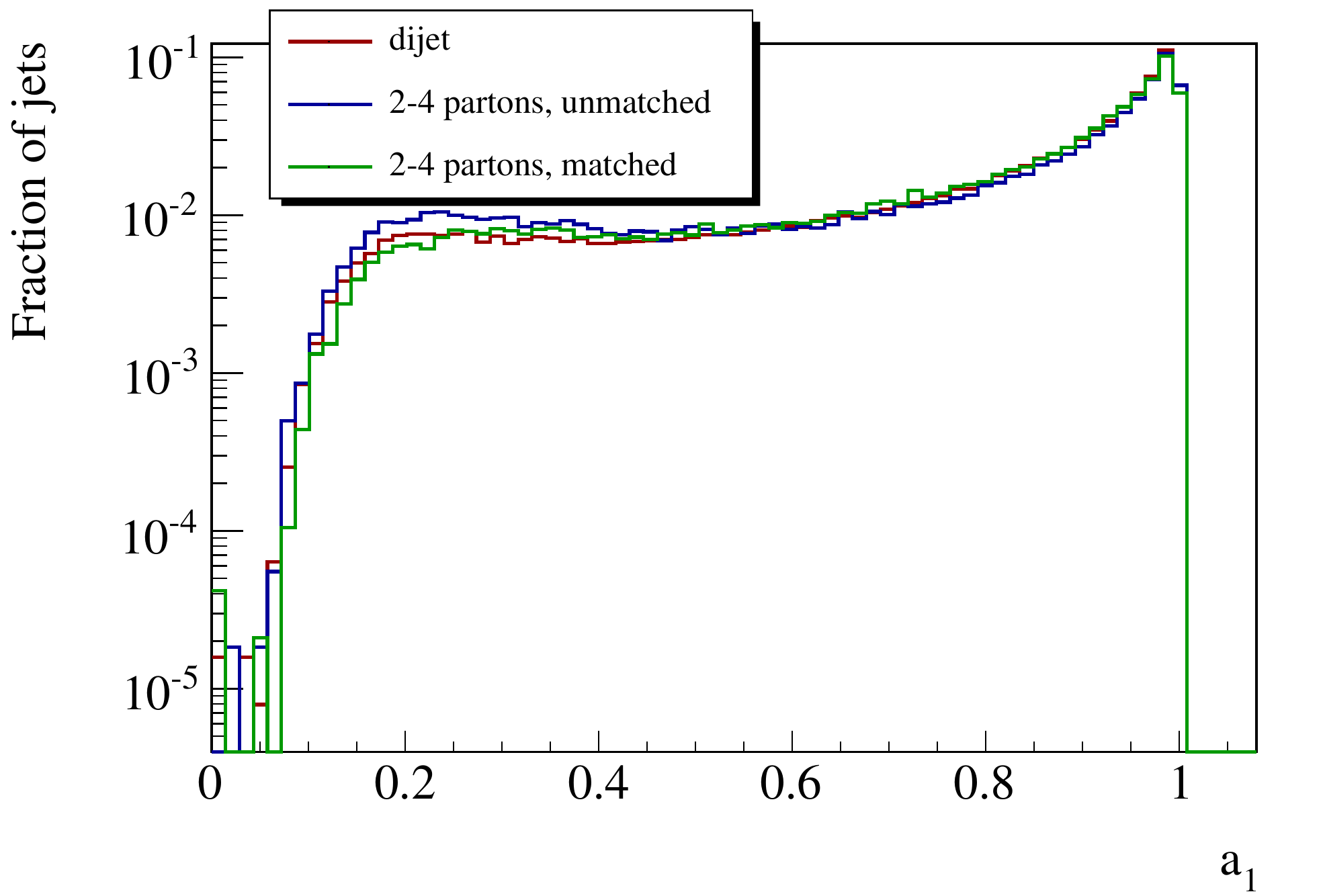}}
\subfloat[Minimum subjet mass] {\includegraphics[width = .48\columnwidth]{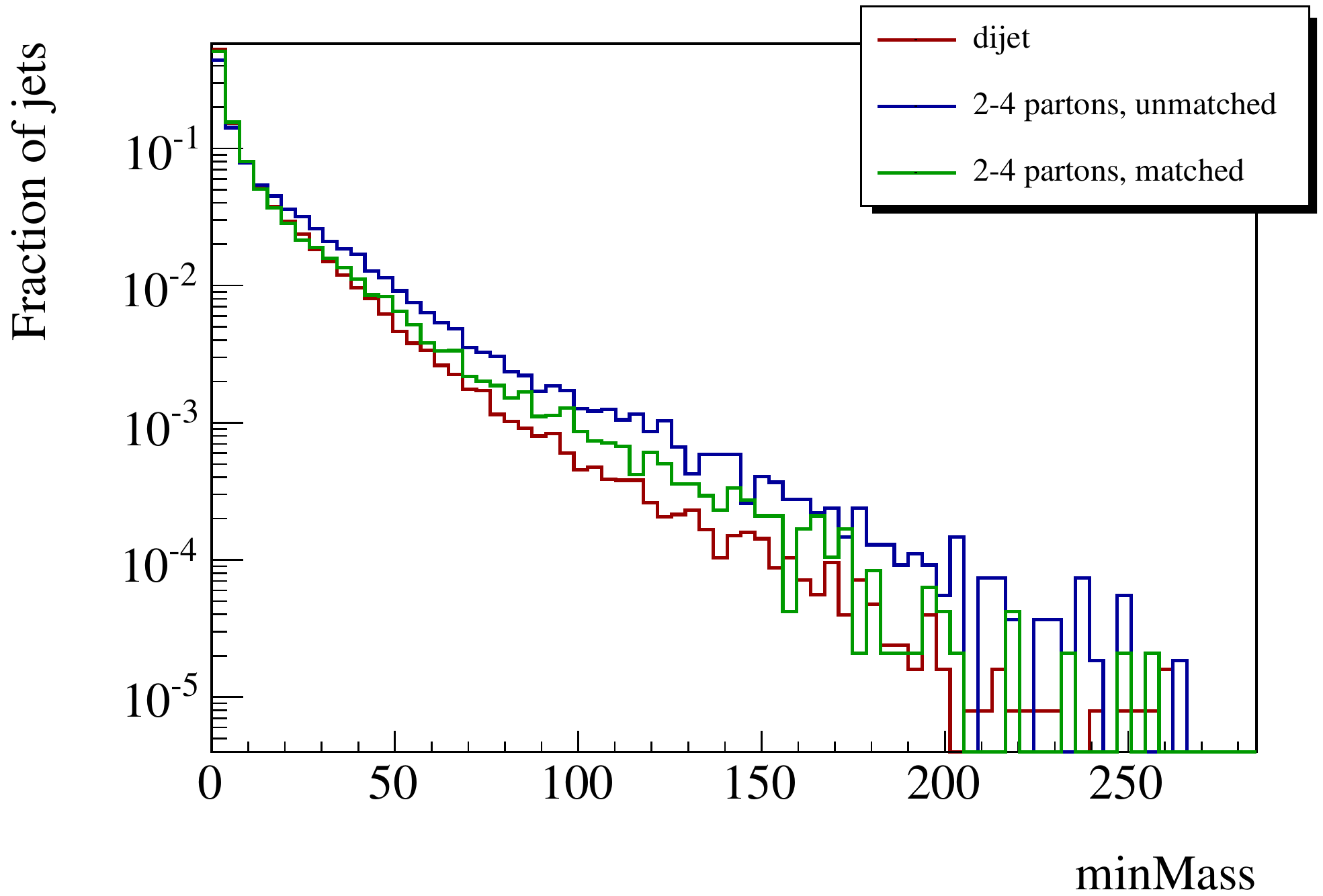}}

\end{center}
\caption[Distribution in jet and subjet masses for matched and unmatched samples]{Distribution in $m_\text{jet}$, $m_\text{jet}/p^T_\text{jet}$, $a_1$, and ``minimum subjet mass''.  $a_1$ is the mass of the heavier subjet scaled to the jet mass;  the ``minimum subjet mass'' is the minimum pairwise mass between subjets if the jet is unclustered to three subjets.  Jets have $p_T > 500$ GeV.}
\label{fig:match_compare}
\end{figure}

As an example, consider the plots in Fig.~\ref{fig:match_compare}.  Three samples are compared: ``dijet'' refers to showered $2\to2$ processes.  The ``matched'' sample is the sample used throughout the paper and described above, with matrix elements for two, three, and four hard partons.  The``unmatched'' sample has the same set of matrix elements, but with no matching --- i.e., no attempt is made to remove double counting.  That the mass spectrum is much harder than either of the other samples suggests that the double counting is significant.

All three samples use the same \texttt{MadGraph} phase space cuts: \{\texttt{xqcut} $>$ 50 GeV, \texttt{htjmin} $>$ 950 GeV\}, corresponding to the second $p_T$ bin of the top quark background samples.  The first cut requires partons to be separated by 50 GeV in $\kt$ distance, and to each have $p_T > 50$ GeV as well.  The second requires that $\sum |p_T^i| > 950$ GeV, where the sum is over all partons.  For the dijet sample the first cut has no effect.

The distributions in Fig.~\ref{fig:match_compare} are individually normalized to unit integral.  The leading order cross sections are given in Table \ref{table:matched_xsec}.

\begin{table}[htbp]
\begin{center}
\begin{tabular}{|l|c|}
\hline
Sample & LO cross section (pb) \\
\hline
\hline
dijet & $788.5 \pm 0.5$ \\
\hline
2--4 parton unmatched & $3424 \pm 2$\\
\hline
2--4 parton matched & $964 \pm 2$\\
\hline
\end{tabular}
\end{center}

\caption[Leading order cross sections for matched and unmatched samples]{Leading order cross sections for the three samples in Fig.~\ref{fig:match_compare}.  Note the extreme overcounting if we include additional hard partons but do not match.  The cross sections are taken from \prog{MadGraph} and include statistical errors.}
\label{table:matched_xsec}
\end{table}

The important comparison is between the dijet sample and the matched sample.  The matched sample has a slightly harder mass spectrum, even more noticeable when we scale by $p^T_\text{jet}$.  In the lower left we see that the distribution in $a_1$, the measure of subjet mass used repeatedly in this thesis does not change much.  However, in the lower left I show another variable, inspired by the CMS top tagger \cite{CMSTopTagging}.  The ``minimum subjet mass'' is defined to be the minimum pairwise mass between subjets if the jet is unclustered to three subjets (by undoing the last two clustering steps).  In addition to the CMS top tagger, this variable is used in the pruning top tagger described in \cite{BOOST2010}.  We see that the matched sample has significantly more jets with large minimum subjet mass.

The lesson is clear: the details of jet substructure seen in simulated events depend heavily on the details of the Monte Carlo modeling.  Since jet substructure is fundamentally a higher-order effect, it is natural that higher-order simulation makes a difference.


 \graphicspath{{appendixE/graphics/}}

\chapter{\texorpdfstring{\SJ}{SpartyJet} example}
\label{app:example}

In this appendix I give two brief examples of \SJ analyses, with the goal of comparing pruning to top-tagging for top finding and comparing pruning to the mass-drop filter method in $W$ finding.  I will first walk through the implementation to demonstrate the construction of a \SJ analysis, then show some results.

\section{Implementation}

Both analyses use the following simple wrapper function that handles input and output, setting up a few input selector tools:

\begin{lstlisting}[language=Python]
def RunAlgorithms(infile, outfile, jetAlgs, pTCut = 50, N = -1):
	"""
	This function wraps the algorithm-running functionality of SpartyJet.
	
	Infile is assumed to be in 'HuskyInput' format.  N events are processed;
	  N = -1 is all events.
	Output is stored in outfile.root.
	jetTools must be a list (or iterable container) of SpartyJet JetTools;
	  specifically, these should be jet finders.
	pTCut is the final pT cut on jets.
	"""

	
	# Create a jet builder---------------------------
	builder = SJ.JetBuilder()
	builder.silent_mode() # turns off debugging information
	
	# Configure input -------------------------------
	if(infile.find('UW') != -1):
		input = SJ.HuskyInput(infile)
	elif(infile.find('hep') != -1):
		input = SJ.StdHepInput(infile)
	else:
		print 'Unrecognized input format in', infile
		exit(1)
	builder.configure_input(input)
	
	# Configure output--------------------------------
	#builder.add_text_output(outfile+".dat")
	builder.configure_output("SpartyJet_Tree",outfile);
	builder.output_var_style.array_type = "vector" # output as "array" or "vector"
	builder.output_var_style.base_type = "float"  # output as "float" or "double"
	
	for t in jetAlgs: builder.add_custom_alg(t)
	
	# Add input cuts
	builder.add_jetTool_front(SJ.JetPtSelectorTool(0.5))
	builder.add_jetTool_front(SJ.JetEtaCentralSelectorTool(-4.9,4.9))
	
	# Add output cuts
	builder.add_jetTool(SJ.JetPtSelectorTool(pTCut))
	builder.add_jetTool(SJ.JetEtaCentralSelectorTool(-2.5,2.5))
	
	# Add jet moments
	SubjetMassMoment = SJ.HeavierSubjetMass('subjetM')
	builder.add_jetTool(SJ.JetMomentTool('subjetM', SubjetMassMoment))
	SubjetMassMoment = SJ.HeavierSubjetMass('a1', True) # scale to jet mass
	builder.add_jetTool(SJ.JetMomentTool('a1', SubjetMassMoment))
	zMoment = SJ.zMoment('z')
	builder.add_jetTool(SJ.JetMomentTool('z', zMoment))
	DeltaRMoment = SJ.DeltaRMoment('DeltaR')
	builder.add_jetTool(SJ.JetMomentTool('DeltaR', DeltaRMoment))
		
	# Run SpartyJet
	if N > 0:
		builder.print_event_every(max(1,N/20))
	else: # process all is N = -1
		builder.print_event_every(1000)
	builder.process_events(N)
\end{lstlisting}

The main input is a set of \codes{JetAlgorithm}.  These are defined for the top and $W$ analyses by the following functions:

\begin{lstlisting}[language=Python]
def TopCompareAnalysis(infile, outfile, N=-1):	
	algs = {}
	
	# set up initial antikt
	algs['AntiKt10'] = SJ.JetAlgorithm(SJ.FastJet.FastJetFinder('AntiKt10', fj.antikt_algorithm, 1.0, False))
	algs['AntiKt10'].addTool(SJ.JetPtSelectorTool(500))
	initialJets = SJ.ForkToolParent('AntiKt10Parent')
	algs['AntiKt10'].addTool(initialJets)
	
	# recluster with CA, fork again
	algs['CA10'] = SJ.JetAlgorithm(SJ.ForkToolChild(initialJets, 'CA10'))
	algs['CA10'].addTool(SJ.FastJet.FastJetRecluster('CA10cluster', fj.cambridge_algorithm, 1.5, False))
	CAjets = SJ.ForkToolParent('CA10Parent')
	algs['CA10'].addTool(CAjets)
	
	# JH tagger
	algs['CA10JH'] = SJ.JetAlgorithm(SJ.ForkToolChild(CAjets, 'CA10JH'))
	algs['CA10JH'].addTool(SJ.FastJet.TopTaggerTool(fj.JHTopTagger)(fj.JHTopTagger(0.1, 0.19, 81.0)))
	
	# Alternative, more aggressive JH tagger
	algs['CA10JH2'] = SJ.JetAlgorithm(SJ.ForkToolChild(CAjets, 'CA10JH2'))
	JHPrune = SJ.JHPruneTool(0.1, 0.19, 2)
	algs['CA10JH2'].addTool(JHPrune)
	algs['CA10JH2'].addTool(SJ.SubjetCutTool(JHPrune, 3, True))
	algs['CA10JH2'].addTool(SJ.MinMassTool())
	
	# pruning
	algs['CA10prune'] = SJ.JetAlgorithm(SJ.ForkToolChild(CAjets, 'CA10prune'))
	big_CA_def = fj.JetDefinition(fj.cambridge_algorithm, 3.14*0.5)
	algs['CA10prune'].addTool(SJ.FastJet.FastPruneTool(big_CA_def))
	
	RunAlgorithms(infile, outfile, algs.values(), 500, N)
	
def WCompareAnalysis(infile, outfile, N=-1):
	algs = {}
	
	# set up initial antikt
	algs['AntiKt10'] = SJ.JetAlgorithm(SJ.FastJet.FastJetFinder('AntiKt10', fj.antikt_algorithm, 1.0, False))
	algs['AntiKt10'].addTool(SJ.JetPtSelectorTool(200))
	initialJets = SJ.ForkToolParent('AntiKt10Parent')
	algs['AntiKt10'].addTool(initialJets)
	
	# recluster with CA, fork again
	algs['CA10'] = SJ.JetAlgorithm(SJ.ForkToolChild(initialJets, 'CA10'))
	algs['CA10'].addTool(SJ.FastJet.FastJetRecluster('CA10cluster', fj.cambridge_algorithm, 1.5, False))
	CAjets = SJ.ForkToolParent('CA10Parent')
	algs['CA10'].addTool(CAjets)
	
	# MDF analysis
	algs['CA10MDF'] = SJ.JetAlgorithm(SJ.ForkToolChild(CAjets, 'CA10MDF'))
	subjetFinder = SJ.MassDropTool(0.67, 0.09, 1, 'MassDrop')
	algs['CA10MDF'].addTool(subjetFinder)
	algs['CA10MDF'].addTool(SJ.SubjetCutTool(subjetFinder, 2))
	algs['CA10MDF'].addTool(SJ.FastJet.BDRSFilterTool(1.2, 0.3, 3))
	
	# pruning
	algs['CA10prune'] = SJ.JetAlgorithm(SJ.ForkToolChild(CAjets, 'CA10prune'))
	big_CA_def = fj.JetDefinition(fj.cambridge_algorithm, 3.14*0.5)Ch
	algs['CA10prune'].addTool(SJ.FastJet.FastPruneTool(big_CA_def))
		
	RunAlgorithms(infile, outfile, algs.values(), 200, N)
\end{lstlisting}

The new plots in Chapters \ref{sec:sub} and \ref{sec:prune} were generated with similar functions, not given here.  The input file must be in ``UW'' or \prog{StdHEP} format (the former is a simple text format); the output is a \SJ\,\prog{ROOT} file with all jet information stored, including measured values of $z$, $\Delta R$, $a_1$, and $m_1$ (the last two both look for the heavier subjet; $a_1 \equiv m_1/m_J$).

The $W$ analysis compares initial anti-$\kt$ jets, jets reclustered with CA (identical contents but different substructure), and CA jets with pruning or mass-drop filtering \cite{FilteringHiggs} applied.  The top analysis compares the same initial jets with pruned or top-tagged \cite{TopTagging} jets.  This analysis also includes an additional top-tagging implementation I have set up with a set of \SJ tools.  This version discards asymmetric branchings even for subjets that do not eventually split, so is somewhat more aggressive.\footnote{To illustrate the difference, consider their action on a putative top jet.  Both will remove from the jet soft, wide-angle splittings until a top-level splitting is found.  Both will then repeat this procedure on the two subjets.  Consider then that for one subjet, several soft protojets are discarded before finding an irreducible splitting --- the subjet does not split.  The original top-tagger (at least as implemented by Gavin Salam's \code{JHTopTagger.hh} \cite{FastJetWebsite}) keeps an entire subjet; my implementation will discard the soft protojets and keeps only the subjet formed at the irreducible splitting.}  In addition, my implementation finds the $W$ subjet by unclustering the top jet to three subjets, then merging the pair with minimum combined mass --- as in the CMS top tagging implementation \cite{CMSTopTagging}.  The original implementation simply takes the pair with combined mass closest to $m_W$.  No attempt has been made to optimize the parameters of this modified top tagger; it is included as an example of a \SJ tool implementation and a foil for the other methods.

\section{Top quark results}

\begin{figure}
\begin{center}
\includegraphics[width=.7\textwidth]{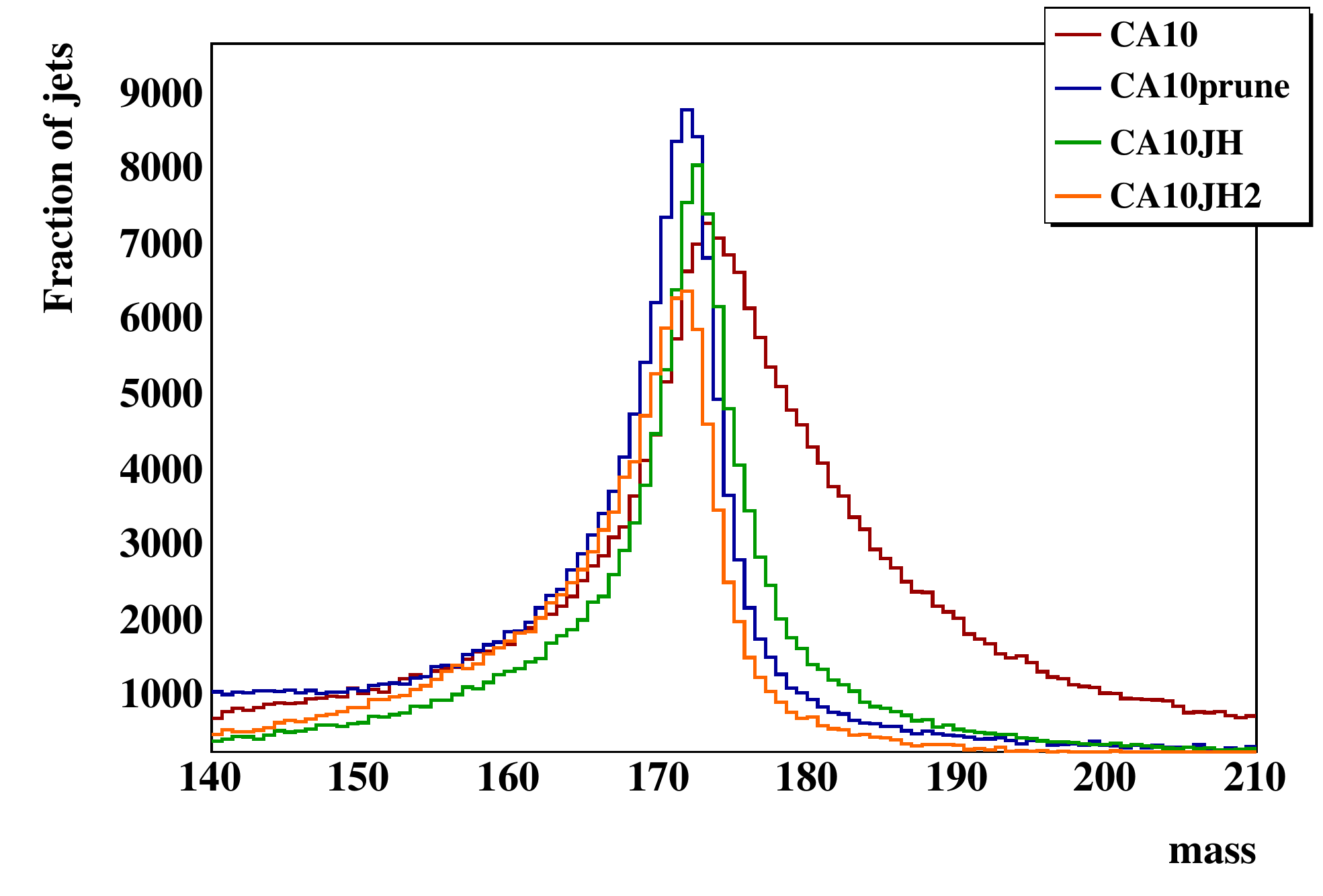}
\end{center}
\caption[Distribution in $m_J$ for anti-$\kt$ jets reclustered with CA, then pruned or top-tagged]{Distribution in $m_J$ for anti-$\kt$ jets reclustered with CA, then pruned or top-tagged.  ``CA10JH'' is the original Johns Hopkins tagger; ``CA10JH2'' is my variant.  Jets have $p_T > 500$ GeV; the initial $D = 1.0$.}
\label{fig:topcompare_mjet}
\end{figure}

The jet mass distribution for each method is shown in Fig.~\ref{fig:topcompare_mjet}.  The events are the same as in the second $p_T$ bin studied in Chapter \ref{sec:prune}, with jets have $p_T > 500$ GeV.  All three substructure methods improve on plain anti-$\kt$ jets.  As expected, pruning removes more soft radiation than top-tagging, since pruning is applied to the whole jet; the result is a mass peak that is slightly higher but shifted slightly lower.  The second implementation of top tagging is shifted even further lower but is clearly over-grooming --- the $z$ and $\Delta R$ criteria used by top tagging are both looser than for pruning, resulting in more vetoed mergings for the ``JH2'' sample.  The jet mass windows, found as described in Sec. \ref{sec:prune:study:metrics}, are given in Table \ref{table:topwindows}.

\begin{table}[htbp]
\begin{center}
\begin{tabular}{|l|c|c||c|c|}
\hline
Method & $m_\text{jet}^\text{low}$ & $m_\text{jet}^\text{high}$ & $m_\text{subjet}^\text{low}$ & $m_\text{subjet}^\text{high}$ \\
\hline \hline
CA10 & 160.3 & 187.9 & 72.6 & 84.6 \\
\hline
CA10 + pruning & 165.7 & 178.3 & 73.8 & 83.8 \\
\hline
CA10 + JH tagger & 165.4 & 180.3 & 73.5 & 85.1 \\
\hline
CA10 + JH2 tagger & 163.7 & 179.6 & 72.4 & 85.1 \\
\hline
\end{tabular}
\end{center}
\caption[Jet mass and subjet mass windows for each top-finding method]{Jet mass and subjet mass windows for each top-finding method.}
\label{table:topwindows}
\end{table}

\begin{figure}
\begin{center}
\includegraphics[width=.7\textwidth]{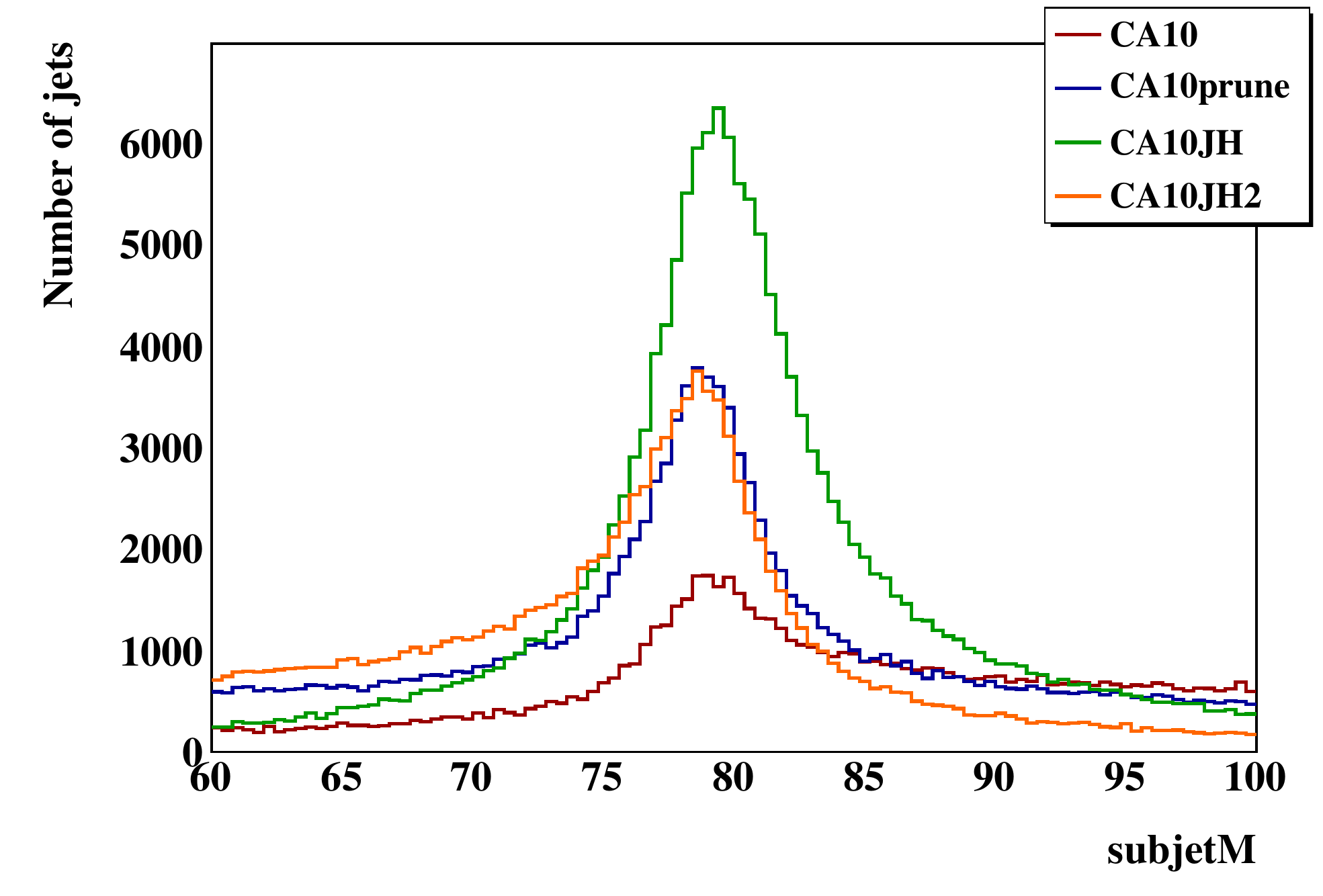}
\end{center}
\caption[Distribution in subjet mass for anti-$\kt$ jets reclustered with CA, then pruned or top-tagged]{Distribution in subjet mass for anti-$\kt$ jets reclustered with CA, then pruned or top-tagged.  ``CA10JH'' is the original Johns Hopkins tagger; ``CA10JH2'' is my variant.  For the JH tagger the identified $W$ subjet is used; for the others I take the heavier subjet.  Jets have $p_T > 500$ GeV; the initial $D = 1.0$.}
\label{fig:topcompare_mW}
\end{figure}

After restricting jets to lie in the mass windows given in Table \ref{table:topwindows}, we can look for evidence of the $W$ mass.  In Fig.~\ref{fig:topcompare_mW} we plot the found subjet mass; for the JH tagger we use the identified $W$; for the other three we take the heavier subjet.  The results are broadly similar, with pruning giving a slightly narrow peak and ``JH2'' a slightly wider peak than the JH tagger.  Again using the methods of Sec.~\ref{sec:prune:study:metrics} we can find the subjet mass windows, also given in Table \ref{table:topwindows}.

\begin{figure}
\begin{center}
\includegraphics[width=.45\textwidth]{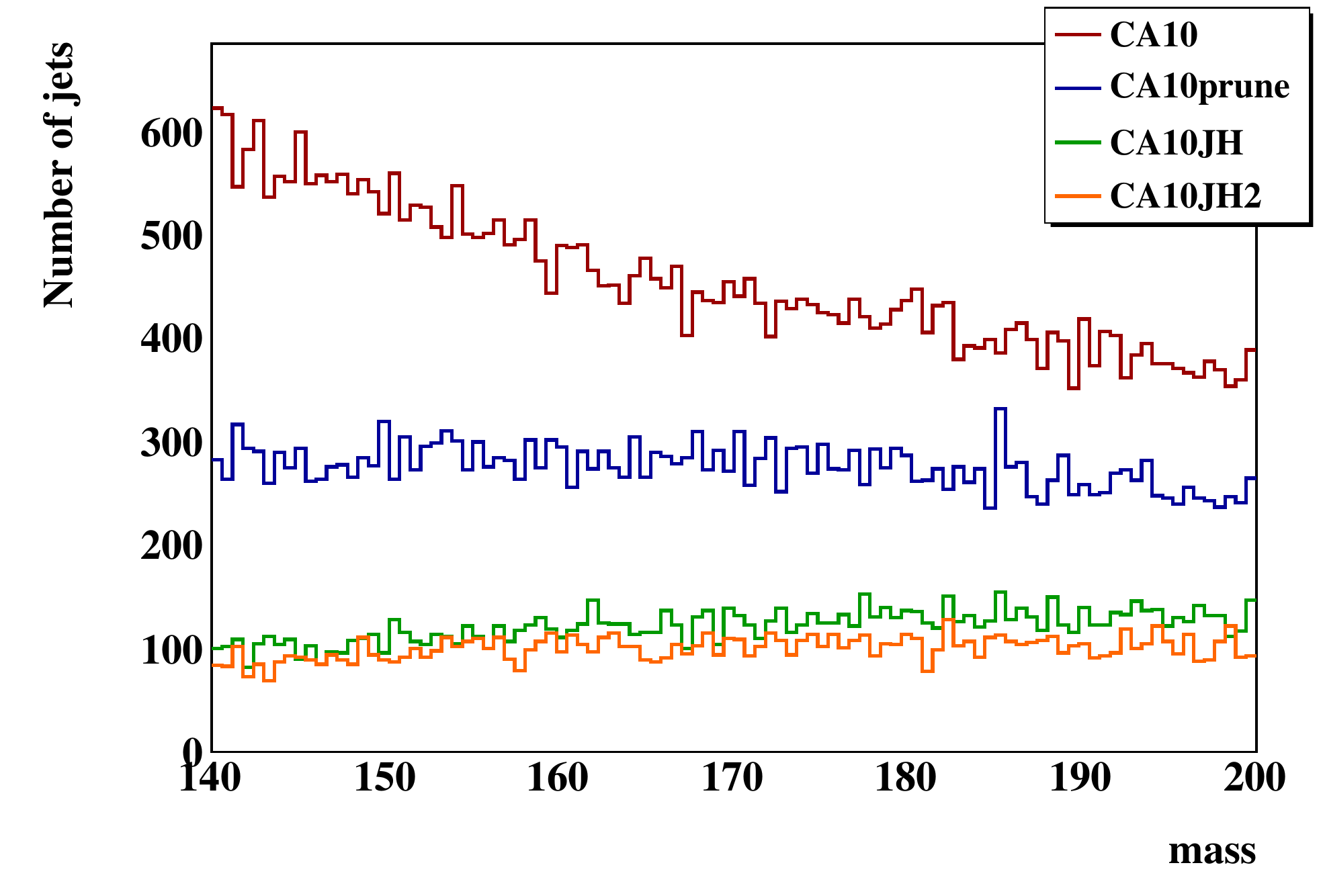}
\includegraphics[width=.45\textwidth]{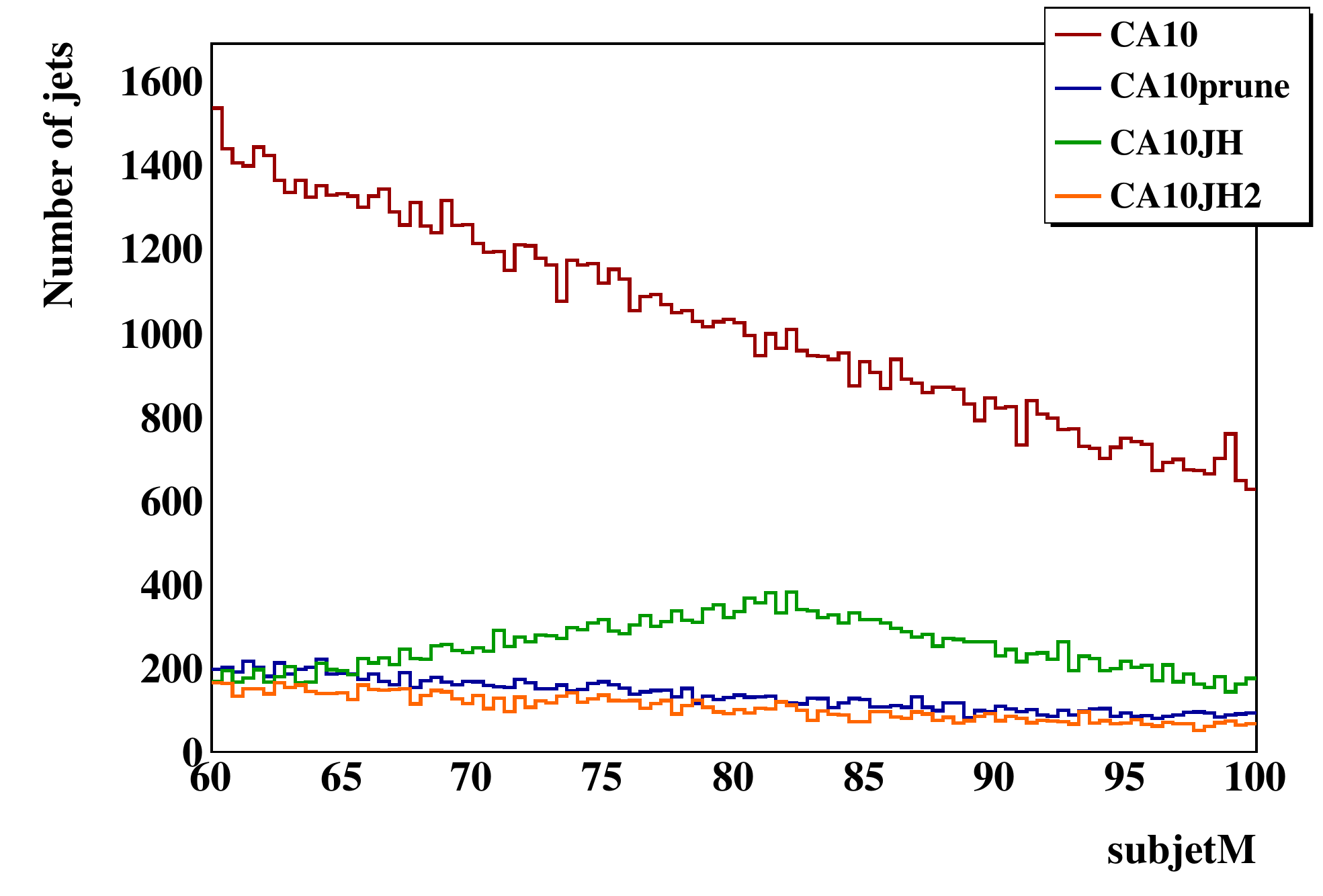}
\end{center}
\caption[Background distribution in jet and subjet mass for anti-$\kt$ jets reclustered with CA, then pruned or top-tagged]{Background distribution in jet and subjet mass for anti-$\kt$ jets reclustered with CA, then pruned or top-tagged.  ``CA10JH'' is the original Johns Hopkins tagger; ``CA10JH2'' is my variant.  For the JH tagger the identified $W$ subjet is used; for the others I take the heavier subjet.  Jets have $p_T > 500$ GeV; the initial $D = 1.0$.}
\label{fig:topcompare_BG}
\end{figure}

In Fig.~\ref{fig:topcompare_BG} we give the jet and subjet mass distributions for the background sample (the same matched multijet as in Chapter \ref{sec:prune}, $p_T$ bin 2).  Note that the JH tagger takes three or four subjets and merges the two closest in combined mass to $m_W$, producing a peak in the background subjet mass distribution.  The ``minimum mass'' taken in JH2, and the CMS implementation of the JH tagger, does not share this feature.

The tagging efficiencies and mis-tag rates for each method are given in Table \ref{table:topefficiencies}.  The efficiency (mis-tag rate) for each method is the number of jets in the signal (background) sample that survive after all cuts, divided by the number of initial jets that pass the $p_T$ cut.  Only mass cuts are imposed, unlike in the original top-tagging analysis which also used a cut on the cosine of the helicity angle, $\cos \theta_h$.

\begin{table}[htbp]
\begin{center}
\begin{tabular}{|l|c|c||c|c|}
\hline
 & \multicolumn{2}{c||}{Signal} & \multicolumn{2}{c|}{Background} \\
\hline
Method & $m_\text{jet}$ cut & $m_\text{jet}$ and $m_\text{subjet}$ cuts & $m_\text{jet}$ cut & $m_\text{jet}$ and $m_\text{subjet}$ cuts \\
\hline \hline
CA10 & 0.49 & 0.05 & 0.054 & 0.0018 \\
\hline
CA10 + pruning & 0.27 & 0.11 & 0.016 & 0.00075 \\
\hline
CA10 + JH tagger & 0.27 & 0.20 & 0.0086 & 0.0025 \\
\hline
CA10 + JH2 tagger & 0.23 & 0.14 & 0.0074 & 0.0014 \\
\hline
\end{tabular}
\end{center}
\caption[Tagging efficiencies and mis-tag rates for each method]{Tagging efficiencies and mis-tag rates for each method, applied to $t \bar t$ events (Signal) and matched multi-jet events (Background).  Initial jets have $p_T > 500$ GeV and $D = 1.0$.  Efficiencies are relative to initial numbers of jets passing the $p_T$ cut.}
\label{table:topefficiencies}
\end{table}

\section{\texorpdfstring{$W$}{W} results}

\begin{figure}
\begin{center}
\includegraphics[width=.7\textwidth]{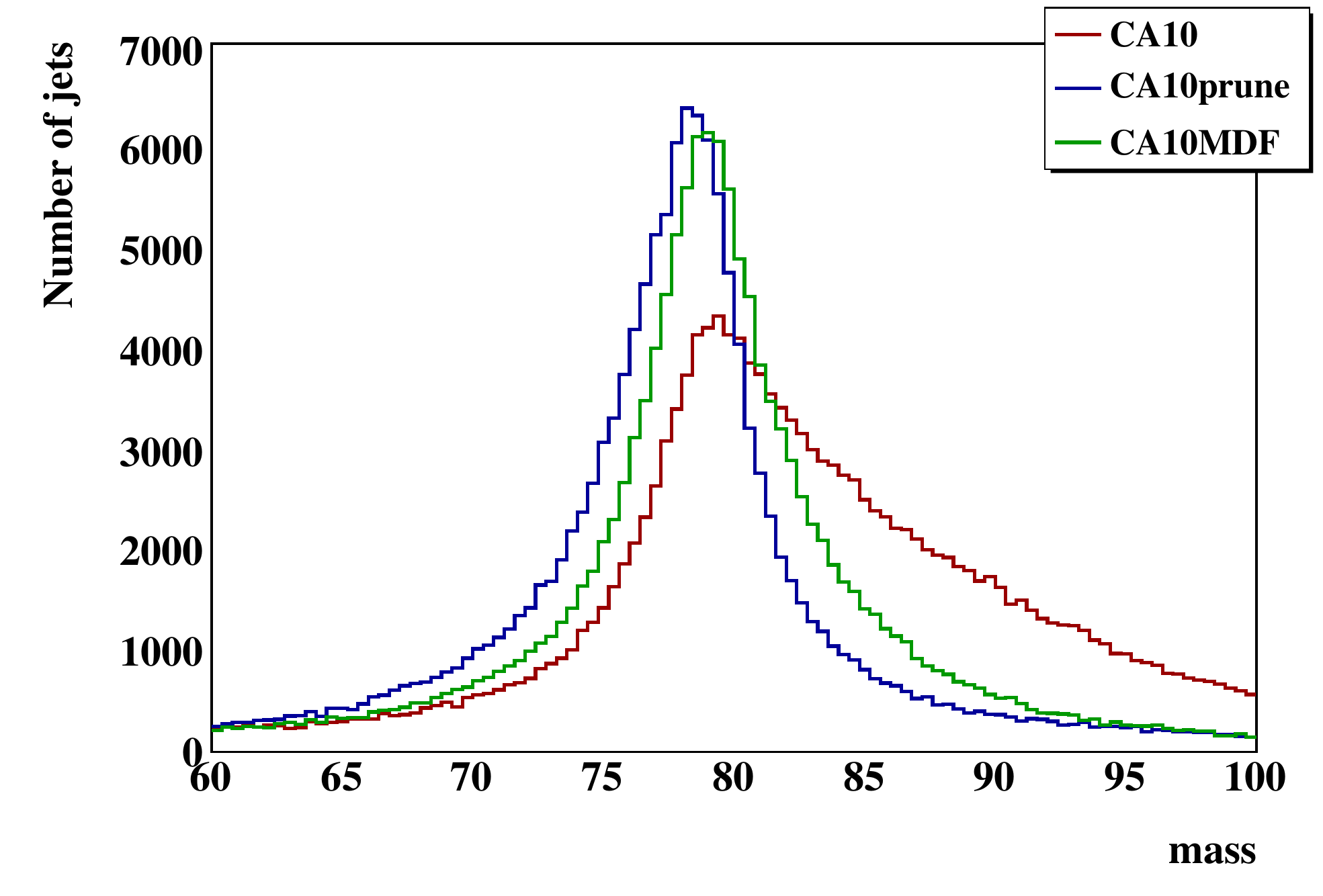}
\end{center}
\caption[Distribution in $m_J$ for anti-$\kt$ jets reclustered with CA, then pruned or mass-drop filtered]{Distribution in $m_J$ for anti-$\kt$ jets reclustered with CA, then pruned or mass-drop filtered.  Jets have $p_T > 200$ GeV; the initial $D = 1.0$.}
\label{fig:Wcompare_mjet}
\end{figure}

\begin{figure}
\begin{center}
\includegraphics[width=.7\textwidth]{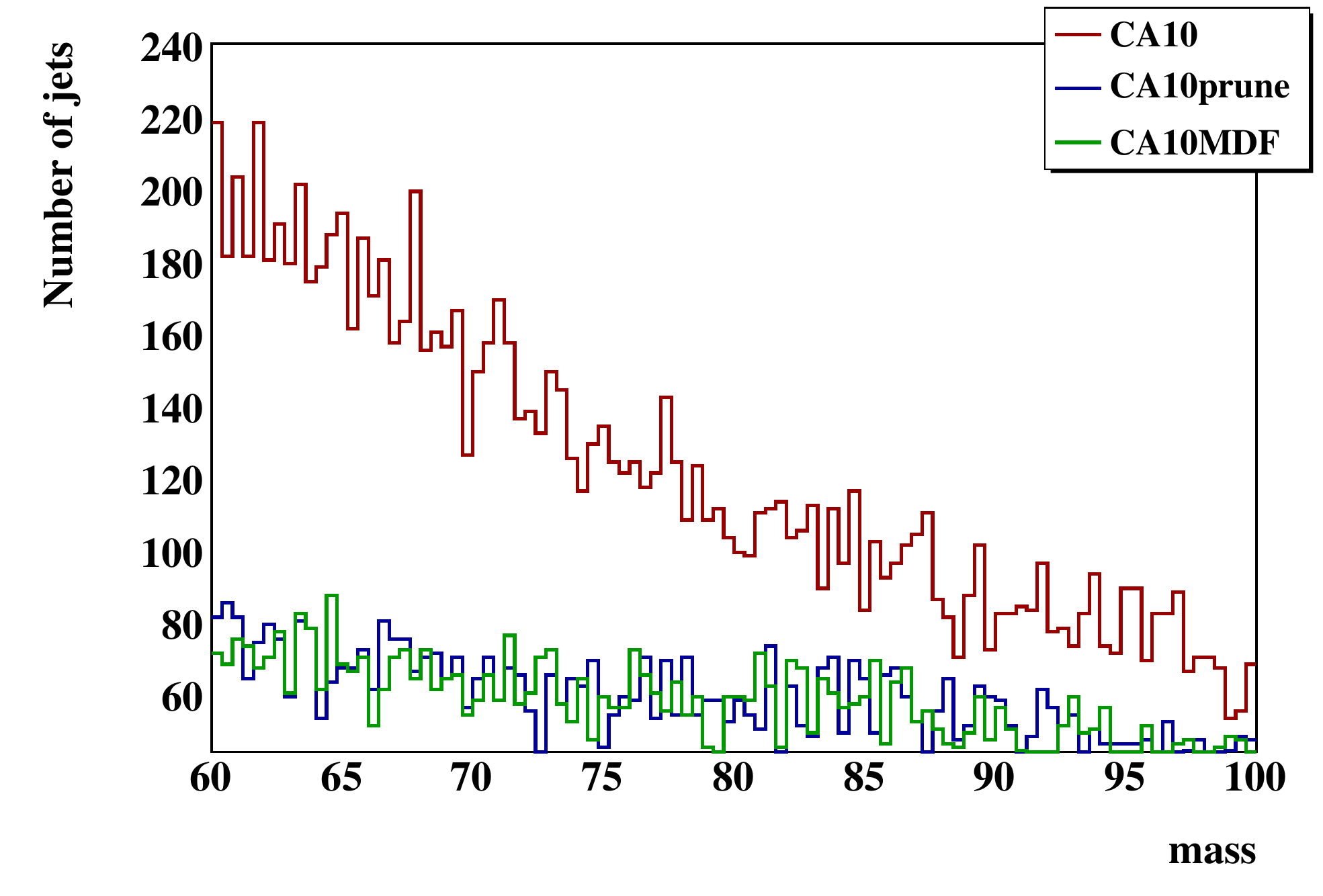}
\end{center}
\caption[Background distribution in $m_J$ for anti-$\kt$ jets reclustered with CA, then pruned or mass-drop filtered]{Background distribution in $m_J$ for anti-$\kt$ jets reclustered with CA, then pruned or mass-drop filtered.  Jets have $p_T > 200$ GeV; the initial $D = 1.0$.}
\label{fig:Wcompare_BG}
\end{figure}

We now turn to $W$ finding, repeating the analysis of the previous section but this time comparing pruning to the mass-drop filter method.  The signal jet mass distributions are shown in Fig.~\ref{fig:Wcompare_mjet}.  We can see that the performance of pruning is quite similar to the mass-drop filter method.  The mass windows for each method are given in Table \ref{table:Wwindows}.  The background jet mass distributions are shown in Fig.~\ref{fig:Wcompare_BG}.  Tagging and mis-tagging efficiencies are given in Table \ref{table:Wefficiencies}.  We can see that pruning and mass-drop filtering are both superior to plain CA, but that they are quite similar in performance.

\begin{table}[htbp]
\begin{center}
\begin{tabular}{|l|c|c|}
\hline
Method & $m_\text{jet}^\text{low}$ & $m_\text{jet}^\text{high}$ \\
\hline \hline
CA10 & 69.0 & 89.9\\
\hline
CA10 + pruning & 71.4 & 84.0 \\
\hline
CA10 + MDF & 71.7 & 86..4 \\
\hline
\end{tabular}
\end{center}
\caption[Jet mass windows for each $W$-finding method]{Jet mass windows for each $W$-finding method.}
\label{table:Wwindows}
\end{table}

\begin{table}[htbp]
\begin{center}
\begin{tabular}{|l|c||c|}
\hline
Method & Signal & Background \\
\hline \hline
CA10 & 0.62 & 0.117 \\
\hline
CA10 + pruning & 0.54 & 0.036 \\
\hline
CA10 + MDF & 0.57 & 0.042 \\
\hline
\end{tabular}
\end{center}
\caption[Tagging efficiencies and mis-tag rates for each method]{Tagging efficiencies and mis-tag rates for each method after a jet mass cut, applied to semileptonic $WW$ events (Signal) and matched $W+$ jets events (Background).  Initial jets have $p_T > 200$ GeV and $D = 1.0$.  Efficiencies are relative to initial numbers of jets passing the $p_T$ cut.}
\label{table:Wefficiencies}
\end{table}

\vita{Christopher Vermilion was born September 5, 1984 in Seattle, WA.  A short time later he attended Boston University, receiving a Bachelor of Science degree in Electrical Engineering and a Bachelor of Arts in Physics, both in May 2006.  In September 2006 he arrived at the physics department of the University of Washington, from whom he received a Master of Arts degree in 2007.  If you are reading this vita, his dissertation was approved and he received a Doctor of Philosophy in 2010.
}

\end{document}